\documentclass[a4paper,fleqn,usenatbib]{mnras}
\usepackage{newtxtext,newtxmath}
\usepackage[T1]{fontenc}
\usepackage{ae,aecompl}

\usepackage{graphicx,graphics}
\usepackage{epstopdf}
\usepackage{amssymb}	% Extra maths symbols

\usepackage[export]{adjustbox}
\title[2D maps for galaxy clusters]{Physical
properties of the X-ray gas as a dynamical diagnosis for galaxy clusters}

\author[Lagan\'{a}, Durret \& Lopes ]{
T. F. Lagan\'{a},$^{1}$\thanks{E-mail: tatiana.lagana@cruzeirodosul.edu.br}
F. Durret$^{2}$
and P. A. A. Lopes$^{3}$
\\
$^{1}$NAT, Universidade Cruzeiro do Sul, Rua Galv\~{a}o Bueno, 868, CEP:01506-000, S\~{a}o Paulo-SP, Brazil\\
$^{2}$ Sorbonne Universit\'{e}, CNRS, UMR 7095, Institut d'Astrophysique de Paris, 98bis Bd Arago, F-75014 Paris, France. \\
$^{3}$  Observat\'{o}rio do Valongo, Universidade Federal do Rio de Janeiro, Ladeira do Pedro Ant\^{o}nio 43, Rio de Janeiro, RJ, 20080-090, Brazil\\
}

\date{Accepted 2018 December 19. Received 2018 December 17; in original form 2018 November 29.}

\pubyear{2019}

\begin{document}
\label{firstpage}
\pagerange{\pageref{firstpage}--\pageref{lastpage}}
\maketitle

\begin{abstract}
%CONTEXT
%GENERAL SENTENCE ABOUT CLUSTERS; LARGE STRUCTURES GRAV. BOUND BUT STILL IN THE PROCESS OF FORMATION.

%Frequently cool-core cluster are chosen for cosmological studies since they are believed to be dynamically relaxed. 
%There are several parameters used to distinguish cool-core from non cool-core clusters, with the aim of physically
%classify them into relaxed and disturbed systems. 

We analysed XMM-{\it Newton}  EPIC data for 53 galaxy clusters. 
Through 2D spectral maps, we provide the most detailed and extended view of the spatial 
distribution of temperature (kT), pressure (P), entropy (S) 
and metallicity (Z) of galaxy clusters to date with the aim of correlating the dynamical state 
of the system to six cool-core diagnoses from the literature.\\
With the objective of building 2D maps and resolving structures in kT, P, S and Z, we divide the data 
in small regions from which spectra can be extracted. 
Our analysis shows that when clusters are spherically symmetric the
cool-cores (CC) are preserved, the systems are relaxed with little
signs of perturbation, and most of the CC criteria agree. The
disturbed clusters are elongated, show clear signs of interaction in
the 2D maps, and most do not have a cool-core. 
However, 16 well studied clusters classified as CC by at least four criteria
show spectral maps that appear disturbed. 
All of these clusters but one 
show clear signs of recent mergers, with a complex structure and geometry but with
a cool-core that remains preserved.
Thus, although very useful for CC characterization, most diagnoses are too simplistic to reproduce the overall 
structure and dynamics of galaxy clusters, and therefore the selection of relaxed systems 
according to these criteria may affect mass estimates. The complex structure of galaxy clusters can be 
reliably assessed through the 2D maps presented here.

%CONCLUSIONS

\end{abstract}


%%%%%%%%%%%%%%%%%%%%%%%%%%%%%%%%%%%%%%%%%%%%%%%%%%

%%%%%%%%%%%%%%%%% BODY OF PAPER %%%%%%%%%%%%%%%%%%

\section{Introduction}
%GENERAL PARAGRAPH OF GALAXY CLUSTERS

Clusters of galaxies are the largest bound structures in the Universe,
and in the $\Lambda$CDM model they are formed through the mergers of
smaller structures. However, halos of comparable mass also occasionally
merge.  Simulations of the hierarchical scenario show that
consequently substructures should be common in galaxy clusters.  The
degree of substructure therefore provides information on the dynamical
state and processes that took place during cluster formation.
%CC CLUSTERS

A large fraction of galaxy clusters that have been imaged with X-ray
telescopes show clear evidence of a centrally peaked surface
brightness profile. The central cooling-times for these systems are
much shorter than the Hubble time and a temperature drop is observed
in the centre, typically reaching one-third of the virial temperature
\citep{Peterson03,Vik05}.  These systems,
long known as cooling-flow clusters \citep{Fabian94}, are now
 referred to as cool-core
(CC, hereafter) clusters \citep{Molendi01}. 

A crucial aspect here is a proper definition of a cool-core cluster, since a variety of cooling criteria have been proposed
to separate CC and non-cool core (NCC, hereafter) clusters, and results can occasionally diverge.
In X-rays, classification attempts are generally based on core properties because cluster centres are more easily accessible.
These indicators are typically based on a central temperature
drop \citep[e.g.,][]{Burns08}, a central cooling time shorter than the
Hubble time \citep[e.g.,][]{Bauer05}, or a significant mass deposition
\citep{Chen07}. \citet{Hudson10} compared 16 cool-core criteria
with the aim of determining a physical property that could unambiguously
separate relaxed from disturbed clusters. 
To segregate CC and NCC clusters, 
they found that the cooling time 
($t_{\rm cool}$) is the best parameter for low redshift clusters with high quality data,
and that cuspiness is the best parameter for high redshift clusters. 
Although these different cooling estimators consider the cool-core phenomenon as a bimodal feature (CC or NCC),
several works have distinguished three regimes of cooling \citep[e.g.,][]{Bauer05, Leccardi10}, 
with an intermediate category suggesting a transition from NCC to CC systems.

\citet{Angulo12} showed that cosmological conclusions based on galaxy
cluster surveys depend on the wavelength used for cluster selection.
This issue has been addressed in the last few years with the
availability of cluster samples selected through the
Sunyaev-Zel'dovich effect by the Planck satellite.
The signal of thermal SZ \citep[SZ,][]{SZ72} effect depends on the Compton-y parameter
which is proportional to the electron pressure
($P_{e} \propto n_{e} kT$), while in X-ray selected samples, the
surface brightness depends quadratically on the electron number
density ($n_{e}$).  Hence,  the  fraction  of  CCs  in  X-ray  samples  is  likely
overestimated.
Due to this difference on the gas density, there
is currently a debate on whether the two experiments are detecting the
same population of clusters, since X-ray surveys are more prone to detect
centrally peaked galaxy clusters. Indeed, it has been shown that X-ray
flux limited samples have larger fractions of CC clusters in
comparison to Planck samples \citep[e.g.,][]{Rossetti16}.
This result was interpreted as the CC bias affecting X-ray selection
\citep{Eckert11}. 

For galaxy clusters to be used as cosmological probes, they must be dynamically relaxed,
since non-gravitational processes can result in the under  or overestimation of the mass.
The comparison of core properties with morphological properties shows that
more disturbed systems tend to have less well defined cores \citep{Buote96,Bauer05}.
Thus, there have been attempts to correlate dynamical properties with simple CC diagnoses as those 
mentioned before.

Recently, \citet{Lopes18} combined optical and X-ray
measurements, and showed that the offset between the brightest cluster galaxy (hereafter BCG) and the X-ray centroid or the
magnitude gap between the first and the second BCGs are good and
simple measures to assess cluster dynamical states.
Another way to separate relaxed from disturbed systems is through the
degree of substructures.  The presence of substructures is a clear sign
of incomplete relaxation in a cluster, and X-ray observations of
substructures in galaxy clusters are imprints of the recent and
ongoing formation process 
\citep{JonesForman99,Jeltema05,Lagana10,AndradeSantos12}.  
However, estimates of cluster substructures vary from $\sim$20\% up to
$\sim$80\% depending on the methods employed for substructure detection
\citep[e.g.,][]{Kolokotronis01}.

Although very useful for CC characterisation, the criteria adopted in the literature 
\citep[e.g.,][]{AndradeSantos17,Hudson10}
account for the overall structure of galaxy clusters in a too simplistic way.
Aiming at a full spectral analysis, we show for the first time 2D
maps of the spatial distribution of temperature (kT), pseudo-entropy
(S), pseudo-pressure (P) and metallicity (Z) for a large sample of 53 
galaxy clusters. These clusters belong to SZ and X-ray samples
and have been used as cosmological tools.  Our aim is to obtain a clear dependence
of these maps with the cluster dynamical state, and to correlate these properties with six important CC
diagnoses considered in \citet{Lopes18}.  We highlight that the maps reveal
the detailed and complex structure of galaxy clusters that can be
missed in CC diagnoses tools.
% in particular by comparing a few of
%these maps with hydrodynamical numerical simulations. 
%\pl {should we remove the end of the sentence, about simulations? 
%Or rewrite it? It sounded (for me) as we were talking of a comparison to simulations that 
%we have already performed.}

A large number of simulations are necessary to account for the variety of properties observed in this large sample,
and this will be the topic of a future paper (Machado et al. in
preparation). We are also performing a detailed analysis of the
dynamical properties of a subsample of these clusters with a large
number of galaxy redshifts available (Biviano et al. in preparation).

The paper is organised as follows. In Section \ref{sect:data} we
present the sample and data reduction and the method to construct the
2D maps.  In Section \ref{sect:res} we show our results, and we
discuss them in Section \ref{sect:disc}.  Finally, we conclude on our
findings in Section \ref{sect:conc}. We present notes on individual
clusters in Appendix \ref{sect:apNotes}.  In this work, we assume a
flat $\Lambda$CDM Universe with $\Omega_{M}=0.3$, and
$H_{0} = 71$ ~km~s$^{-1}$~Mpc$^{-1}$. All errors are given at the
$1\sigma$ level.

\section{The data}
\label{sect:data}

\subsection{Sample and data Reduction}

\begin{figure}
\centering
\includegraphics[width=\columnwidth]{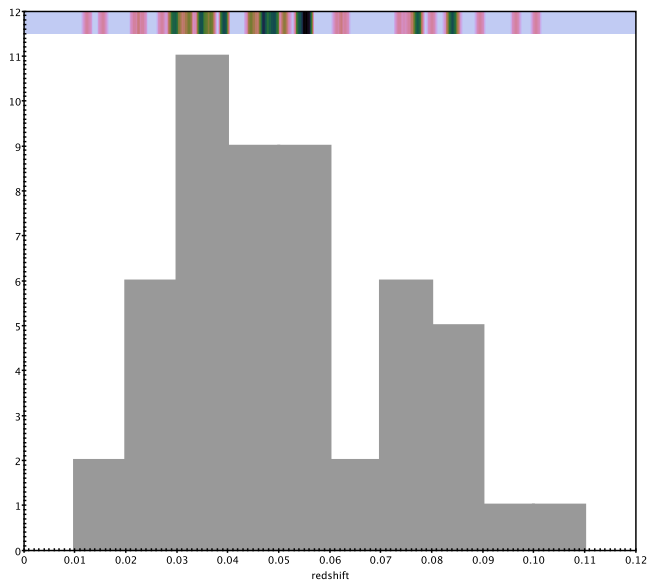}
\caption{Redshift histogram of the sample analysed in this work. In the upper part of the panel we show the
densogram of clusters (values are represented by a fixed-size pixel-width column of a colour from a colour map. 
The darker the color, the higher the number of clusters in that bin).}
\label{fig:histoz}
\end{figure}

This work is based on a sample of clusters studied in \citet{Lopes18},
limited to $z\leq 0.11$, with publicly available XMM-{\it Newton}  data.
\citet{Lopes18} used SZ and X-ray samples presented in
\citet{AndradeSantos17} with optical data available.  
Out of 72 clusters analysed in \citet{Lopes18}, 
55 systems had XMM-{\it Newton}  data
in the archive, but the exposure times for two of them were not sufficient to
produce 2D maps.  Thus, we ended-up with 53 clusters for which we
studied the global spatial distribution of the temperature, metal
abundance, entropy and pressure through 2D spectral maps. The full
list of clusters under study is given in Tab.~\ref{tab:obs} and the
redshift histogram of the sample is shown in Fig.~\ref{fig:histoz}.
We highlight that there are secondary subclusters, that we kept  
separately in Tab.~\ref{tab:obs} as done in \citet{Lopes18}, but most of the times they are in the same X-ray observation as the main cluster and
are shown together with the main cluster in some of the Figs. 1-10.
%Figs.~\ref{fig:CCclusters}, \ref{fig:CCclusters2},\ref{fig:CCclusters3},\ref{fig:CCclusters4},\ref{fig:groupsA85},
%\ref{fig:NCCclusters1}, \ref{fig:NCCclusters2}, \ref{fig:NCCclusters3}, \ref{fig:NCCclusters4}, \ref{fig:NCCclusters5}.}

Data reduction was done with SAS version 16.1.0 (July 2017) and
calibration files updated to 2017 July.  Background flares were
identified and rejected  by applying a 1.8$\sigma$ clipping method to the count rate
histogram of the the light curves in the energy range of [1$-$10] keV.
Point sources were detected by visual inspection, confirmed in the \textit{High Energy Catalogue 2XMMi Source},
and excluded  from our analysis.

To take
into account each detector background contribution, we obtained a
background spectrum in an outer annulus of the observation in the
[10-12] keV energy band. A corresponding background spectrum was also
extracted from the blank sky background file by \citet{ReadPonman03}
and then rescaled to obtain a normalisation parameter that will be
used in the spectral fits.

%We compared these spectra with the one obtained by \citet{ReadPonman03} blank sky in the same region and energy band.
%Then, we rescaled the observation background to the blank sky background to obtain a normalisation parameter that will be used in the spectral fits.

The spectral analysis was performed in the [0.7-7.0] keV energy band and we
also excluded the [1.2-1.9]~keV  band to avoid any
influence from Al and Si lines.  For the spectral fits, the spectra
were grouped to have at least 15 counts per spectral bin and we used
XSPEC v12.10.0 for fitting.  In this study the MEKAL XSPEC thermal
spectral model \citep{KM93} is used to model the emission of an
optically-thin plasma, and WABS \citep{Balucinska92,Morrison83} to
model photoelectric absorption.  We fixed the redshift and the
Galactic absorption values (see Tab.~\ref{tab:obs}), letting all other
parameters (e.g. temperature, metallicity and normalization) vary.
Abundances were measured assuming the ratios from \citet{Asplund09}.

%Using $\chi^{2}$ statistics, we adopted an absorbed  single temperature plasma model 
%Mewe-Kaastra-Liedahl (called MEKAL), absorbed to account
%for Galactic attenuation (i.e. WABS $\times$ MEKAL). 

\begin{table*}
	\centering
	\caption{Description of the cluster sample. The columns list the cluster name,  the
          \textit{Planck} name where the prefix PLCKESZ is omitted for simplicity, RA, DEC, redshift, 
          XMM-{\it Newton} observation identification, exposure time, and Galactic absorption. We note that secondary subclusters (startng with a ``b") are kept separately as done in \citet{Lopes18}.} 
	\label{tab:obs}
	\begin{tabular}{lccccccc} % four columns, alignment for each
		\hline
		Cluster & Planck Name & RA & DEC & z & ObsID & $t_{\rm exp}$ & $n_{\rm H}$ \\
		&  & (J2000) & (J2000) &   & & (s) & ($10^{20} cm^{-3}$) \\
		\hline	
A2734     			&					&	2.84036		&	-28.85404	&	0.0613	&	0675470801	&	17519	&	1.39	\\
A85         			&	 G115.16-72.09		&	10.45996		&	-9.30265	&	0.0555	&	0723802101	&	101400	&	2.78	\\
A119       			&	G125.58-64.14		&	14.06754		&	-1.24972	&	0.0441	&	0505211001	&	13916	&	3.54	\\
EXO0422			&					&	66.46318		&	-8.55954	&	0.0392	&	0300210401	&	40507	&	7.86	\\
A496       			&	G209.56-36.49		&	68.40797		&	-13.26168	&	0.0326	&	0506260401	&	78919	&	3.78	\\
S0540     			&					&	85.02782		&	-40.83657	&	0.0365	&	0149420101	&	18217	&	3.16	\\
A3376     			&	G246.52-26.05		&	90.42327		&	-39.98886	&	0.0461	&	0151900101	&	47206	&	4.82	\\
A3391     			&					&	96.58534		&	-53.69329	&	0.0554	&	0505210401	&	27915	&	5.65	\\
A3395     			&	G263.20-25.21		&	96.70078		&	-54.54923	&	0.0514	&	0400010301	&	29910	&	6.70	\\
bA3395   			& 					& 	96.90003 		& 	-54.44629&	0.0505	&	0400010301	& 	29910	&	6.70	 \\
A754       			&					&	137.33525	&	-9.68431	&	0.0543	&	0556200501	&	92270	&	4.84	\\
G269.51+26.42	&						&	159.17049	&	-27.52653	&	0.0124	&	0206230101	&	68554	&	4.90	\\
USGCS152		&					&	162.60879	&	-12.84501	&	0.0154	&	0146510301	&	40921	&	3.87	\\
A1367 			&	G234.59+73.01		&	176.18399	&	19.70444	&	0.0216	&	0061740101	&	33208	&	1.77	\\
ZwCl1215			&	G282.49+65.17		&	184.42165	&	3.65612	&	0.0768	&	0300211401	&	29215	&	1.72	\\
A3528n   			&	G303.75+33.65		&	193.59252	&	-29.01299	&	0.0539	&	0030140101	&	17205	&	6.38	\\
A3528s			&					&	193.66960	&	-29.22793	&	0.0539	&	0030140101	&	17205	&	6.38	\\
A1644n   			&					&	194.45546	&	-17.27307	&	0.0471	&	0010420201	&	22808	&	4.17	\\
A1644s   			&					&	194.29821	&	-17.40932 &	0.0471	&	0010420201	&	22808	&	4.17	\\
A3532     			&					&	194.34180	&	-30.36371	&	0.0558	&	0030140301	&	16895	&	6.43	\\
A1650     			&	G306.68+61.06		&	194.67264	&	-1.76207	&	0.0842	&	0093200101	&	43103	&	1.35	\\
A1651     			&	G306.80+58.60		&	194.84308	&	-4.19592	&	0.0848	&	0203020101	&	26646	&	1.52	\\
Coma			&	G057.33+88.01		&	194.94856	&	27.95189	&	0.0232	&	0300530301	&	31015	&	0.85	\\
A3558     			&	G311.99+30.71		&	201.98677	&	-31.49551	&	0.0475	&	0107260101	&	44615	&	4.05	\\
bA3558   			&					&	202.44904	&	-31.60724&	0.0490	&	0651590201	&	26917	&	4.05	\\
A3560     			&					&	203.11565		&	-33.14266	&	0.0489	&	0205450201	&	45522	&	4.27	\\
A3562     			&					&	203.39470	&	-31.67291	&	0.0489	&	0105261301	&	47166	&	3.76	\\
A1775     			&					&	205.45360	&	26.37219	&	0.0754	&	0108460101	&	33021	&	1.04	\\
A3571     			&	G316.34+28.54		&	206.86713	&	-32.86611	&	0.0393	&	0086950201	&	33642	&	4.25	\\
A1795     			&	G033.78+77.16		&	207.21963	&	26.59200	&	0.0629	&	0097820101	&	66559	&	1.19	\\
MKW8    			&					&	220.16445	&	3.47031	&	0.0269	&	0300210701	&	23610	&	2.40	\\
A2029     			&	G006.47+50.54		&	227.73382	&	5.74455	&	0.0774	&	0551780301	&	46815	&	3.25	\\
A2052     			&					&	229.18537	&	7.02162	&	0.0349	&	0401520501	&	22356	&	2.70	\\
A2061     			&					&	230.30289	&	30.63350	&	0.0773	&	0721740101	&	50000	&	1.69	\\
MKW3s			&					&	230.46593	&	7.70879	&	0.0448	&	0723801501	&	119600	&		\\
A2065     			&	G042.82+56.61		&	230.62280	&	27.70521	&	0.0735	&	0202080201	&	34110	&	3.03	\\
A2063     			&					&	230.77138	&	8.60957	&	0.0344	&	0550360101	&	28615	&	2.66	\\
A2142     			&	G044.22+48.68		&	239.58792	&	27.22996	&	0.0894	&	0674560201	&	59440	&	3.78	\\
A2147     			&					&	240.55862	&	15.97118	&	0.0365	&	0505210601	&	11911	&	3.39	\\
A2151     			&					&	241.14913	&	17.72143	&	0.0349	&	0147210101	&	29946	&	3.34	\\
AWM4    			&					&	241.23602	&	23.93268	&	0.0320	&	0093060401	&	21125	&	5.14	\\
G049.33+44.38	&						&	245.12627	&	29.89331	&	0.0964	&	0692930901	&	11917	&	2.58	\\
A2199     			&	G062.92+43.70		&	247.15930	&	39.55093	&	0.0306	&	0723801101	&	57000	&	0.89	\\
A2244     			&	G056.81+36.31		&	255.67738	&	34.06060	&	0.1004	&	0740900101	&	28000	&	1.88	\\
A2249			&	G057.61+34.94		& 	257.44080 	&      34.45566	& 	0.0838	&      0827010501	&	27000	&	2.21 \\
A2255     			&	G093.91+34.90		&	258.18160	&	64.06303	&	0.0799	&	0112260801	&	21051	&	2.50	\\
NGC6338			&					&	258.84579	&	57.41119	&	0.0290	&	0792790101	&	13600	&	2.23	\\
bNGC6338		&					&	258.84653	&	57.43462	&	0.0291	&	0792790101 	&	13600	& 	2.23	\\
A2572     			&					&	349.30335	&	18.70286	&	0.0392	&	0762950201	&	26000	&	4.03	\\
A2597     			&					&	351.33235	&	-12.12388	&	0.0831	&	0723801701	&	113170	&	2.48	\\
A2626     			&					&	354.12631	&	21.14673	&	0.0558	&	0148310101	&	41364	&	5.50	\\
A4038     			&					&	356.92869	&	-28.14290	&	0.0298	&	0723800801	&	67000	&	4.25	\\
A4059     			&					&	359.25423	&	-34.75899	&	0.0494	&	0723800901	&	98093	&	1.21	\\
\hline
\end{tabular}
\end{table*}

\begin{figure*}
\centering
\includegraphics[scale=0.22]{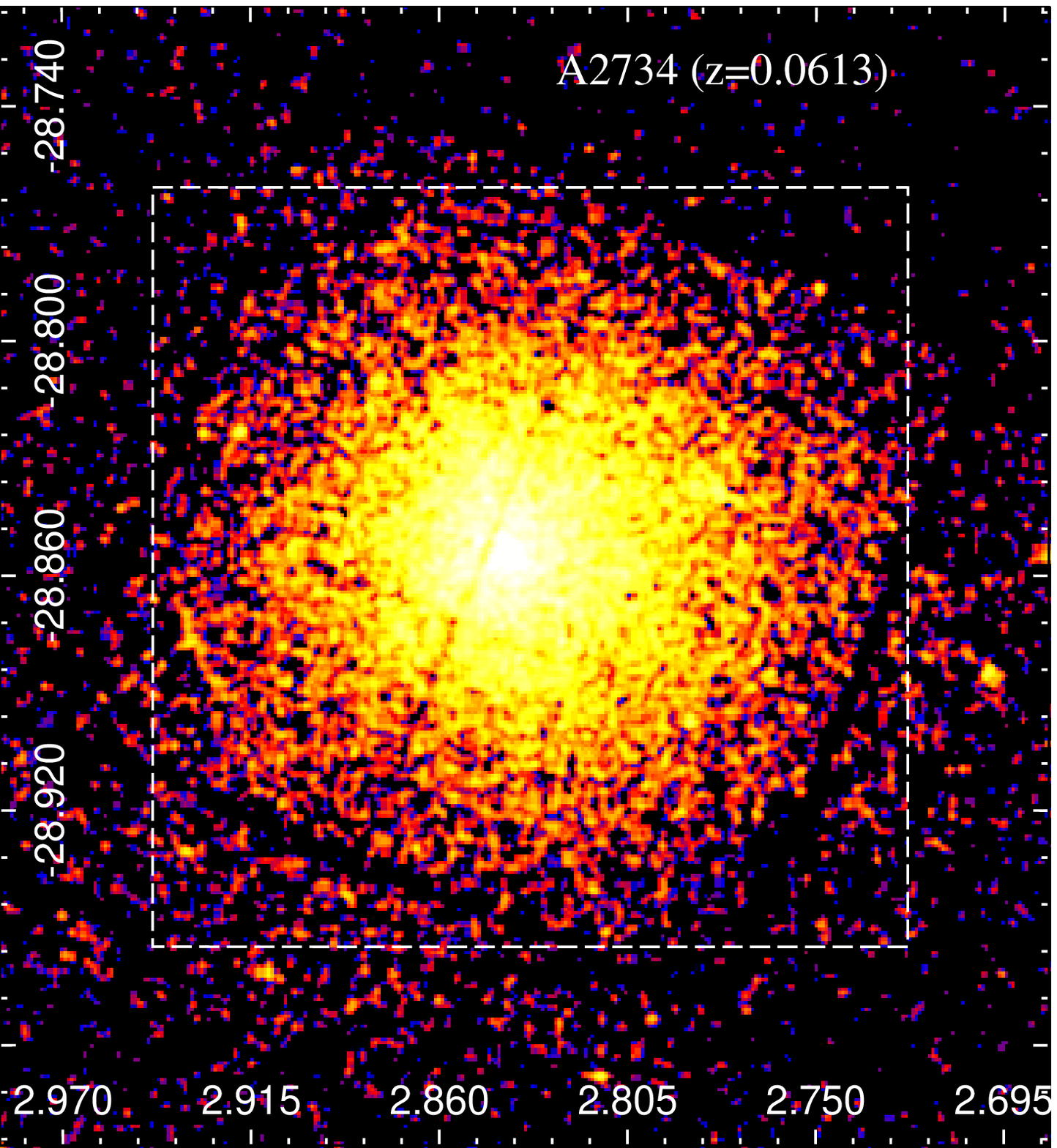}
\includegraphics[scale=0.22]{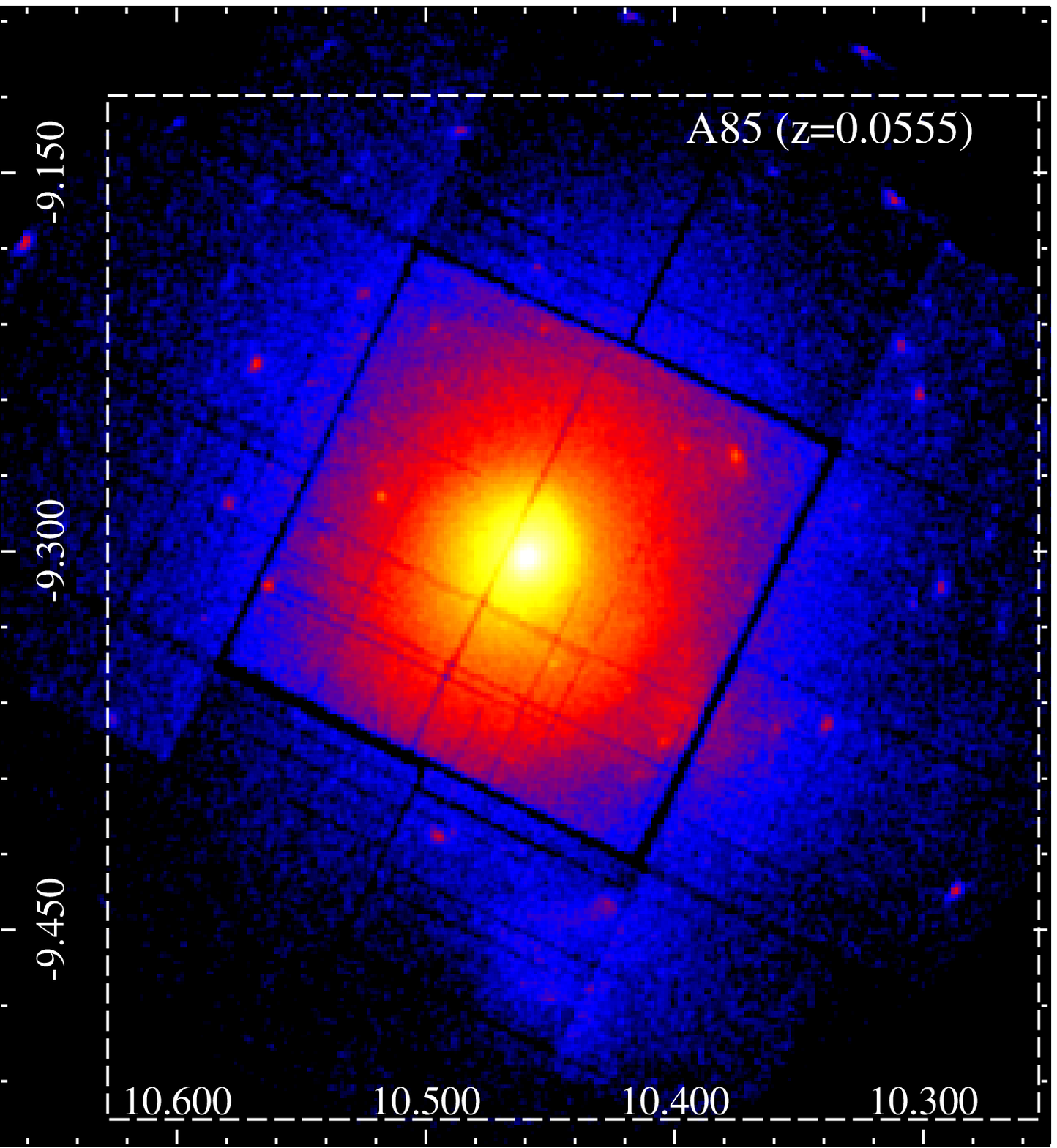}
\includegraphics[scale=0.22]{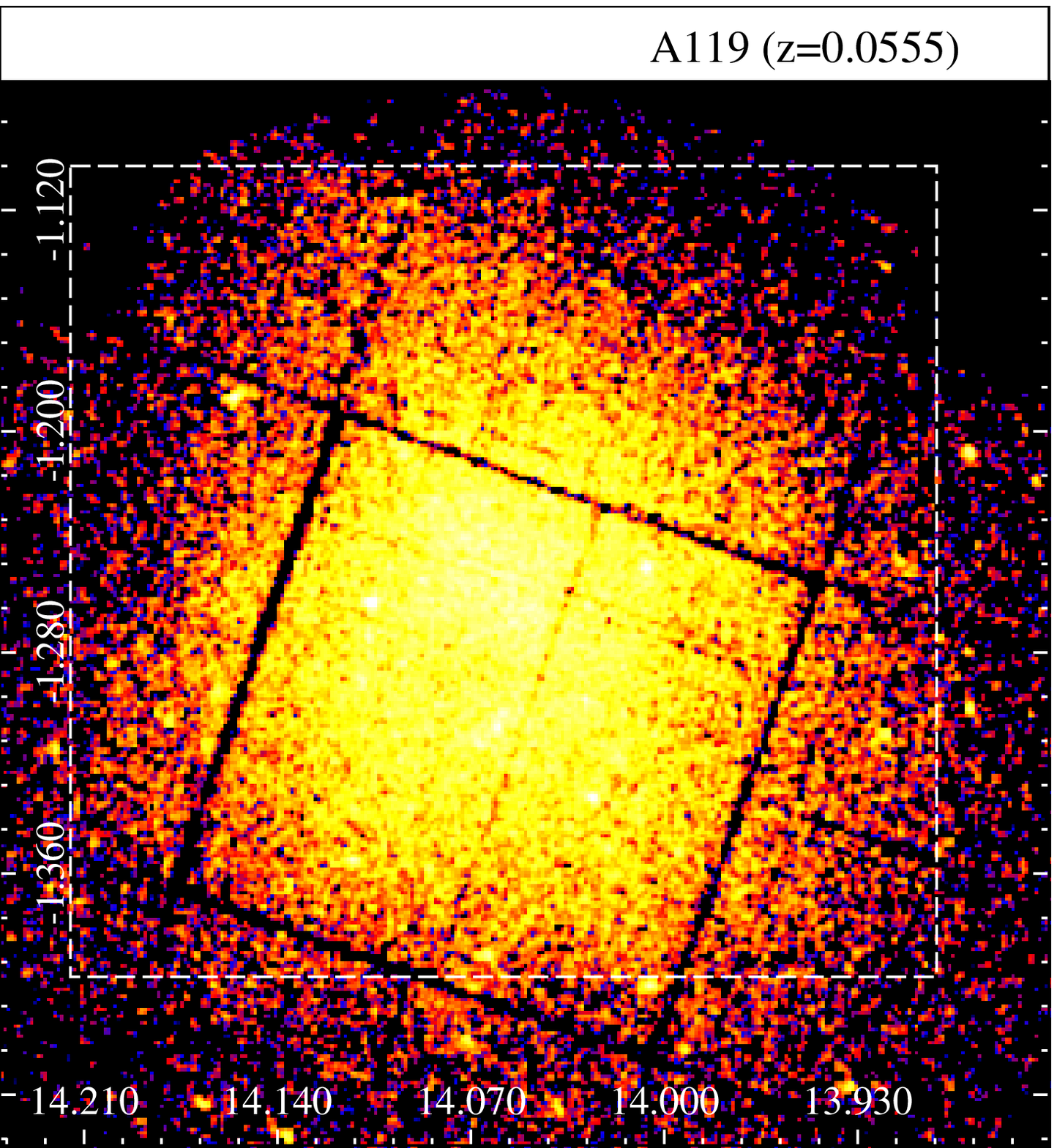}
\includegraphics[scale=0.22]{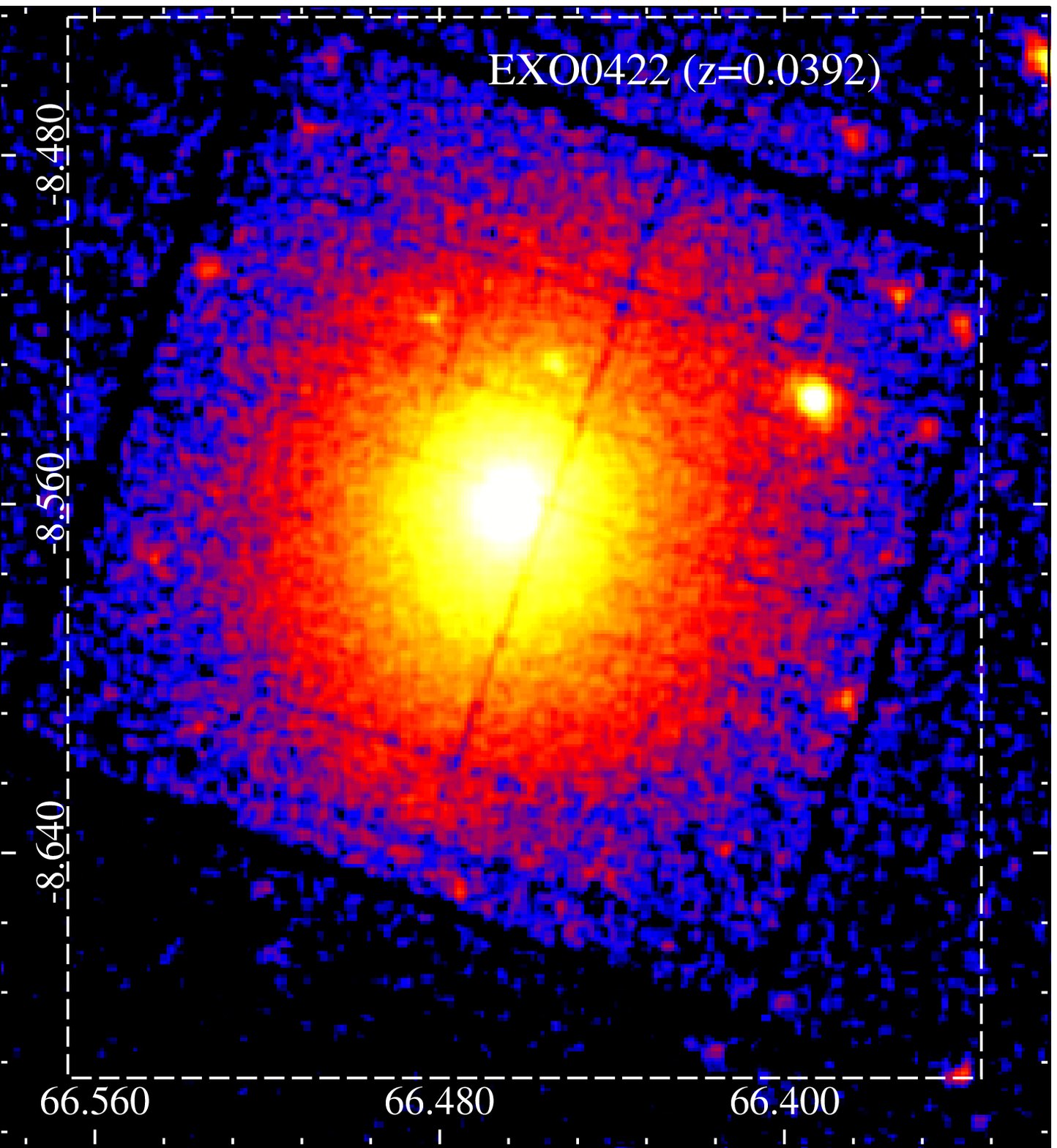}
\includegraphics[scale=0.22]{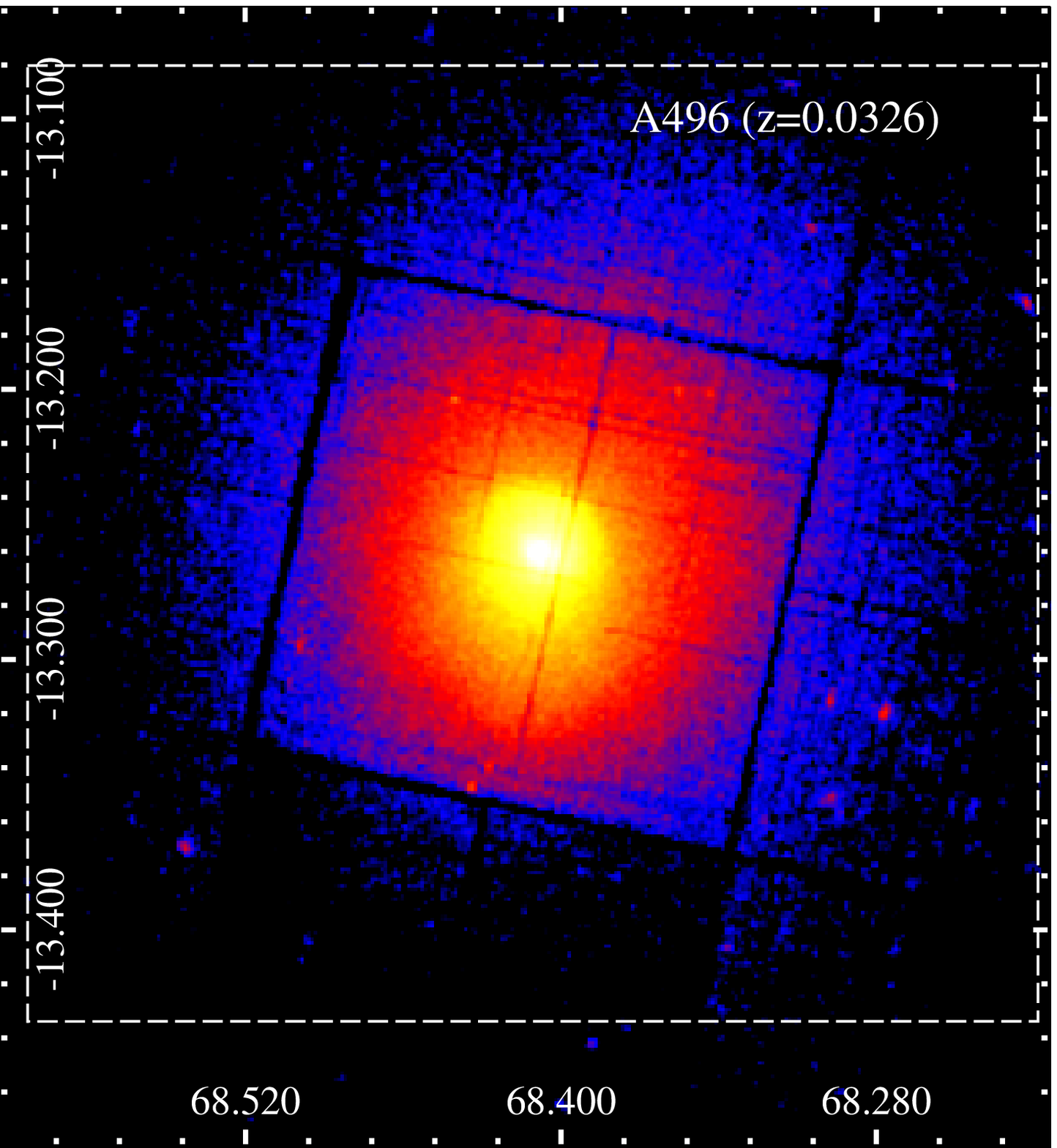}

\includegraphics[scale=0.22]{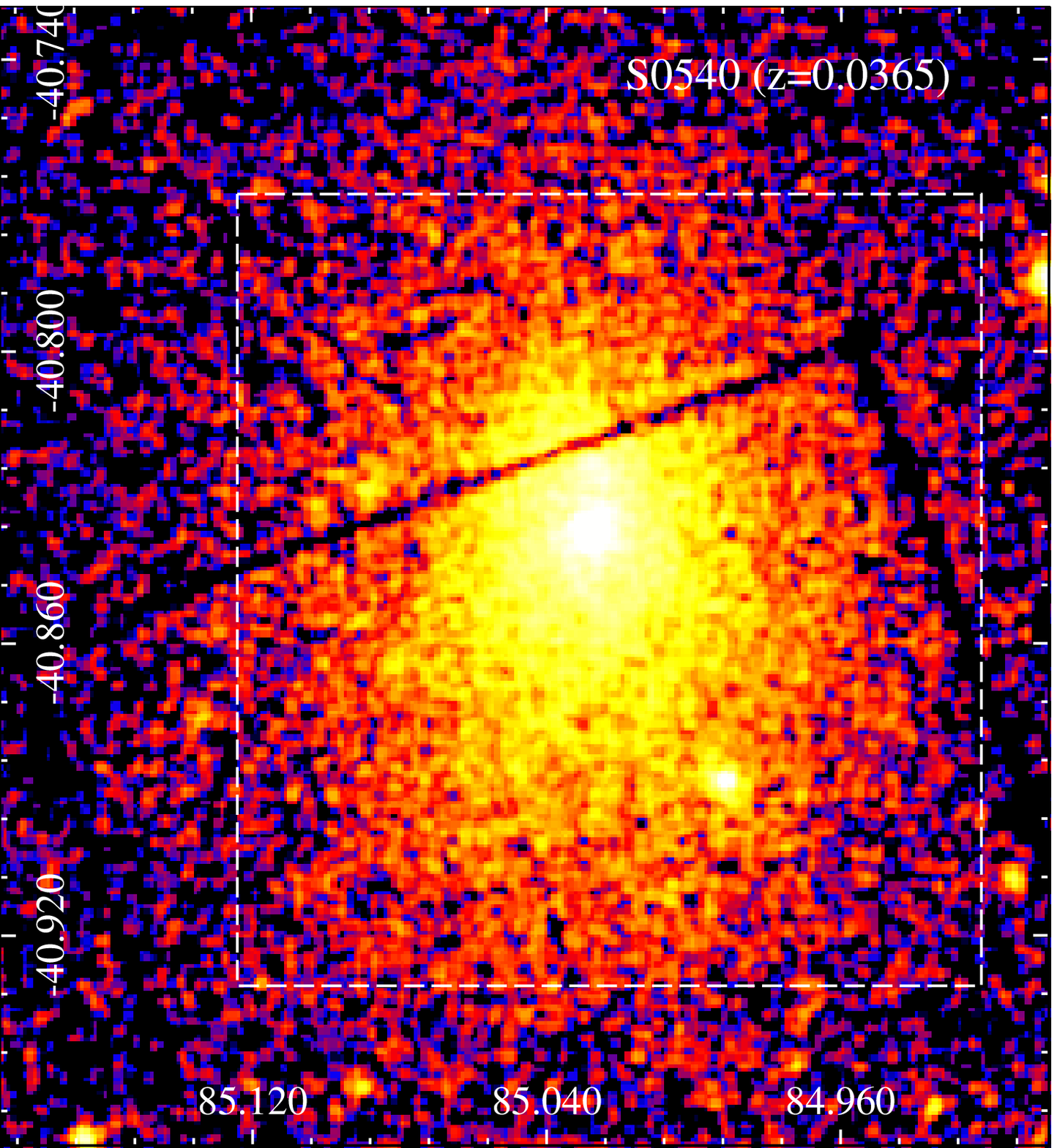}
\includegraphics[scale=0.22]{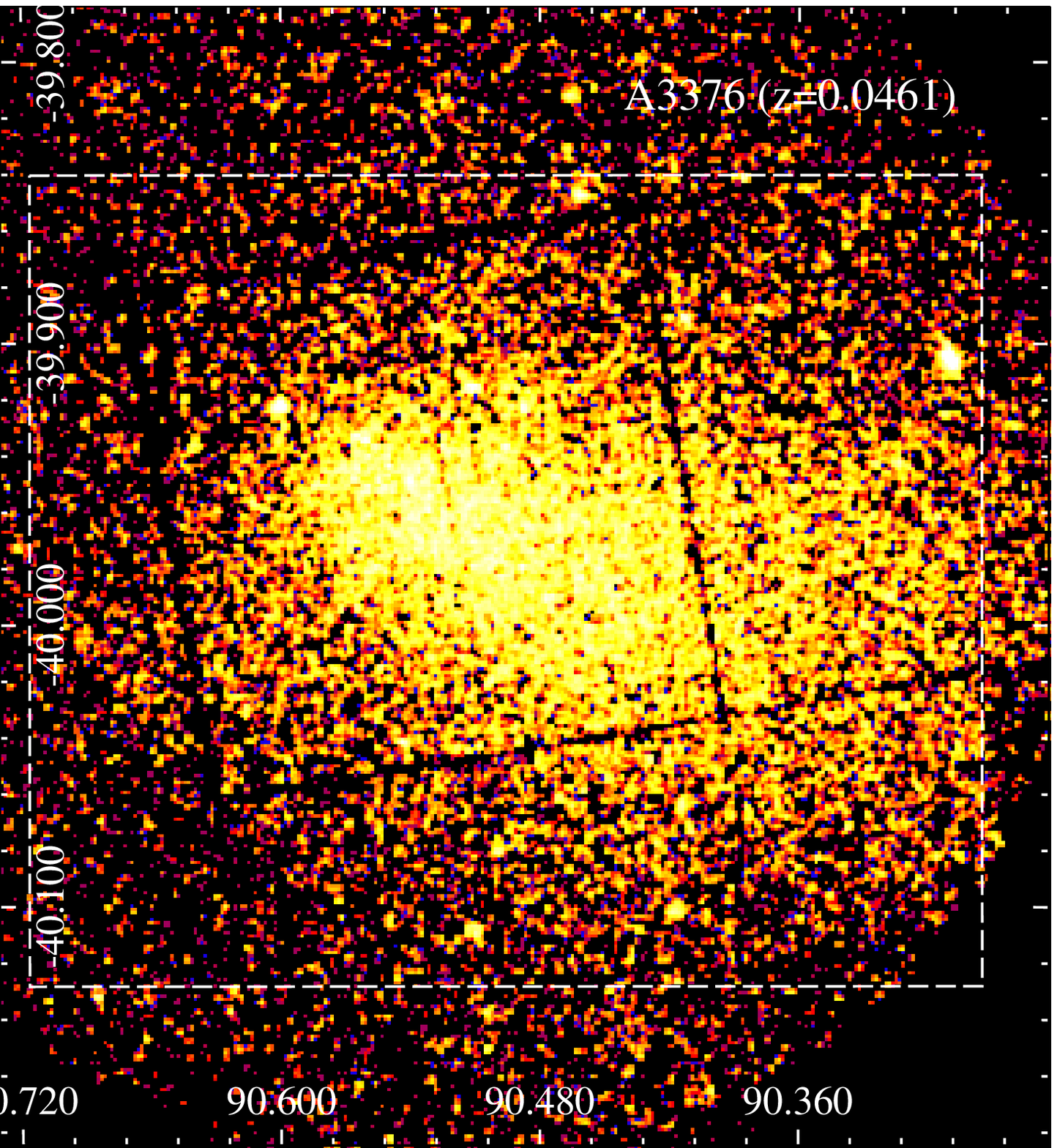}
\includegraphics[scale=0.22]{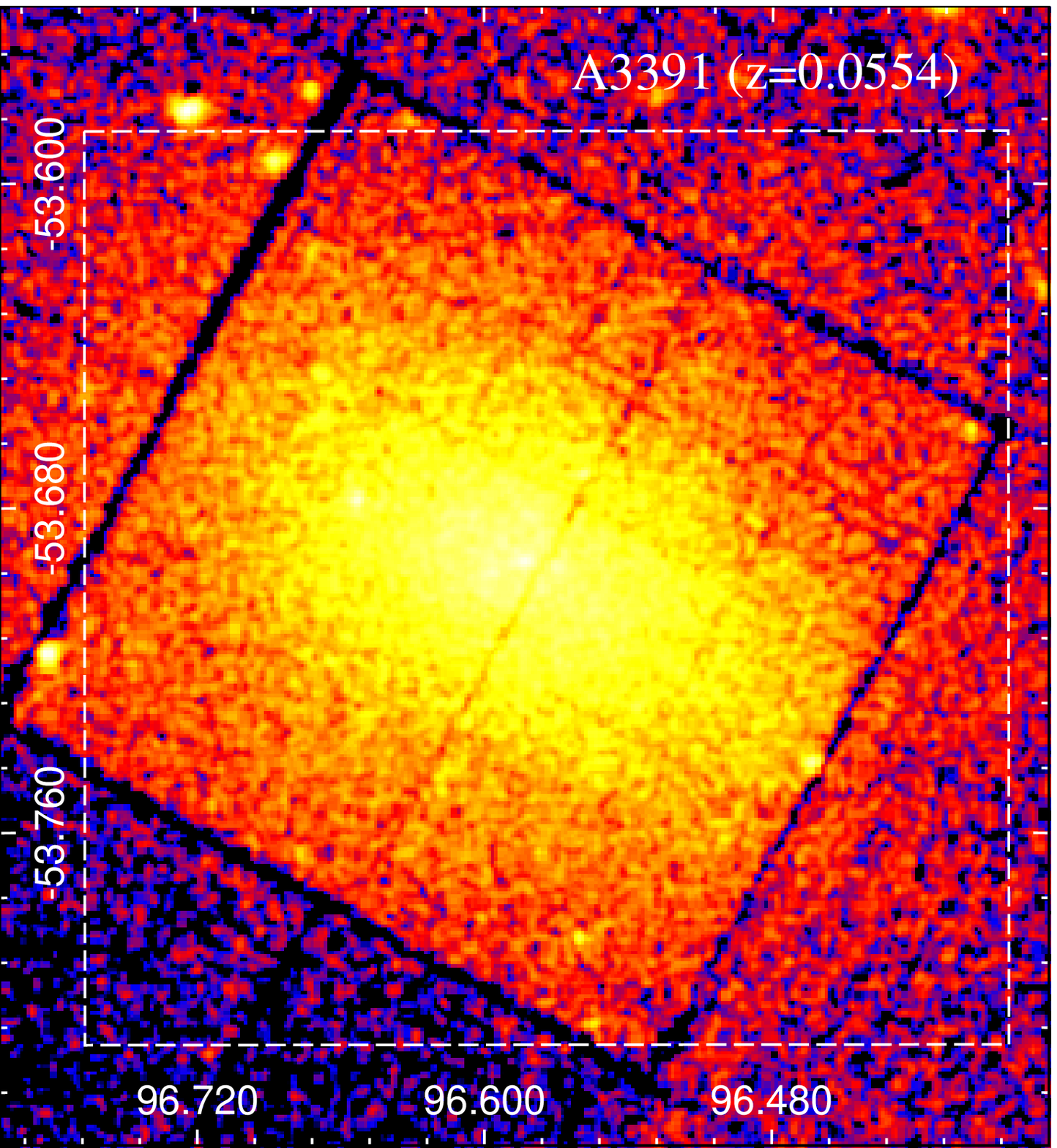}
\includegraphics[scale=0.22]{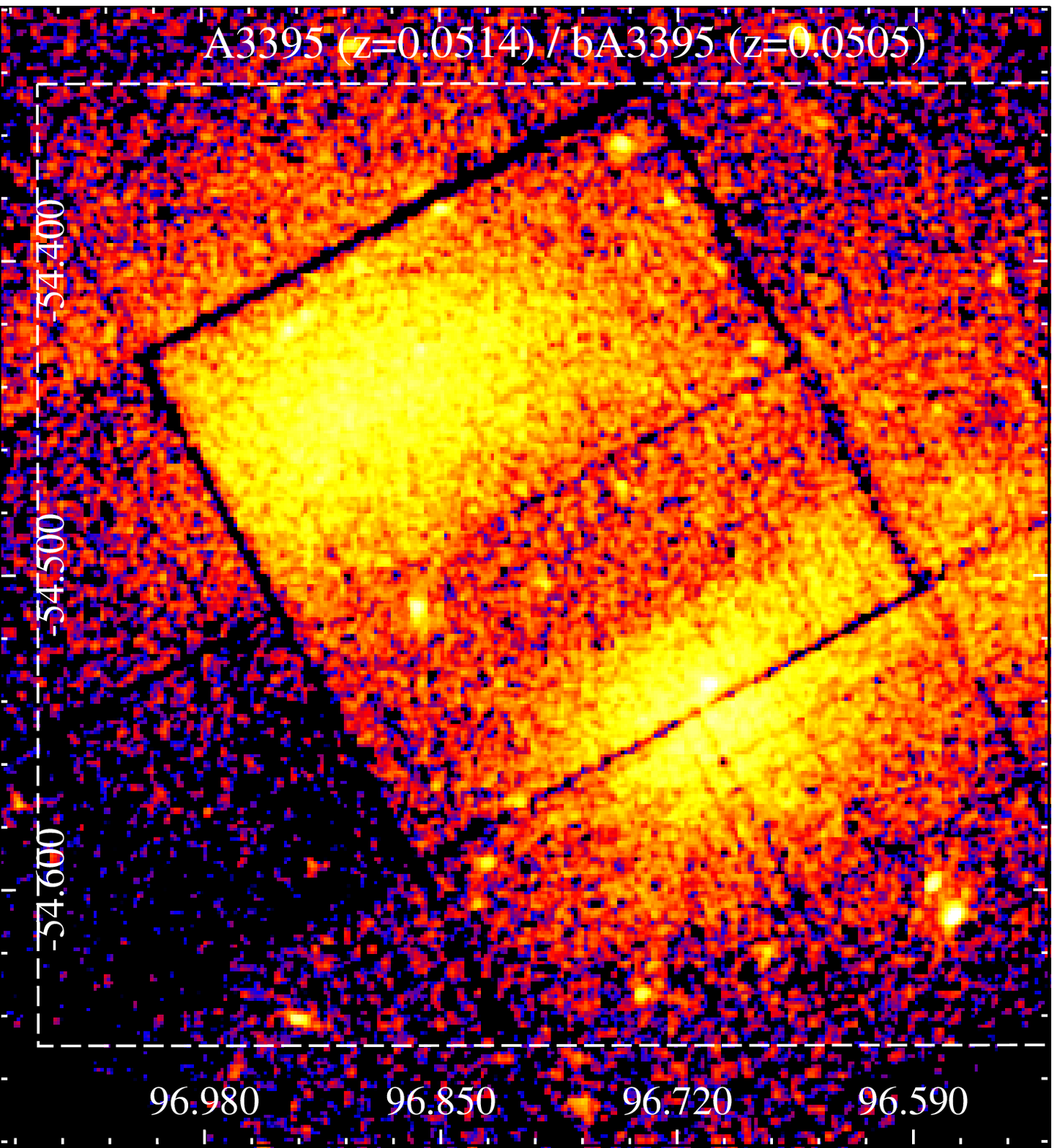}
\includegraphics[scale=0.22]{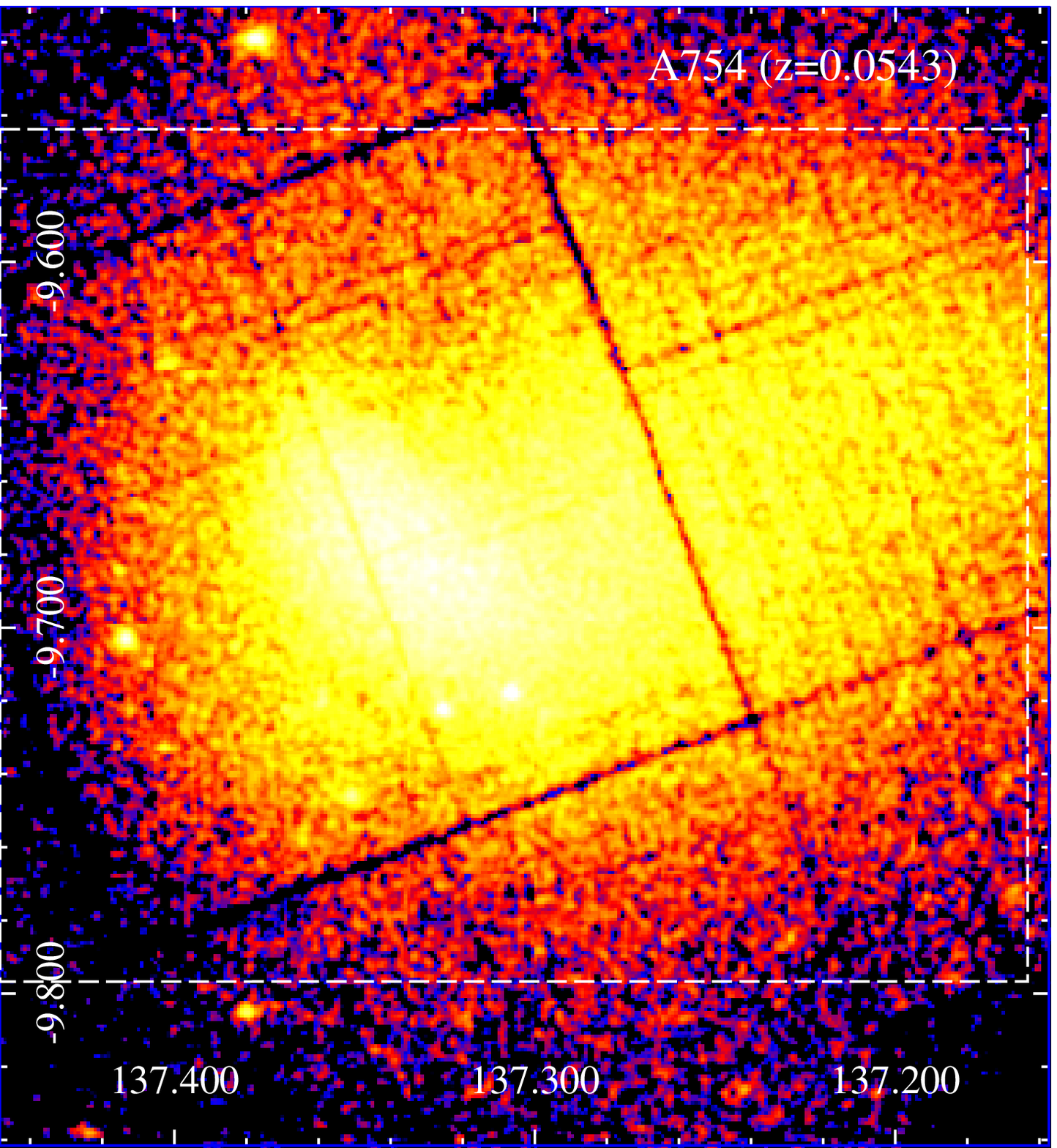}

\includegraphics[scale=0.22]{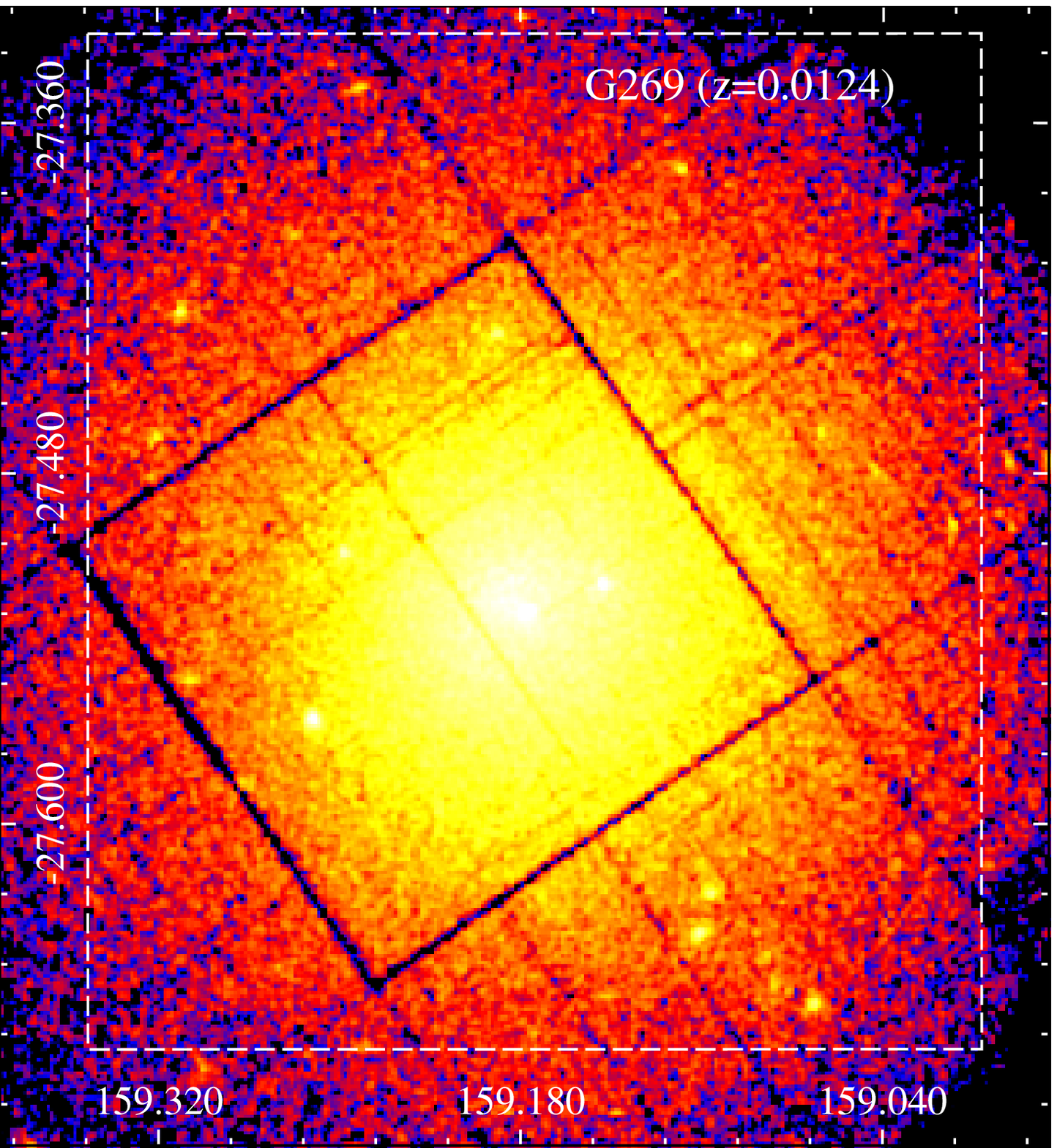}
\includegraphics[scale=0.22]{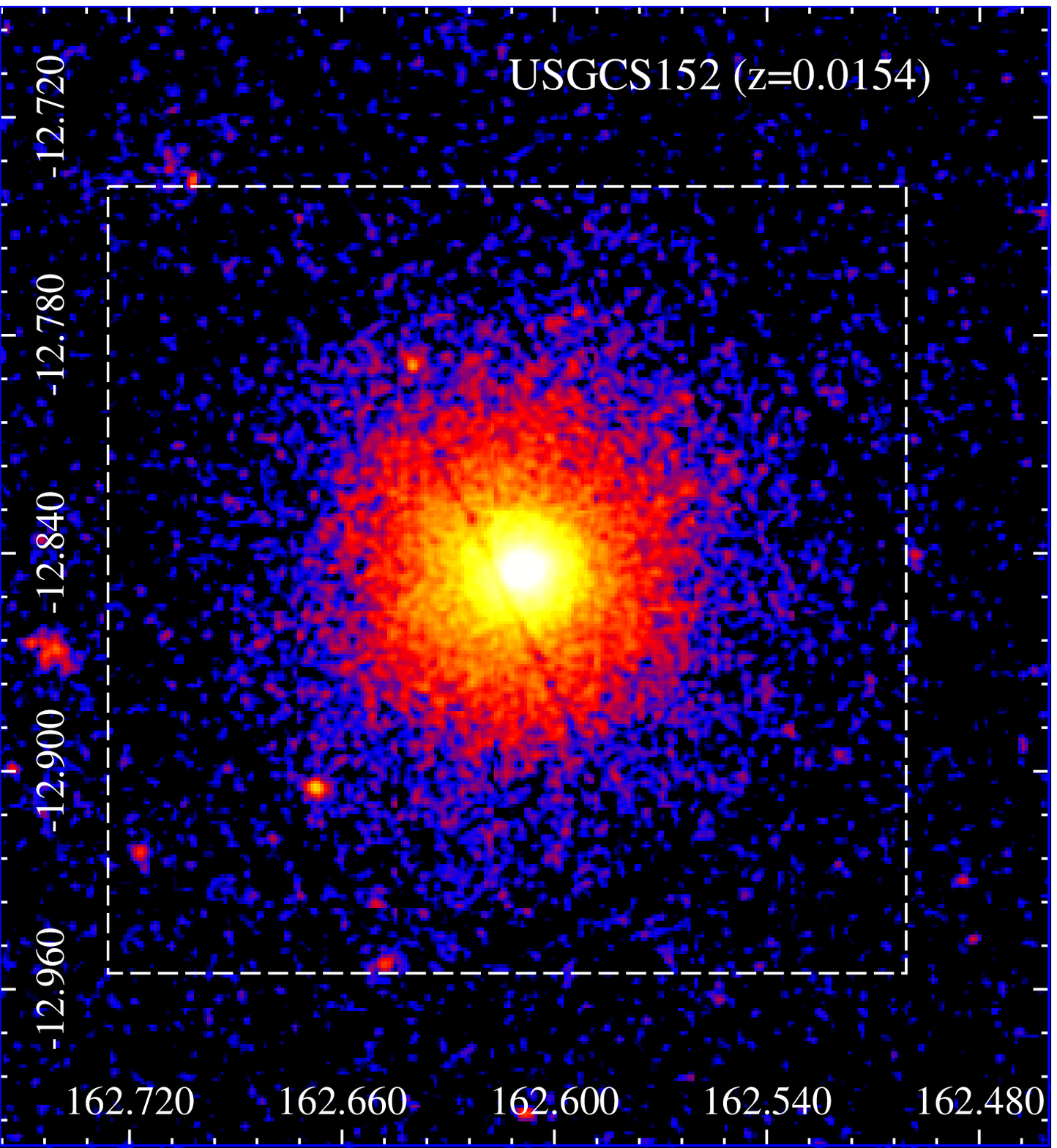}
\includegraphics[scale=0.22]{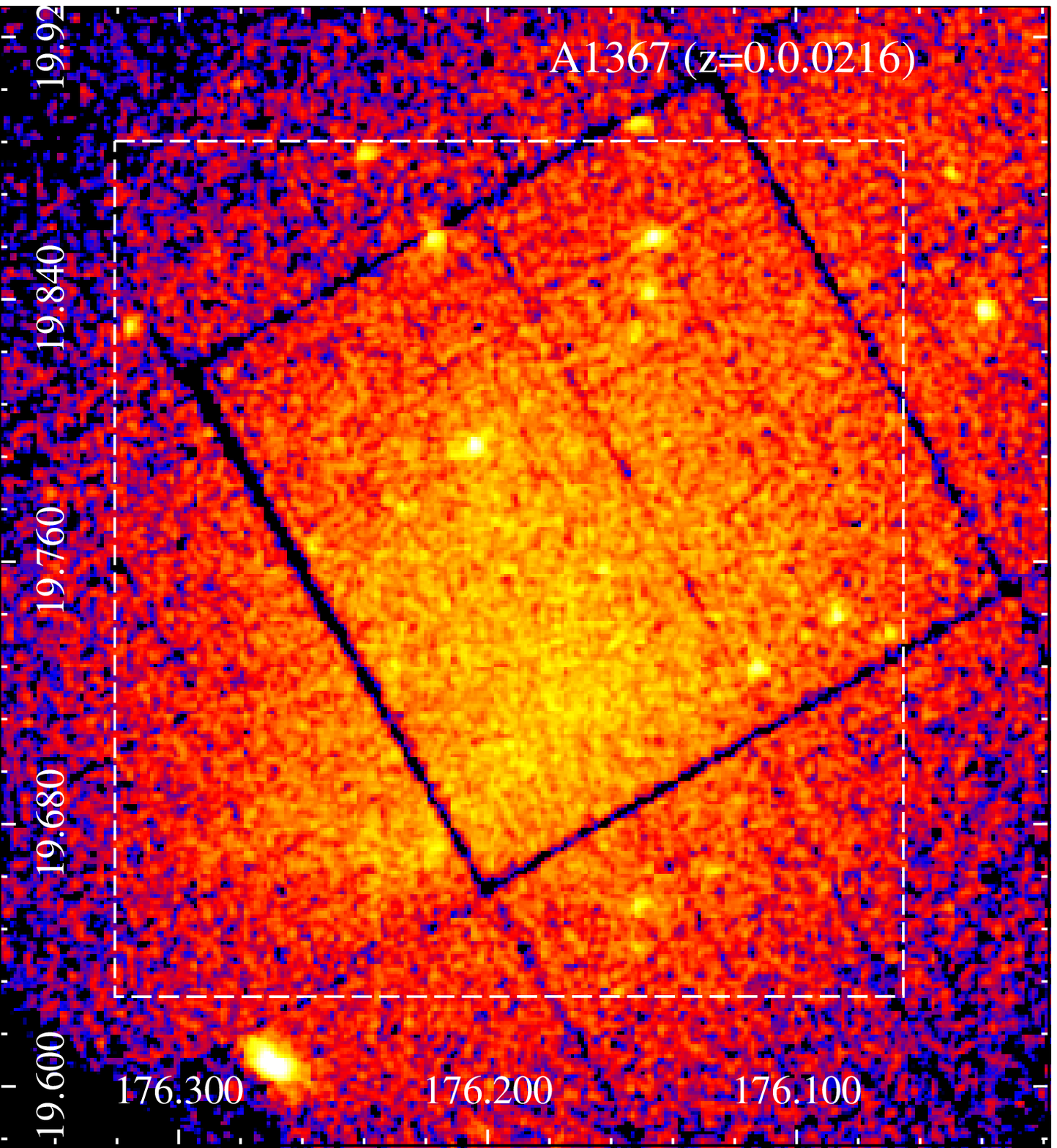}
\includegraphics[scale=0.22]{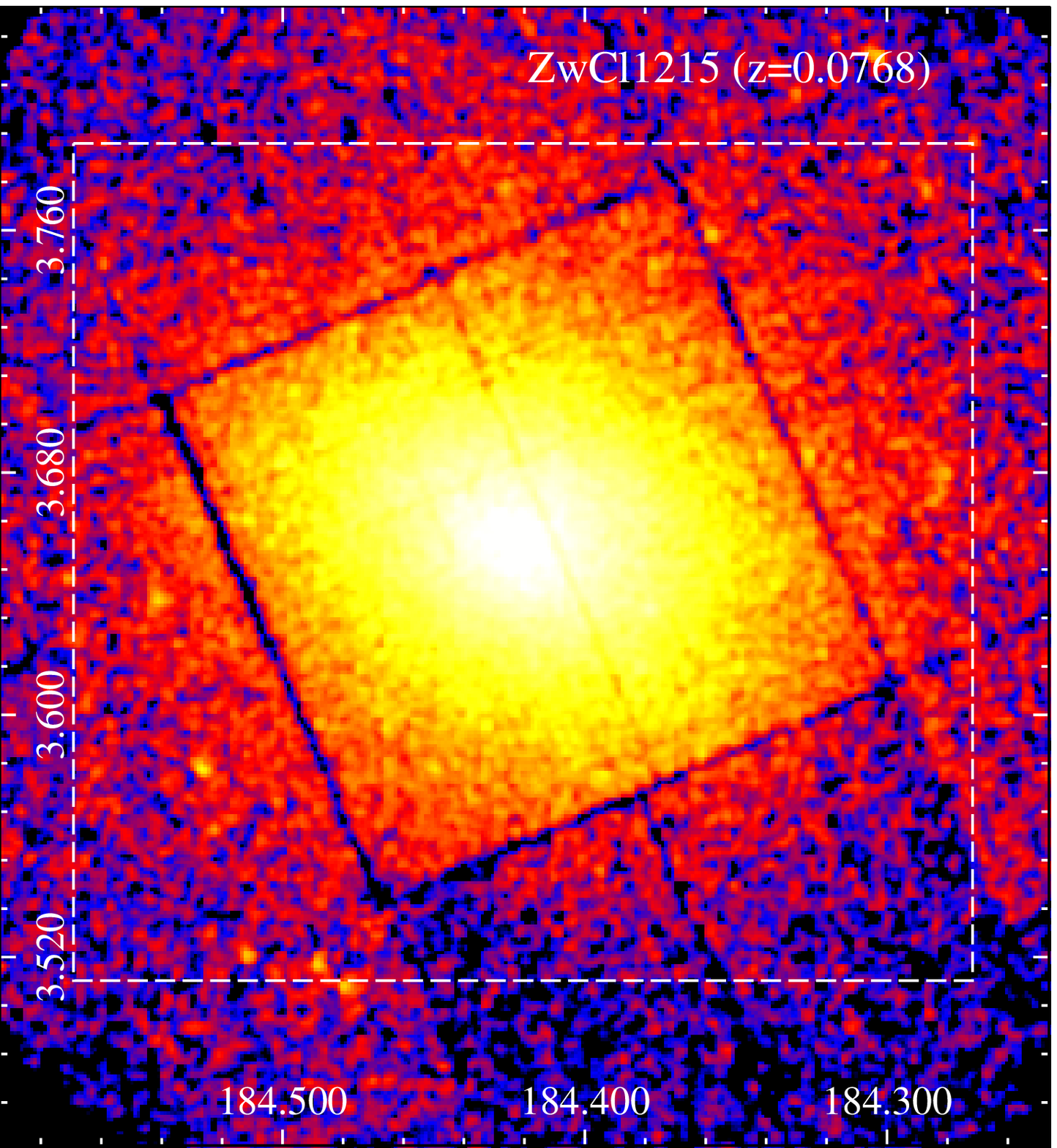}
\includegraphics[scale=0.22]{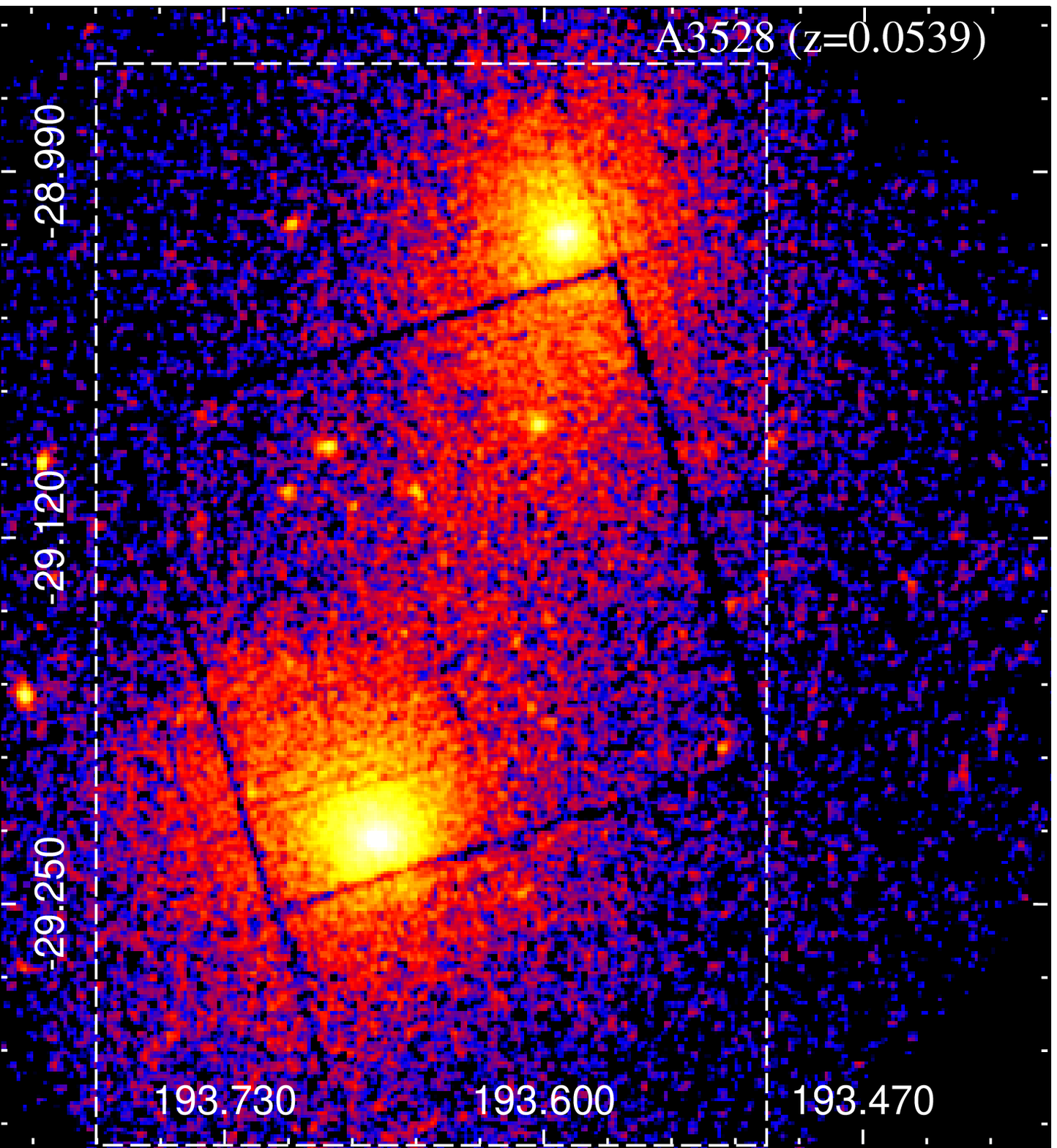}

\includegraphics[scale=0.22]{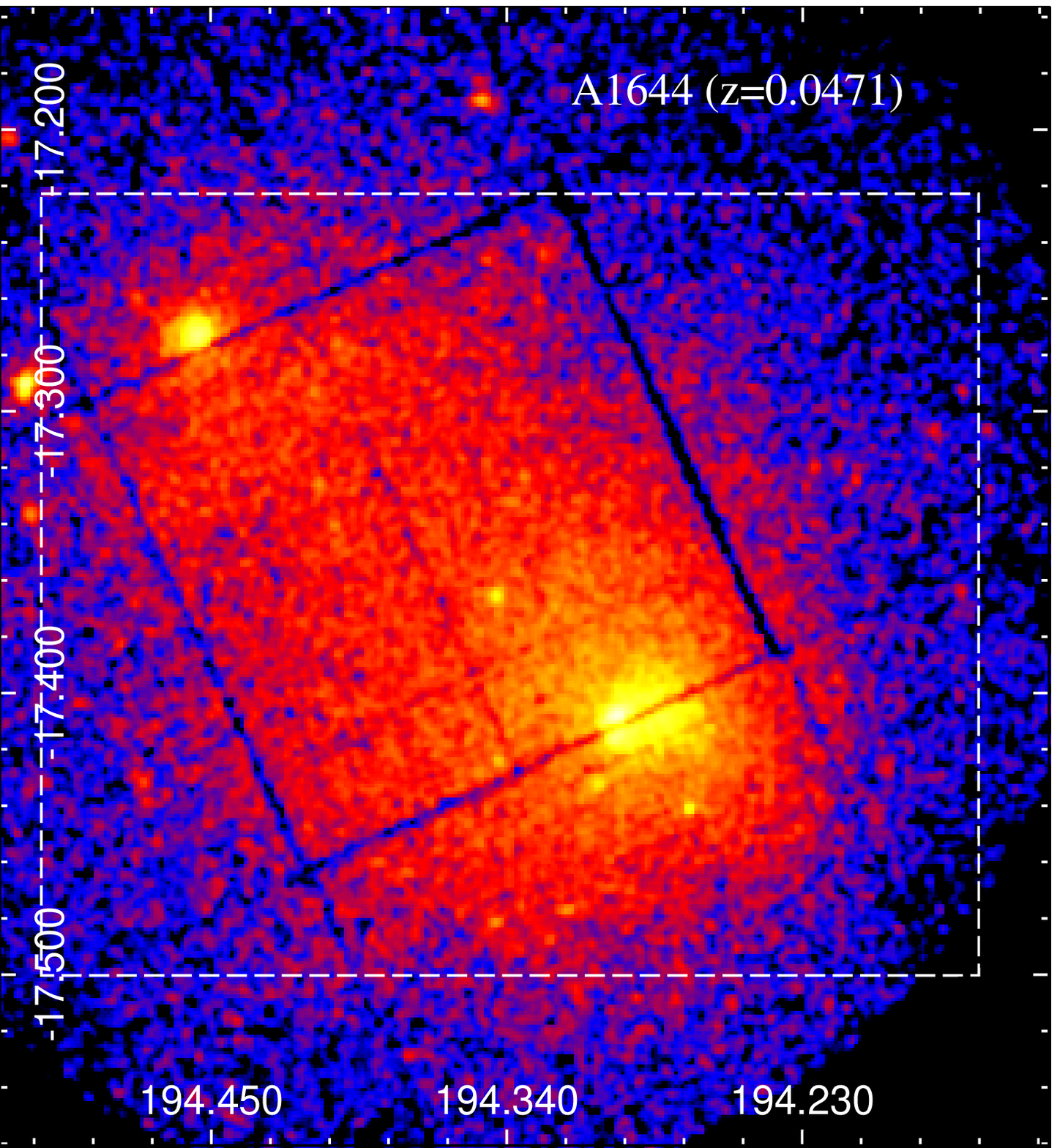}
\includegraphics[scale=0.22]{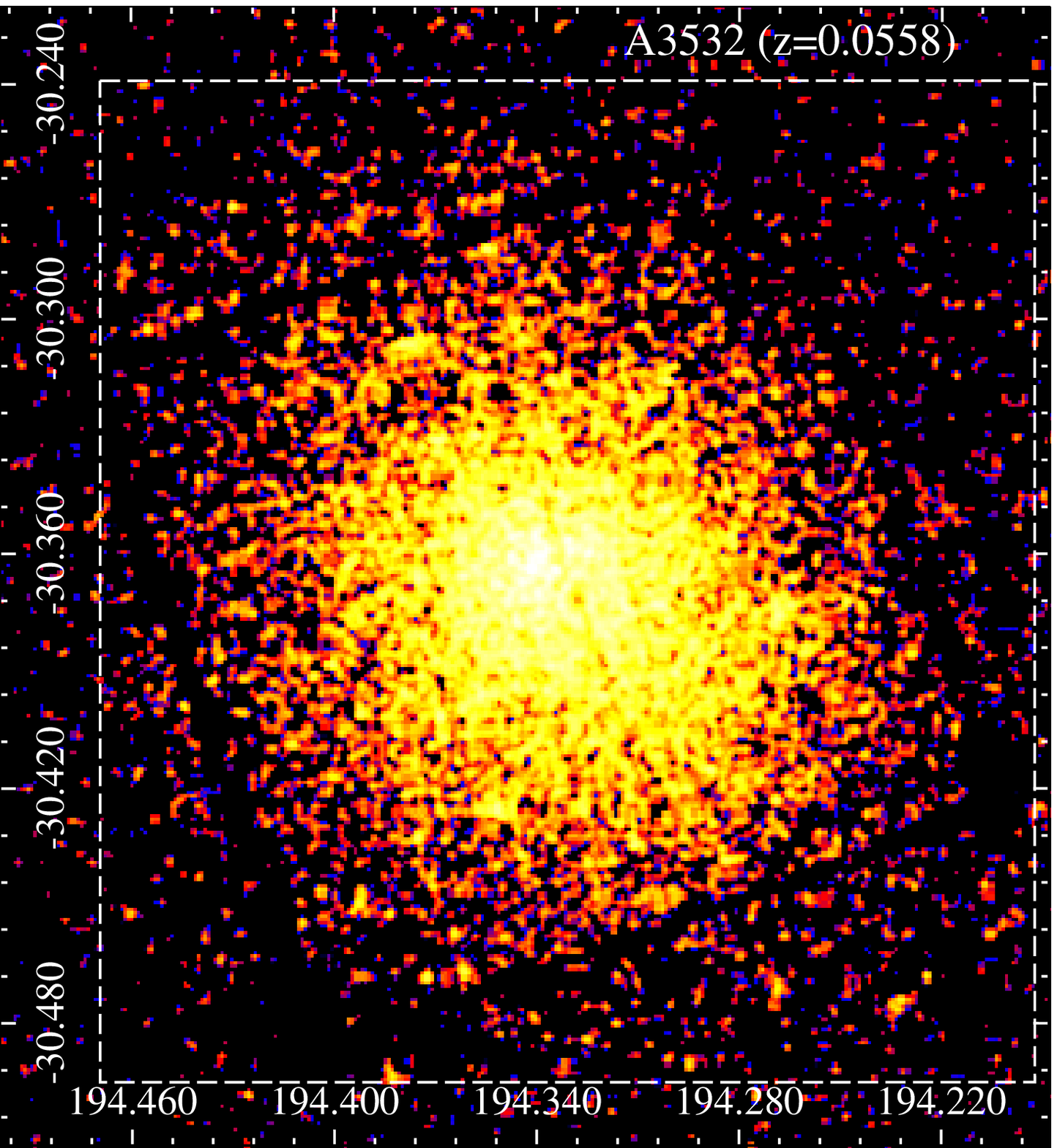}
\includegraphics[scale=0.22]{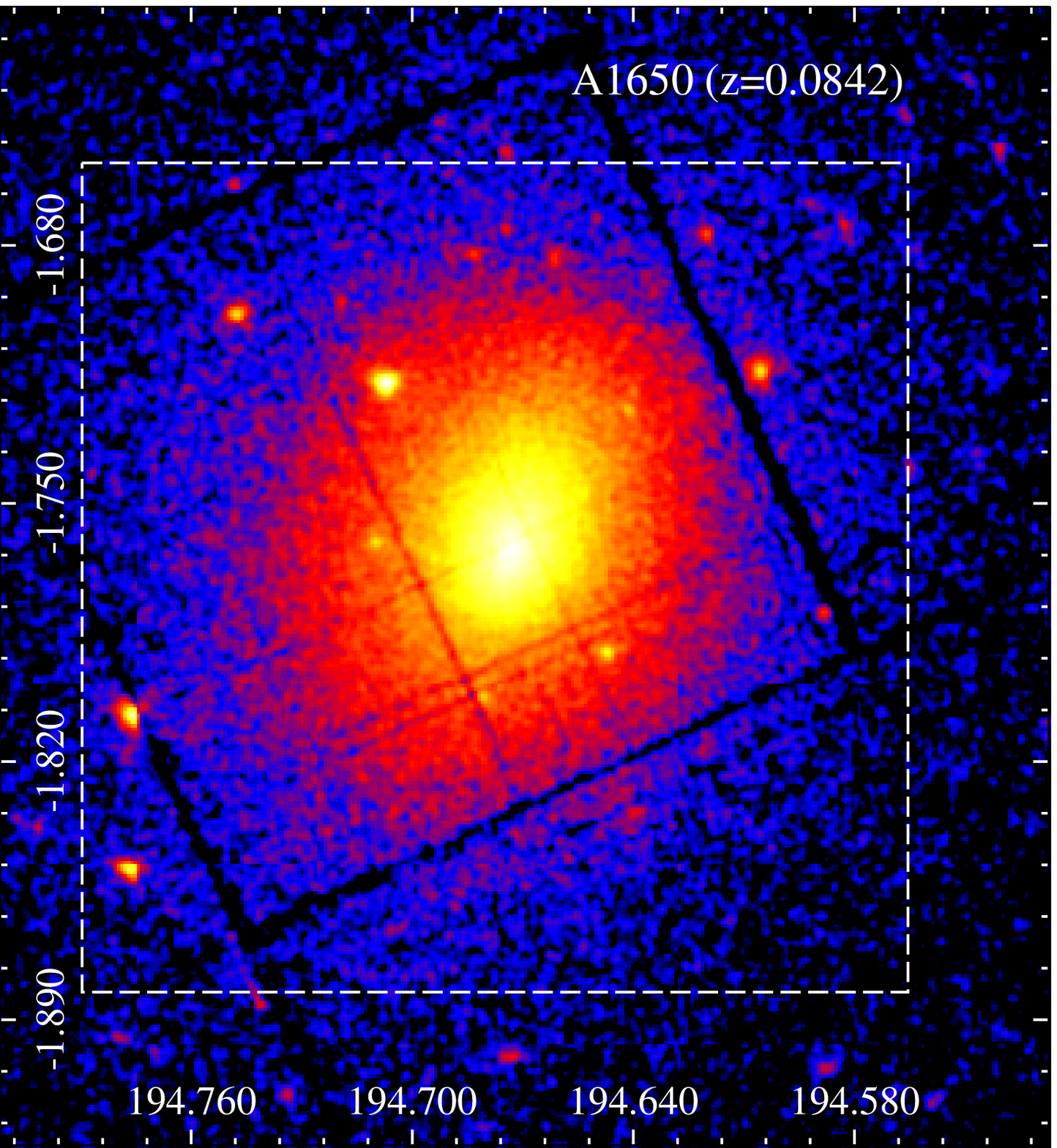}
\includegraphics[scale=0.22]{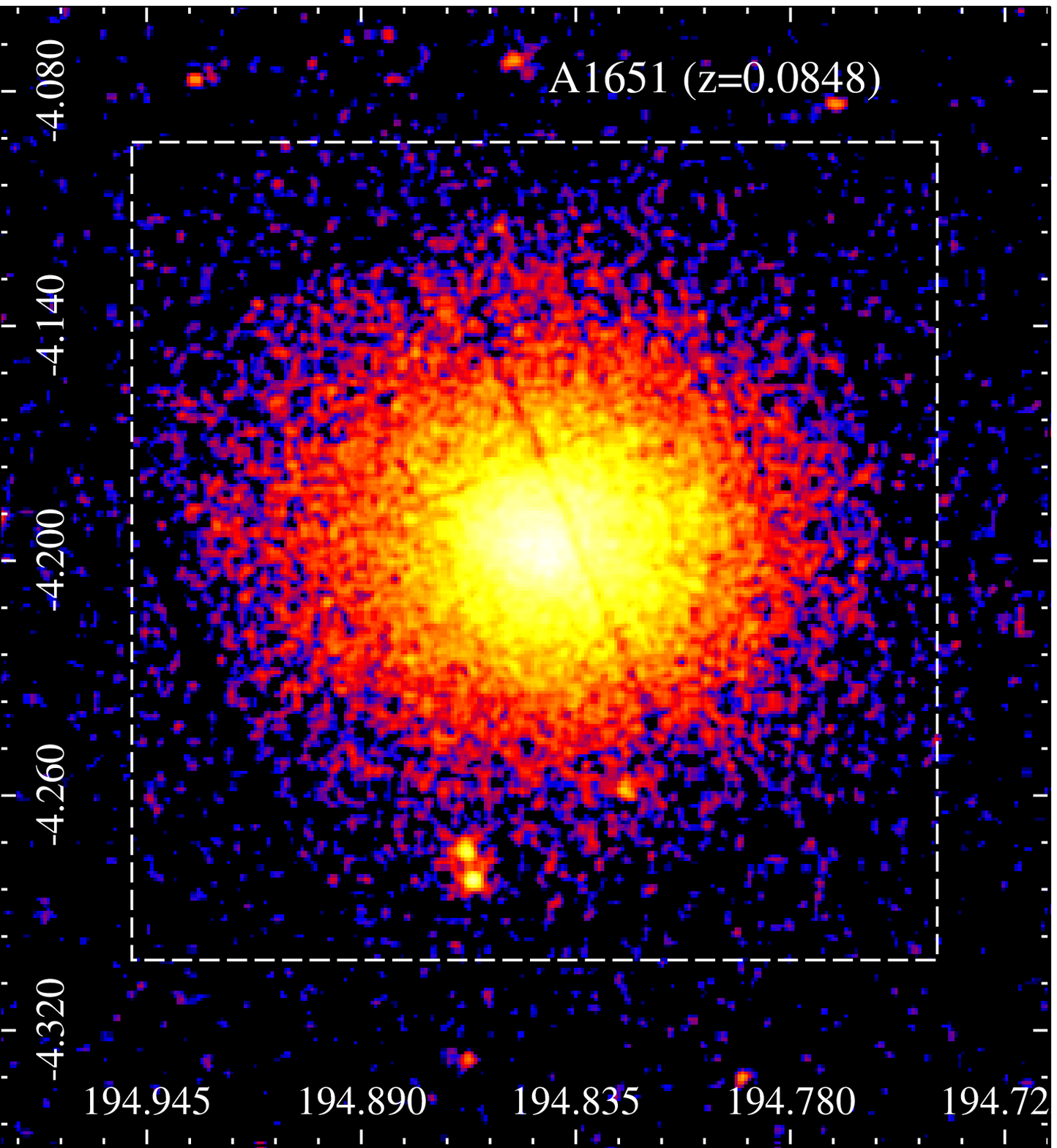}
\includegraphics[scale=0.22]{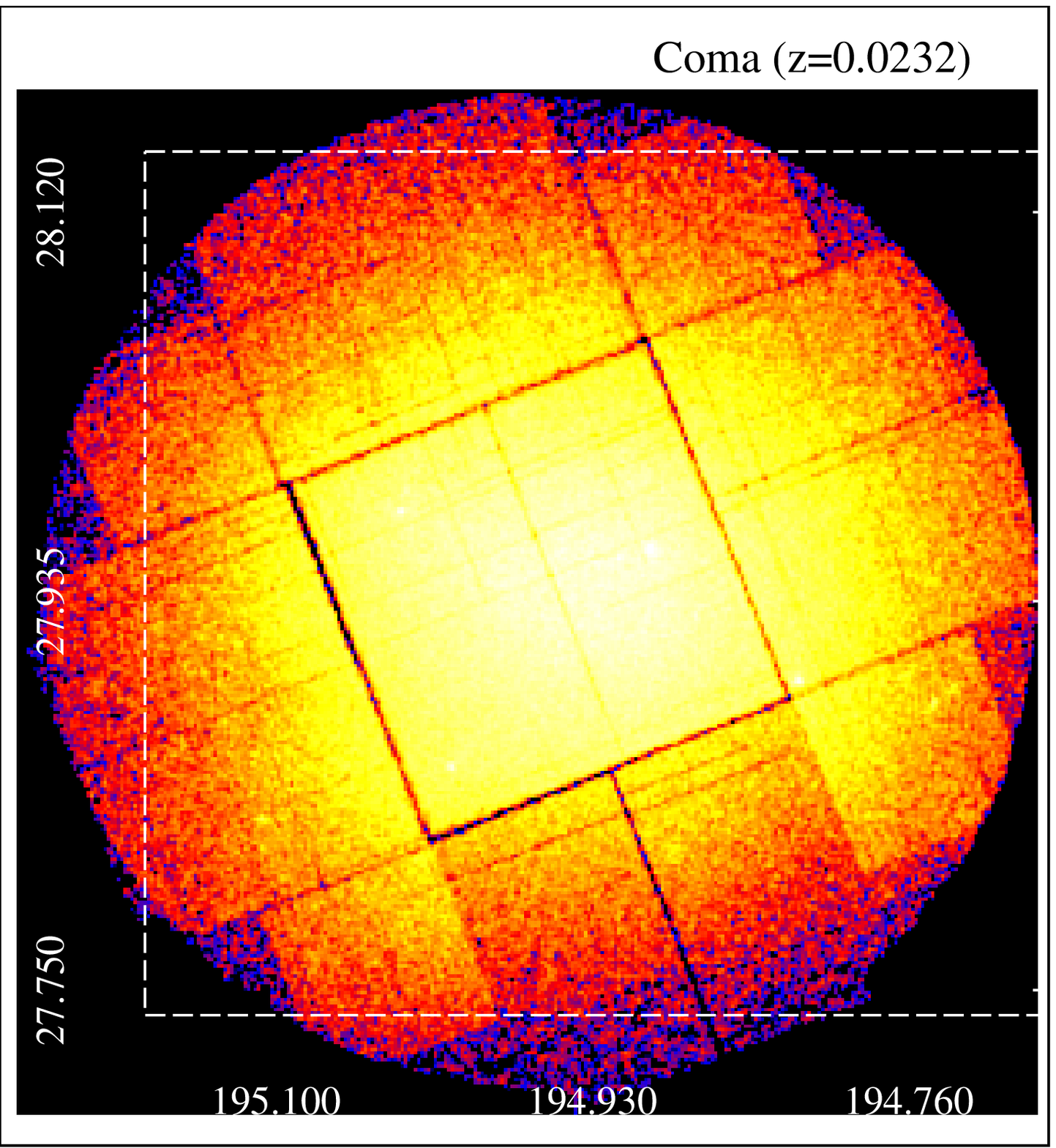}

\includegraphics[scale=0.22]{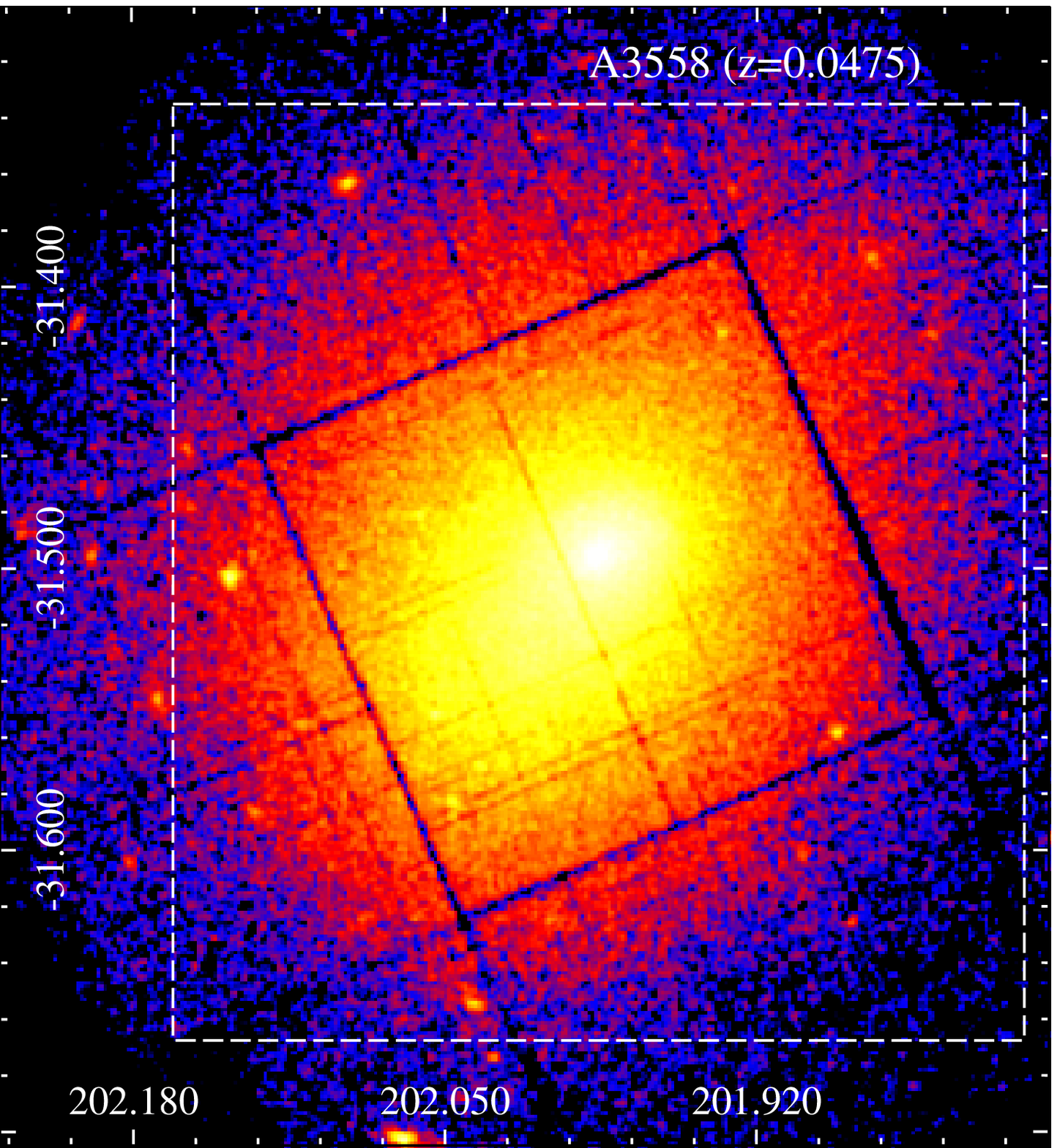}
\includegraphics[scale=0.22]{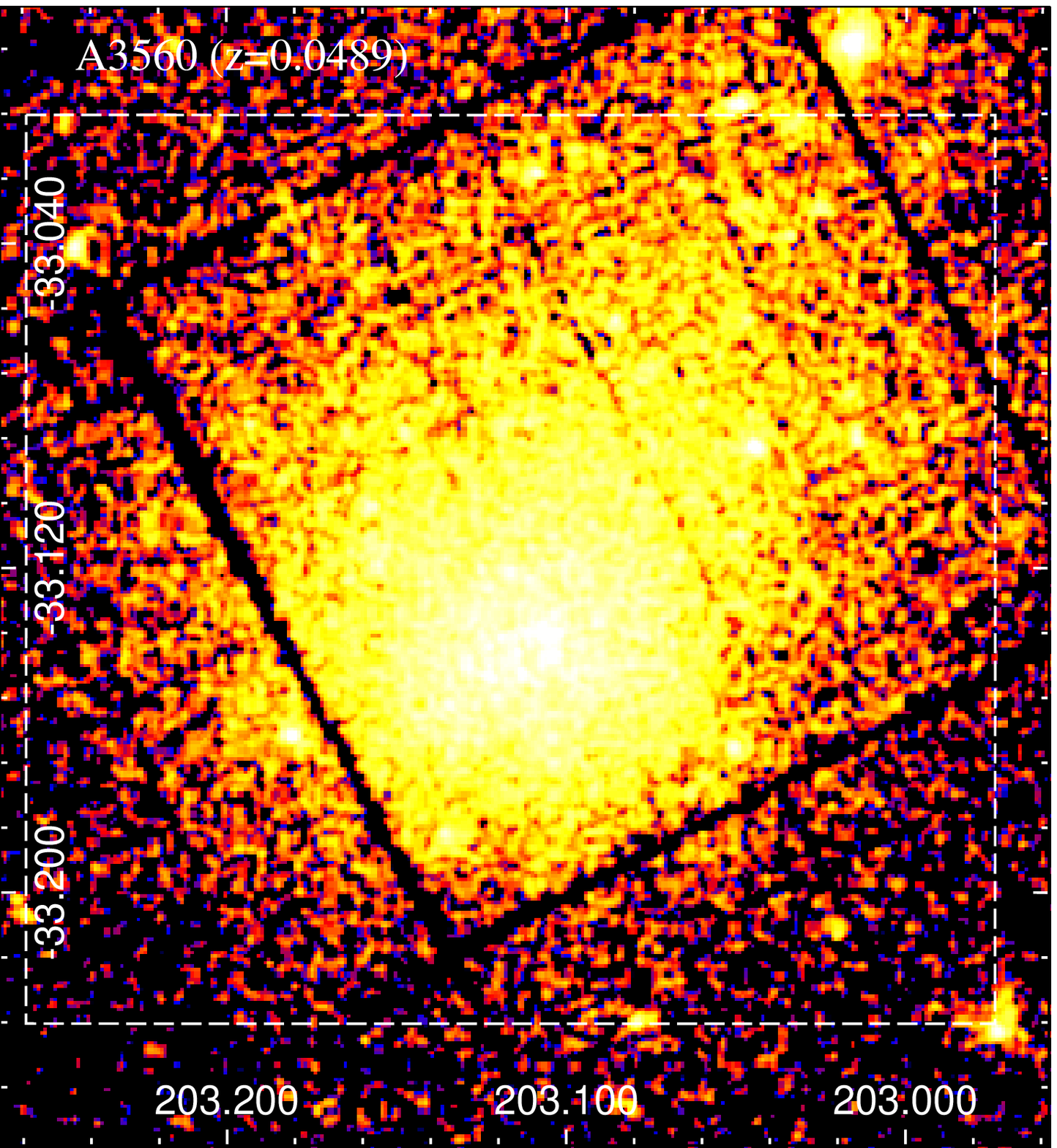}
\includegraphics[scale=0.22]{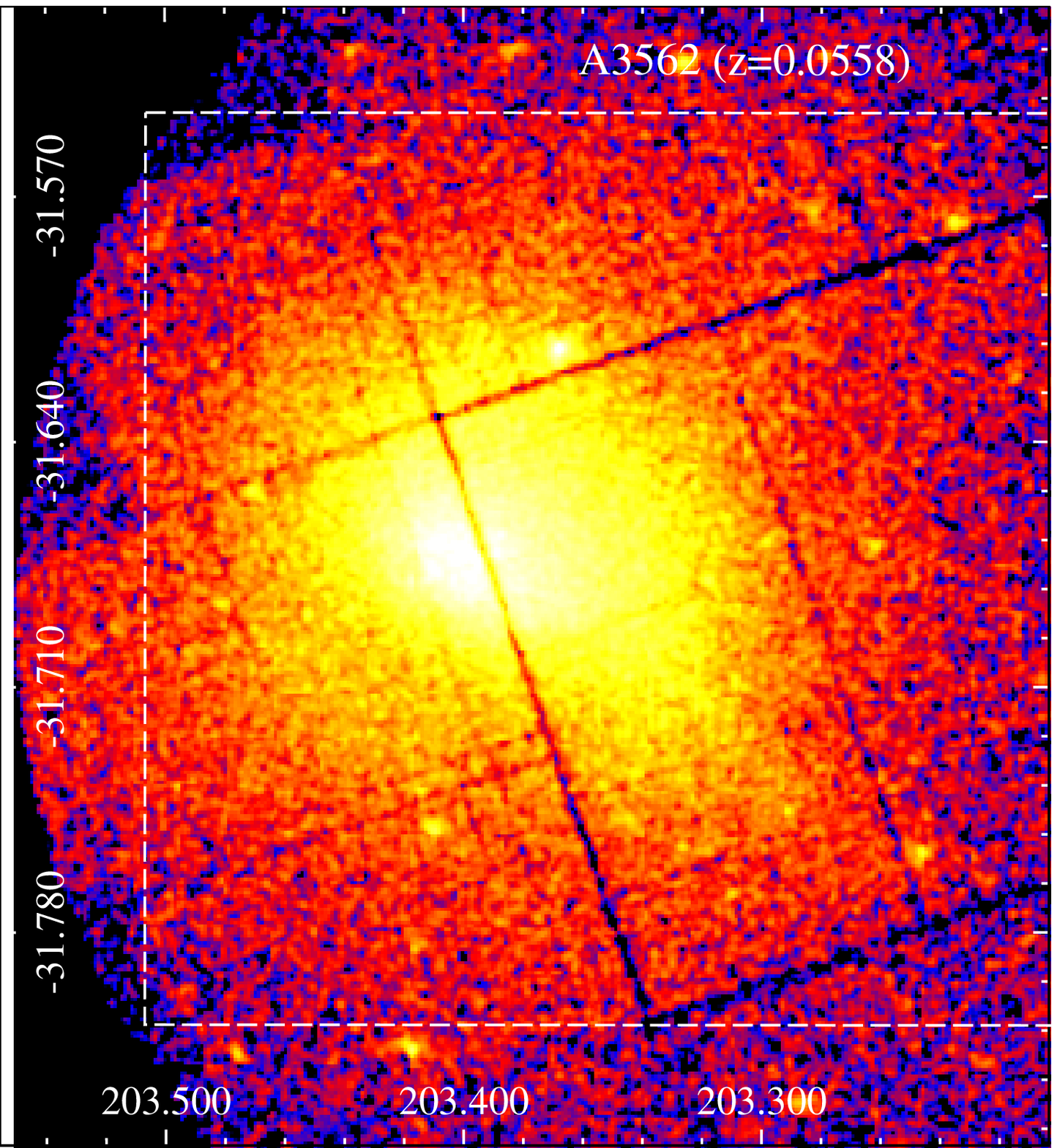}
\includegraphics[scale=0.22]{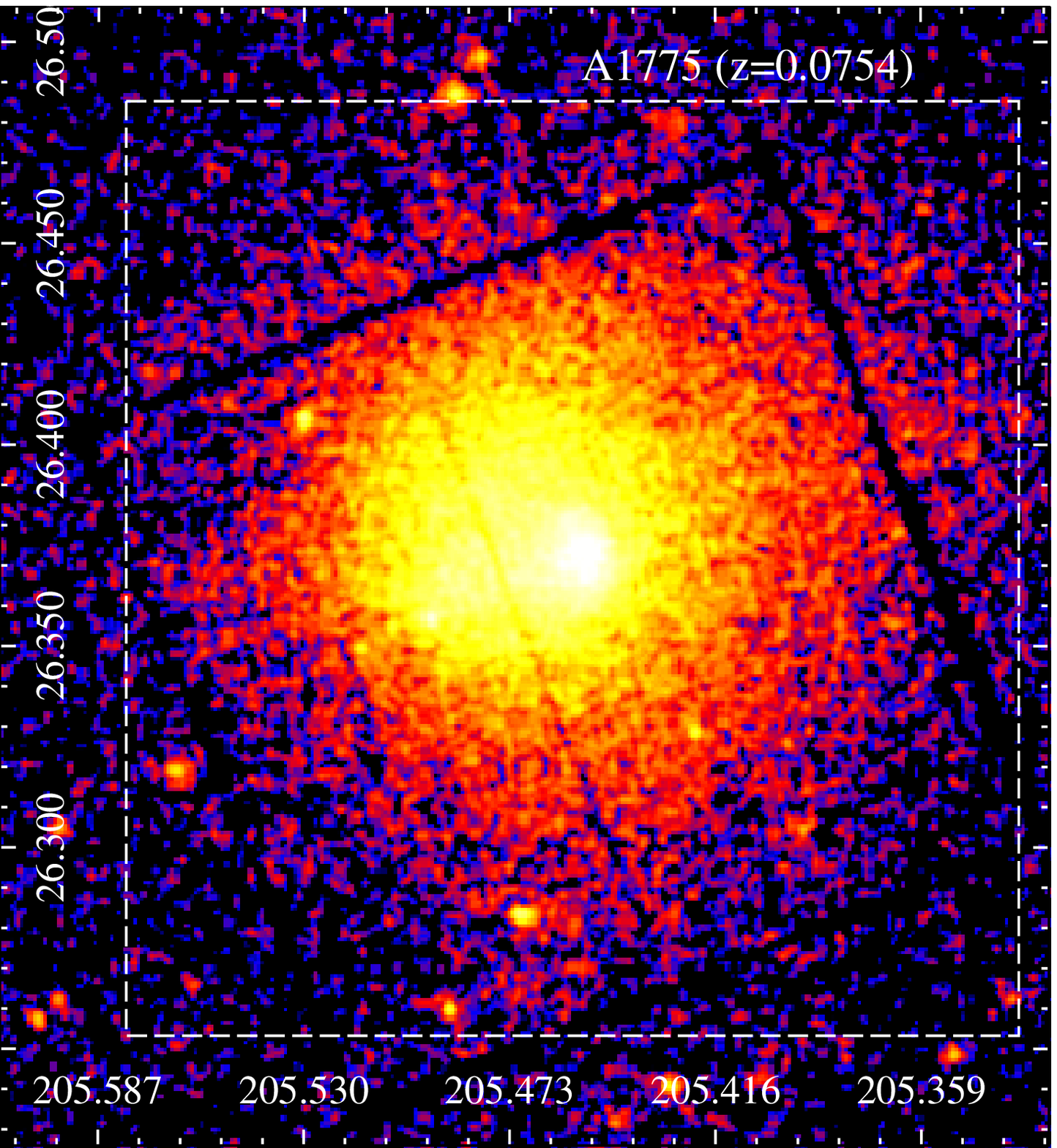}
\includegraphics[scale=0.22]{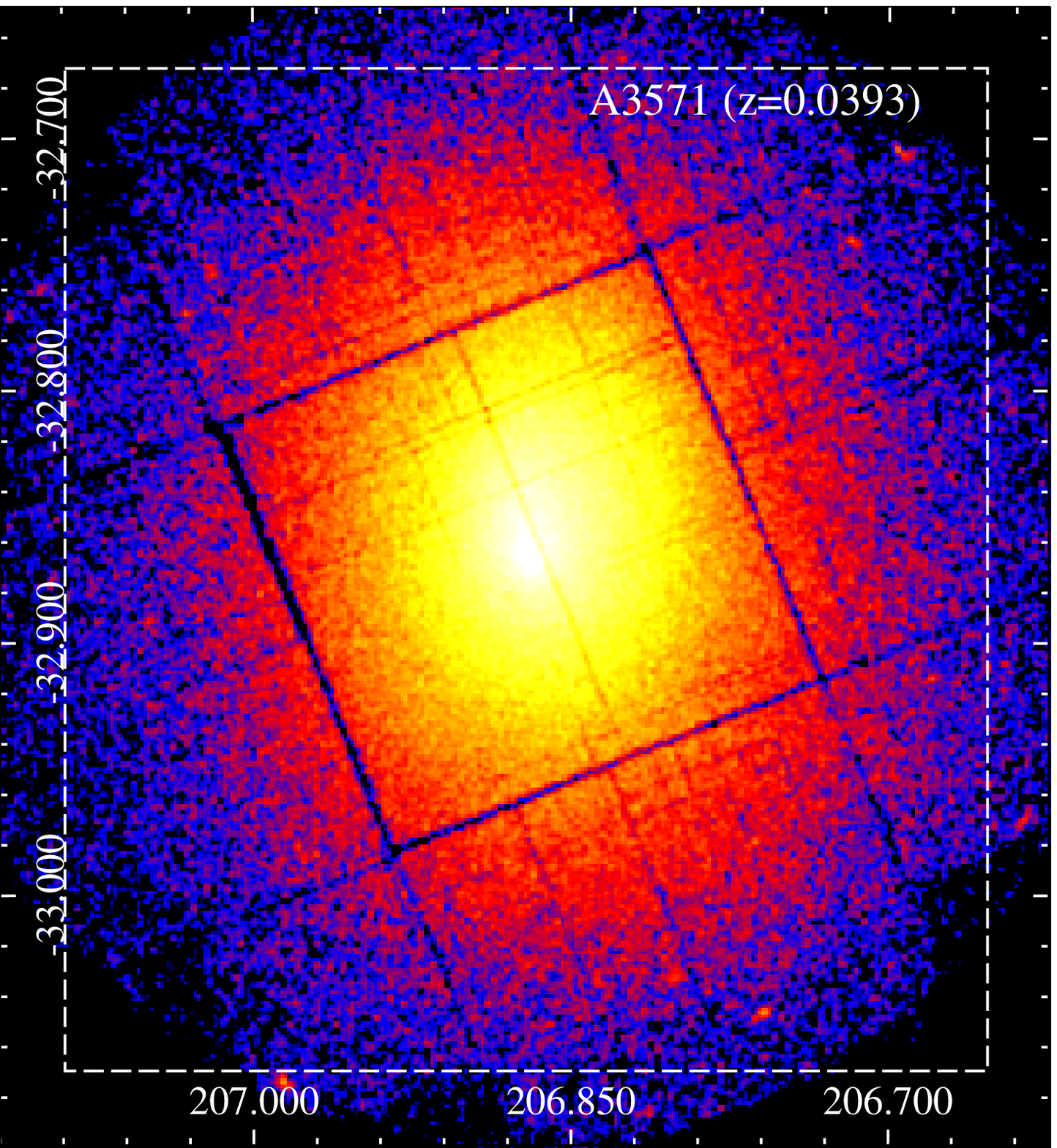}

\includegraphics[scale=0.22]{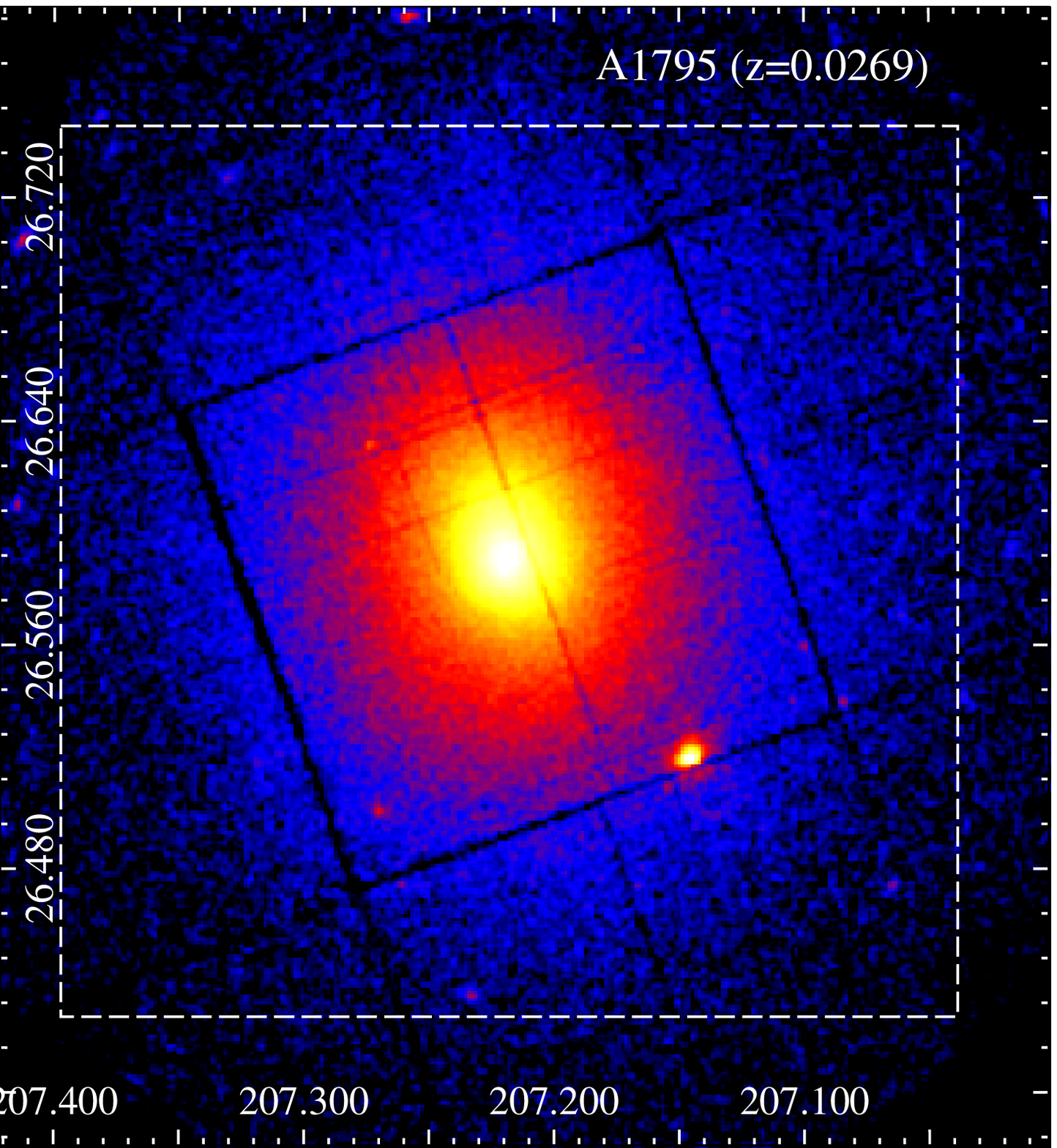}
\includegraphics[scale=0.22]{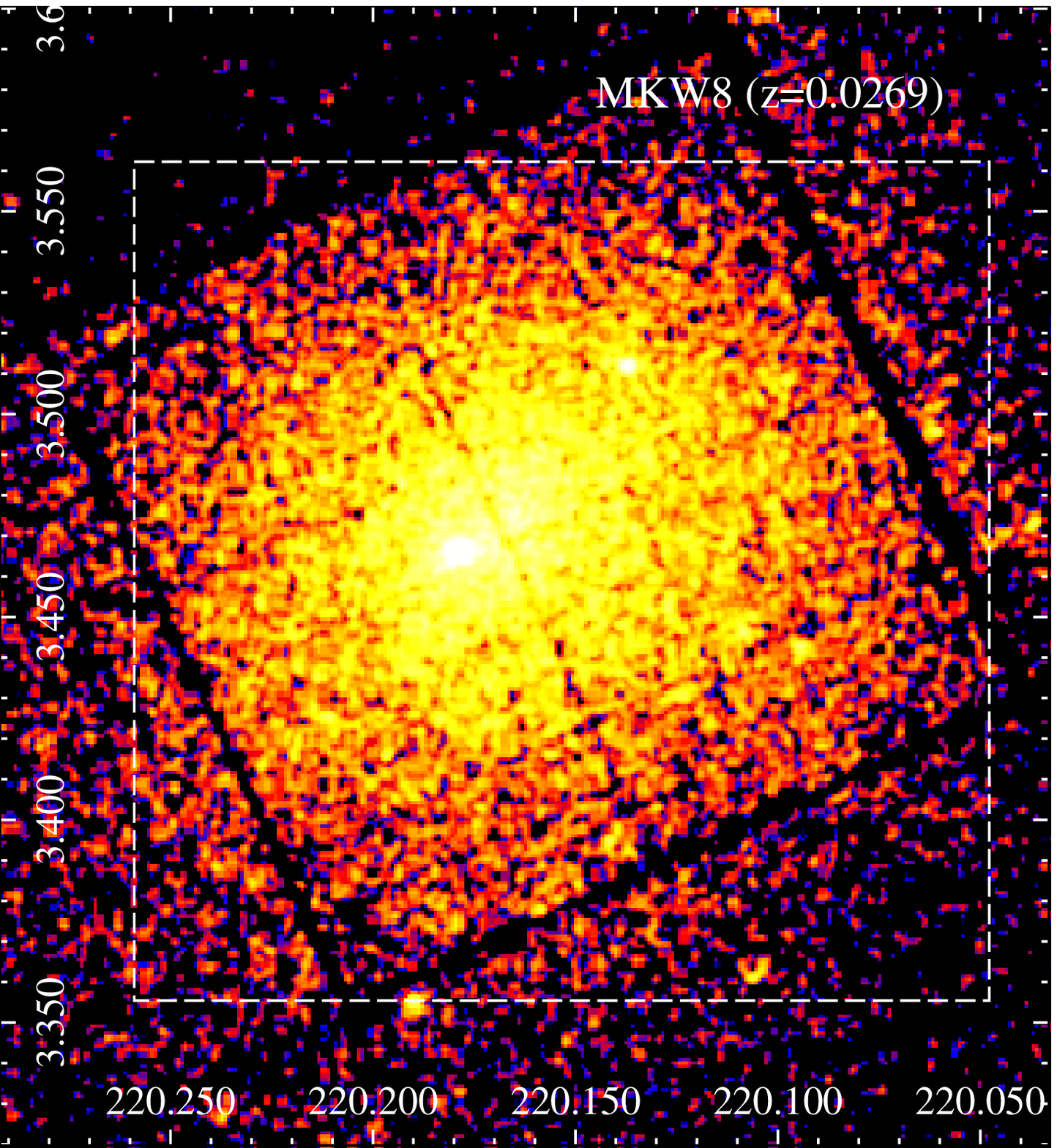}
\includegraphics[scale=0.22]{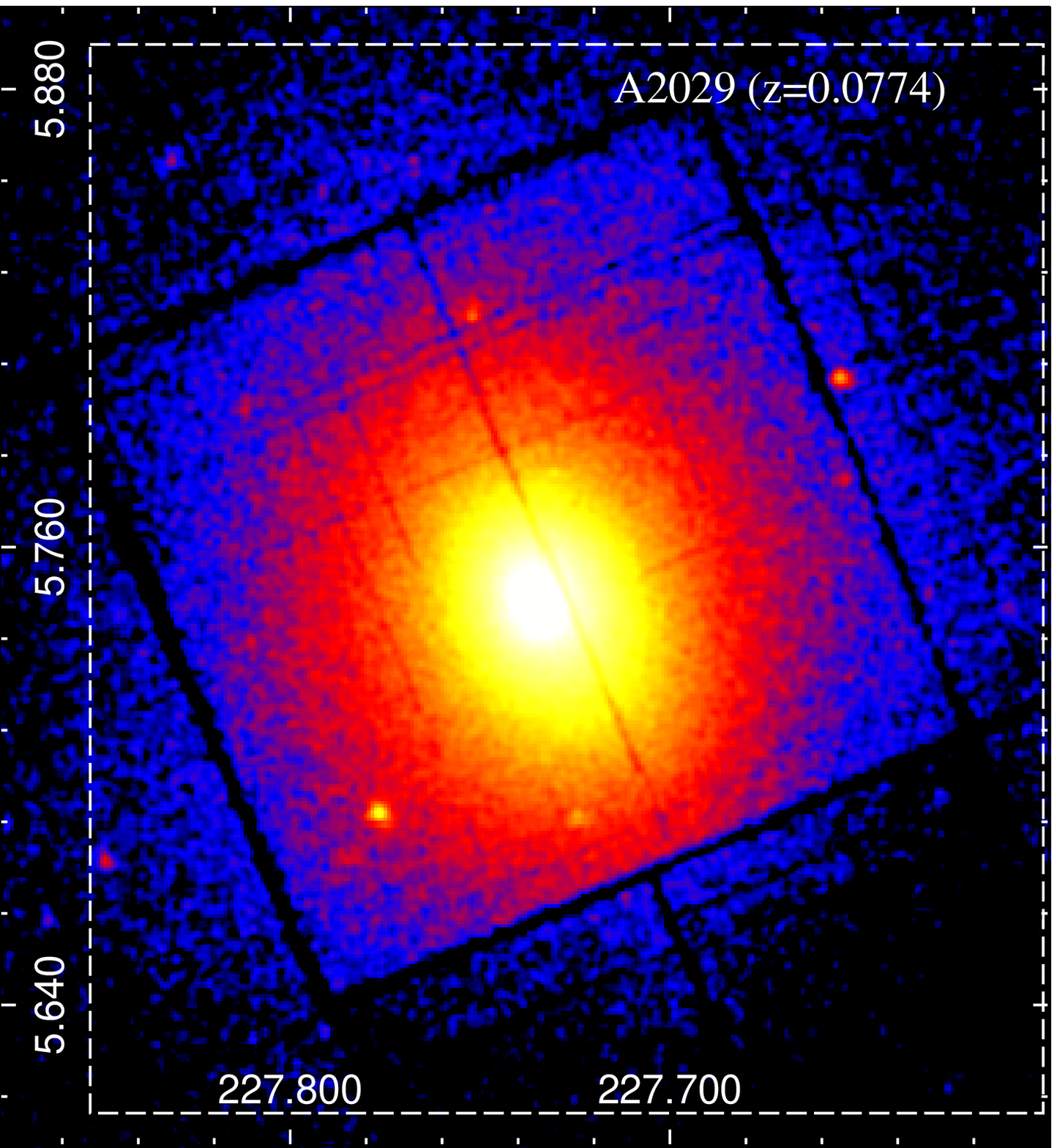}
\includegraphics[scale=0.22]{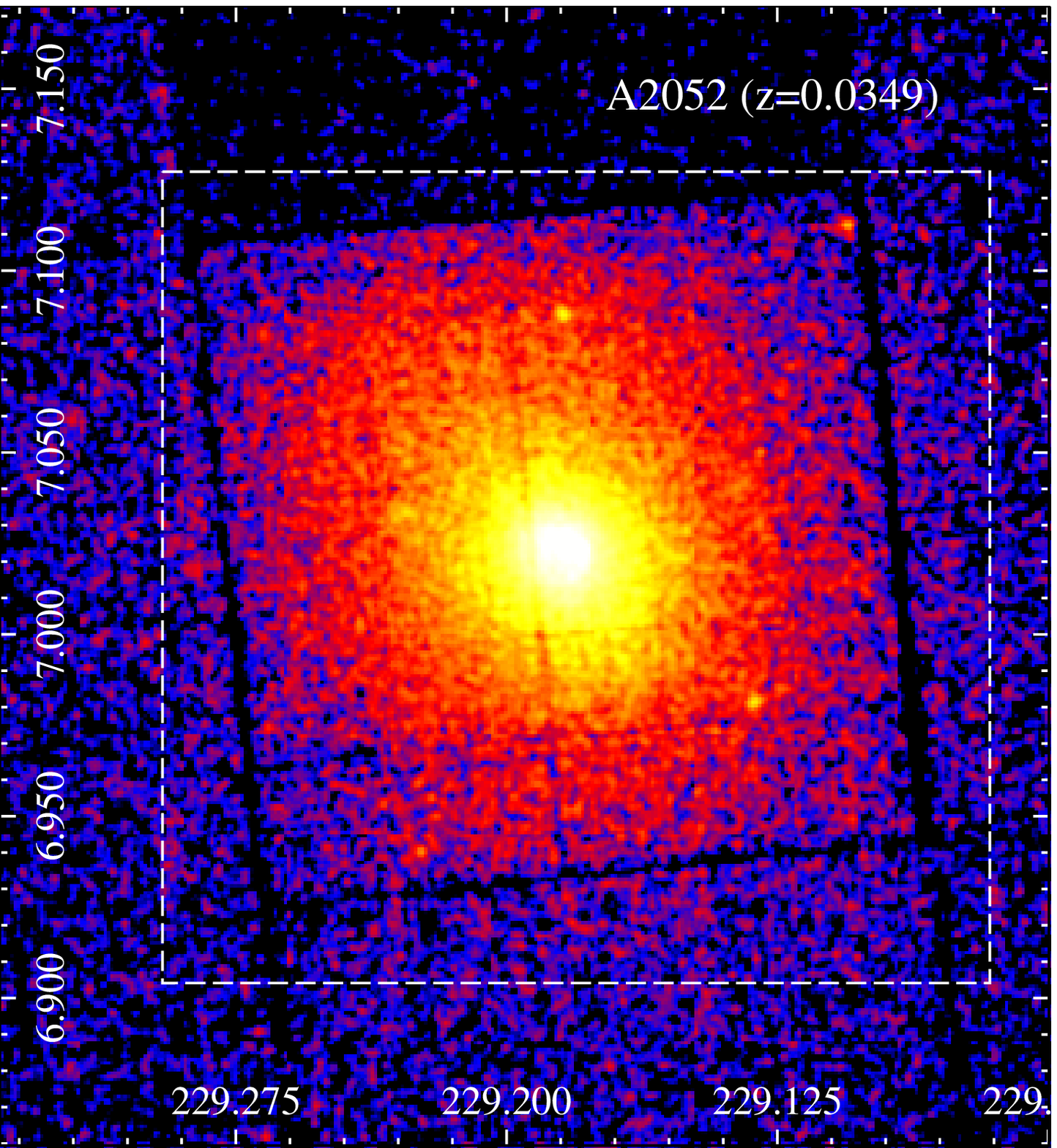}
\includegraphics[scale=0.22]{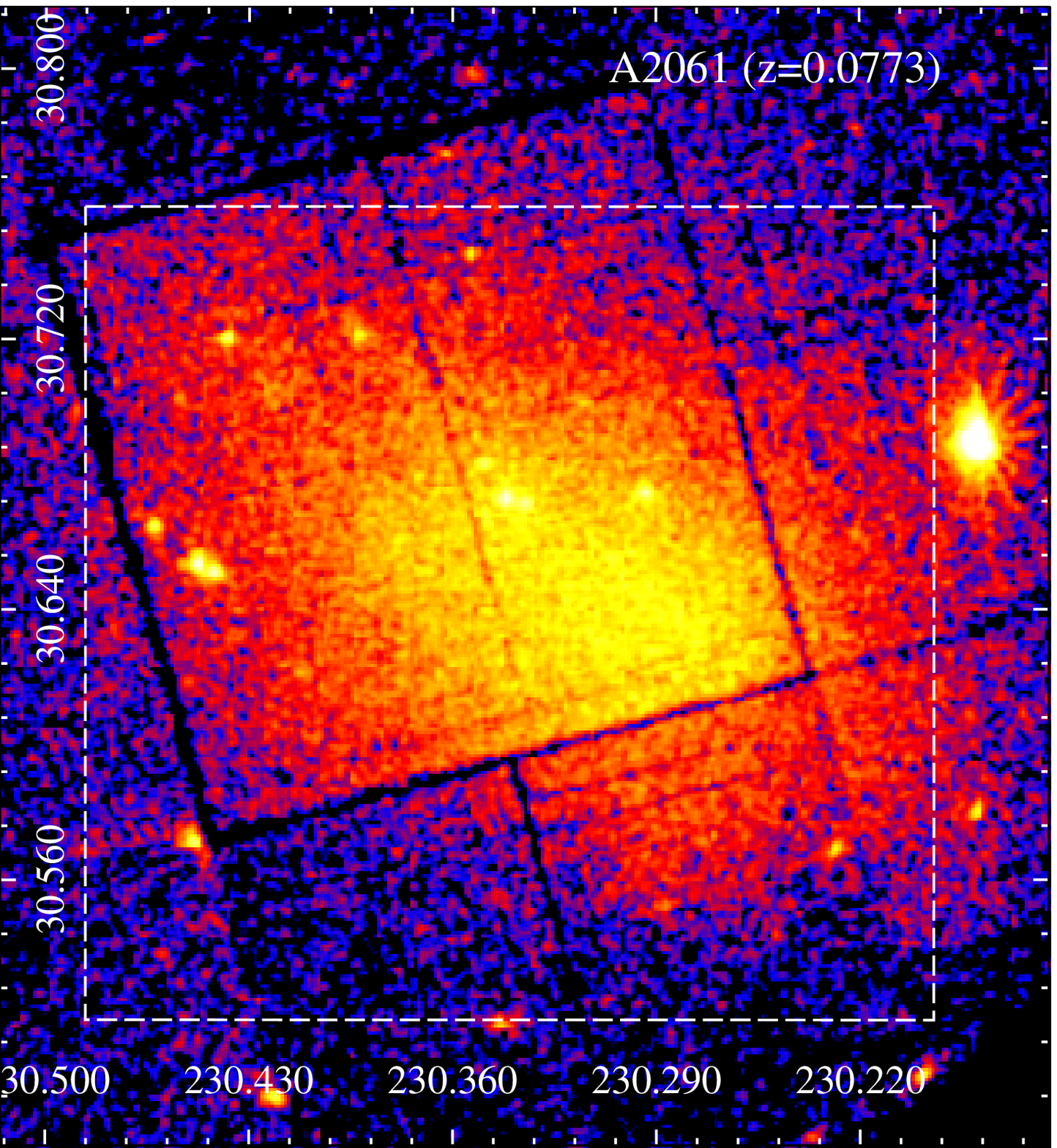}
\caption{X-ray images of the cluster sample in the same order as in Tab.~\ref{tab:obs}. The white dashed boxes represent the 
regions used to construct the 2D maps.}
\end{figure*}

\renewcommand{\thefigure}{\arabic{figure} (Cont.)}
\addtocounter{figure}{-1}

\begin{figure*}
\includegraphics[scale=0.22]{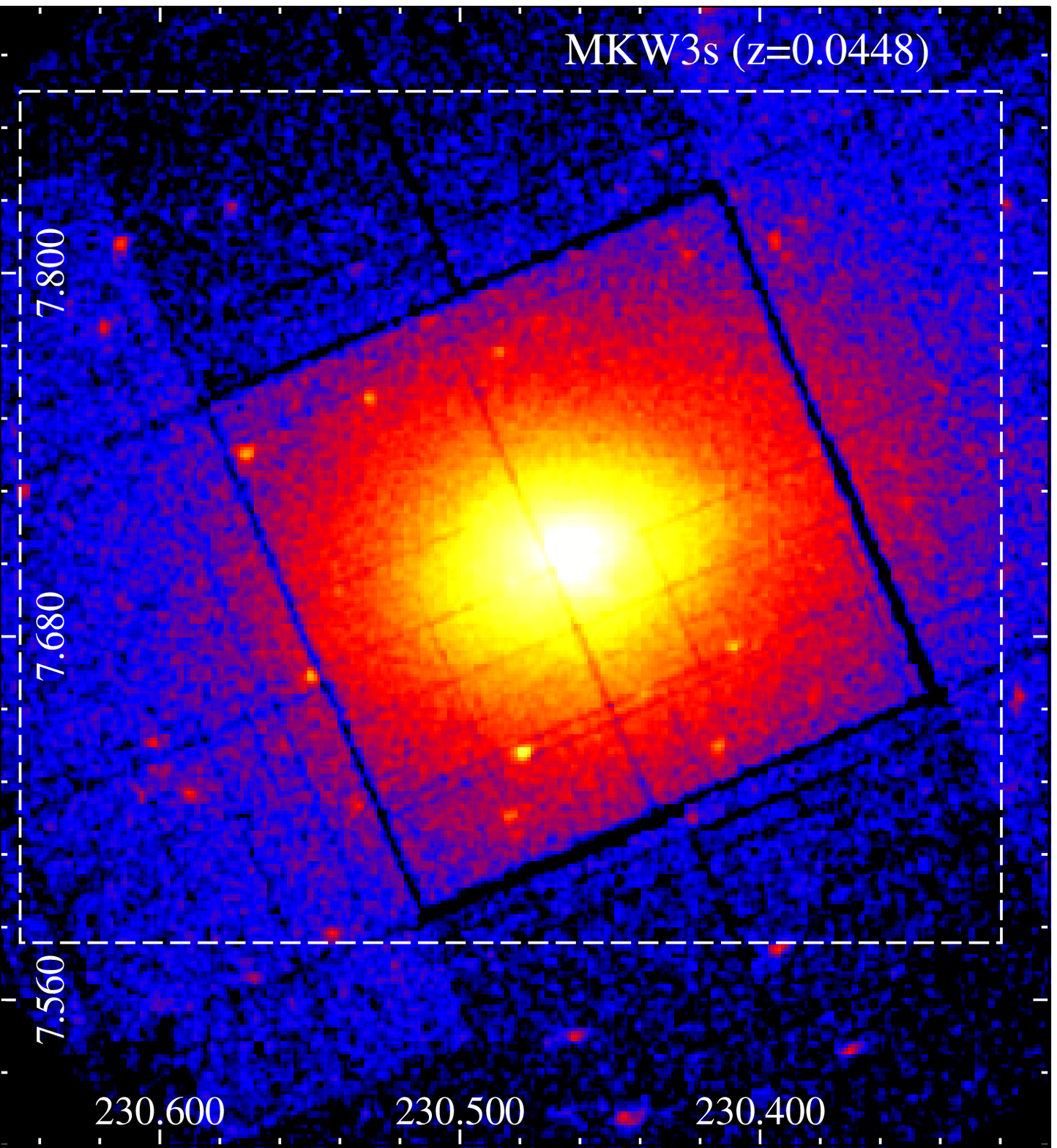}
\includegraphics[scale=0.22]{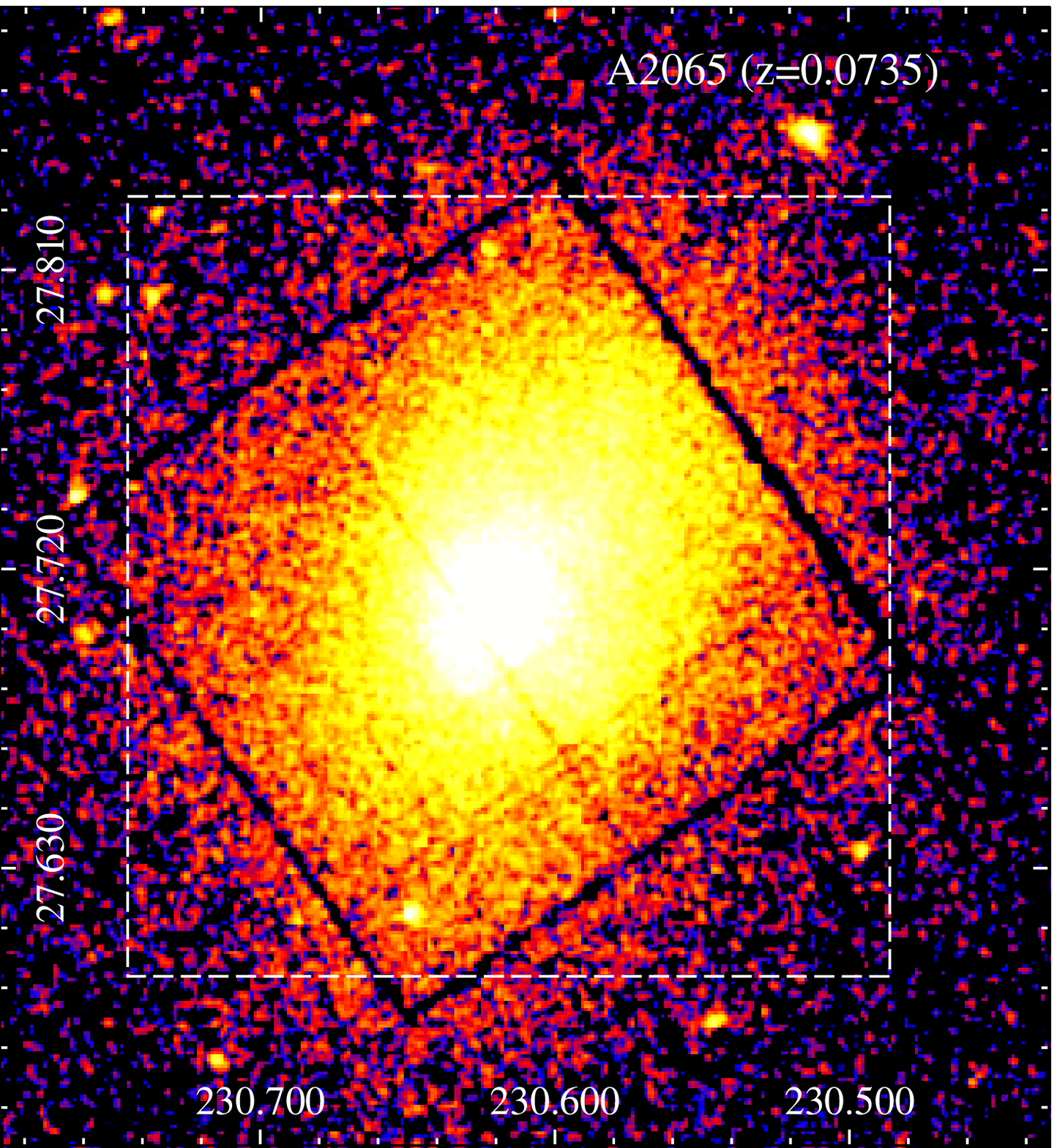}
\includegraphics[scale=0.22]{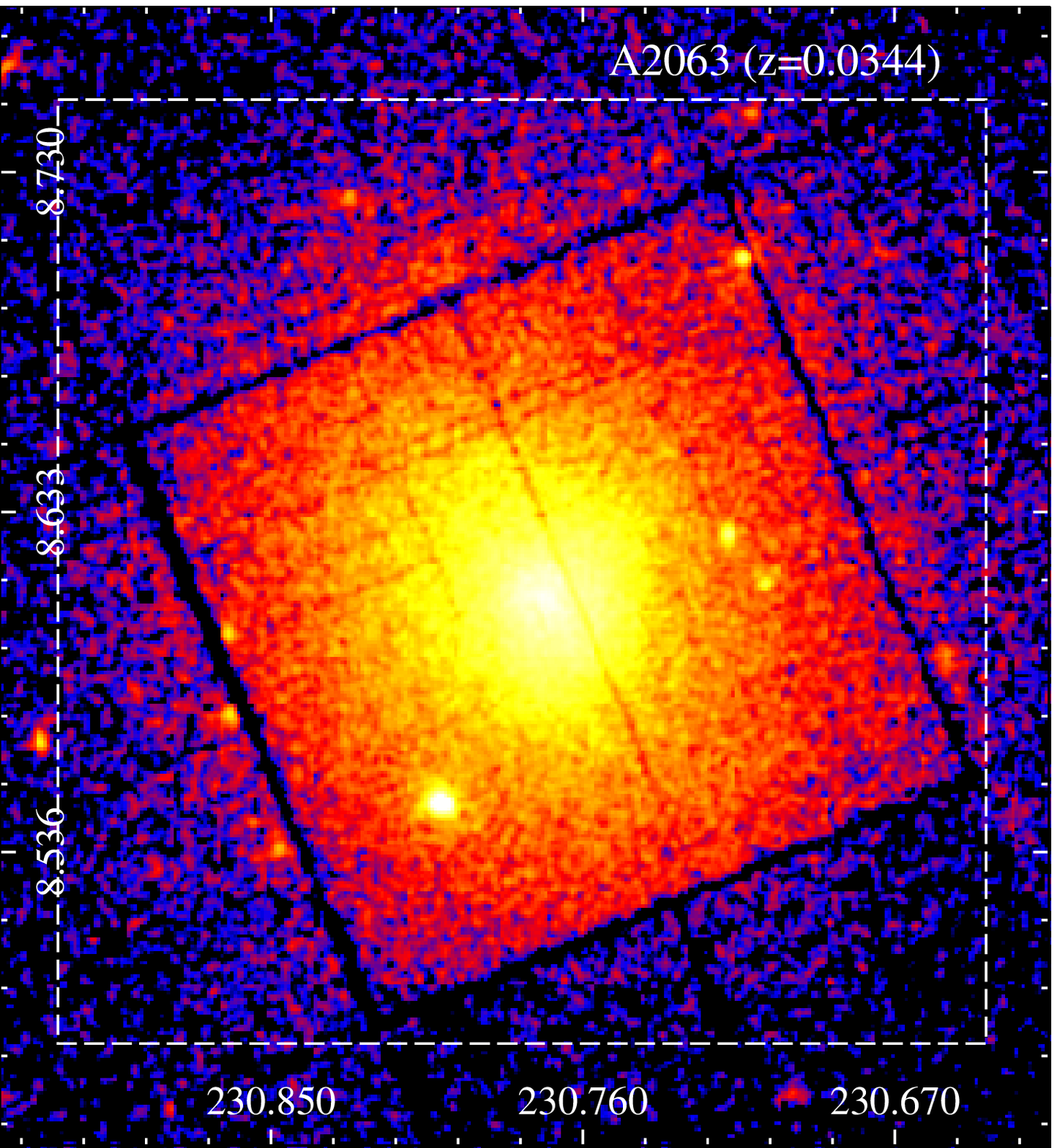}
\includegraphics[scale=0.22]{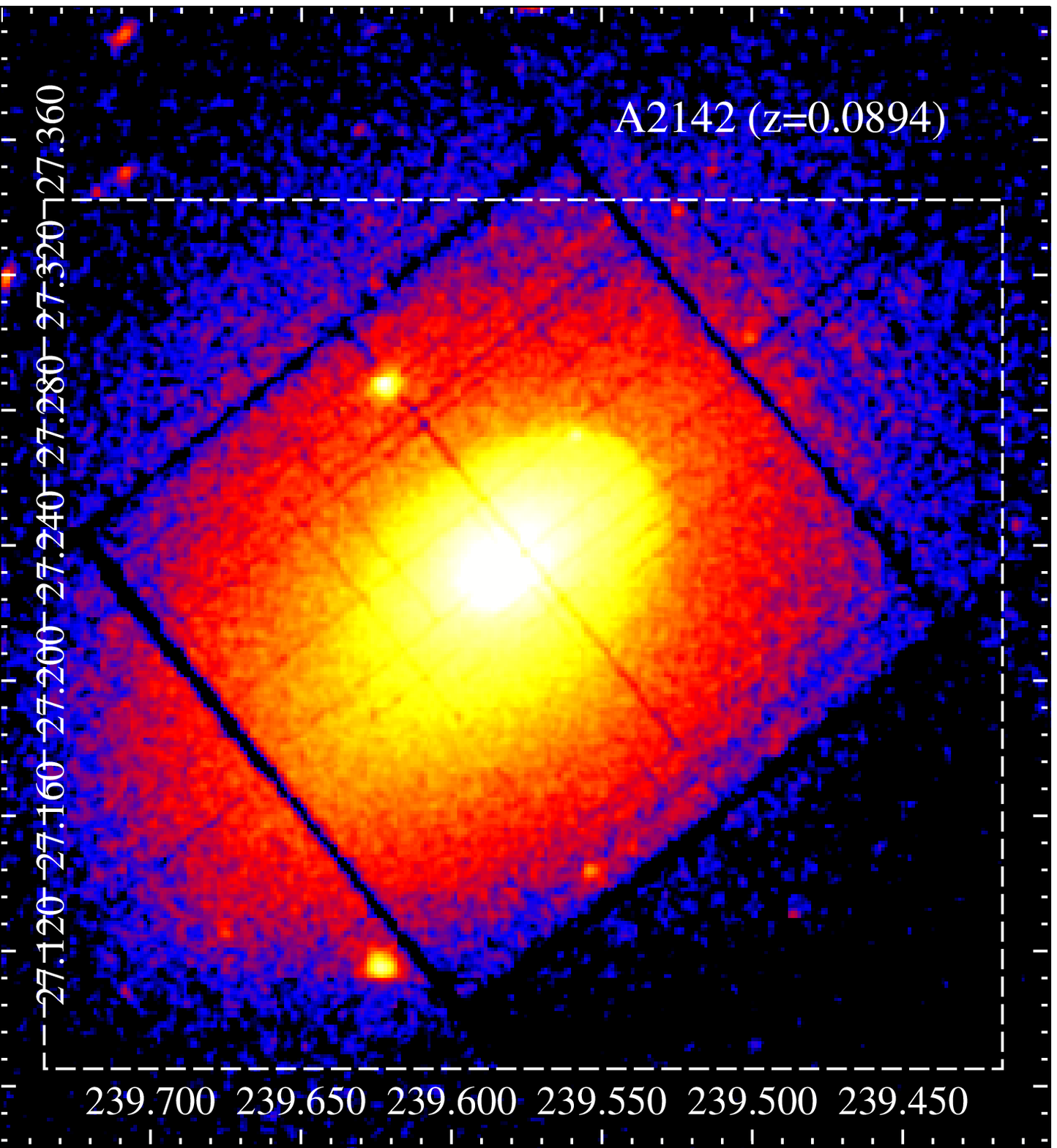}
\includegraphics[scale=0.22]{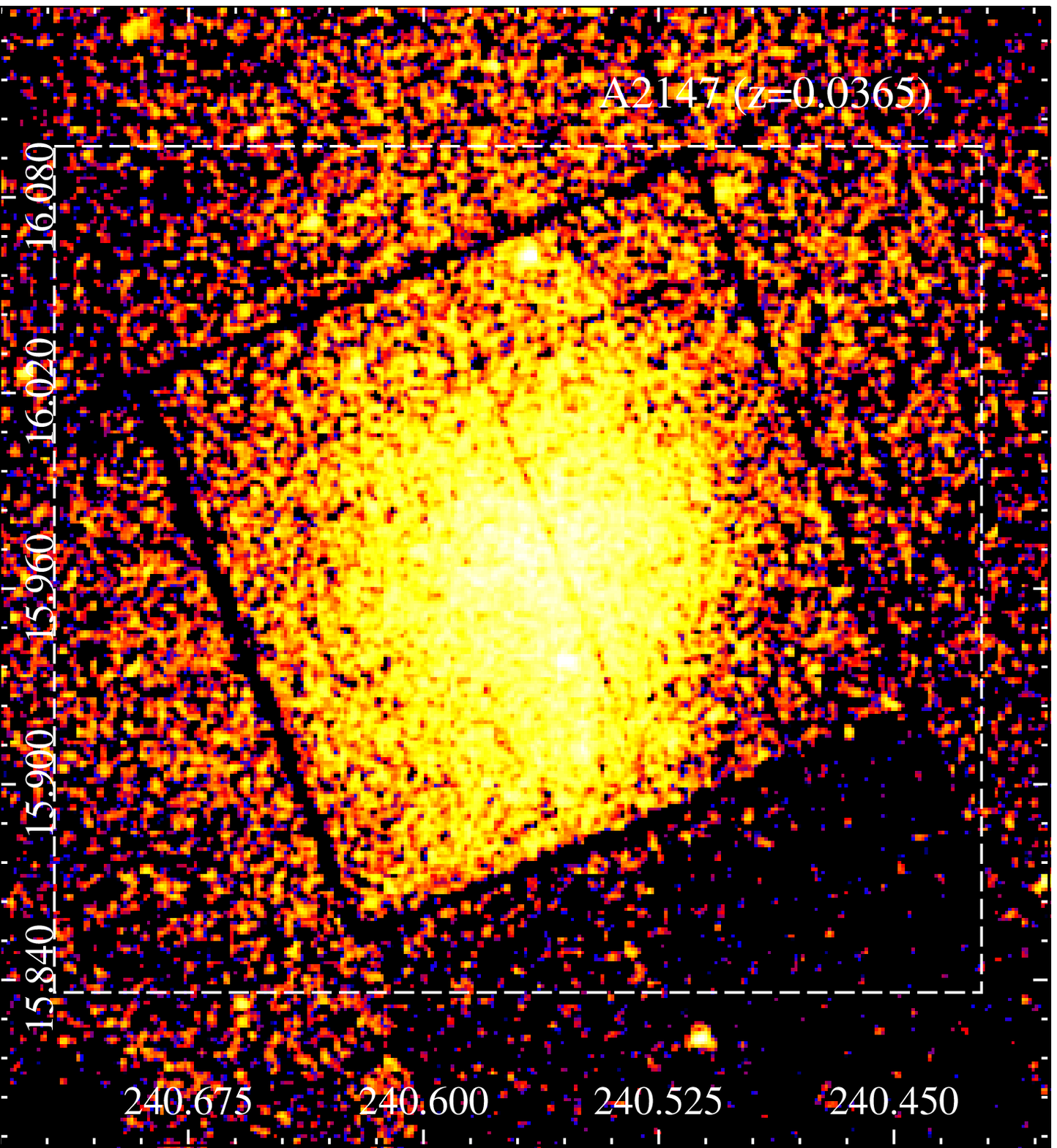}

\includegraphics[scale=0.22]{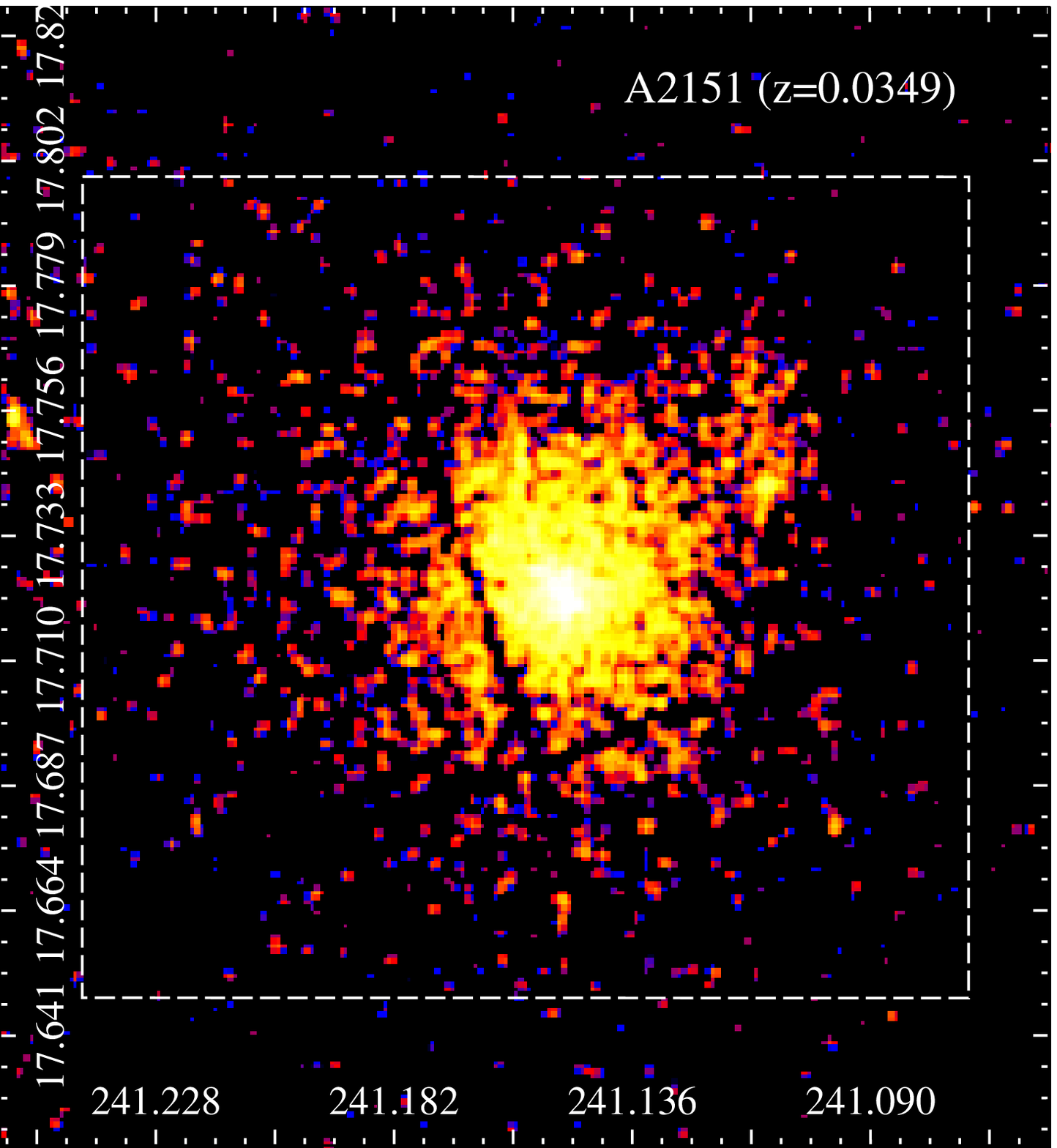}
\includegraphics[scale=0.22]{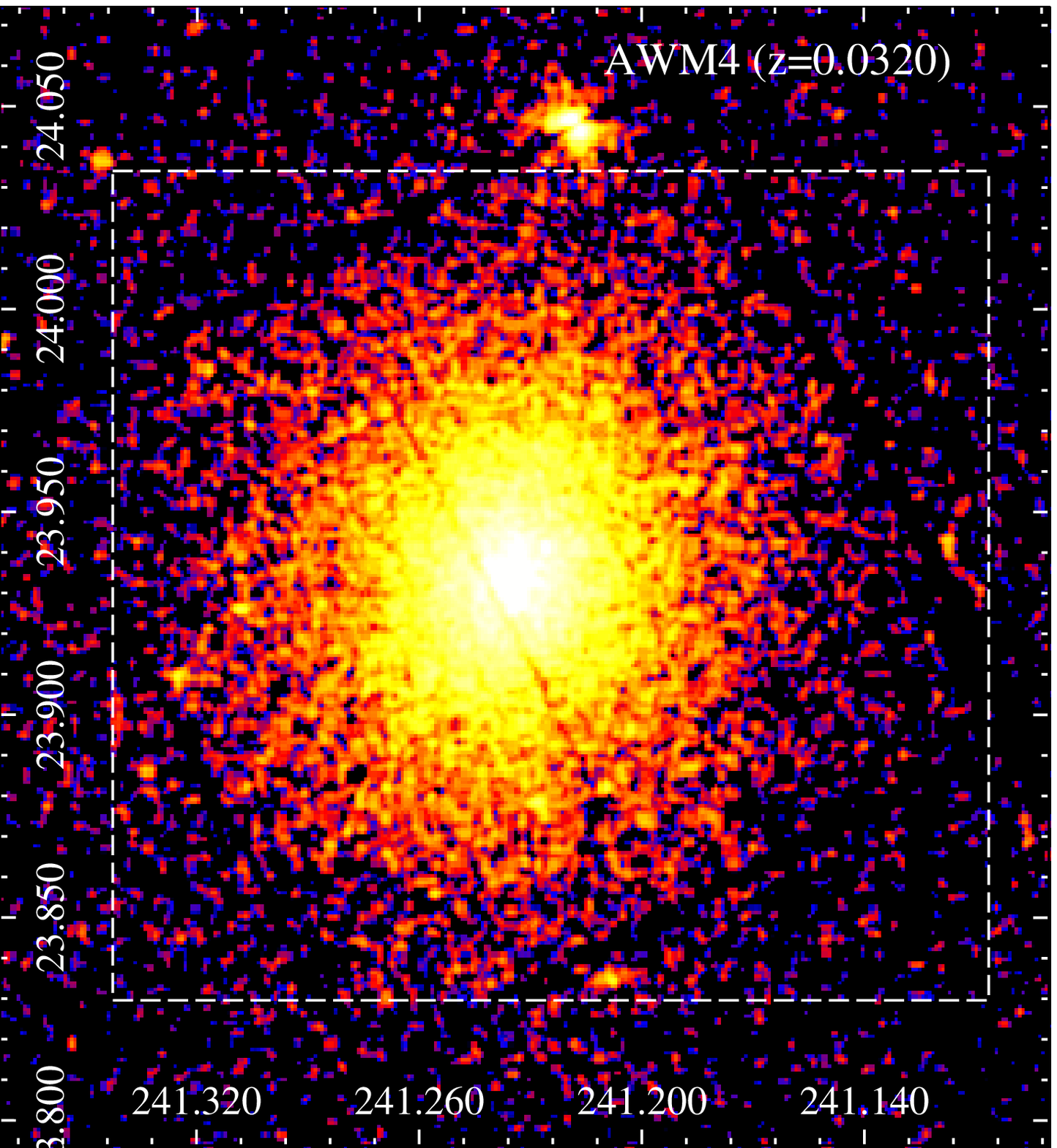}
\includegraphics[scale=0.22]{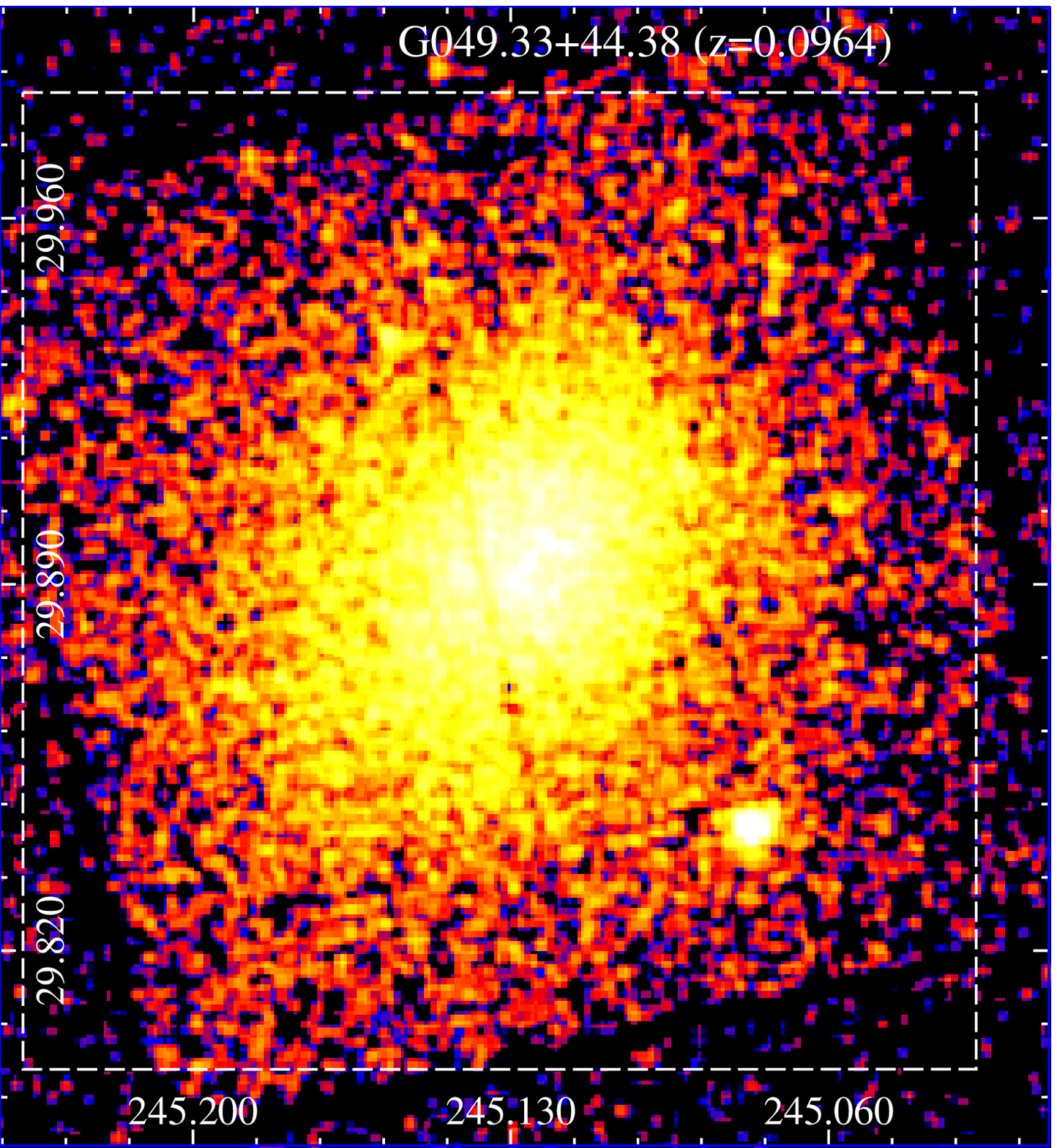}
\includegraphics[scale=0.22]{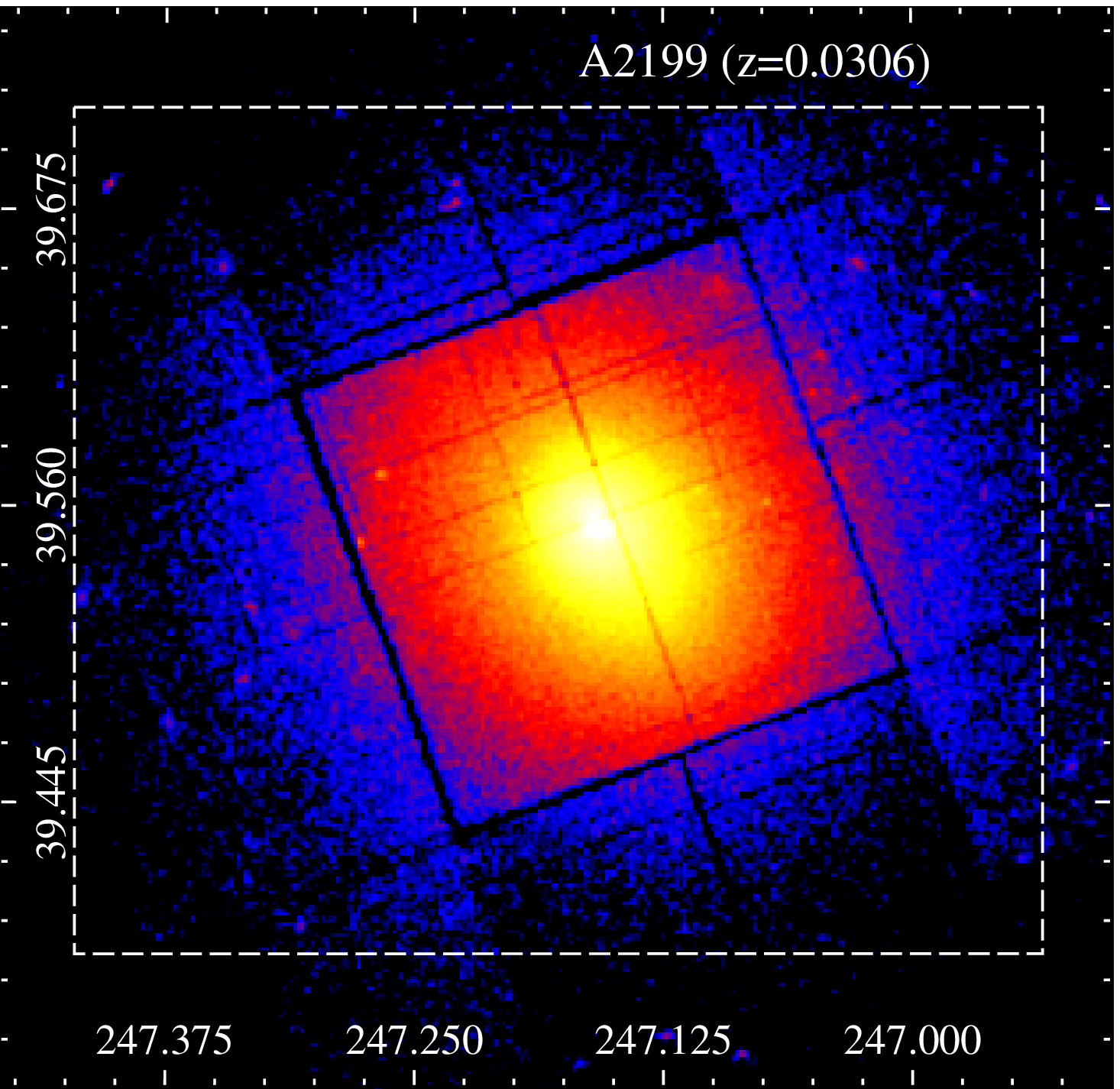}
\includegraphics[scale=0.22]{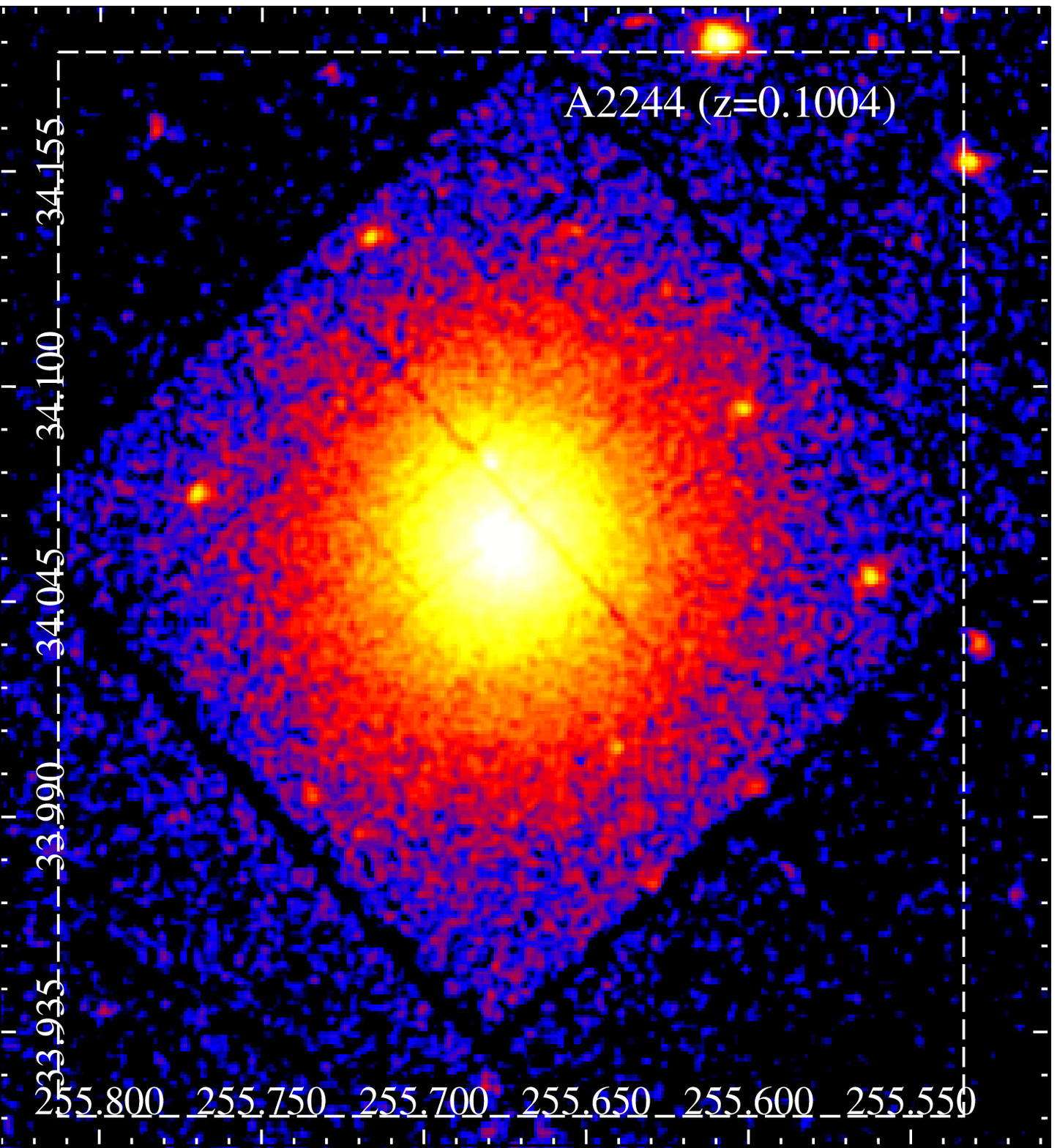}

\includegraphics[scale=0.22]{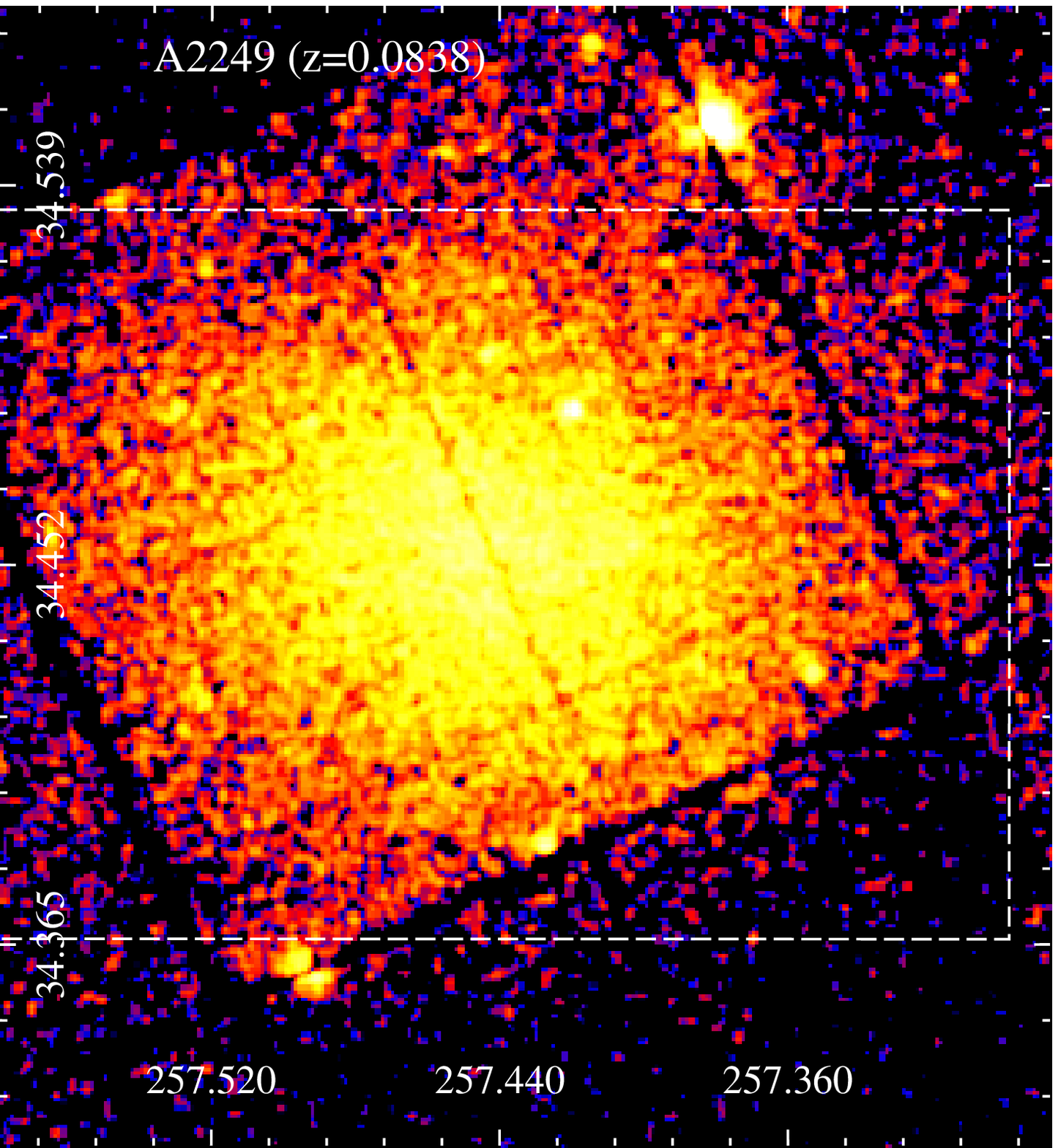}
\includegraphics[scale=0.22]{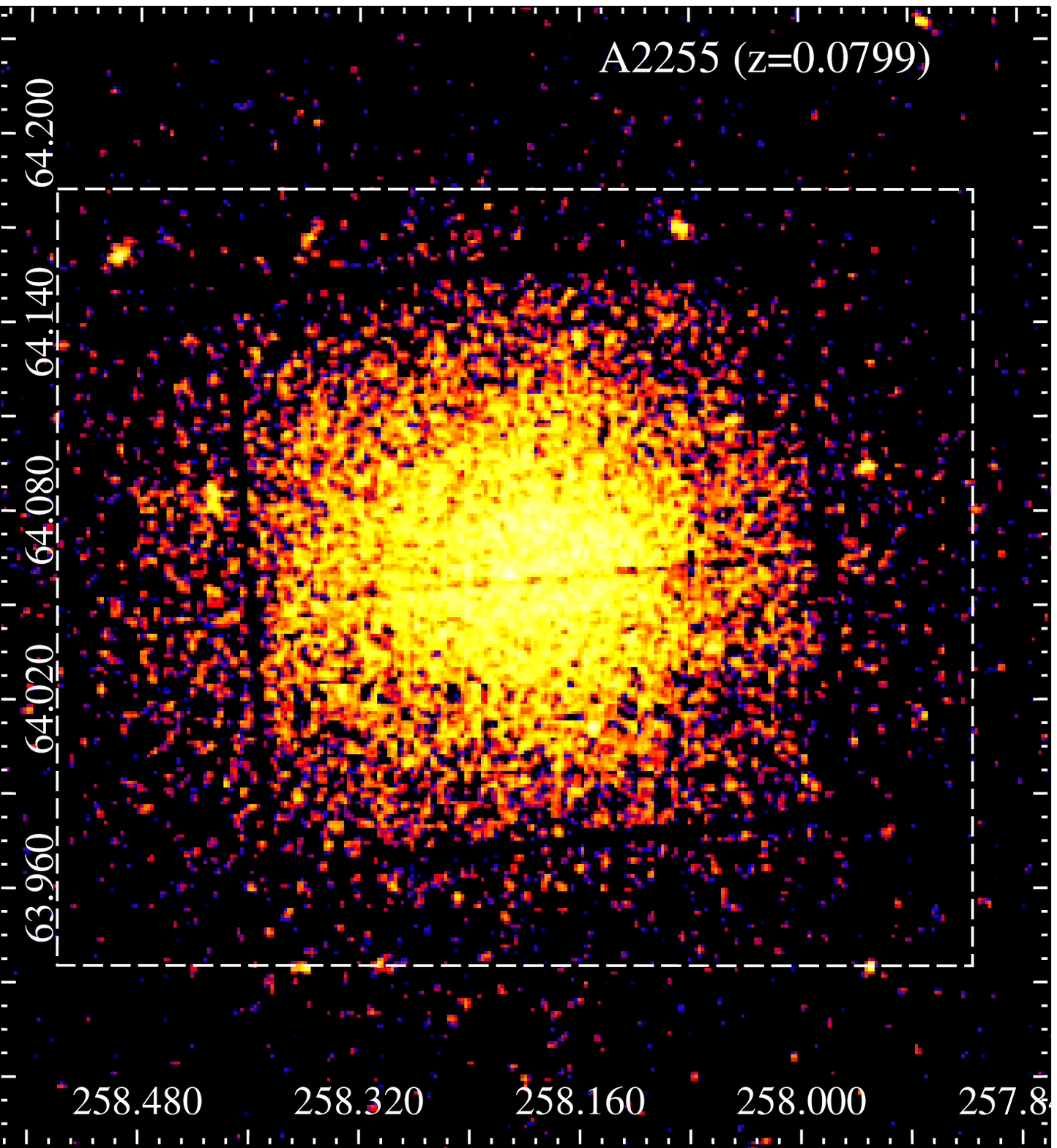}
\includegraphics[scale=0.22]{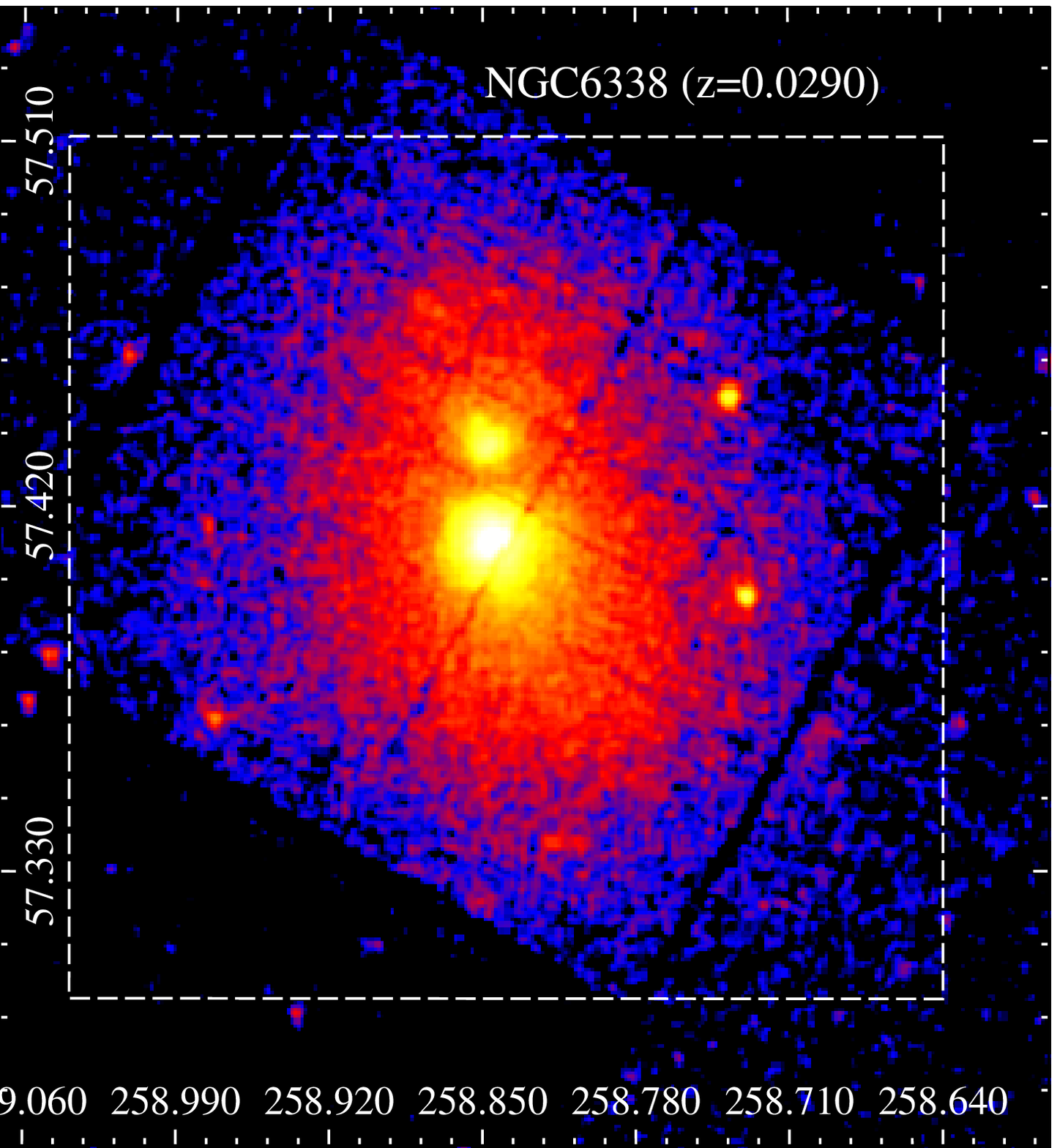}
\includegraphics[scale=0.22]{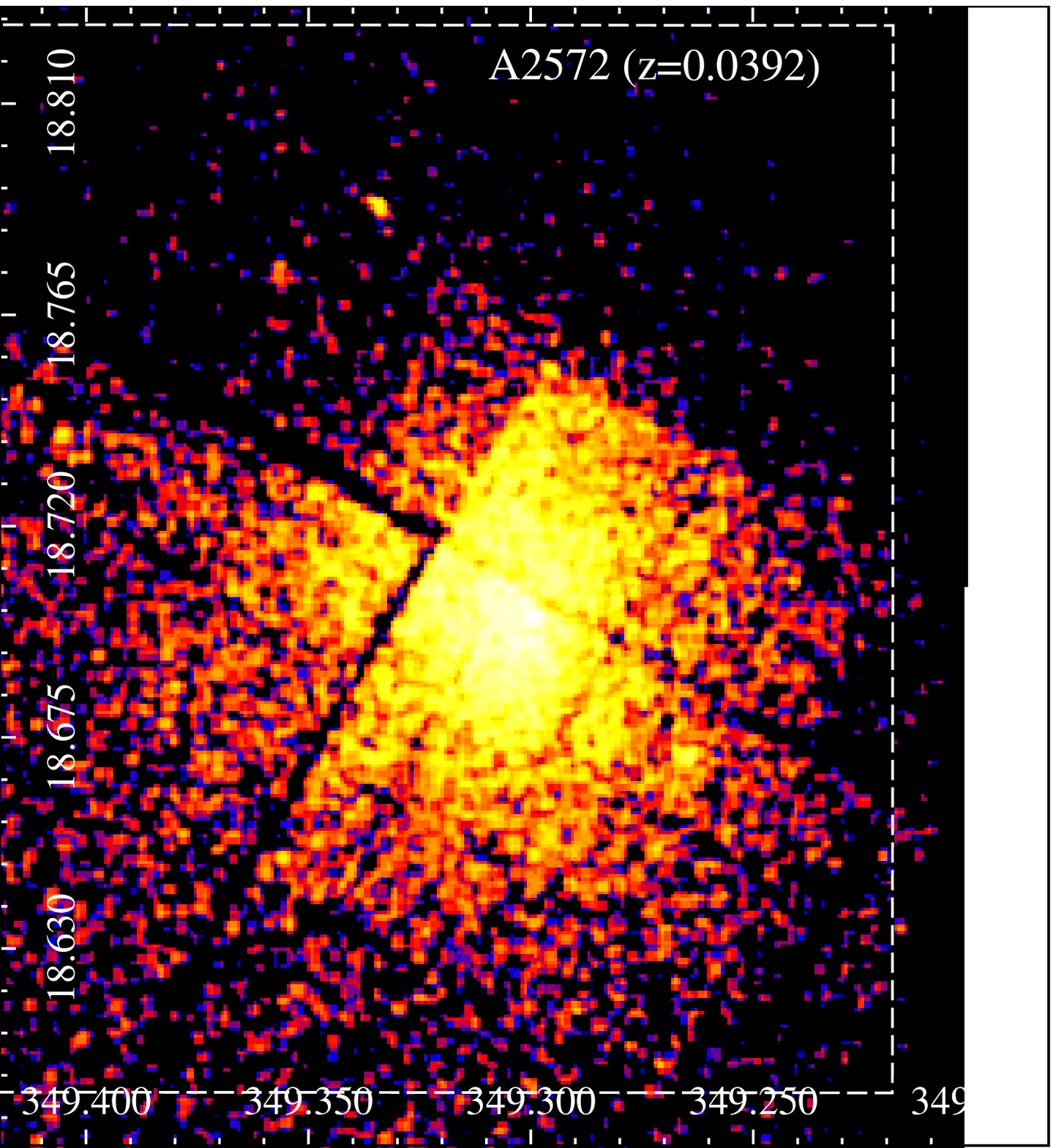}
\includegraphics[scale=0.22]{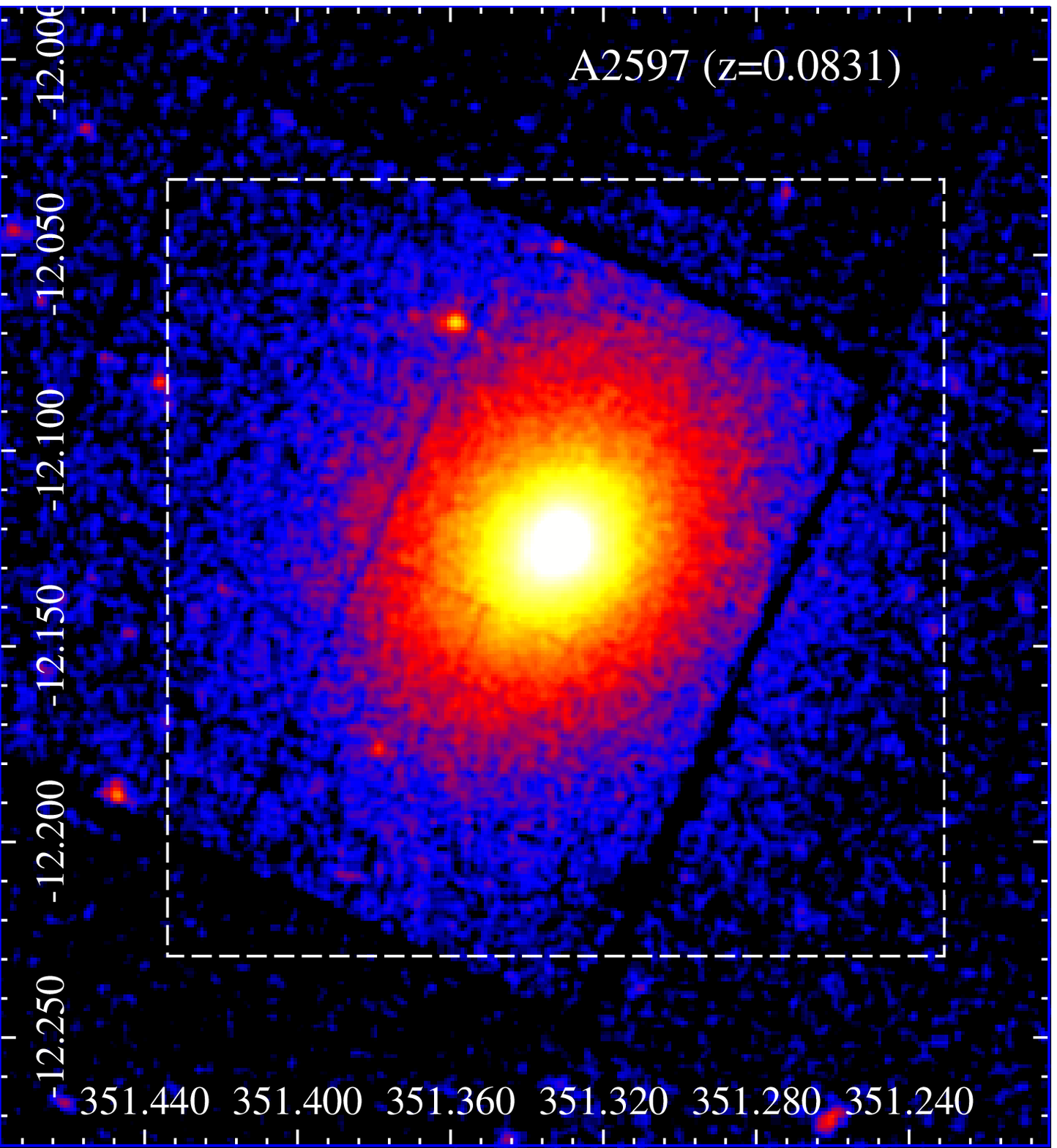}

\includegraphics[scale=0.22]{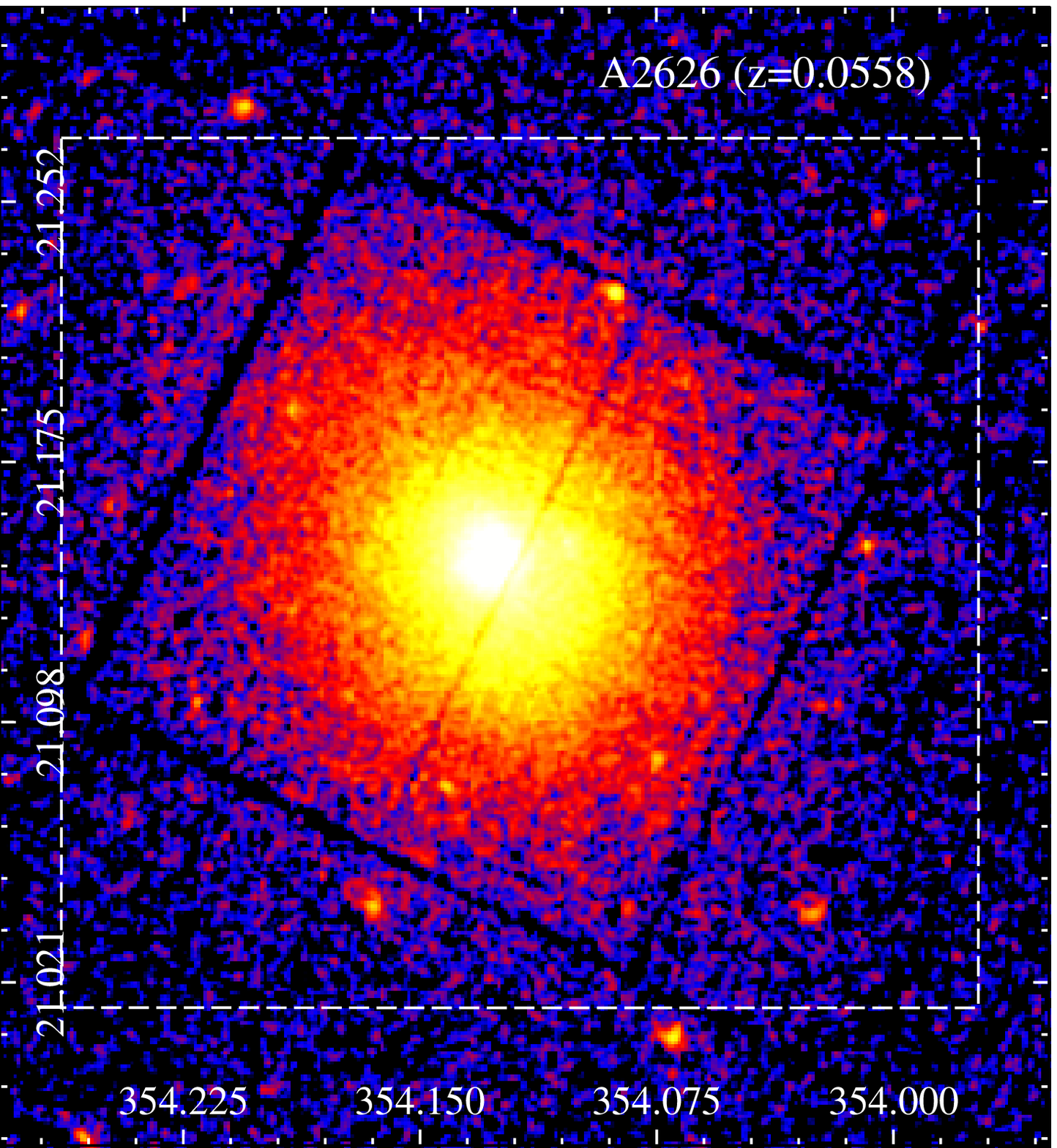}
\includegraphics[scale=0.22]{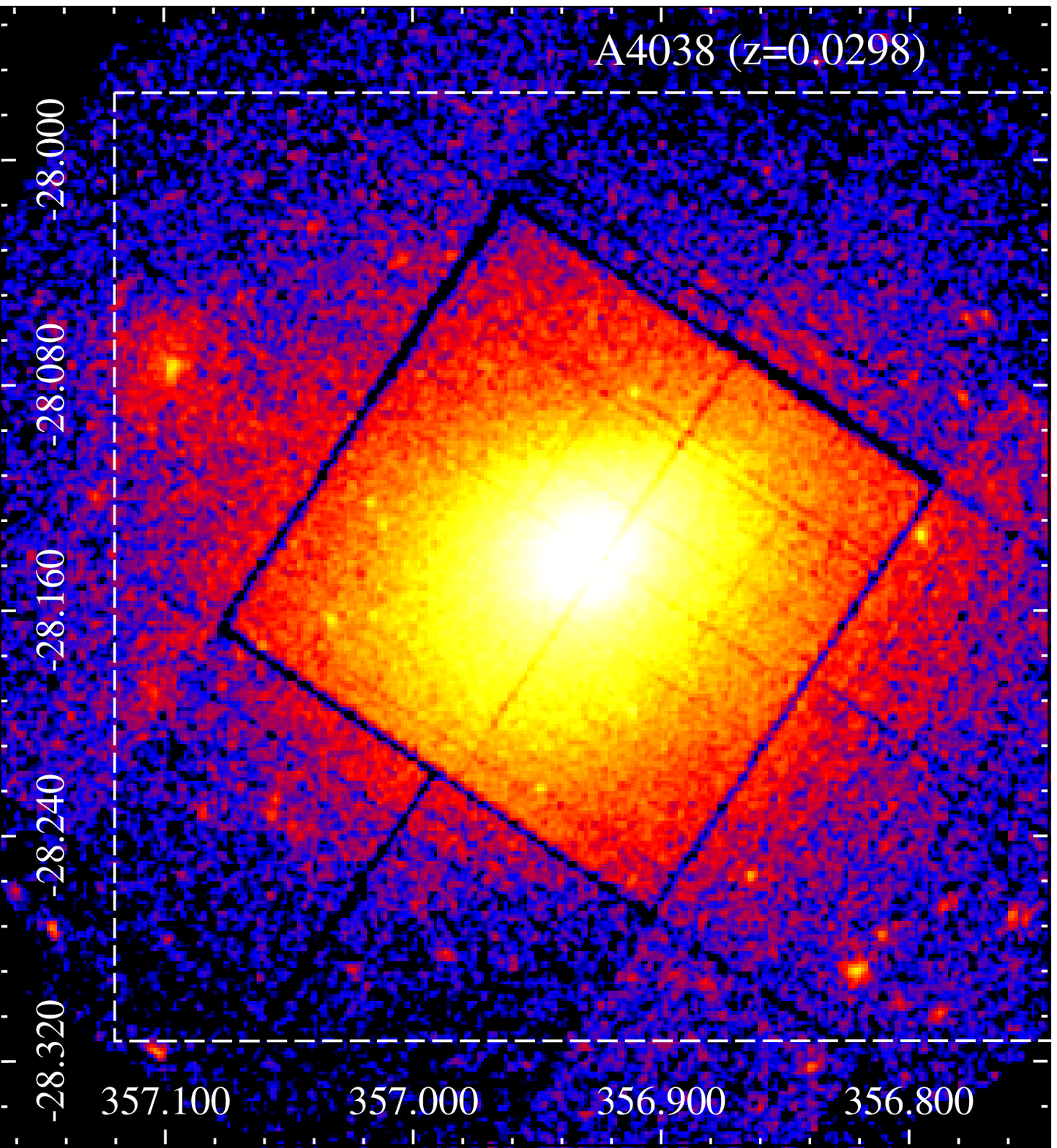}
\includegraphics[scale=0.22]{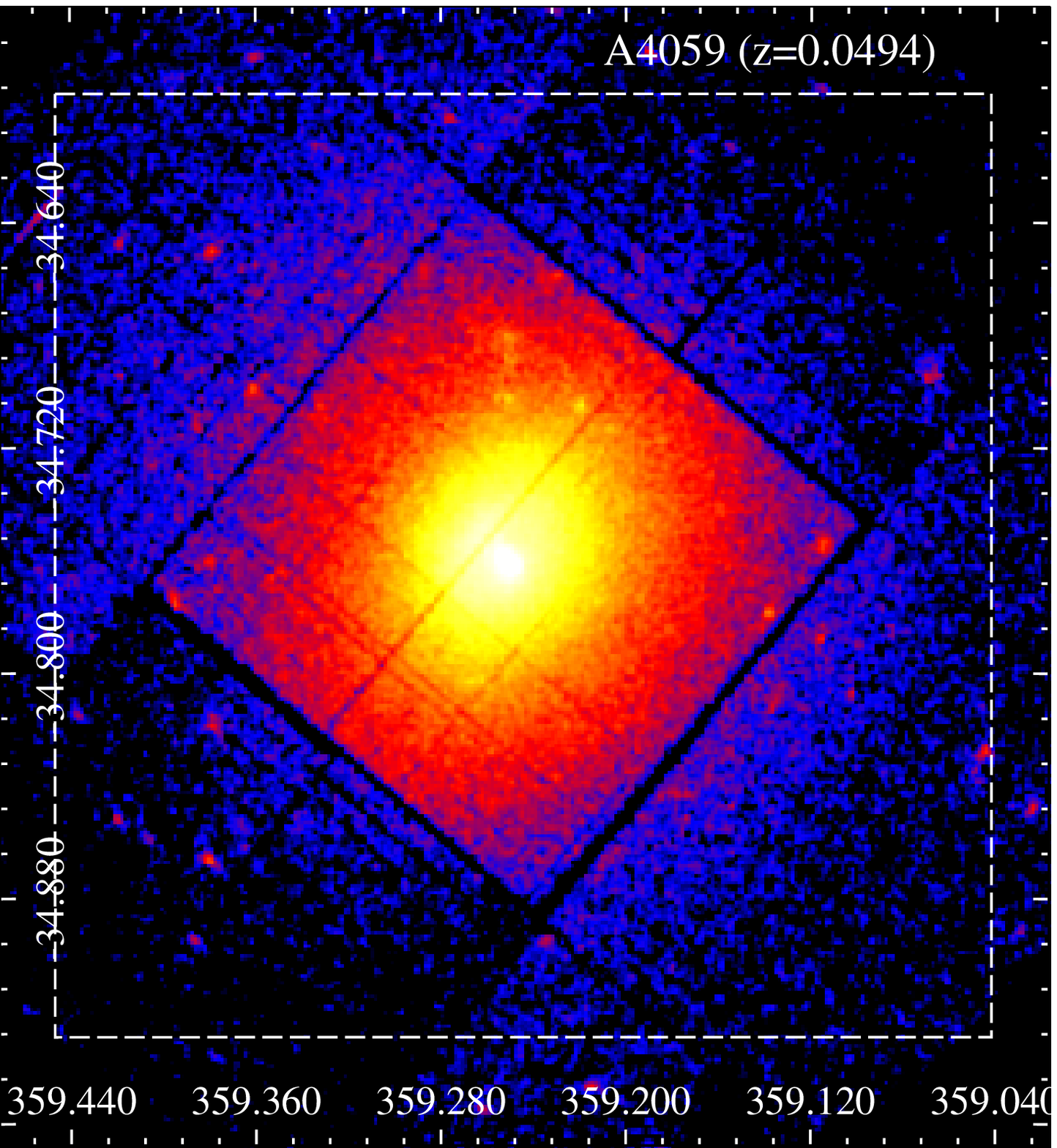}

%\label{fig:sample}
\caption{X-ray images of the cluster sample in the same order as in Tab.~\ref{tab:obs}. The white dashed boxes represent the 
regions used to construct the 2D maps.}
\end{figure*}

\renewcommand{\thefigure}{\arabic{figure}}

\subsection{2D Spectral Maps}

We have derived 2D temperature (kT), metallicity (Z), pseudo-entropy
(${\rm S \propto kT/I^{1/3}}$, where $I$ is the Intensity in net
counts/arcsec$^{2}$) and pseudo-pressure
(${\rm P \propto kT \times I^{1/2}}$) maps for these 53 clusters. This
was achieved by dividing the data into small regions from which
spectra can be extracted.  The 2D maps were made in a grid, where each
pixel is 512 $\times$ 512 XMM-{\it Newton}  EPIC physical pixels.  In each pixel we
set a minimum count number of 1500 (after background subtraction,
corresponding to a S/N of almost 40), necessary for obtaining a
spectral fit.  If we do not reach the minimum count number in a pixel,
we try a square region of 3 $\times$ 3 pixels; and if we still do not
have the minimum number of counts, we try a 5 $\times$ 5 pixel
region. If we still do not have enough counts, the pixel is ignored
and we proceed to the next neighbouring pixel. This is done for all
the pixels in the grid. When we have enough counts, the spectra of the
three EPIC instruments (MOS1, MOS2 and pn) are then simultaneously
fitted as described above, and the best temperature and metallicity
values are attributed to the central pixel.  This procedure
\citep[already described in][]{Durret10, Durret11,Lagana15} allows us
to perform a reliable spectral analysis in each spectral bin, in order
to derive kT, Z, S and P maps.

\section{Results}
\label{sect:res}

Besides possible deviations from symmetry, 2D maps can reveal complex structures.
The temperature maps are commonly used in studies of galaxy clusters
and groups since they trace the recent dynamical history of the
system.
 
The entropy map for a system in equilibrium should be symmetrical
around the centre and exhibit increasing values towards the outskirts
of the cluster. Low-entropy gas can be displaced from the centre
either due to instabilities like cold-fronts \citep[e.g.,][]{Mark07} or
to gas stripping \citep{Finog04}. Areas of high entropy gas are
produced by local heating, most likely AGNs.
 
The pressure structure of the ICM is sensitive to the combined action
of gravitational and non-gravitational physical processes affecting
galaxy clusters.The ICM pressure distribution is directly connected to
the cluster potential well, and thus to the cluster total mass.
Pressure fluctuations trace departure from local equilibrium, such as
that produced by shocks \citep{Mark02,Simionescu09} and pressure waves
\citep{Fabian03,Schu04}.
 
To date, X-ray and SZ observations of galaxy clusters, in line with
the results from numerical simulations \citep[e.g.,][]{Piffaretti08},
suggest that the ICM radial pressure profiles follow a nearly universal
shape within $R_{500}$ \citep{Arnaud10}.

The distribution of heavy elements in the ICM provides important information on the relevance of mergers, 
AGN outbursts and ram-pressure stripping. 
In recent years it has been possible to derive metallicity profiles for a number of clusters and those profiles 
correlate with the CC status. 
It has been found that CC systems display a centrally peaked metallicity distribution, while NCC clusters are 
characterised by a nearly flat distribution
\citep{DeGrandi01, Leccardi10}.

We use four diagnoses from \citet{AndradeSantos17} to identify CC
clusters: (i) the concentration parameter in the (0.15-1.0)$r_{500}$
range (CSB), (ii) the ratio of the integrated projected emissivity
profile within 40~kpc to that within 400 kpc (i. e., the concentration parameter in the 40-400 kpc range, CSB4), 
(iii) the cuspiness of the gas density profile ($\delta$), and (iv) the central gas density
($n_{\rm core}$). We
also include two more criteria from \citet{Lopes18}: (v) the
magnitude gap between the first and second BCG ($\Delta m_{12}$), and
(vi) the BCG offset to the X-ray centre.
 
Adopting the values presented in \citet{Lopes18}, we consider as CC, as shown 
in Tab.~\ref{tab:CC},  the systems with:
\begin{enumerate}
\item{$CSB > 0.26$}
\item{$CSB4 > 0.055$}
\item{$\delta > 0.46$}
\item {$n_{\rm core} > 8 \times 10^{-3} \rm cm^{-3}$}
\item{$\Delta m_{12} > 1.0$}
\item{offset < 0.01 $R_{500}$}
 \end{enumerate}
 
 By visual inspection of our 2D maps, we classify the clusters in
 relaxed versus disturbed systems to correlate this classification
 with the different methods used to separate clusters in CC or NCC
 systems. We defined a relaxed cluster as having round or elliptical
 isophotes, little or no sloshing or cold fronts, and no signs of a
 recent merger.  We thus divided our sample in four different groups,
 as shown in Tab.~\ref{tab:CC}
\begin{itemize}
\item {{\scshape CC-relaxed}: systems where at least four out of six criteria
     classify the cluster as CC and
    which appear relaxed in our 2D map inspection;}
\item{{\scshape CC-disturbed}: systems where at least four out of six criteria
    characterize the system as CC, but with spectral maps
    appearing disturbed;}
\item{{\scshape NCC-relaxed}: systems classified as NCC by at least three of the six criteria and which 
appear relaxed in our 2D map inspection;}   
\item{{\scshape NCC-disturbed}: systems classified as NCC by at least three of the six criteria and showing clear signs
    of perturbation in the maps.}

 \end{itemize}
 
\begin{table*}
	\centering
	\caption{Classification of each cluster in CC or NCC according to four diagnoses from \citet{AndradeSantos17}, two from \citet{Lopes18},
	and in relaxed or disturbed according to our 2D map analysis}. Columns list the cluster name, CSB, CSB4, $\delta$, $n_{\rm core}$, $\Delta m_{12}$, the BCG offset to the X-ray centre (see text), and the dynamical state according to the 2D maps.
	\label{tab:CC}
	\begin{tabular}{lccccccc} % four columns, alignment for each
		\hline
		Cluster &  CSB & CSB4 & $\delta$ & $n_{\rm core}$ &$\Delta m_{12}$ & Offset & 2D maps\\
\hline
 \multicolumn{8}{c} {CC - relaxed}\\
\hline	
A2734        		&	CC	&	CC	&	CC	&	CC	&	CC	&	CC	&	relaxed	\\
EXO0422       		&	CC	&	CC	&	CC	&	CC	&		&	CC	&	relaxed	\\
S0540        		&	CC	&	CC	&	CC	&	CC	&	CC	&		&	relaxed	\\
G269.51+26.42 		&	CC	&	CC	&	CC	&		&		&	CC	&	relaxed	\\
A3528n      	  	&	CC	&	CC	&	CC	&	CC	&	CC	&	CC	&	relaxed	\\
A3528s        		&	CC	&	CC	&	CC	&	CC	&	CC	&	CC	&	relaxed	\\
A1650     	    	&	CC	&	CC	&	CC	&	CC	&	CC	&	CC	&	relaxed	\\
A3571       	 	&	CC	&	CC	&	CC	&	CC	&	CC	&	CC	&	relaxed	\\
A1795       	 	&	CC	&	CC	&	CC	&	CC	&	CC	&	CC	&	relaxed	\\
A2052       	 	&	CC	&	CC	&	CC	&	CC	&	CC	&	CC	&	relaxed	\\
A2063   	     	&	CC	&	CC	&	CC	&	CC	&	CC	&	CC	&	relaxed	\\
A2151         		&	CC	&	CC	&	CC	&	CC	&		&	CC	& 	relaxed	\\
A2244  	      		&	CC	&	CC	&	CC	&	CC	&	CC	&	CC	&	relaxed	\\
A2572    	    	&	CC	&	CC	&	CC	&	CC	&		&		&	relaxed	\\
A2597        	 	&	CC	&	CC	&	CC	&	CC	&	CC	&	CC	&	relaxed	\\
A2626       	  	&	CC	&	CC	&	CC	&	CC	&	CC	&	CC	&	relaxed	\\
A4059      	   		&	CC	&	CC	&	CC	&	CC	&	CC	&	CC	&	relaxed	\\
\hline
\multicolumn{8}{c} {CC - disturbed} \\
\hline
A85           		&	CC	&	CC	&	CC	&	 CC	&	CC	&	CC	& disturbed  \\
A496          		&	CC	&	CC	&	CC	&	CC	&	CC	&	CC	& disturbed  \\
USGCS152      		&	CC	&	CC	&	CC	&	CC	&	CC	&	CC	& disturbed	\\
A1644n         		&		&	CC	&	CC	&	CC	&	CC	&	CC	& disturbed \\
A1644s         		&		&	CC	&	CC	&	CC	&	CC	&	CC	& disturbed \\
A1651         		&	CC	&	CC	&	CC	&		&	CC	&		& disturbed	\\
A3558         		&	CC	&	CC	&	CC	&	CC	&	CC	&	CC	& disturbed	\\
A3562         		&		&	CC	&	CC	&		&	CC	&	CC	& disturbed	\\
A1775         		&	CC	&	CC	&	CC	&	CC	&	CC	&	CC	& disturbed \\
A2029         		&	CC	&	CC	&	CC	&	CC	&	CC	&	CC	& disturbed \\
A2142         		&	CC	&	CC	&	CC	&	CC	&		&		& disturbed	\\
AWM4          		&	CC	&	CC	&	CC	&	CC	&	CC	&	CC	&	disturbed	\\
A2199         		&	CC	&	CC	&	CC	&	CC	&		&	CC	& disturbed	\\
NGC6338       		&	CC	&	CC	&	CC	&	CC	&		&	CC	& disturbed	\\
bNGC6338       		&	CC	&	CC	&	CC	&	CC	&		&	CC	& disturbed	\\
A4038         		&	CC	&	CC	&	CC	&	CC	&	CC	&		& disturbed	\\
\hline
 \multicolumn{8} {c} {NCC - relaxed} \\
\hline
Coma 		&		&		&		& 	CC	&		&		& relaxed	\\
A3532				&		&		&		&		&		&	CC	& relaxed	 \\
bA3558				& 		&  	CC 	& 	CC  &   CC 	&		&		& relaxed	\\
MKW8          		&	CC	&	CC	&		&		&		&	CC		& relaxed	\\
\hline
 \multicolumn{8} {c} {NCC - disturbed} \\
\hline
A119          		&		&		&		&		&		&		& disturbed	\\
A3376         		&		&		&		&		&		&		& disturbed	\\
A3391         		&		&		&		&		&		&	CC	& disturbed	\\
A3395      			&		&		&	CC	&		&		&	CC	& disturbed	\\
bA3395      		&		&		&	CC	&		&		&	CC	& disturbed	\\
A754         		&	CC	&	CC	&	CC	&		&		&		& disturbed	\\
A1367			&		&		&		&		&		&	CC	& disturbed \\
ZwCl1215      		&	CC	&		&		&		&		&	CC	& disturbed	\\
A3560         		&		&		&		&		&		&		& disturbed	\\
A2061         		&		&		&		&		&		&		& disturbed	\\
MKW3s				&	CC	&		&	CC	&	CC	&		&	& disturbed \\
A2065         		&		&		&		&	CC	&		&		& disturbed	\\
A2147         		&		&		&		&		&		&		& disturbed	\\
G049.33+44.38 	&		&		&		&		&	CC	&		& disturbed	\\
A2249			&    	&		&		&		&		&	CC		& disturbed  \\
A2255         		&		&		&		&		&		&		& disturbed	\\
\hline
\end{tabular}
\end{table*}

\begin{figure*}
\includegraphics[scale=0.25]{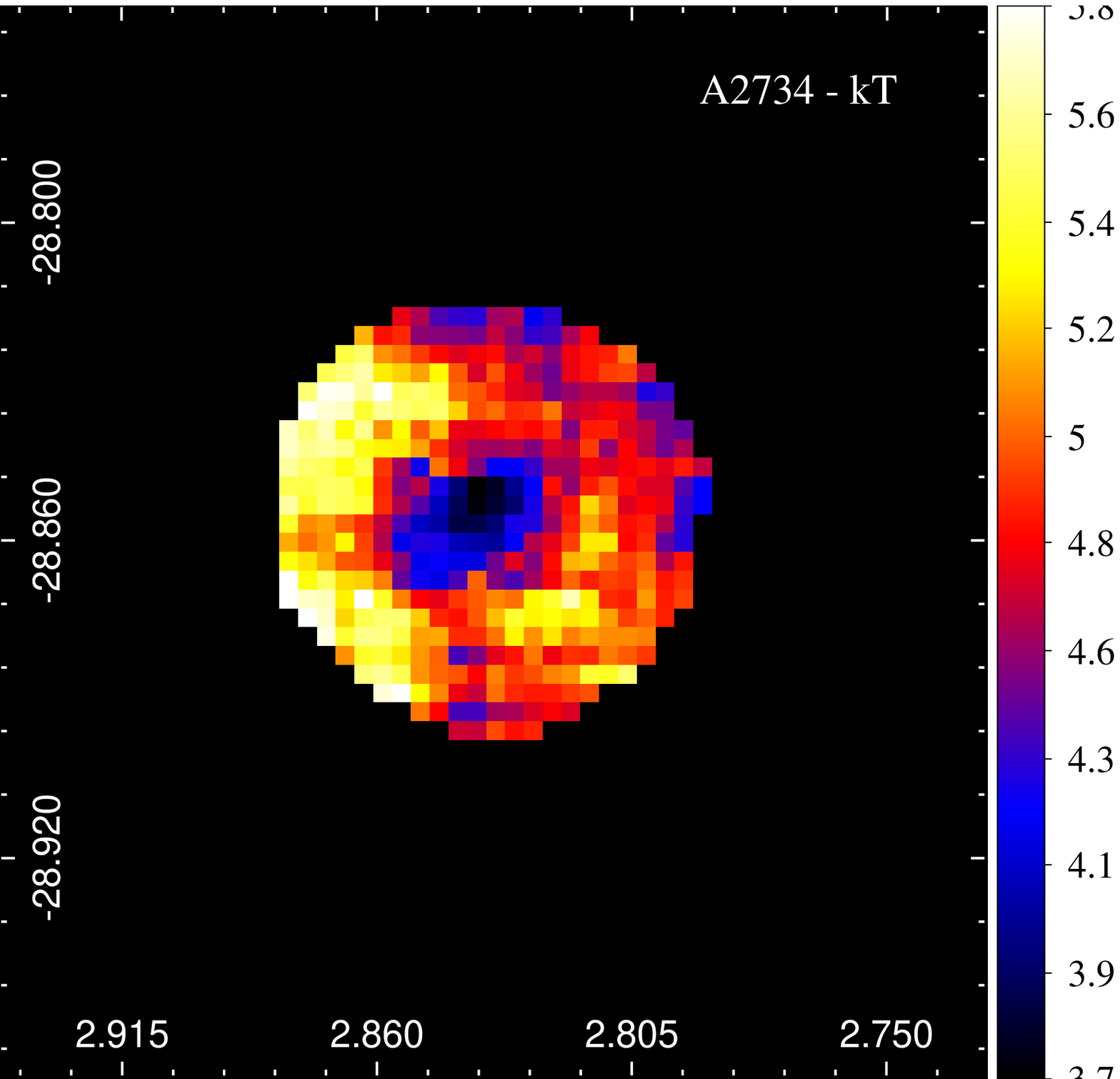}
\includegraphics[scale=0.25]{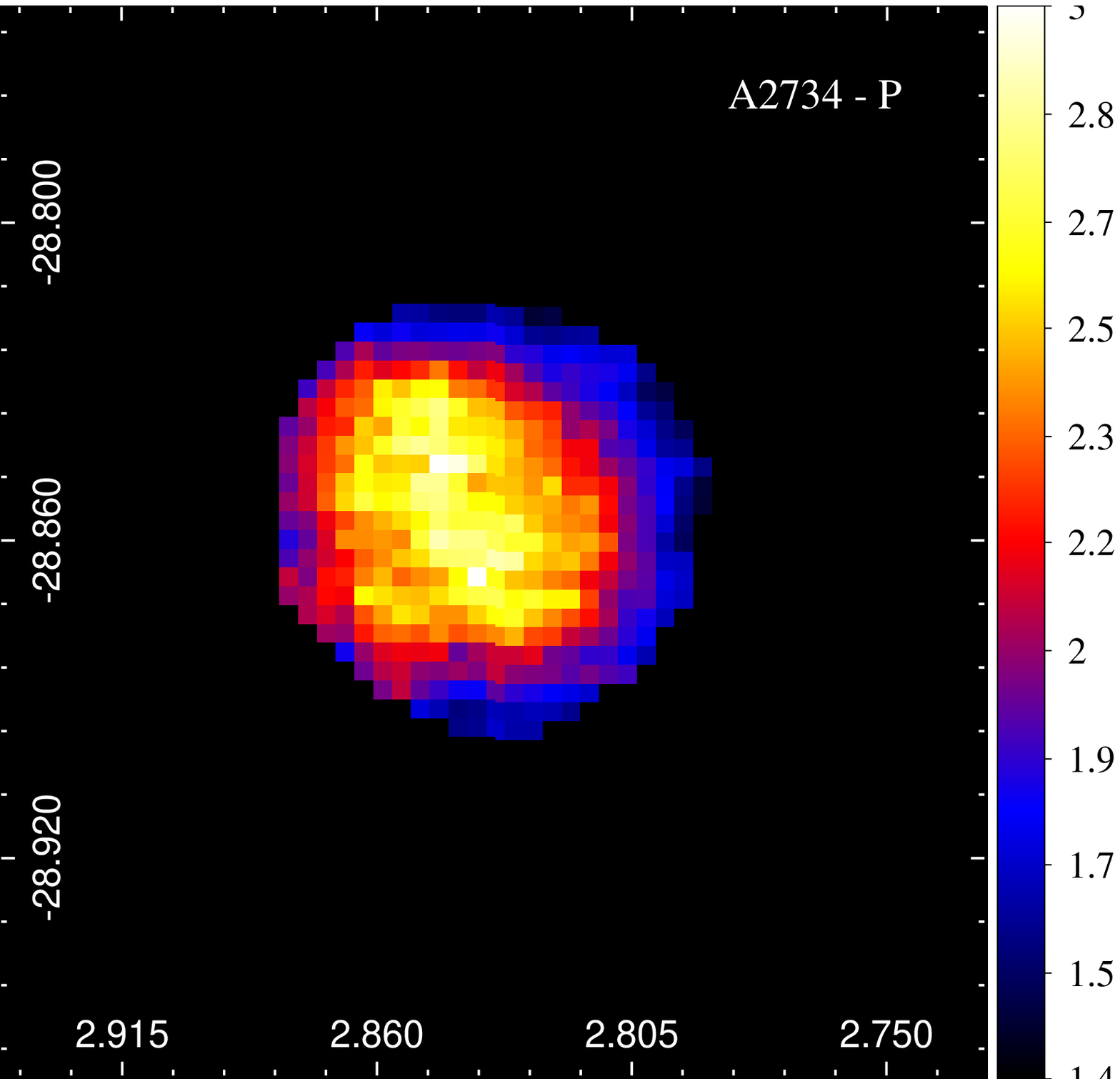}
\includegraphics[scale=0.25]{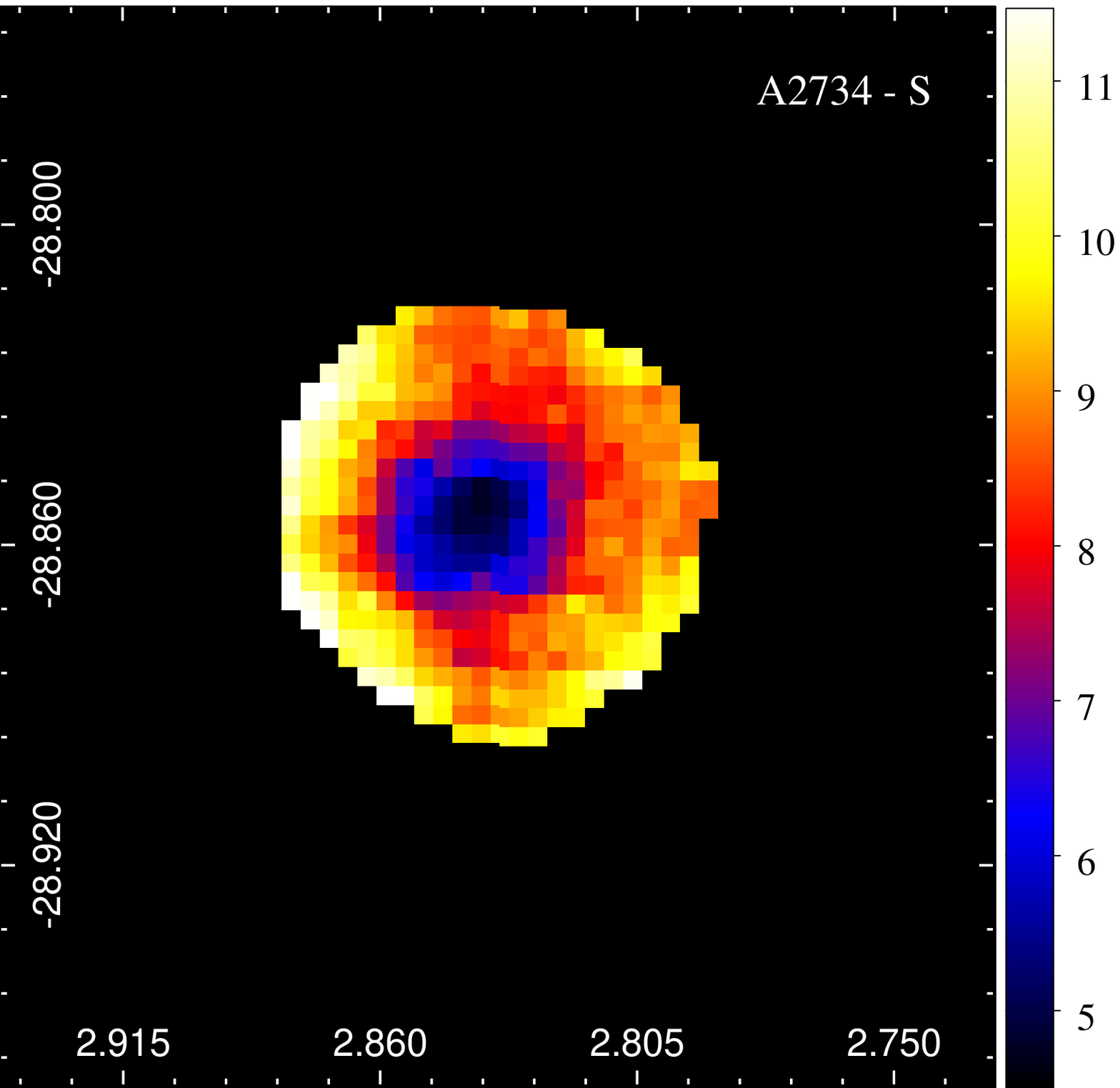}
\includegraphics[scale=0.25]{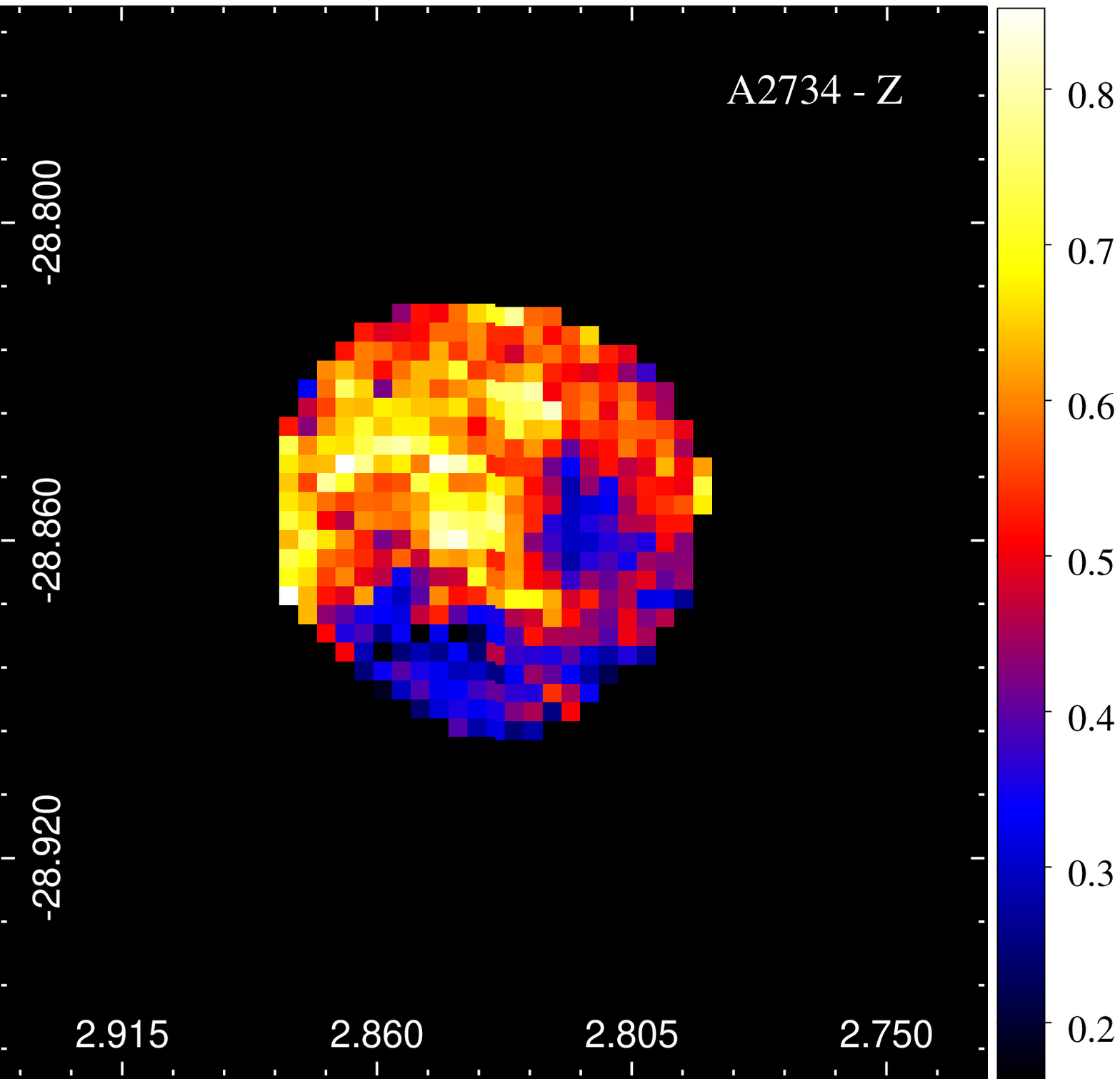}

\includegraphics[scale=0.25]{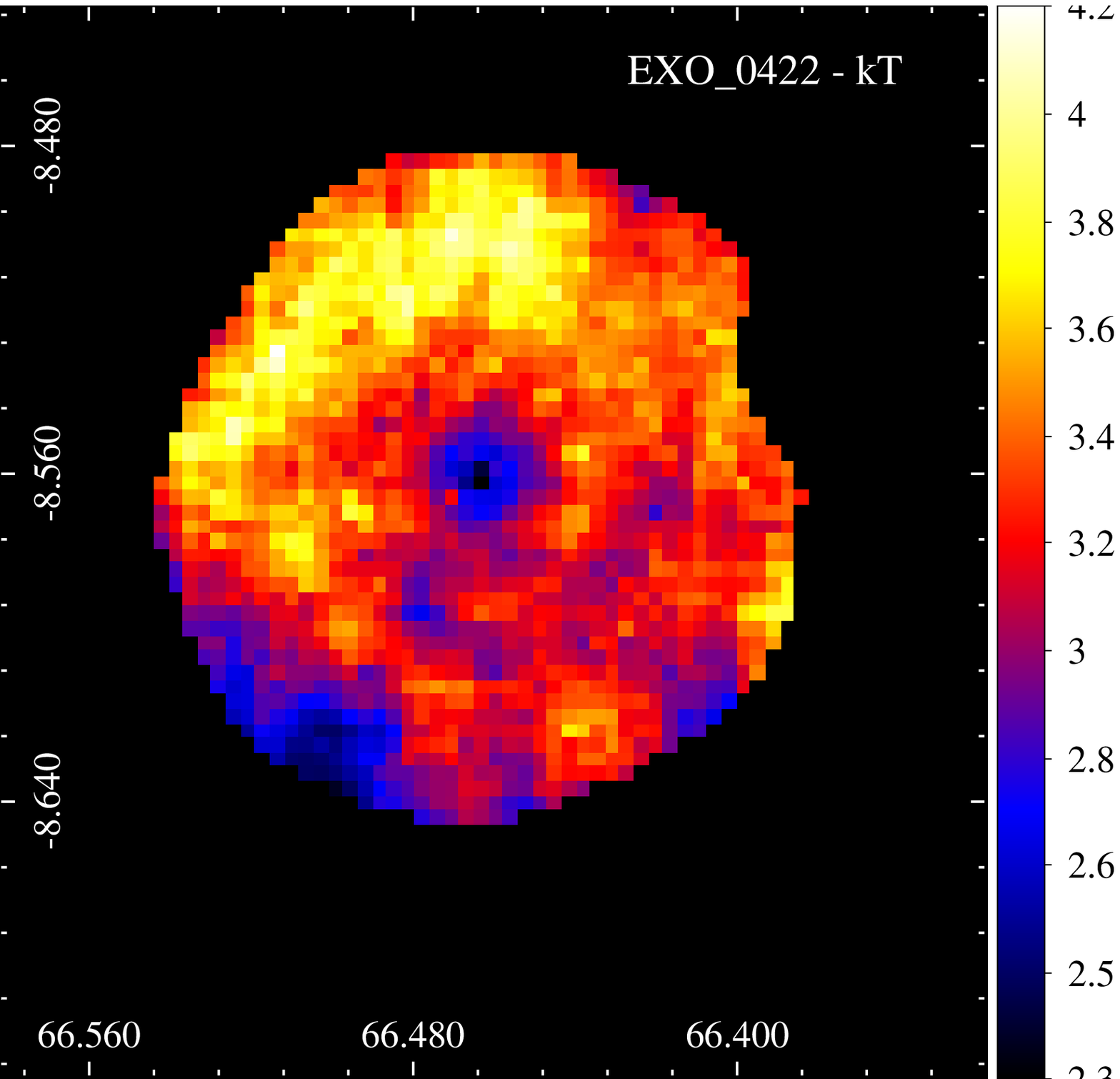}
\includegraphics[scale=0.25]{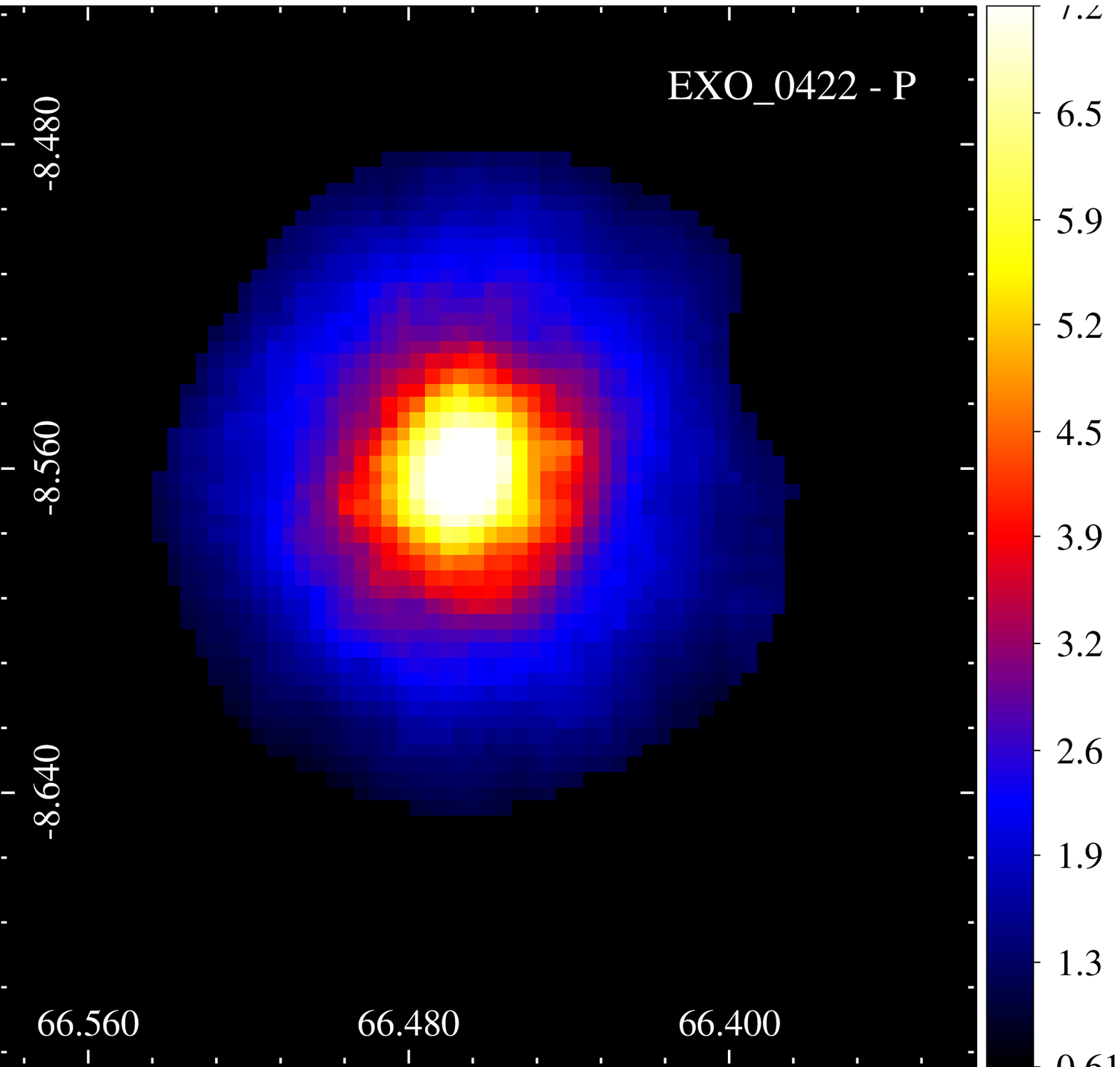}
\includegraphics[scale=0.25]{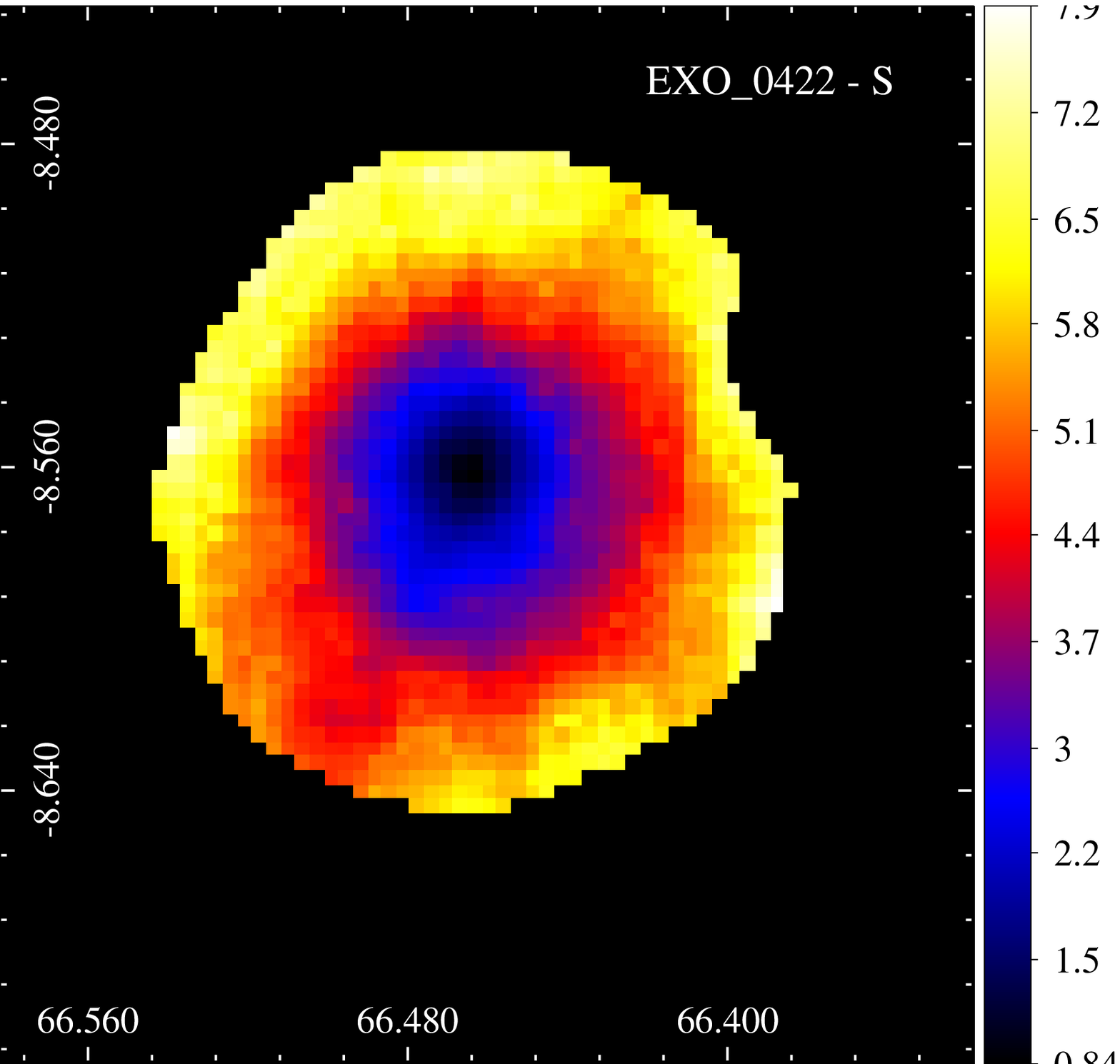}
\includegraphics[scale=0.25]{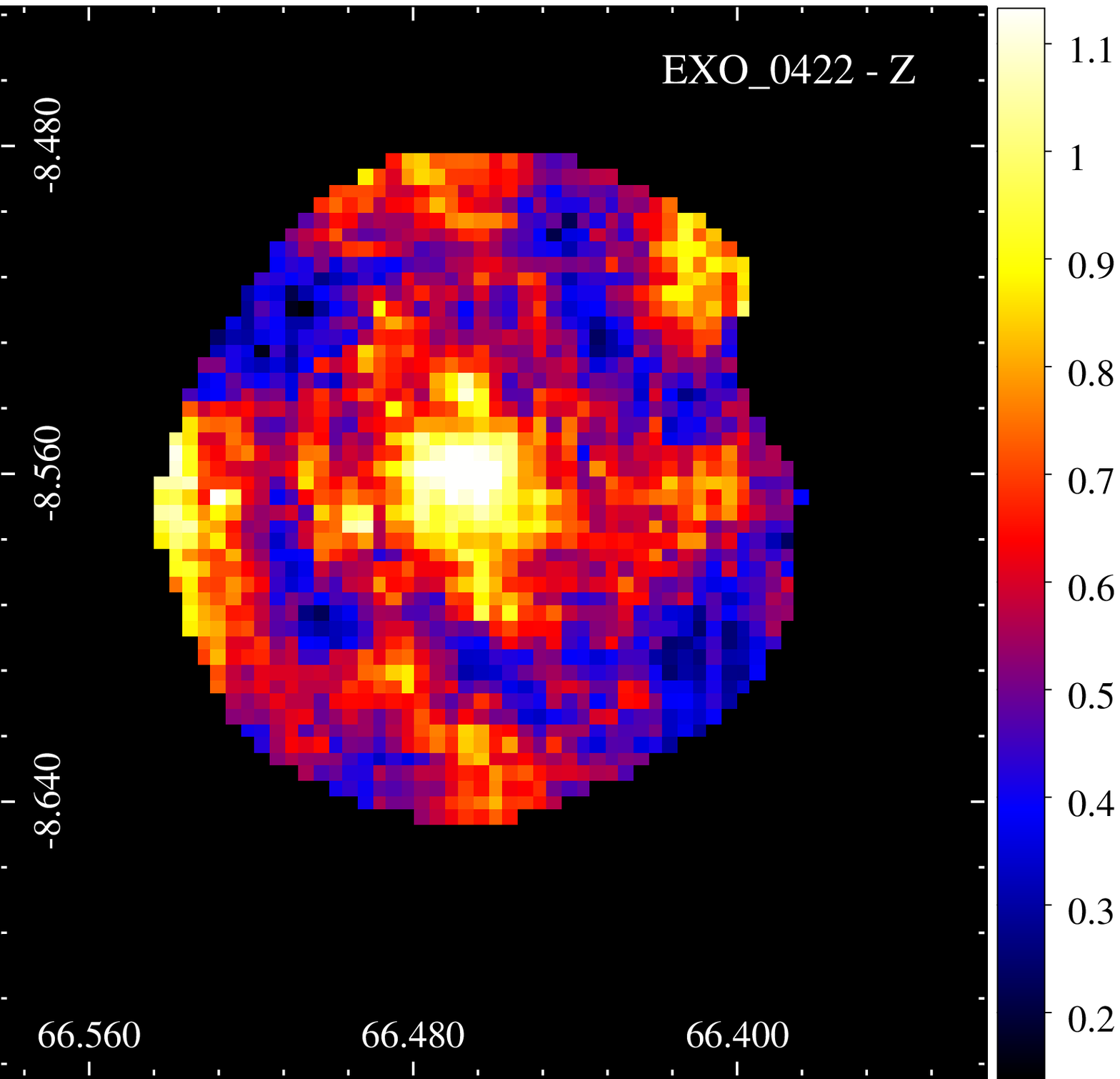}

\includegraphics[scale=0.25]{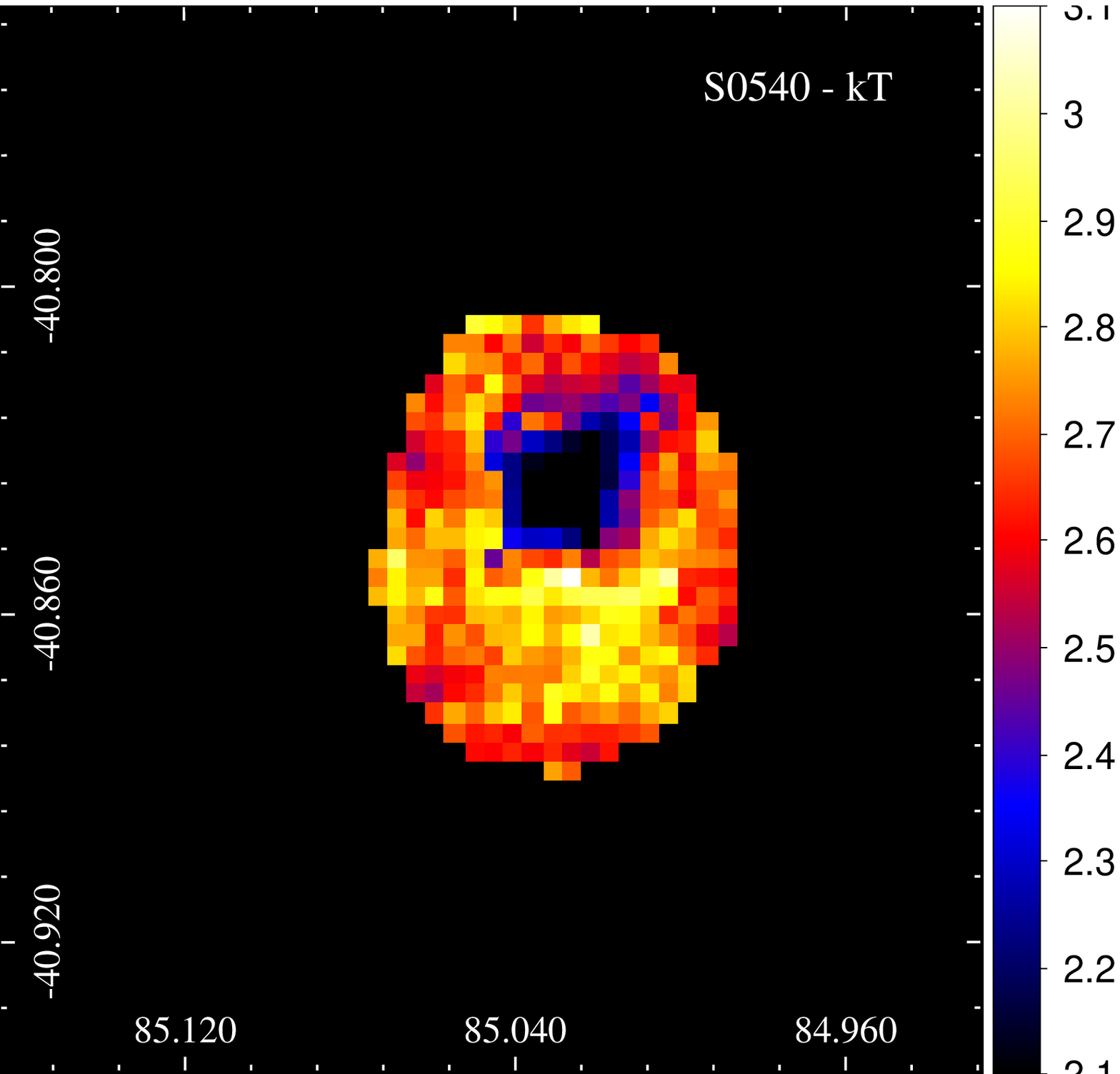}
\includegraphics[scale=0.25]{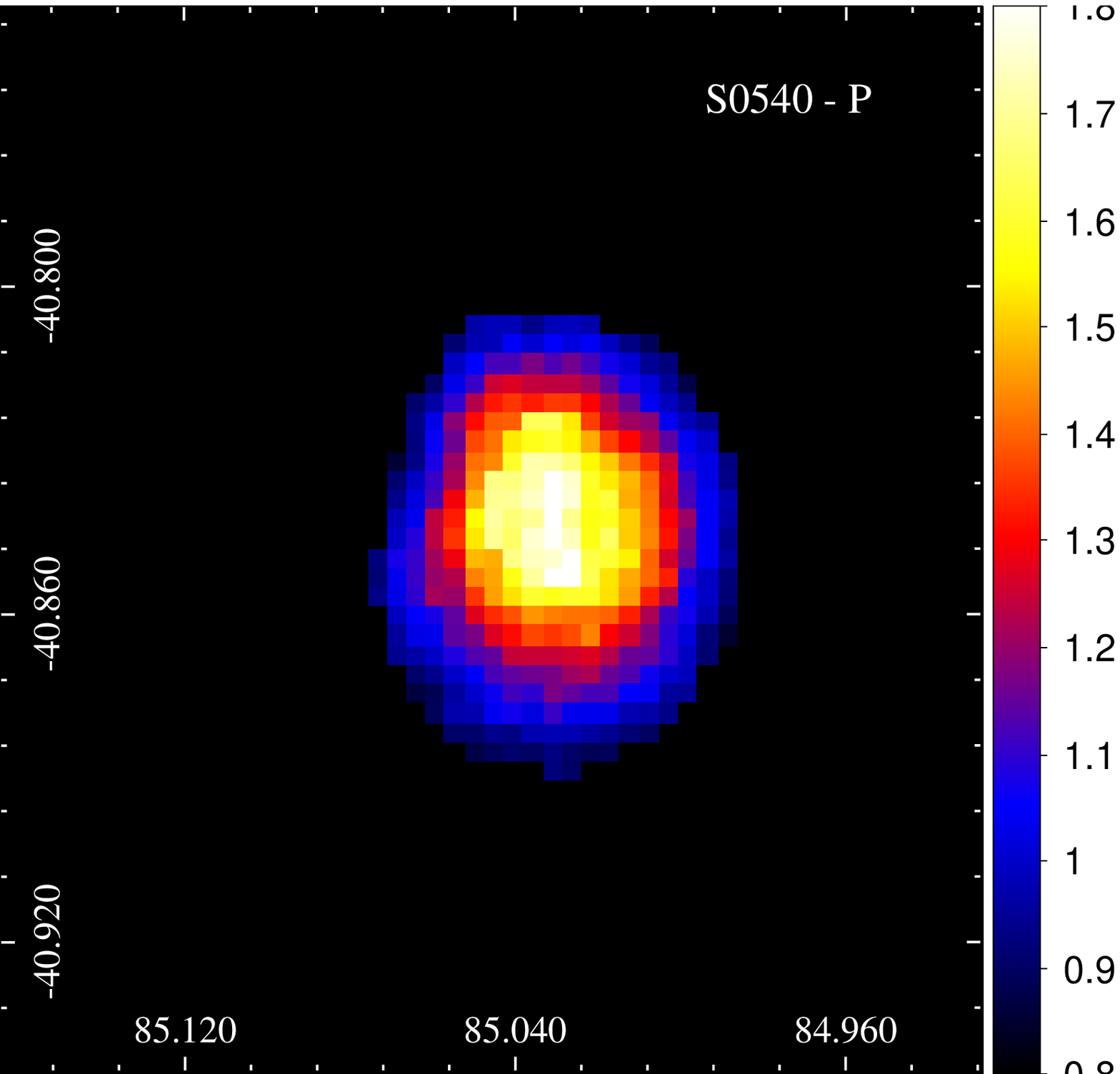}
\includegraphics[scale=0.25]{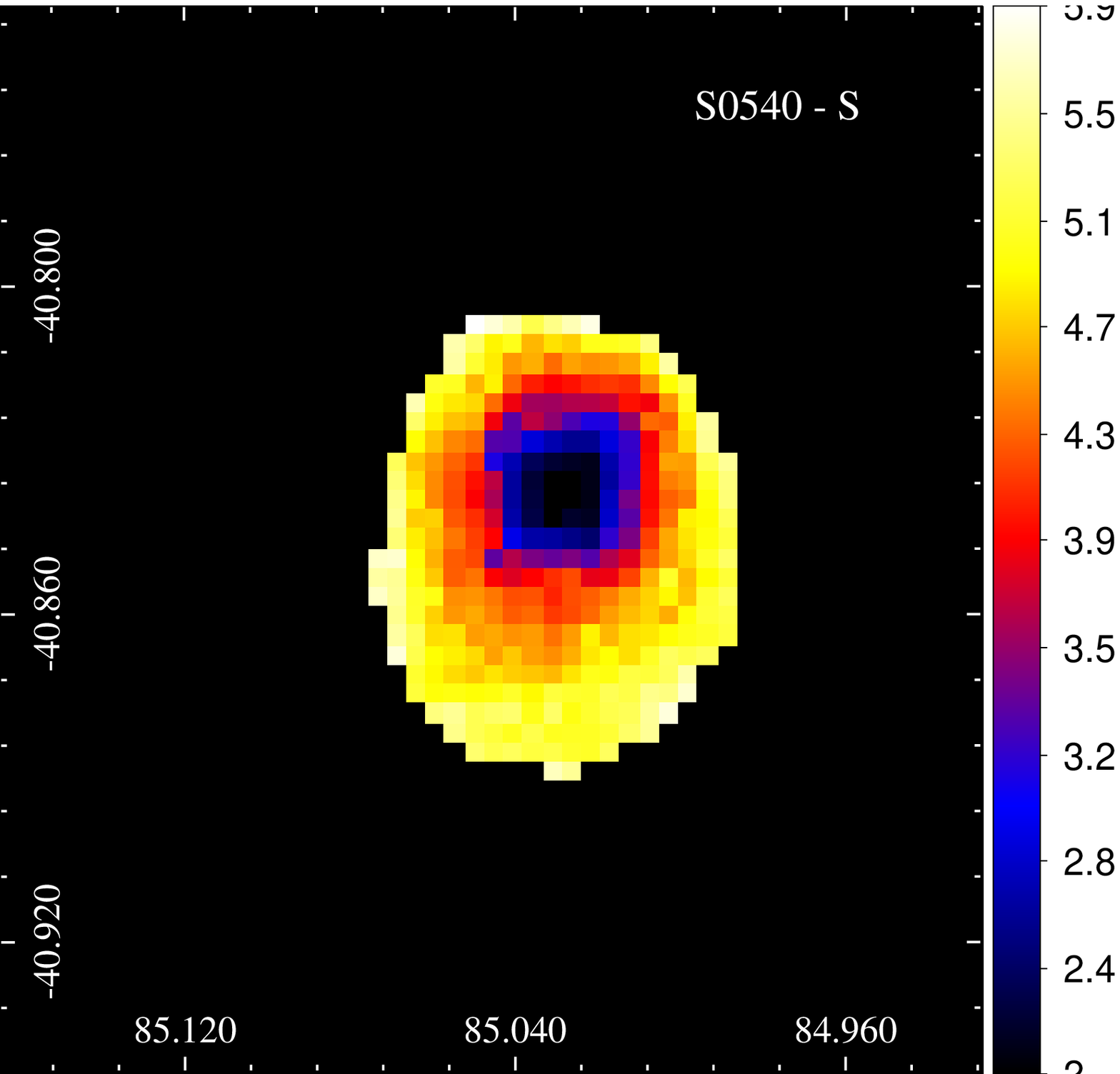}
\includegraphics[scale=0.25]{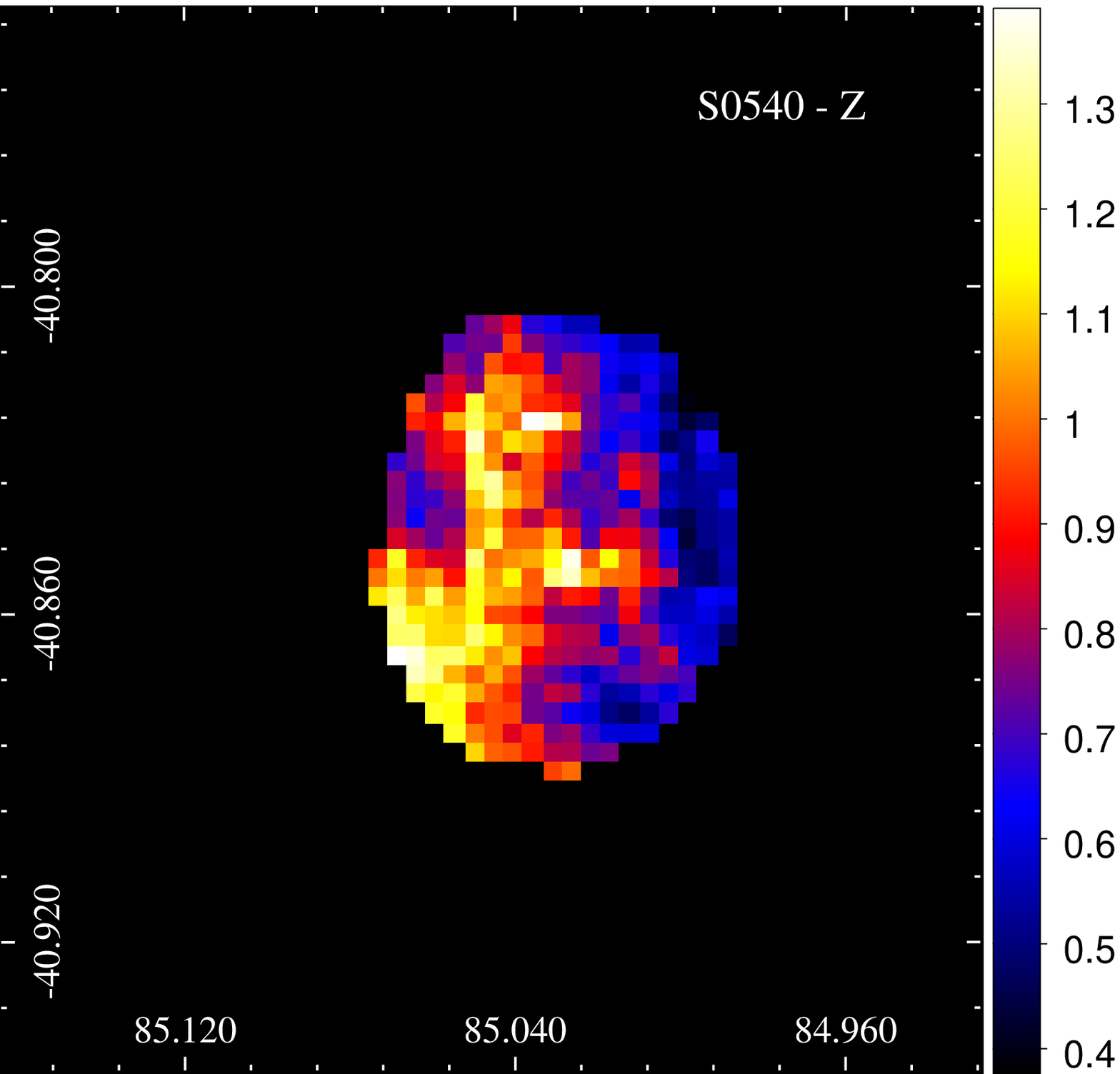}

\includegraphics[scale=0.25]{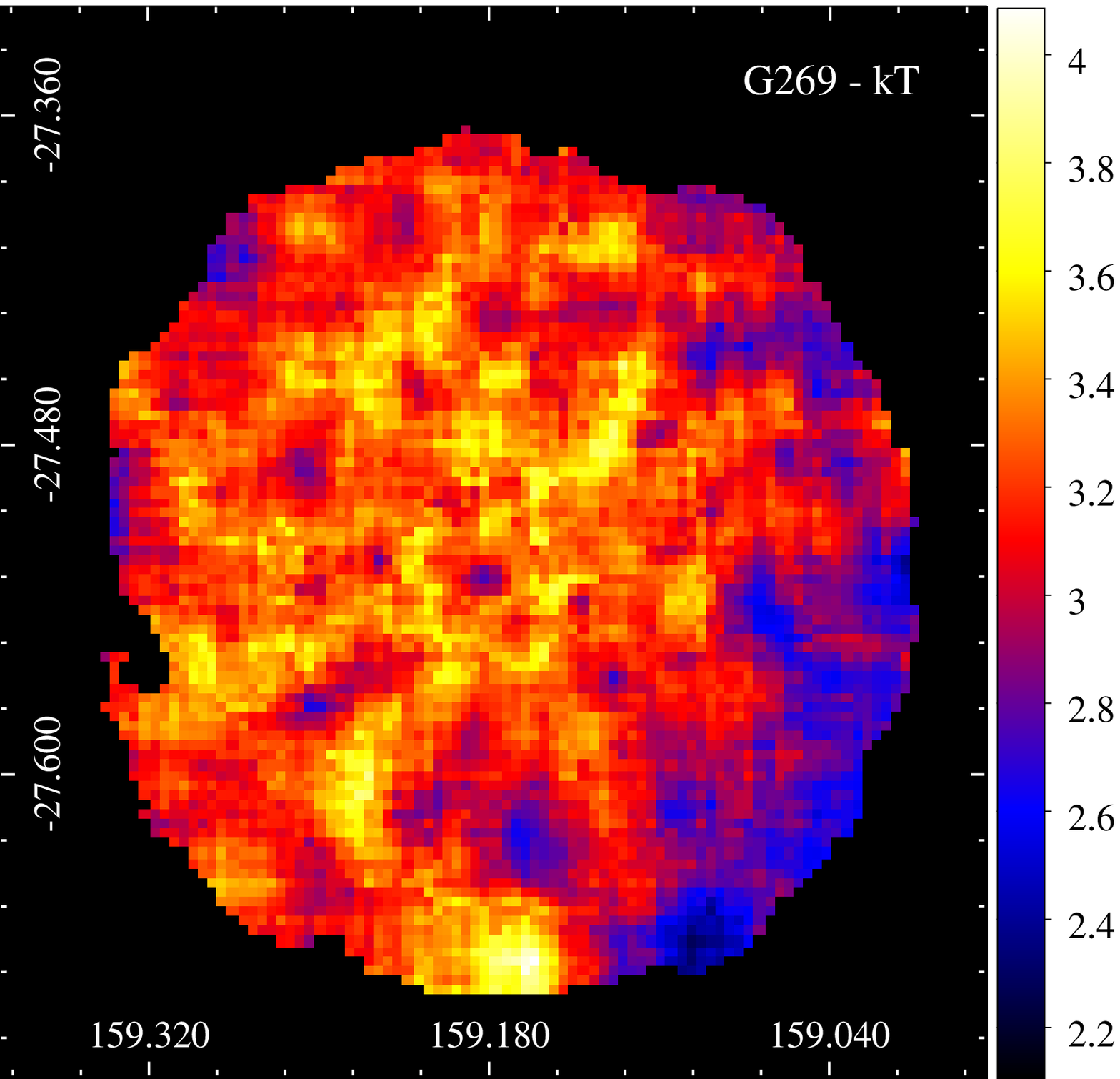}
\includegraphics[scale=0.25]{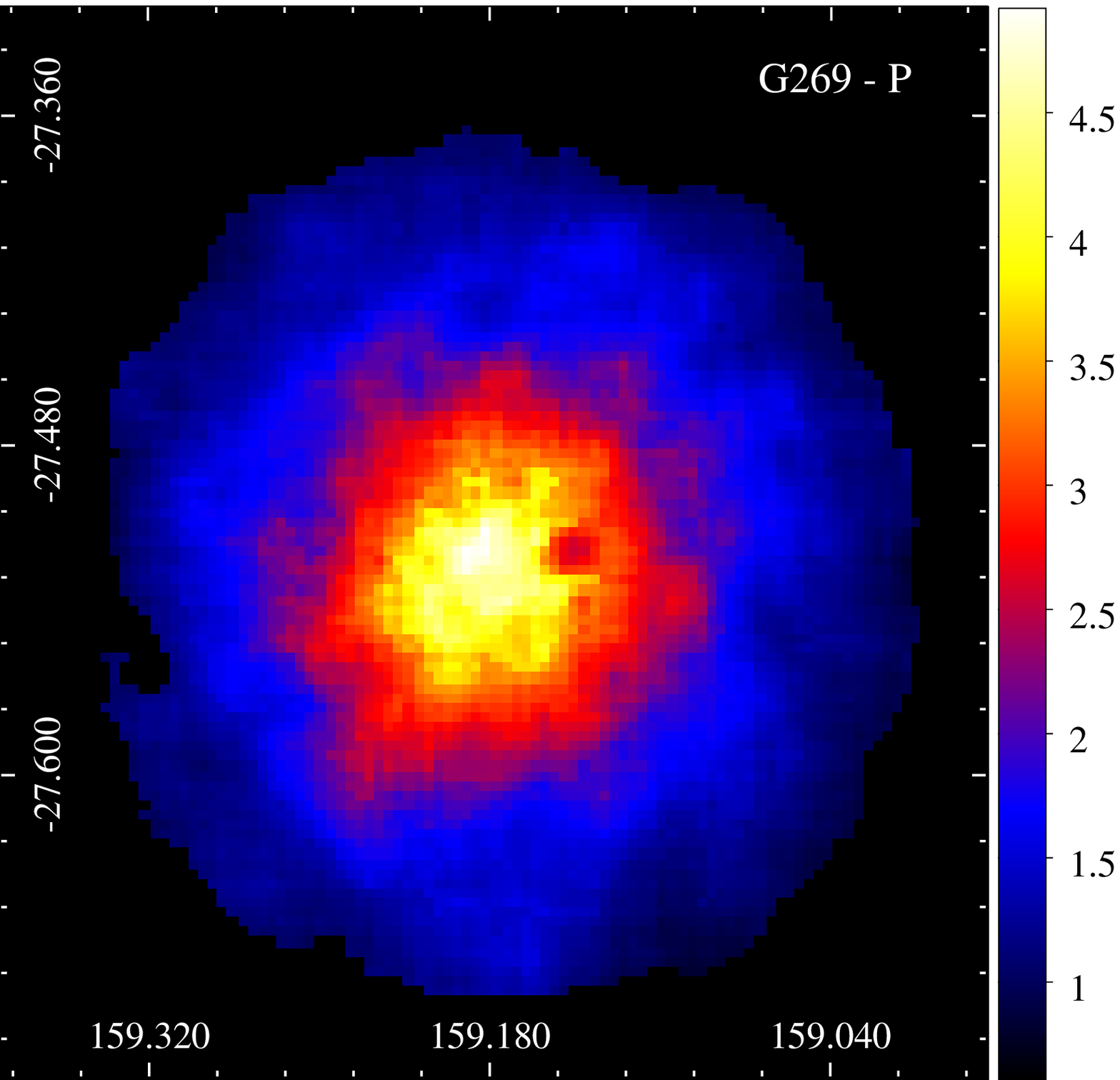}
\includegraphics[scale=0.25]{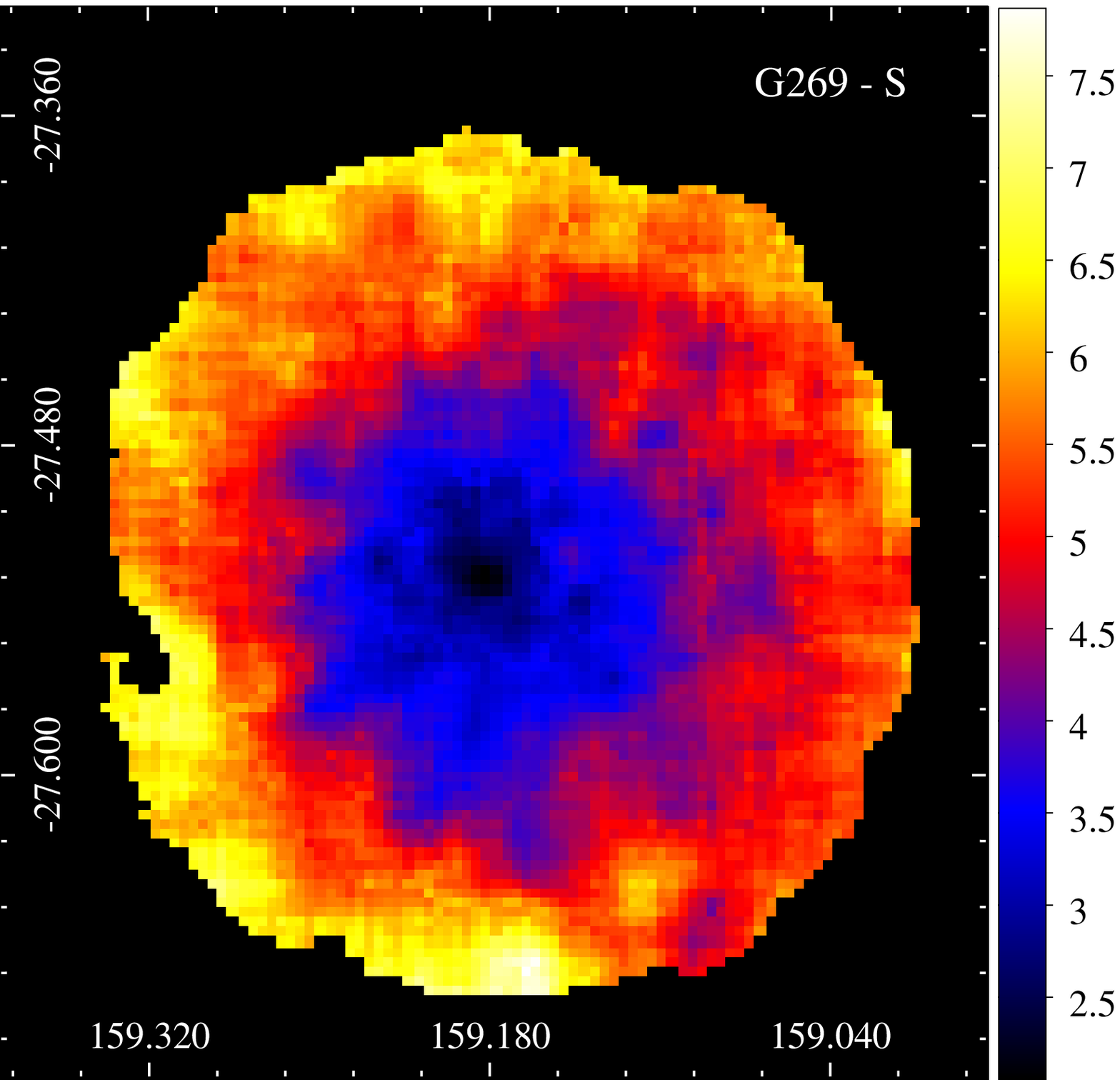}
\includegraphics[scale=0.25]{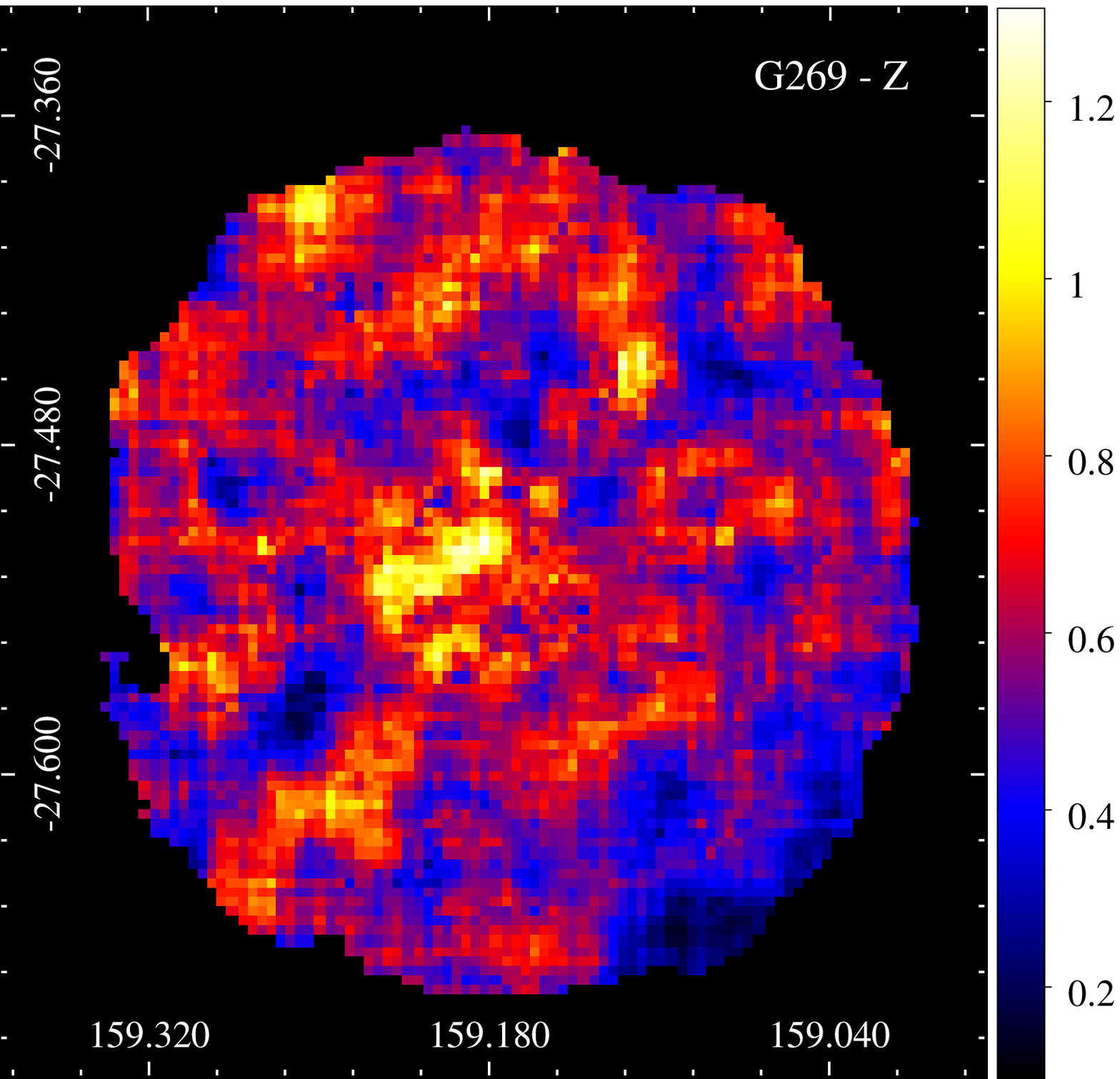}

\includegraphics[scale=0.25]{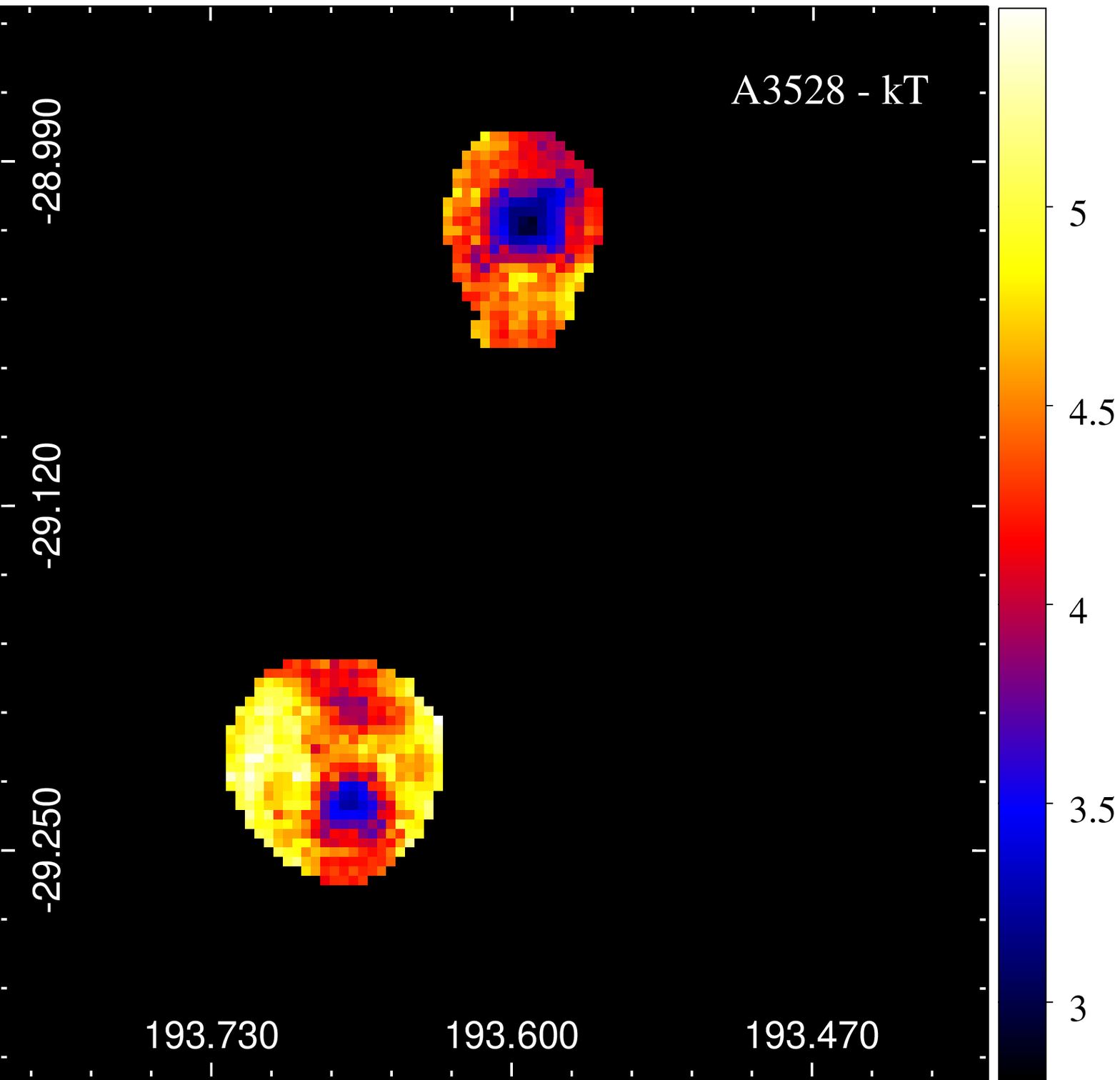}
\includegraphics[scale=0.25]{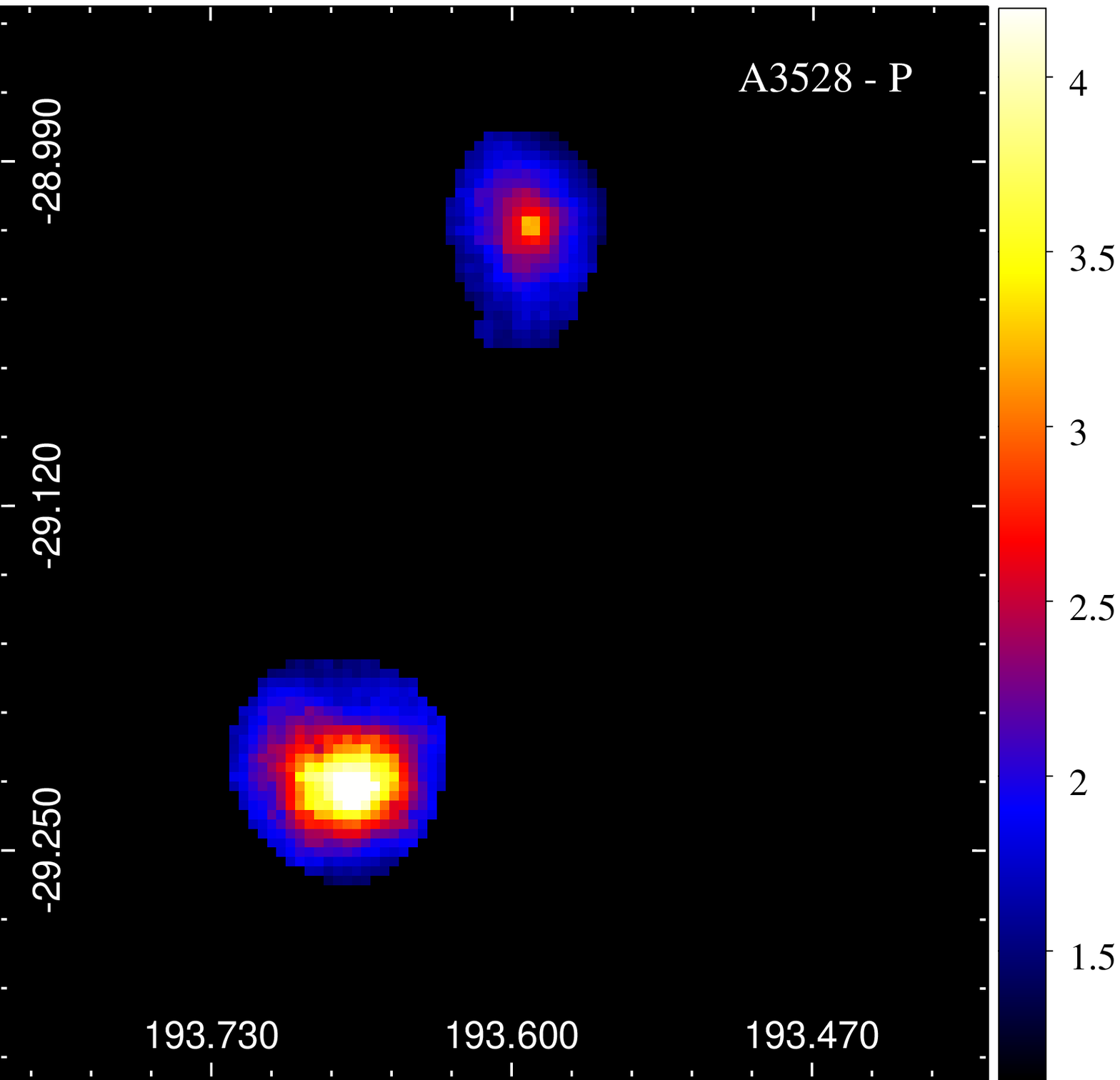}
\includegraphics[scale=0.25]{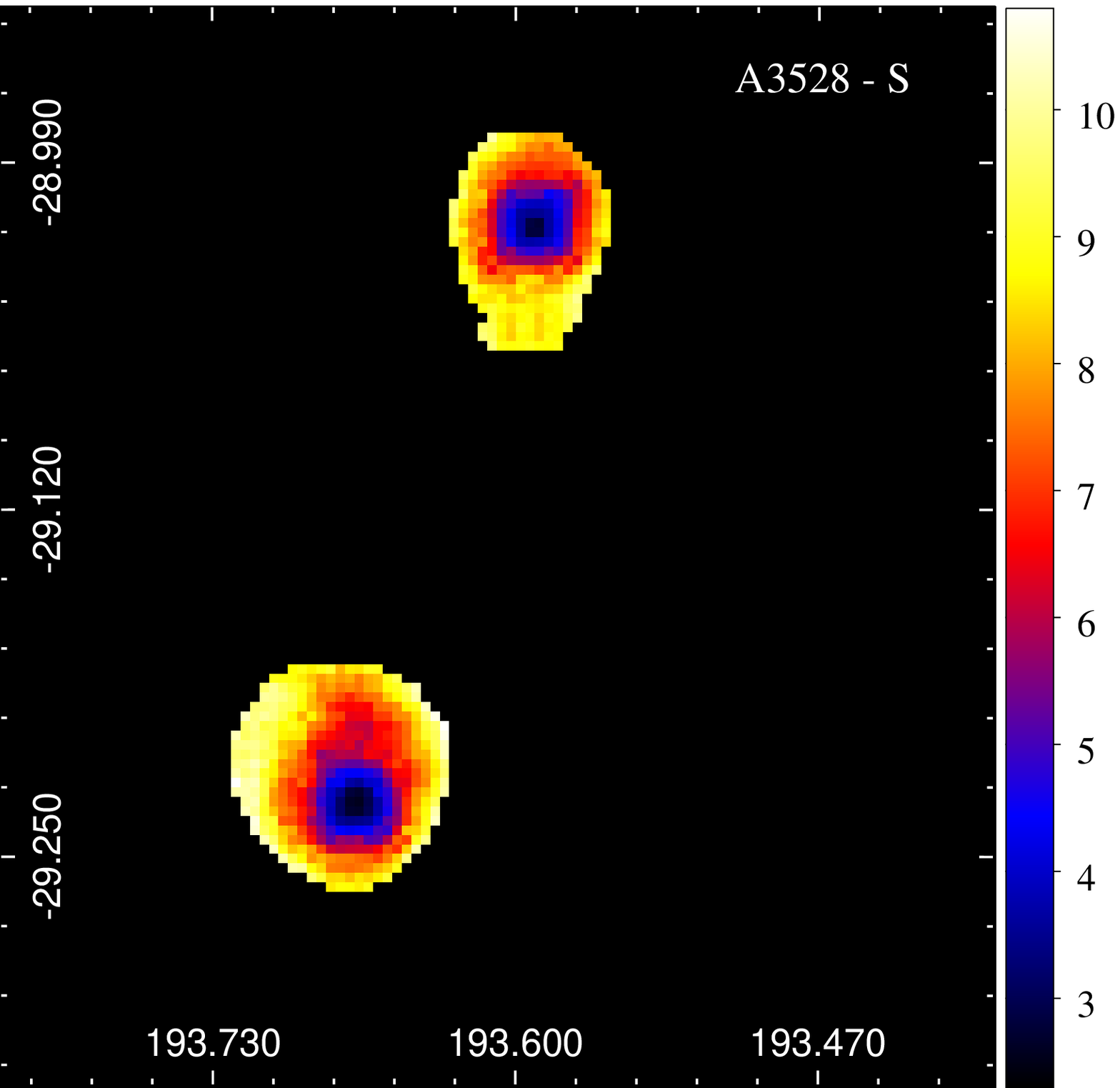}
\includegraphics[scale=0.25]{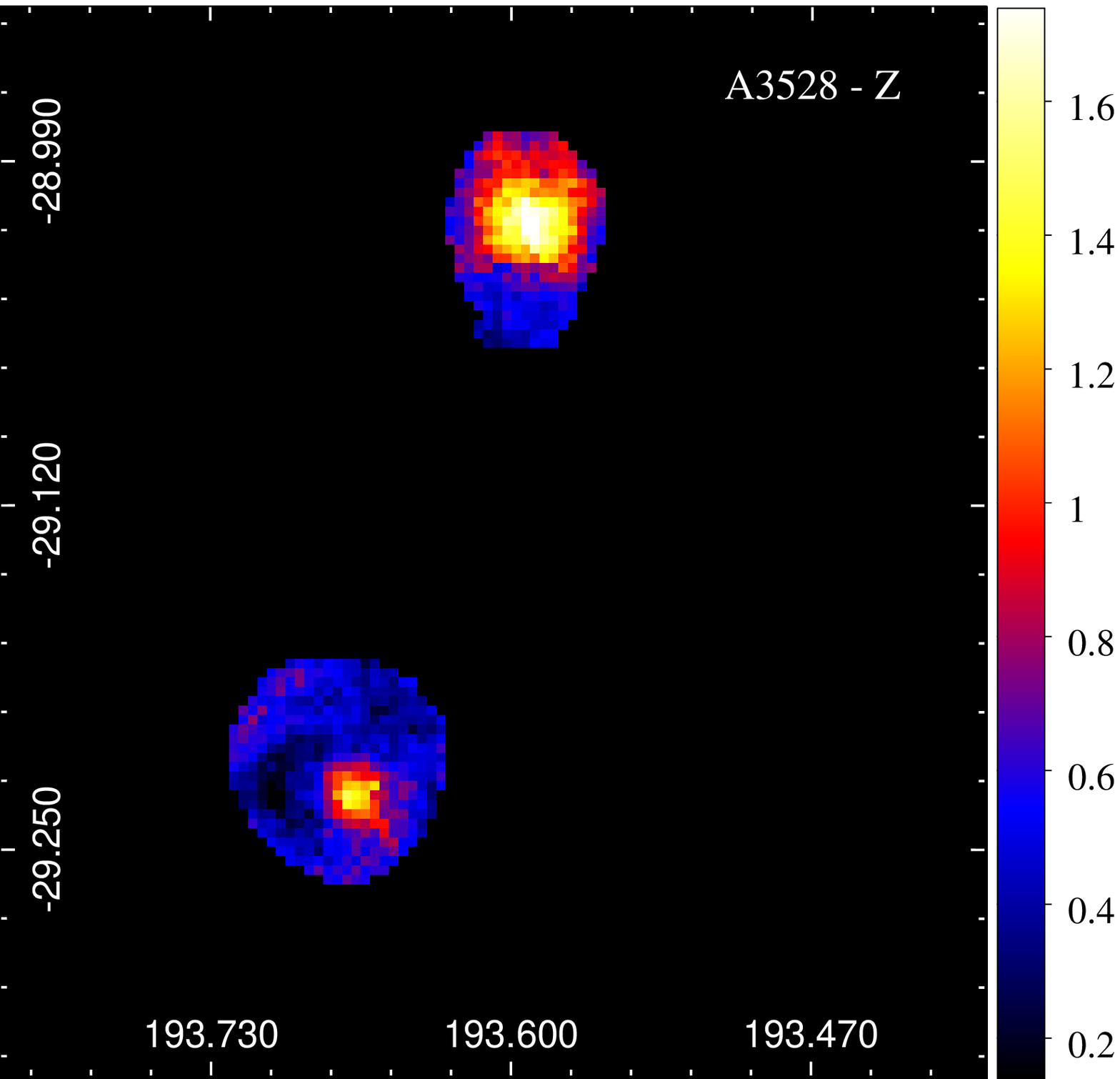}

\caption{CC and relaxed systems. From left to right: temperature,
  pseudo-pressure, pseudo-entropy, and metallicity maps for A2734,
  EXO0422, S0540, G269.51+26.42, and A3528 (A3528n and A3528s).}
\label{fig:CCclusters}
\end{figure*}

\begin{figure*}
\includegraphics[scale=0.25]{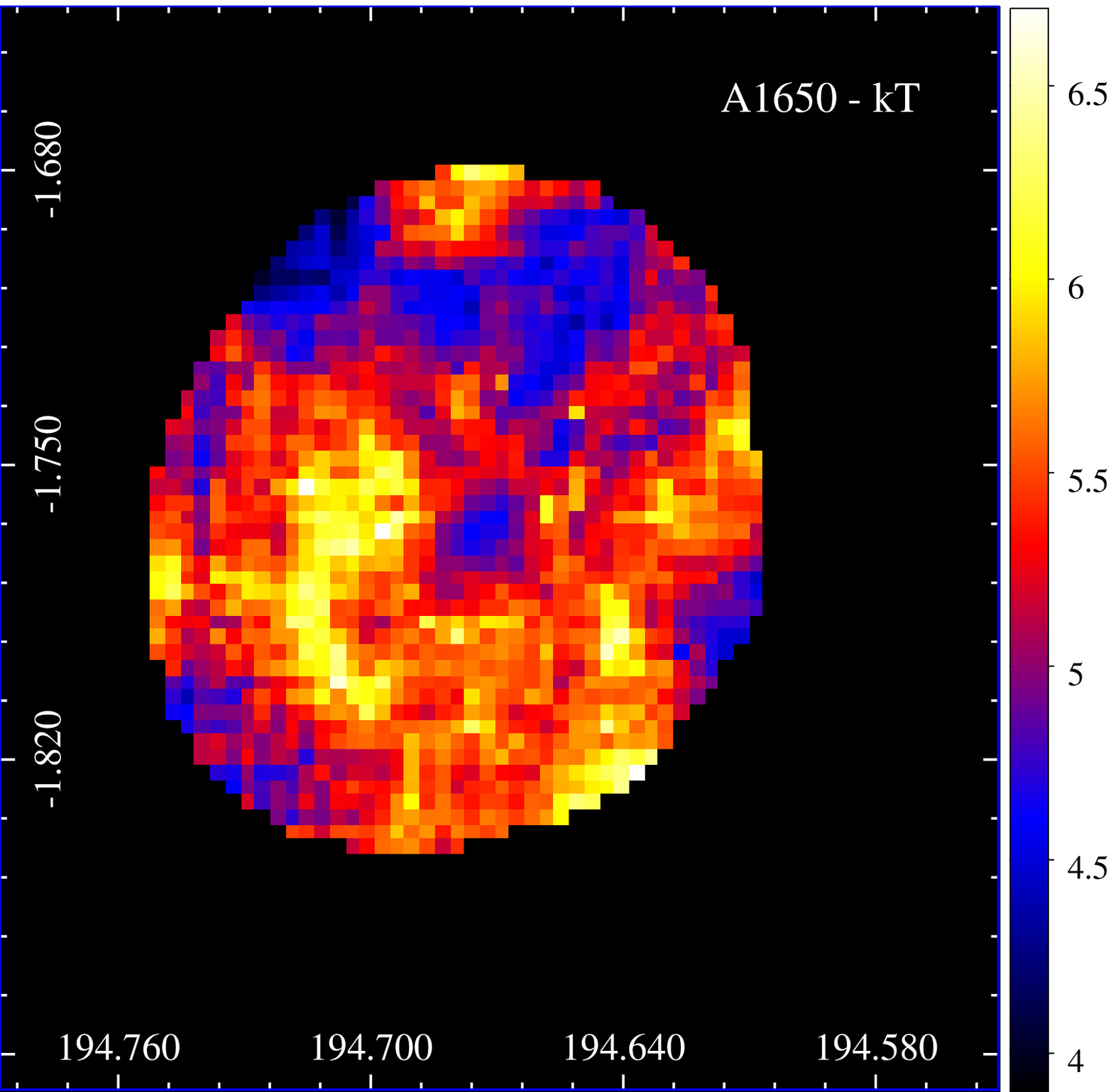}
\includegraphics[scale=0.25]{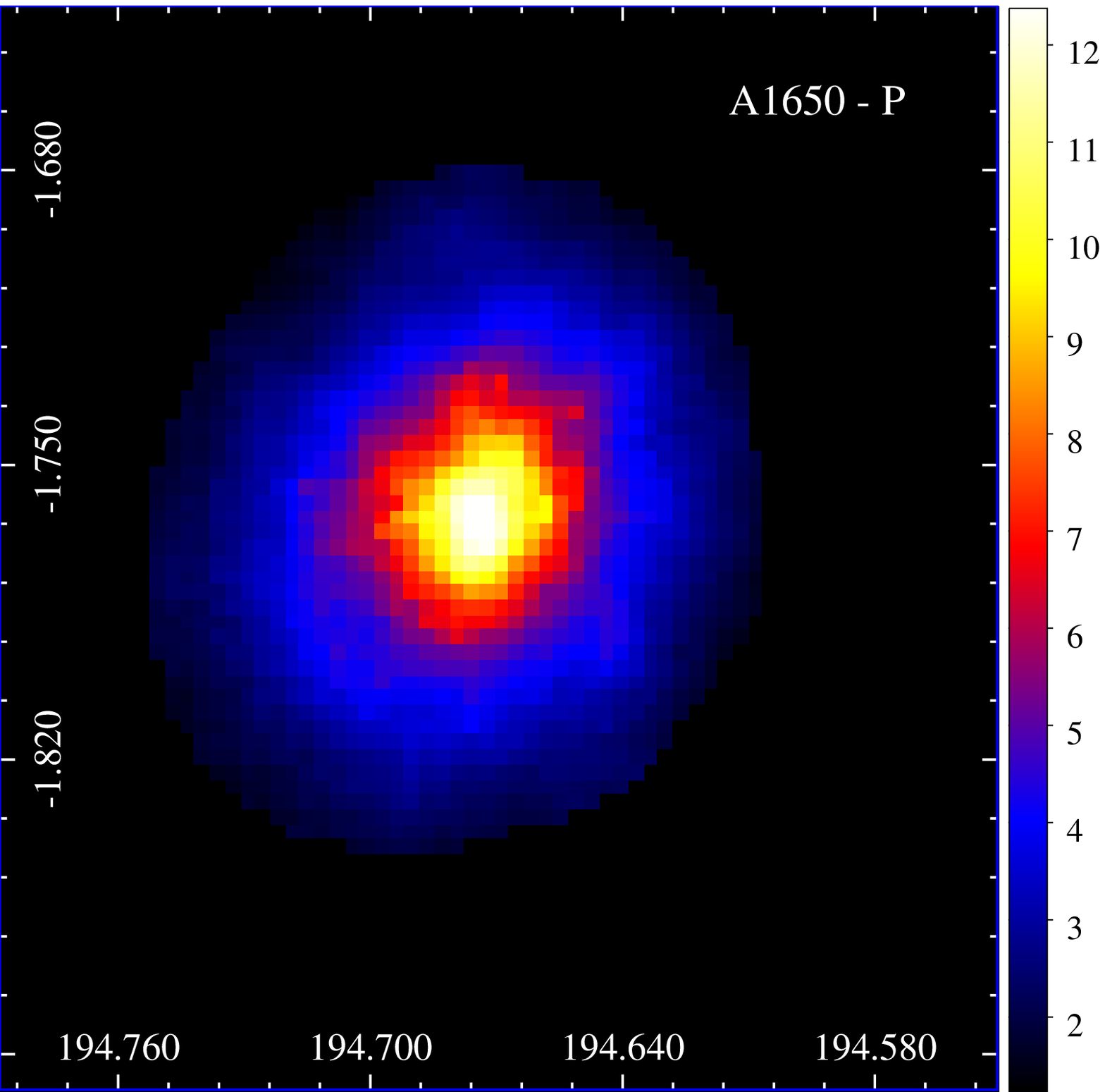}
\includegraphics[scale=0.25]{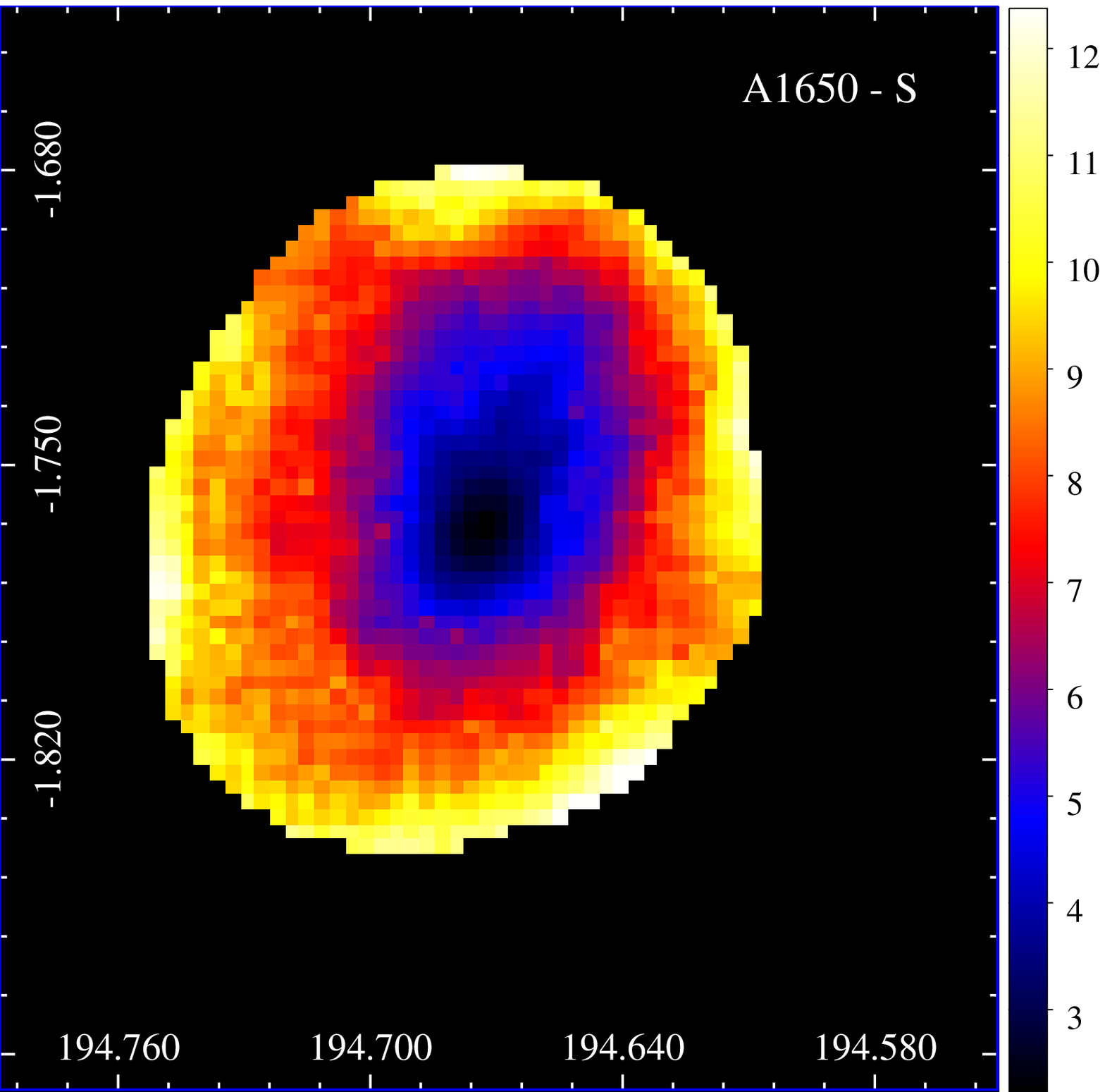}
\includegraphics[scale=0.25]{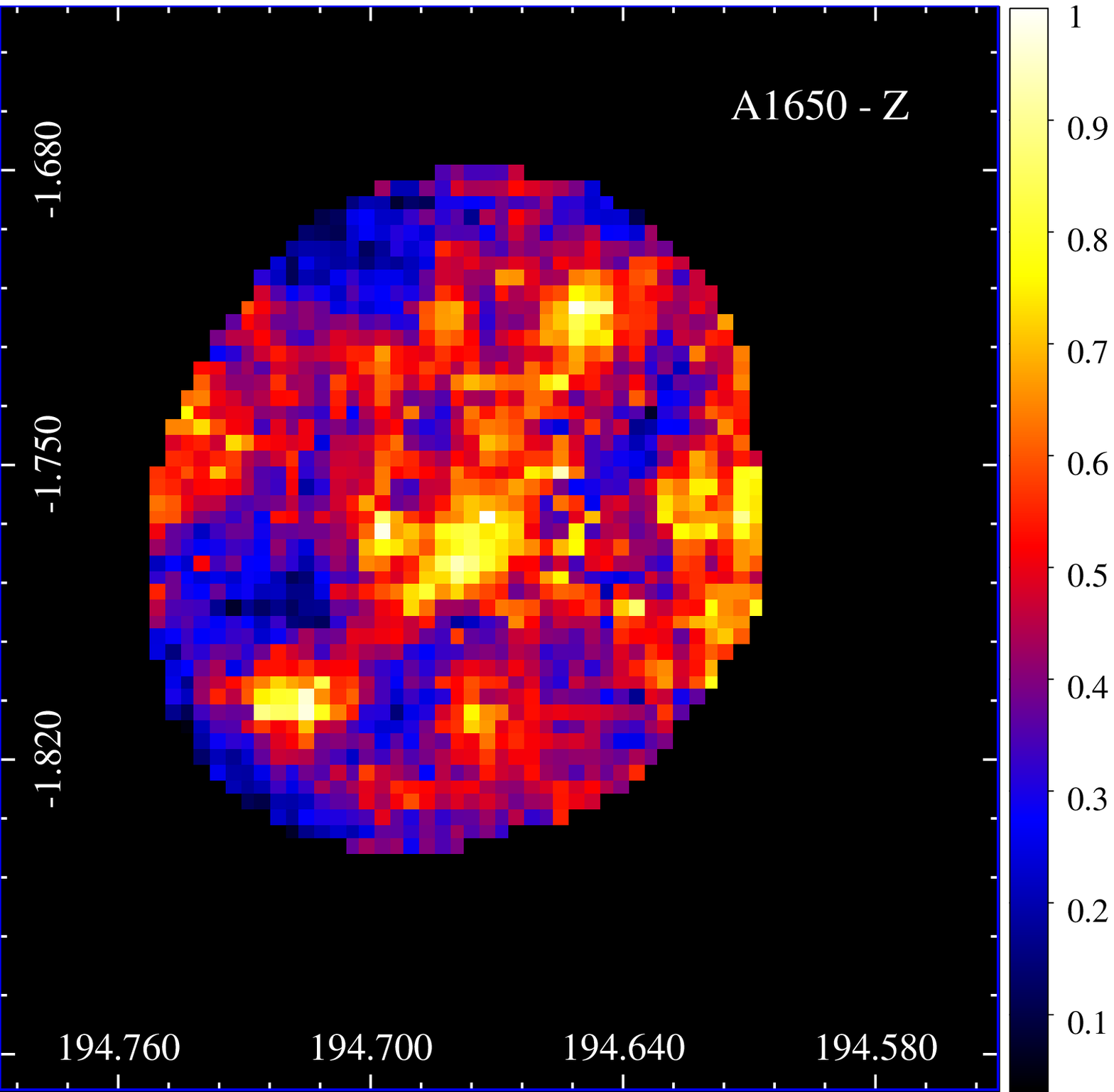}

\includegraphics[scale=0.25]{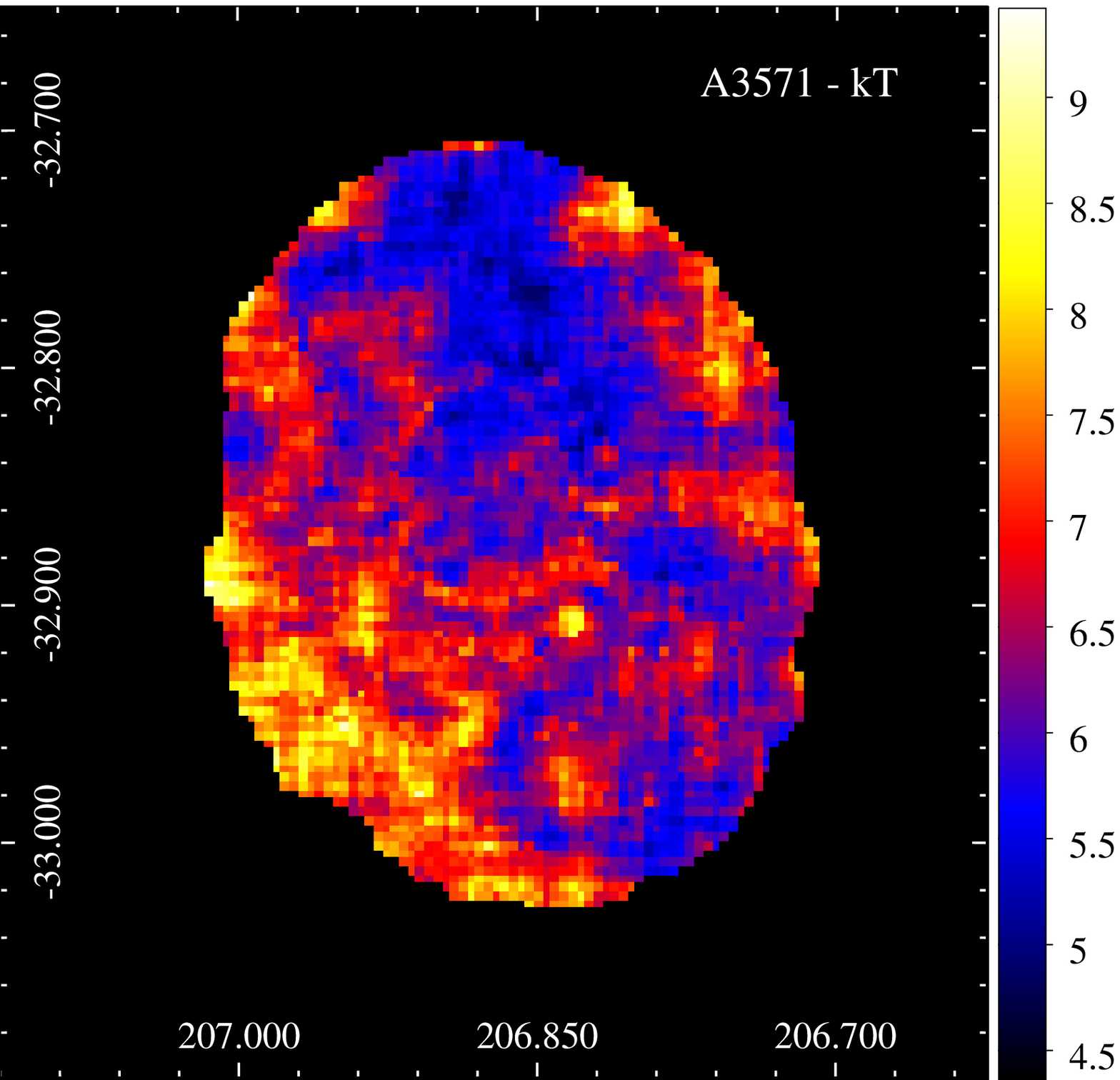}
\includegraphics[scale=0.25]{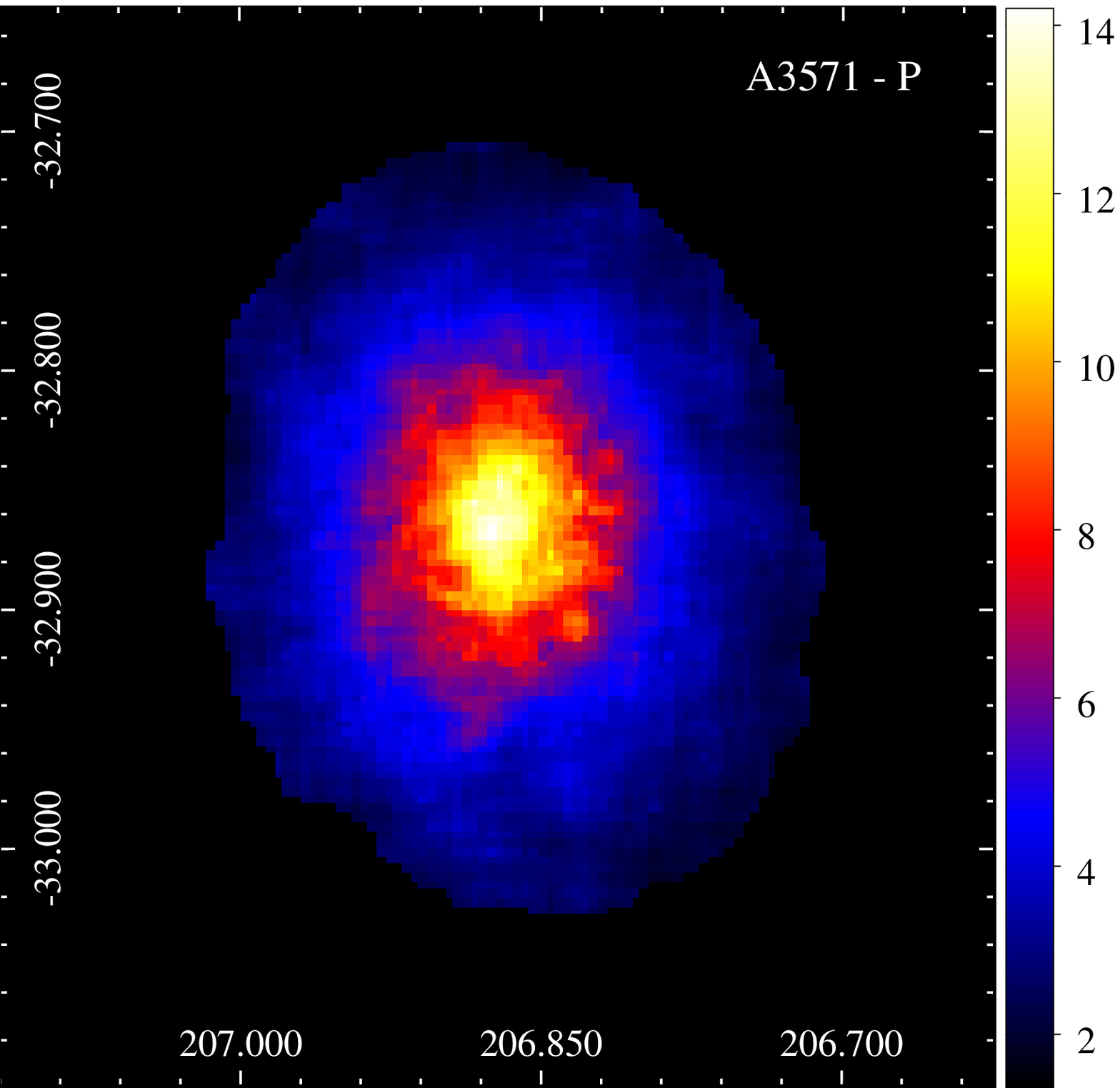}
\includegraphics[scale=0.25]{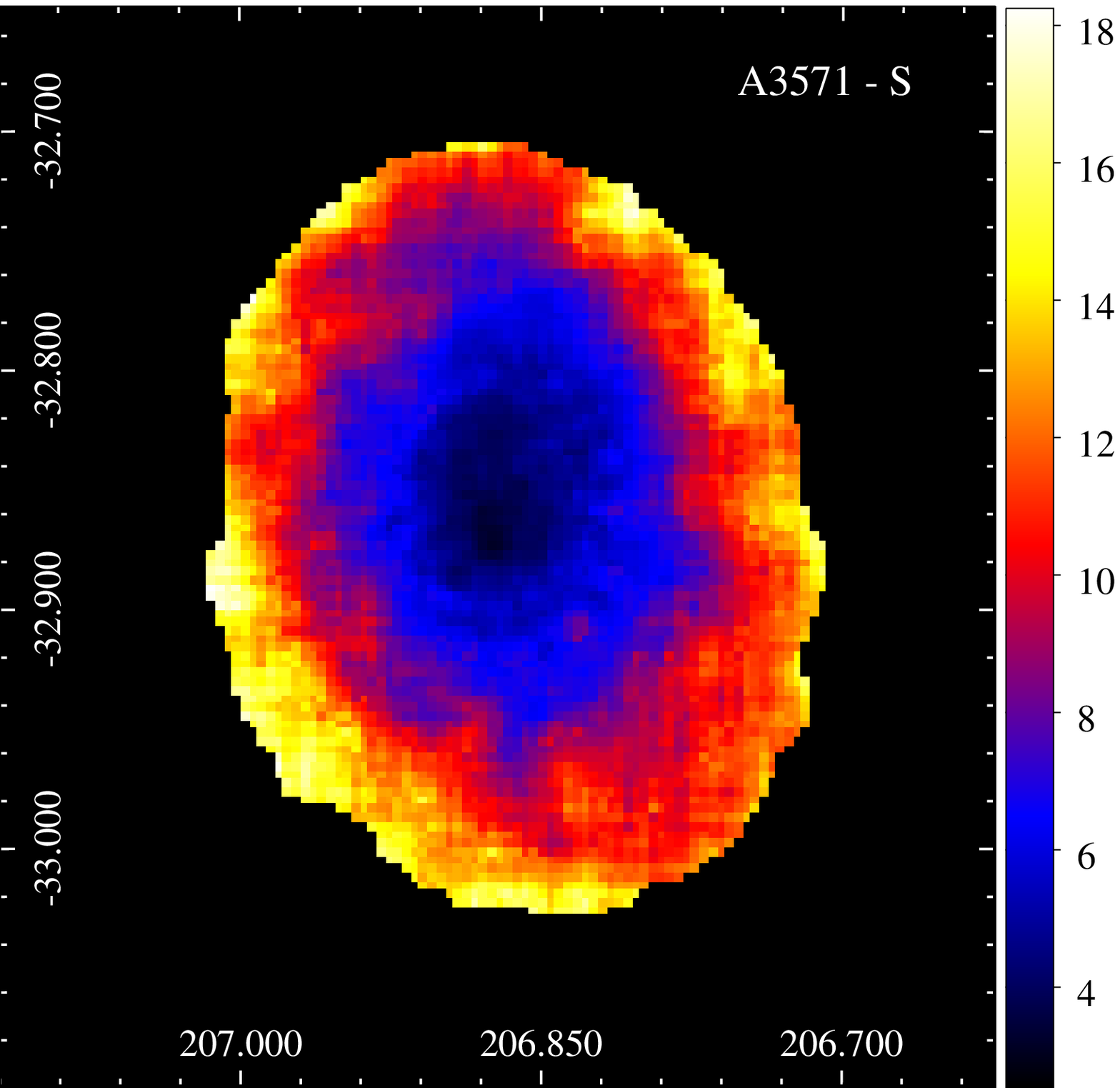}
\includegraphics[scale=0.25]{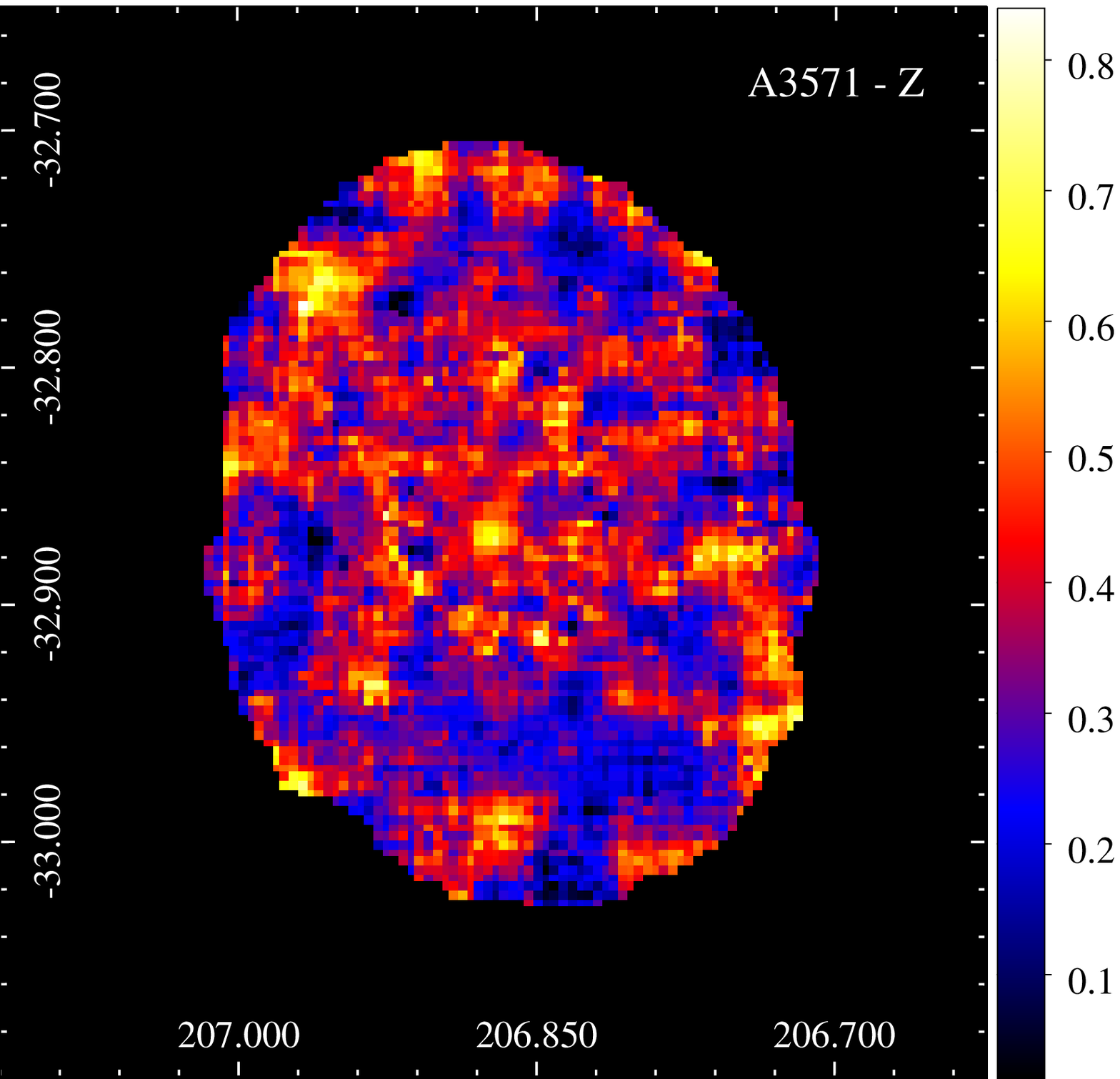}

\includegraphics[scale=0.25]{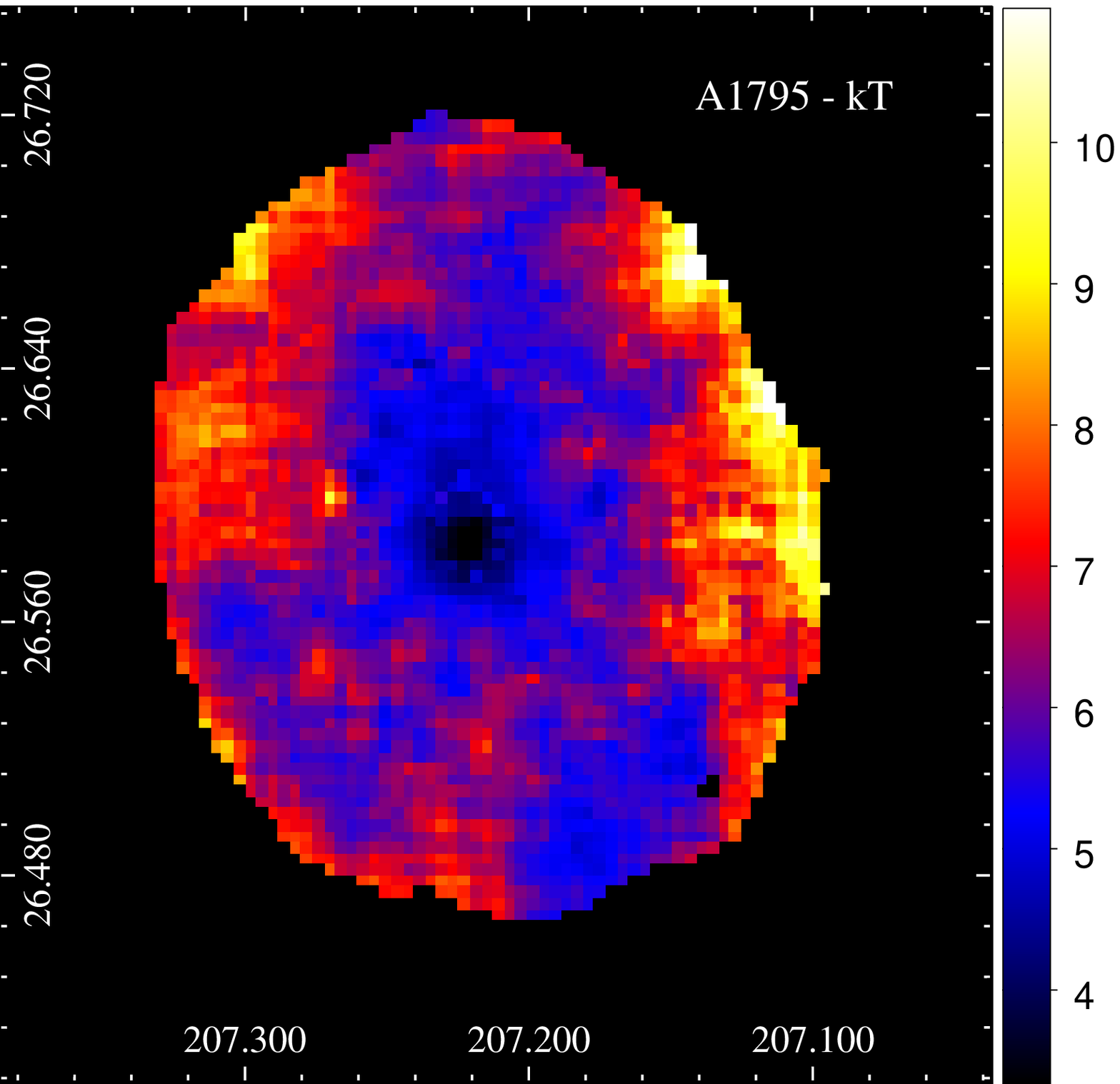}
\includegraphics[scale=0.25]{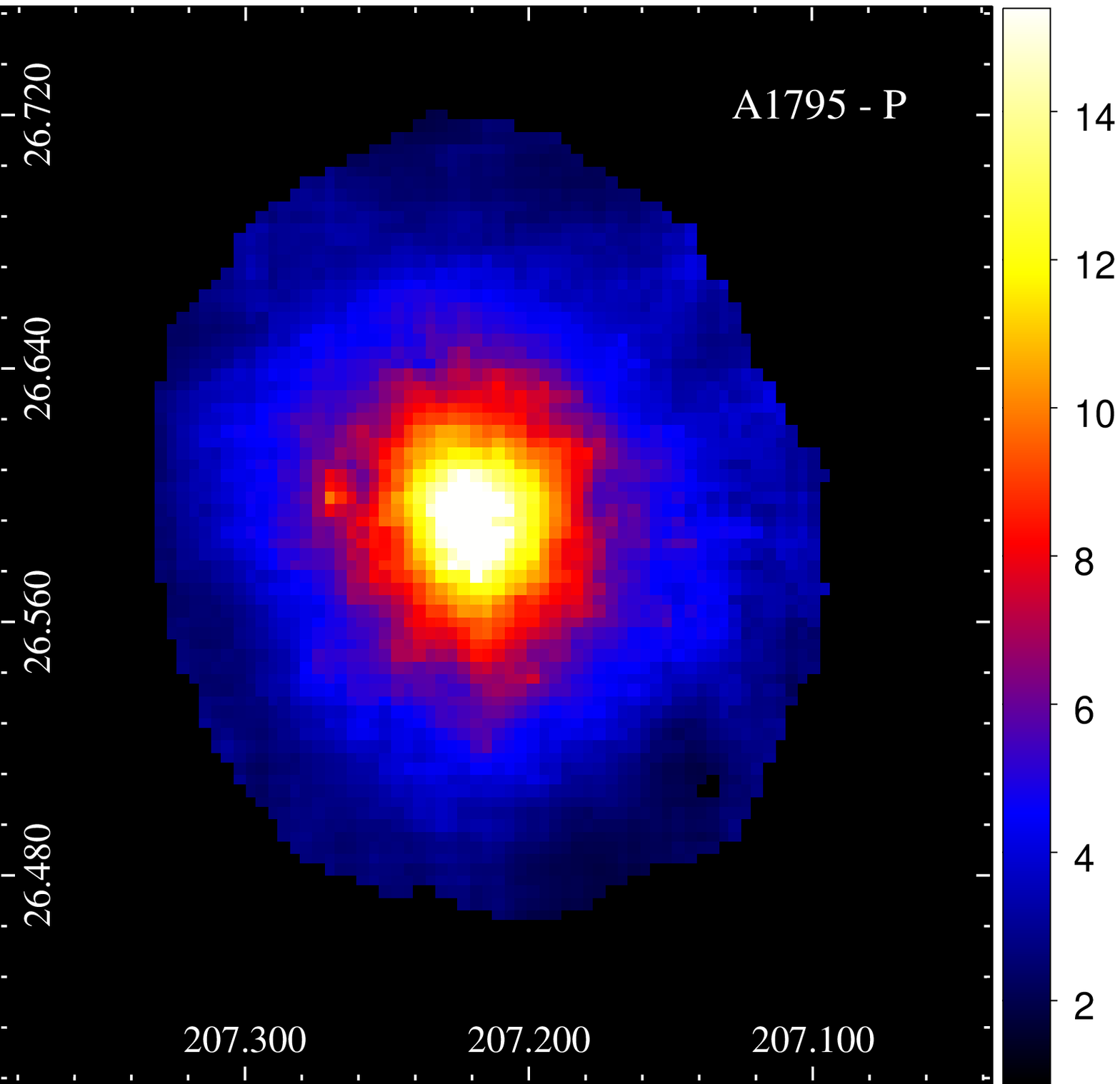}
\includegraphics[scale=0.25]{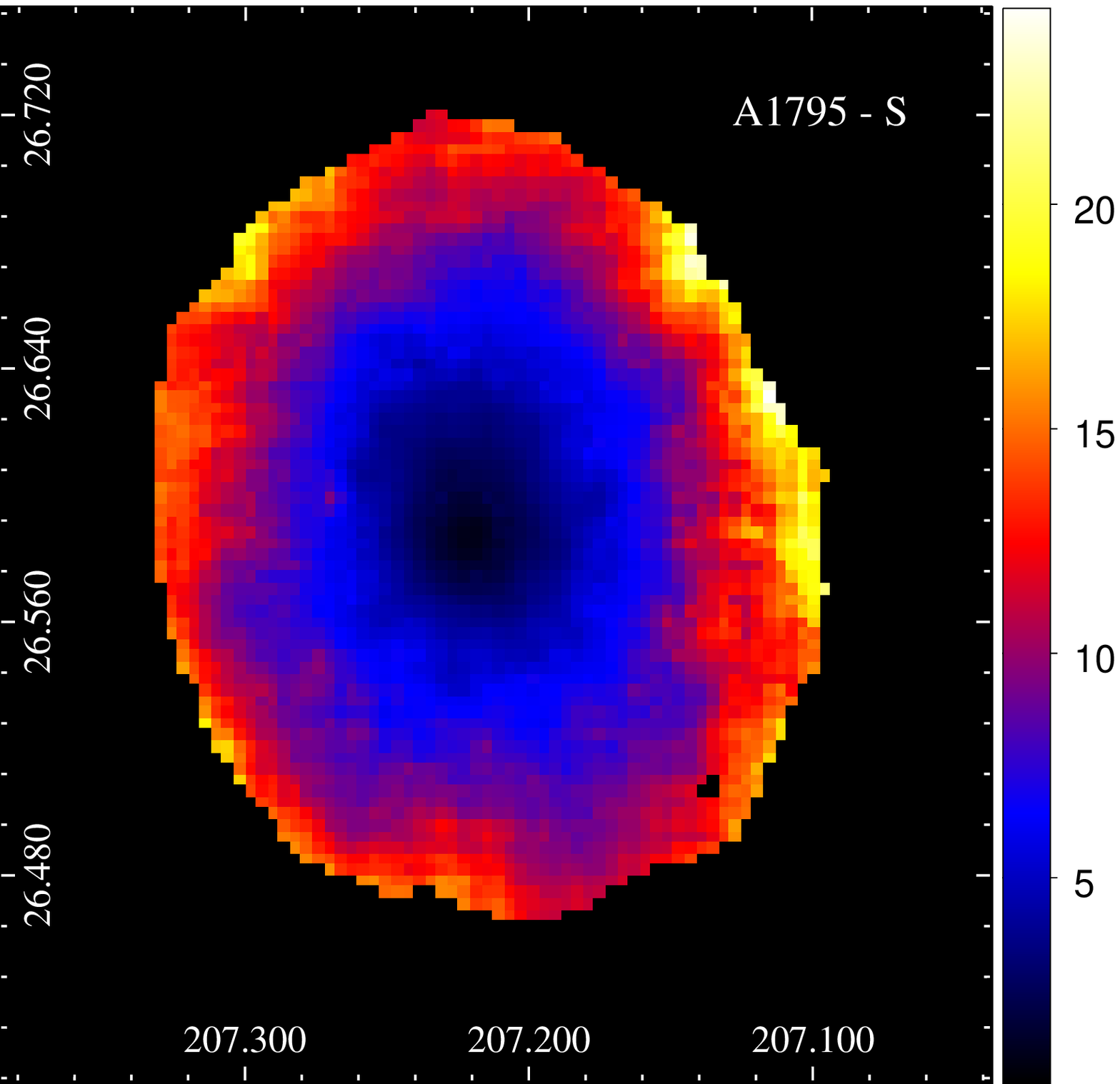}
\includegraphics[scale=0.25]{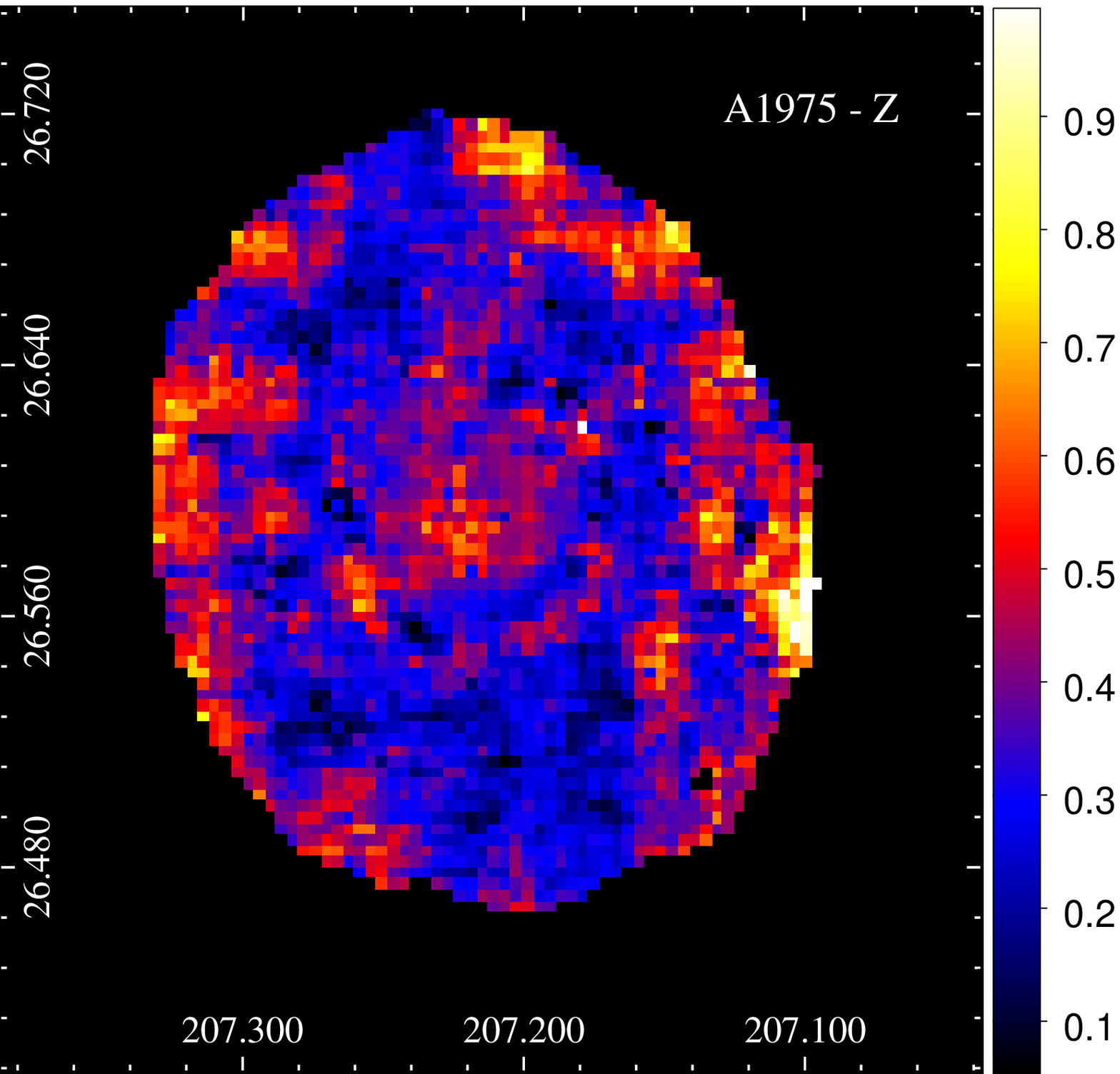}

\includegraphics[scale=0.25]{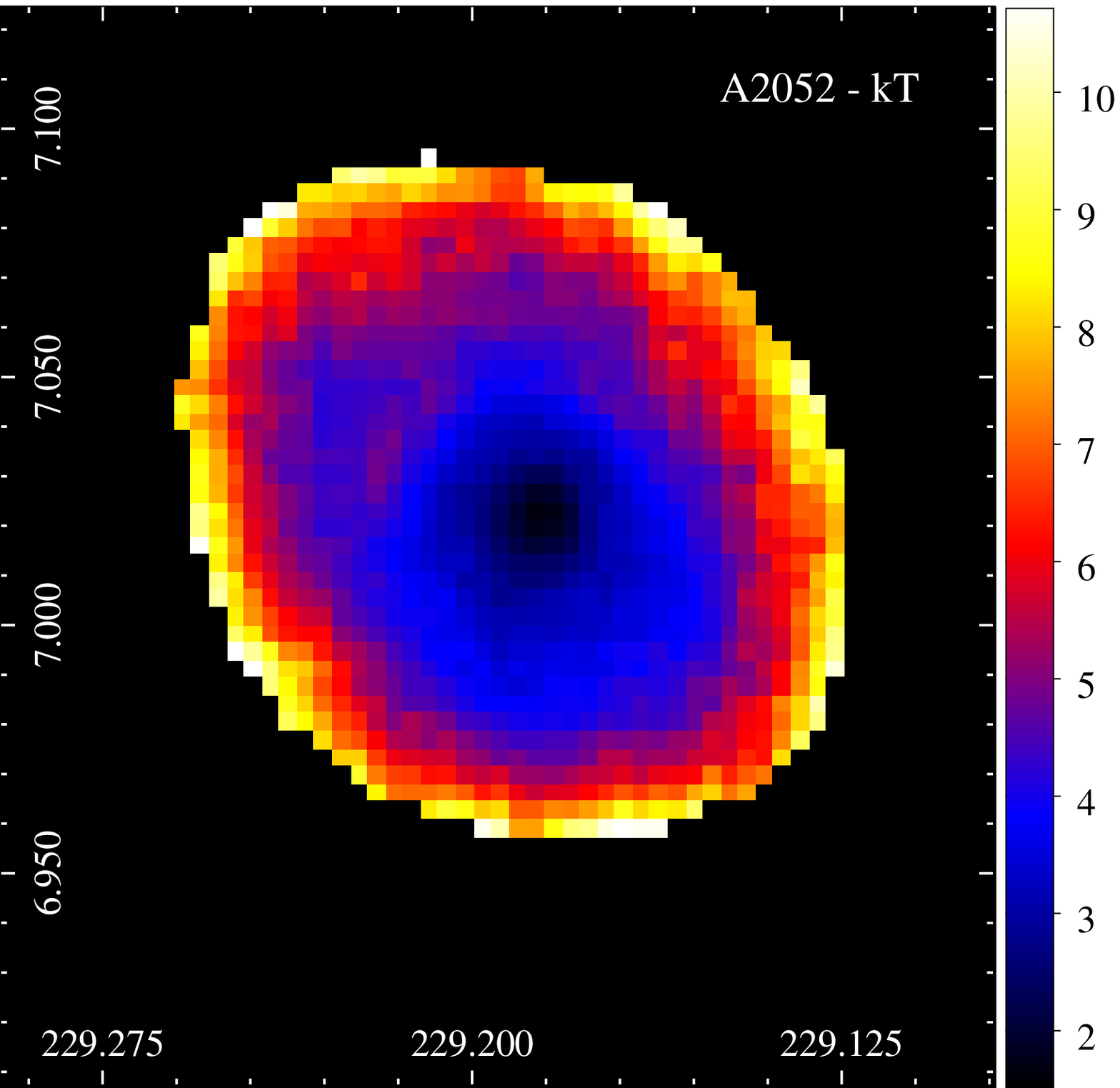}
\includegraphics[scale=0.25]{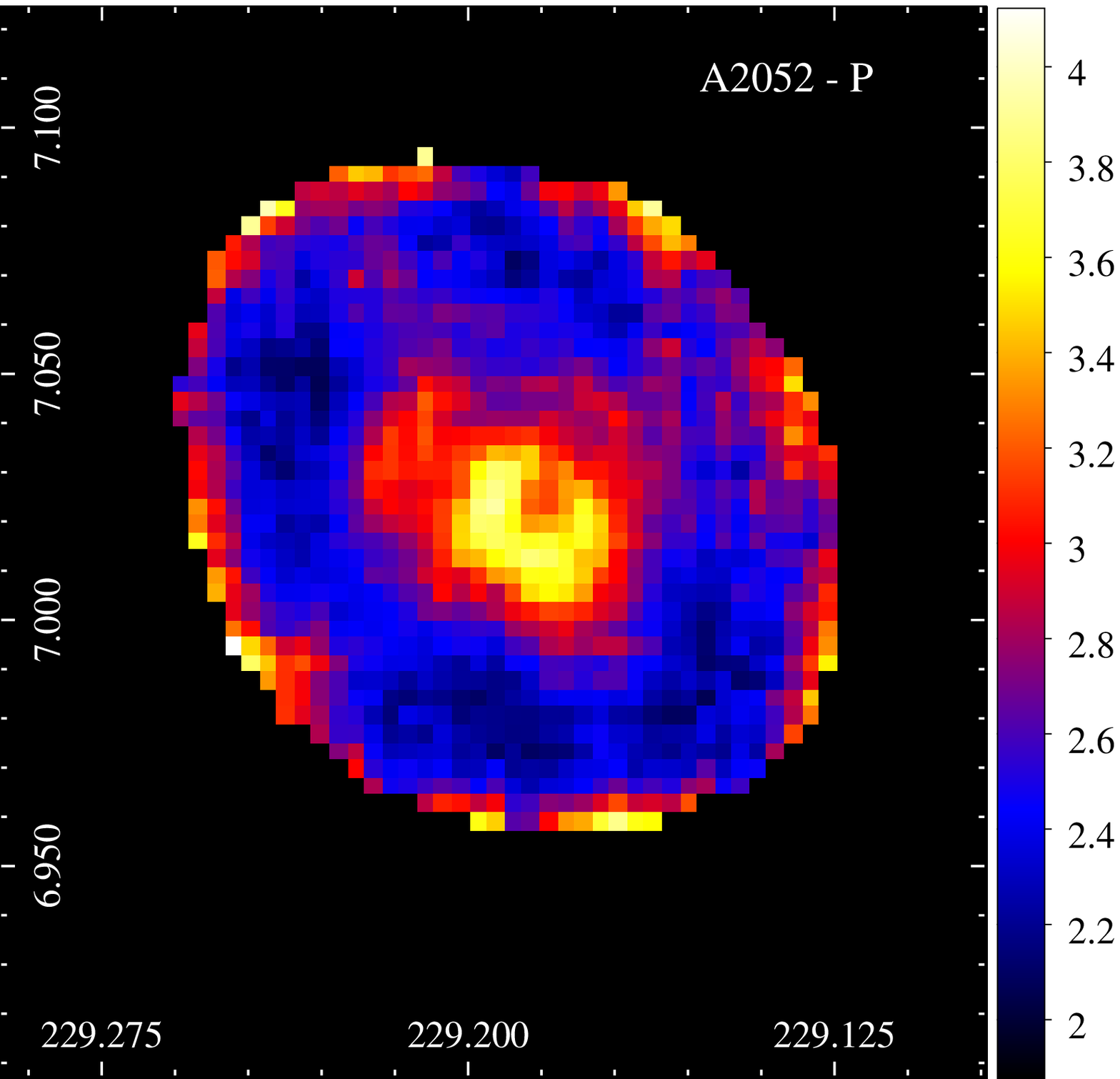}
\includegraphics[scale=0.25]{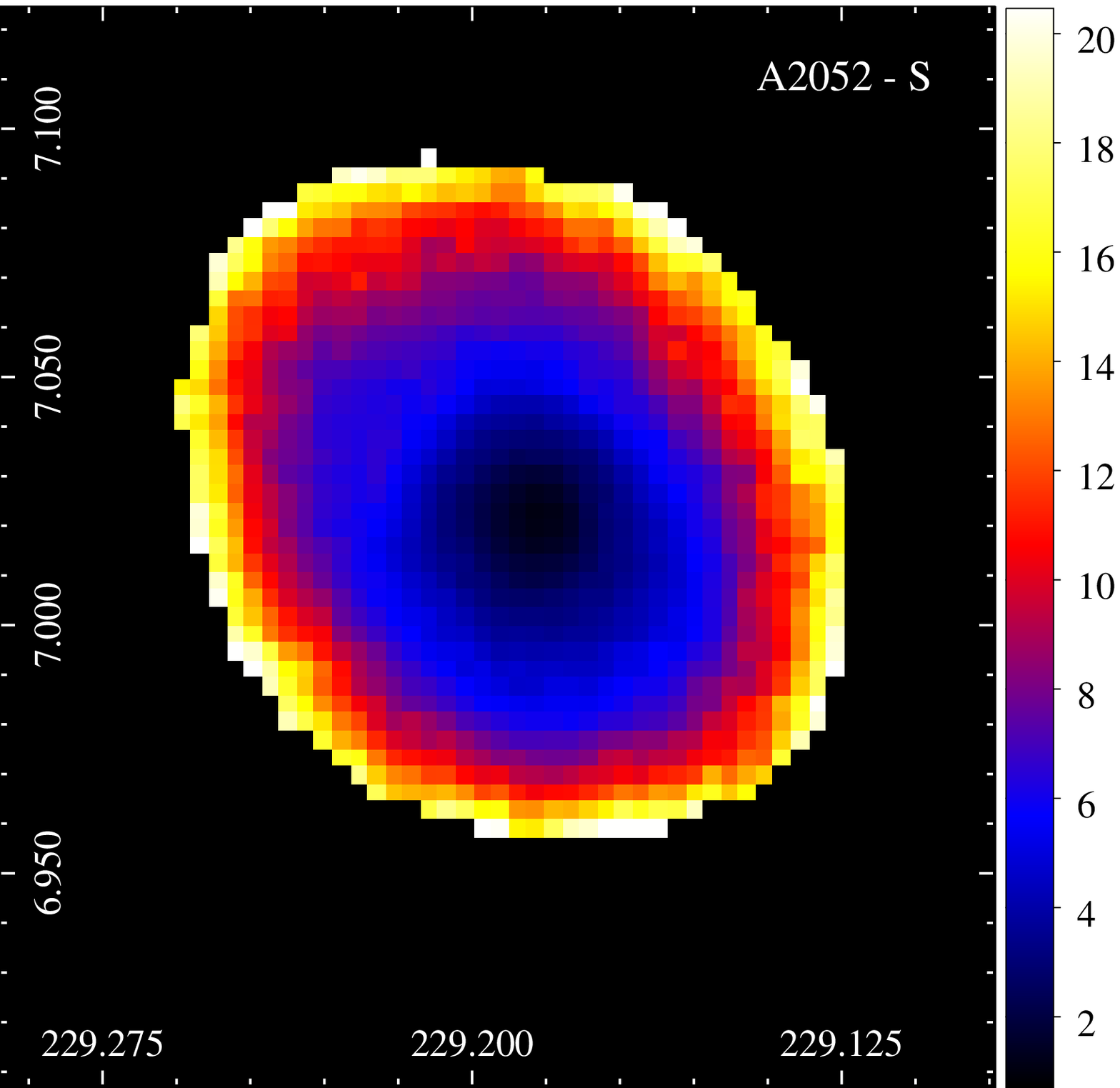}
\includegraphics[scale=0.25]{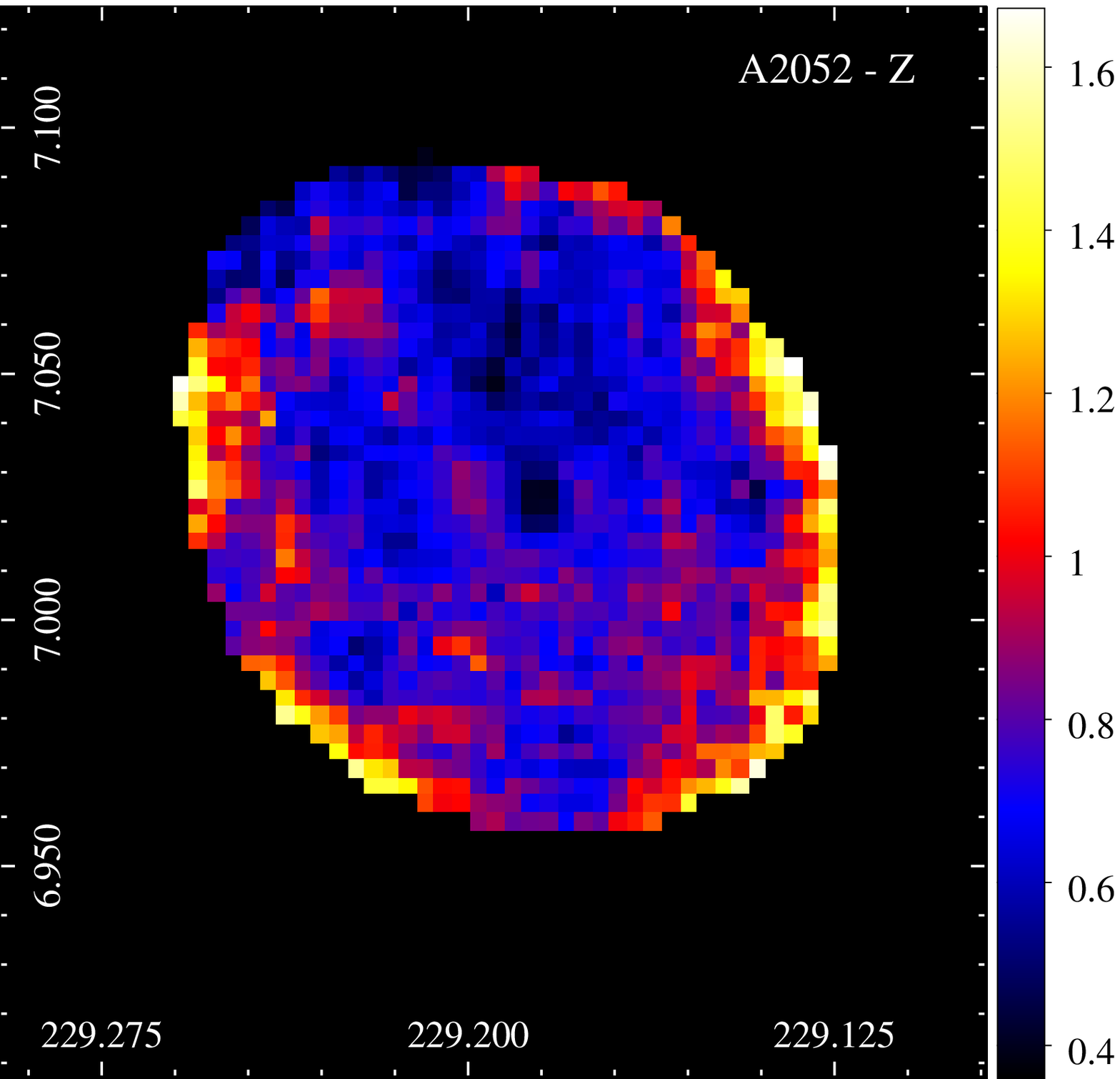}

\includegraphics[scale=0.25]{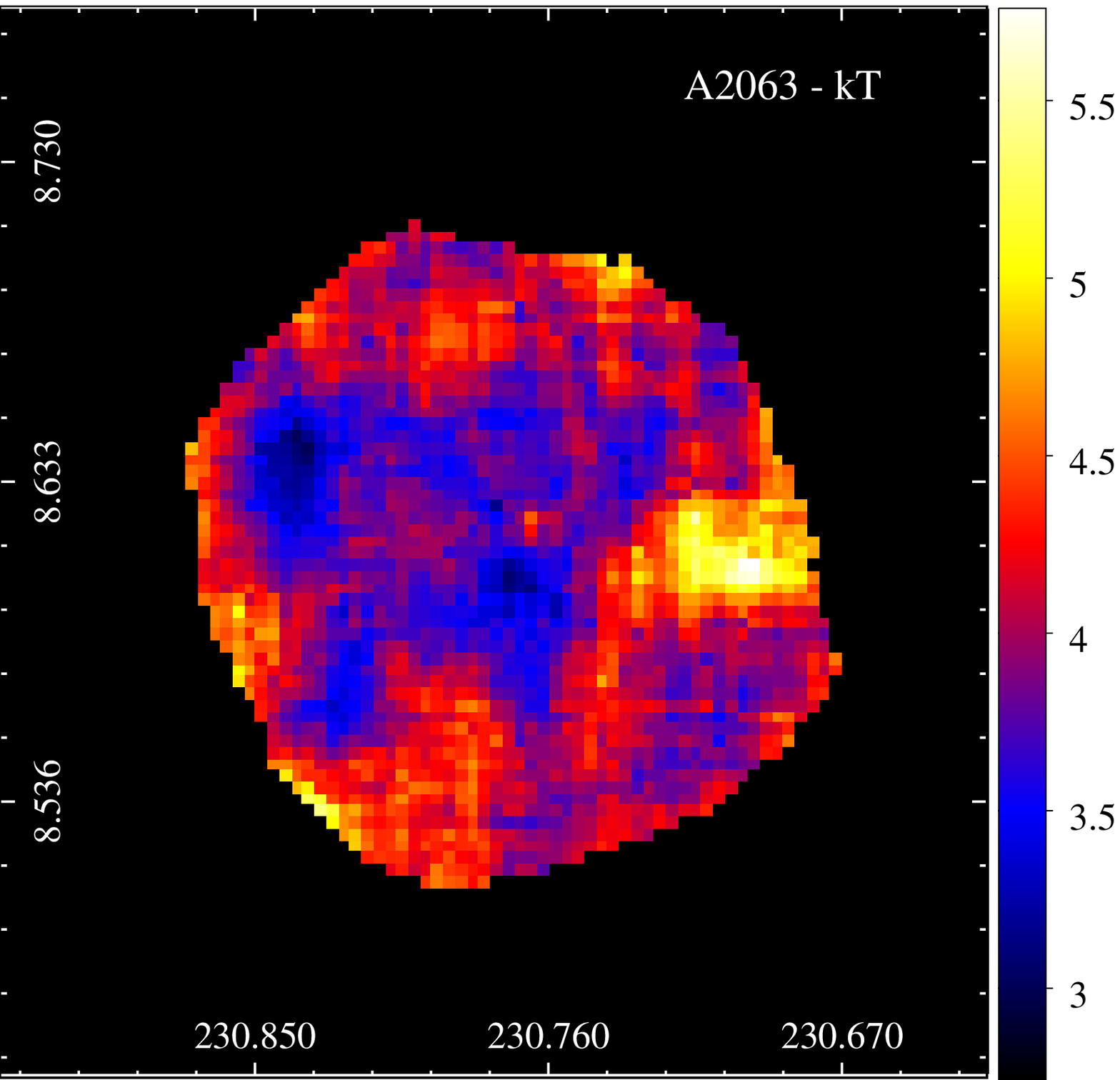}
\includegraphics[scale=0.25]{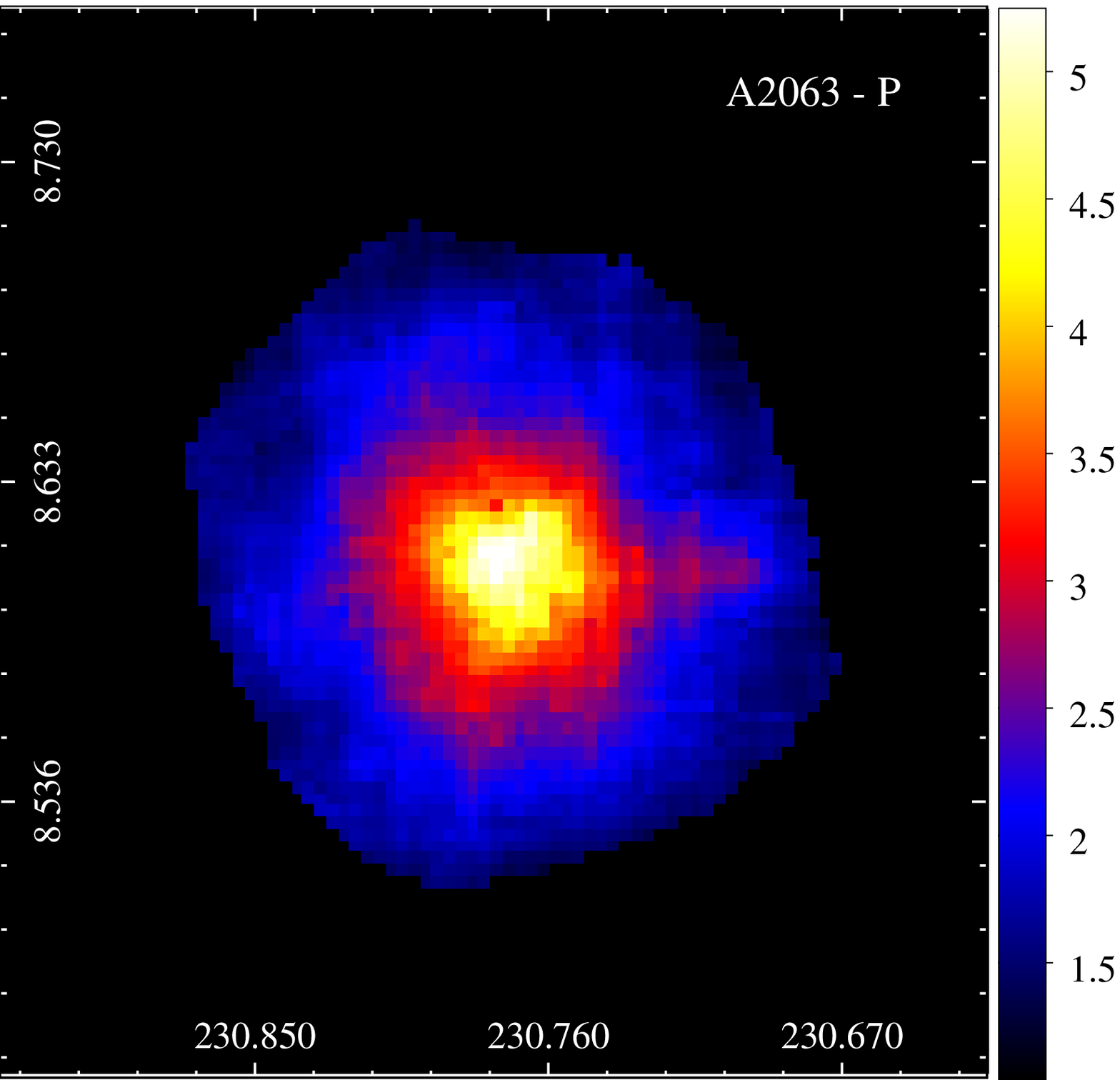}
\includegraphics[scale=0.25]{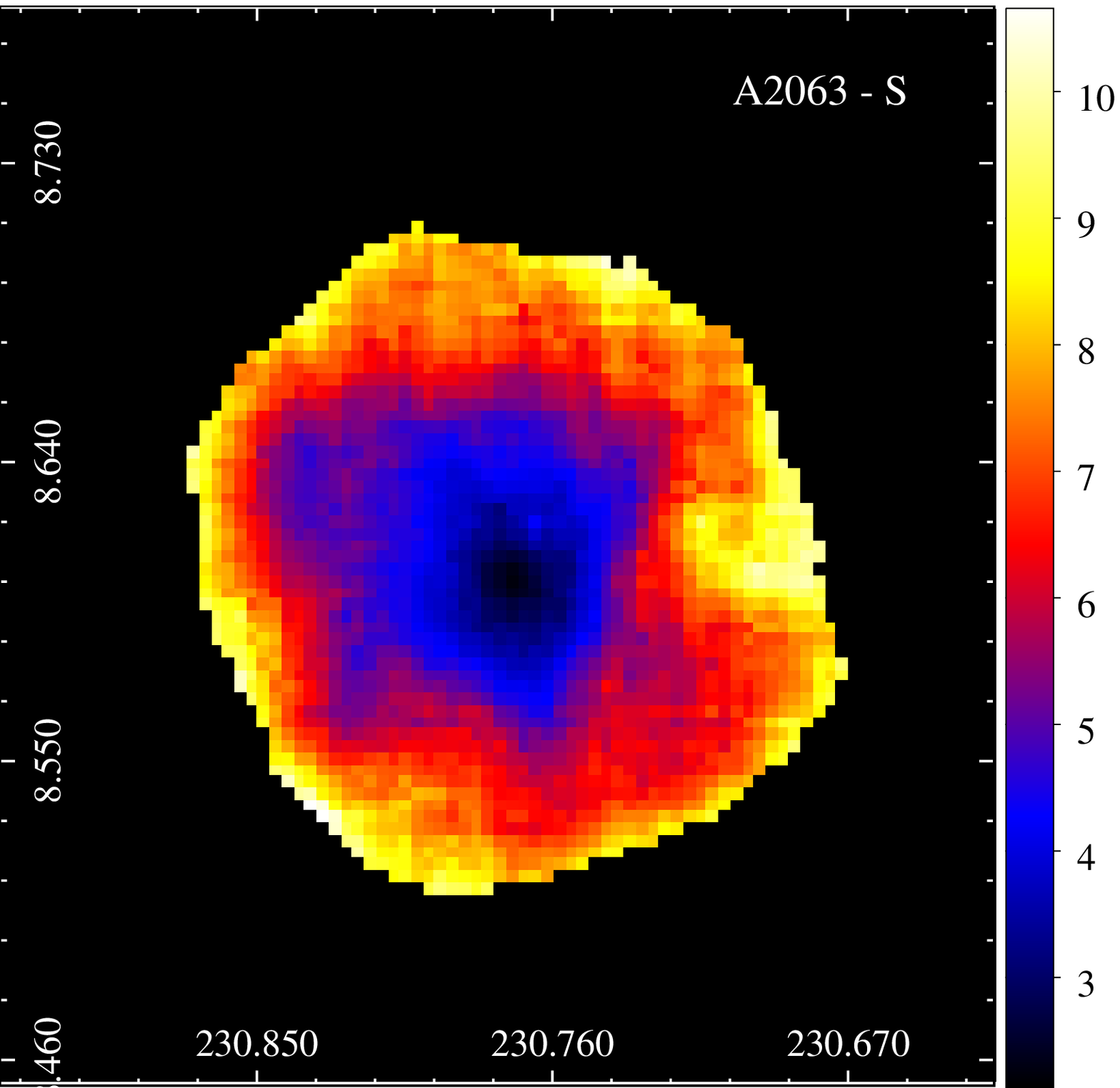}
\includegraphics[scale=0.25]{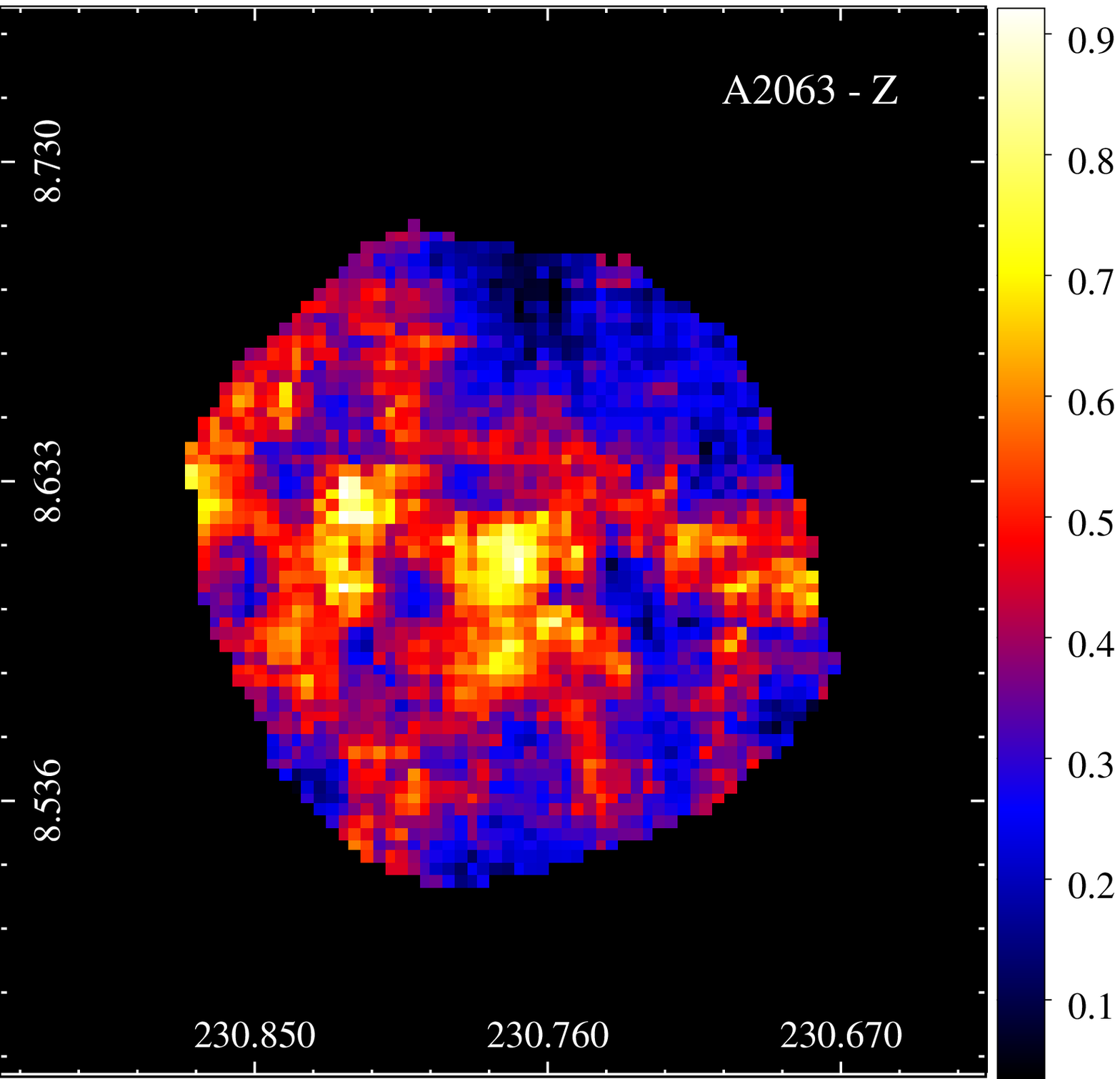}

\caption{CC and relaxed systems. From left to right: temperature,
  pseudo-pressure, pseudo-entropy, and metallicity maps for
  A1650,  A3571,  A1795, A2052, and A2063.}
\label{fig:CCclusters2}
\end{figure*}

\begin{figure*}

\includegraphics[scale=0.235]{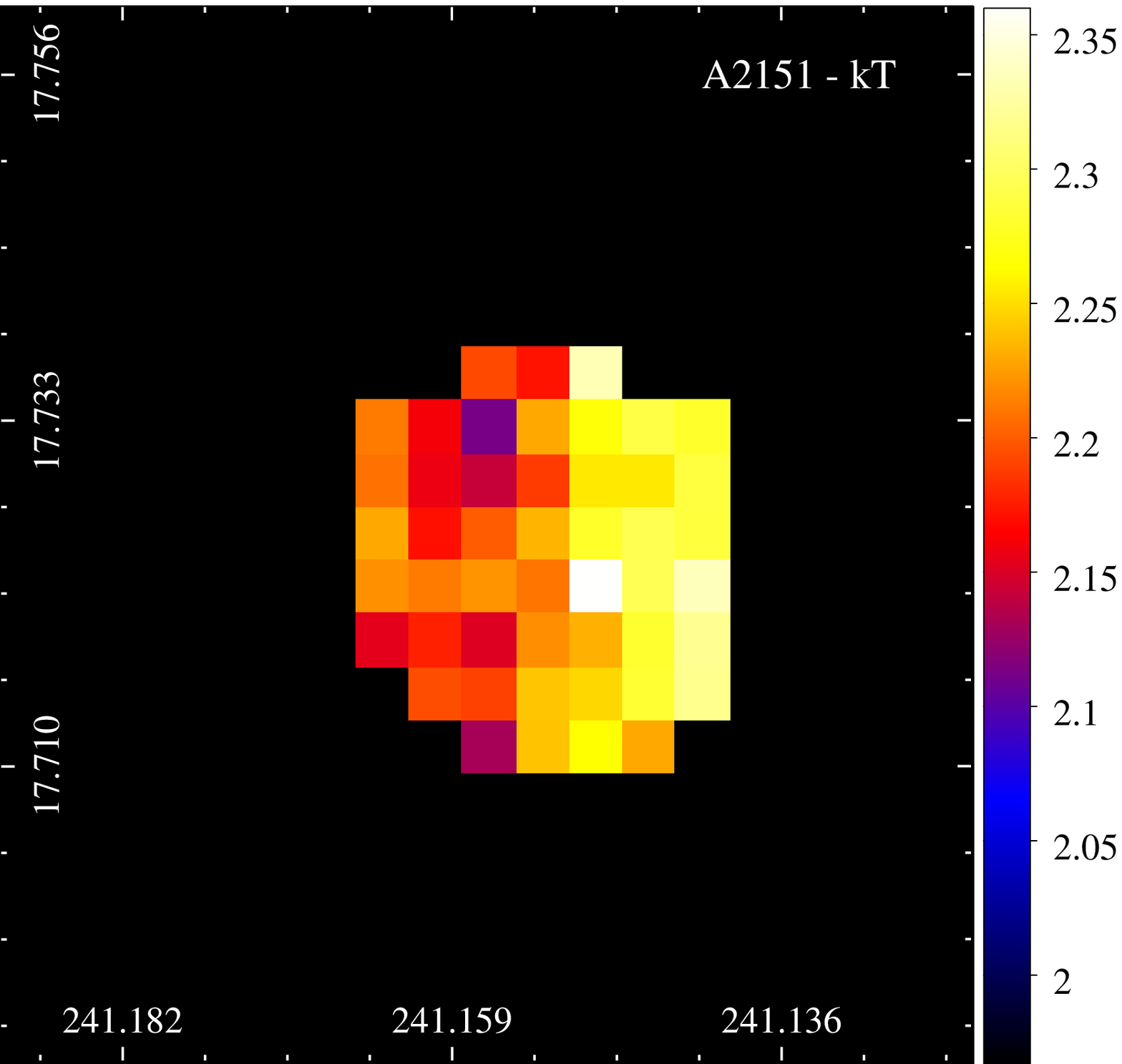}
\includegraphics[scale=0.235]{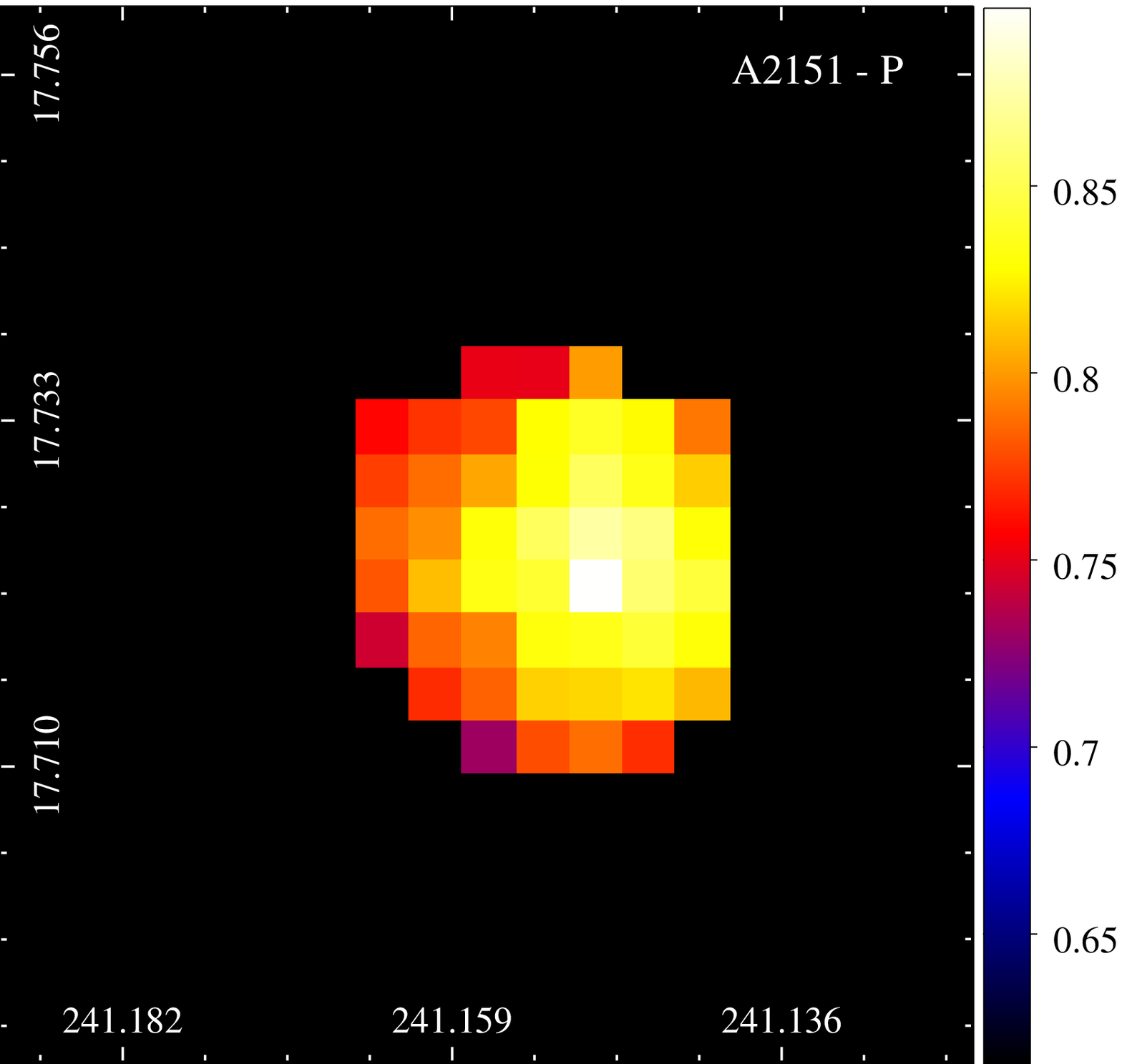}
\includegraphics[scale=0.235]{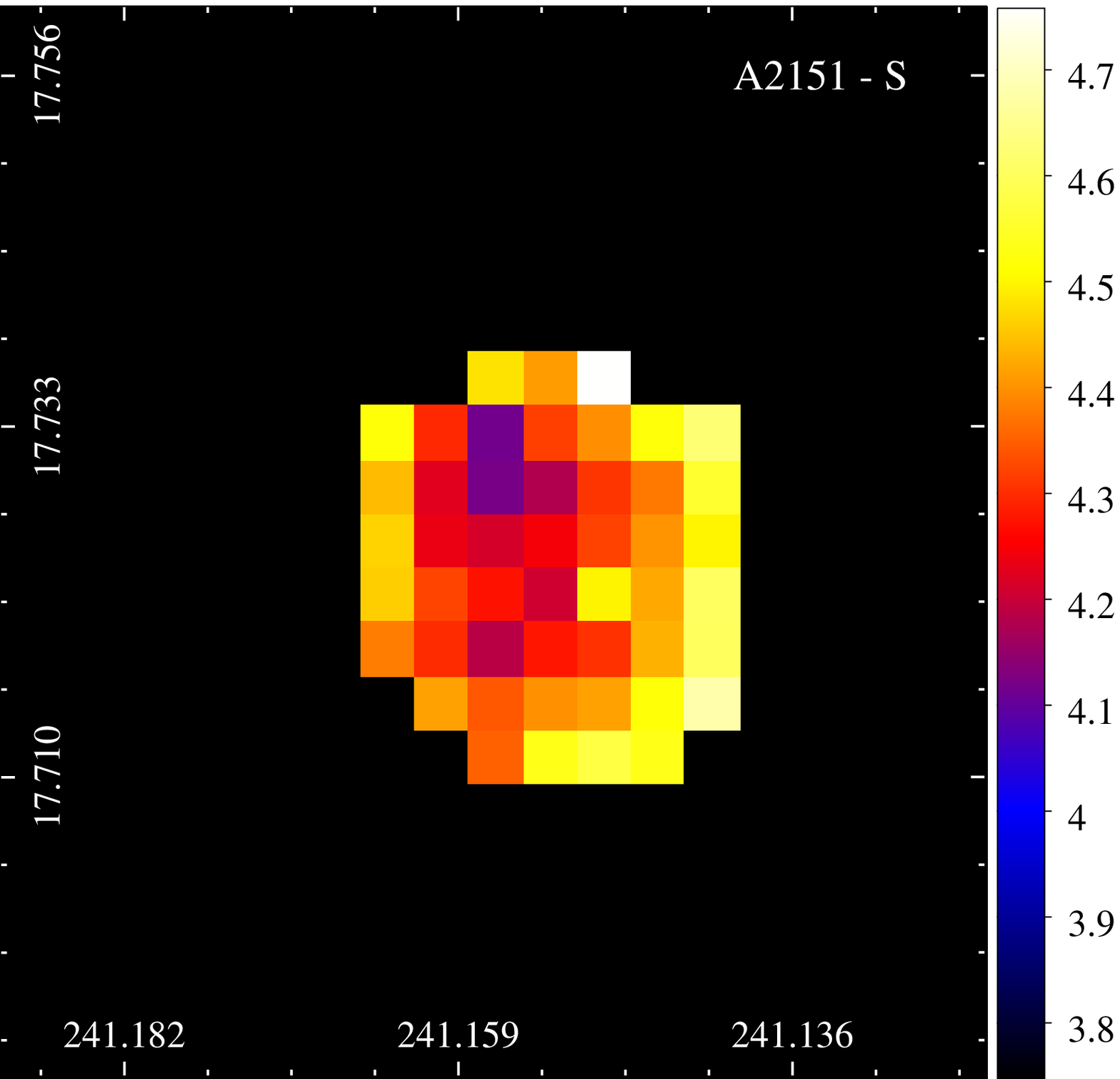}
\includegraphics[scale=0.235]{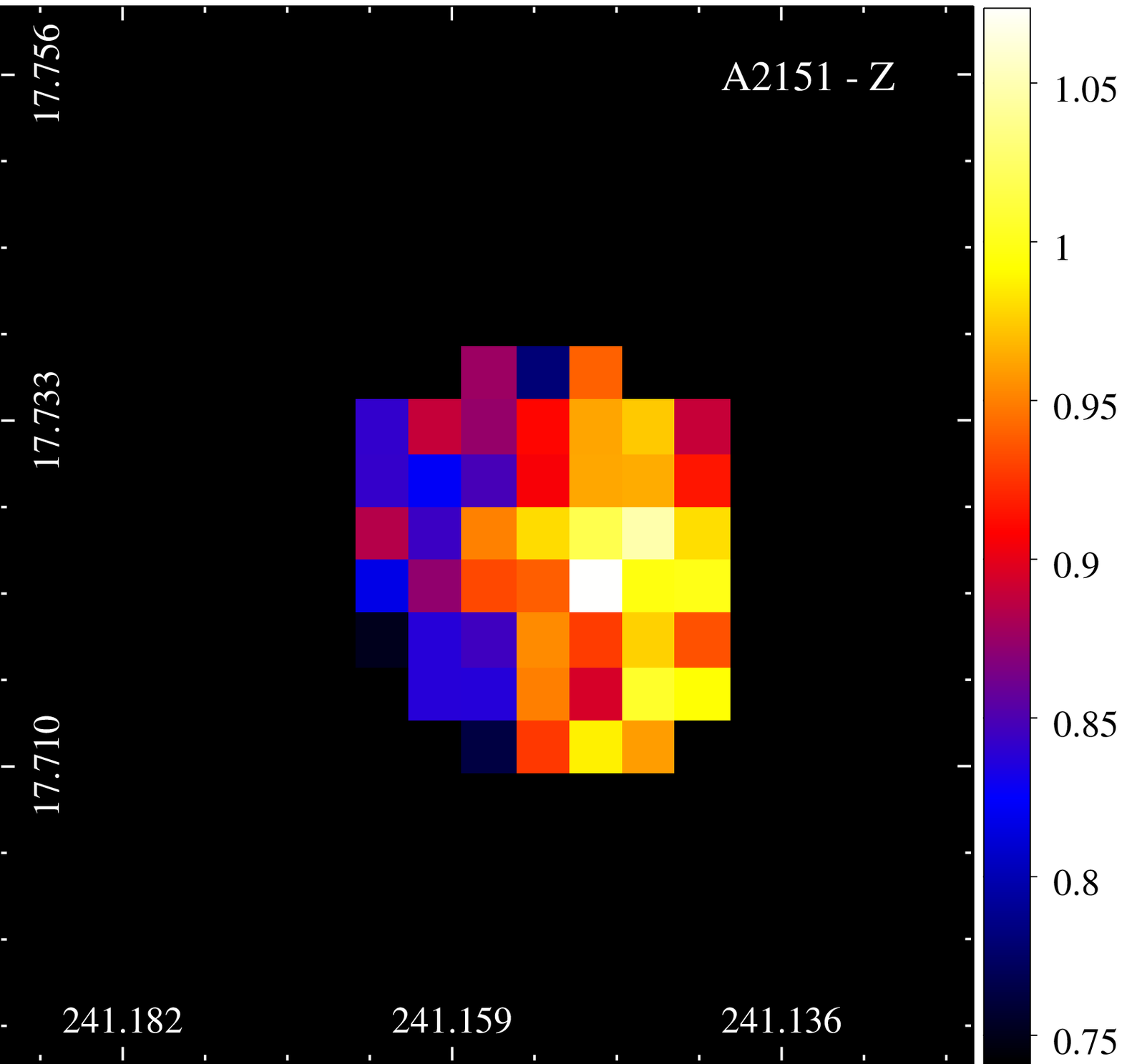}

\includegraphics[scale=0.235]{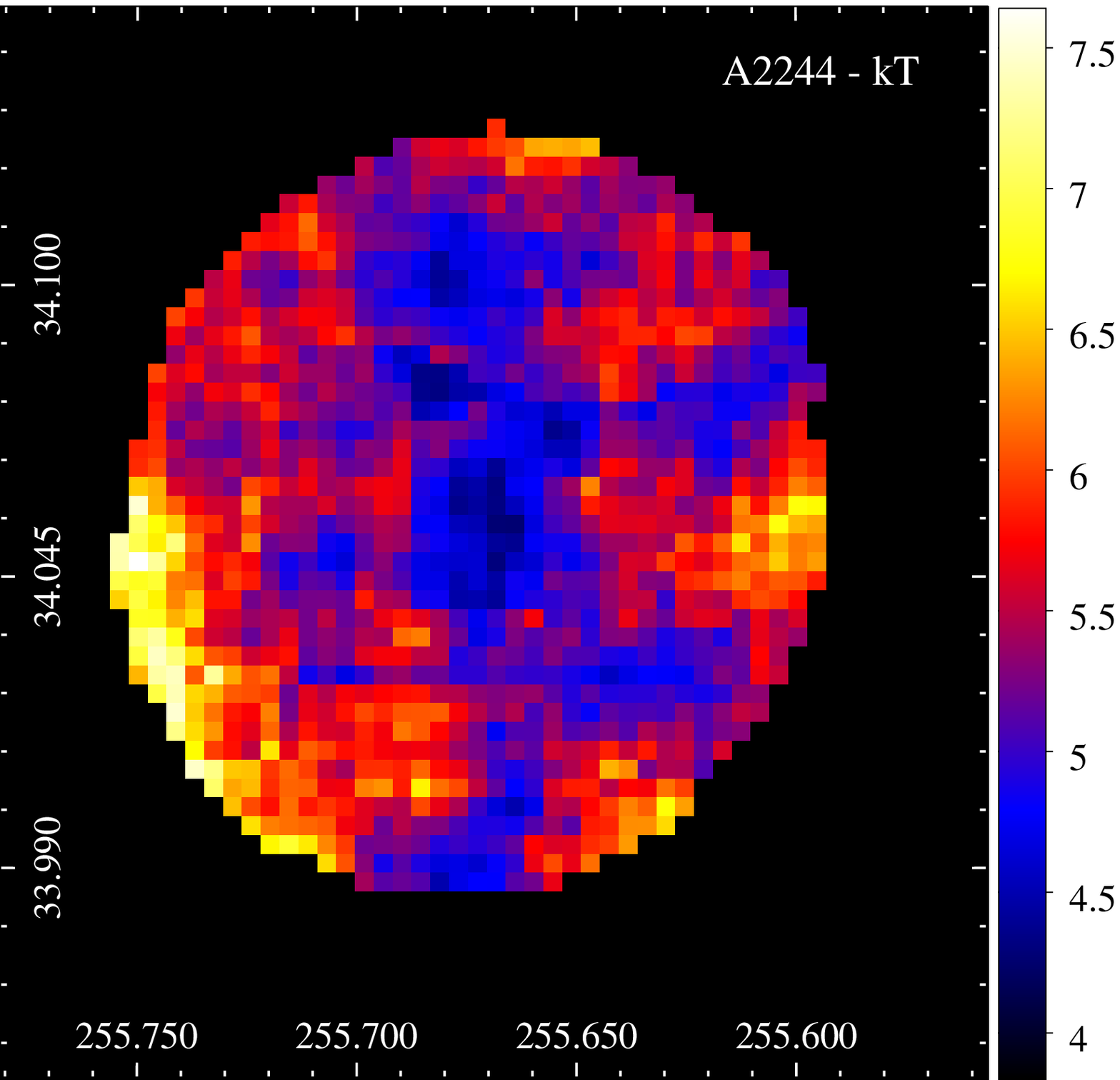}
\includegraphics[scale=0.235]{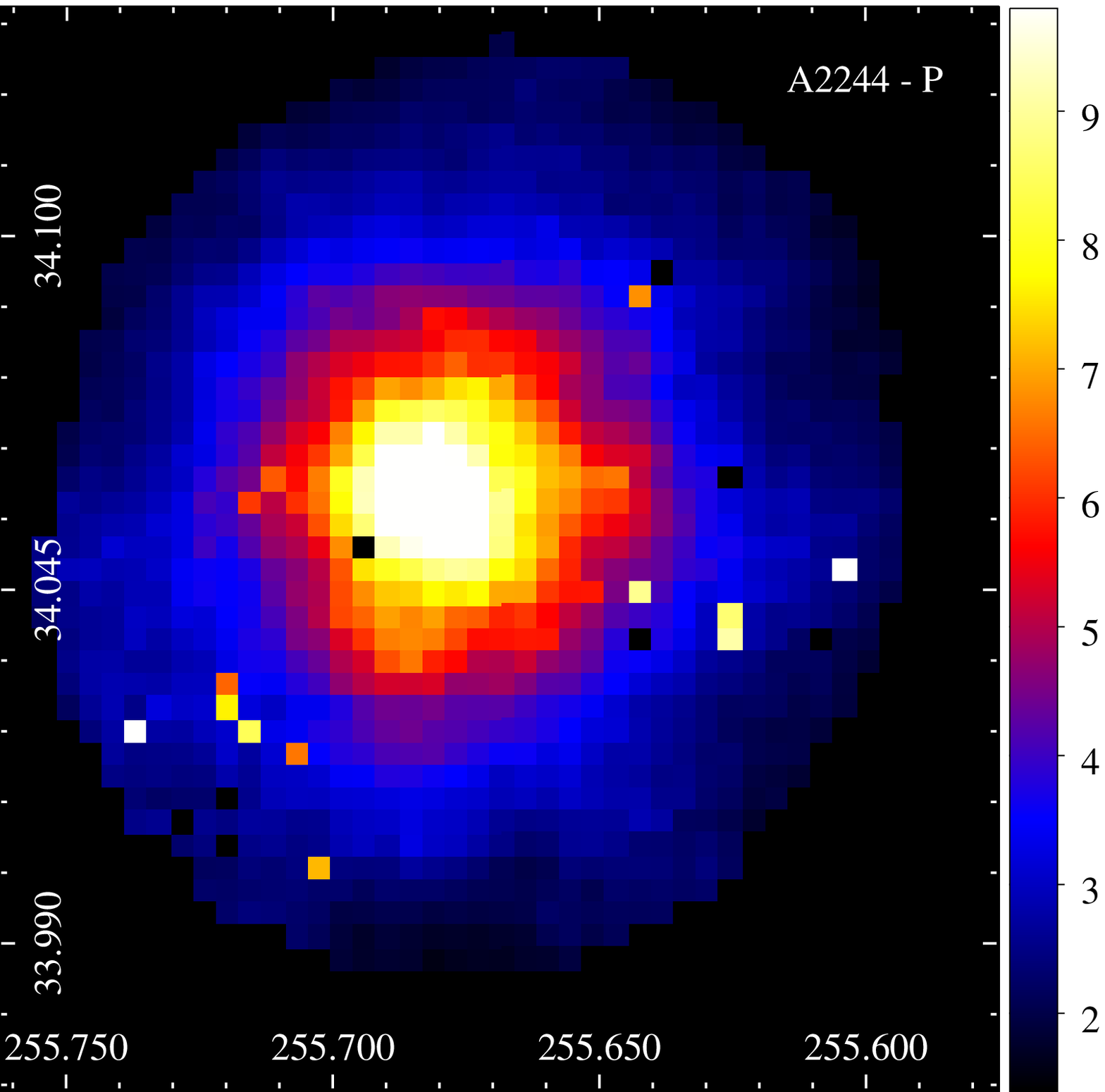}
\includegraphics[scale=0.235]{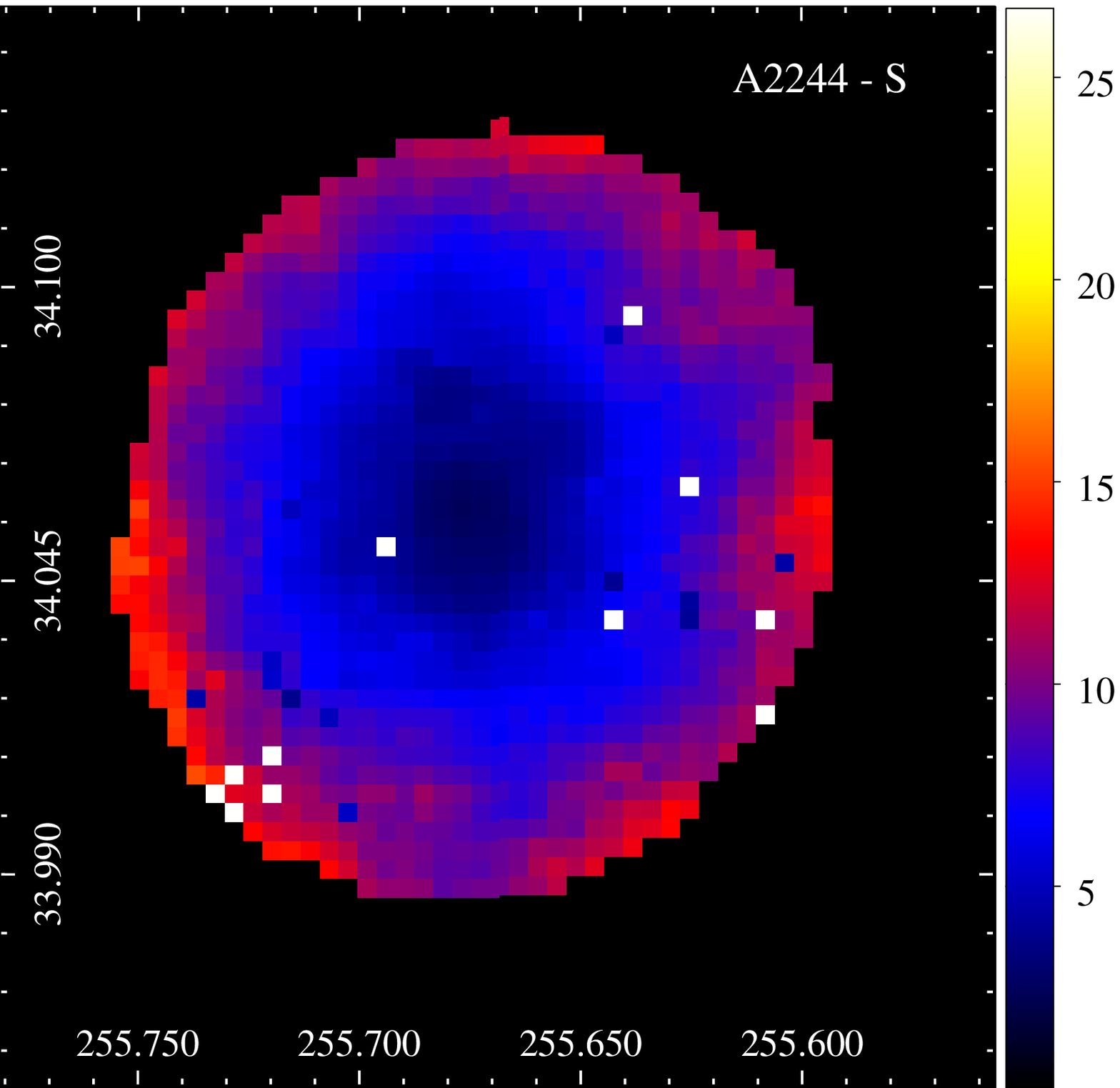}
\includegraphics[scale=0.235]{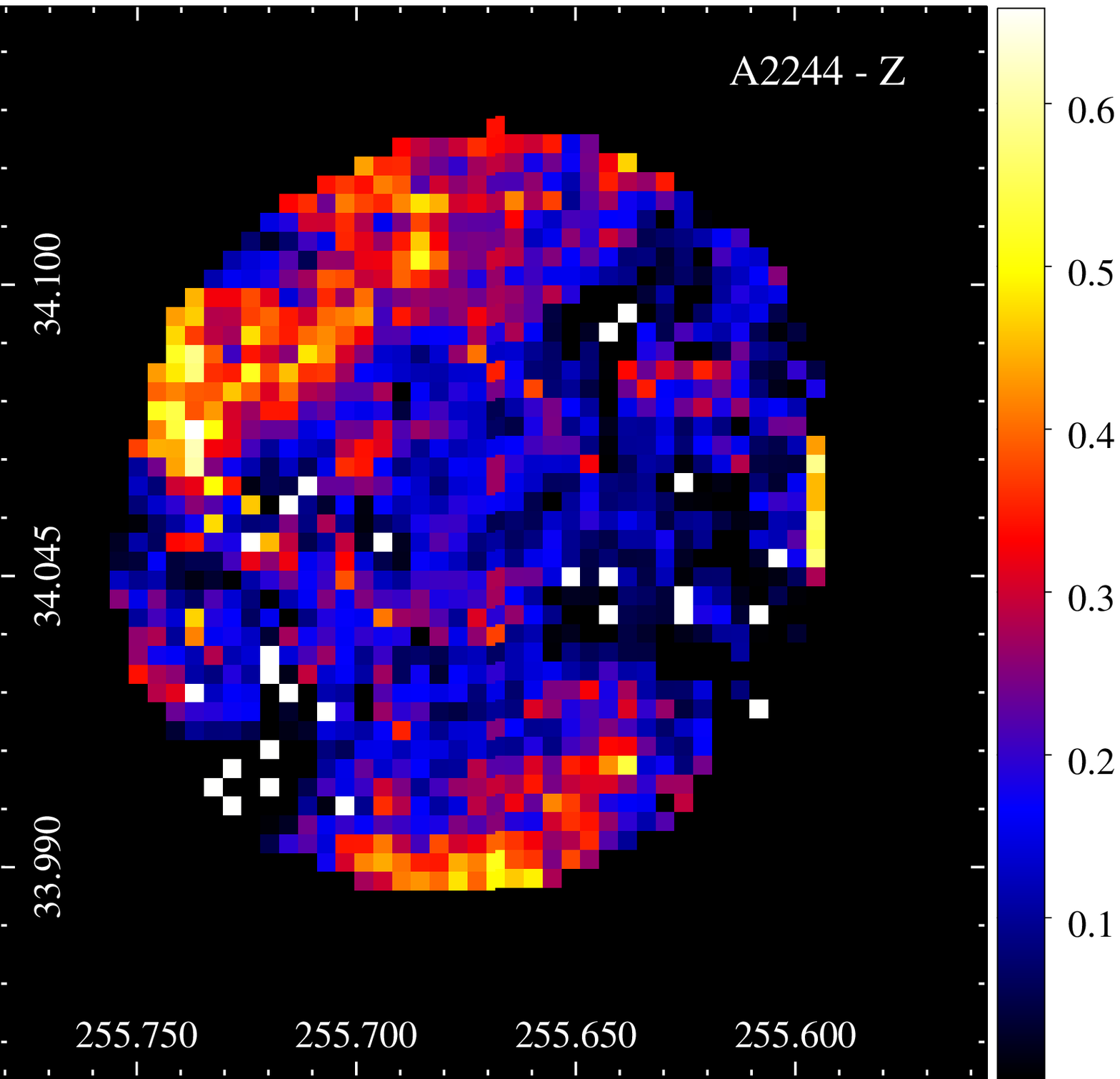}

\includegraphics[scale=0.235]{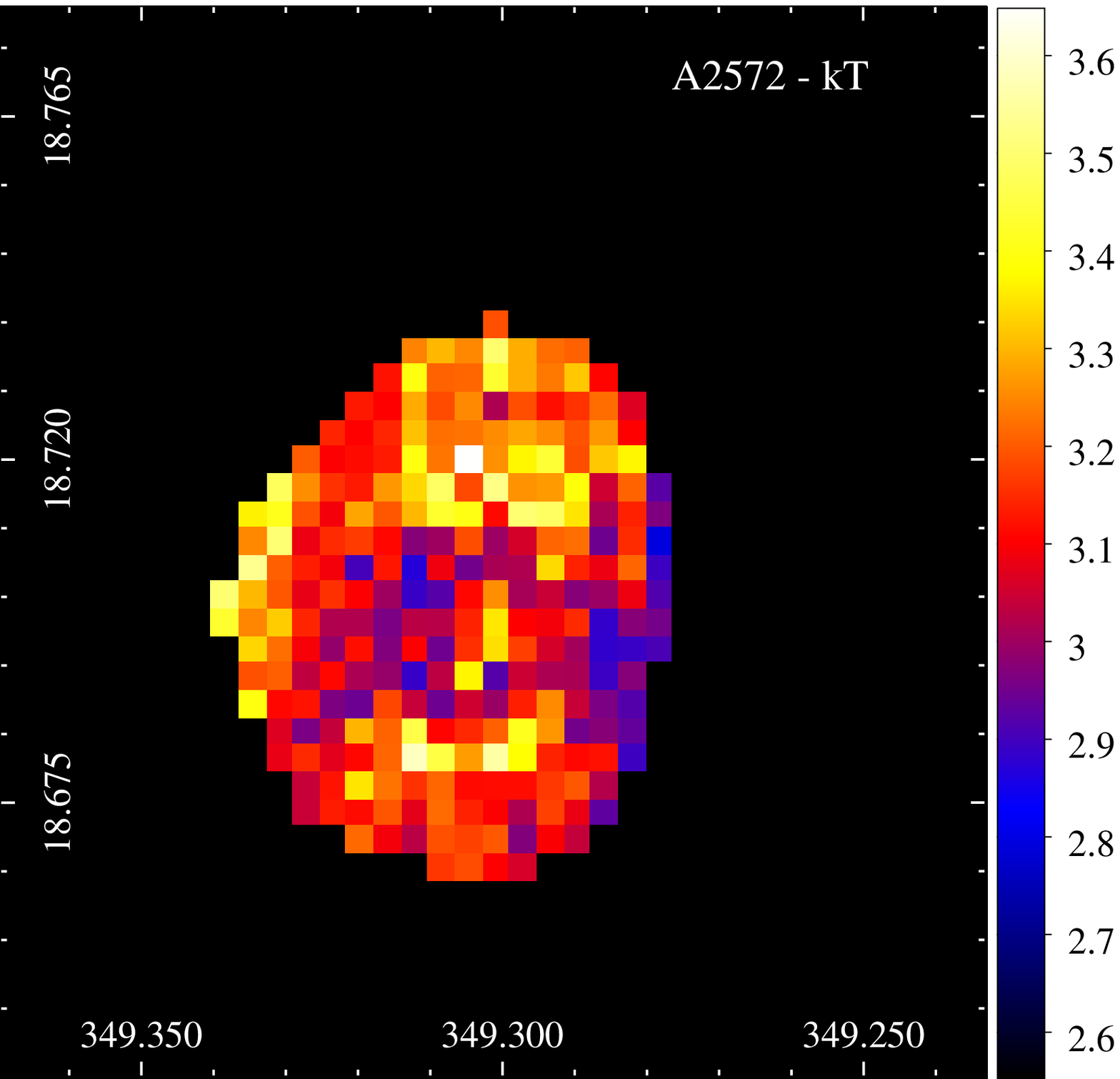}
\includegraphics[scale=0.235]{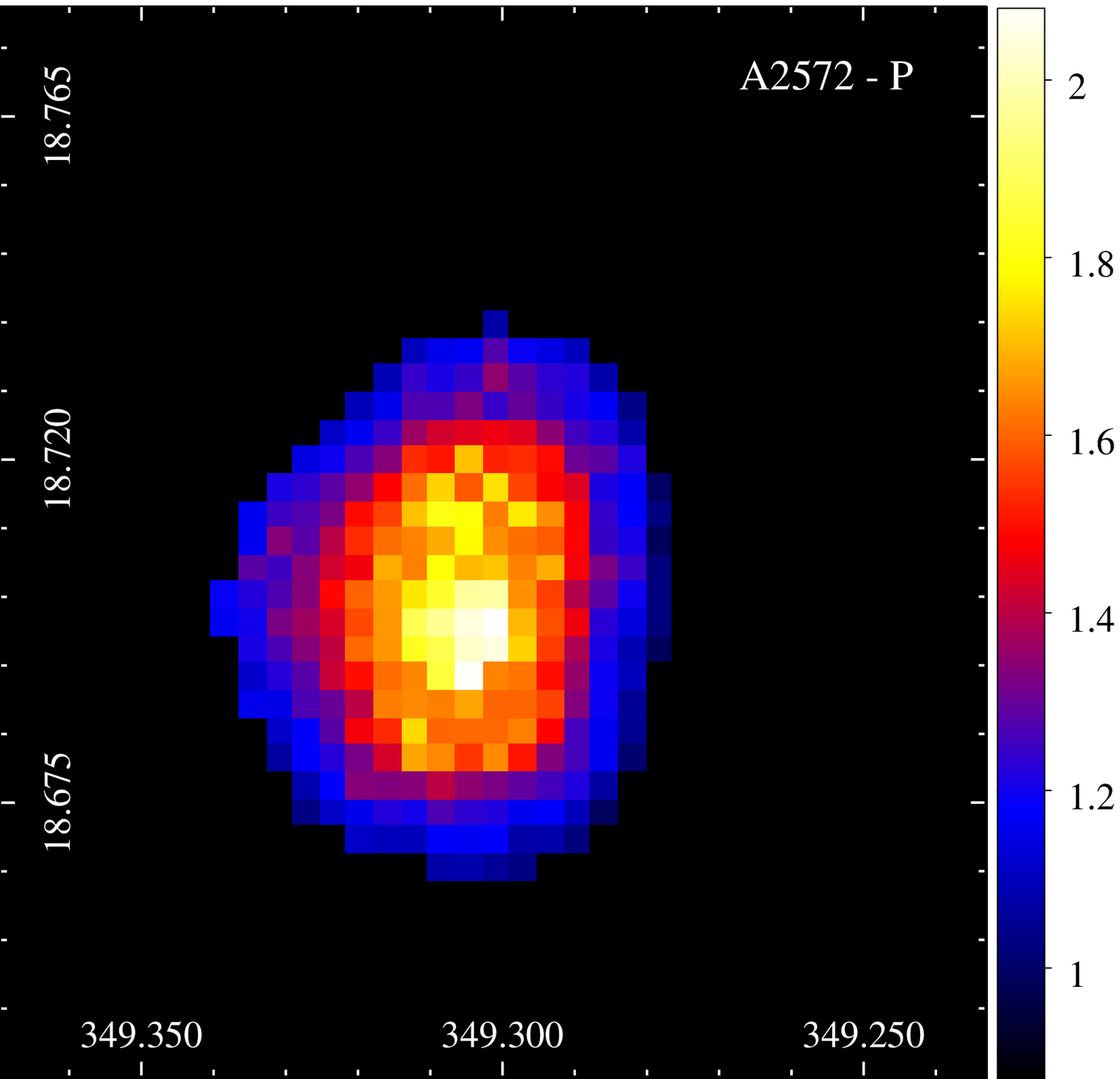}
\includegraphics[scale=0.235]{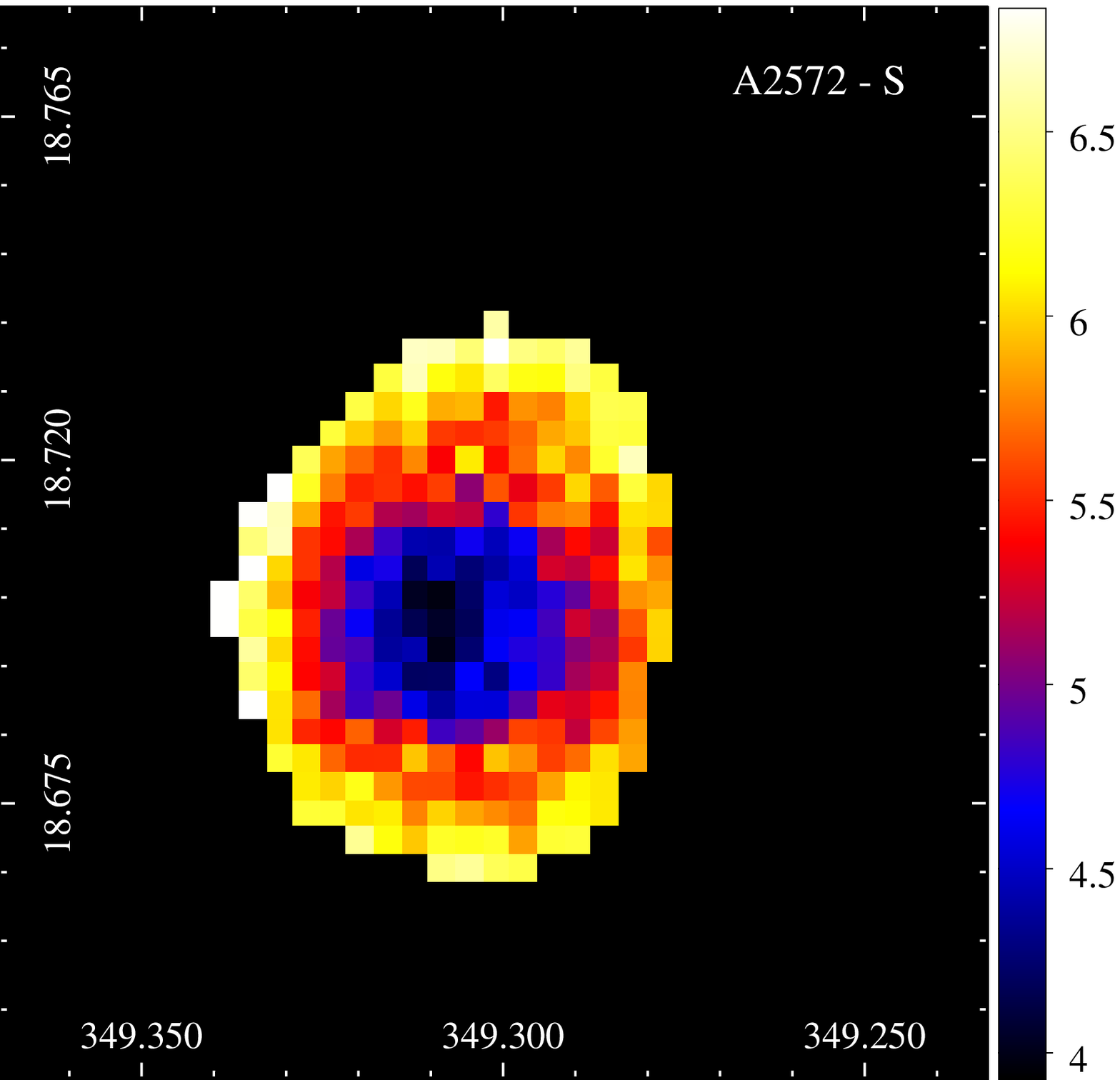}
\includegraphics[scale=0.235]{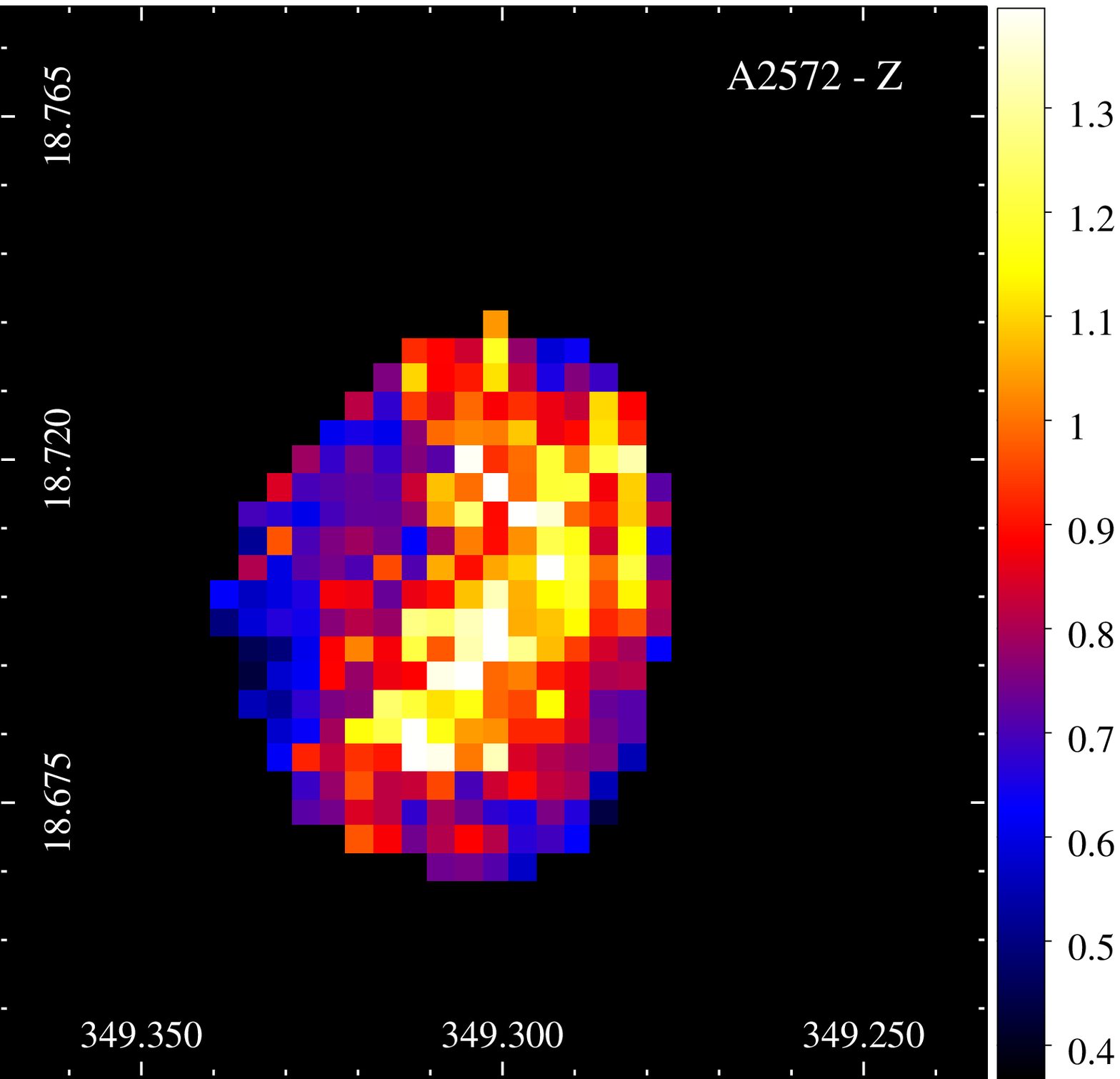}

\includegraphics[scale=0.235]{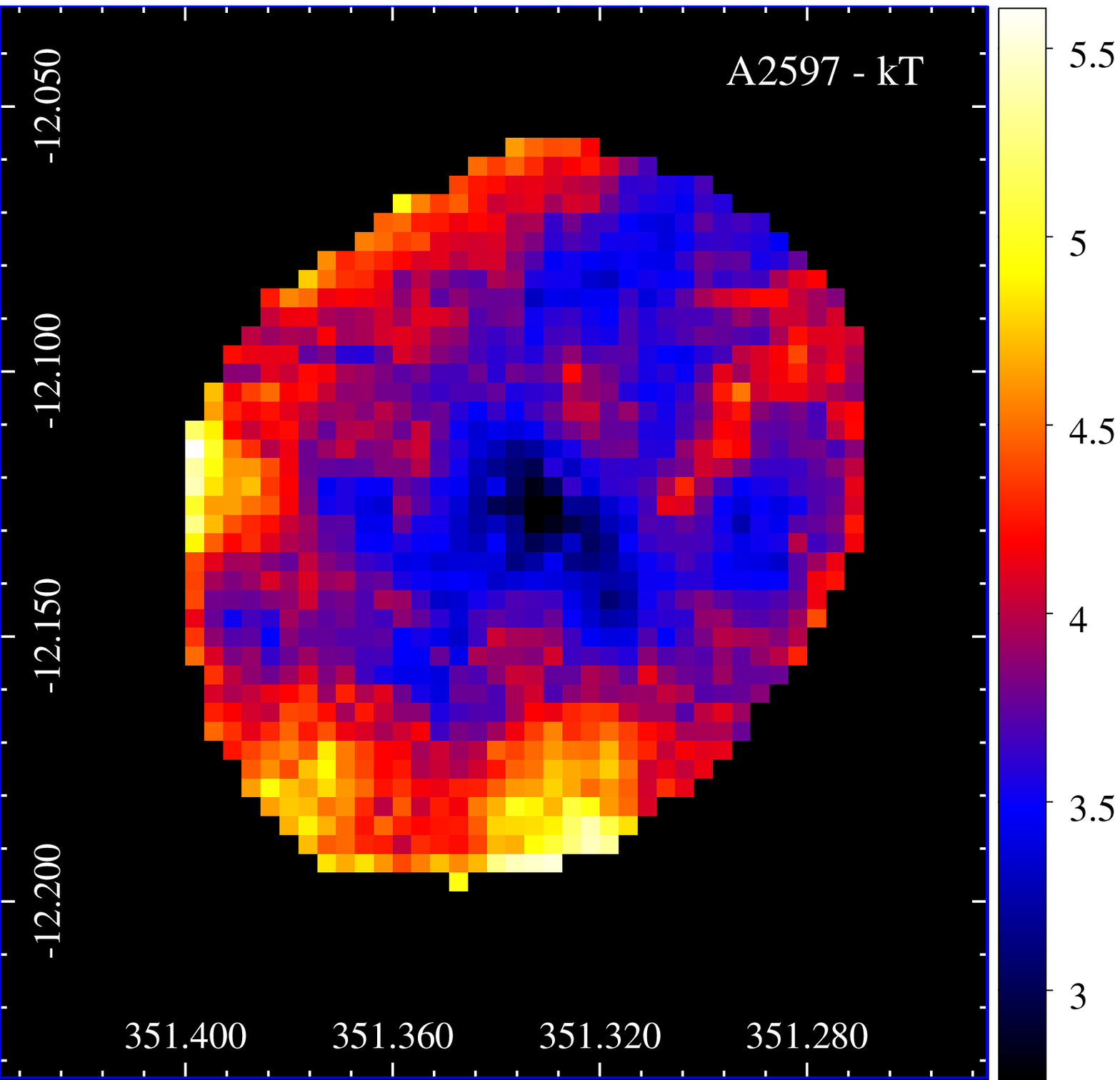}
\includegraphics[scale=0.235]{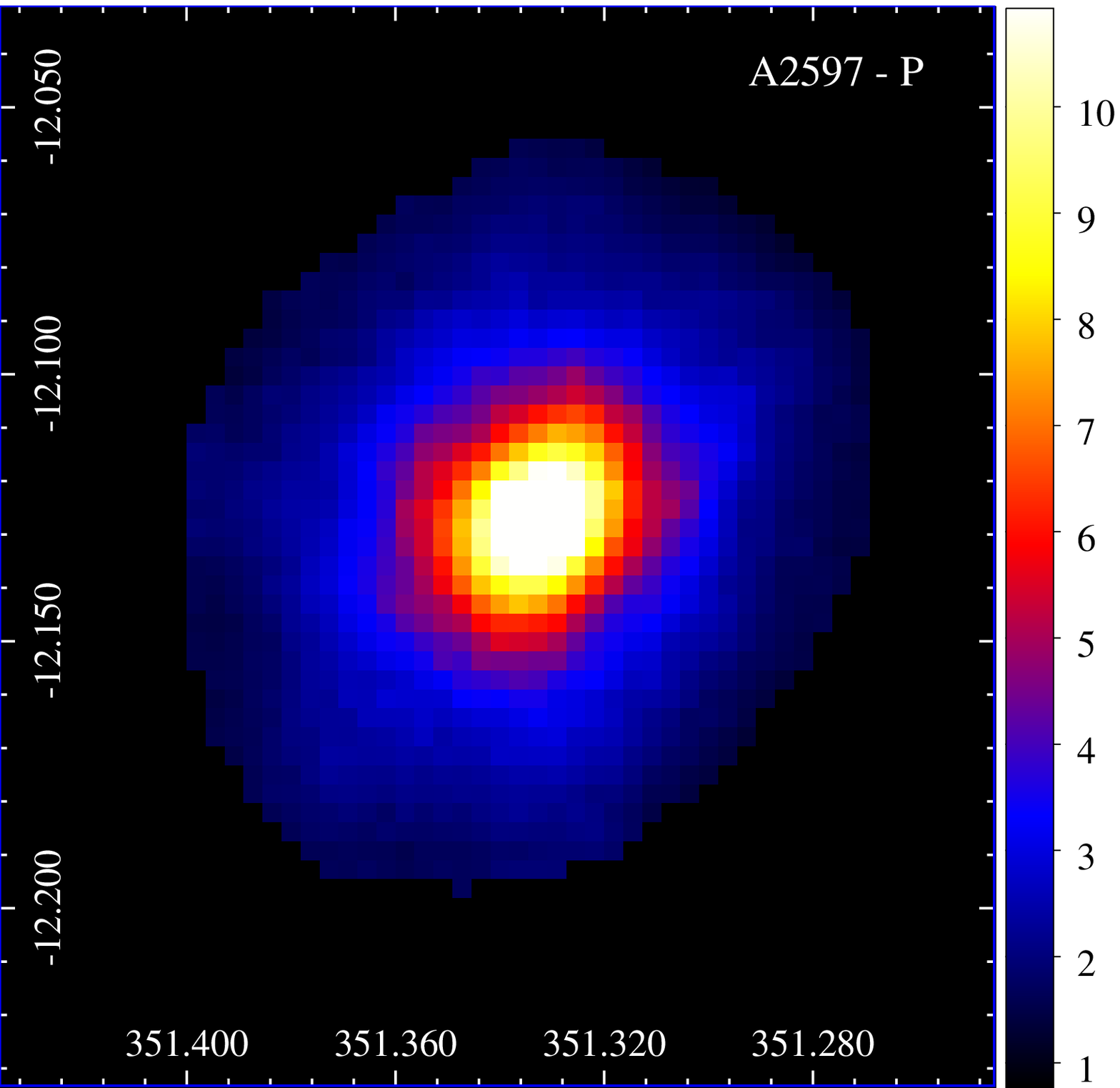}
\includegraphics[scale=0.235]{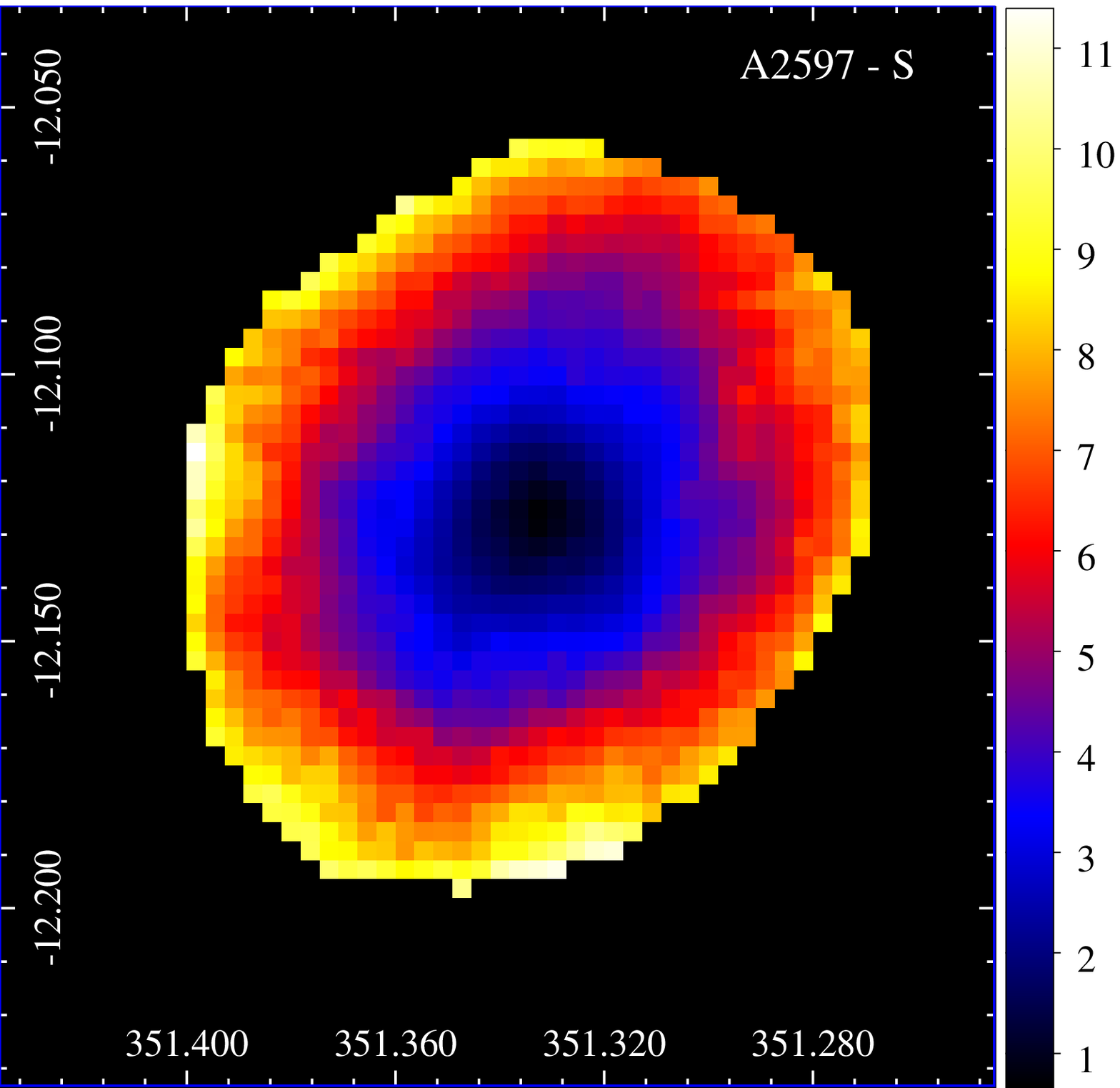}
\includegraphics[scale=0.235]{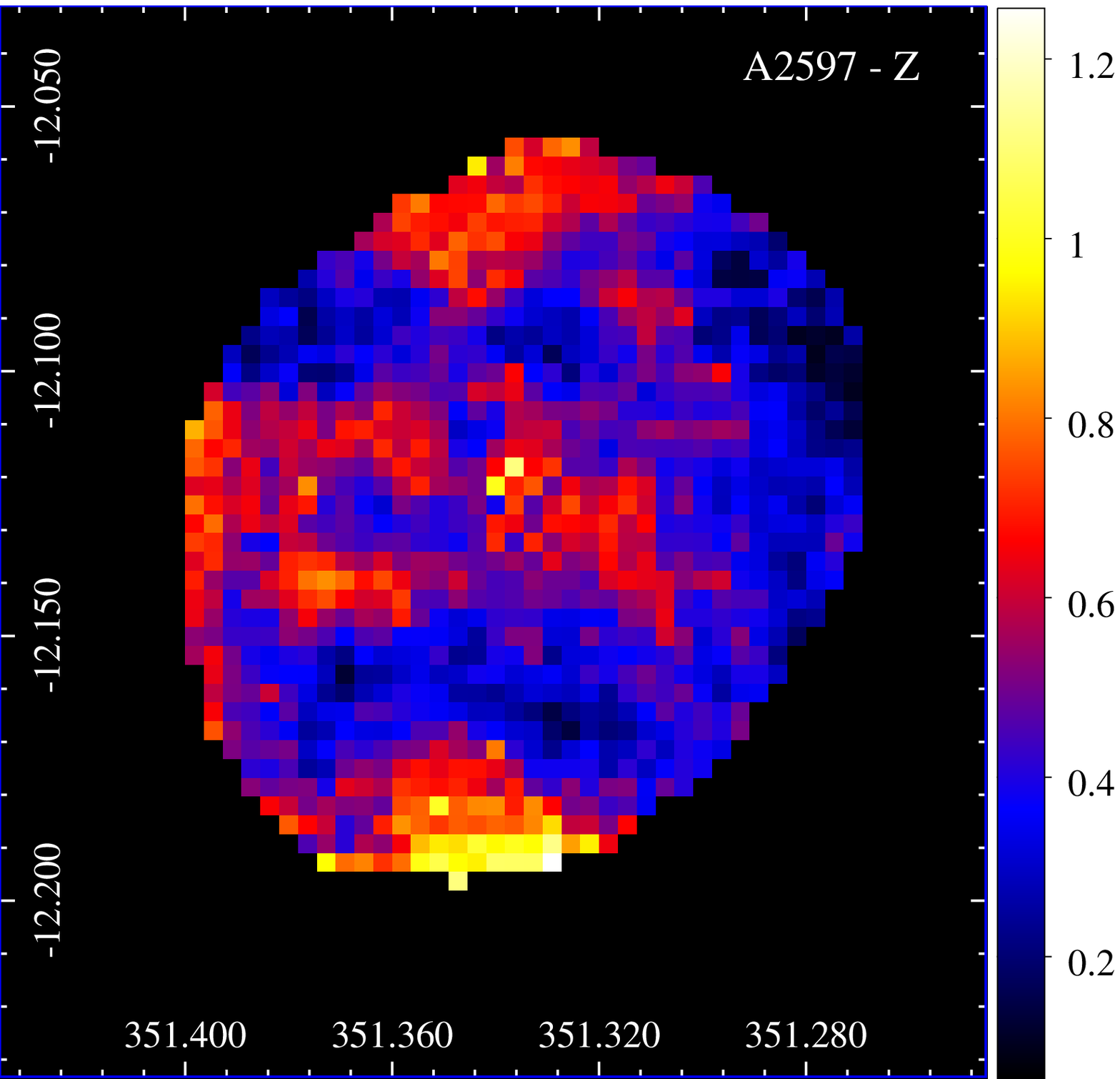}

\includegraphics[scale=0.235]{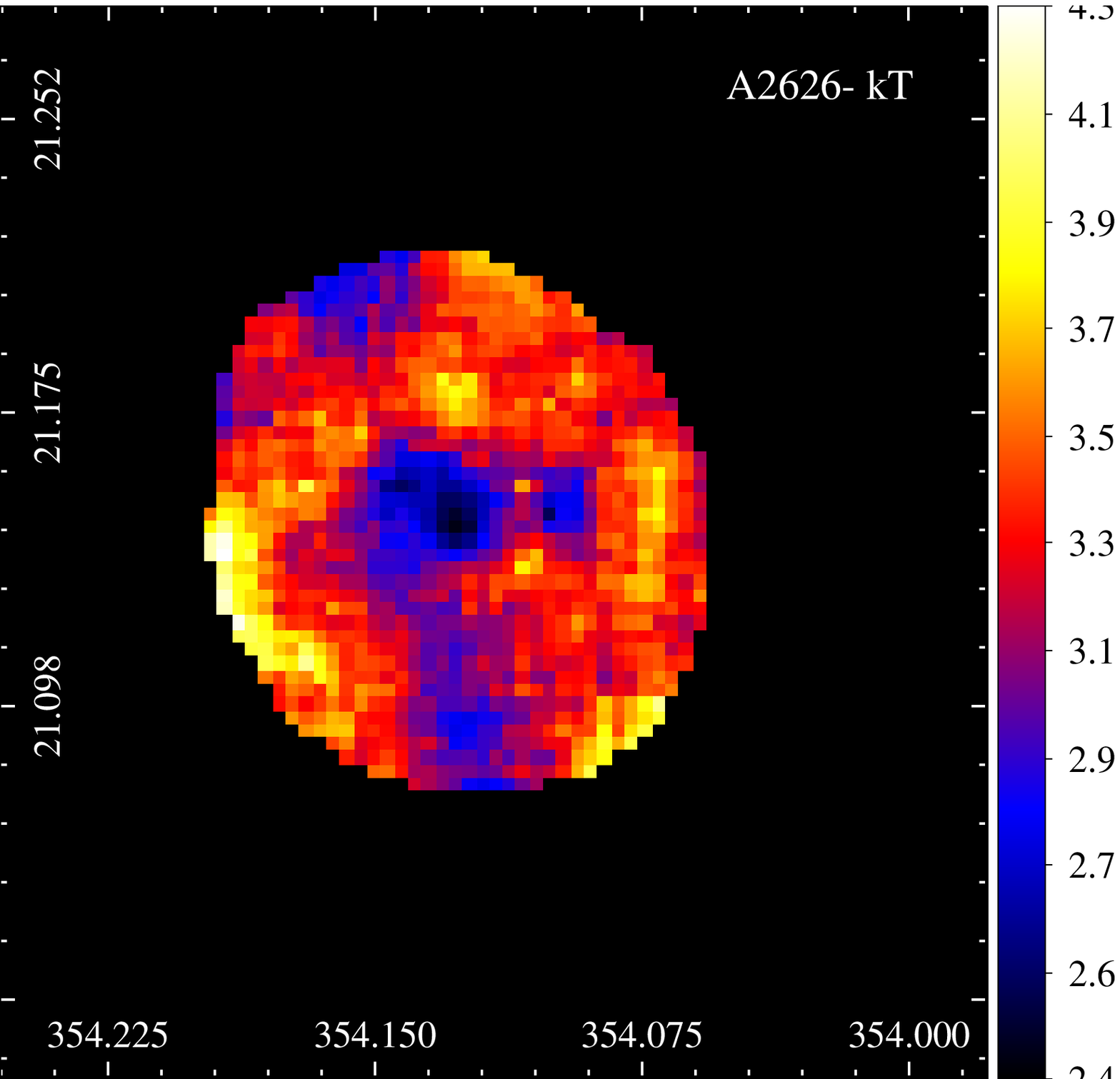}
\includegraphics[scale=0.235]{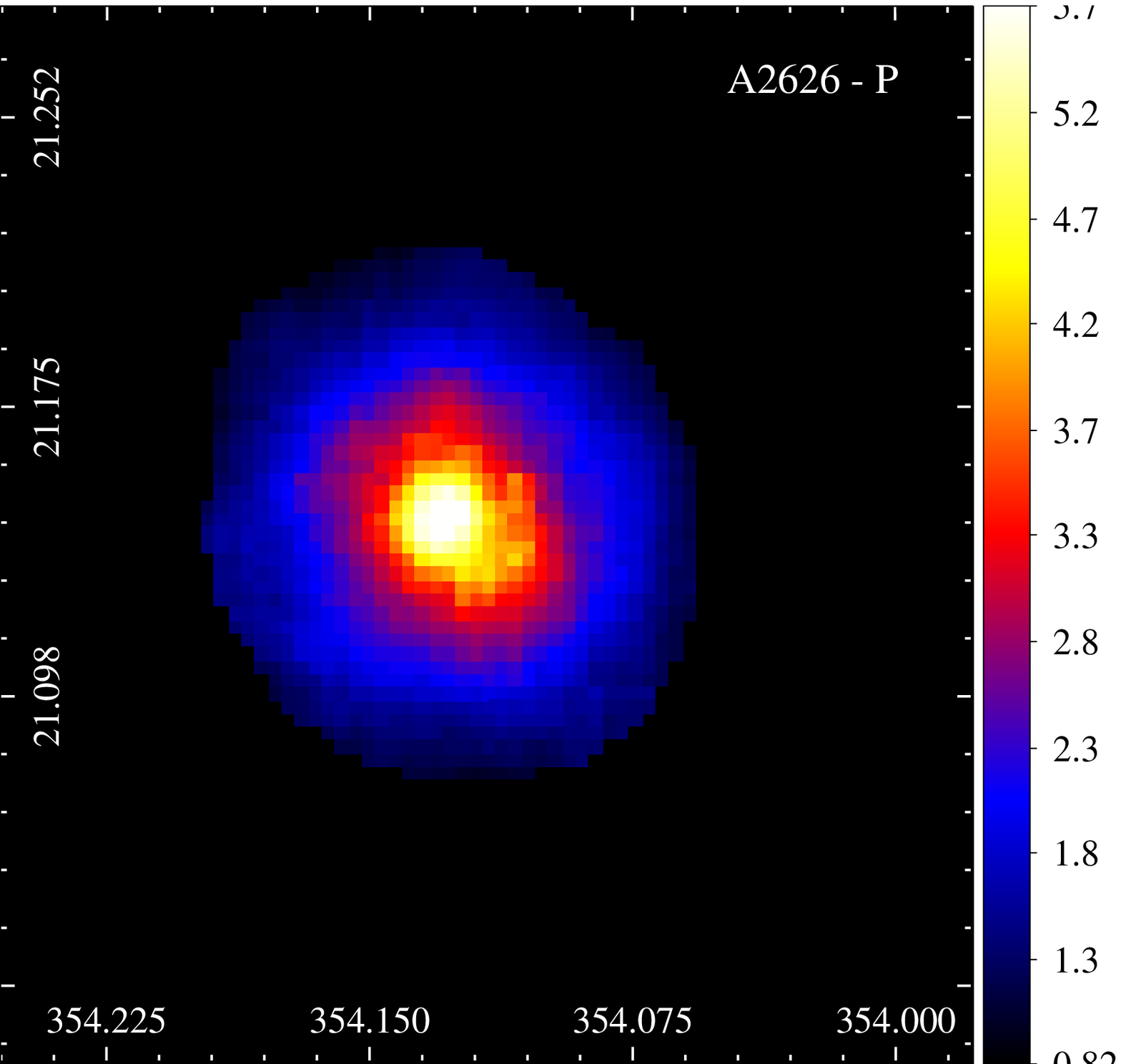}
\includegraphics[scale=0.235]{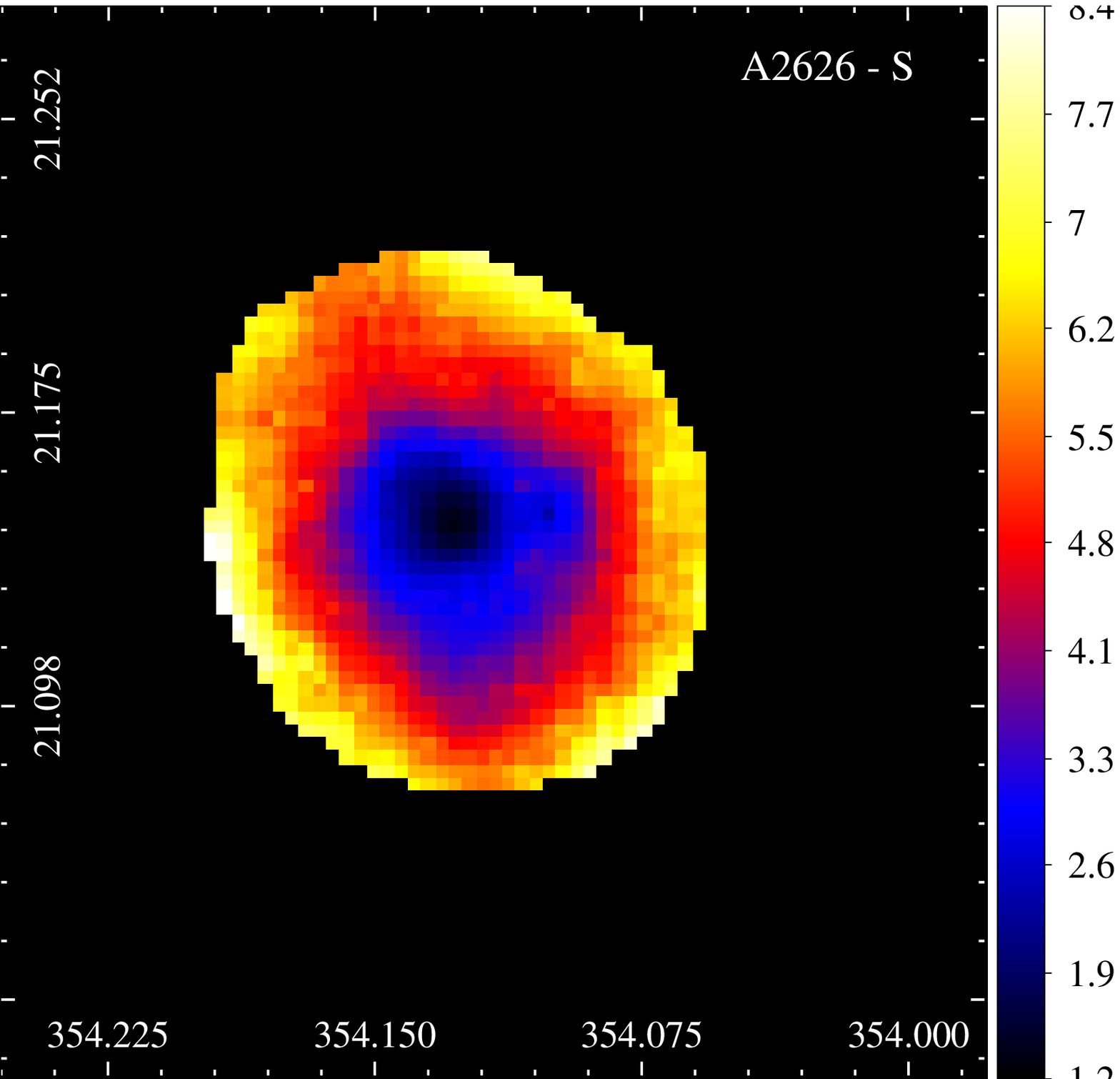}
\includegraphics[scale=0.235]{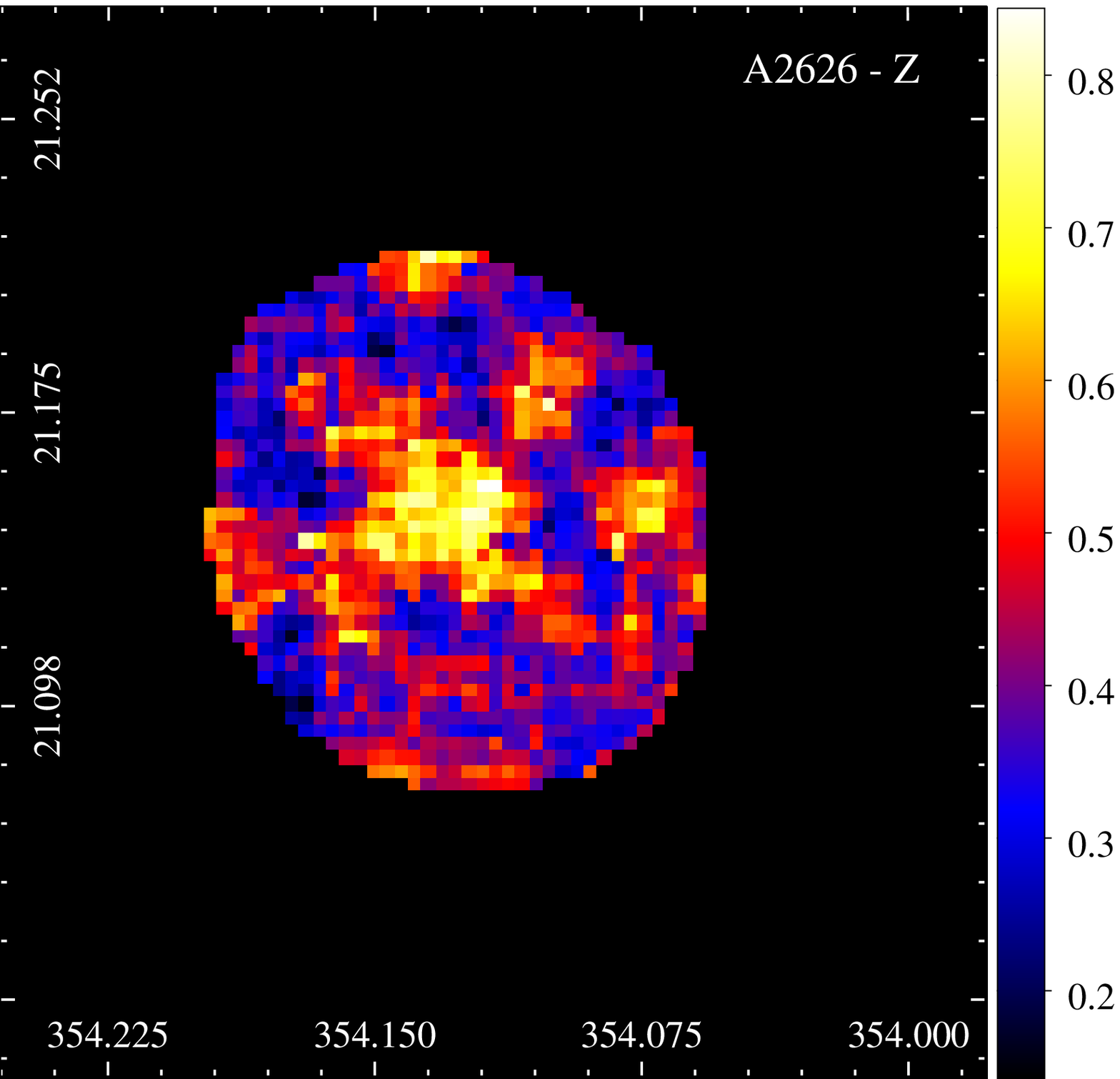}

\includegraphics[scale=0.235]{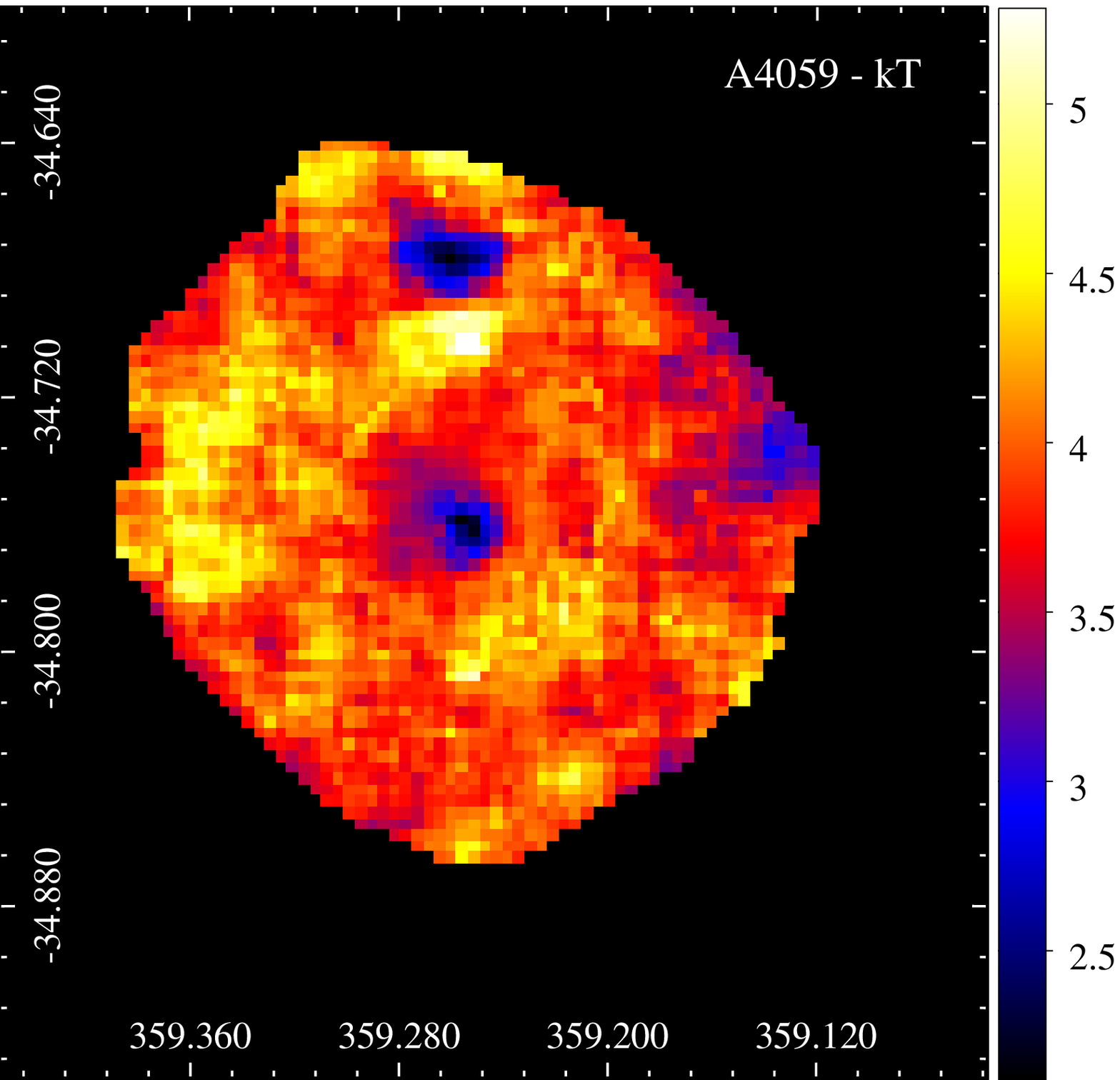}
\includegraphics[scale=0.235]{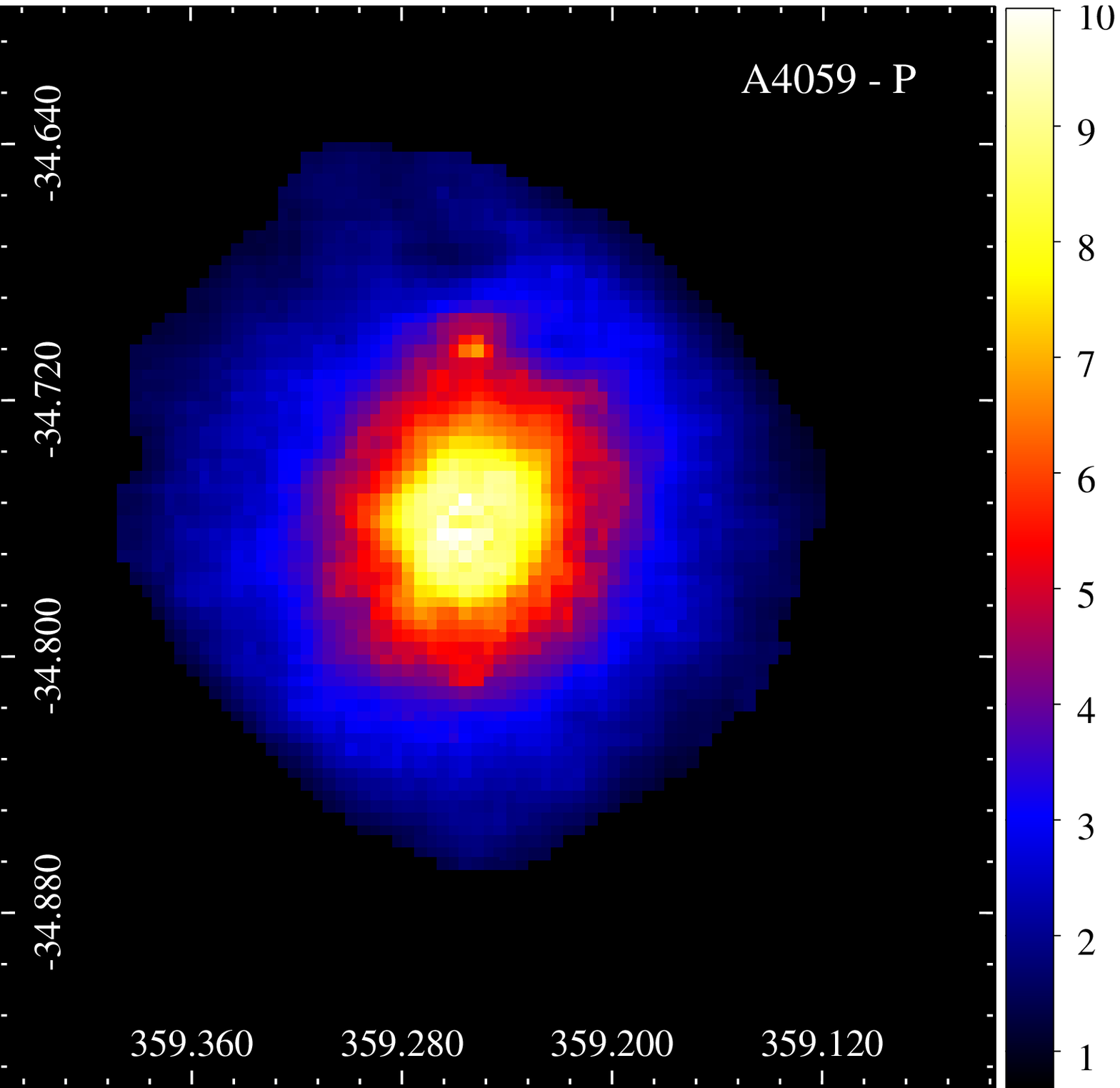}
\includegraphics[scale=0.235]{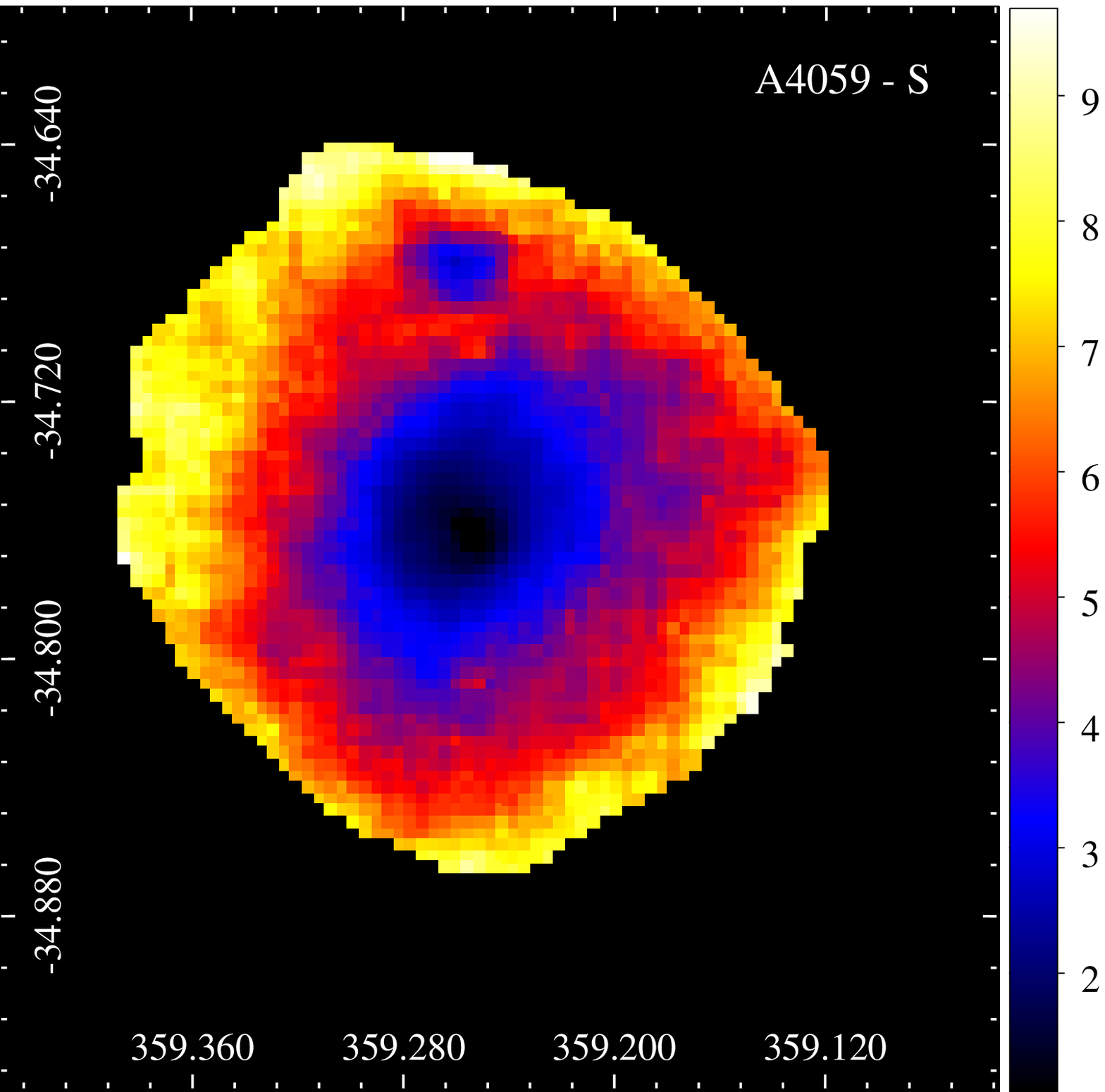}
\includegraphics[scale=0.235]{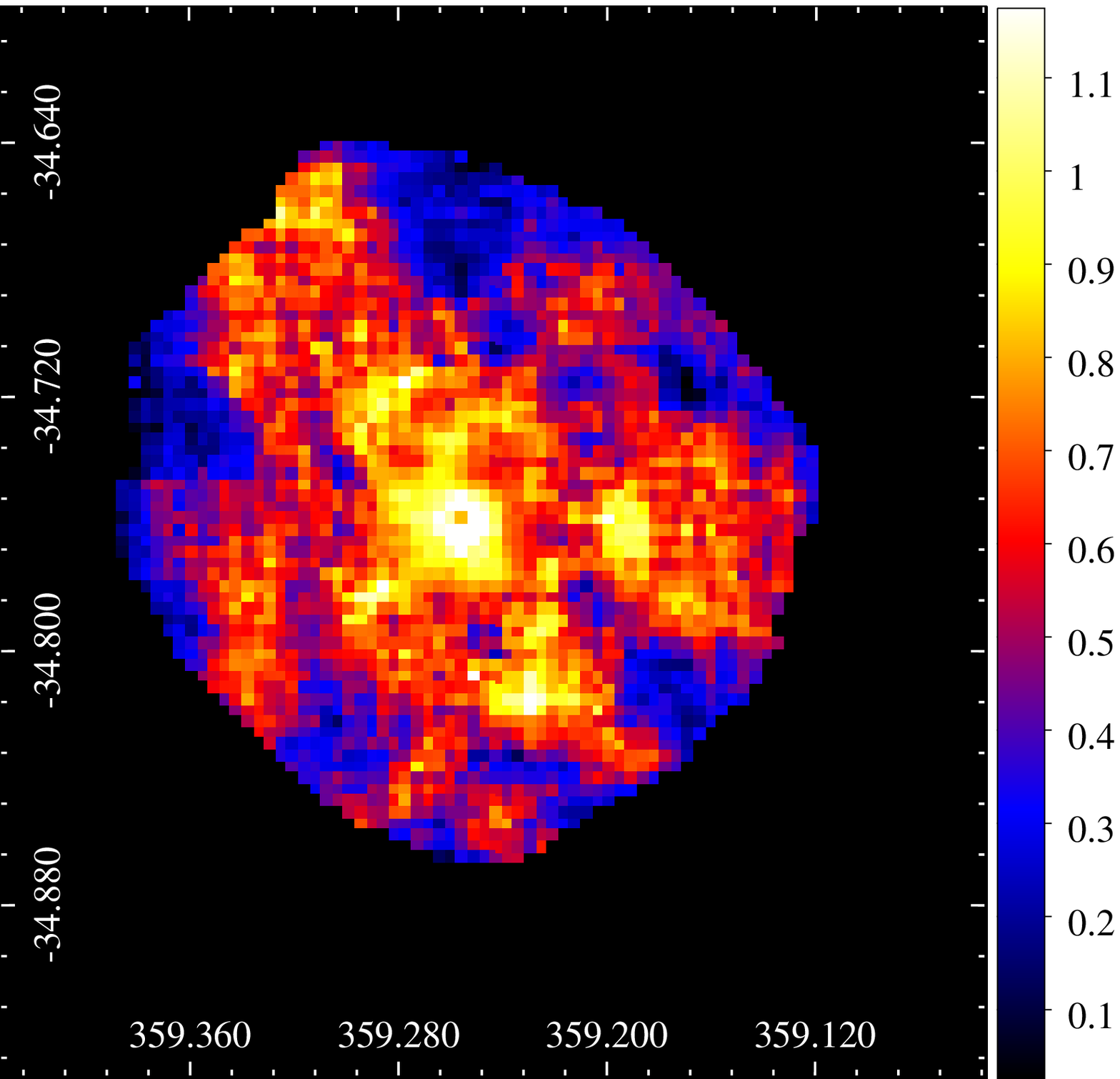}

\caption{CC and relaxed systems. From left to right: temperature,
  pseudo-pressure, pseudo-entropy, and metallicity maps for
   A2151, A2244, A2572, A2597, A2626, and A4059.}
\label{fig:CCclusters3}
\end{figure*}

%%%############ CC-disturbed 

\begin{figure*}
\includegraphics[scale=0.25]{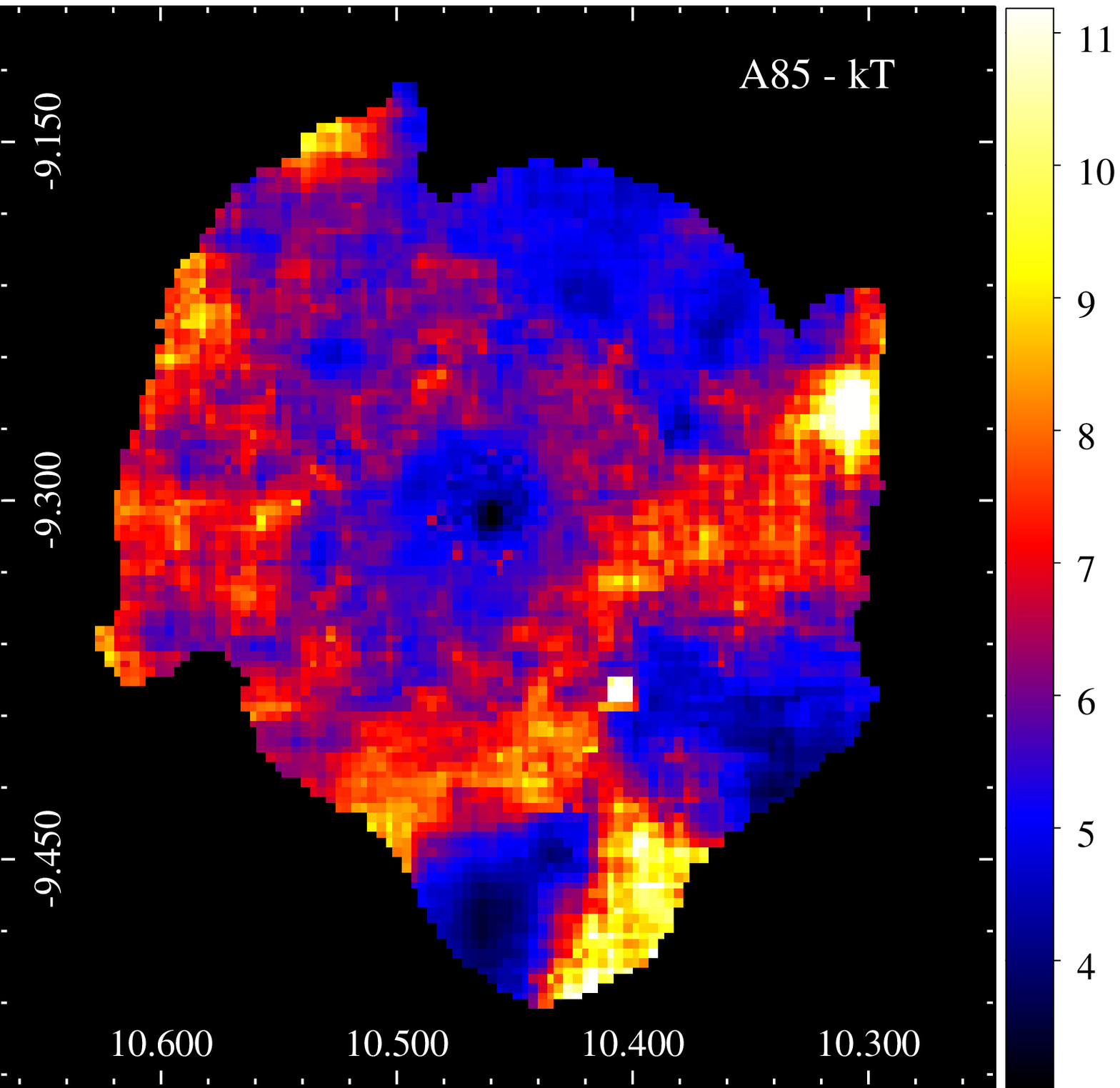}
\includegraphics[scale=0.25]{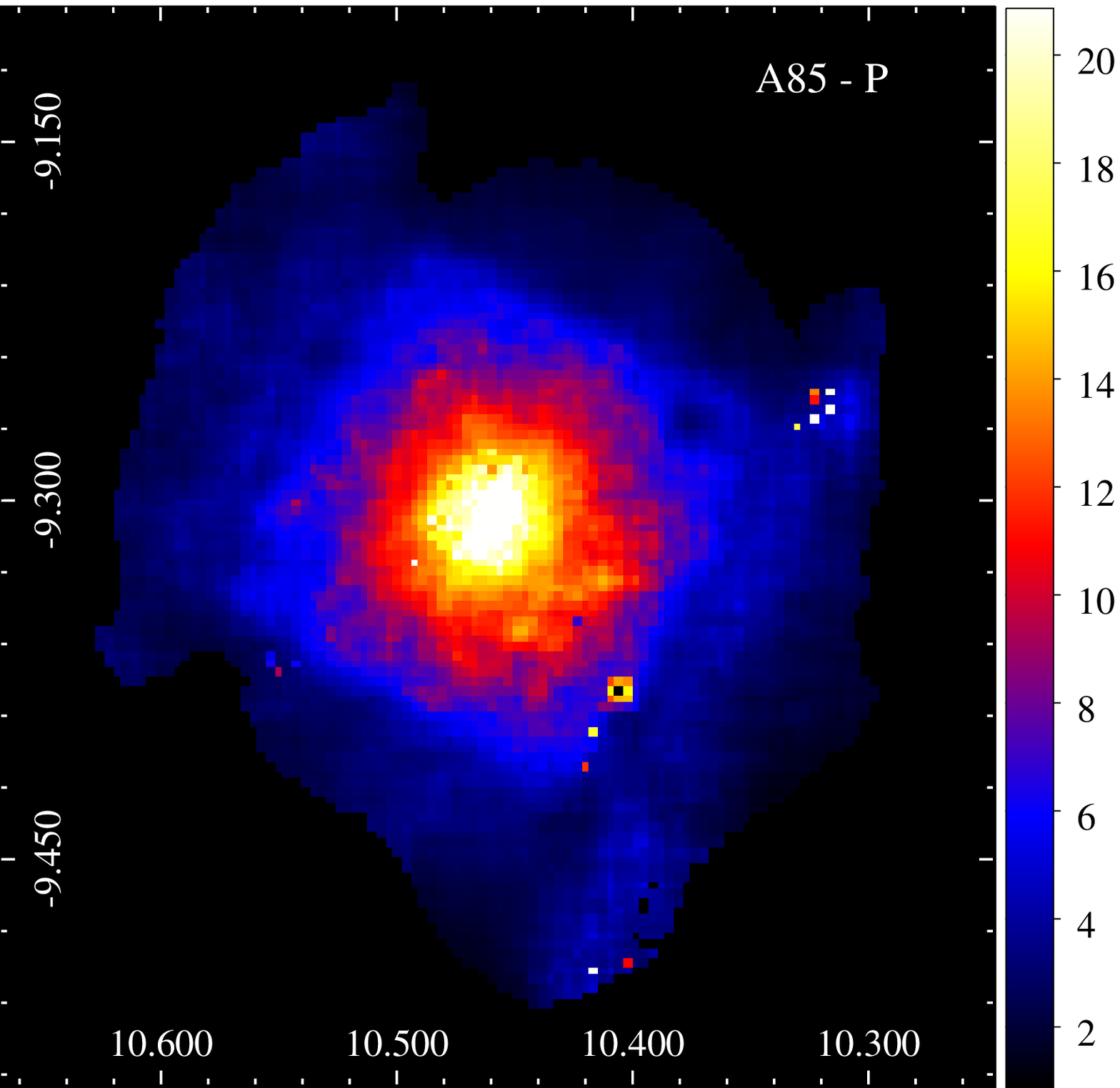}
\includegraphics[scale=0.25]{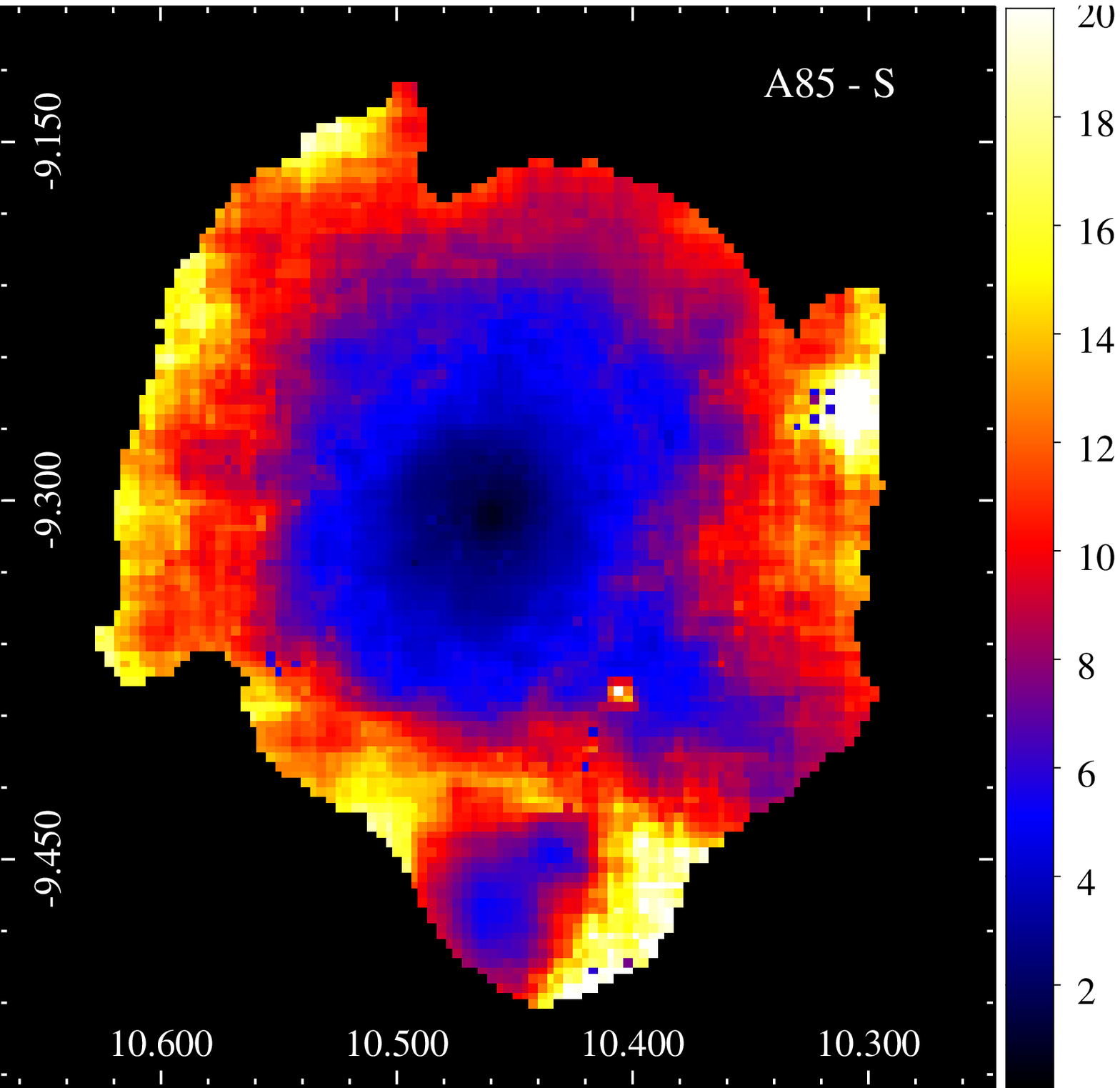}
\includegraphics[scale=0.25]{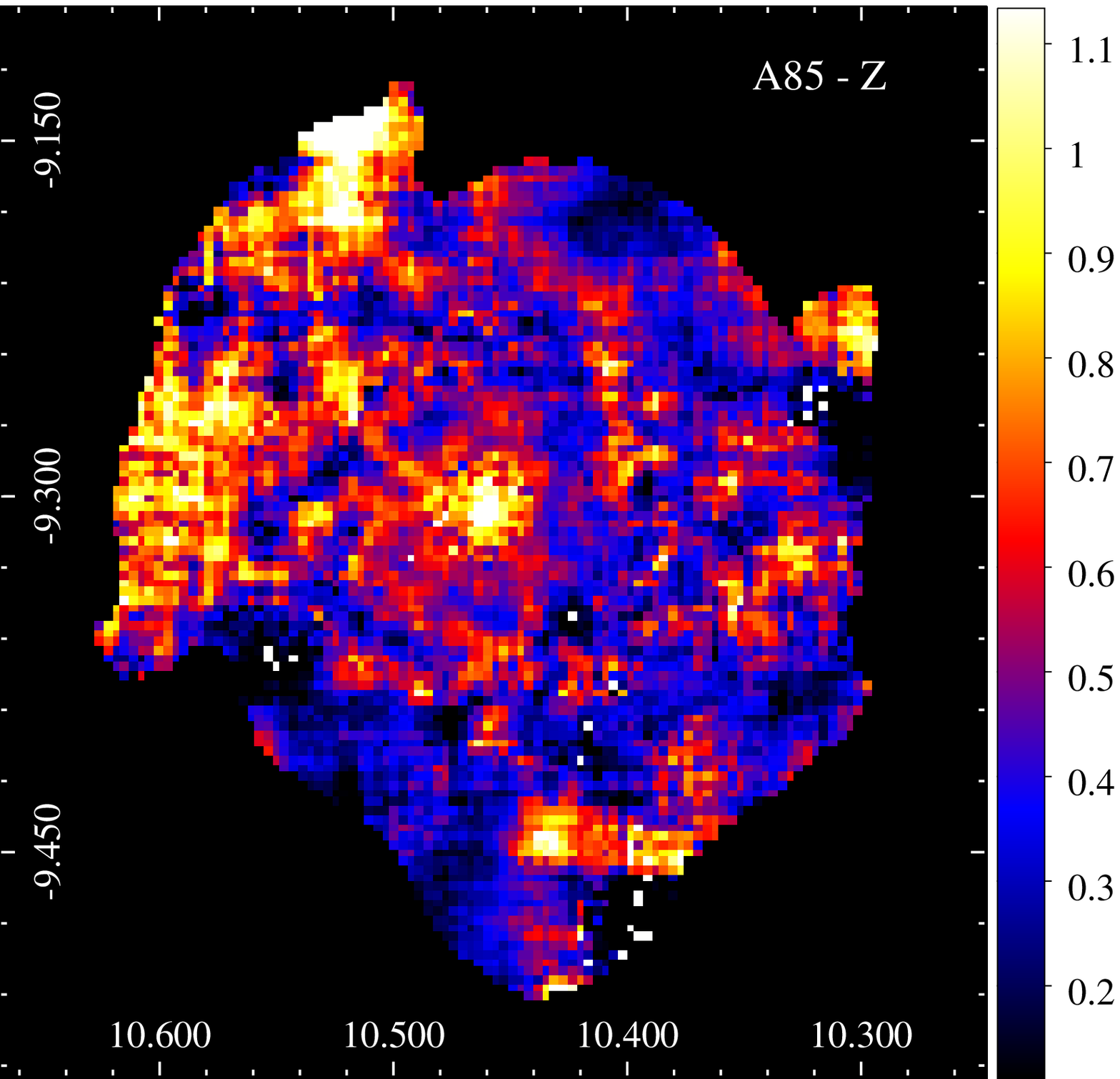}

\includegraphics[scale=0.25]{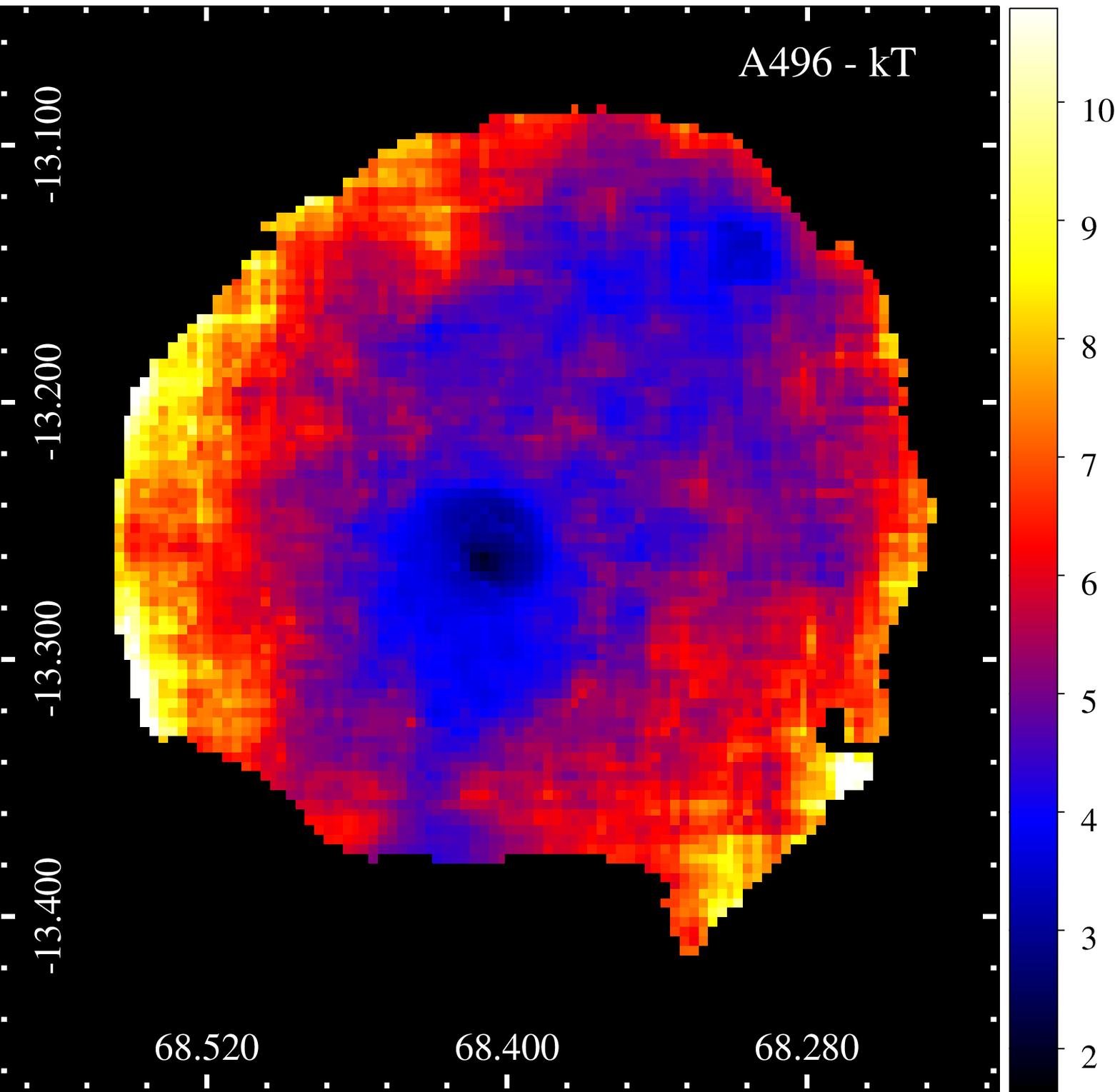}
\includegraphics[scale=0.25]{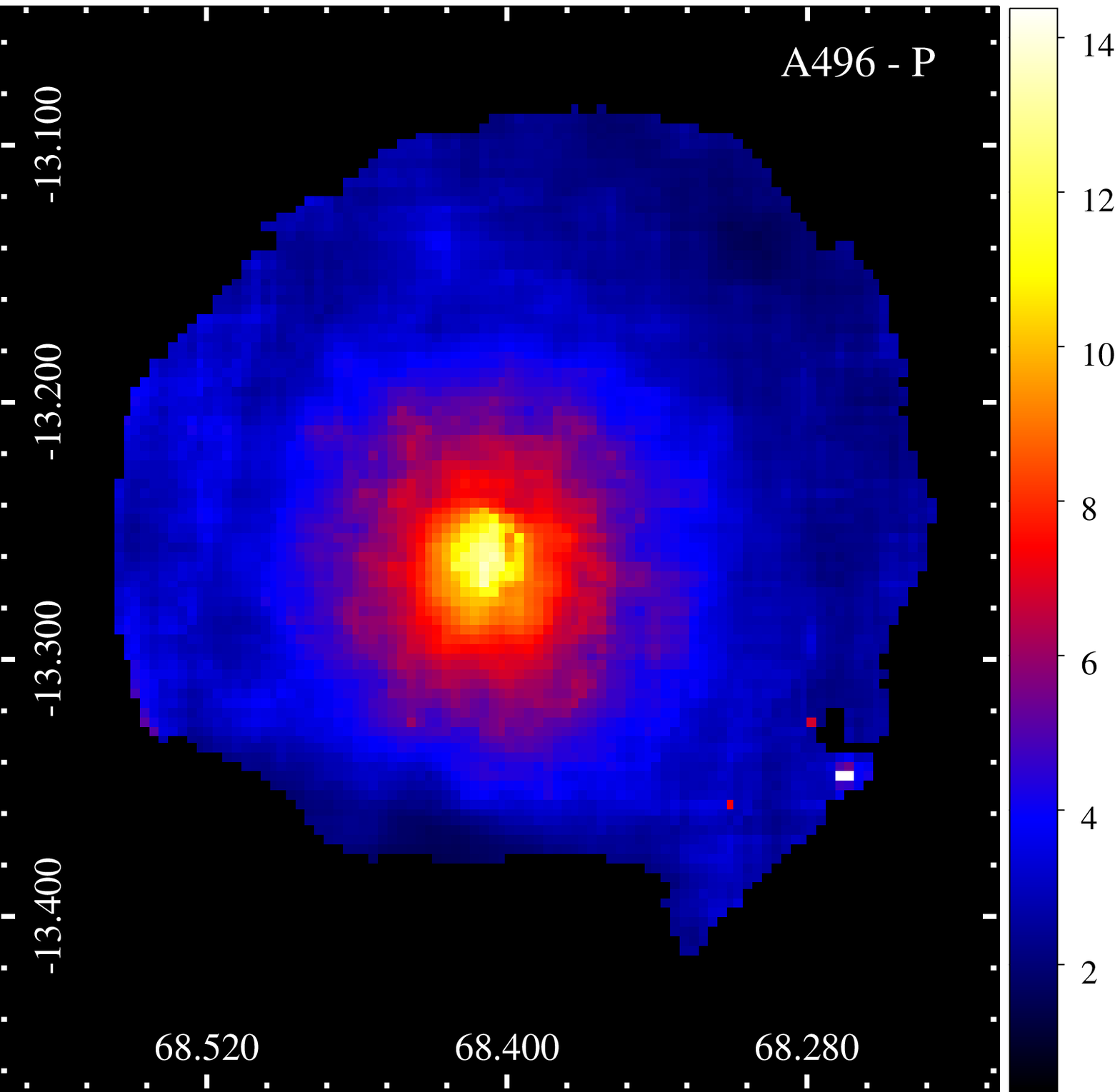}
\includegraphics[scale=0.25]{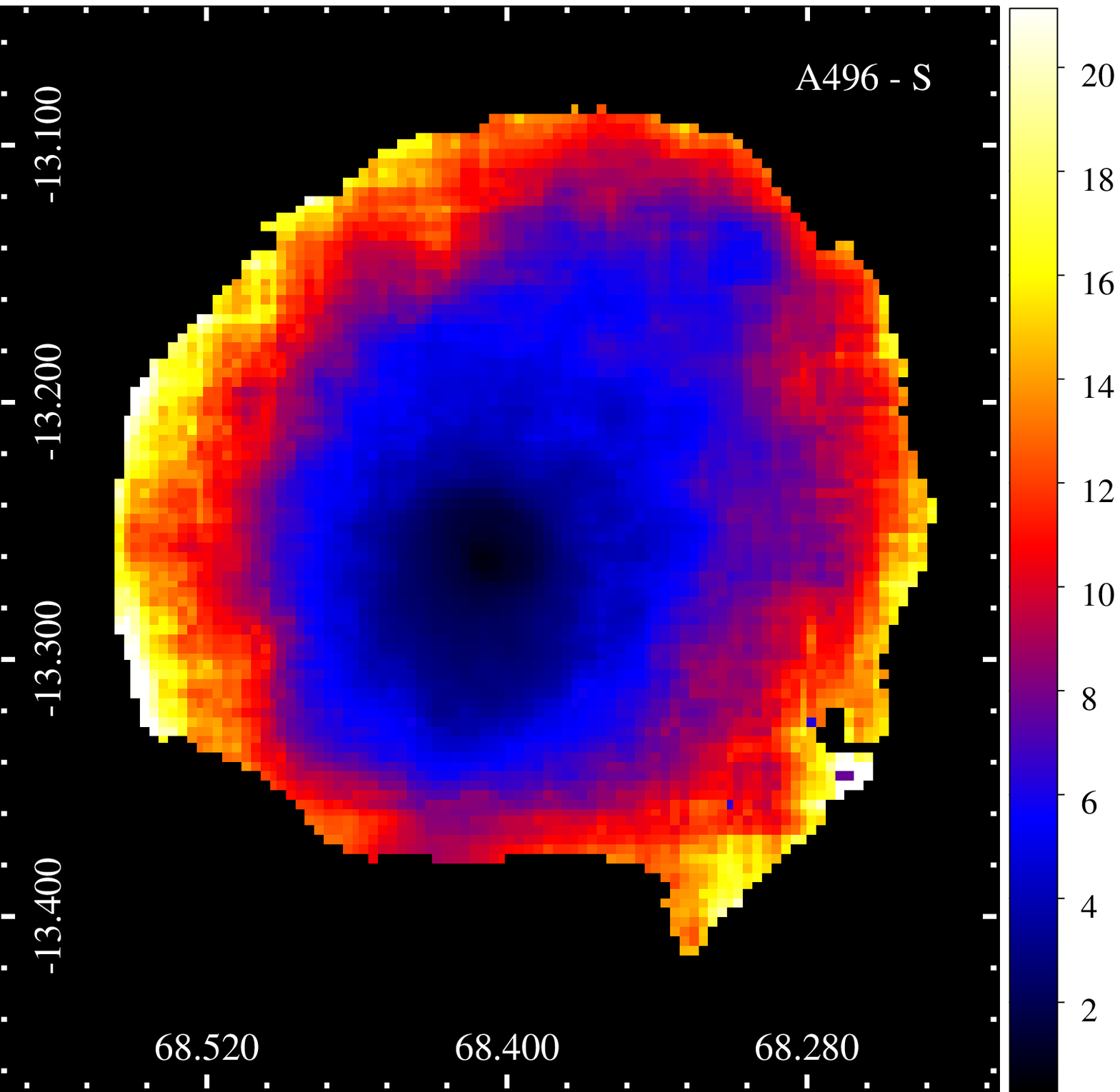}
\includegraphics[scale=0.25]{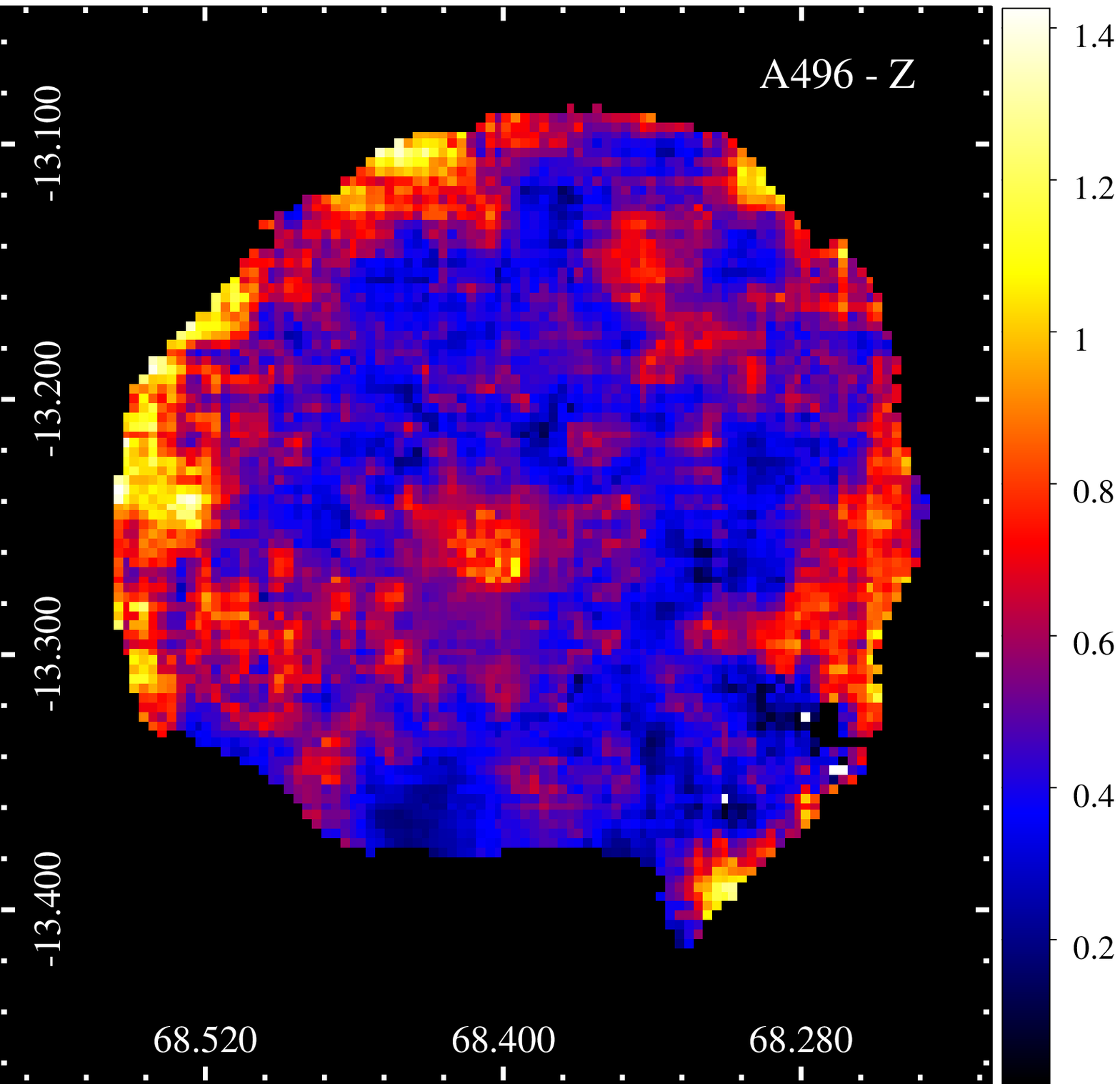}

\includegraphics[scale=0.25]{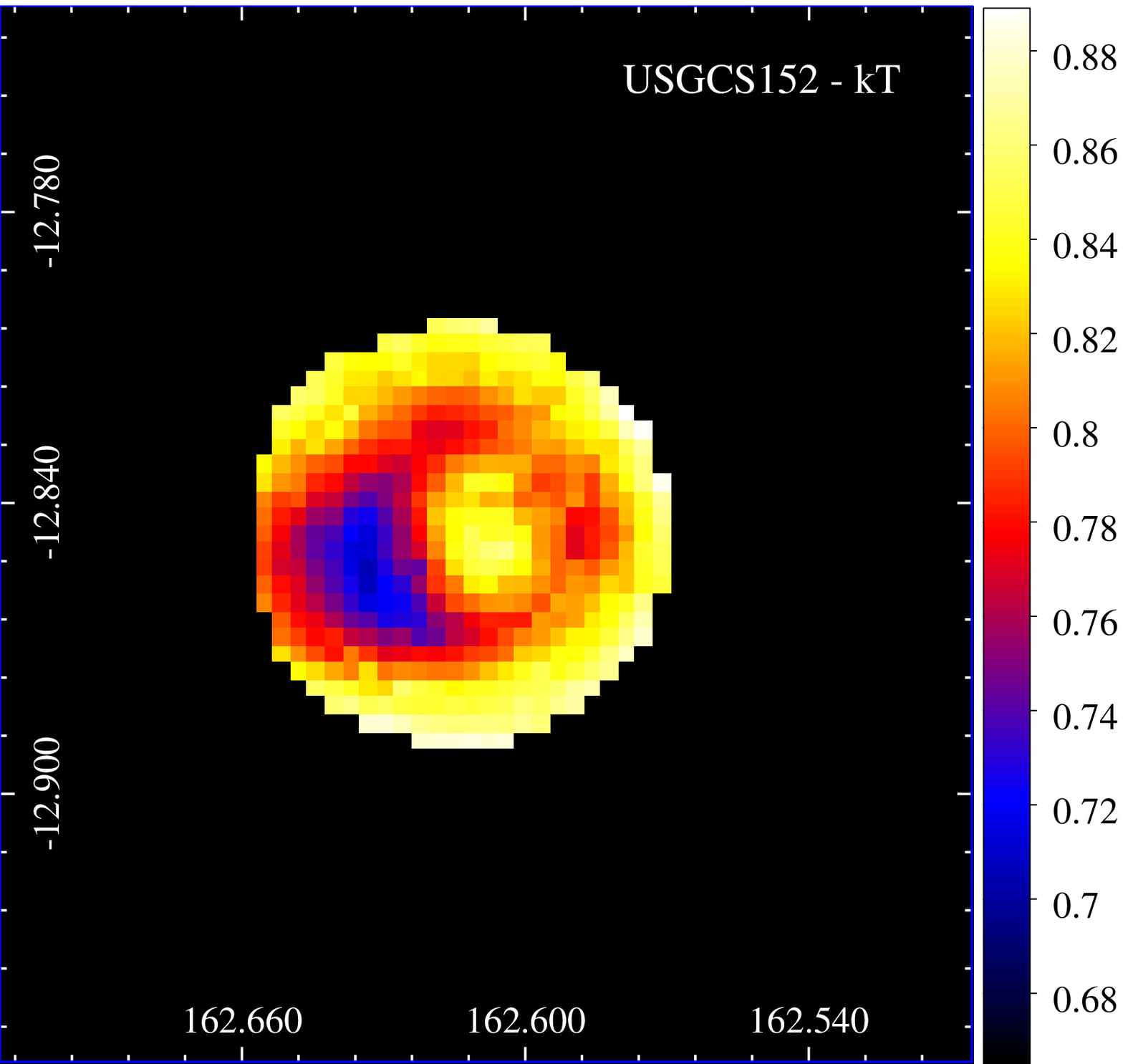}
\includegraphics[scale=0.25]{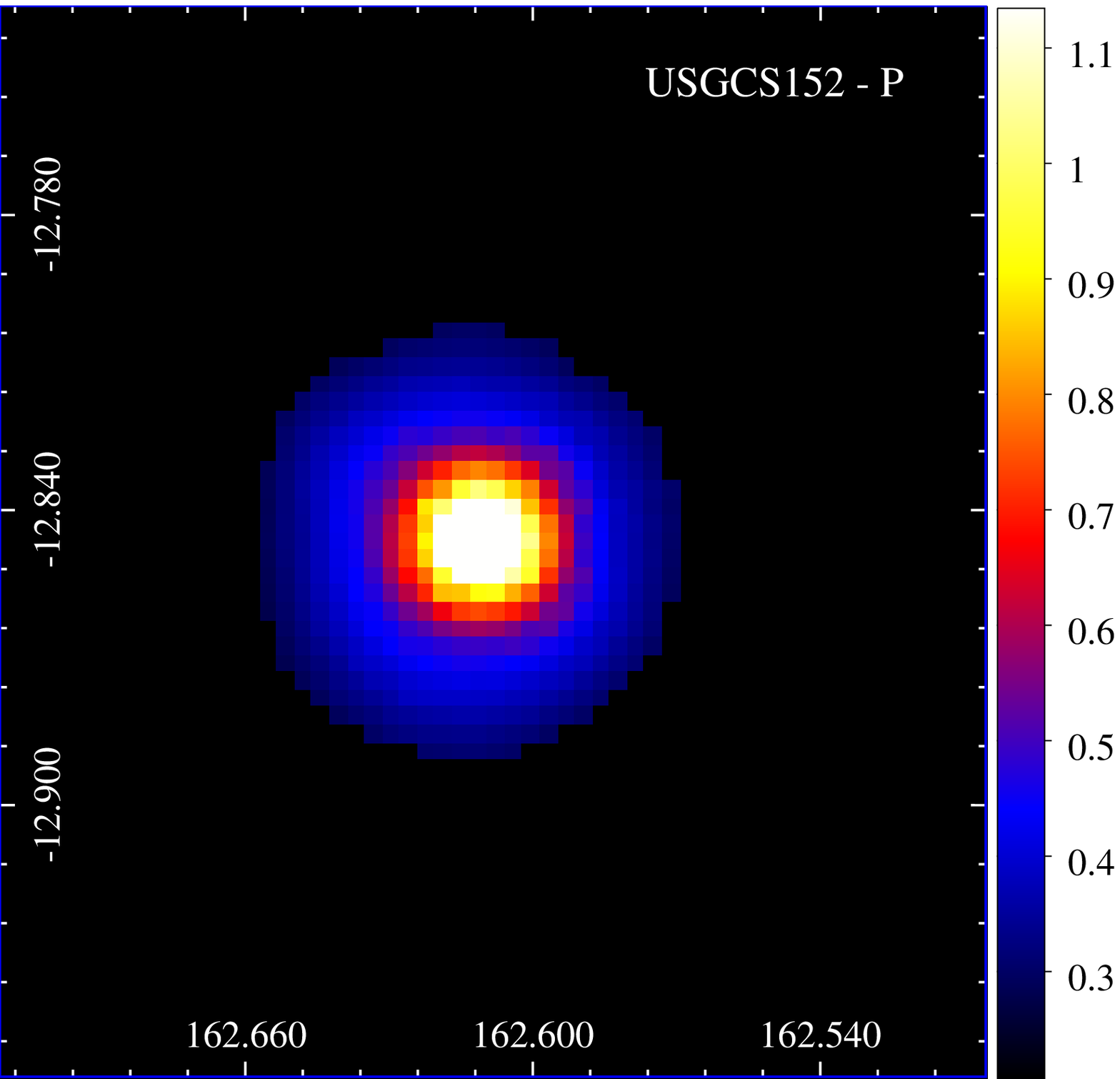}
\includegraphics[scale=0.25]{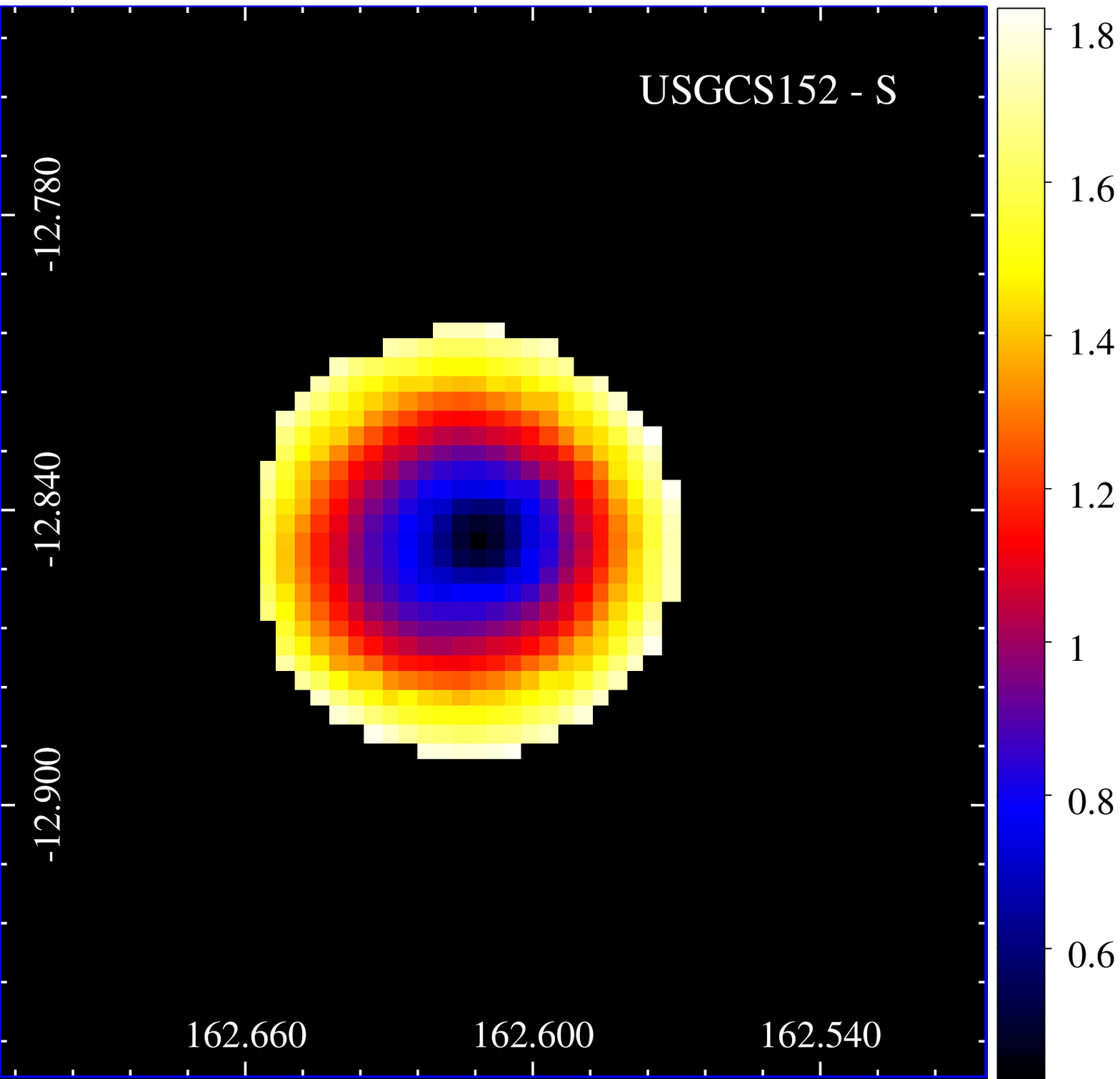}
\includegraphics[scale=0.25]{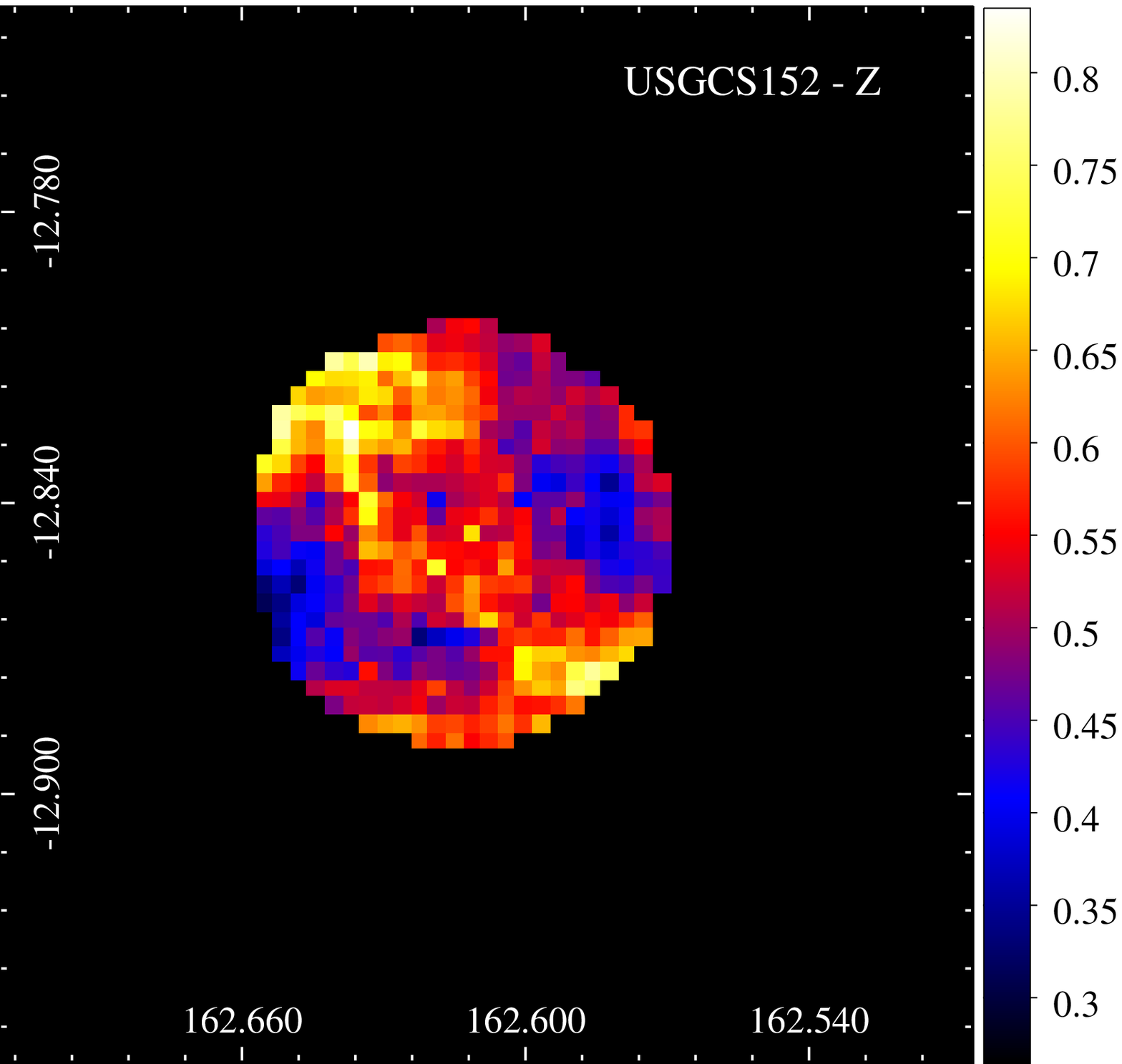}

\includegraphics[scale=0.25]{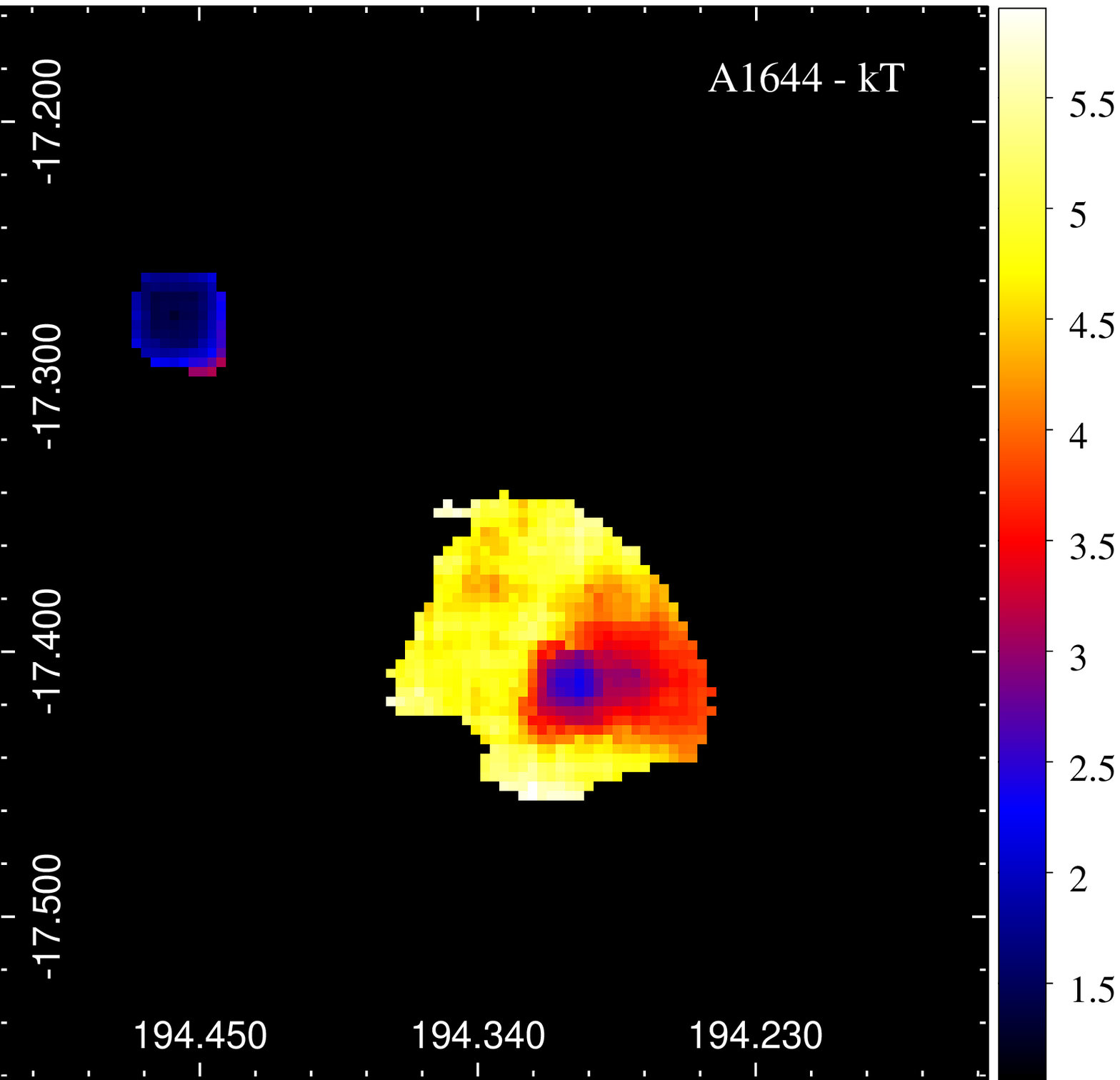}
\includegraphics[scale=0.25]{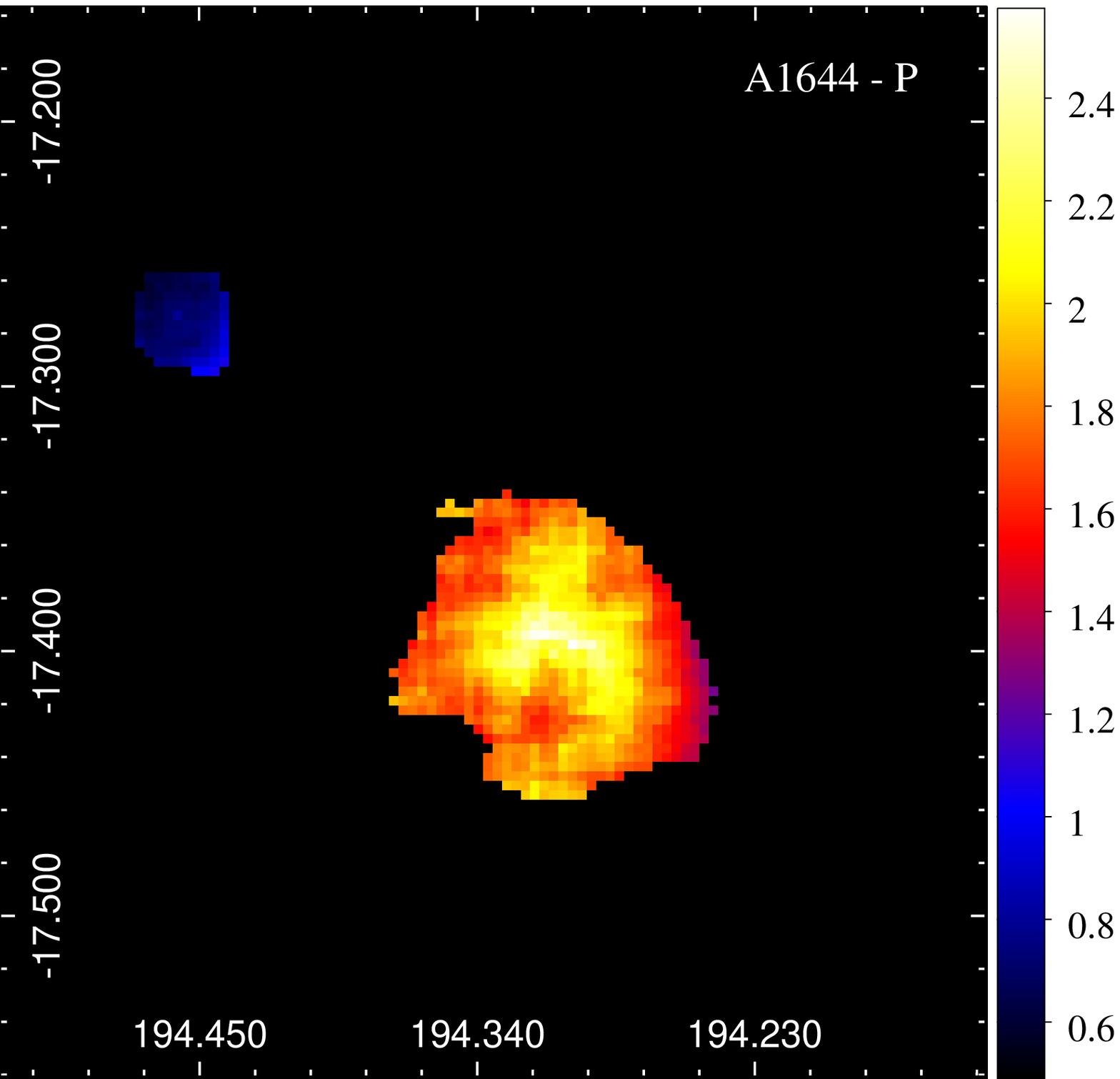}
\includegraphics[scale=0.25]{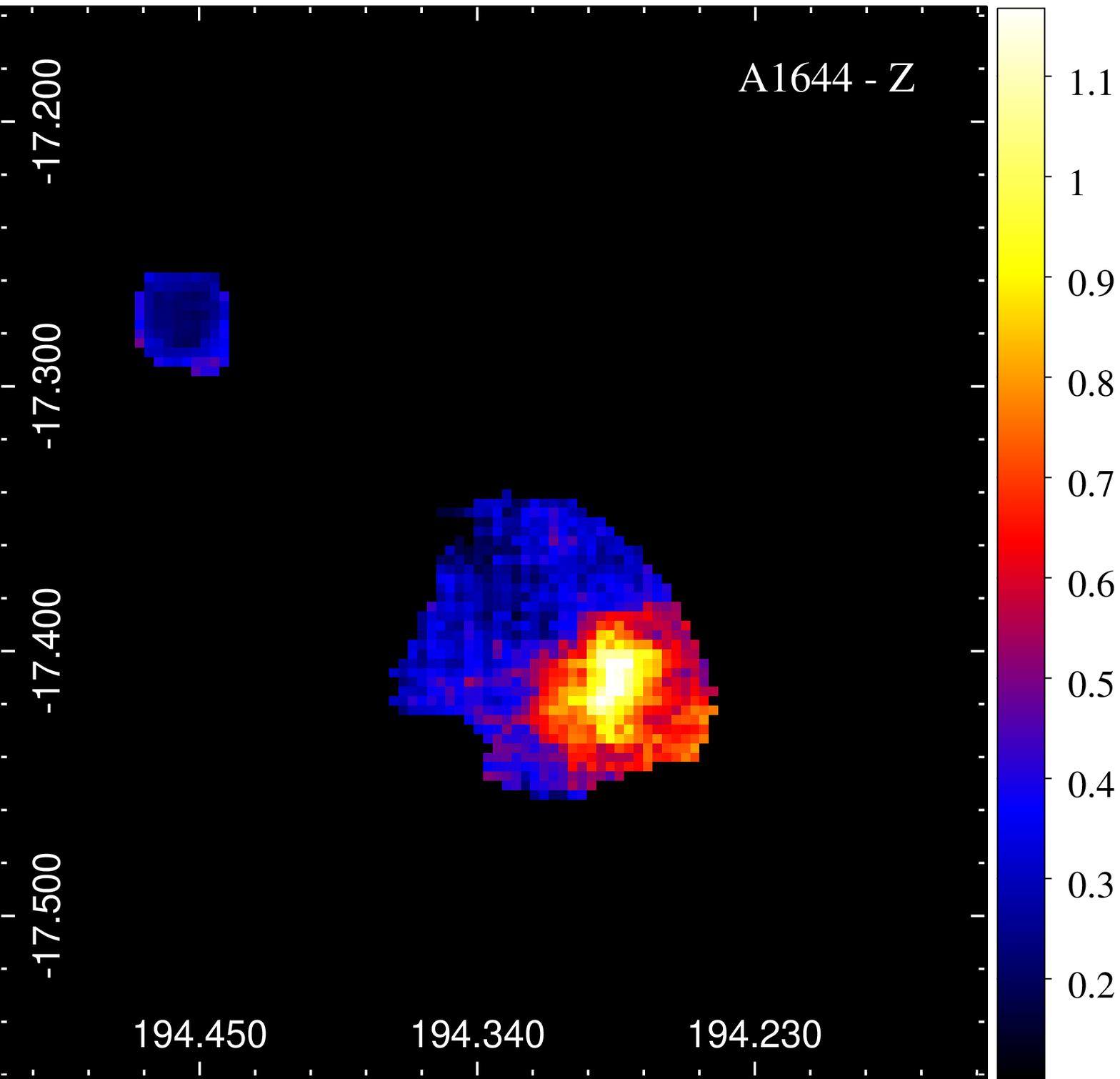}
\includegraphics[scale=0.25]{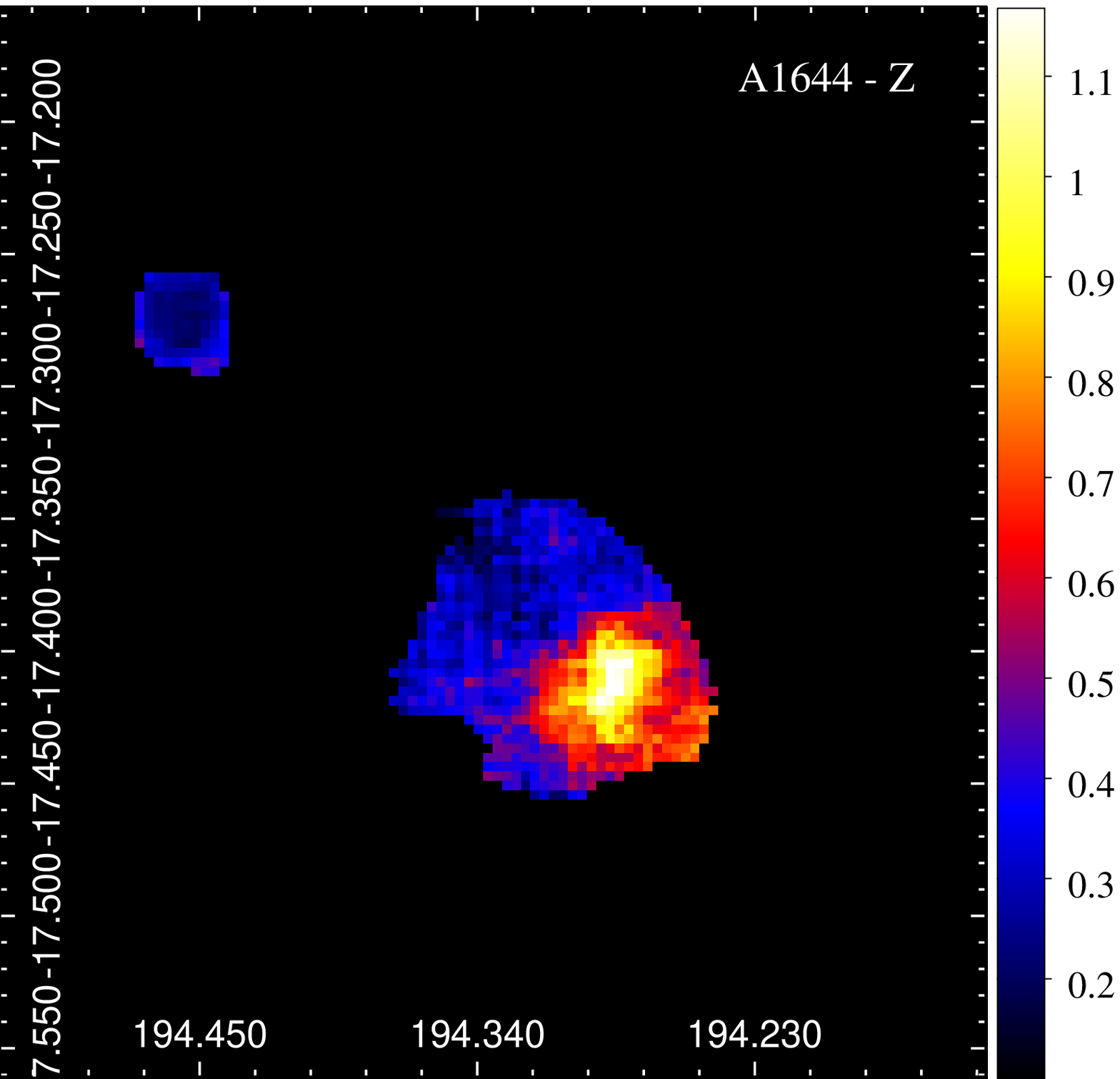}

\includegraphics[scale=0.25]{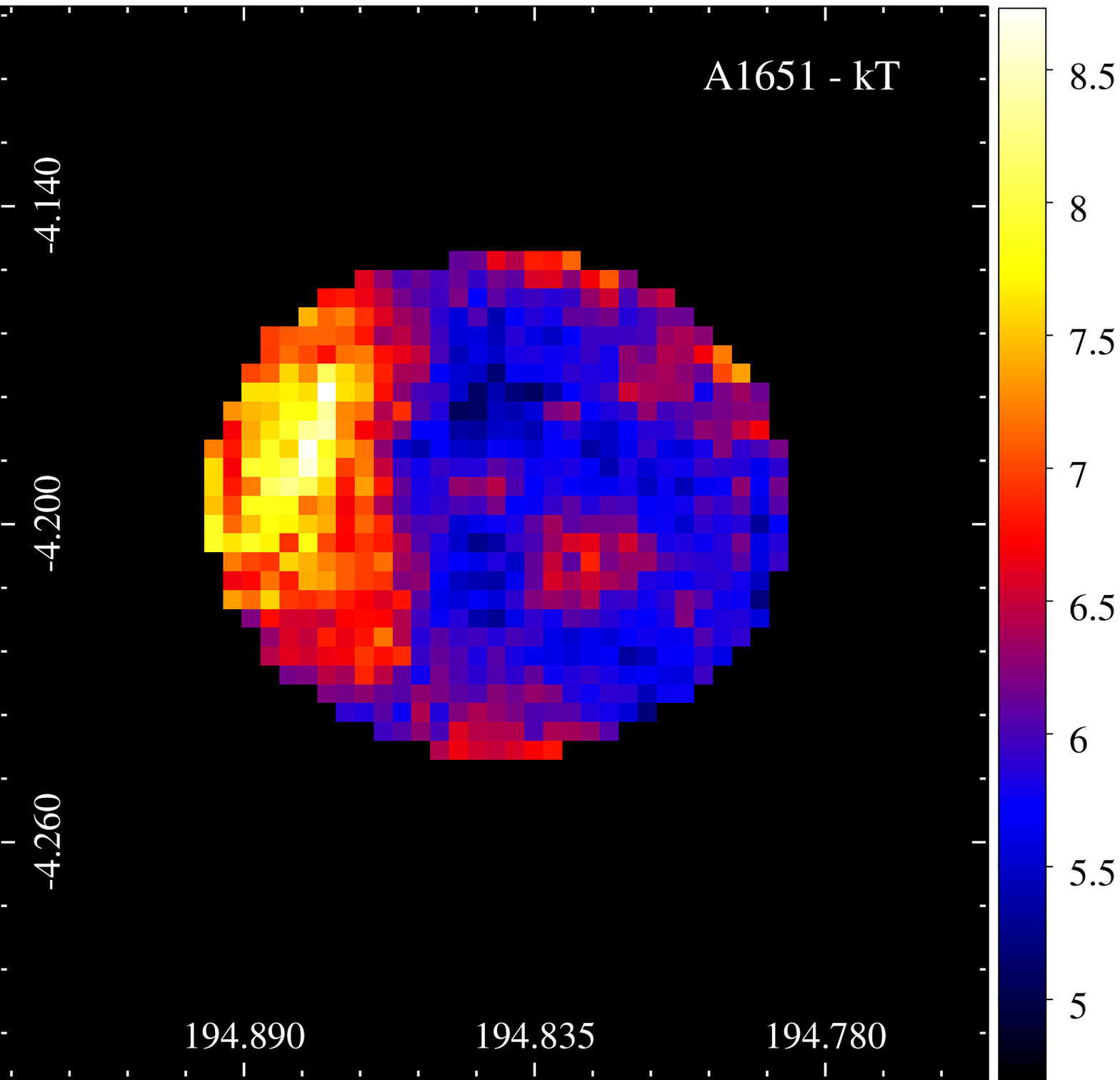}
\includegraphics[scale=0.25]{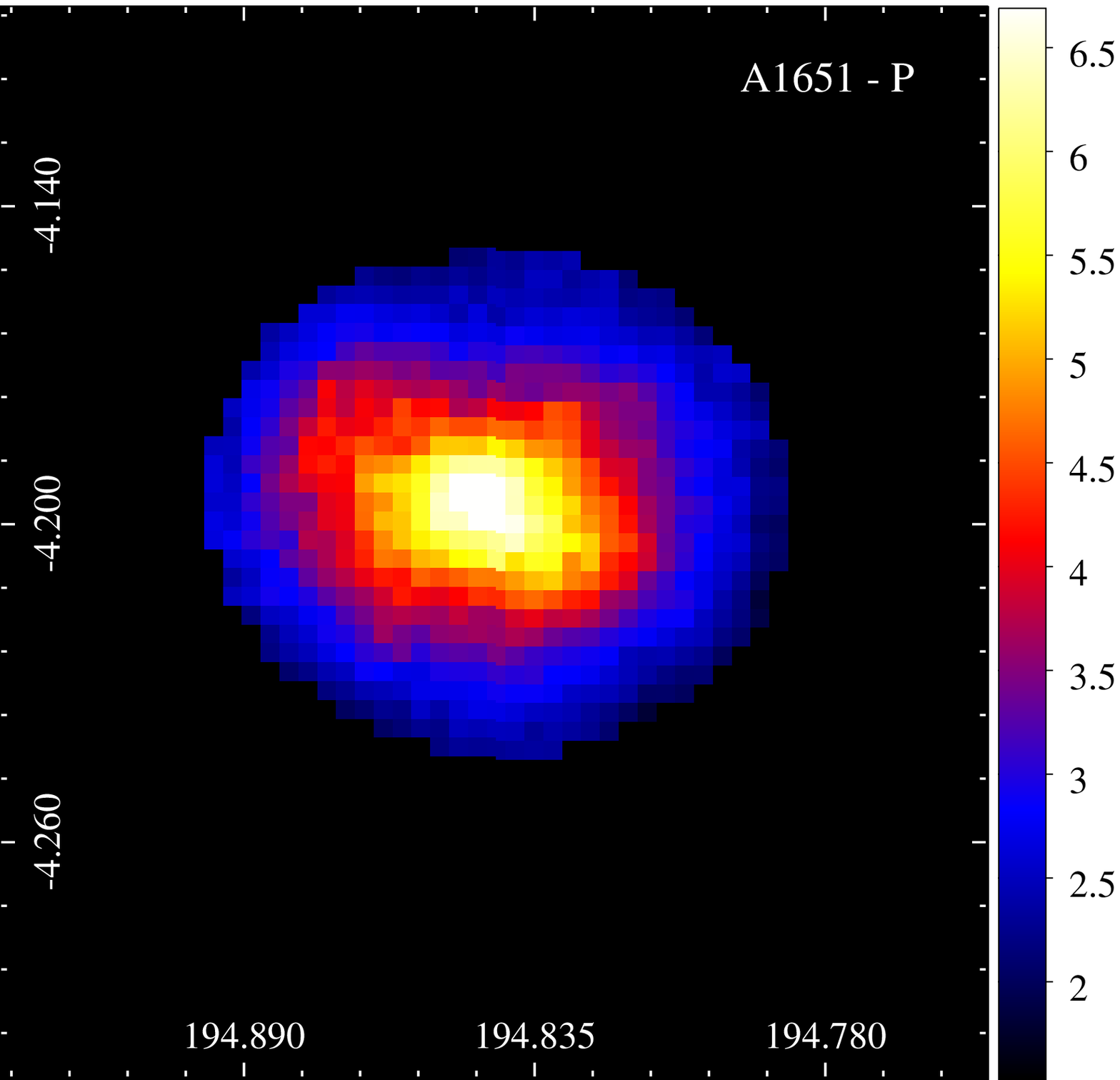}
\includegraphics[scale=0.25]{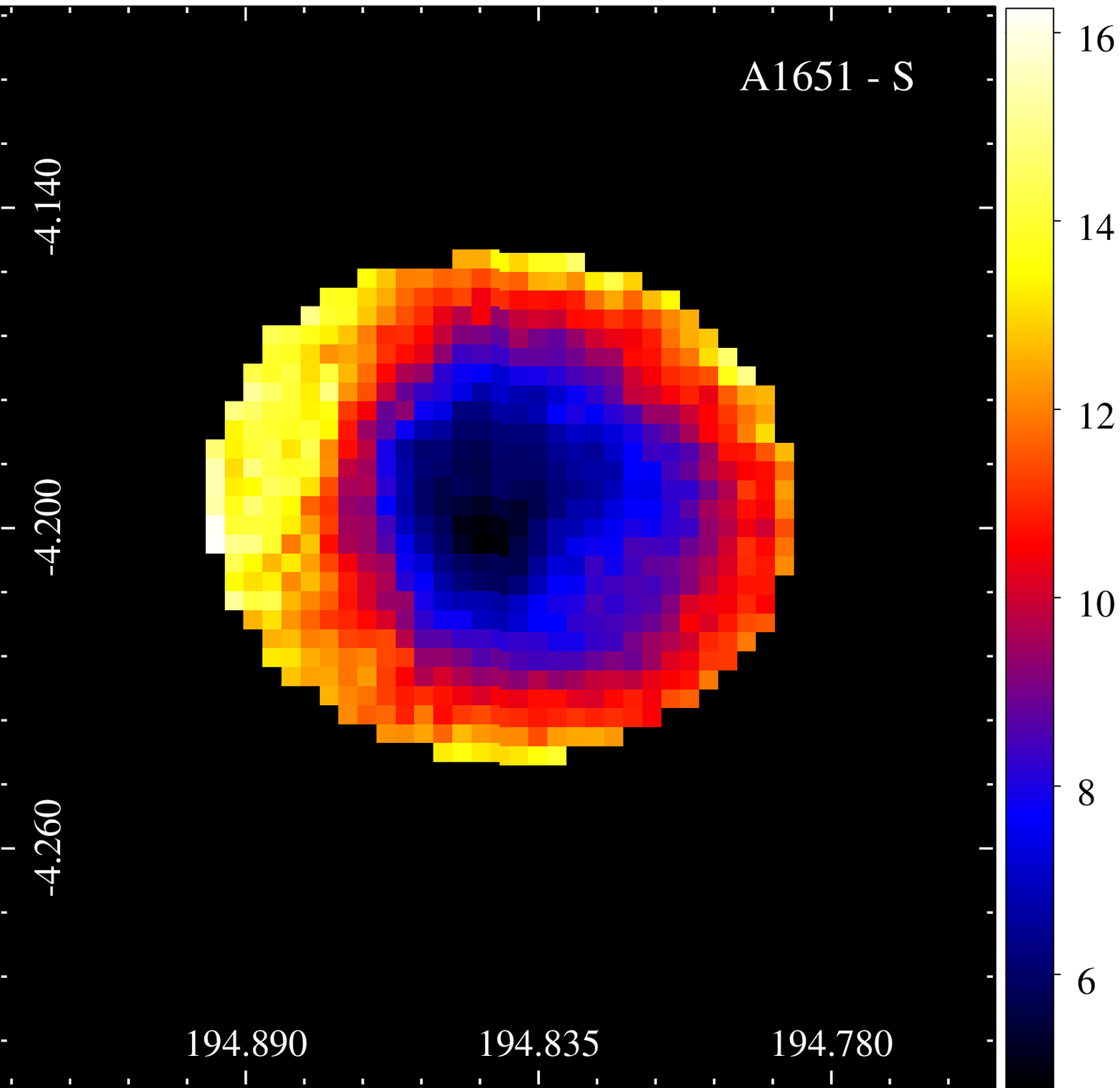}
\includegraphics[scale=0.25]{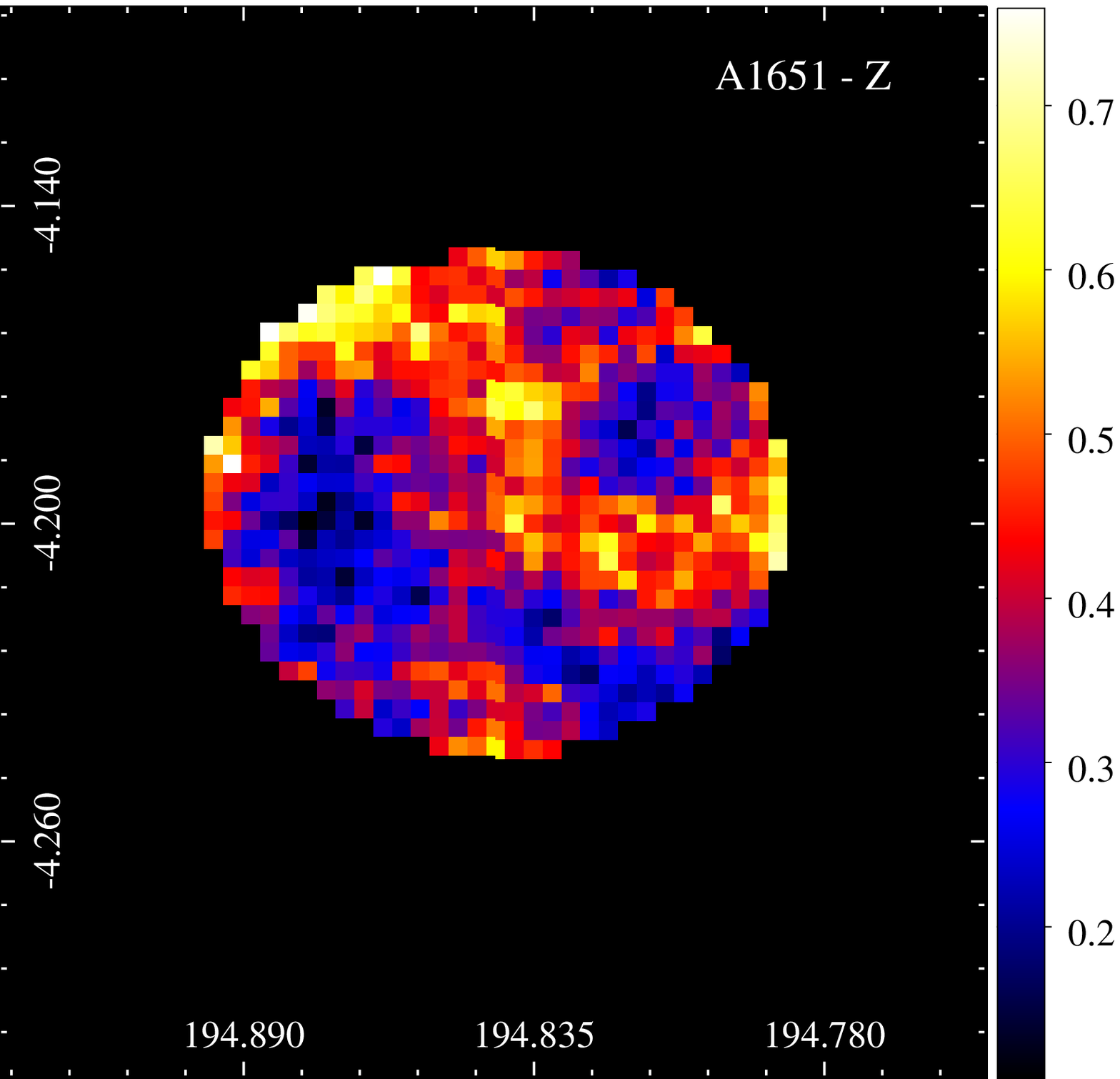}

\caption{CC and disturbed systems. From left to right: temperature, pseudo-pressure, pseudo-entropy, and metallicity map for A85, A496, 
USGC S152, A1644 (A1644n and A1644s in the same map), and A1651.}
\label{fig:groupsA85}
\end{figure*}

\begin{figure*}

\includegraphics[scale=0.25]{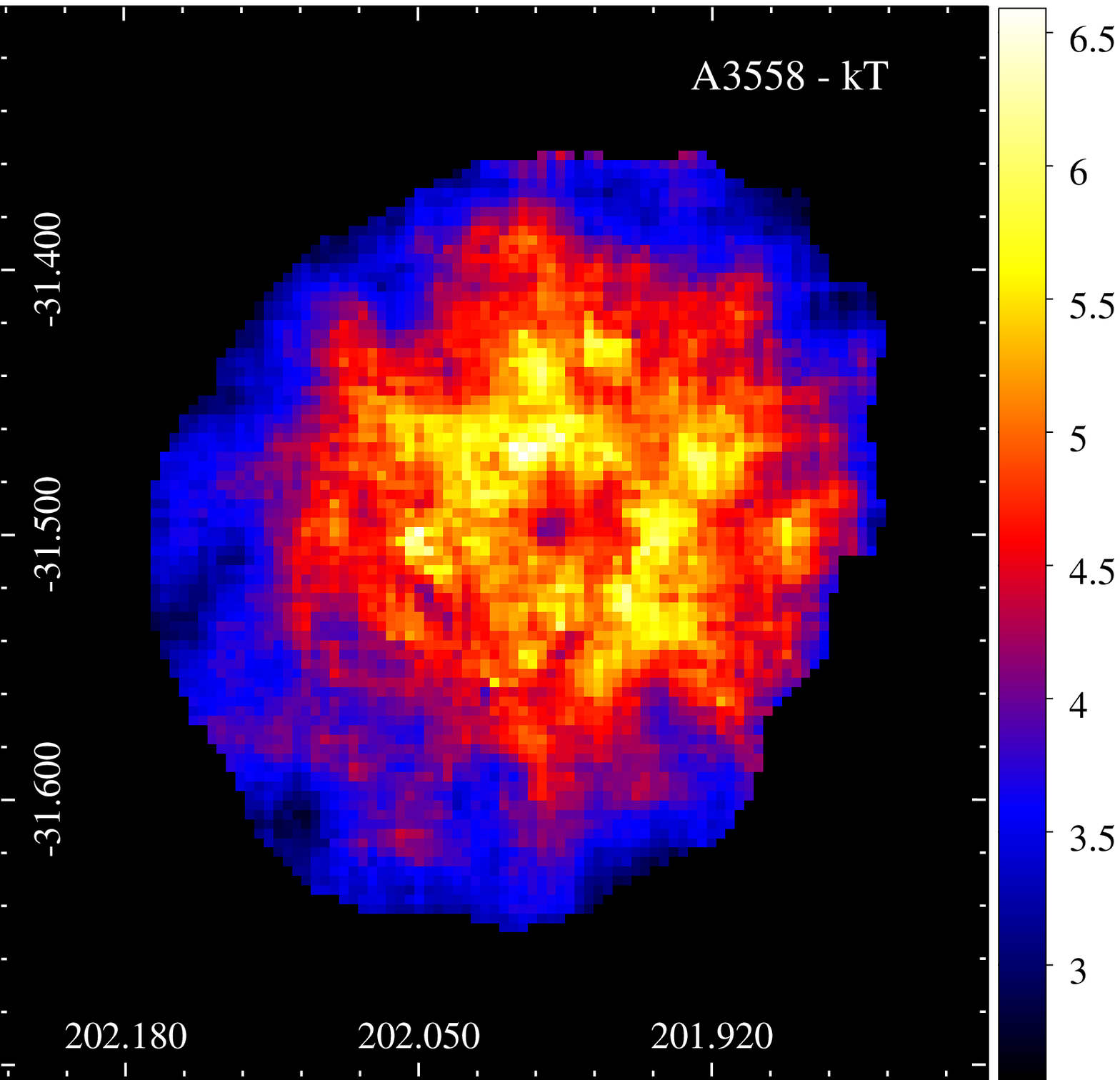}
\includegraphics[scale=0.25]{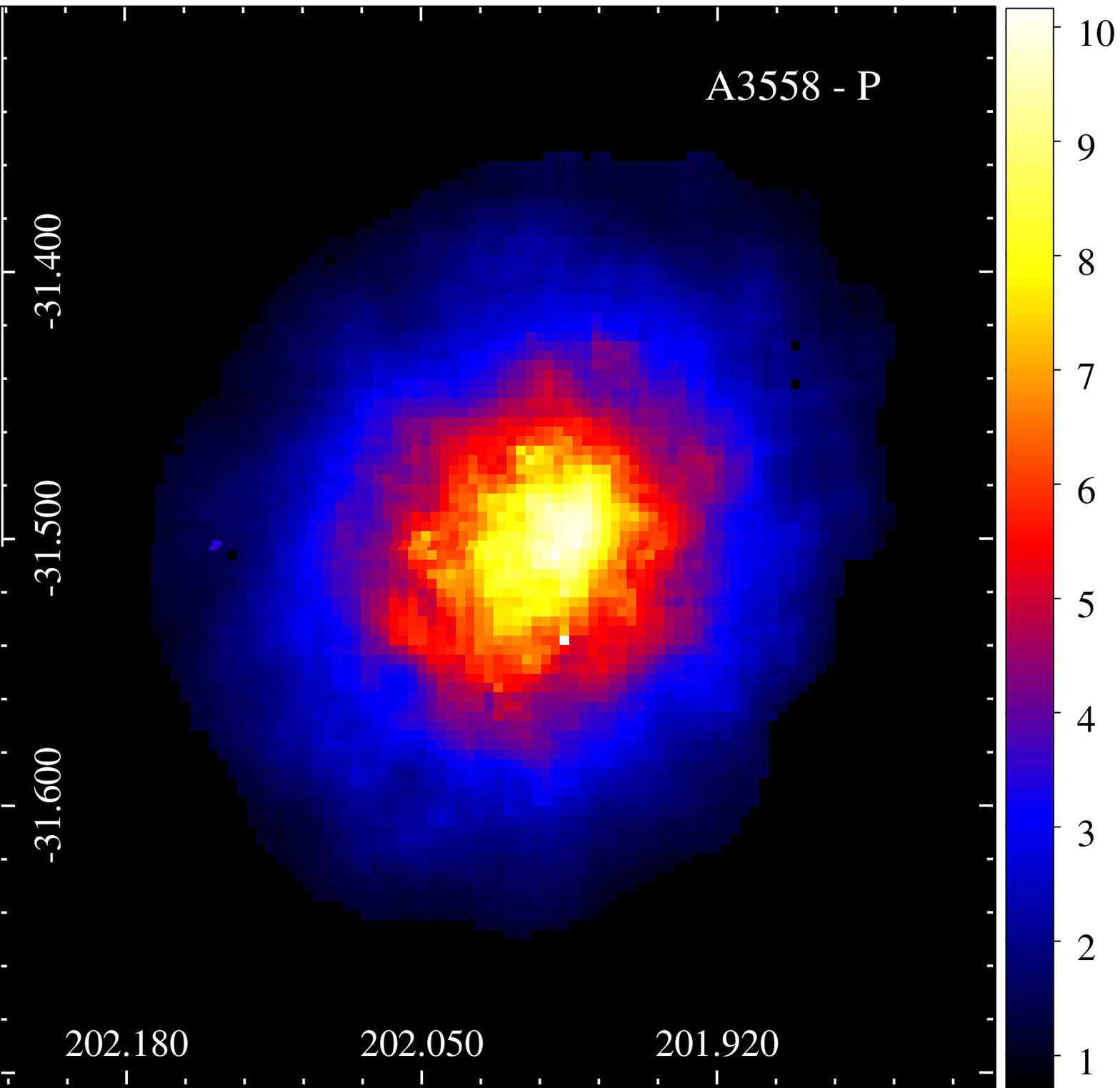}
\includegraphics[scale=0.25]{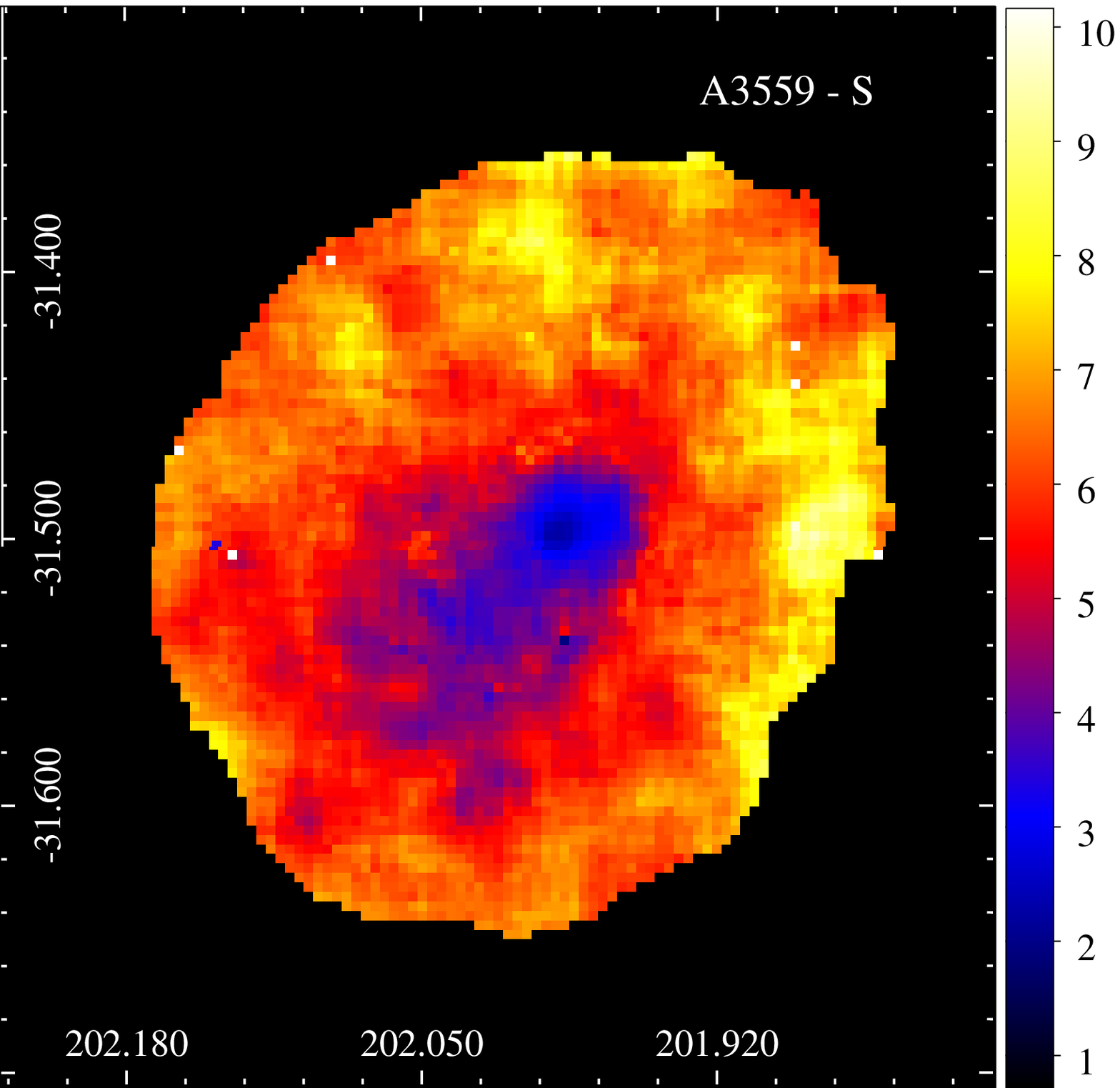}
\includegraphics[scale=0.25]{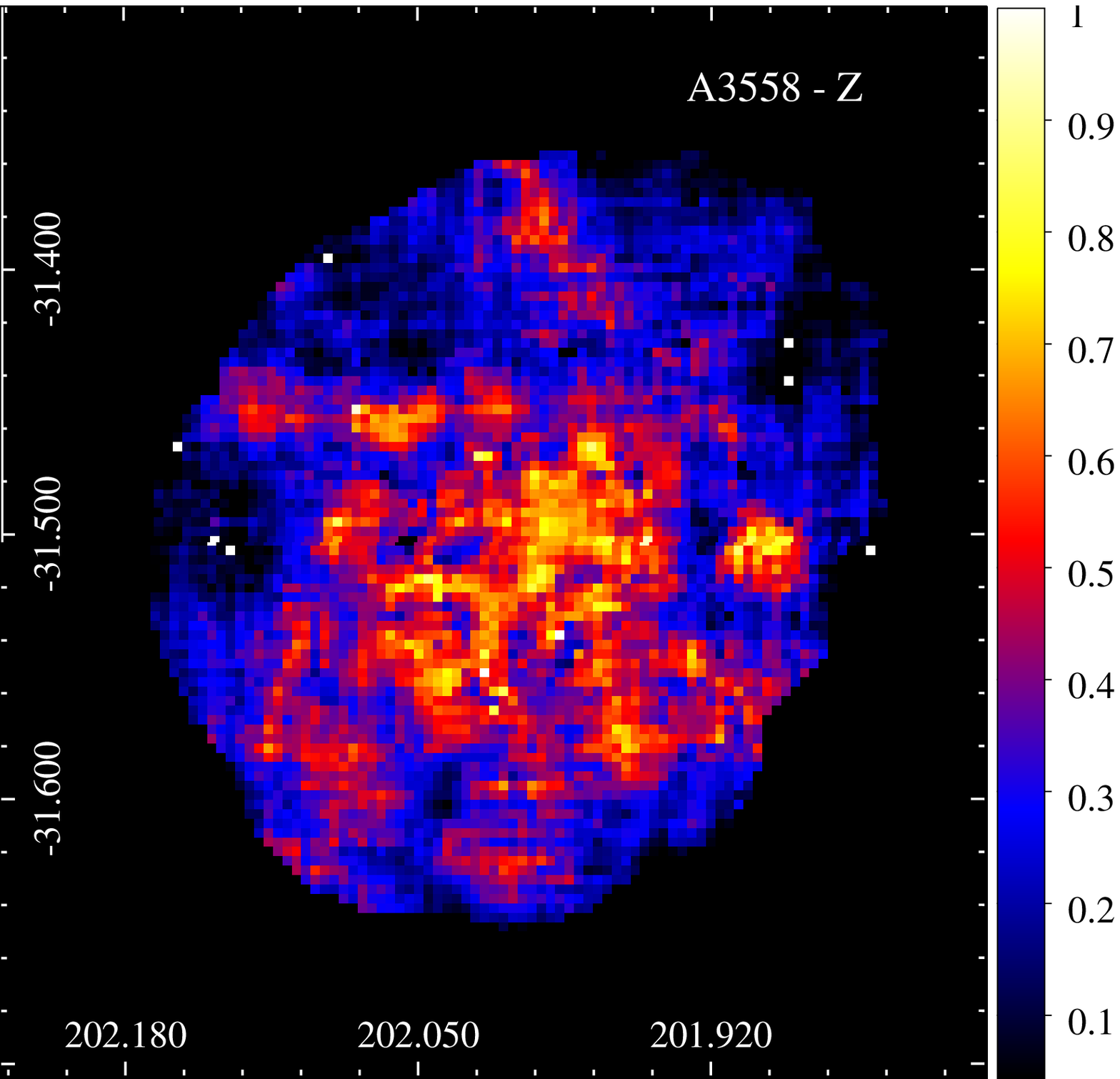}

\includegraphics[scale=0.25]{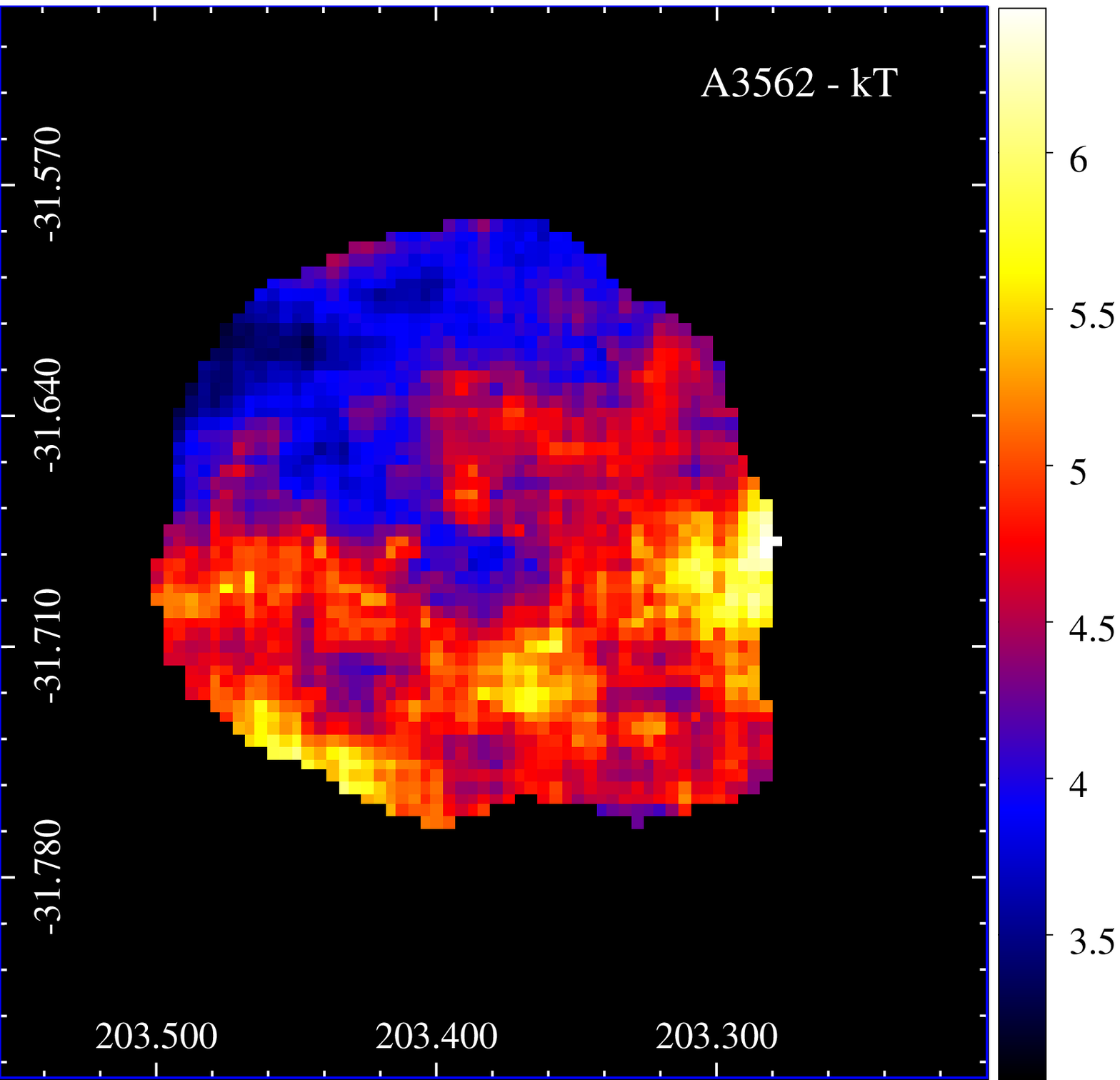}
\includegraphics[scale=0.25]{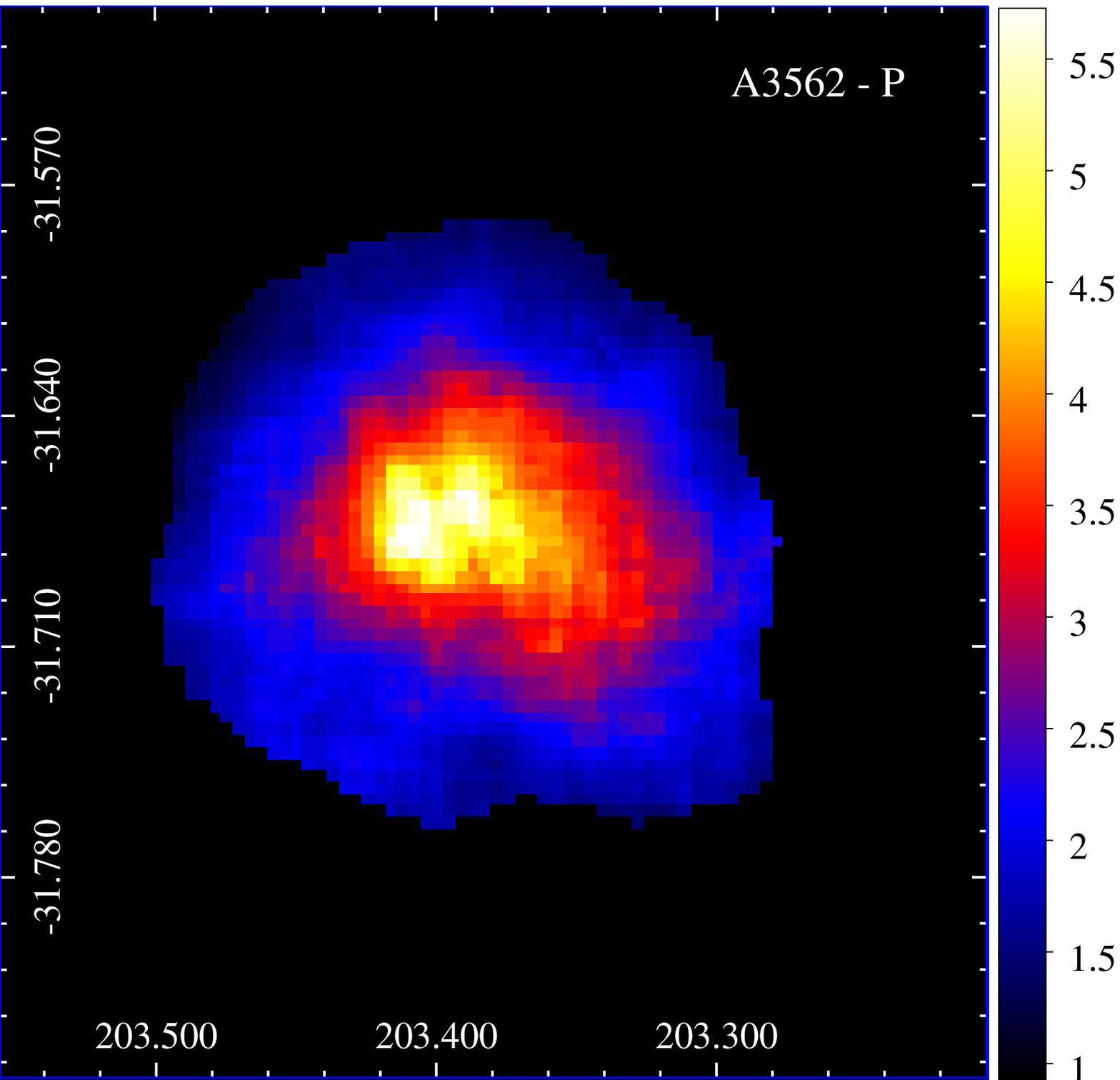}
\includegraphics[scale=0.25]{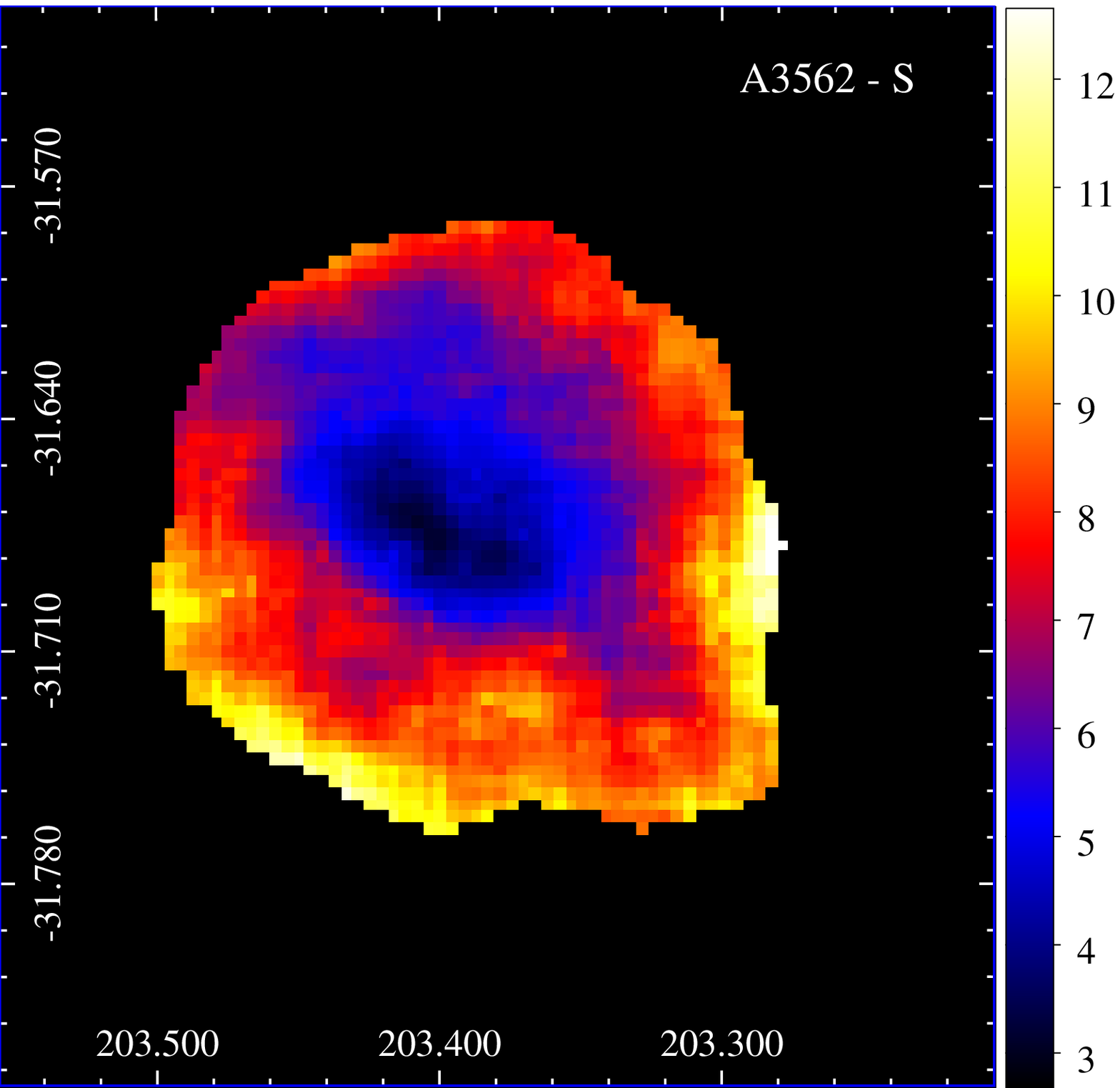}
\includegraphics[scale=0.25]{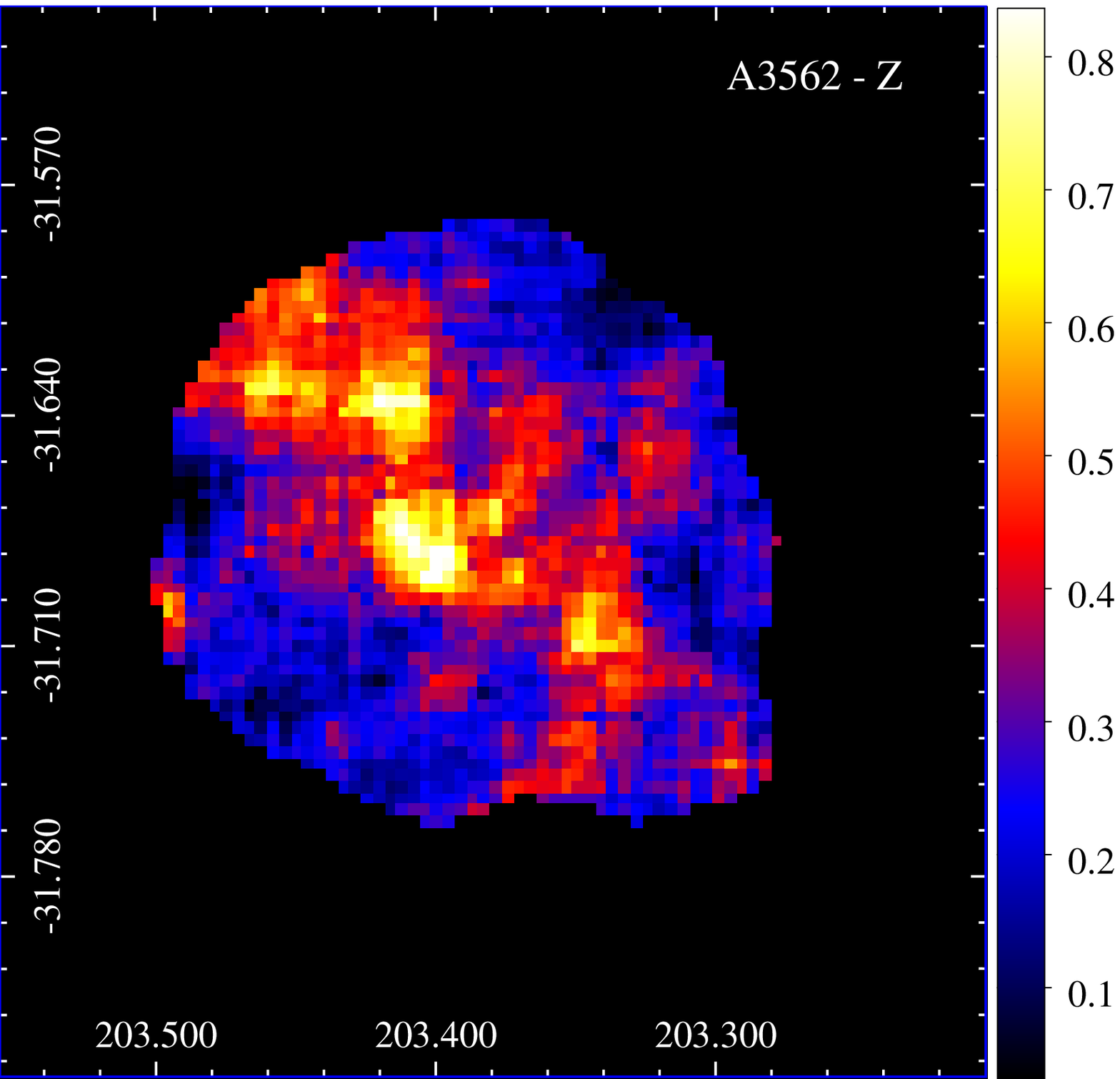}

\includegraphics[scale=0.25]{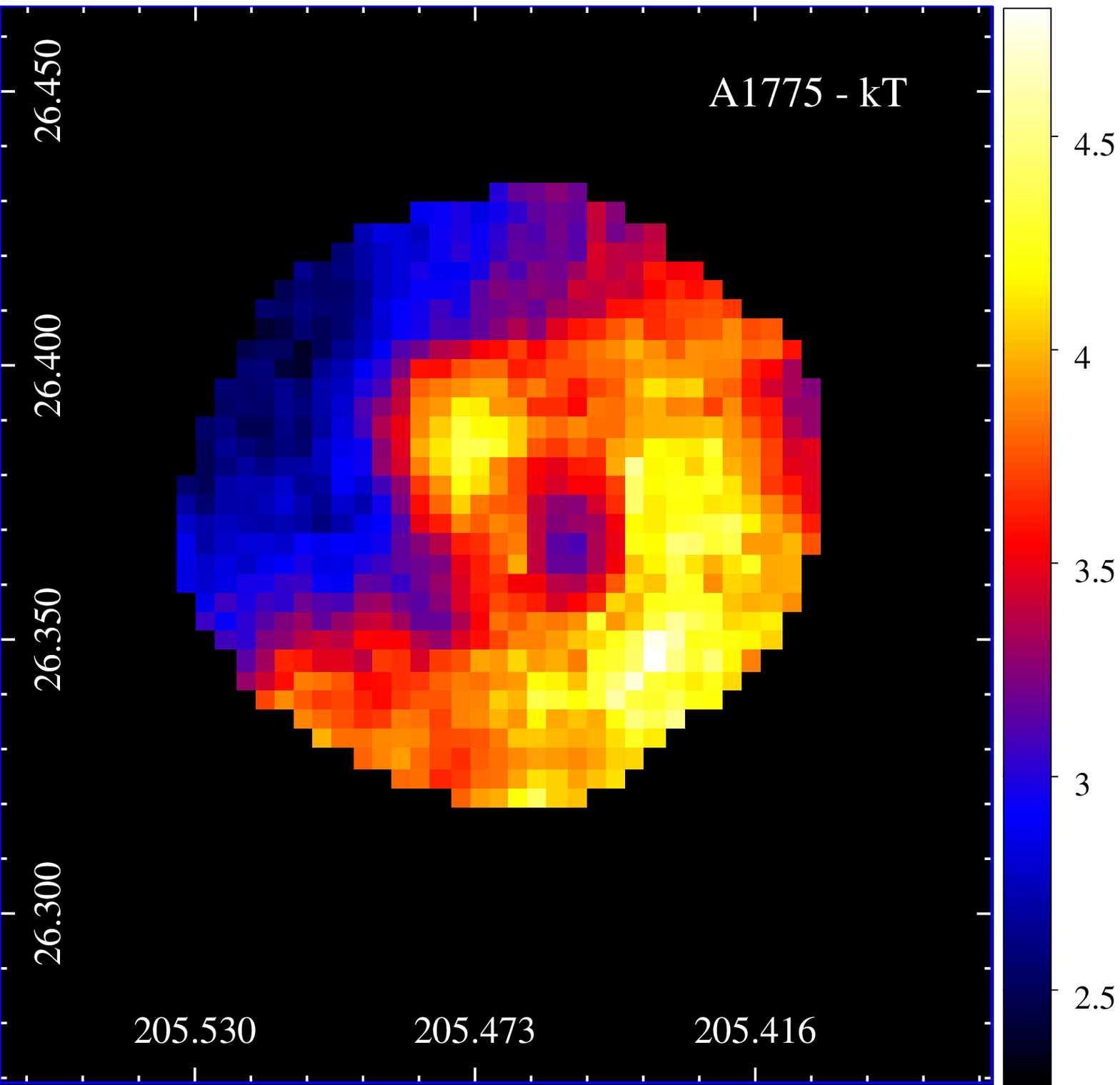}
\includegraphics[scale=0.25]{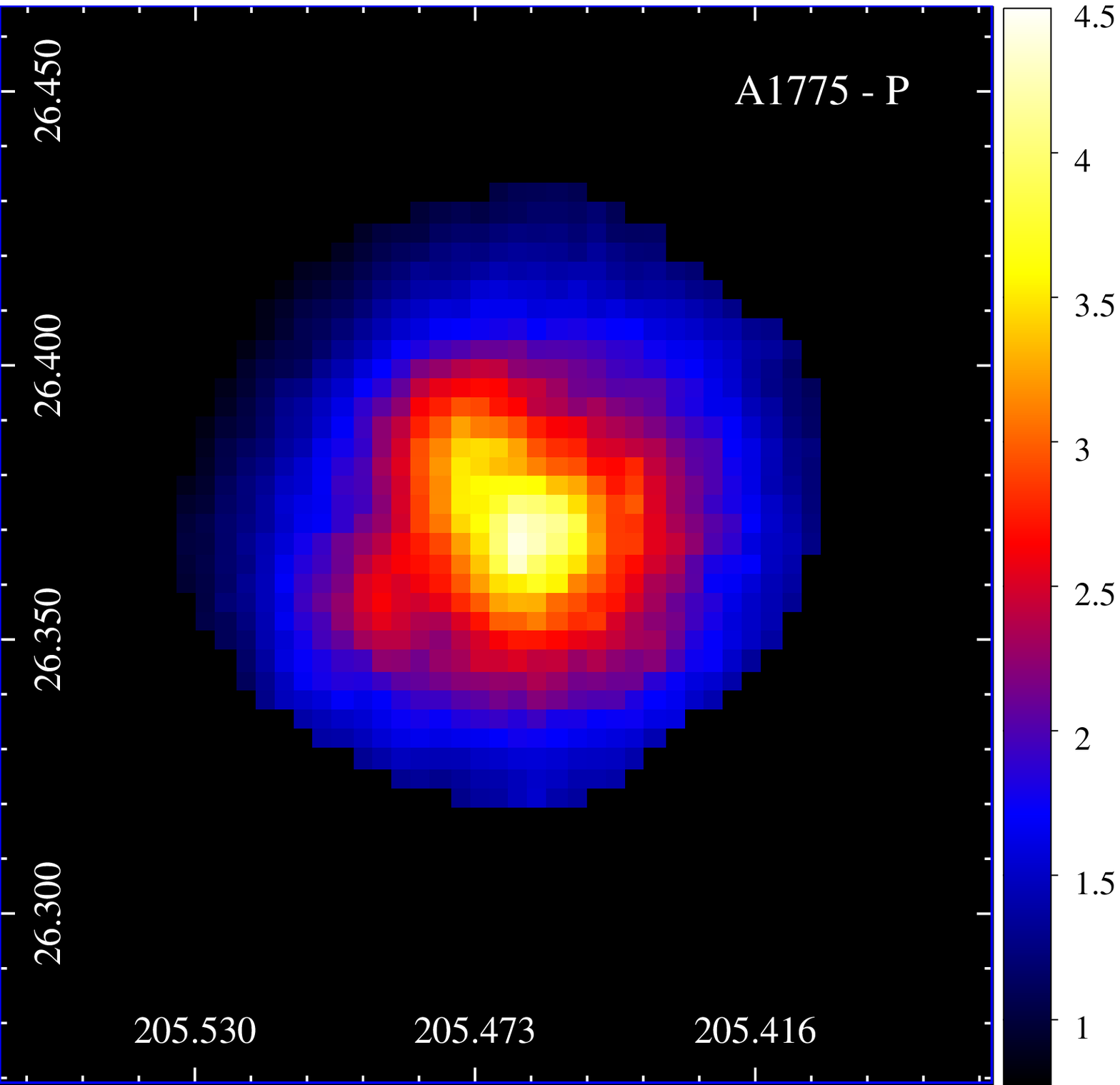}
\includegraphics[scale=0.25]{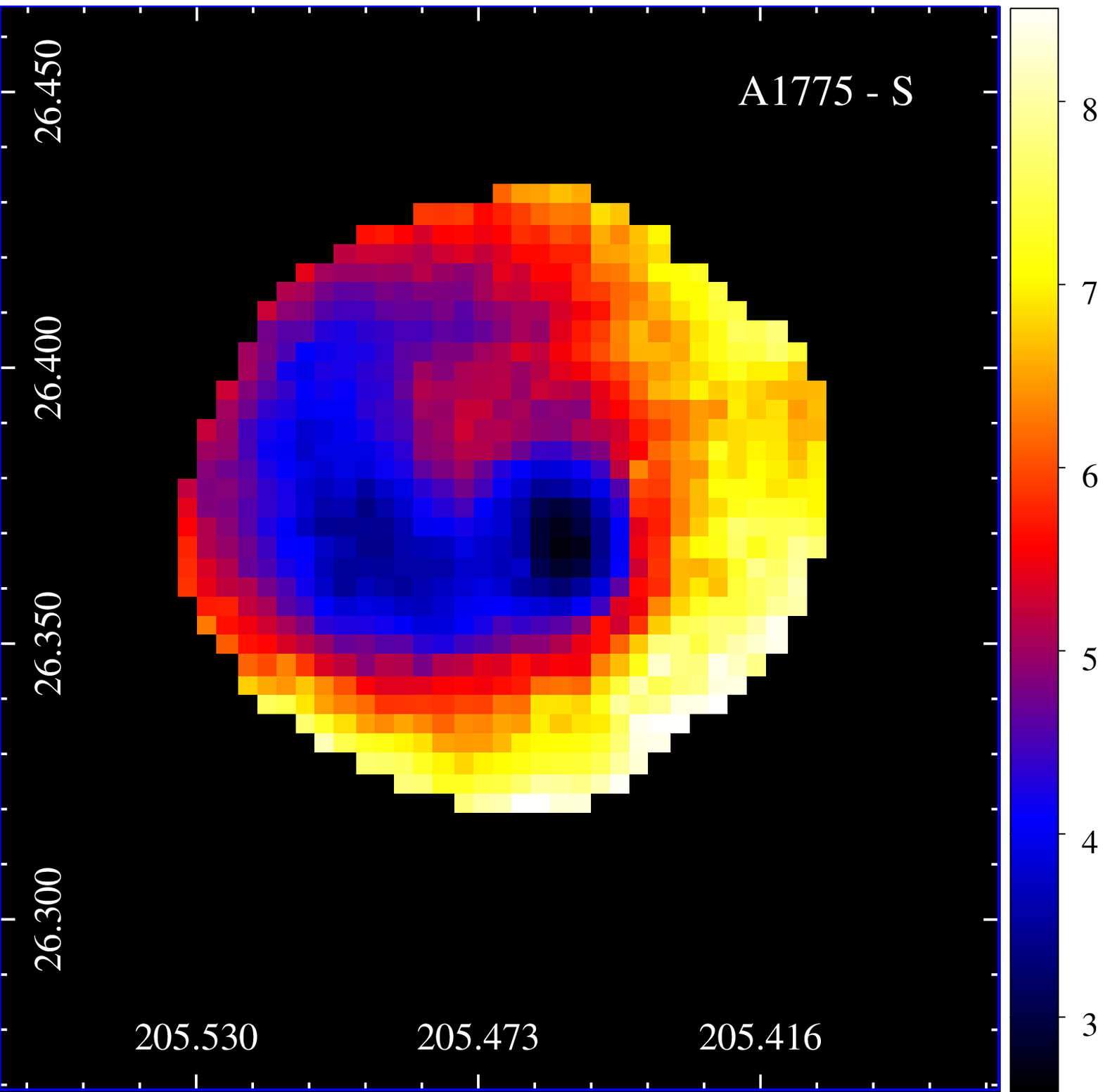}
\includegraphics[scale=0.25]{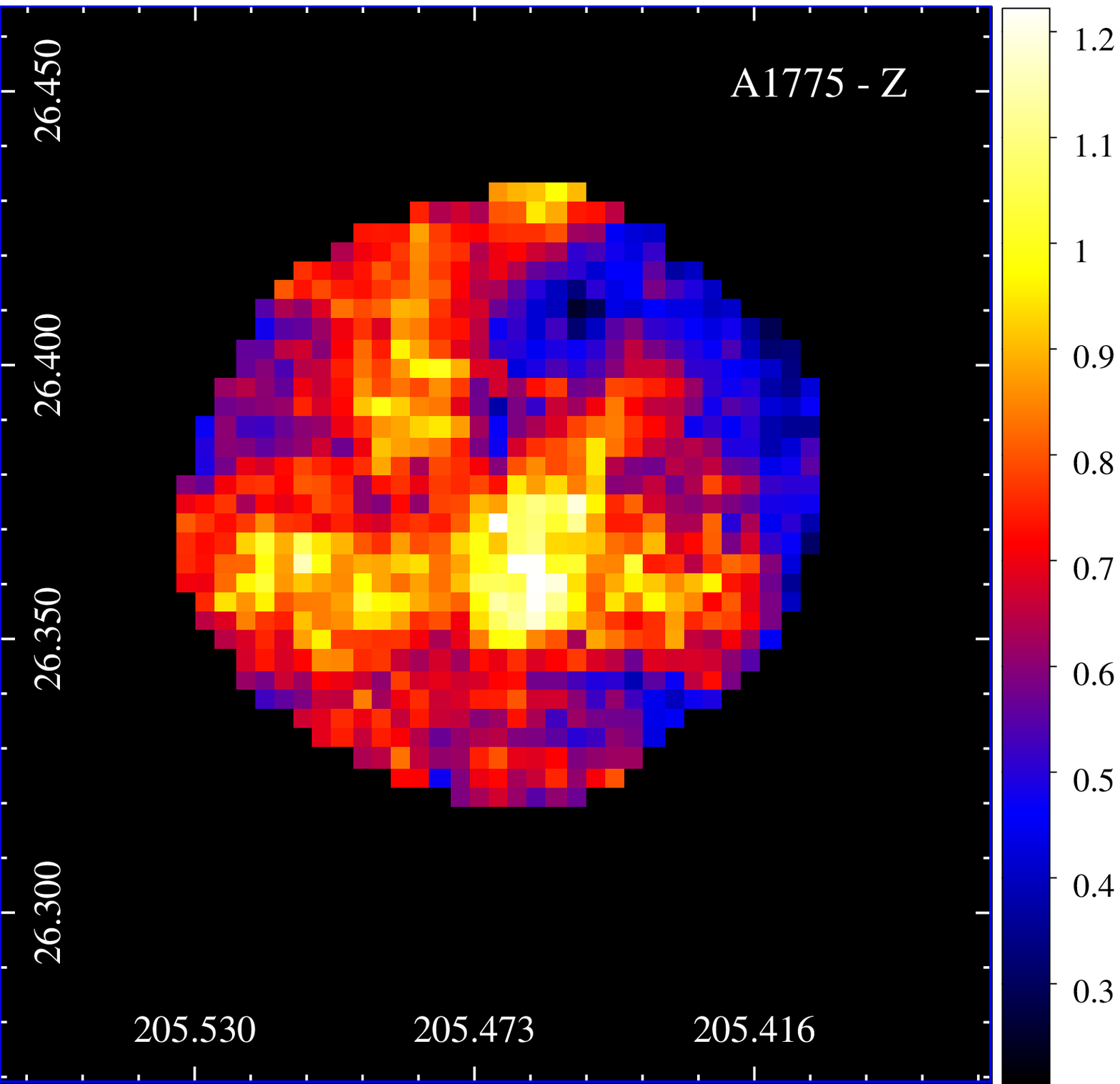}

\includegraphics[scale=0.25]{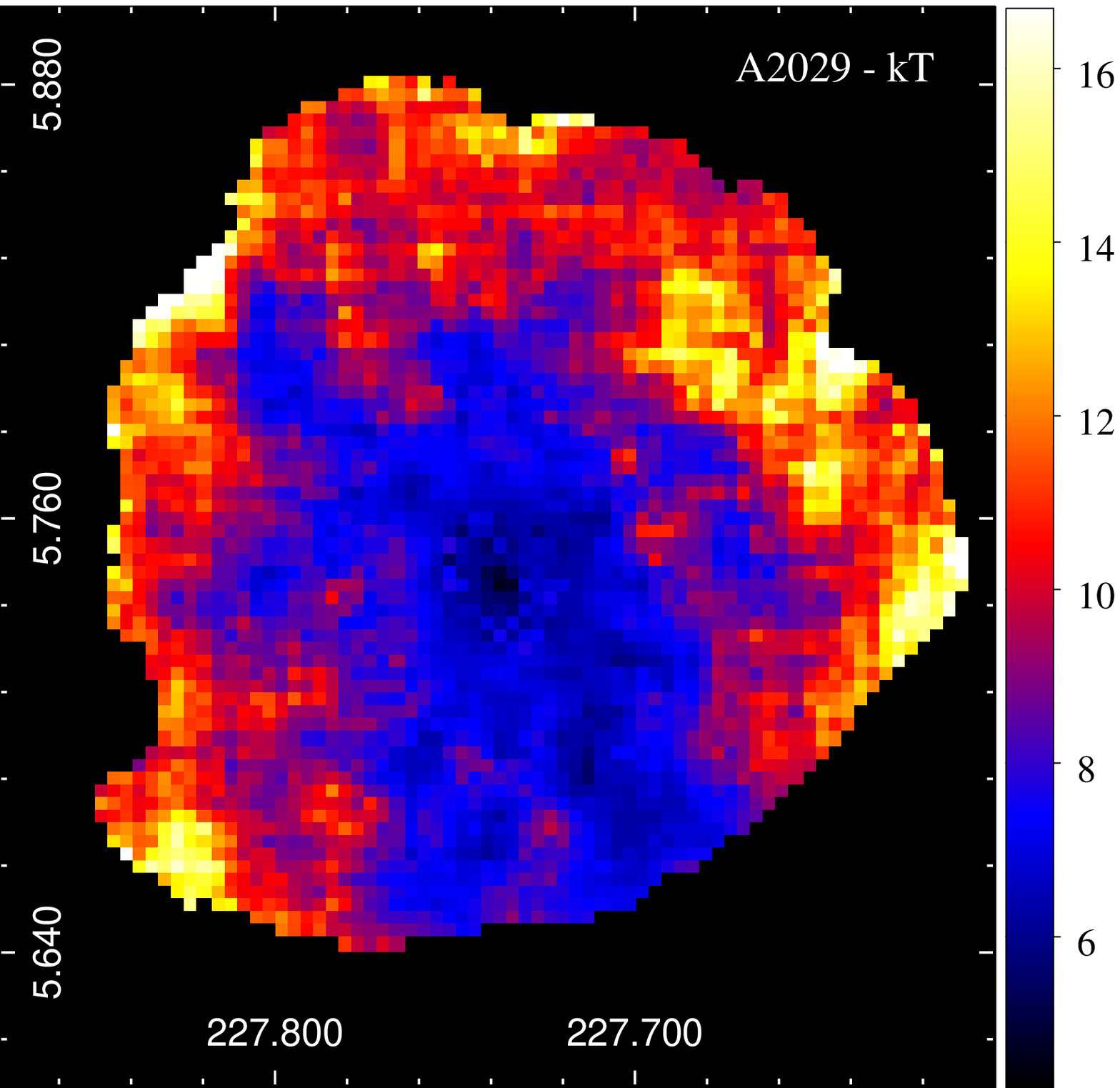}
\includegraphics[scale=0.25]{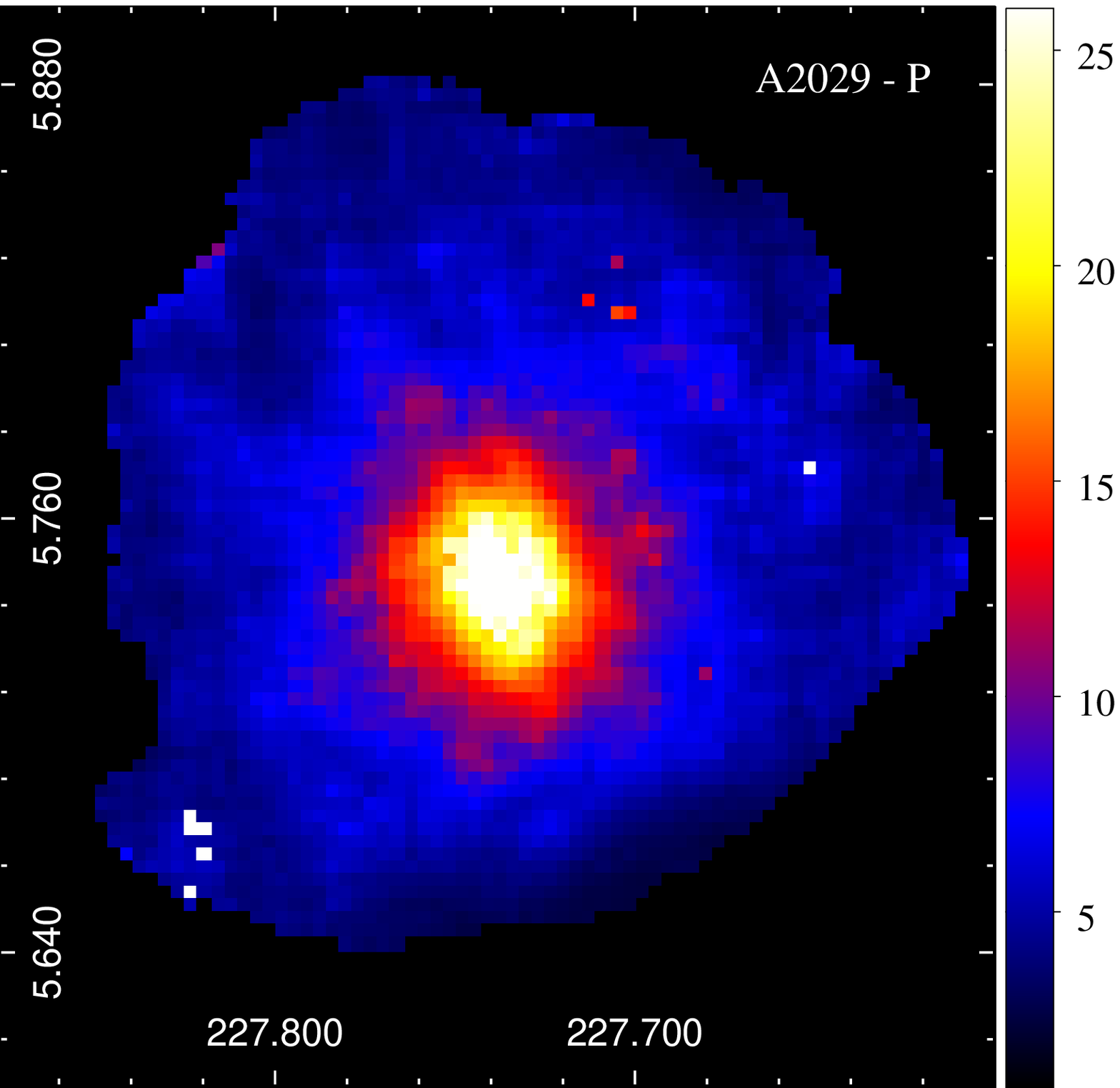}
\includegraphics[scale=0.25]{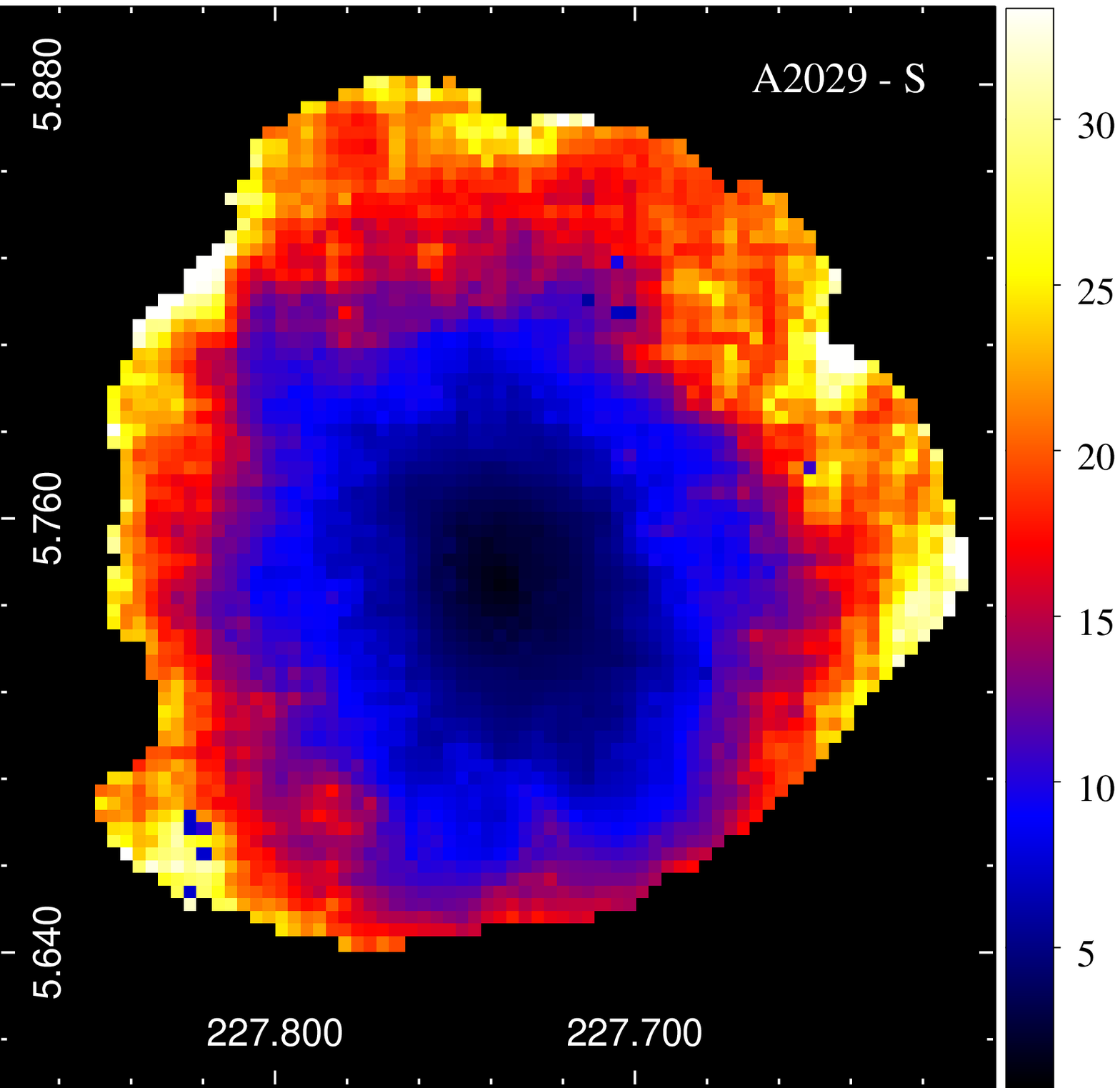}
\includegraphics[scale=0.25]{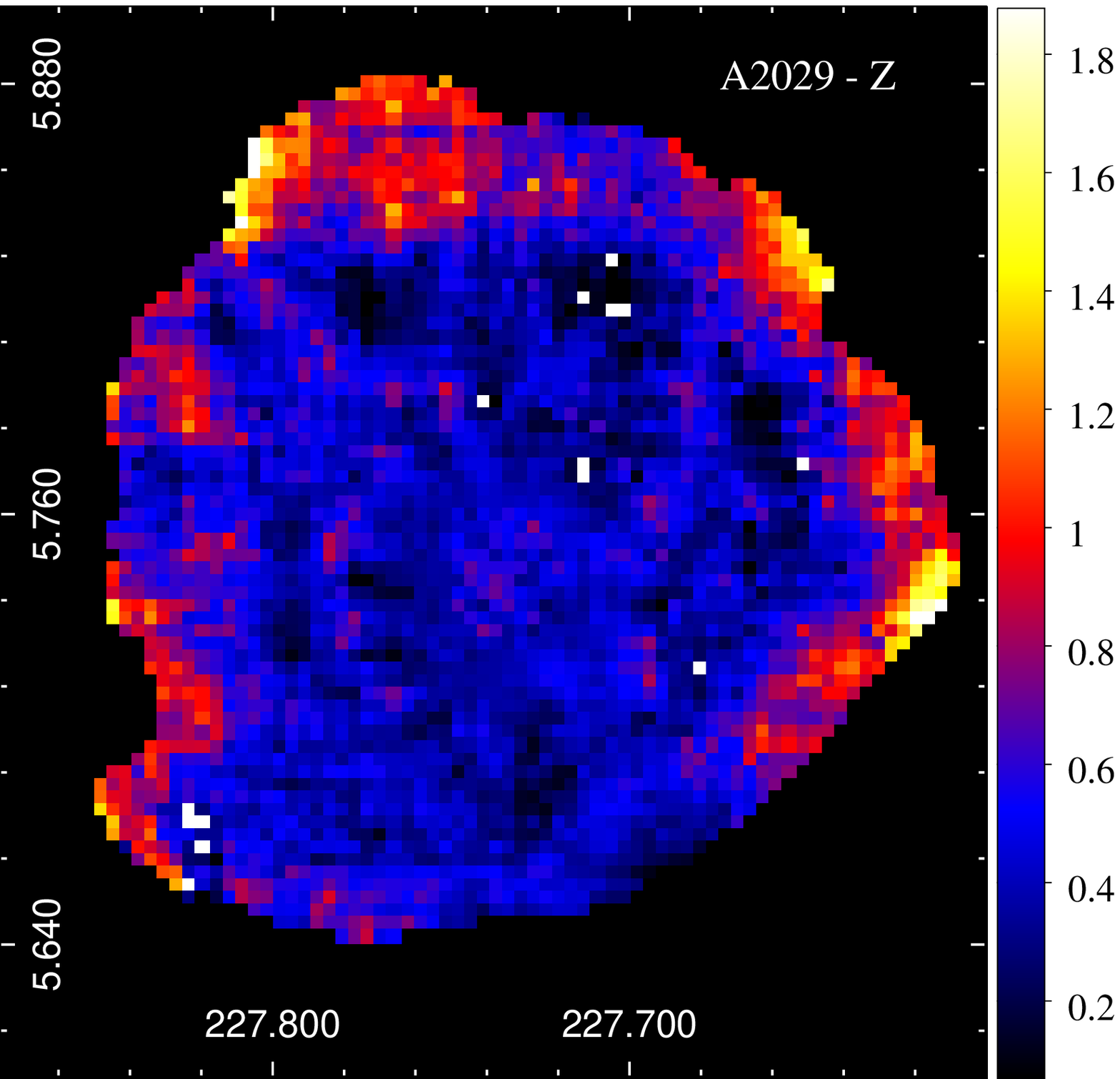}

\includegraphics[scale=0.25]{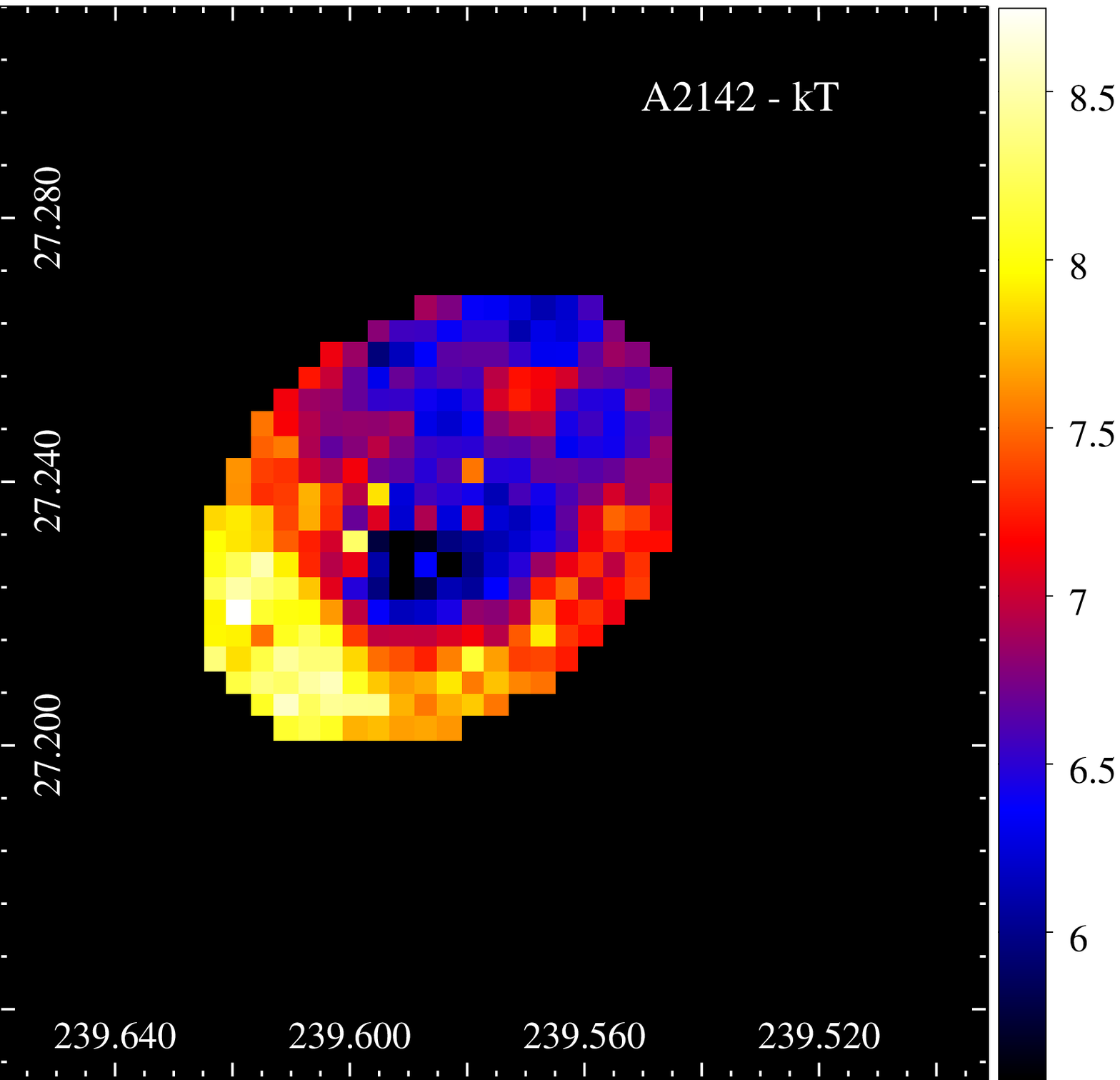}
\includegraphics[scale=0.25]{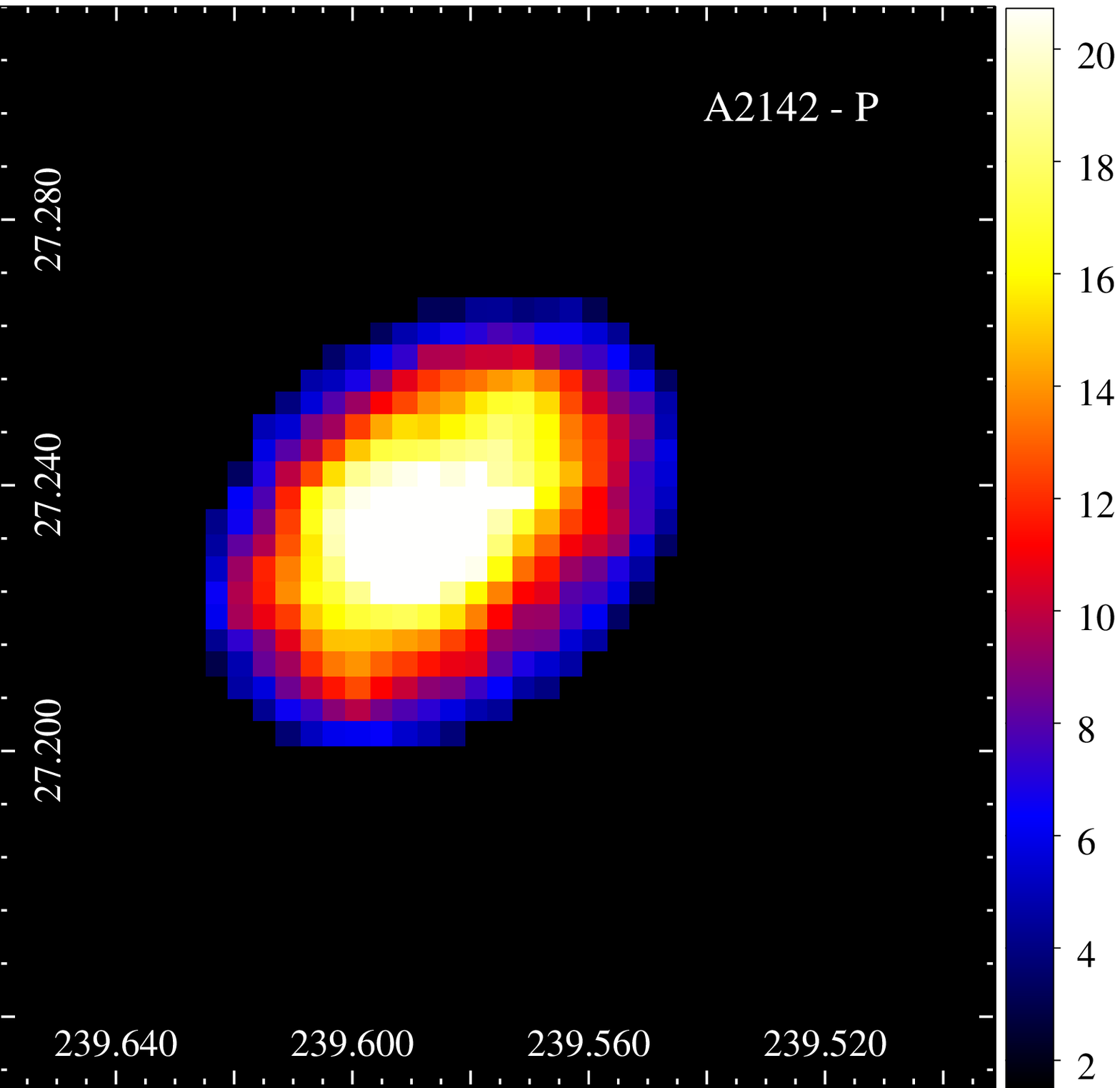}
\includegraphics[scale=0.25]{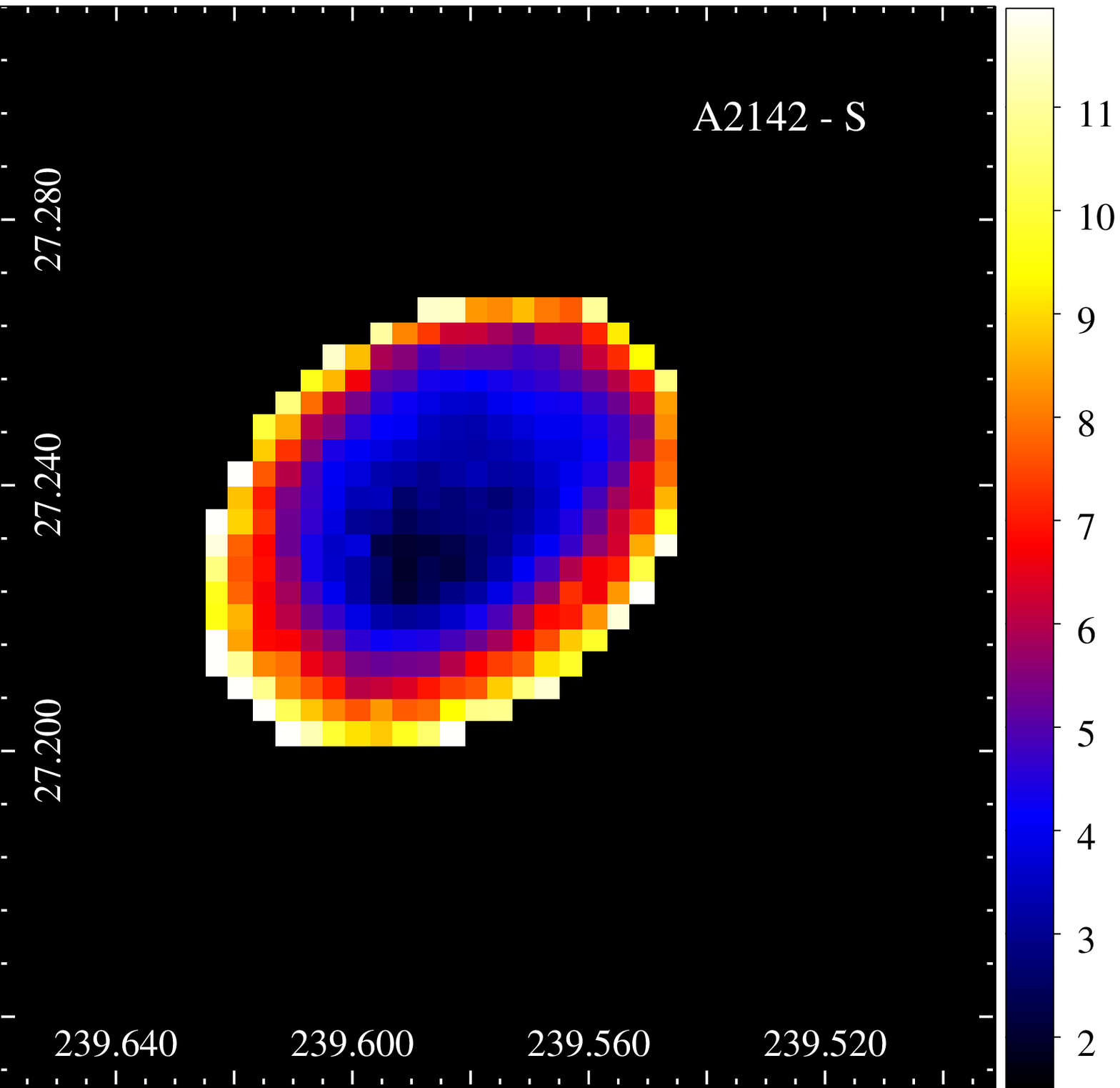}
\includegraphics[scale=0.25]{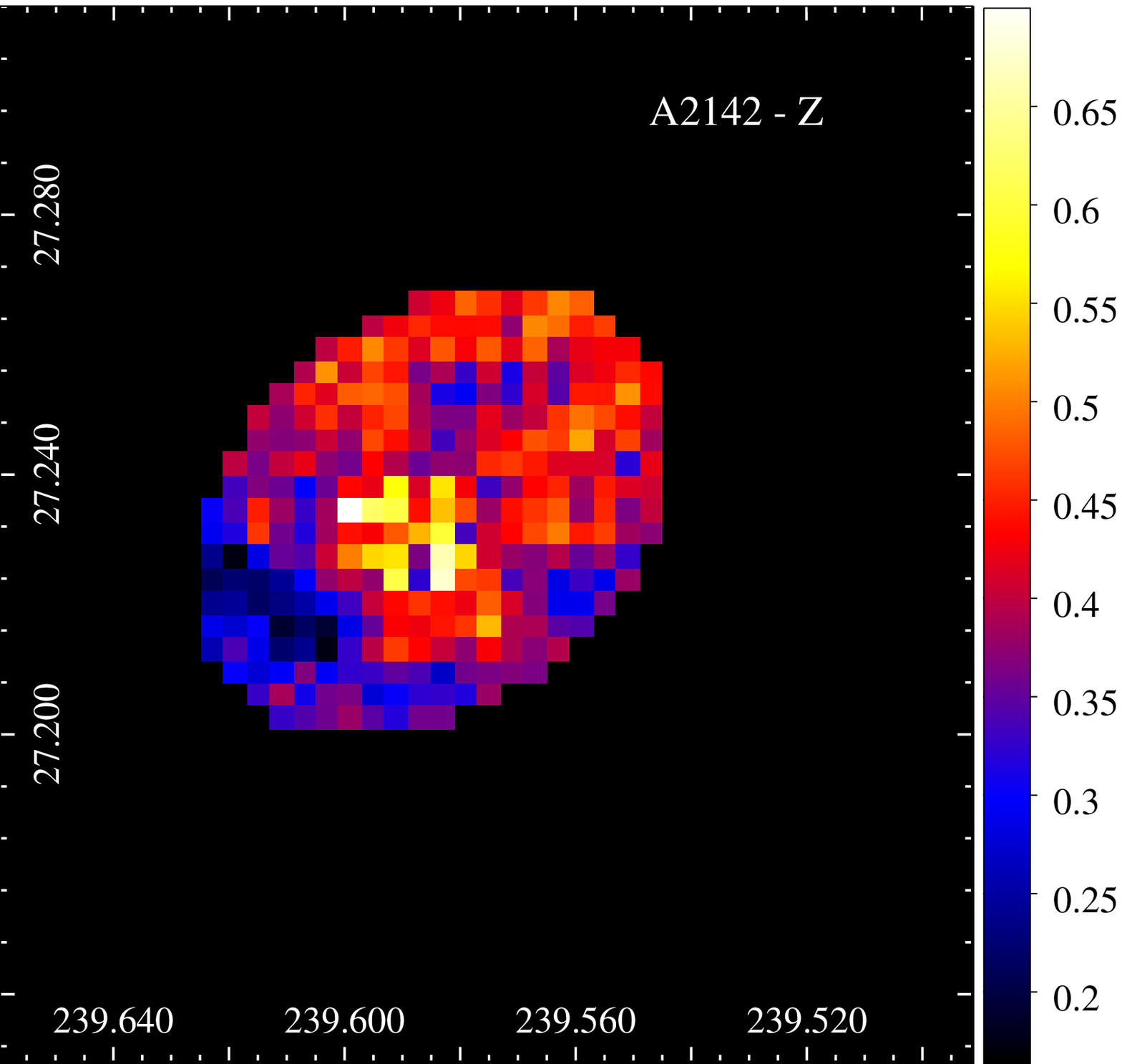}

\caption{CC and disturbed systems. From left to right: temperature, pseudo-pressure, pseudo-entropy, and 
metallicity map for A3558 (A3558 and bA3558 in the same map), A3562, A1775, A2029, and A2142.}
\label{fig:groupsA85b}
\end{figure*}

\begin{figure*}

\includegraphics[scale=0.25]{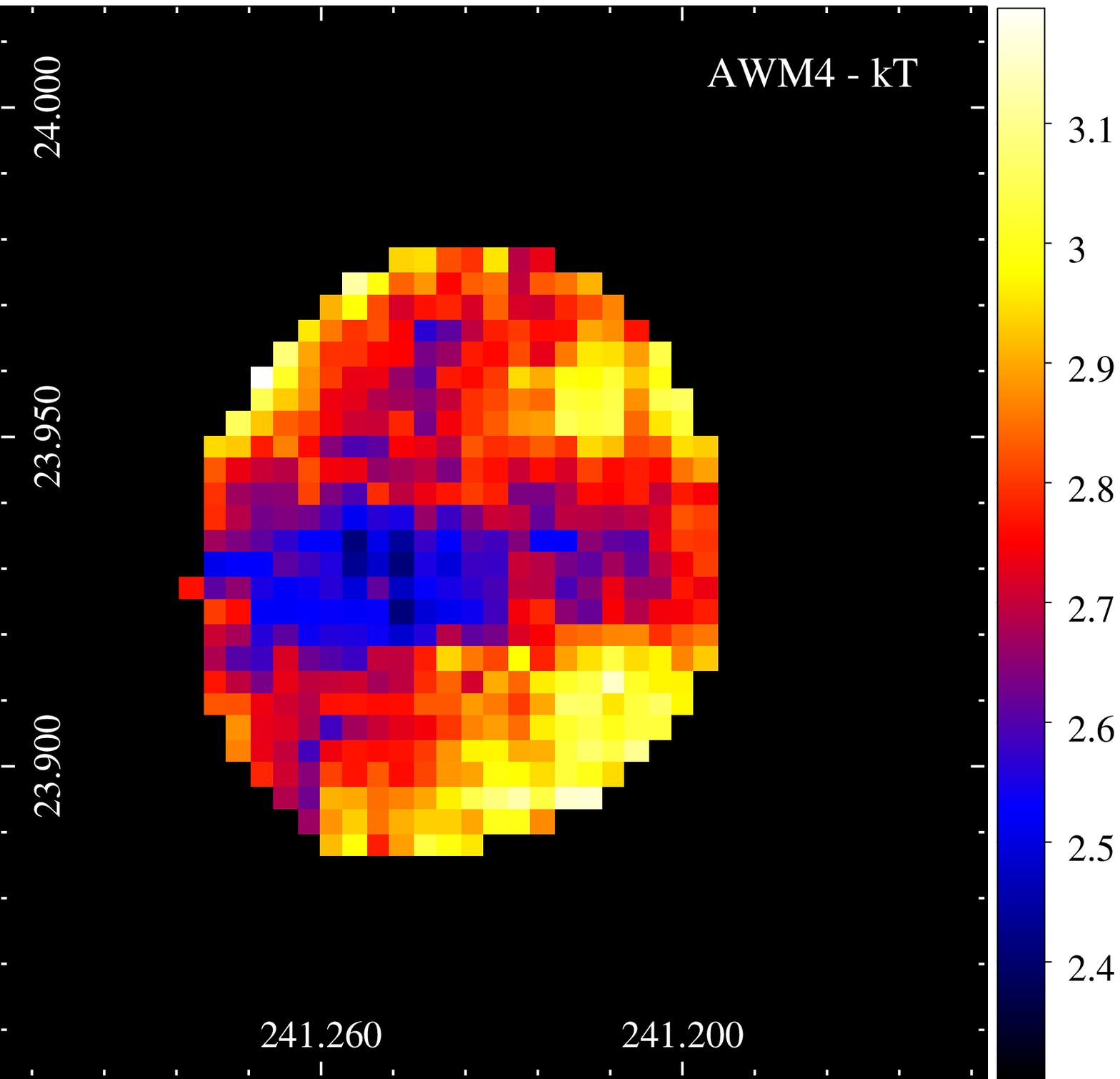}
\includegraphics[scale=0.25]{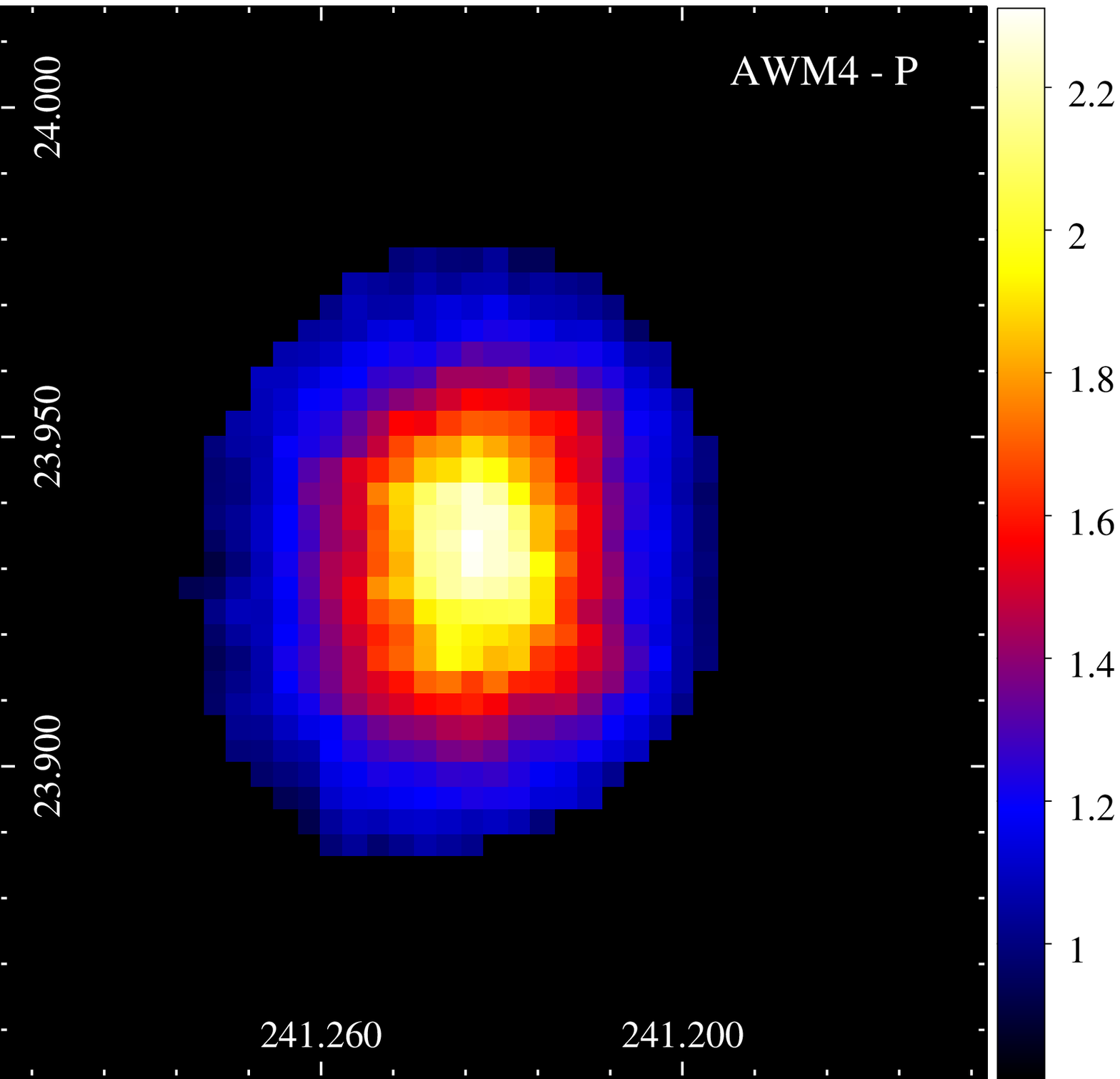}
\includegraphics[scale=0.25]{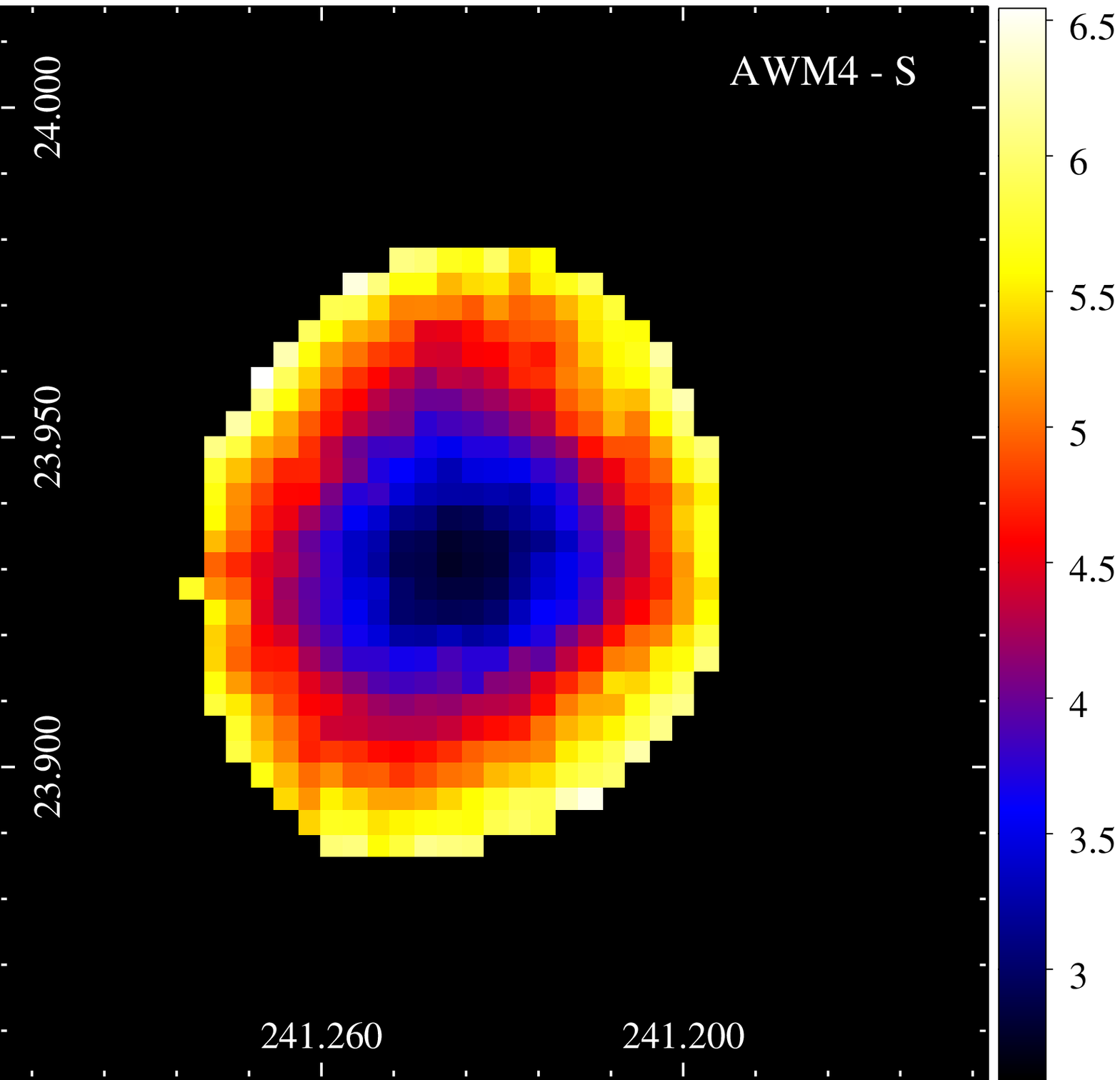}
\includegraphics[scale=0.25]{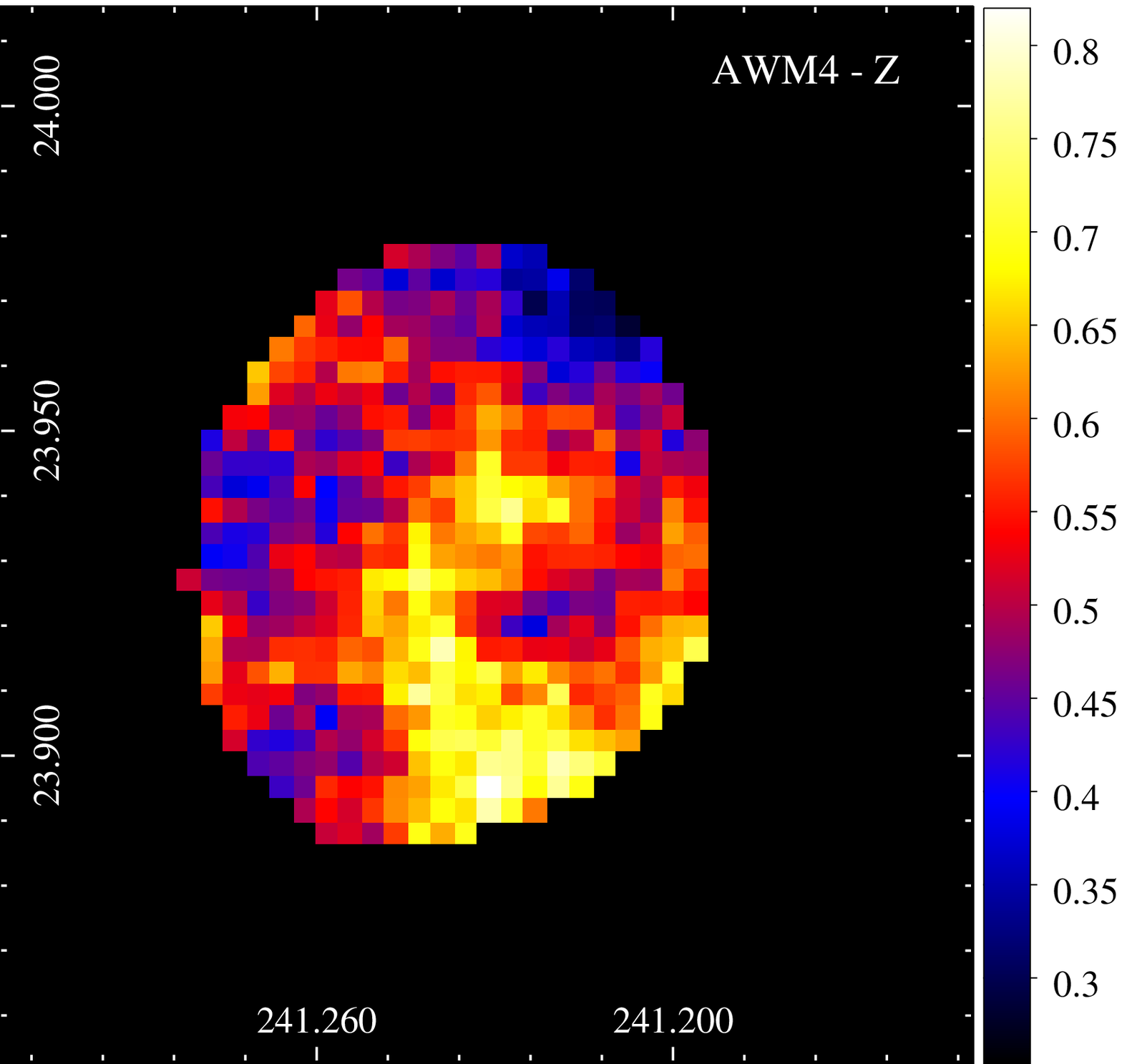}

\includegraphics[scale=0.25]{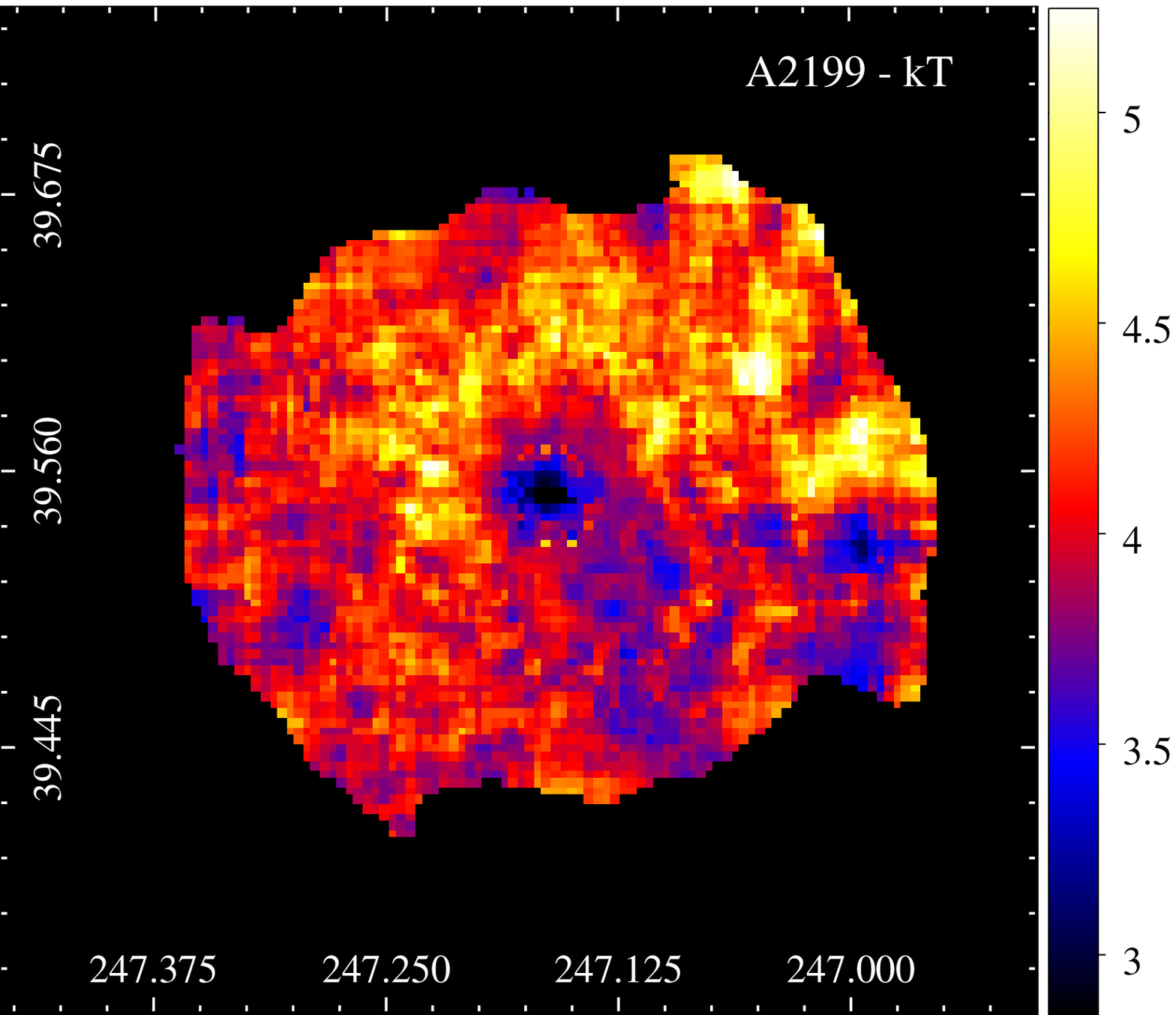}
\includegraphics[scale=0.25]{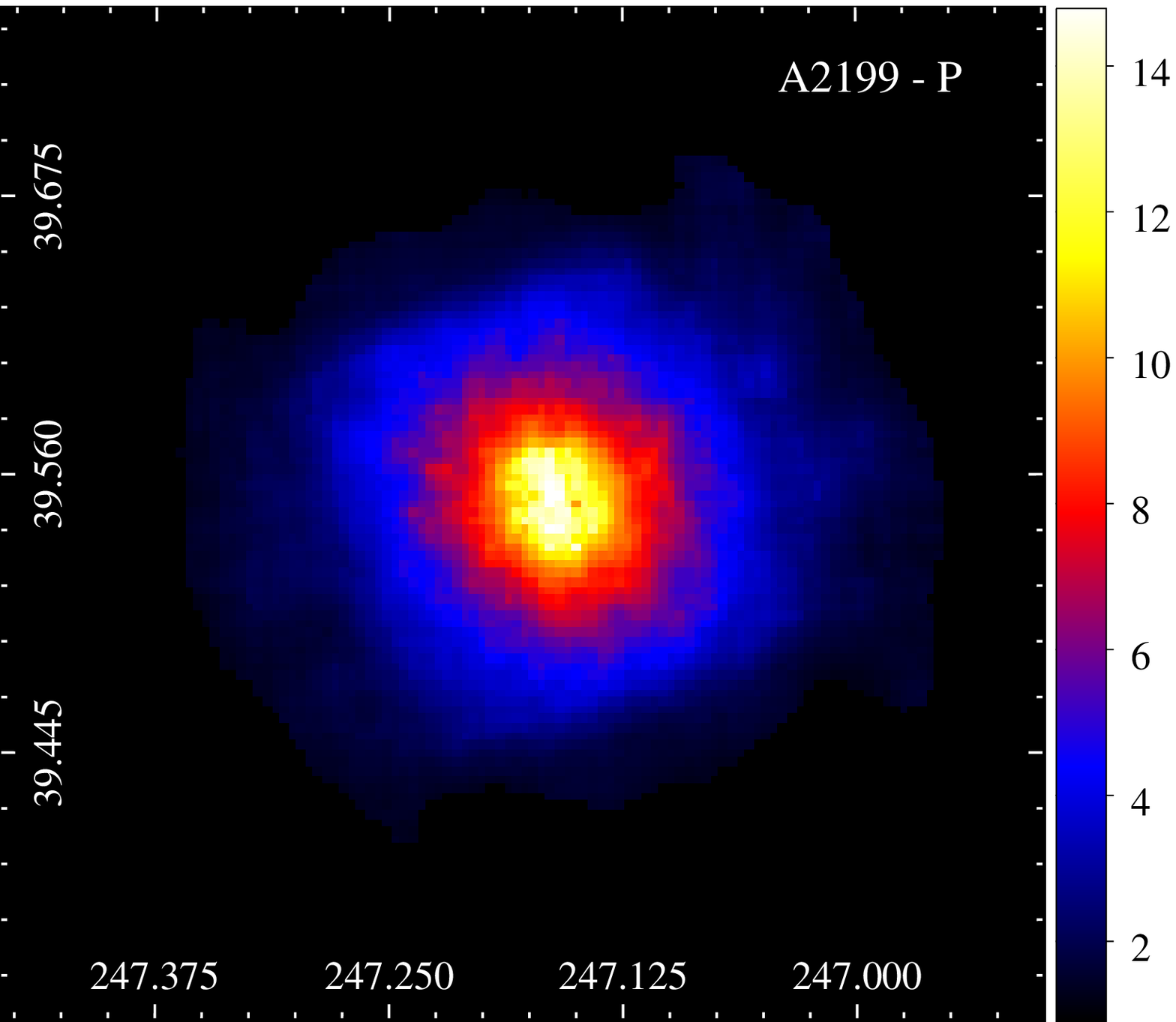}
\includegraphics[scale=0.25]{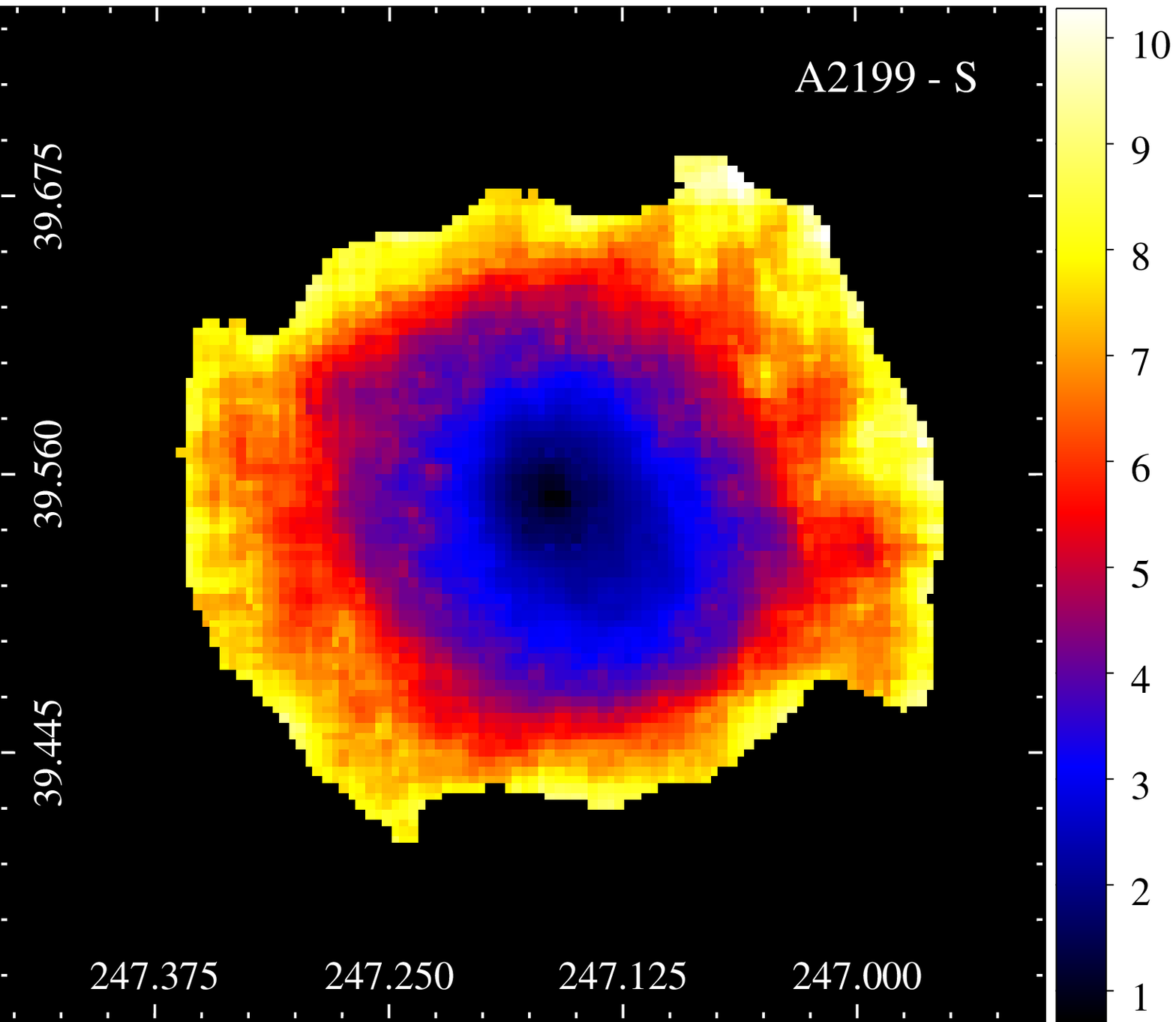}
\includegraphics[scale=0.25]{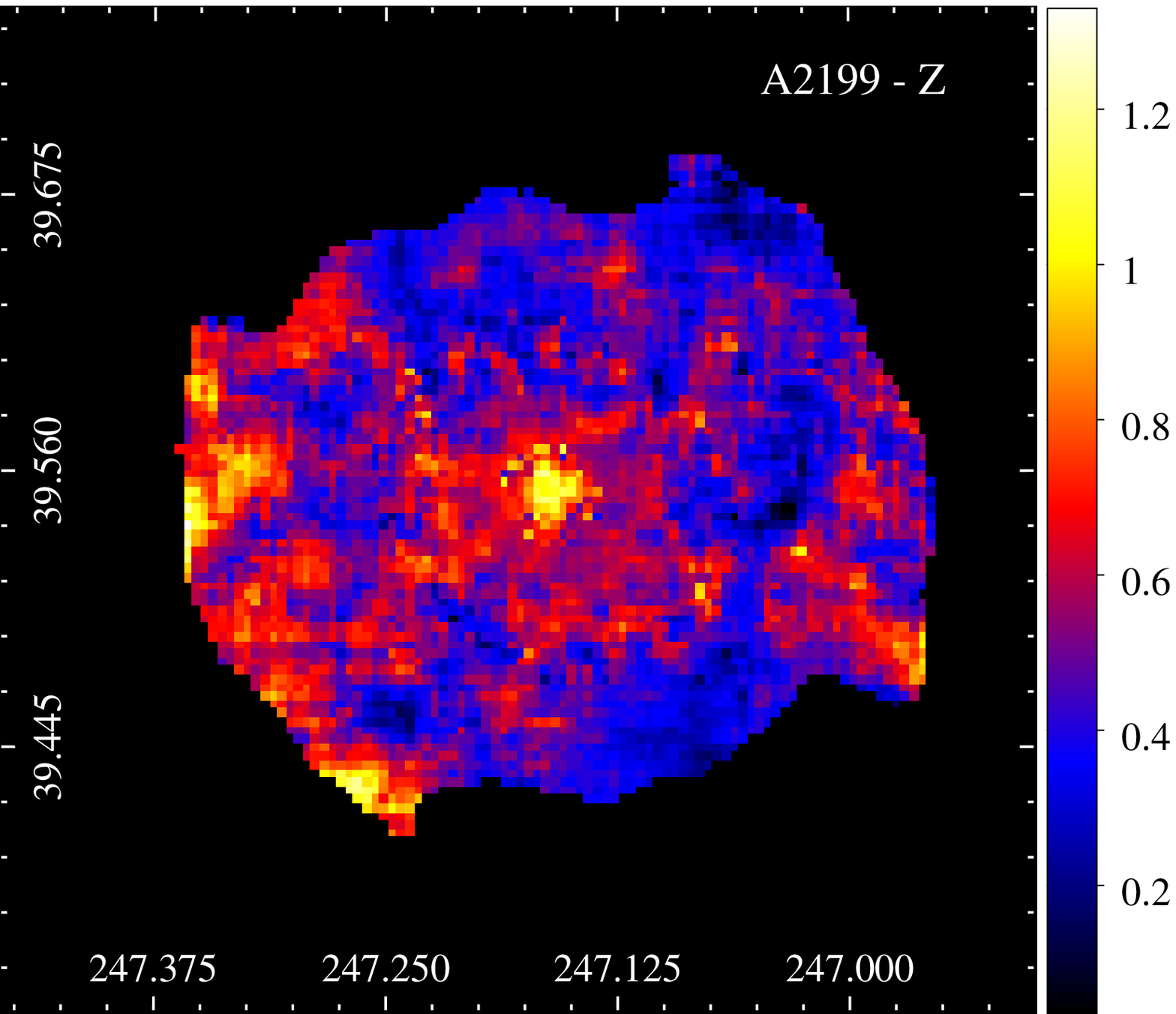}

\includegraphics[scale=0.25]{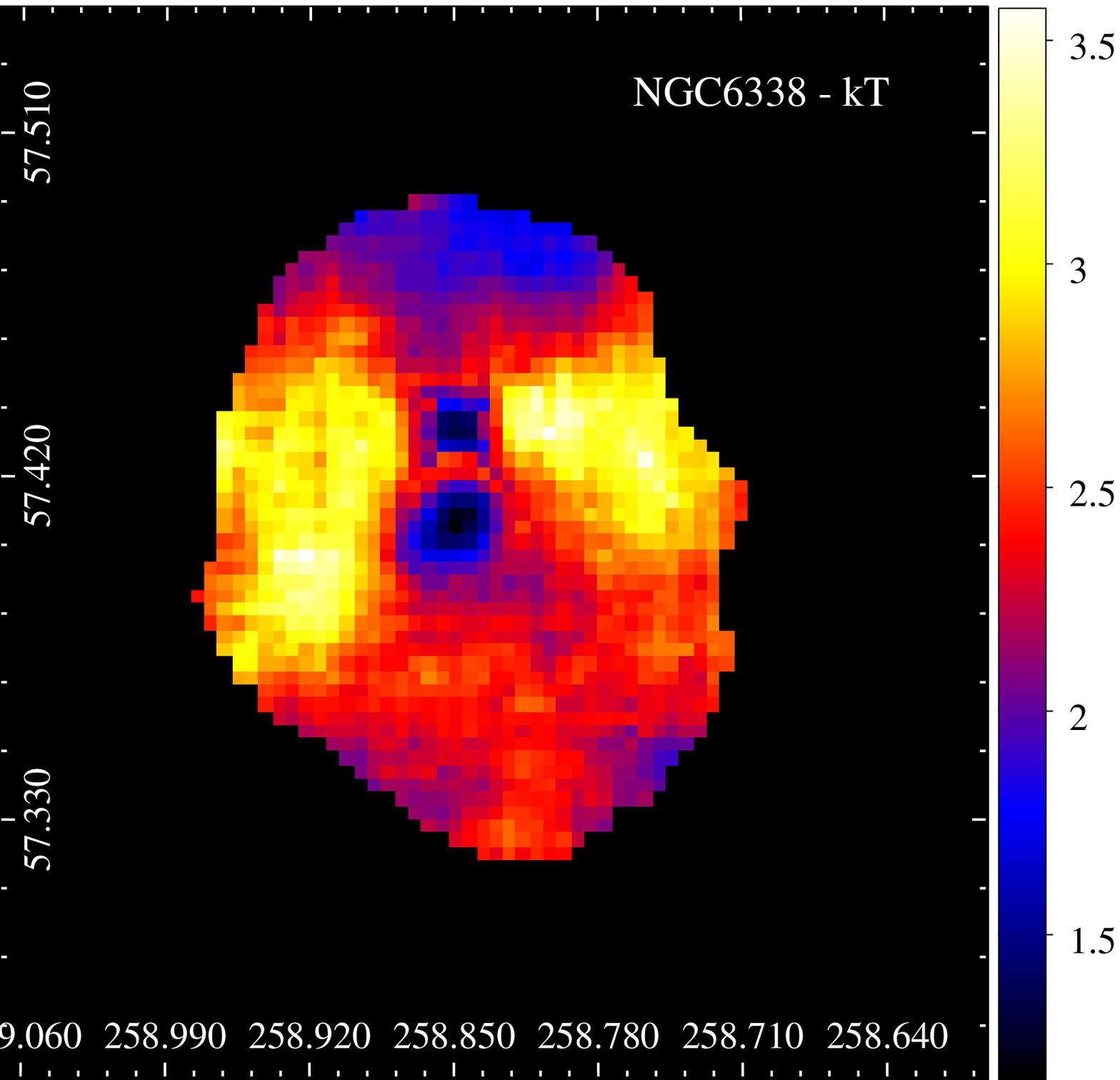}
\includegraphics[scale=0.25]{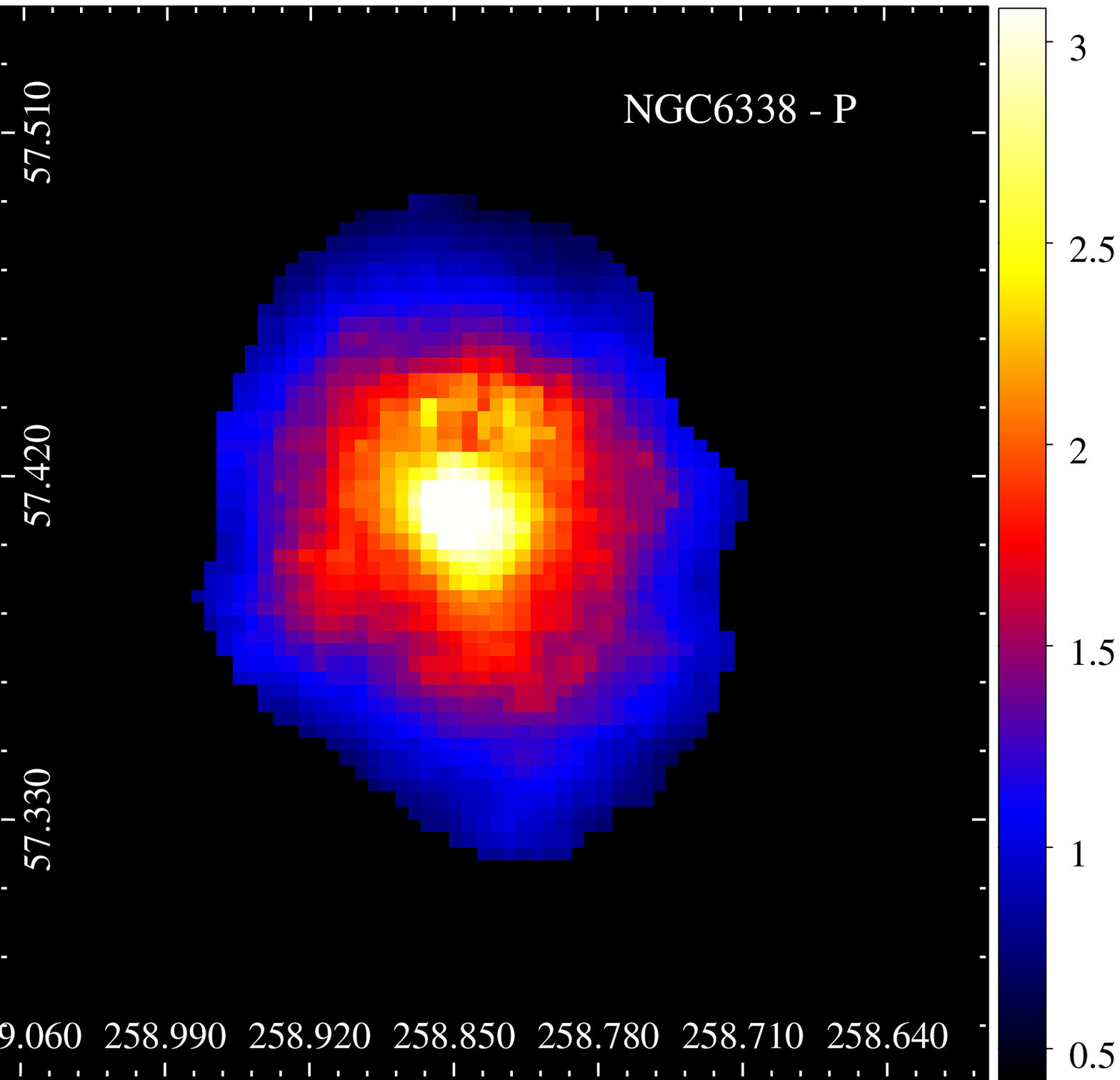}
\includegraphics[scale=0.25]{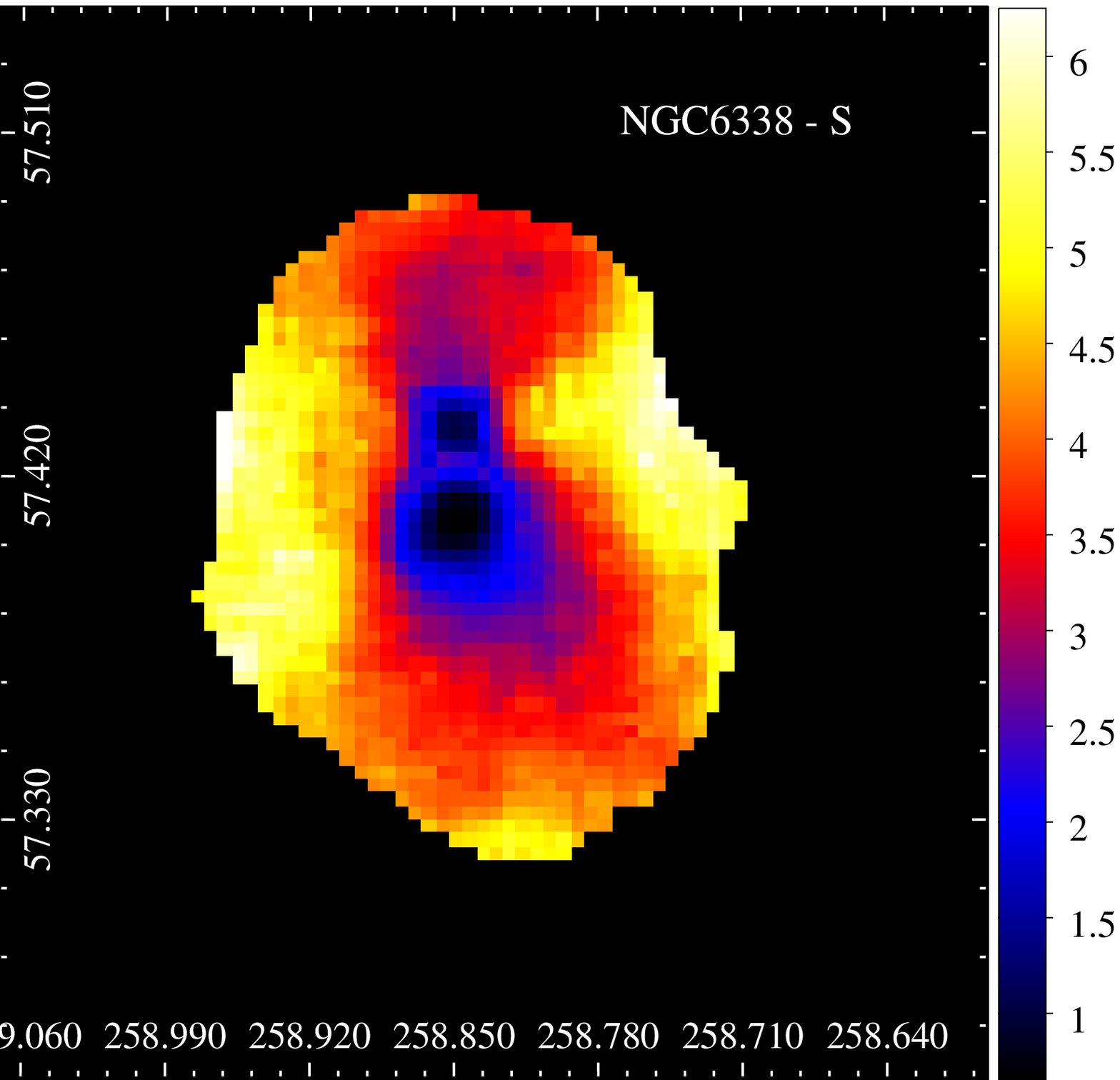}
\includegraphics[scale=0.25]{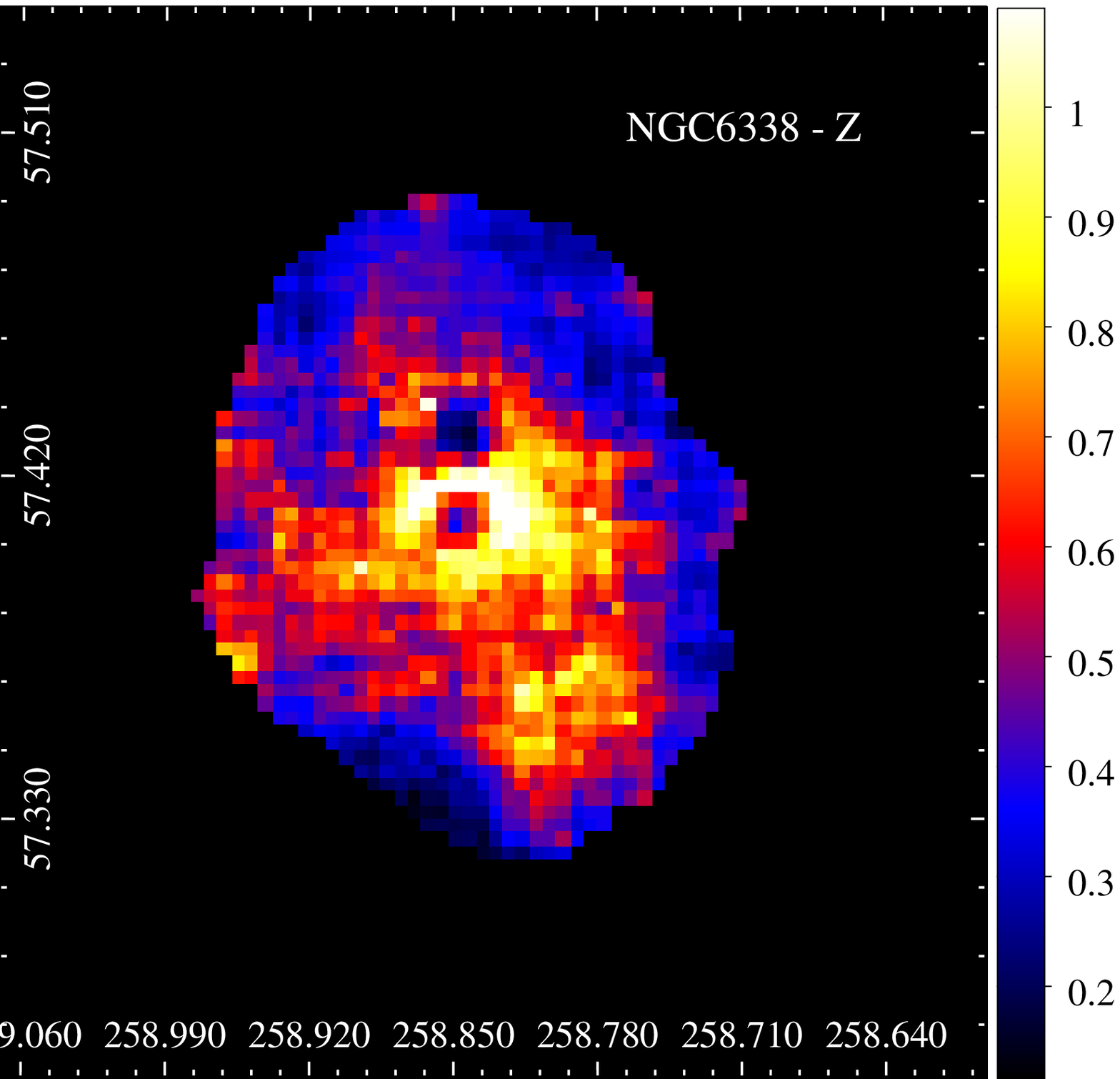}

\includegraphics[scale=0.25]{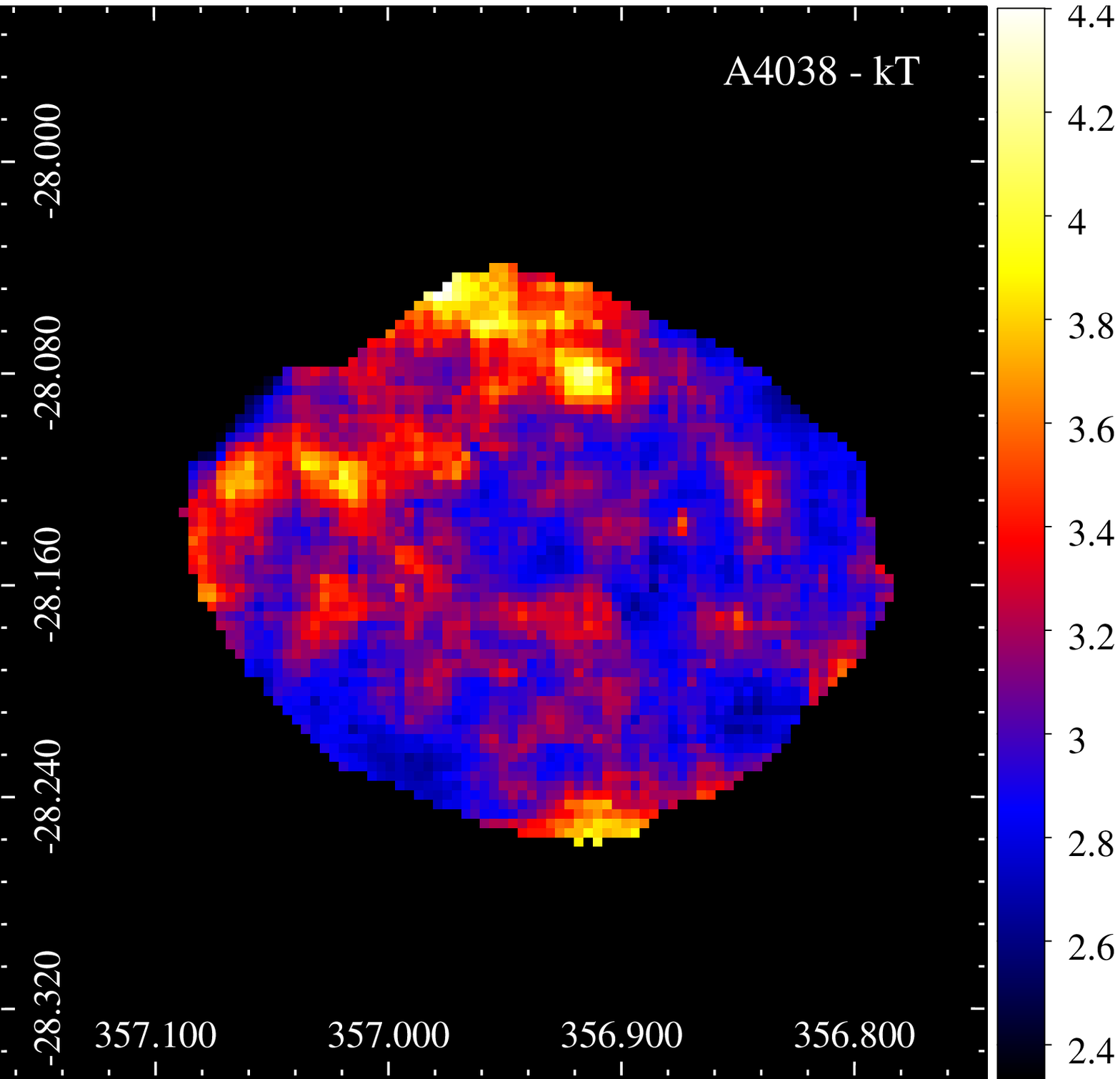}
\includegraphics[scale=0.25]{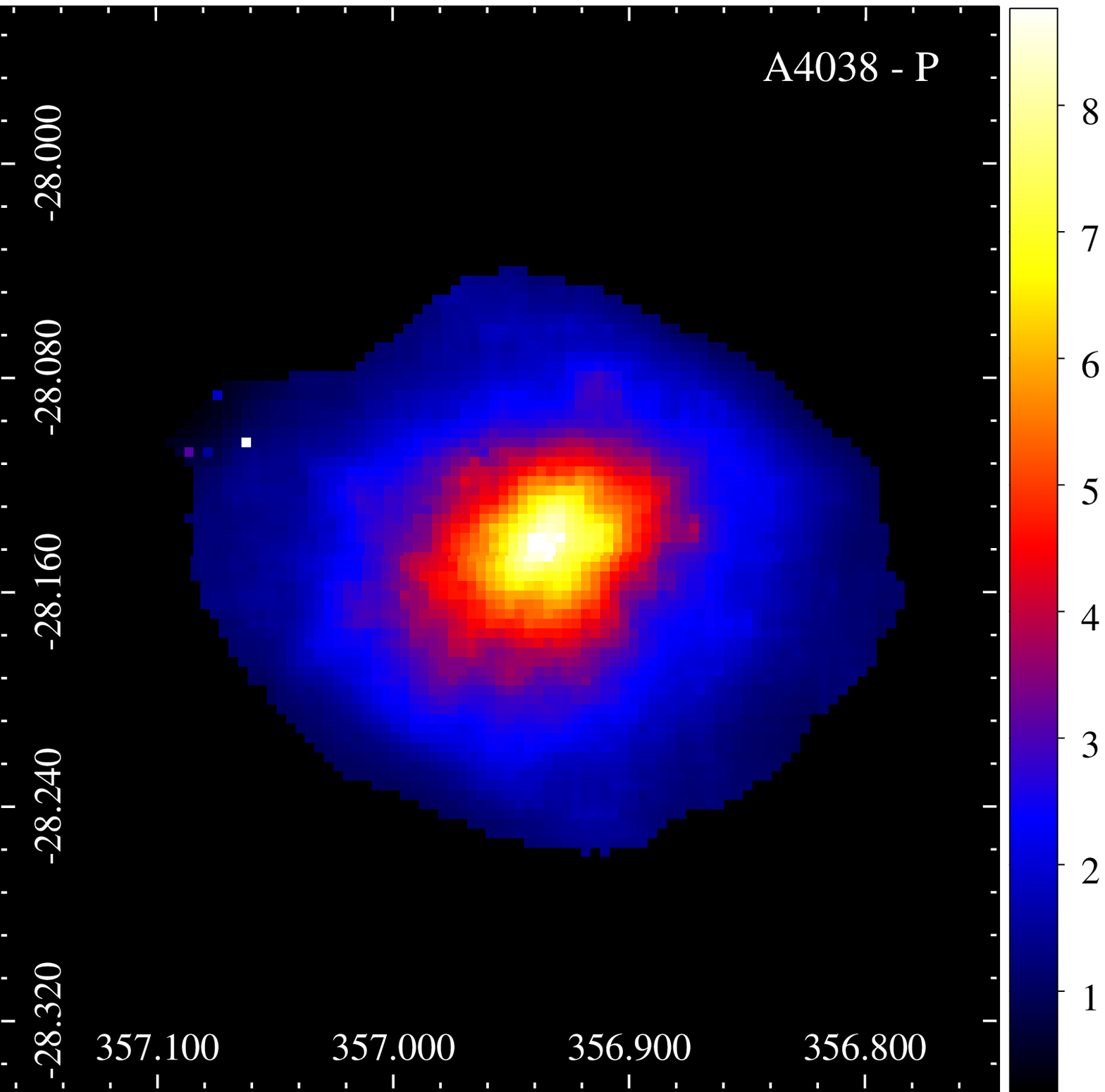}
\includegraphics[scale=0.25]{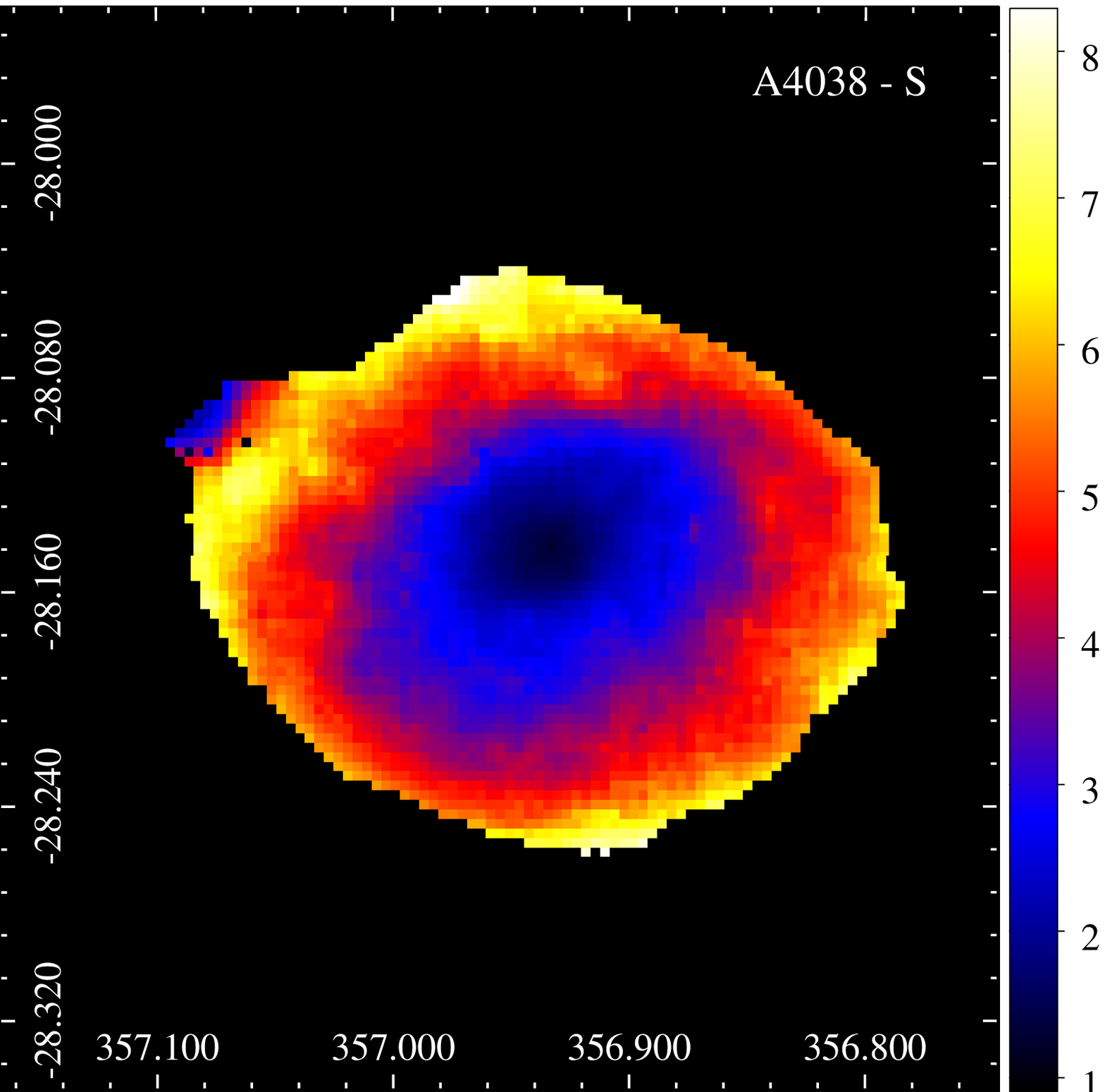}
\includegraphics[scale=0.25]{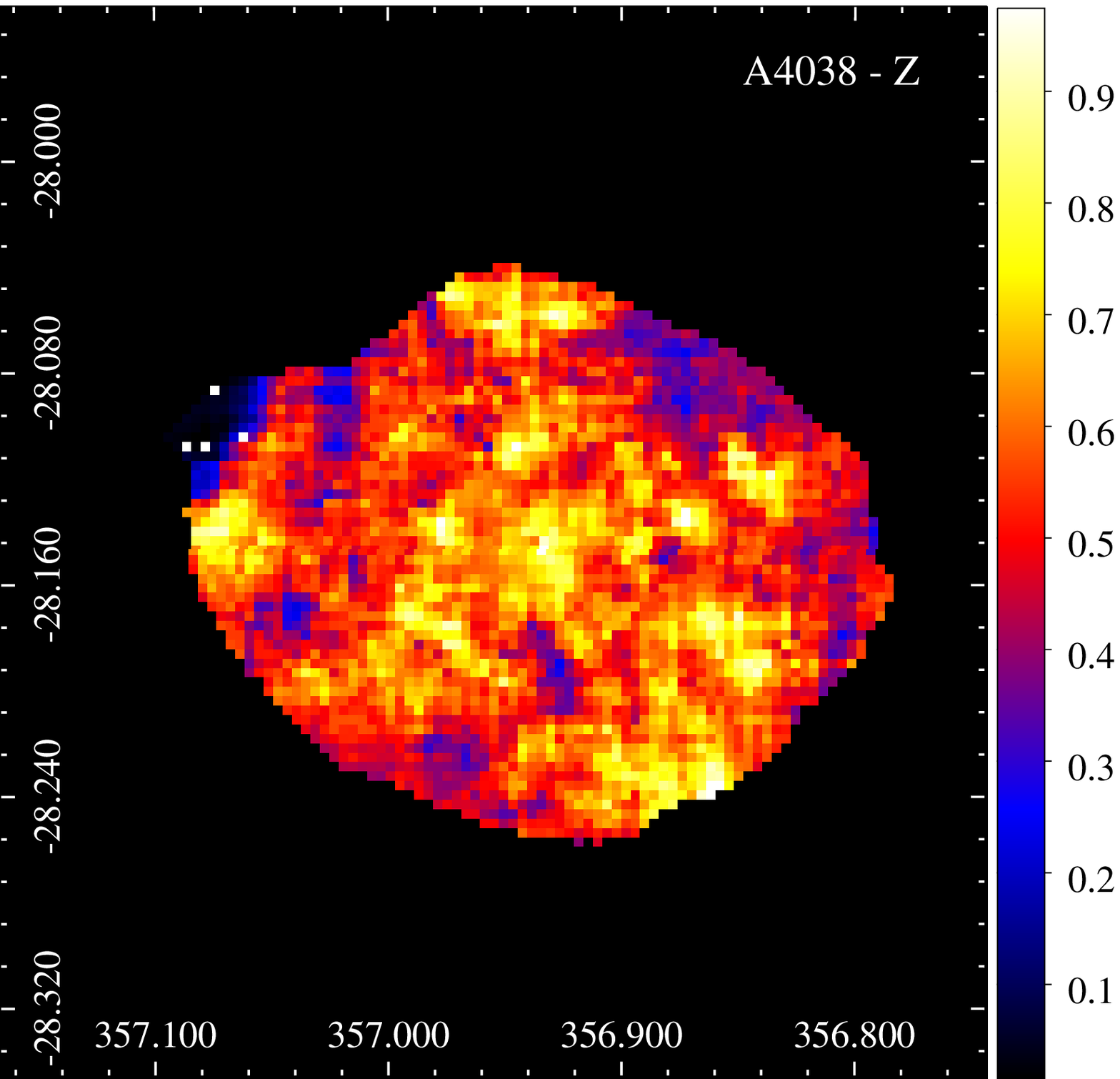}

\caption{CC and disturbed systems. From left to right: temperature, pseudo-pressure, pseudo-entropy, and 
metallicity map for AWM4, A2199, NGC6336 (NGC6338 and bNGC6338), and A4038.}
\label{fig:groupsA85c}
\end{figure*}

%###### NCC-relaxed 

\begin{figure*}

\includegraphics[scale=0.25]{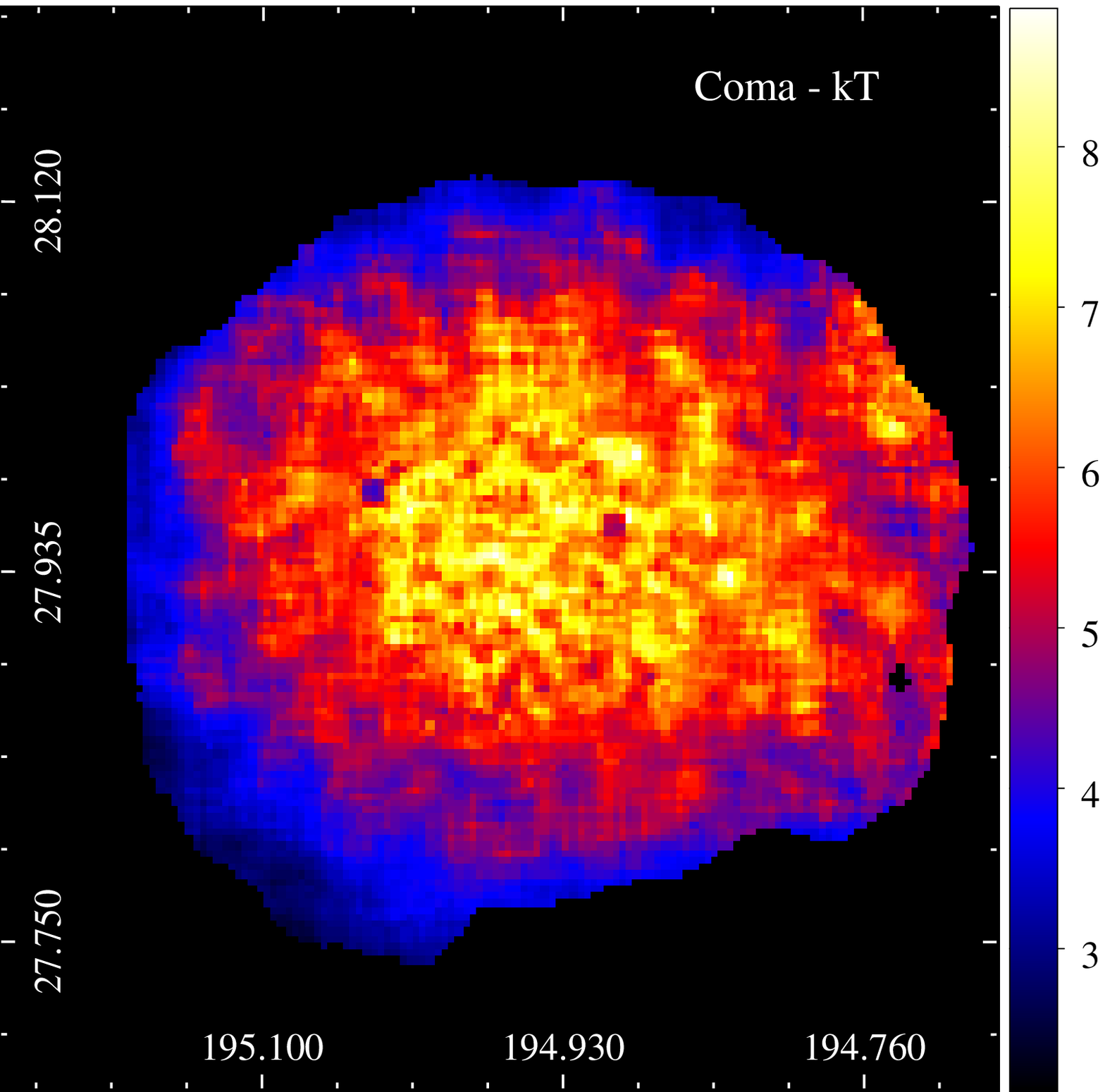}
\includegraphics[scale=0.25]{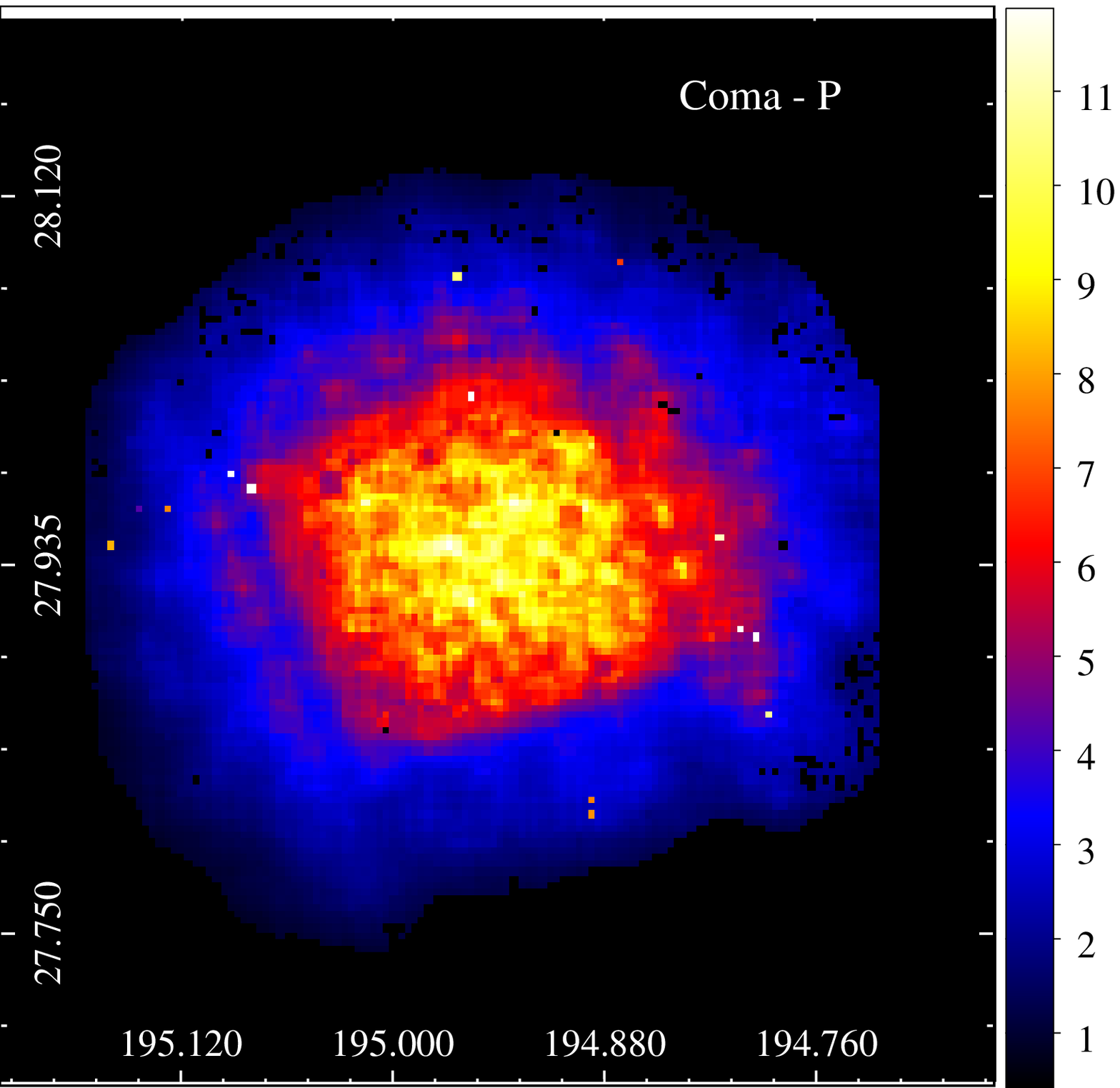}
\includegraphics[scale=0.25]{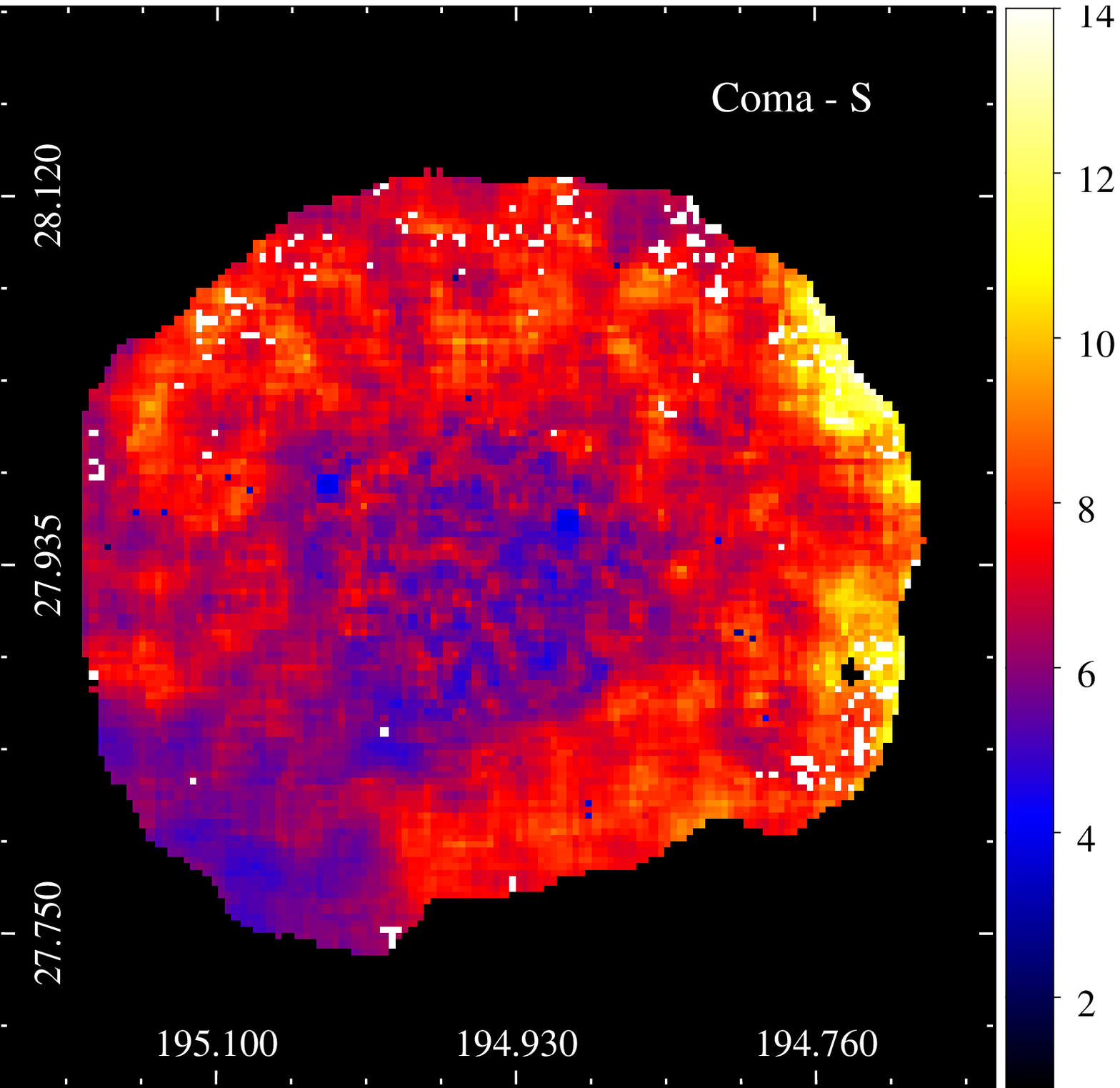}
\includegraphics[scale=0.25]{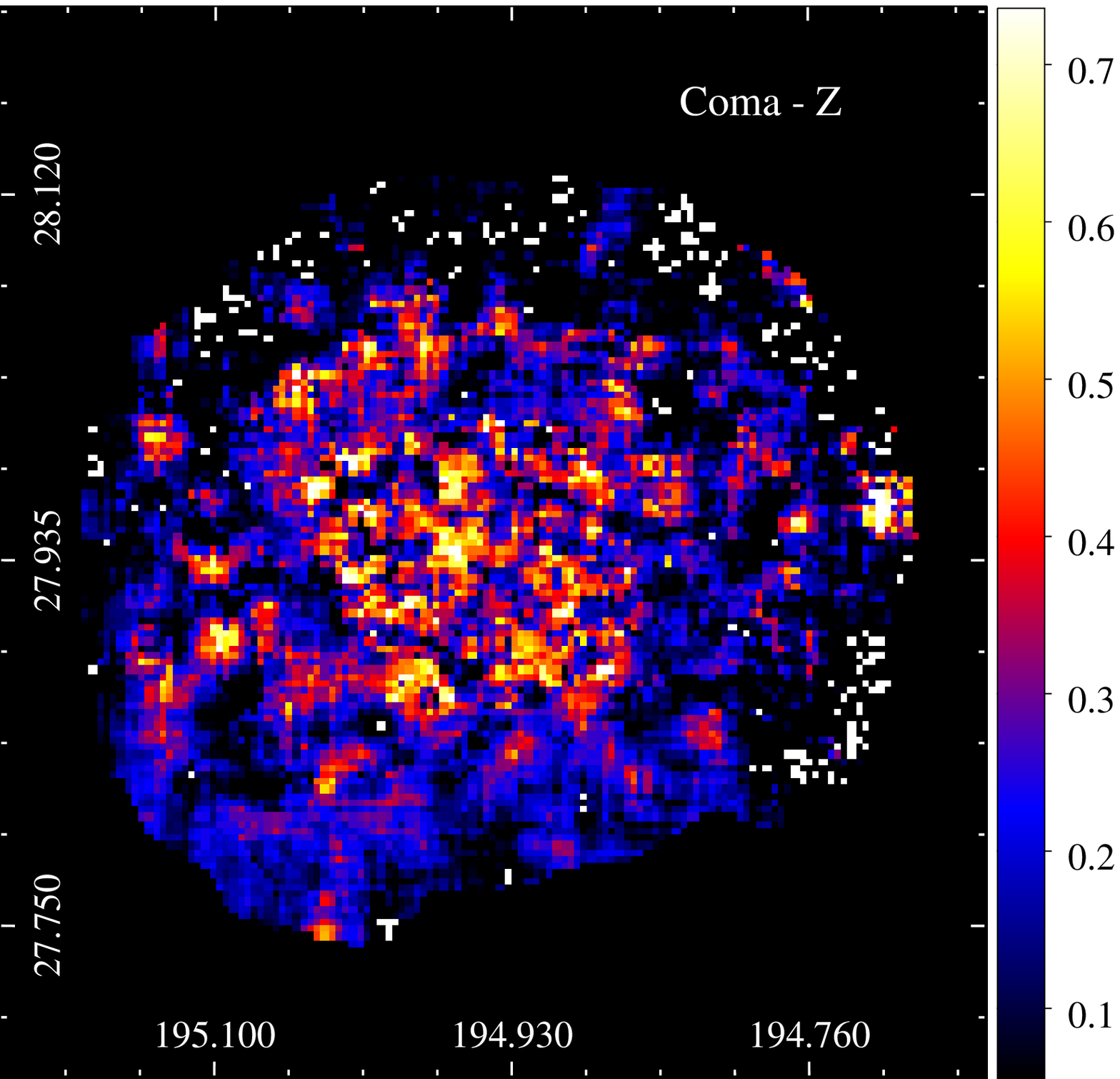}

\includegraphics[scale=0.25]{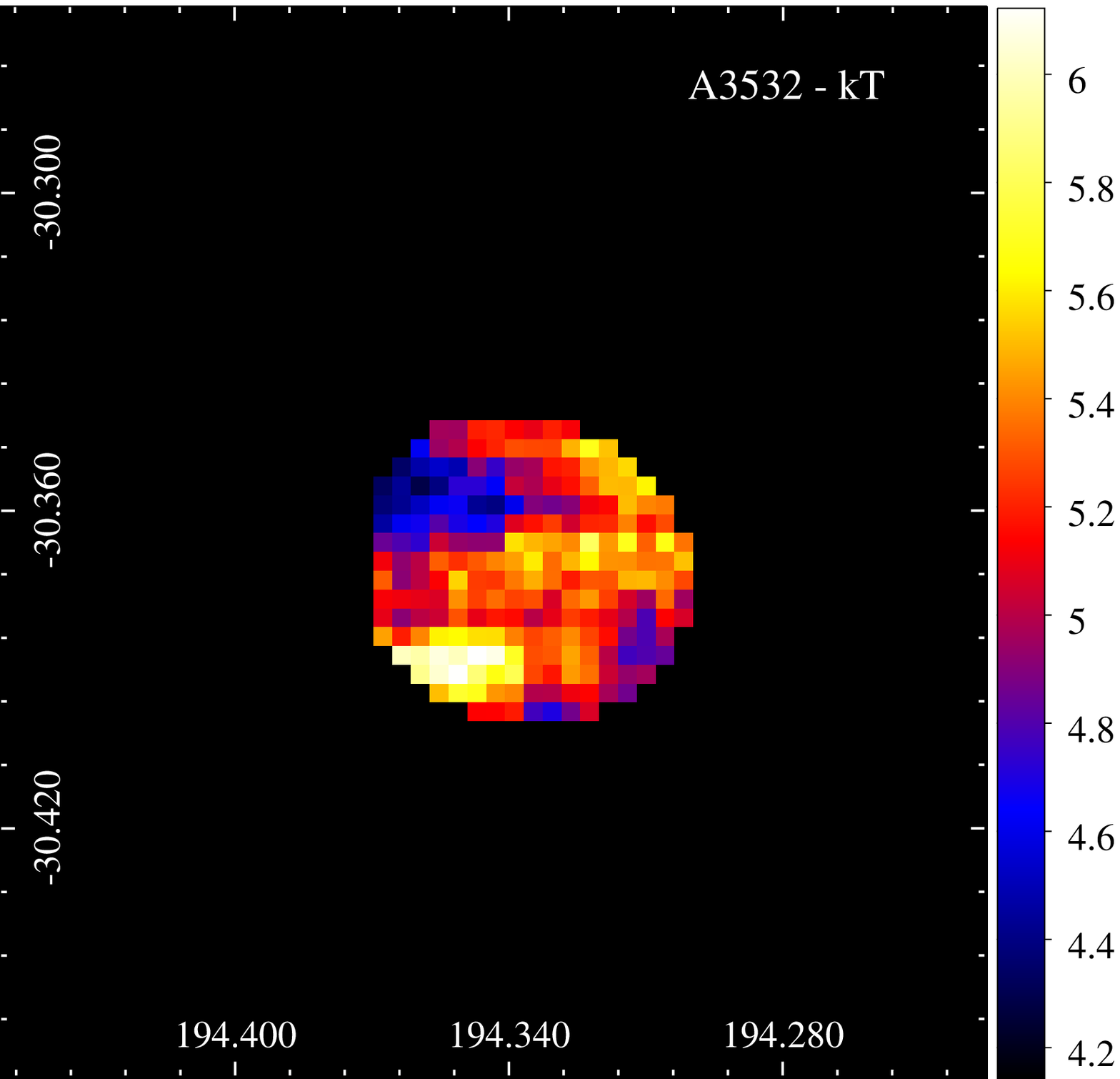}
\includegraphics[scale=0.25]{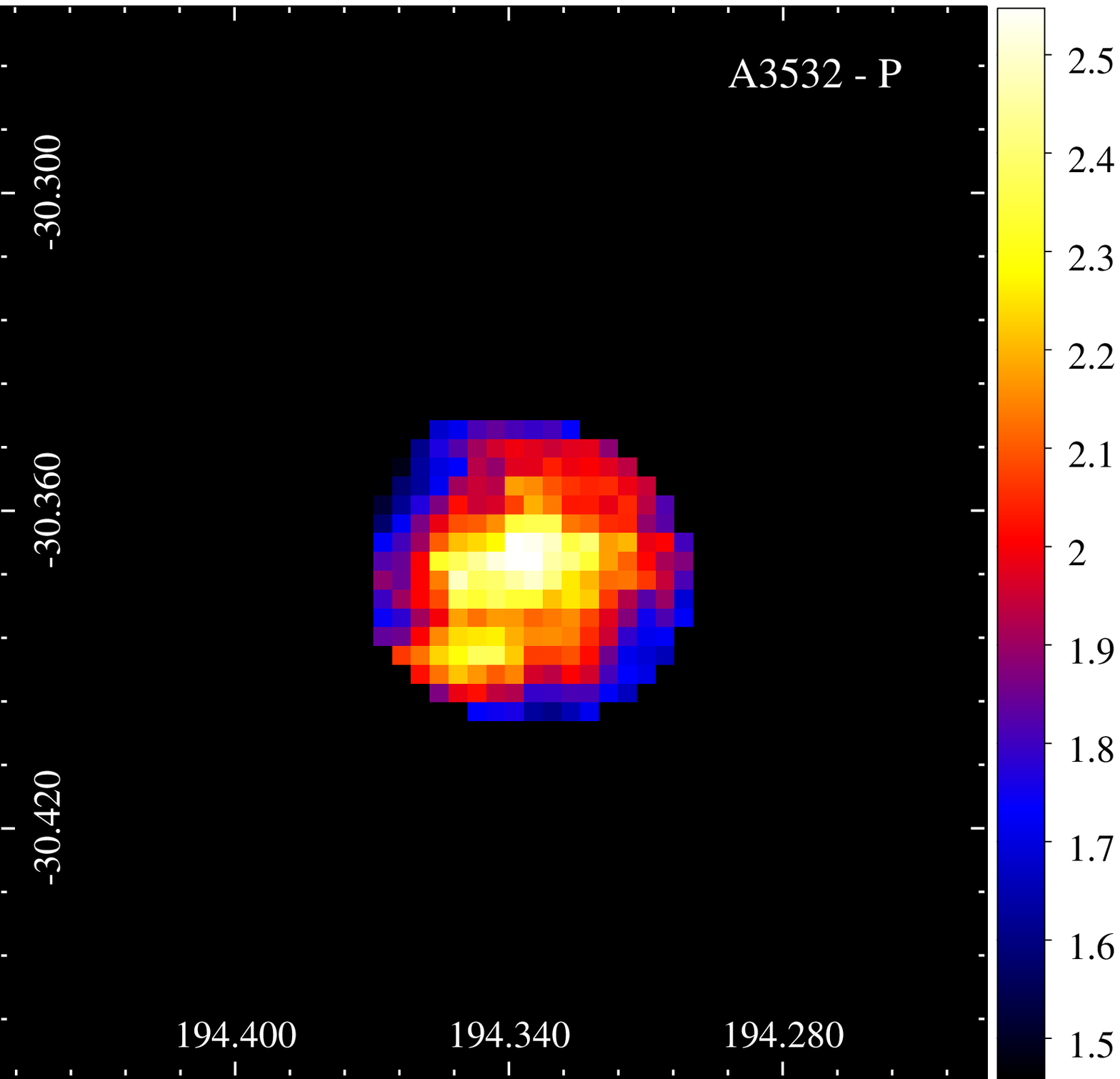}
\includegraphics[scale=0.25]{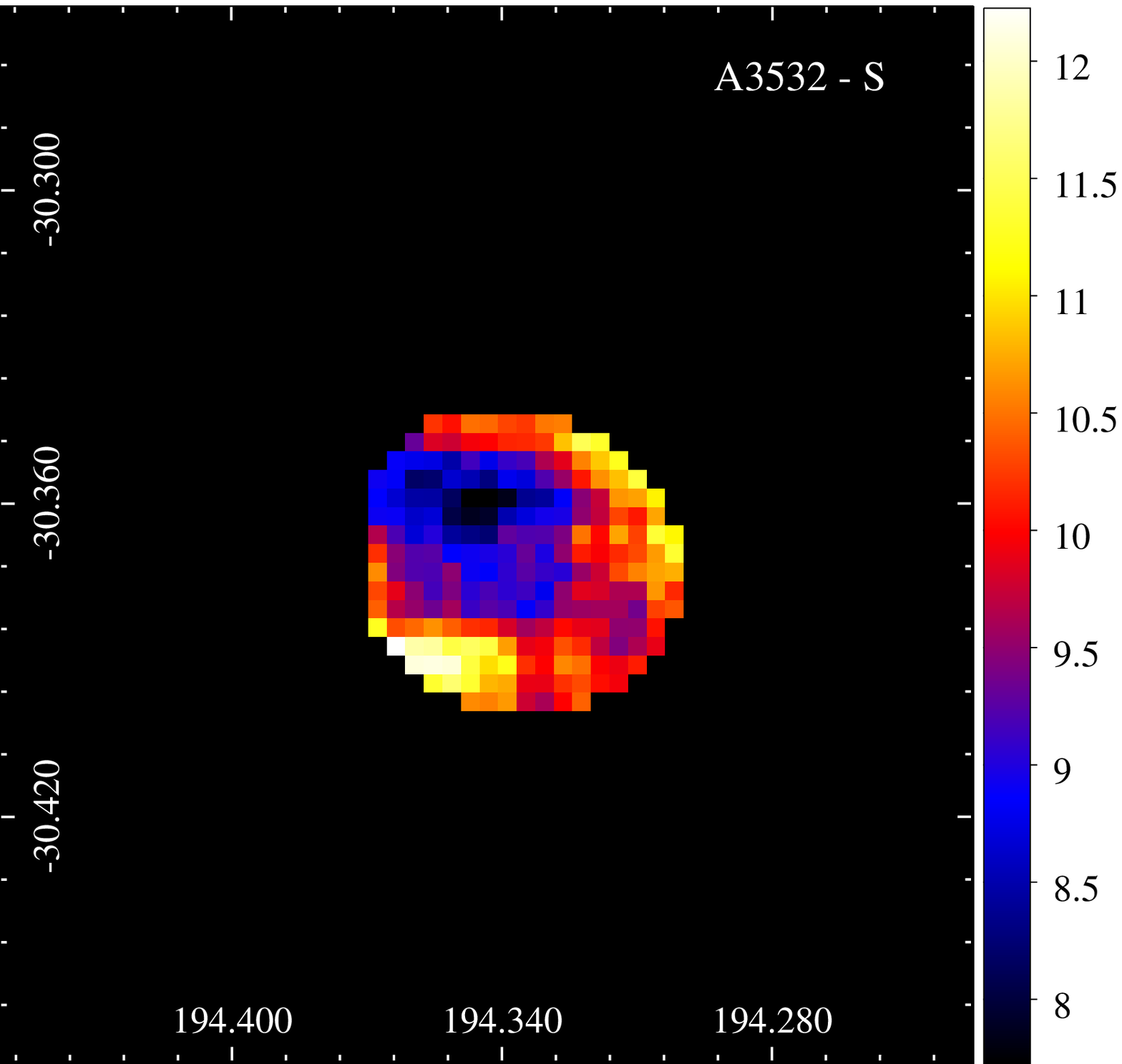}
\includegraphics[scale=0.25]{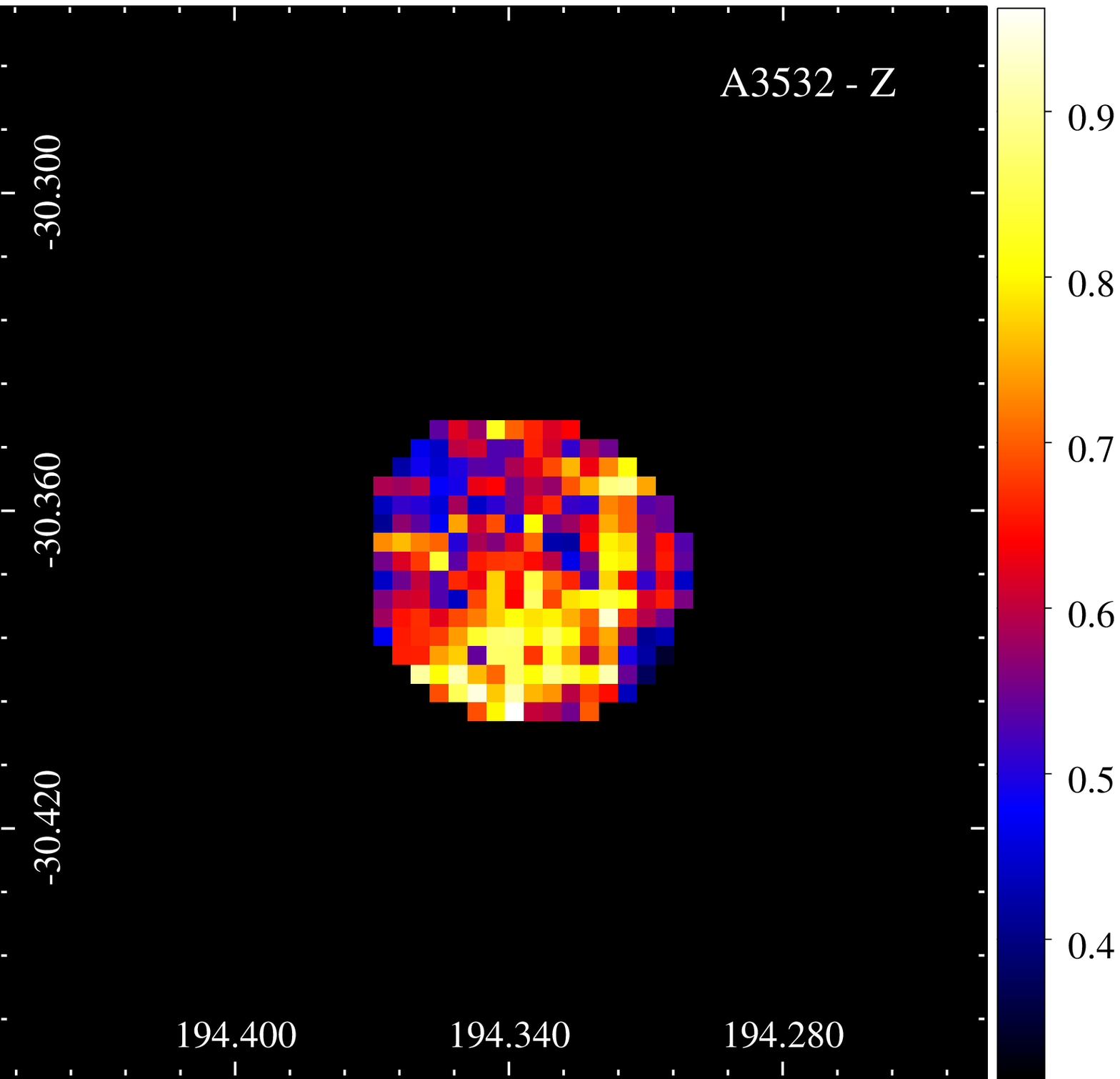}

\includegraphics[scale=0.25]{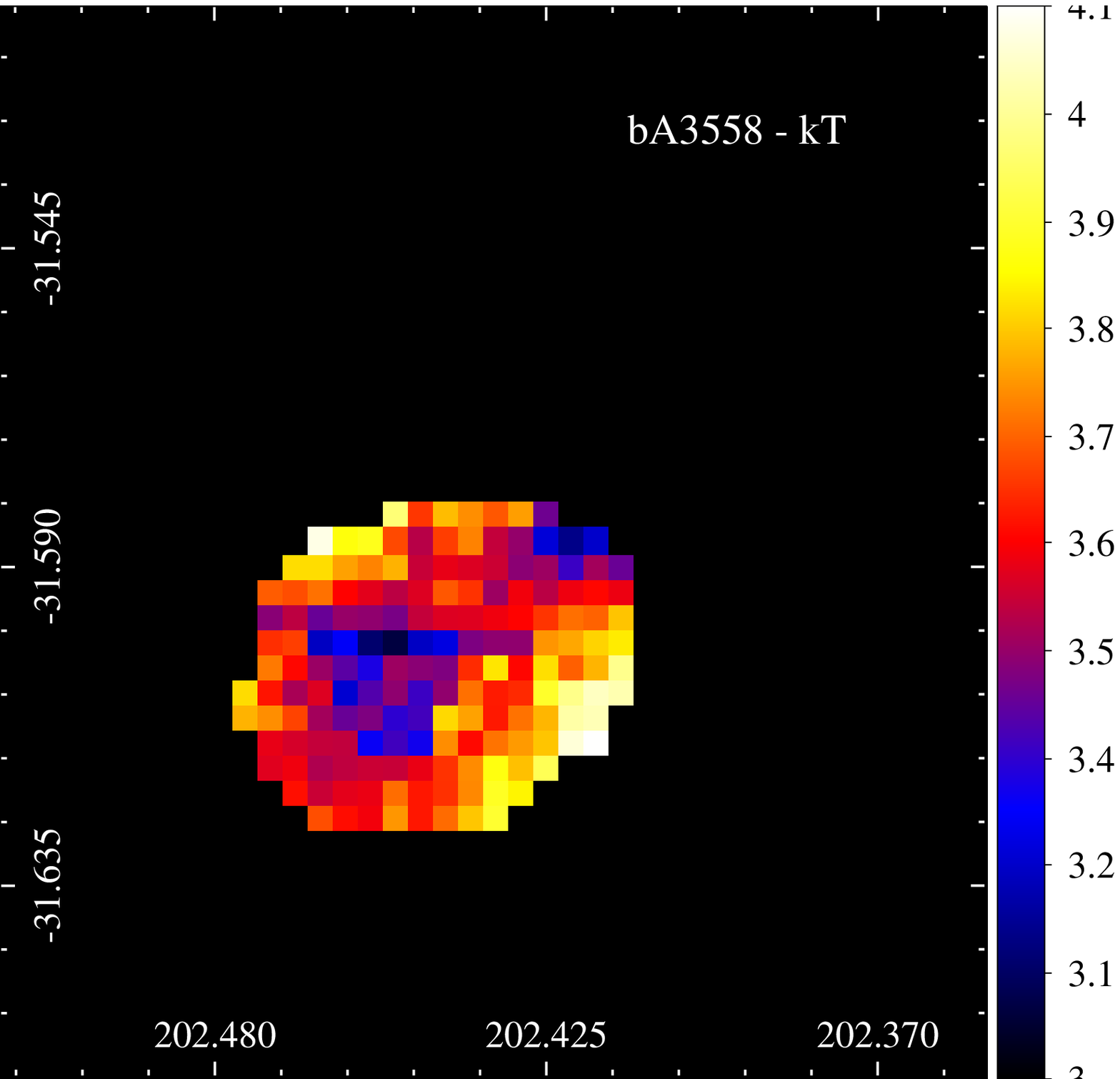}
\includegraphics[scale=0.25]{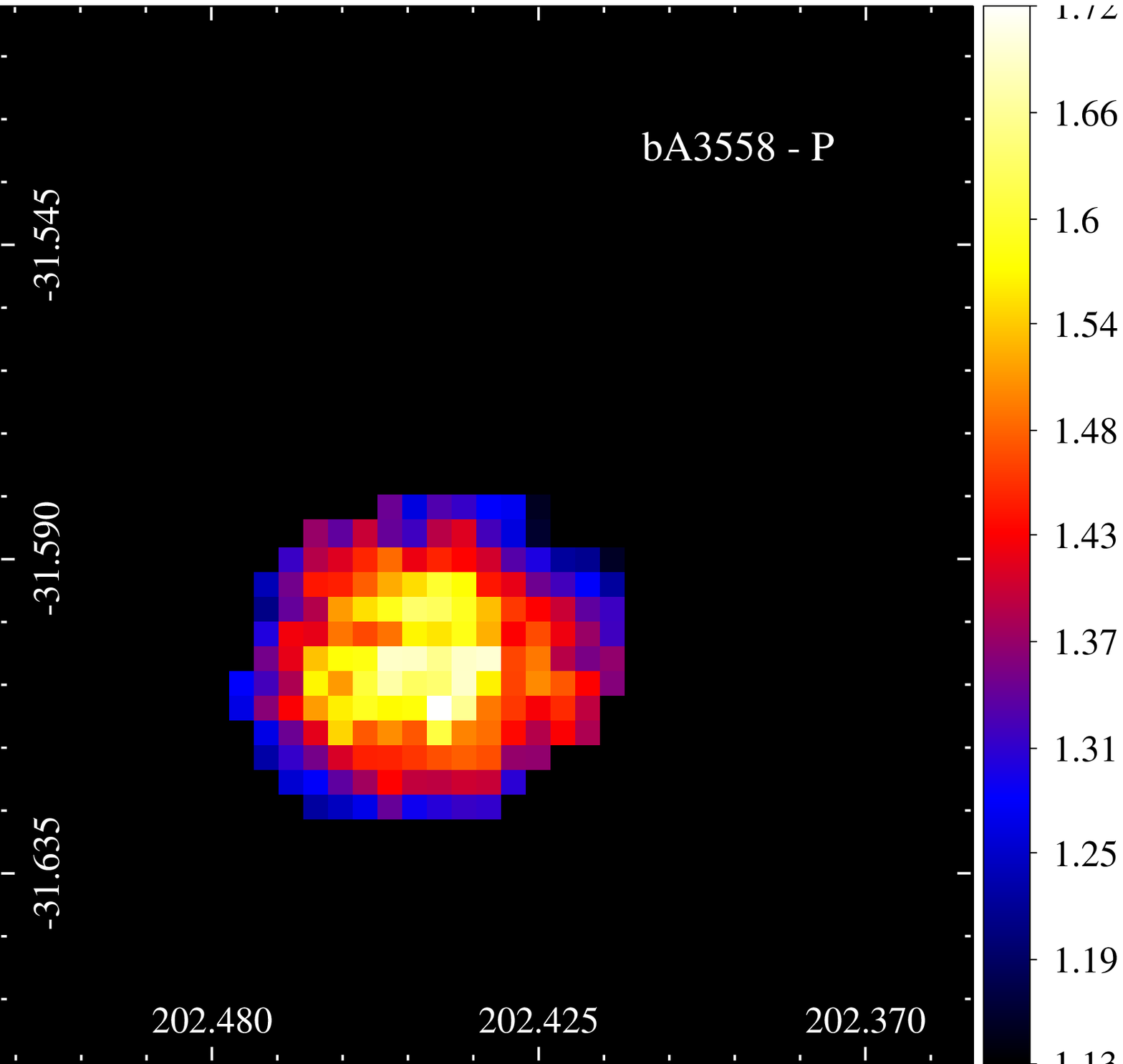}
\includegraphics[scale=0.25]{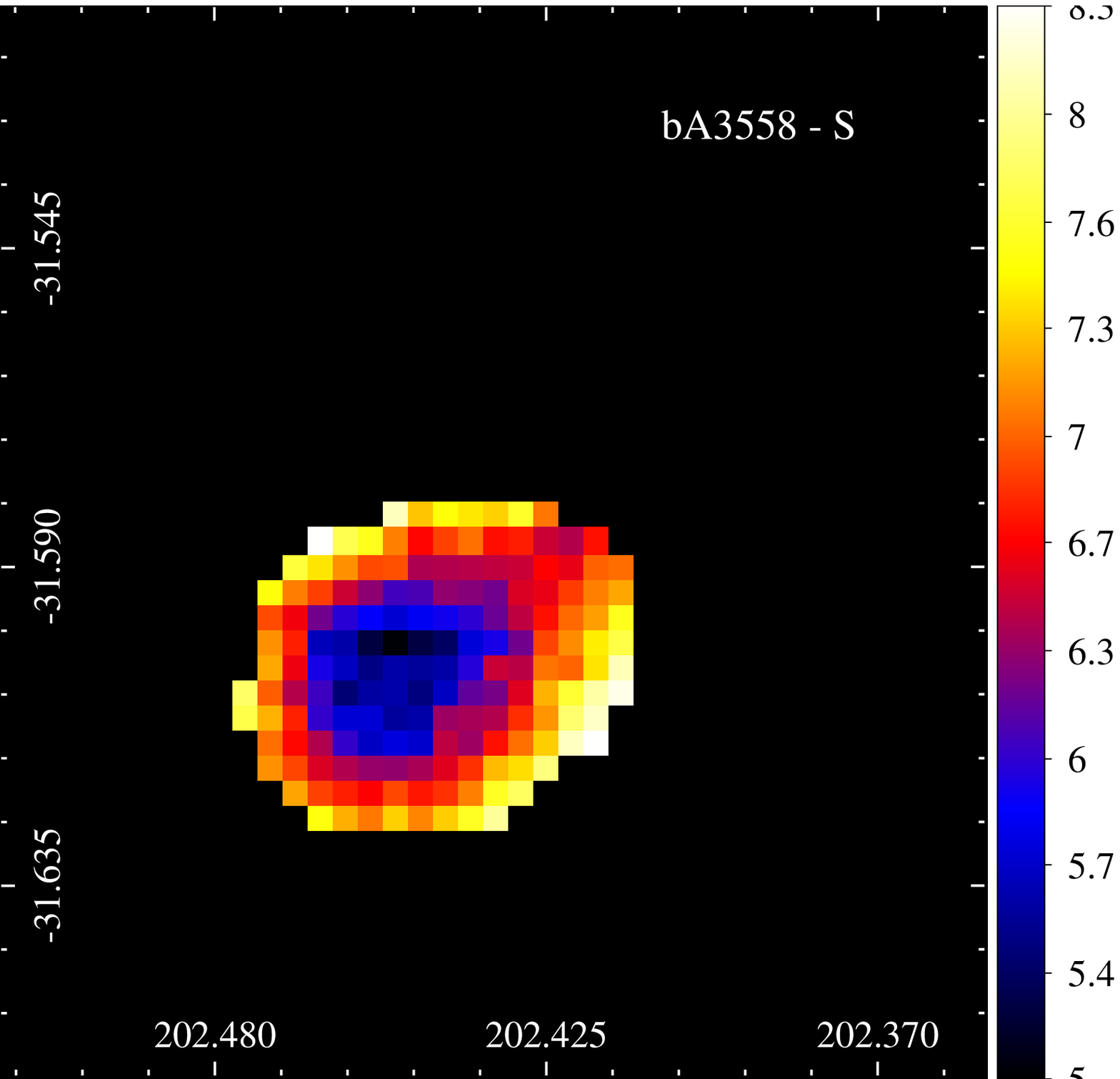}
\includegraphics[scale=0.25]{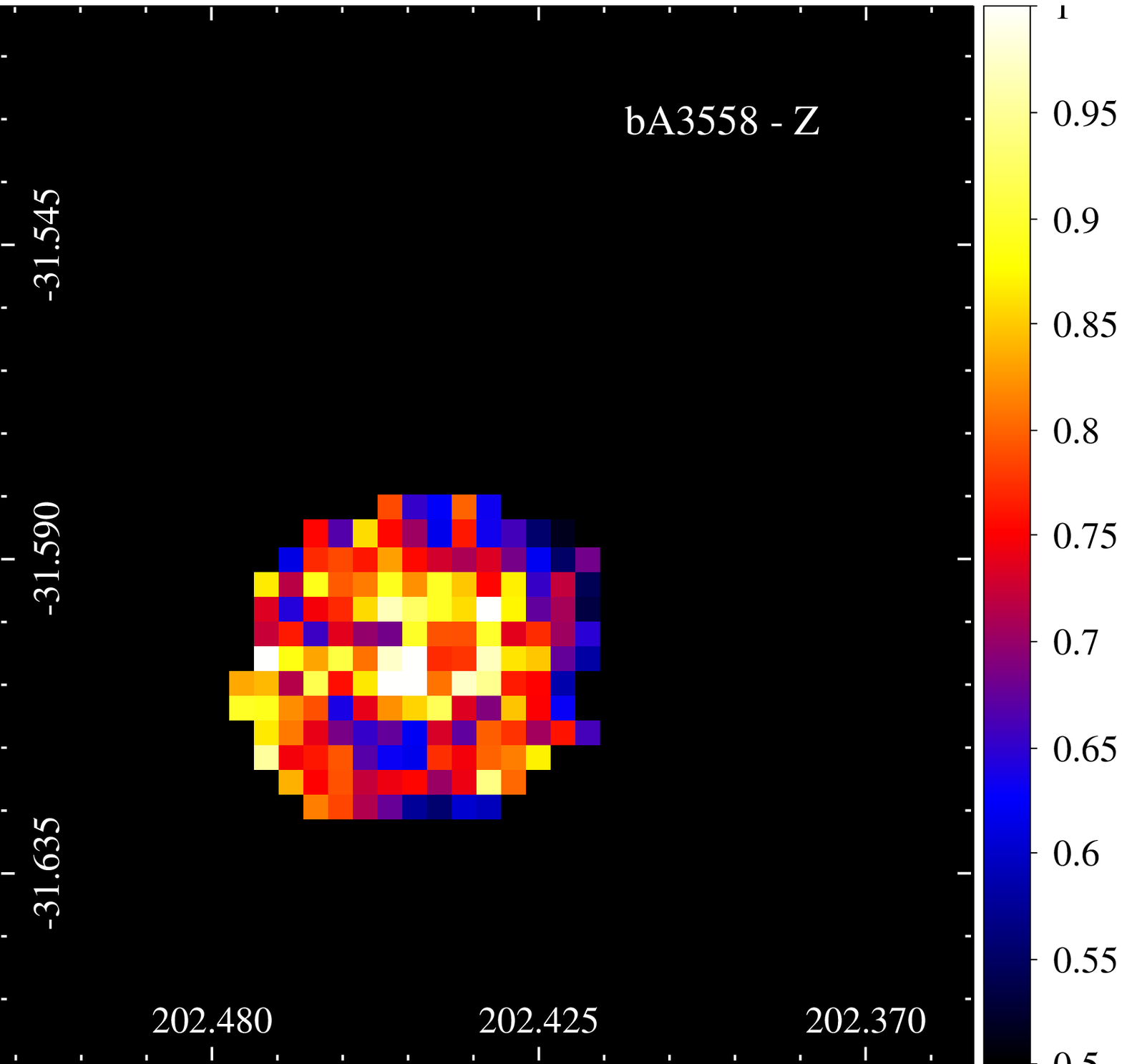}

\includegraphics[scale=0.25]{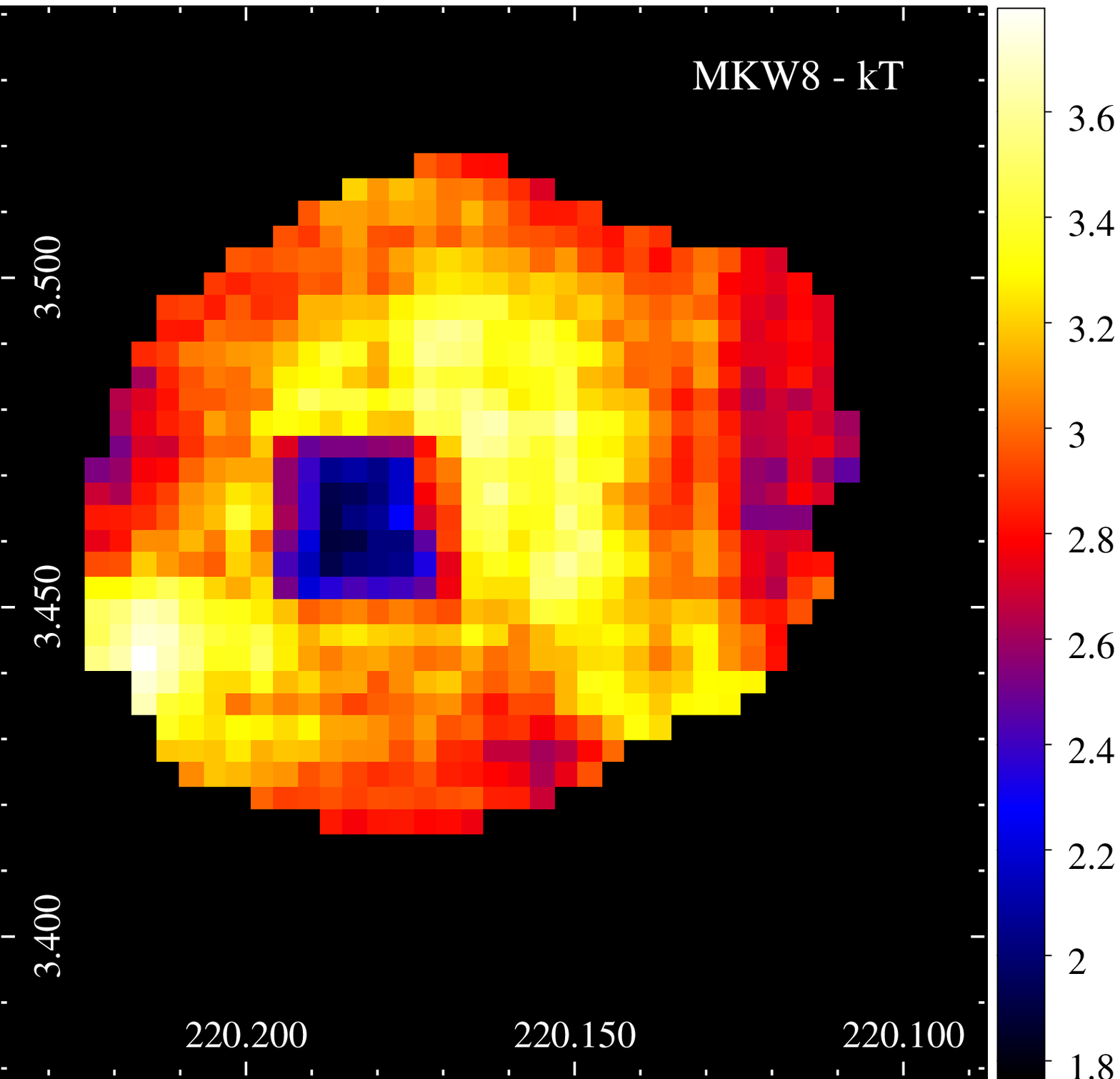}
\includegraphics[scale=0.25]{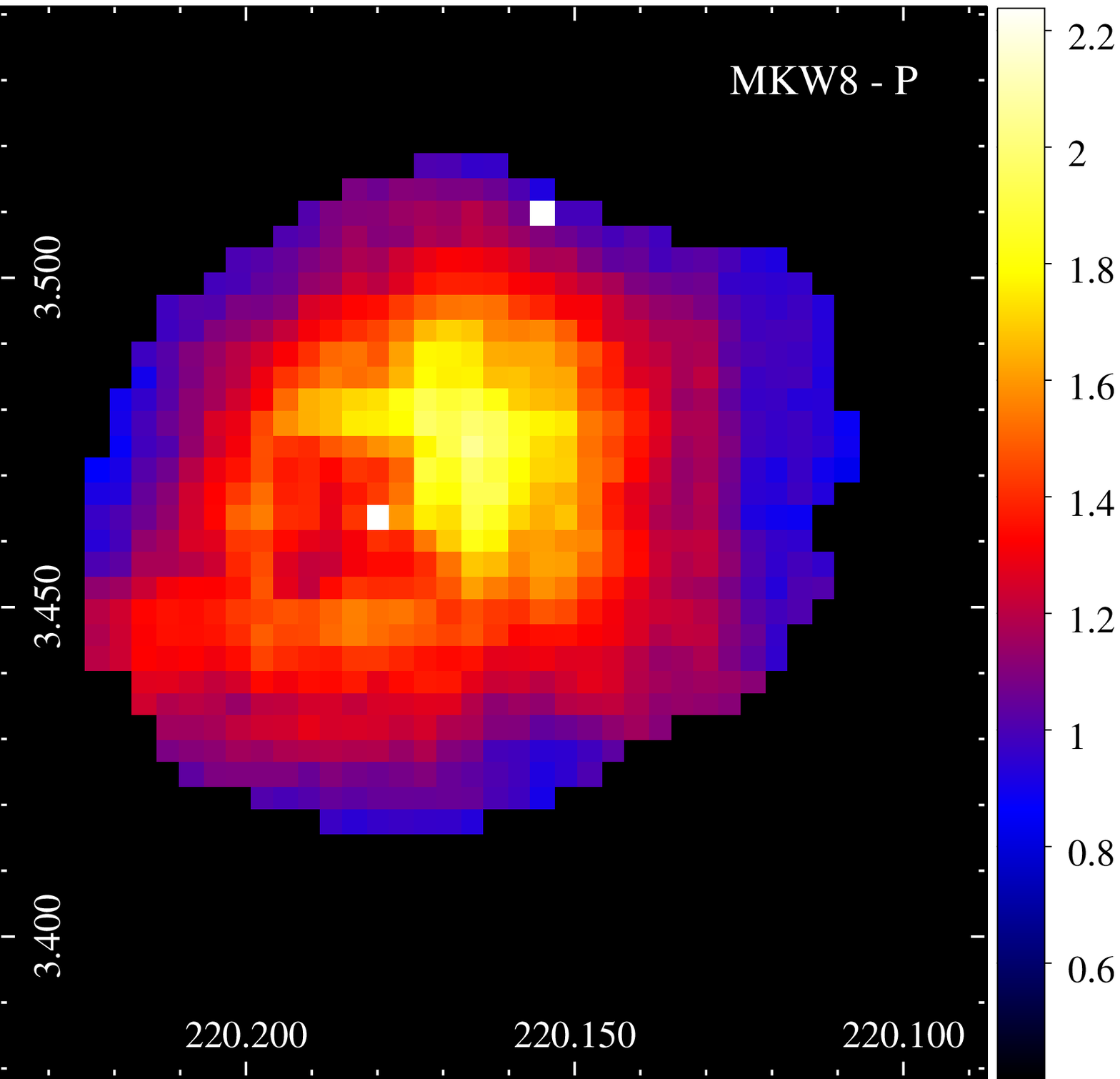}
\includegraphics[scale=0.25]{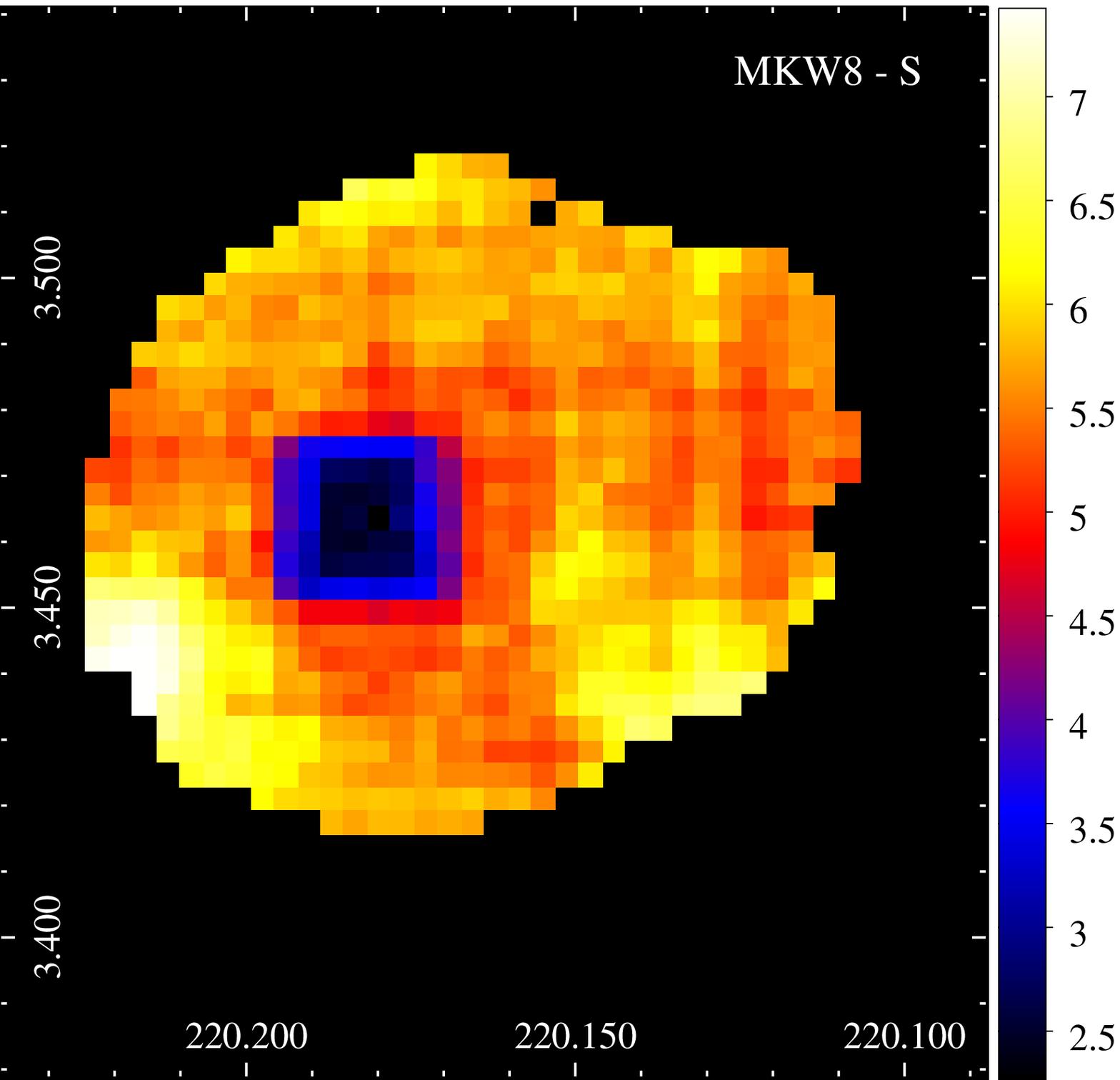}
\includegraphics[scale=0.25]{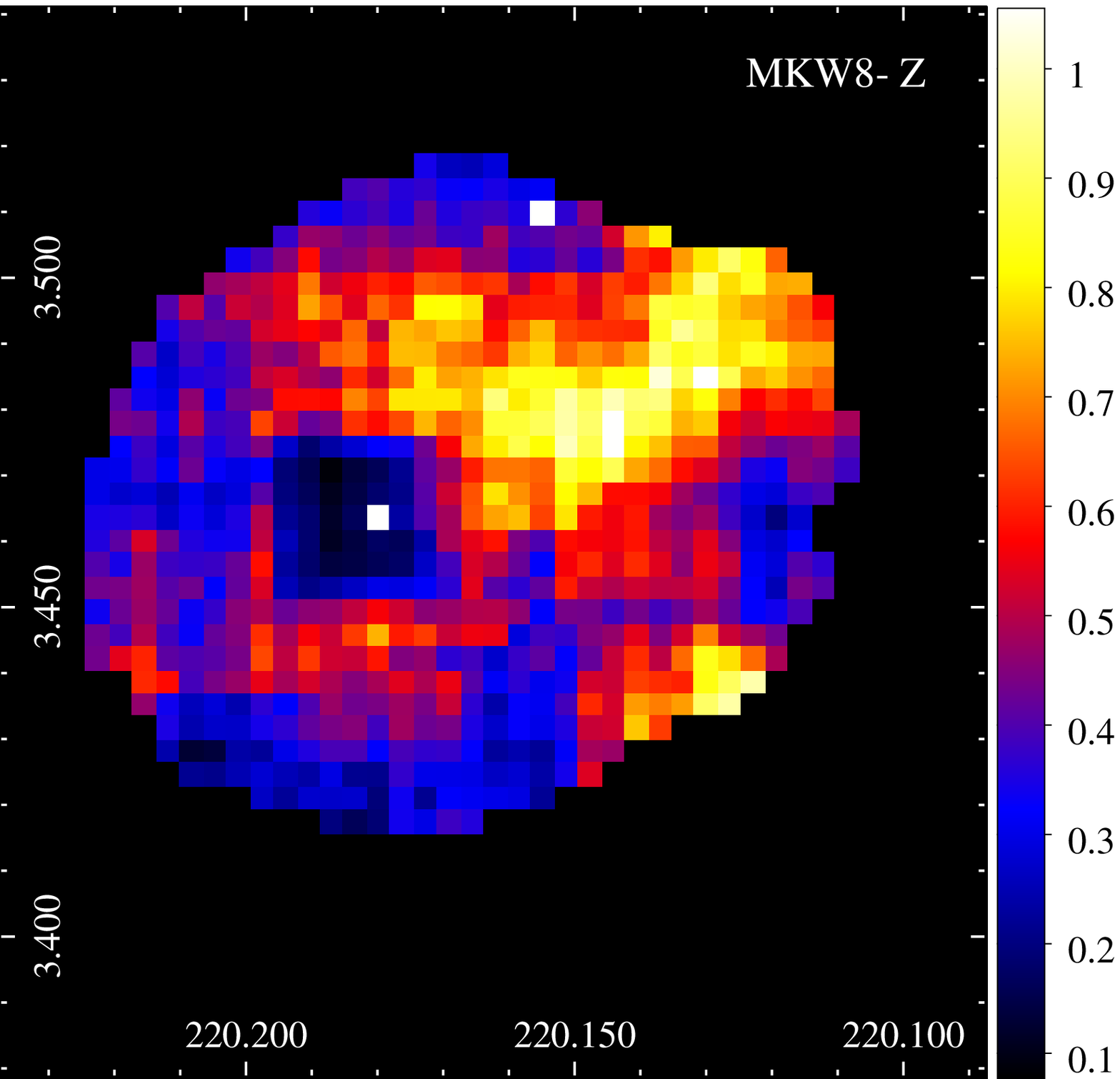}
\caption{NCC and relaxed systems. From left to right: temperature, pseudo-pressure, pseudo-entropy, and 
metallicity map for Coma, A3532, bA3558, and MKW8.}
\label{fig:NCCrelaxed}

\end{figure*}

\begin{figure*}
\includegraphics[scale=0.25]{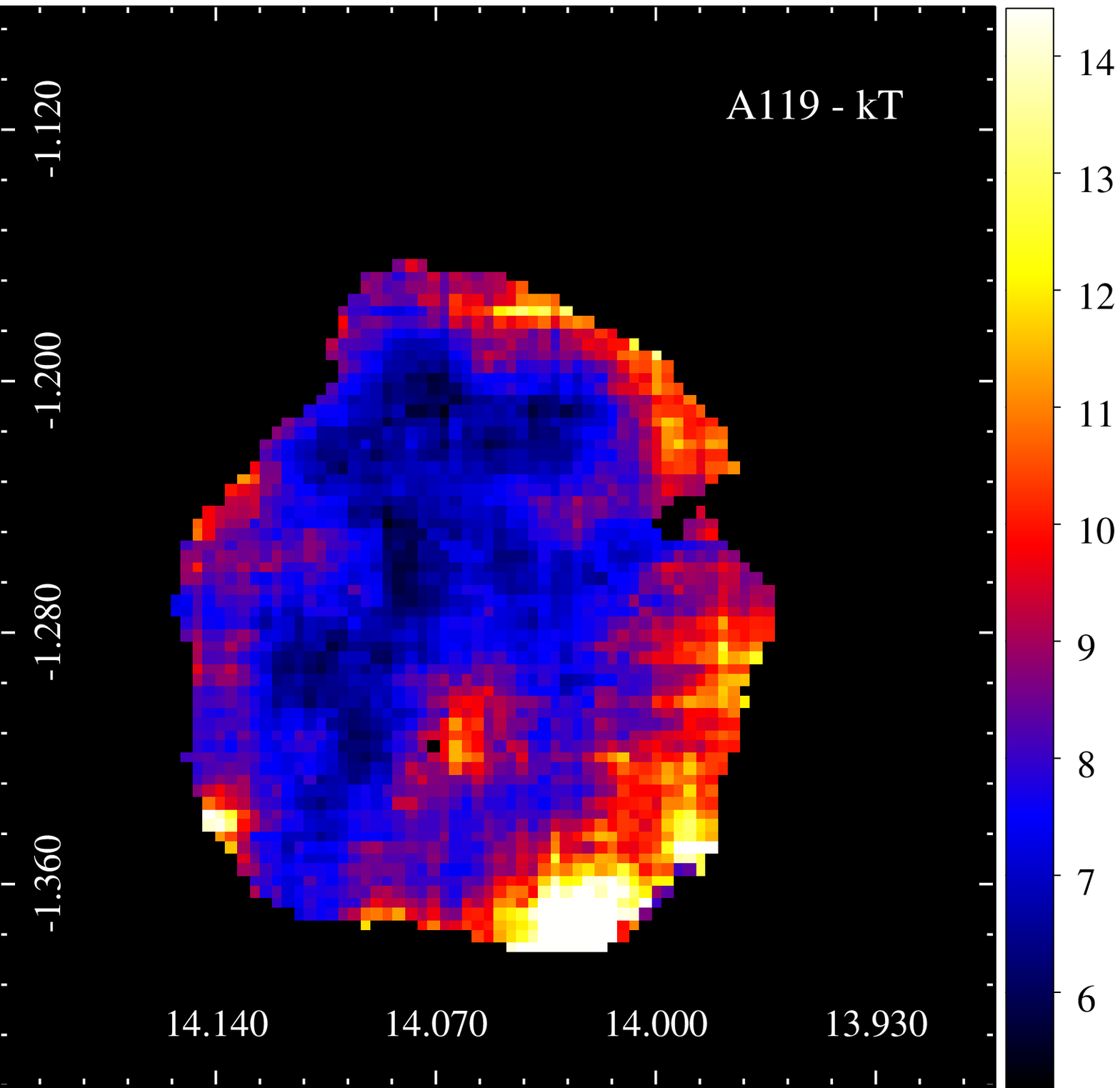}
\includegraphics[scale=0.25]{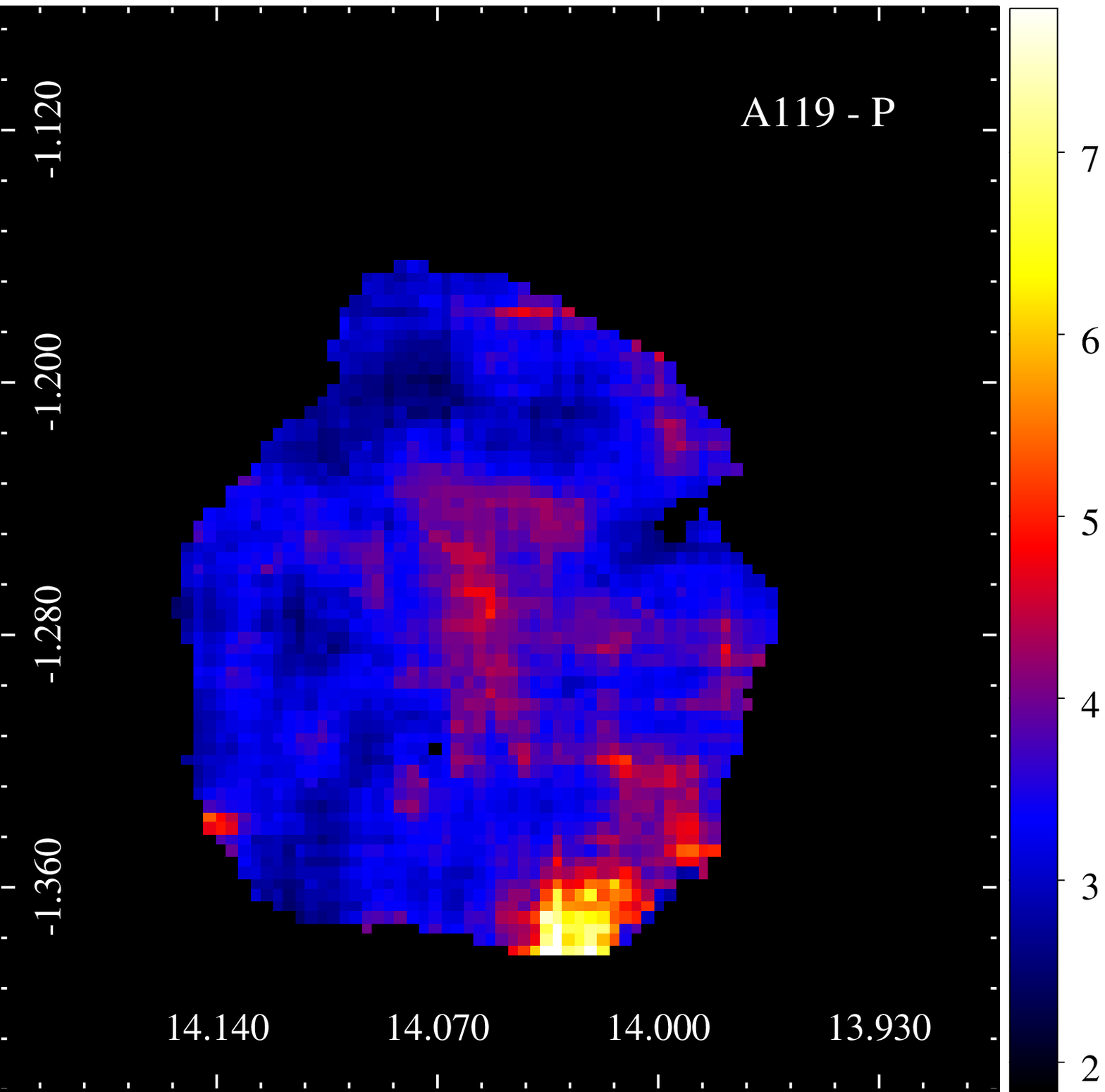}
\includegraphics[scale=0.25]{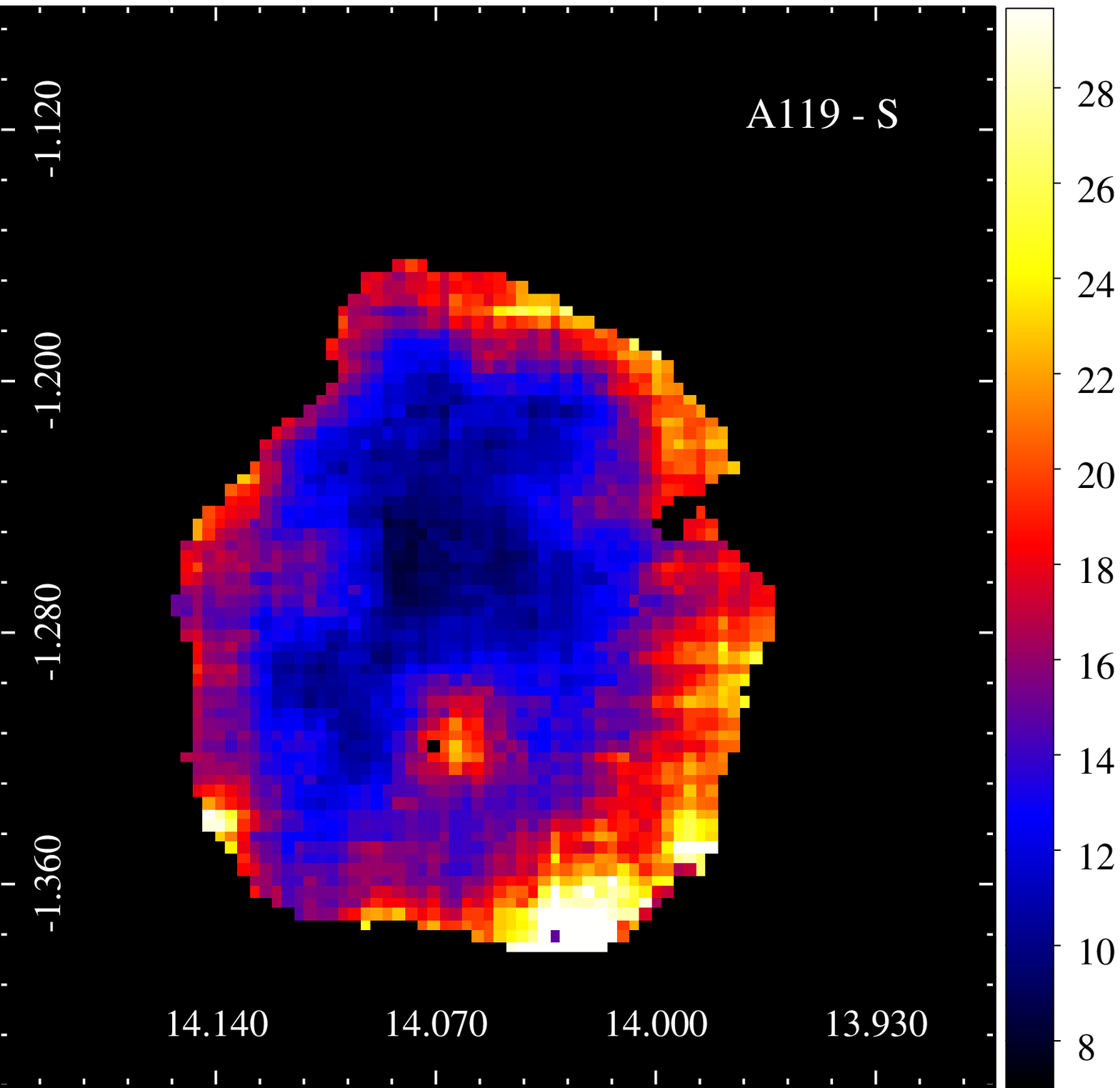}
\includegraphics[scale=0.25]{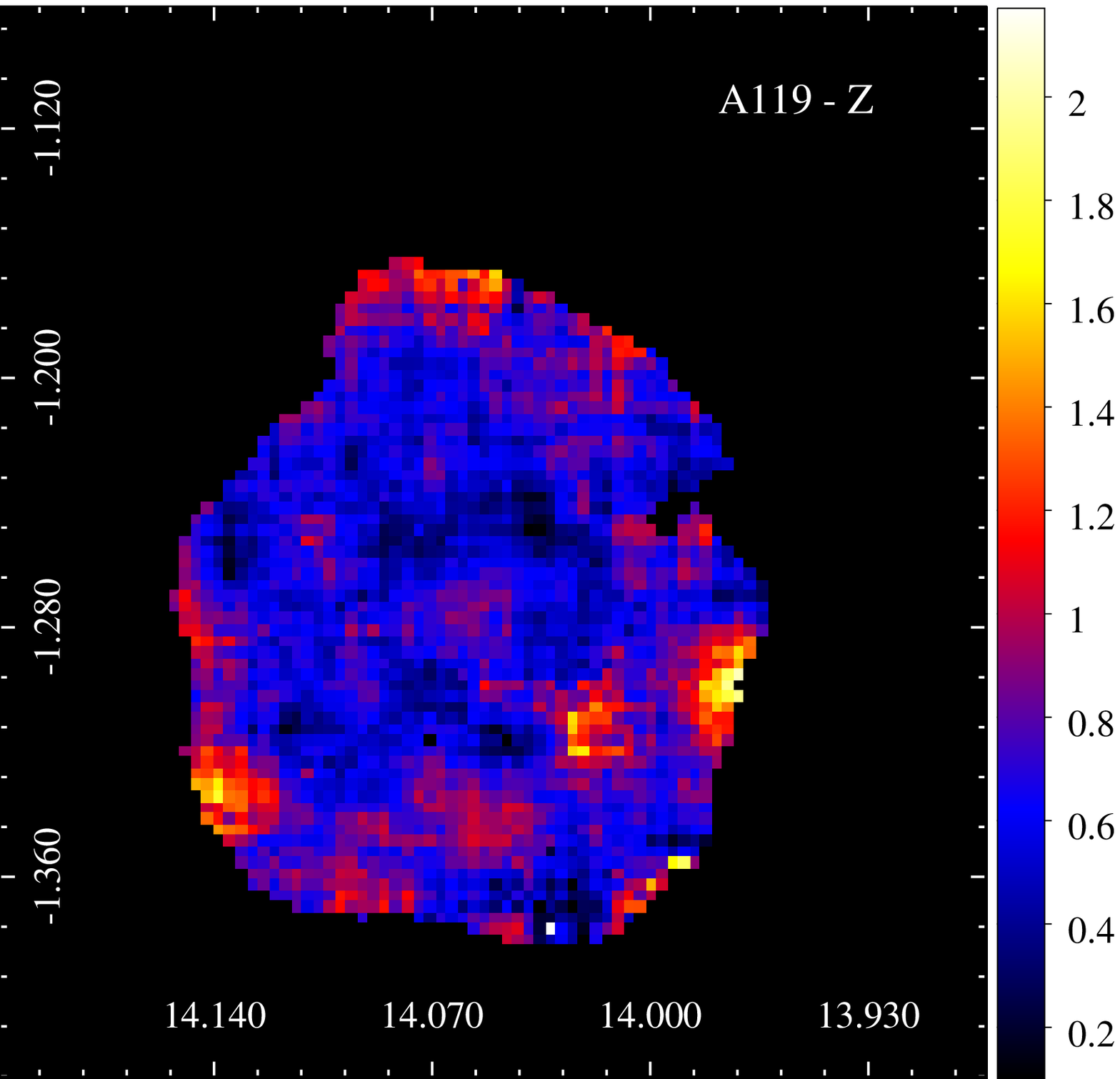}

\includegraphics[scale=0.25]{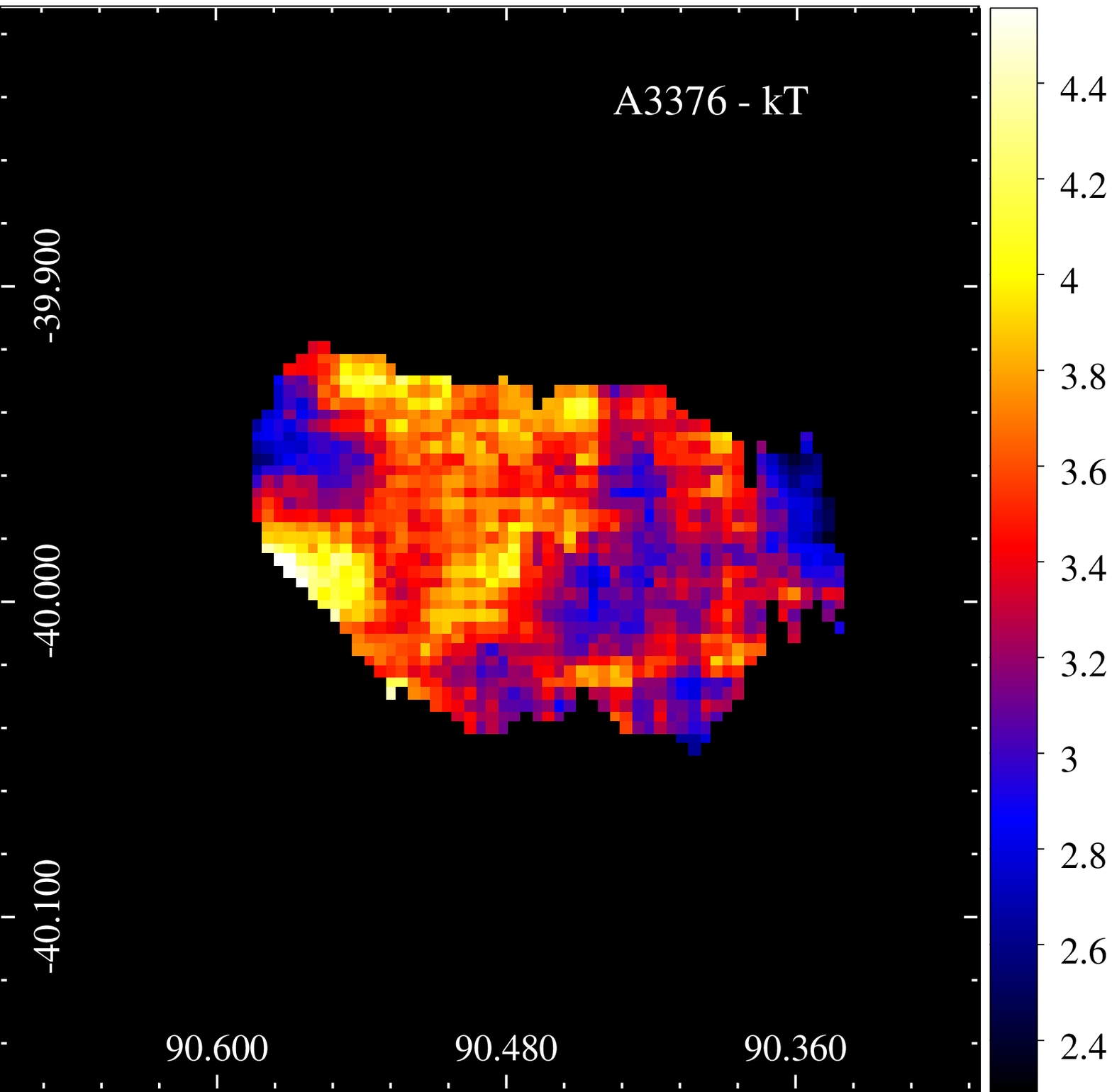}
\includegraphics[scale=0.25]{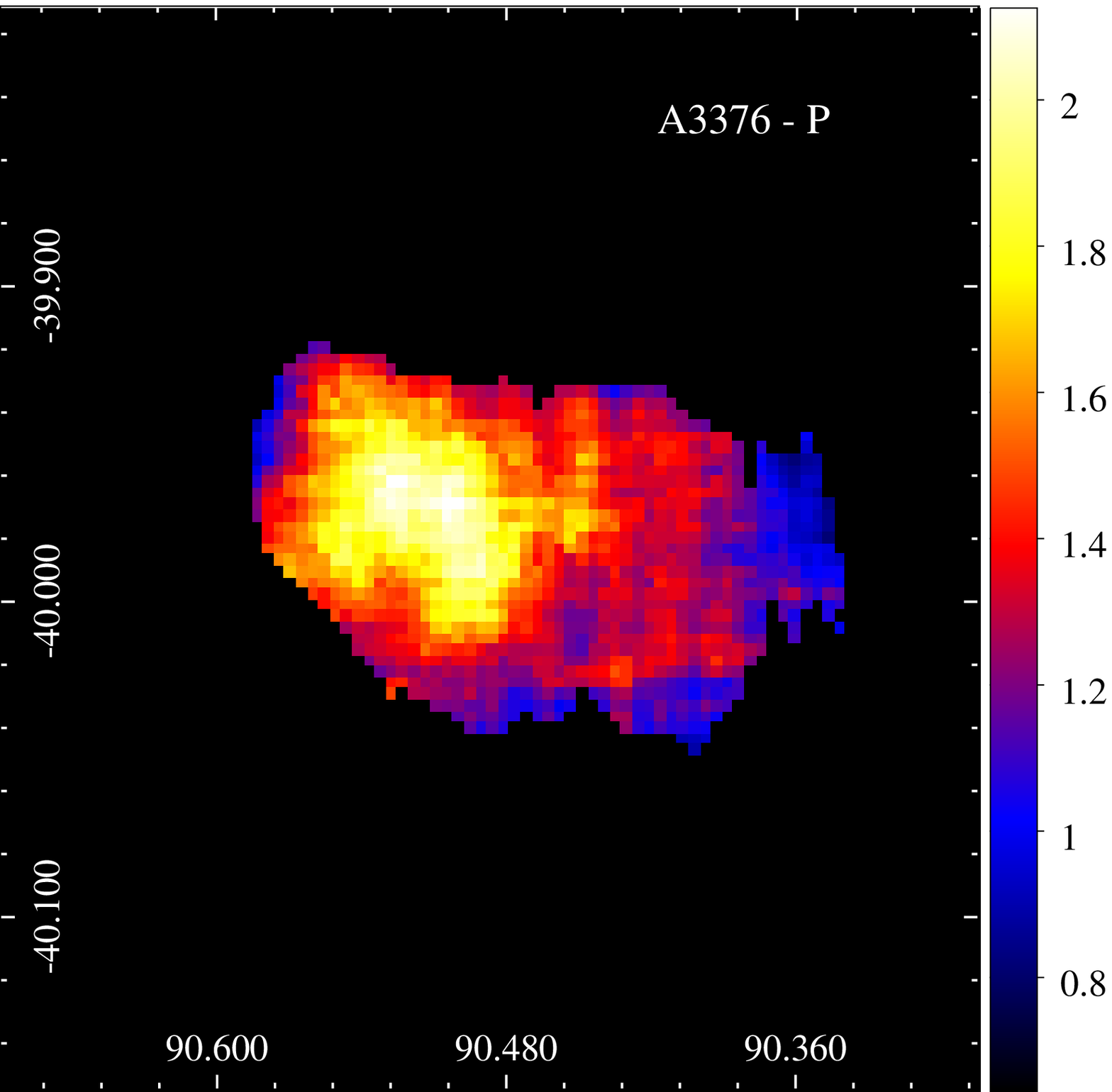}
\includegraphics[scale=0.25]{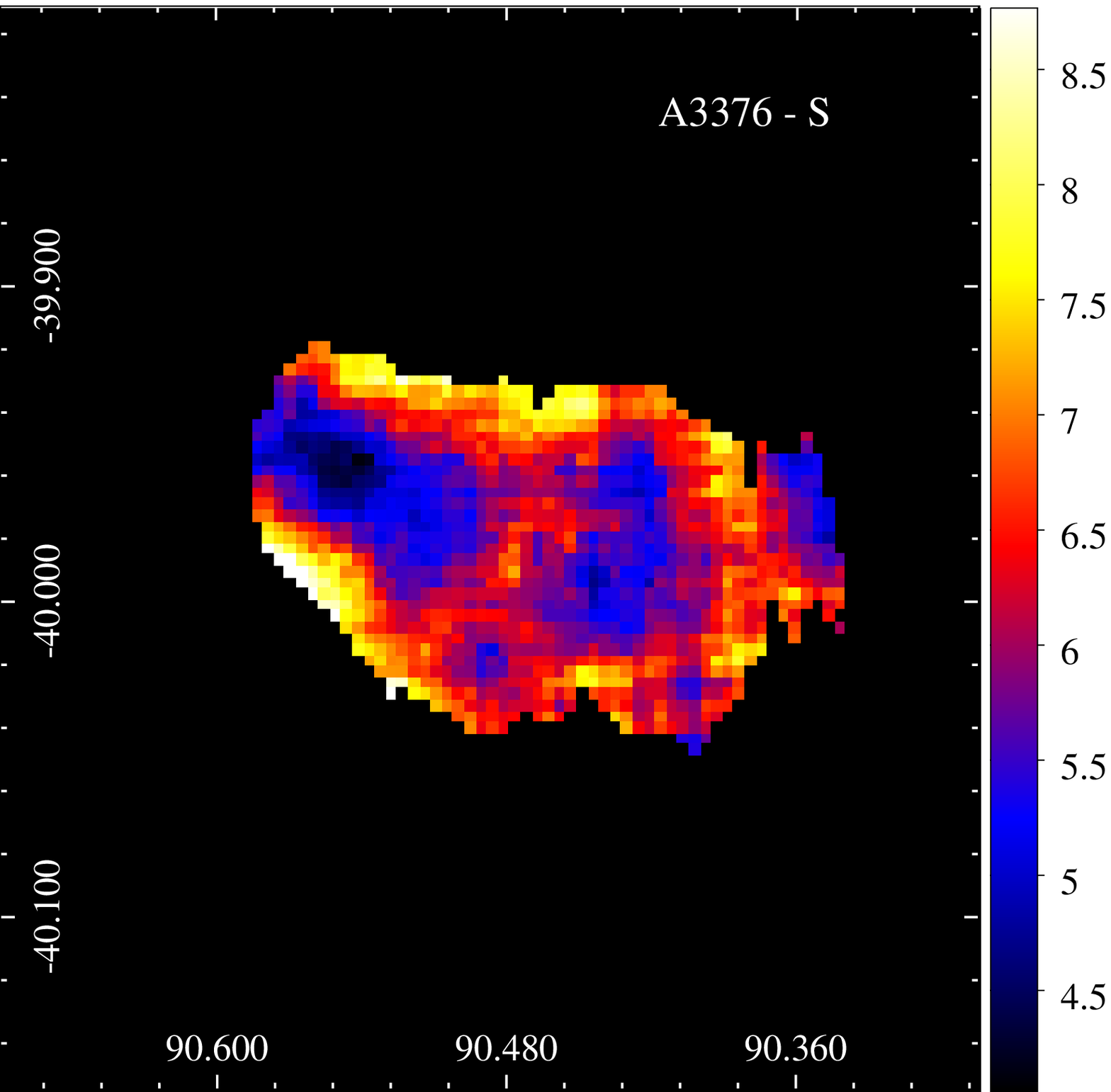}
\includegraphics[scale=0.25]{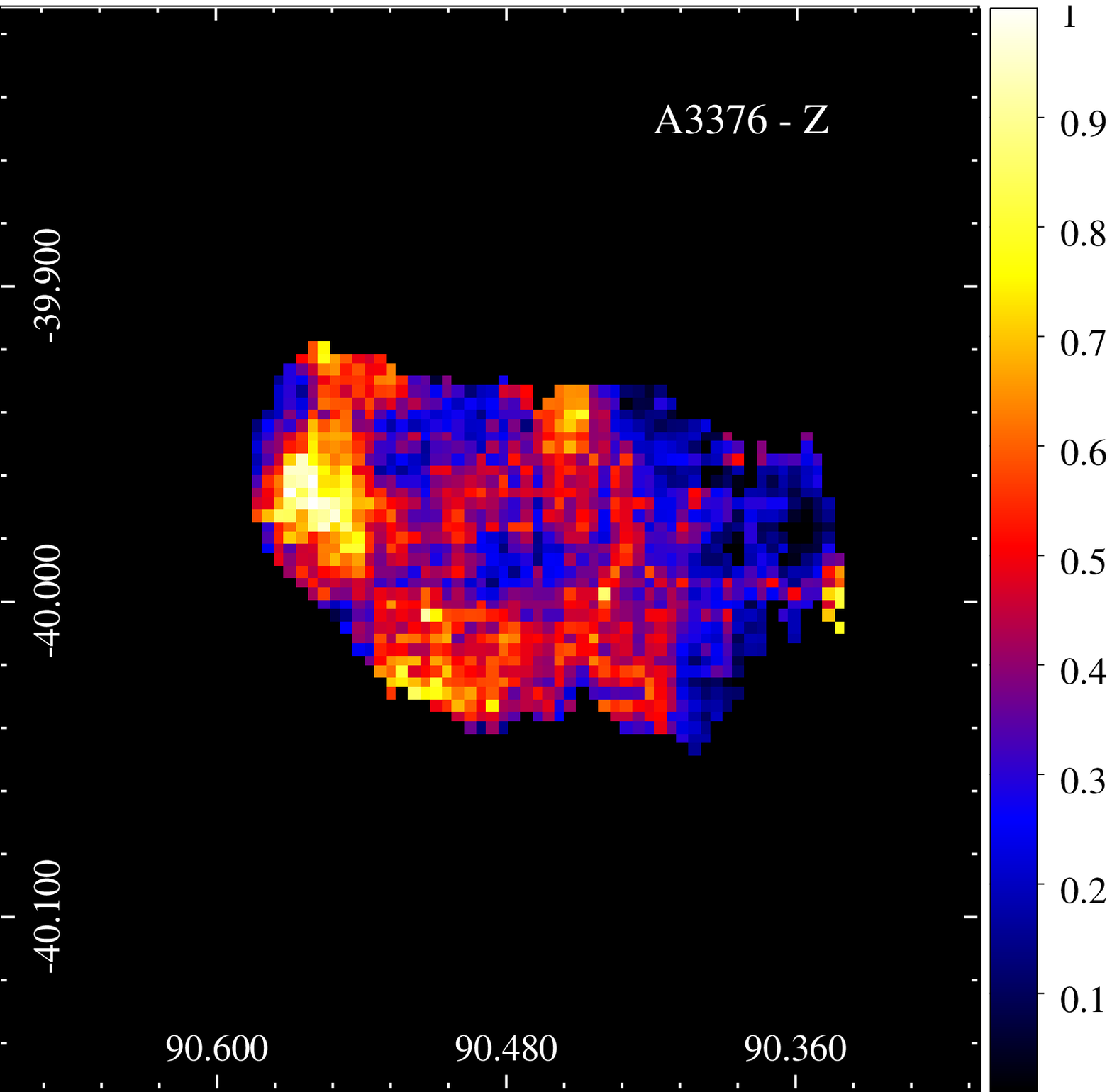}

\includegraphics[scale=0.25]{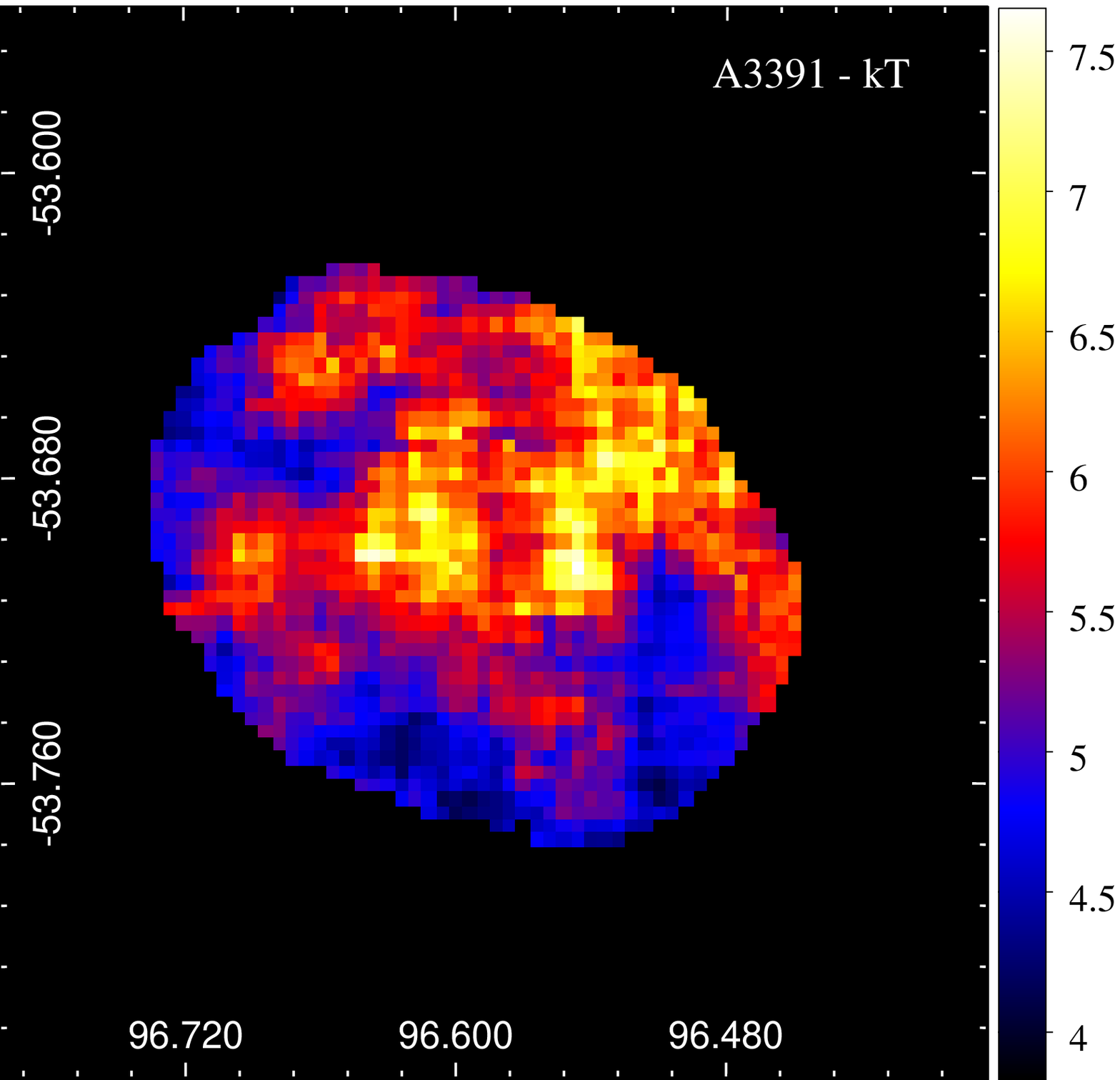}
\includegraphics[scale=0.25]{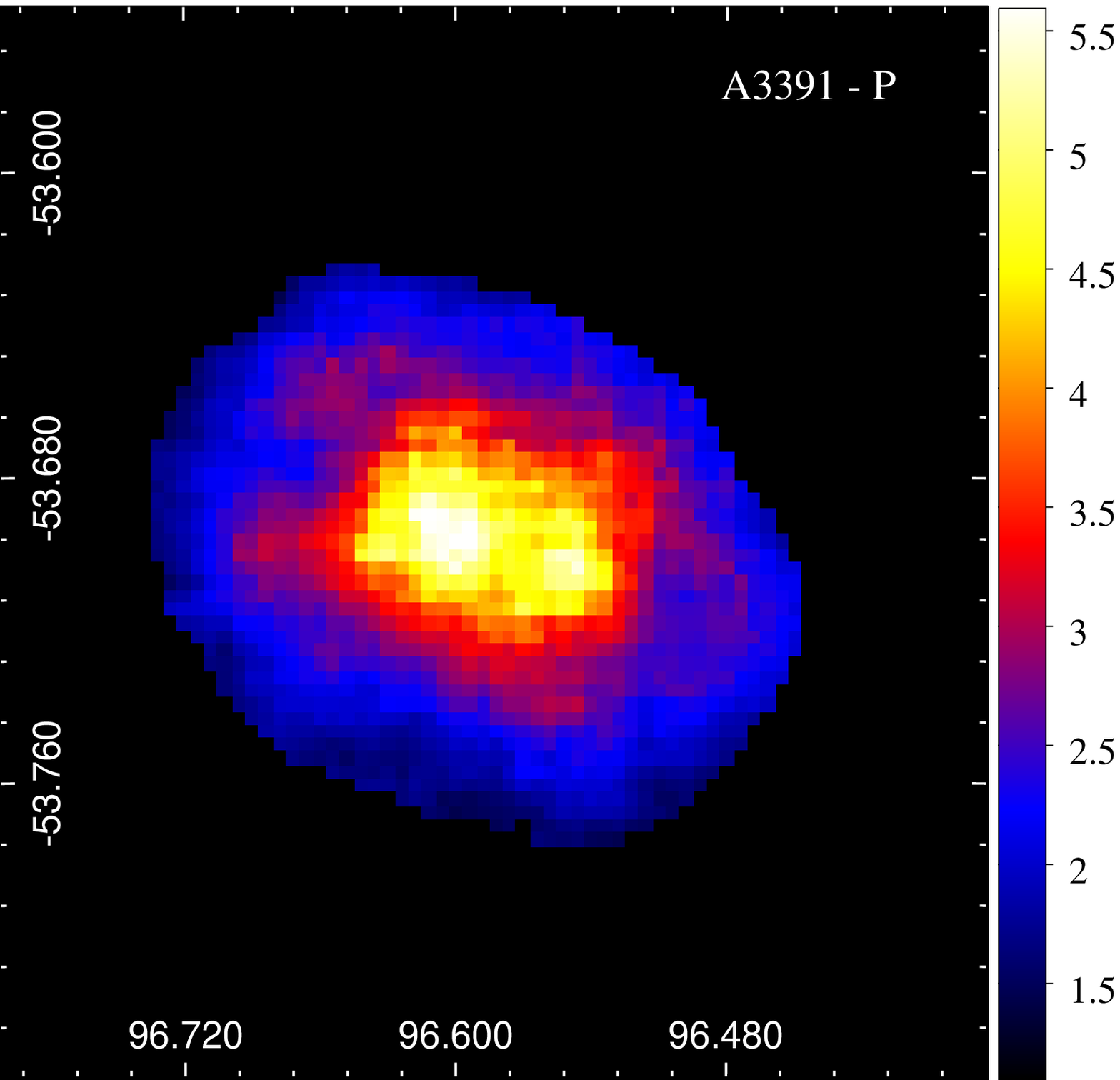}
\includegraphics[scale=0.25]{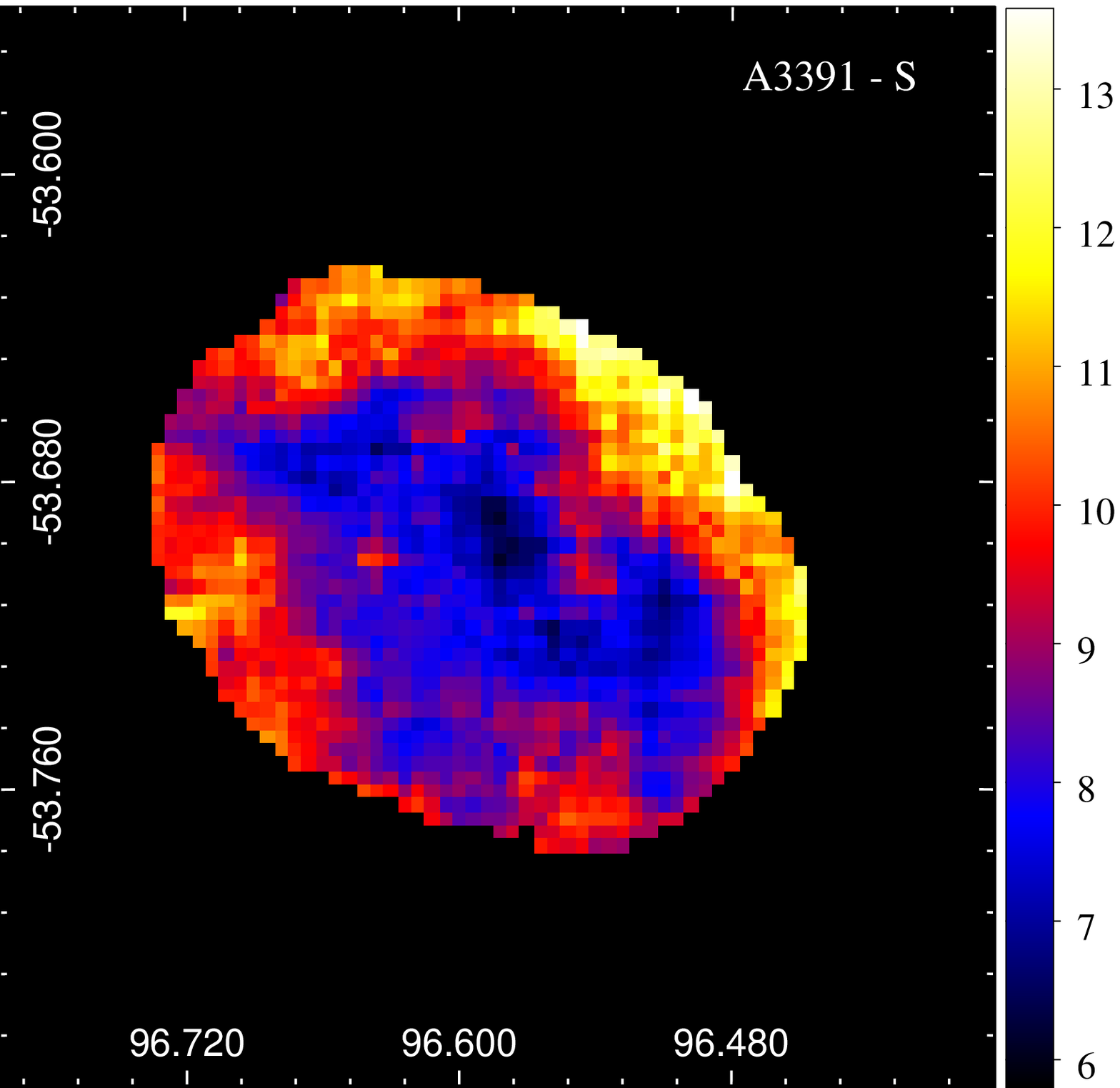}
\includegraphics[scale=0.25]{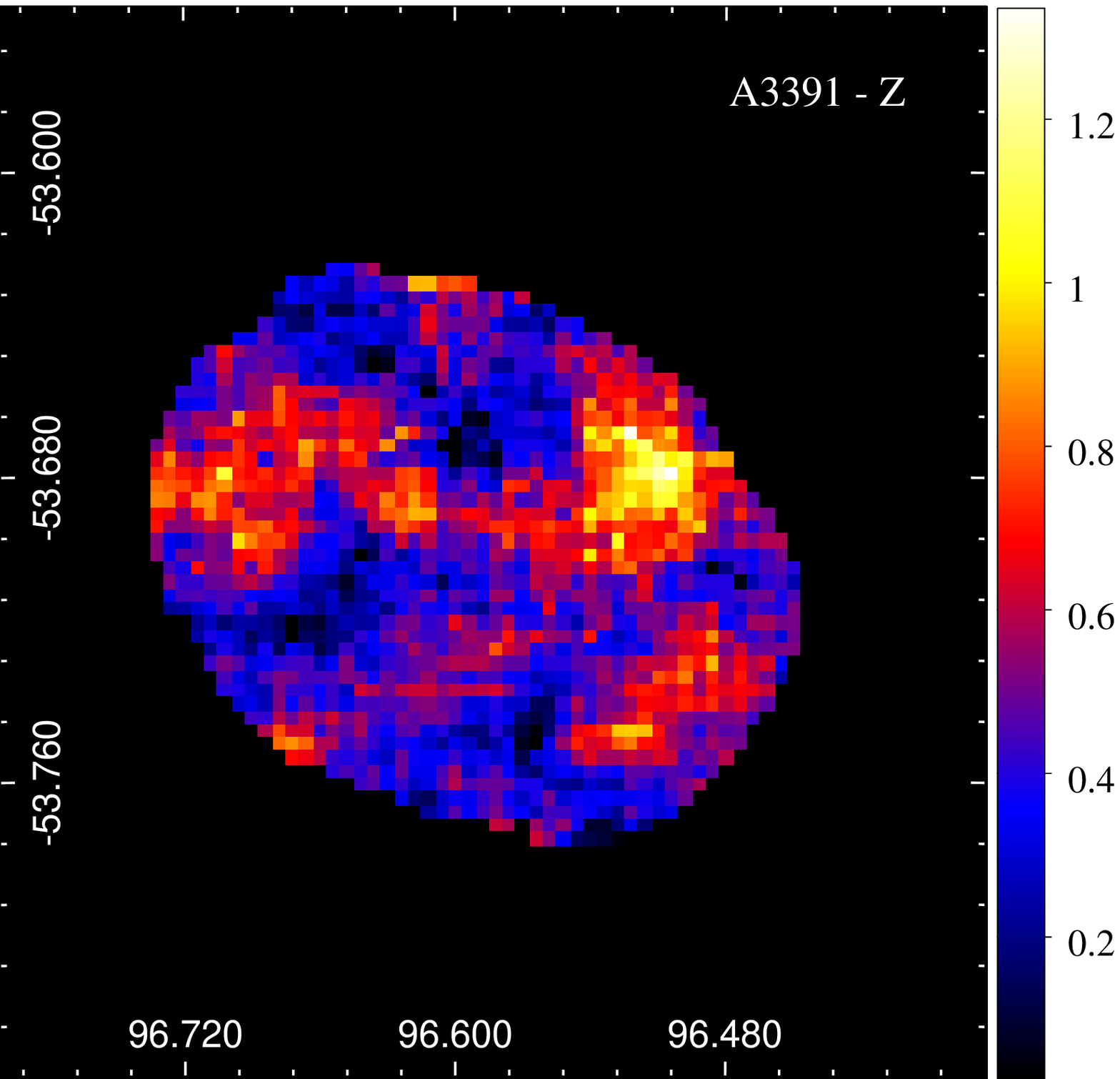}

\includegraphics[scale=0.25]{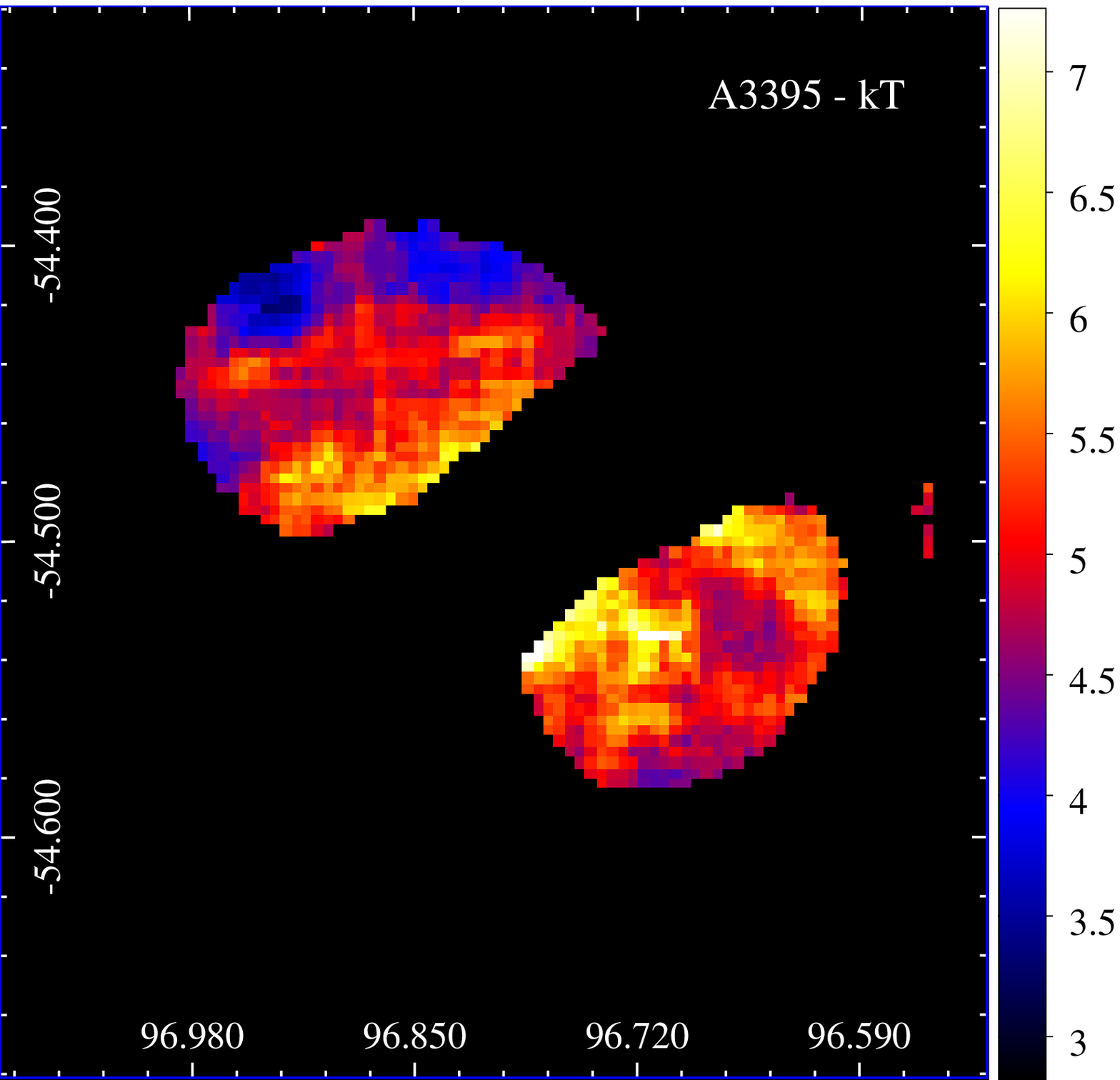}
\includegraphics[scale=0.25]{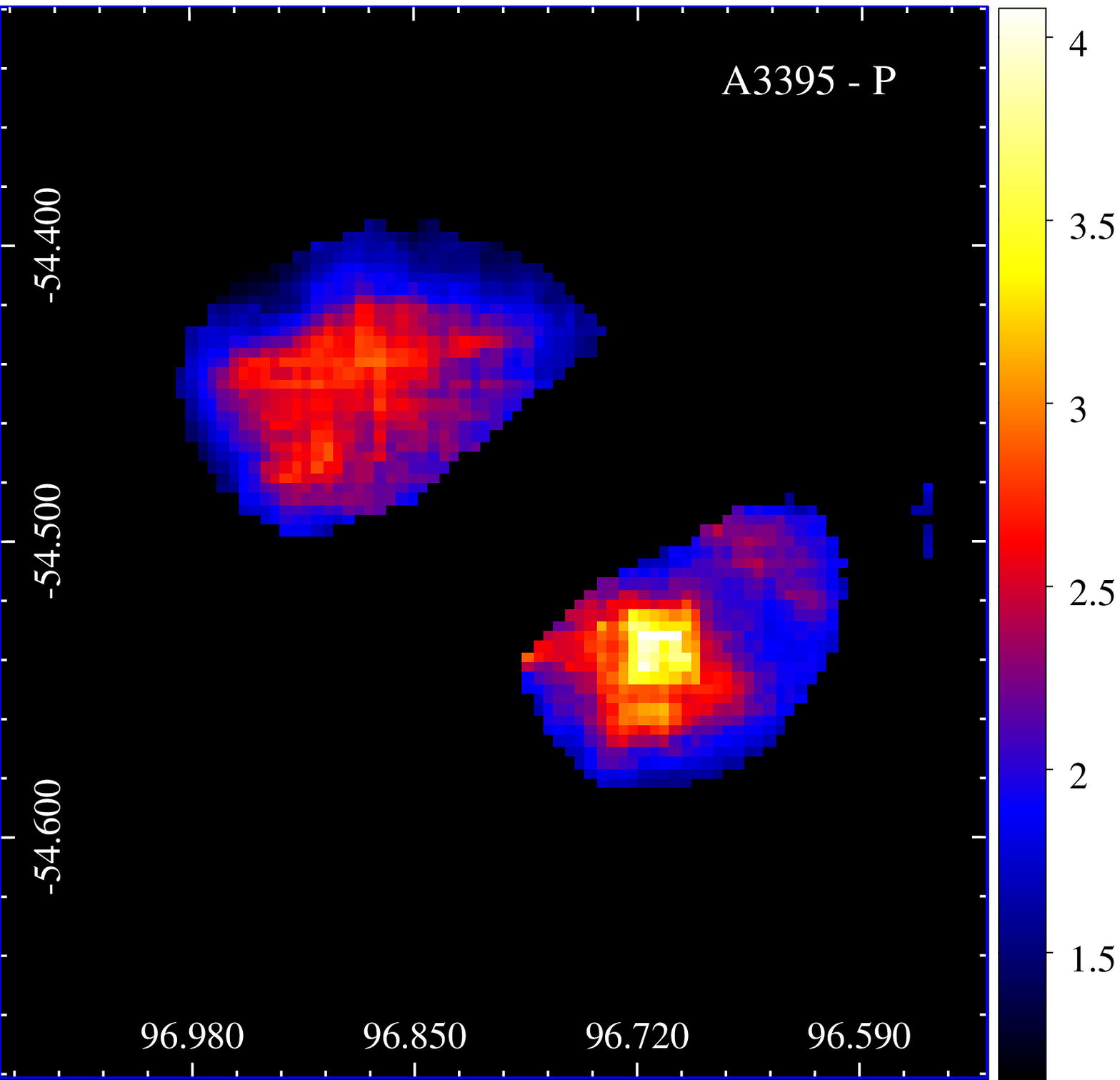}
\includegraphics[scale=0.25]{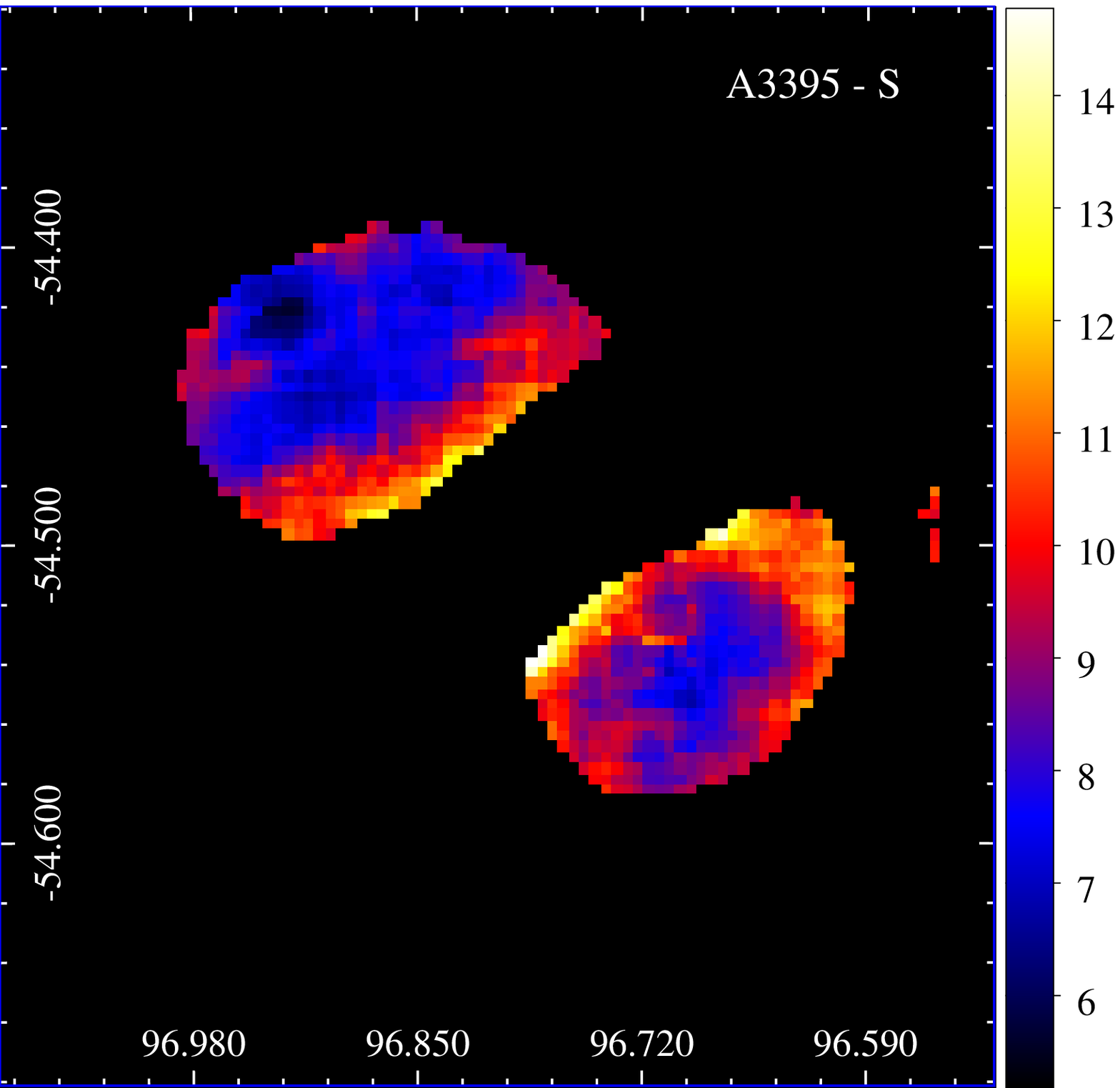}
\includegraphics[scale=0.25]{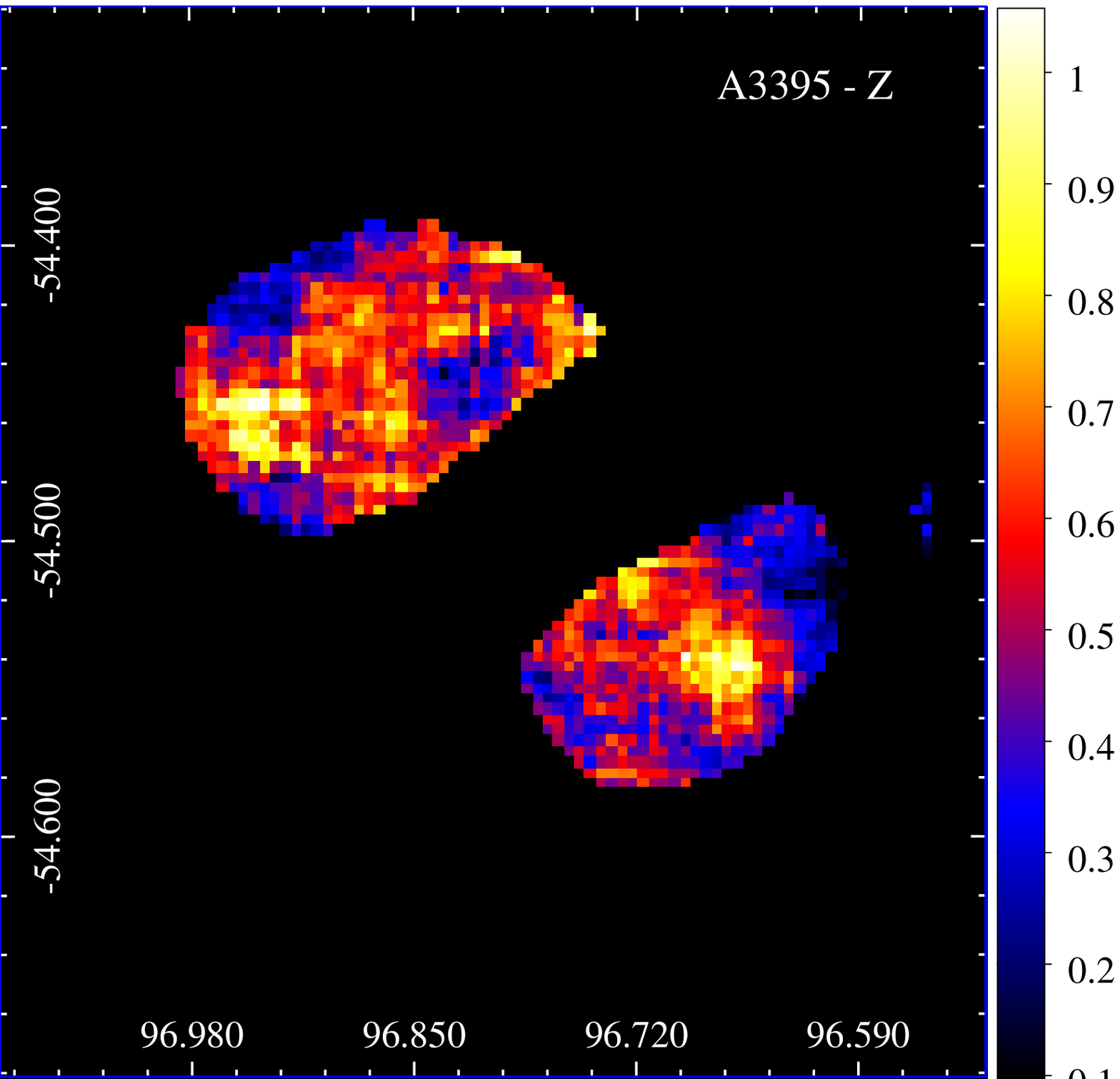}

\includegraphics[scale=0.25]{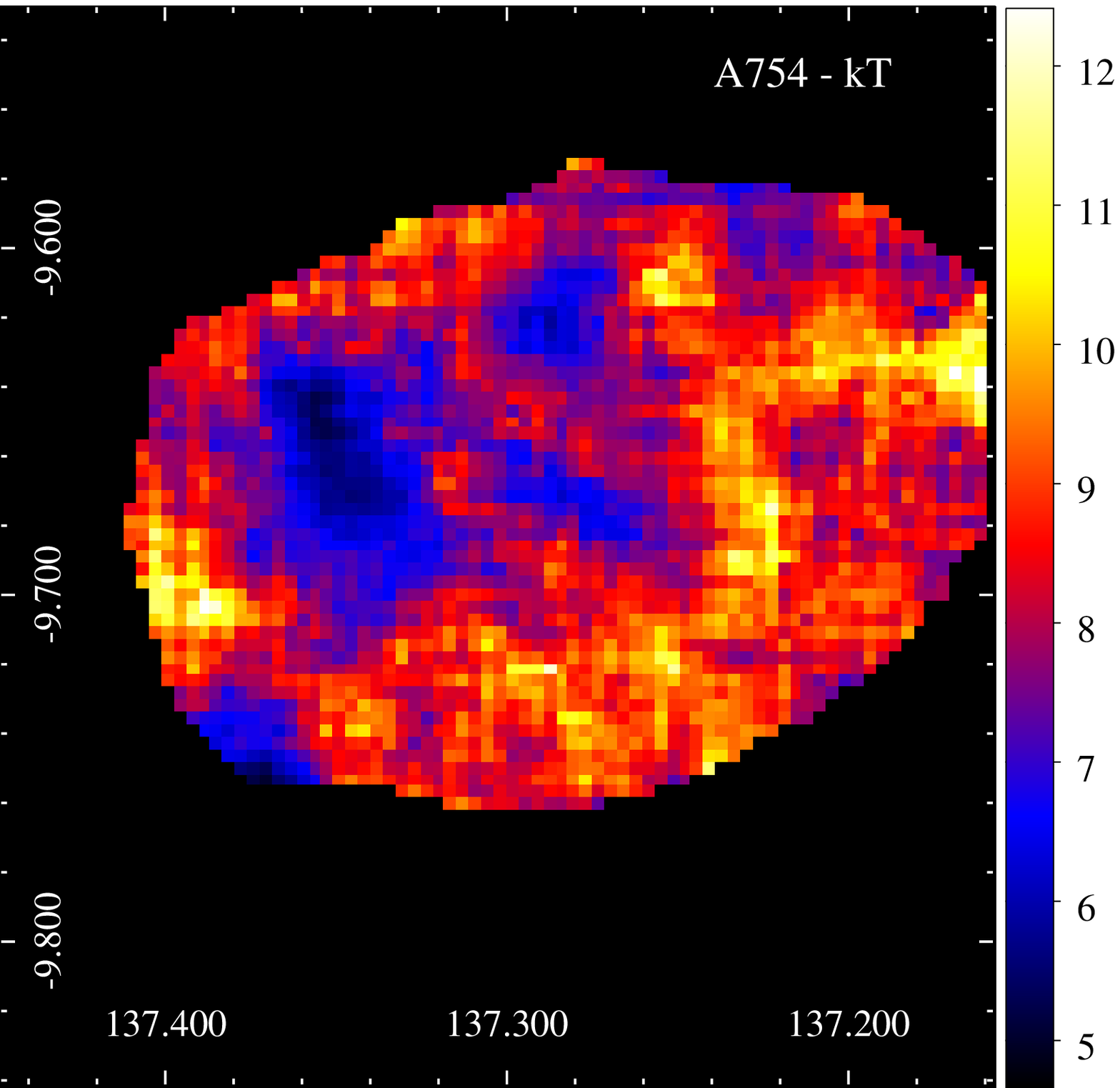}
\includegraphics[scale=0.25]{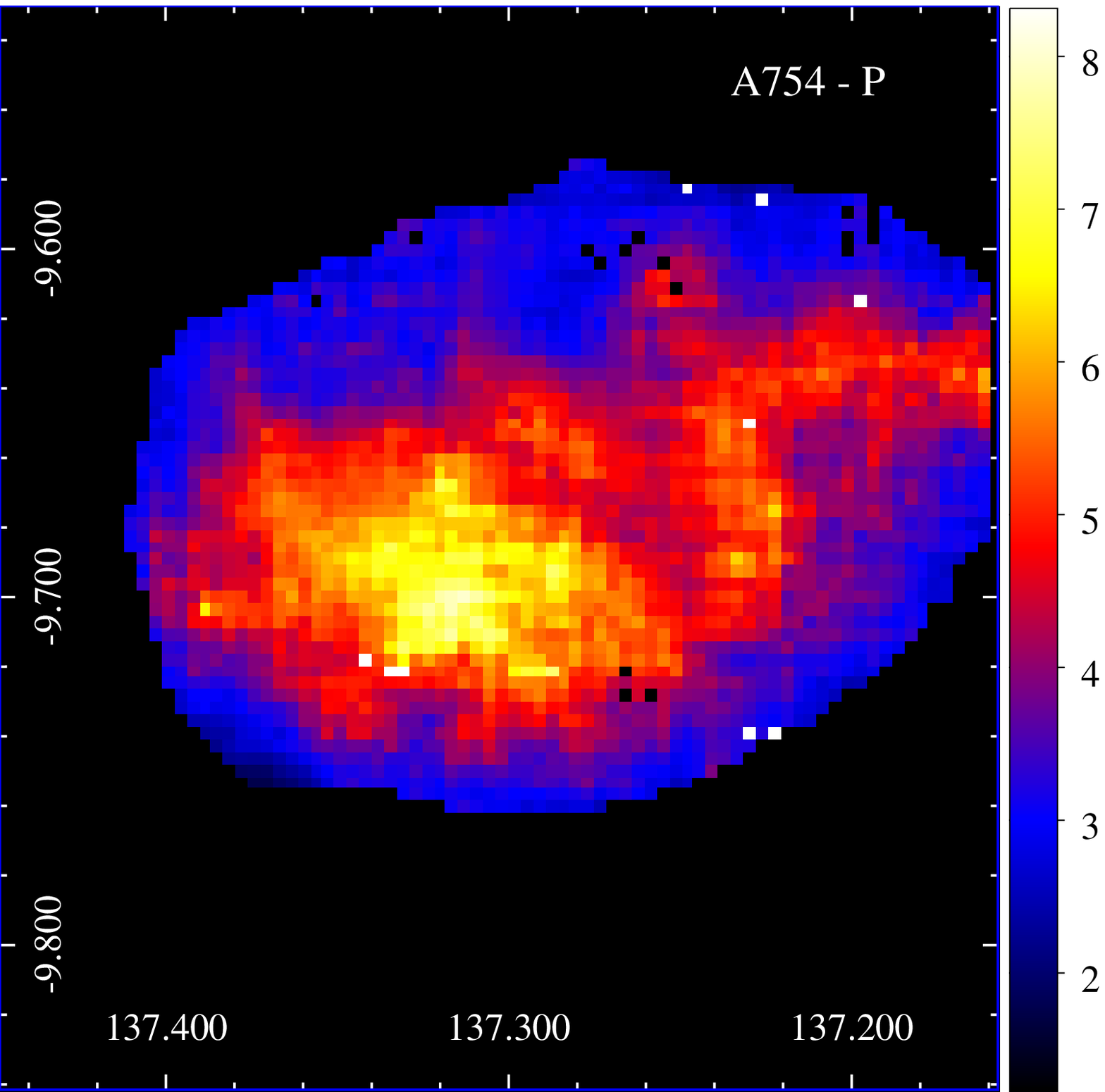}
\includegraphics[scale=0.25]{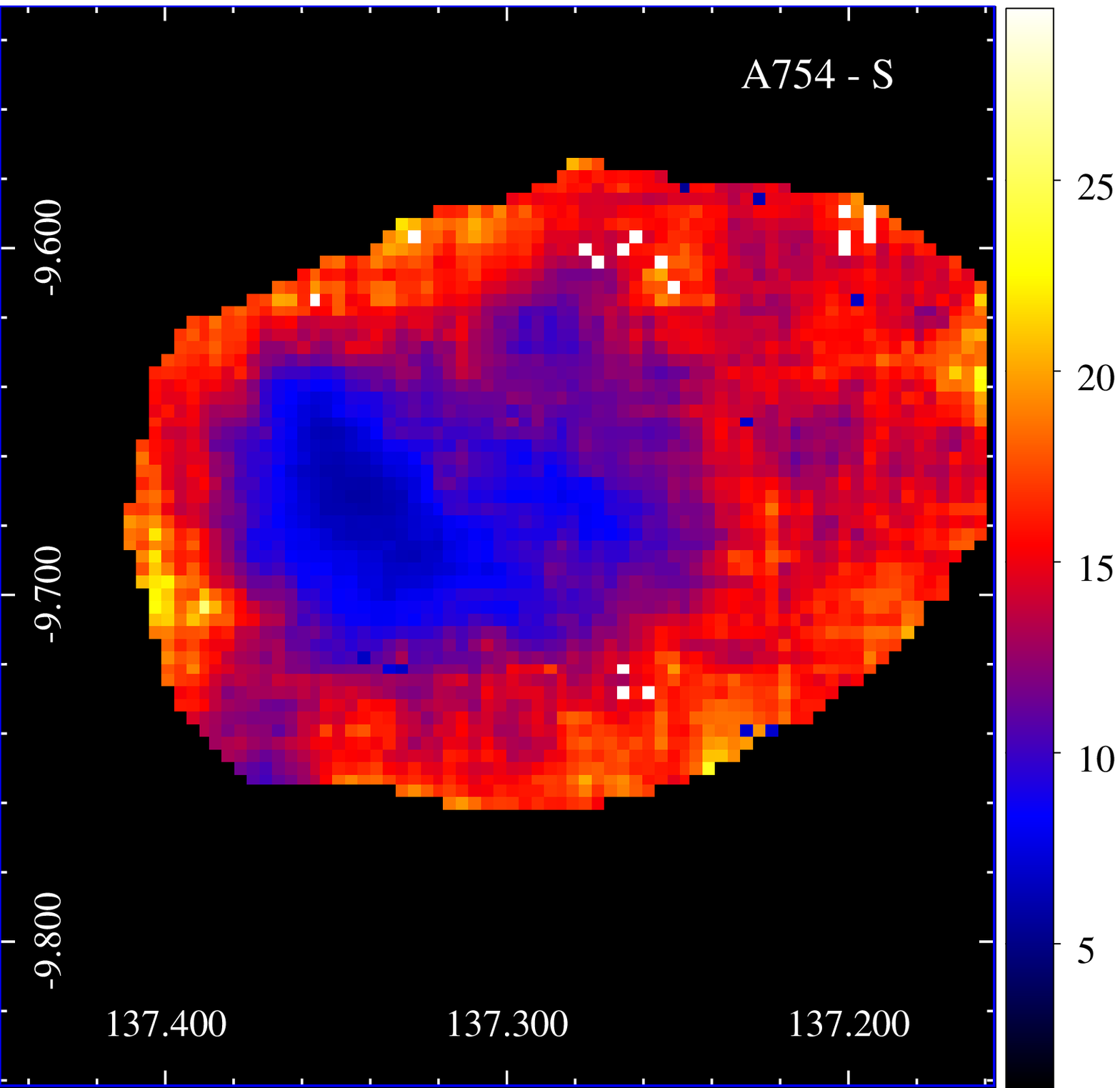}
\includegraphics[scale=0.25]{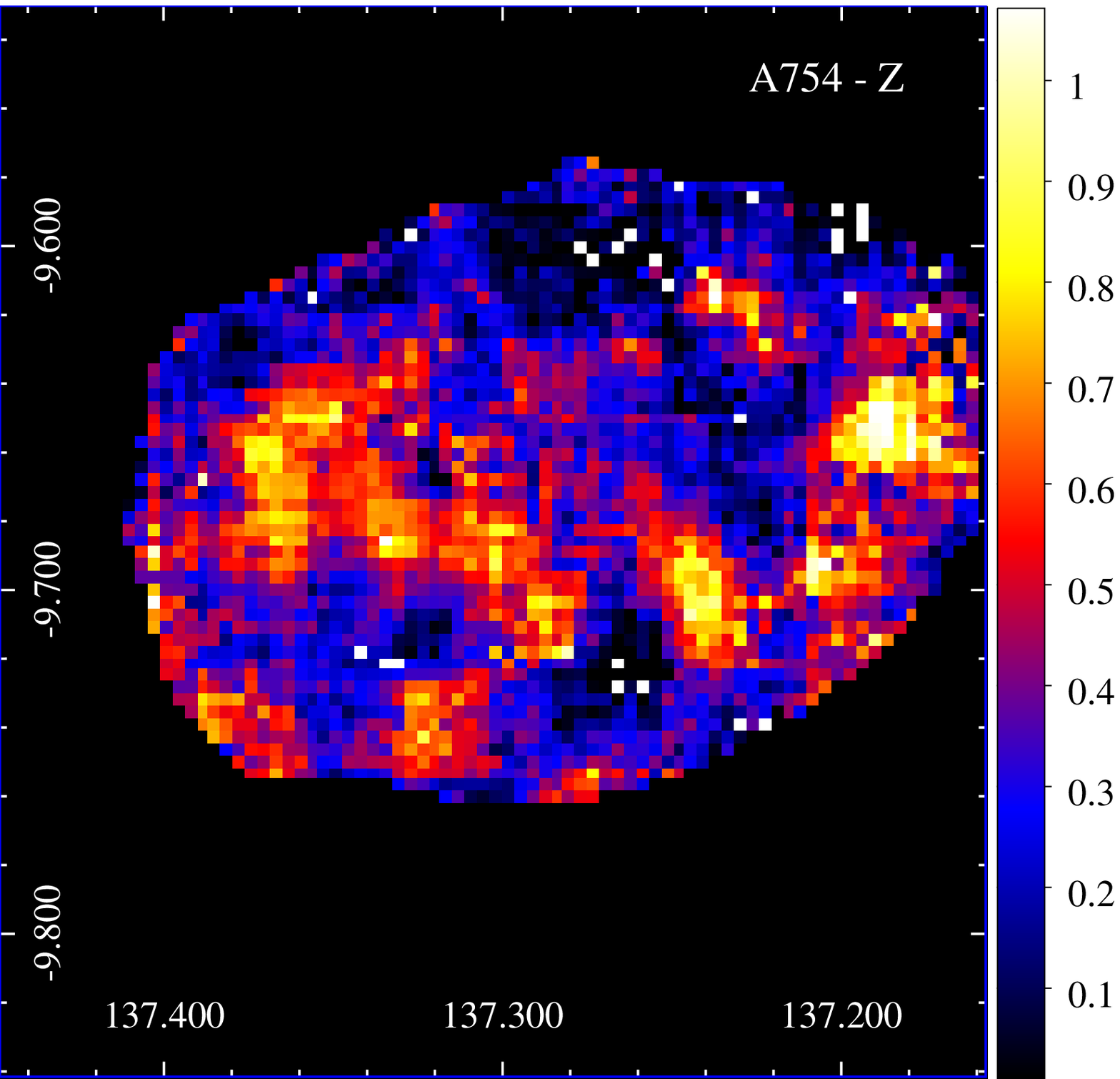}

\caption{NCC and disturbed systems. From left to right: temperature,
  pseudo-pressure, pseudo-entropy, and metallicity maps for A119,
  A3376, A3391, A3395 (A3395 and bA3395 in the same map), and A754.}
\label{fig:NCCclusters1}
\end{figure*}

\begin{figure*}

\includegraphics[scale=0.25]{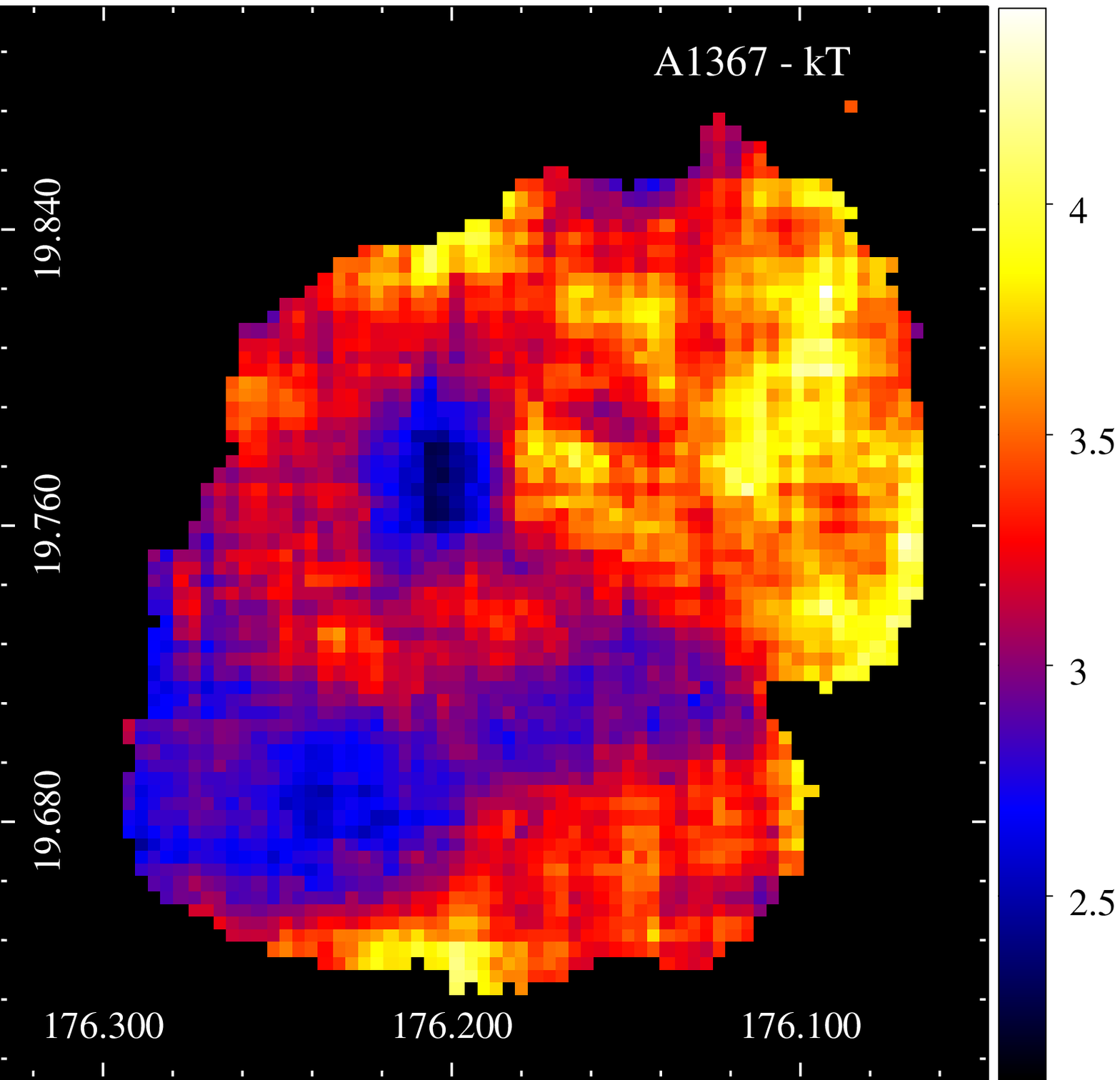}
\includegraphics[scale=0.25]{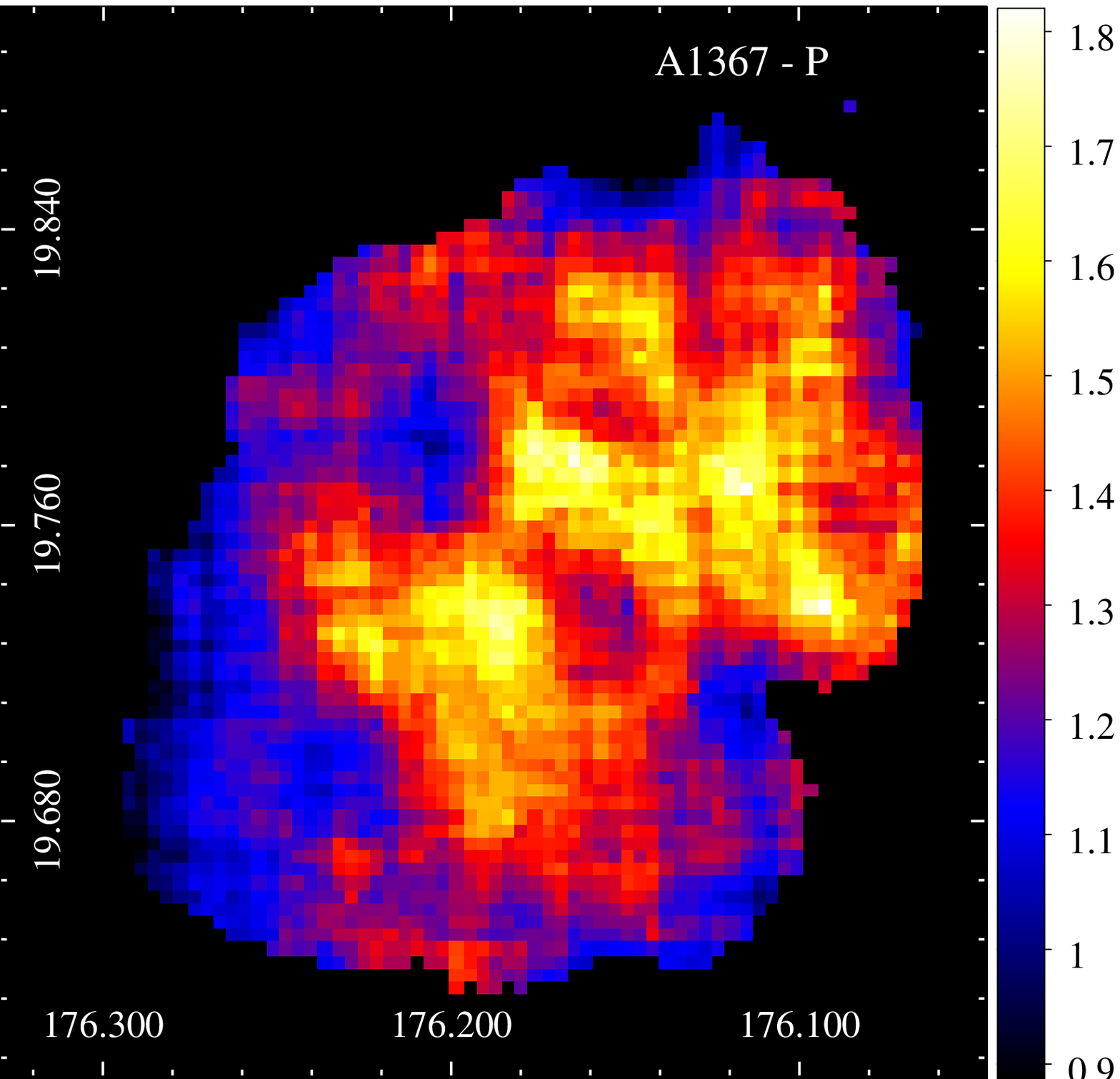}
\includegraphics[scale=0.25]{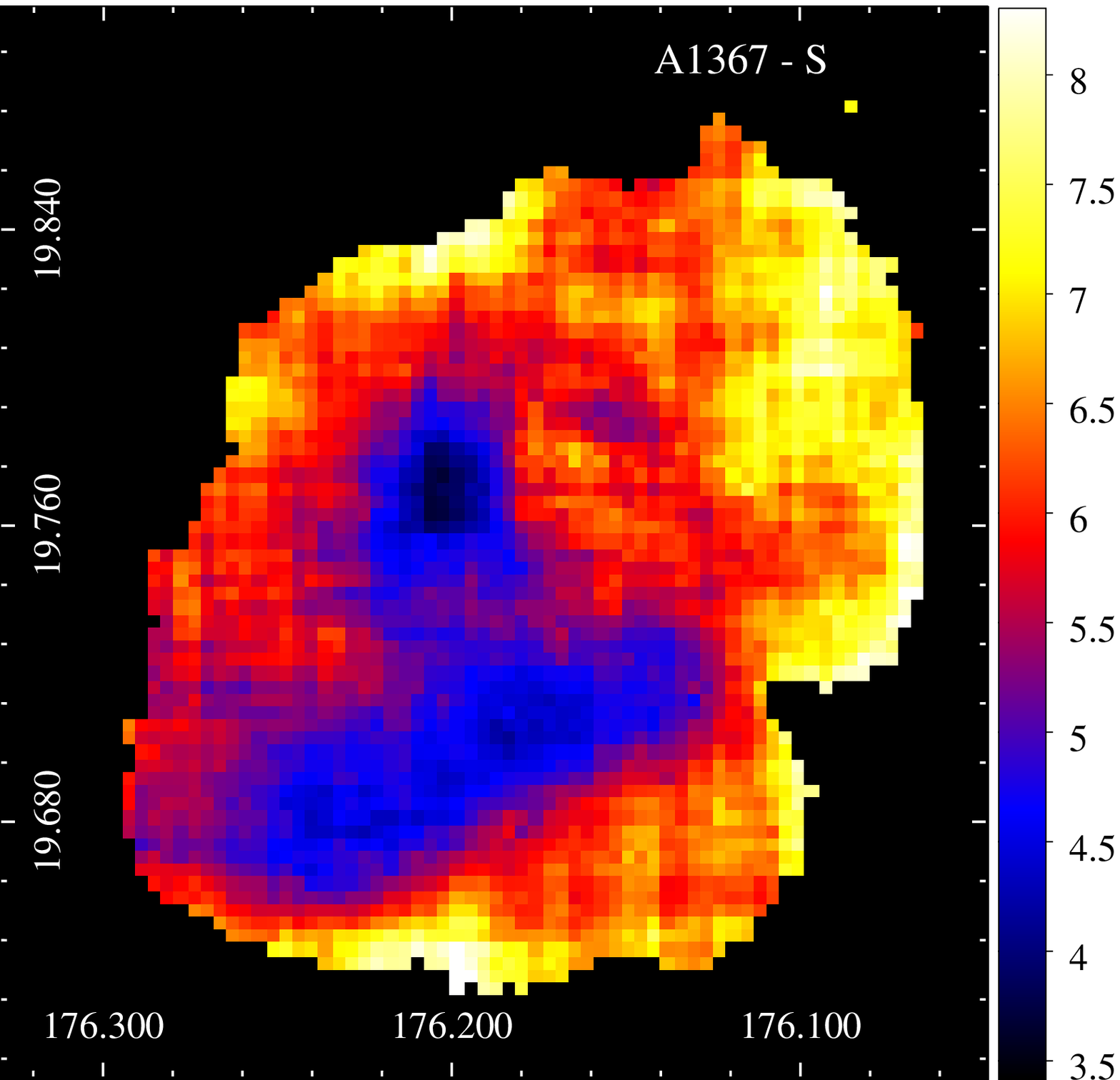}
\includegraphics[scale=0.25]{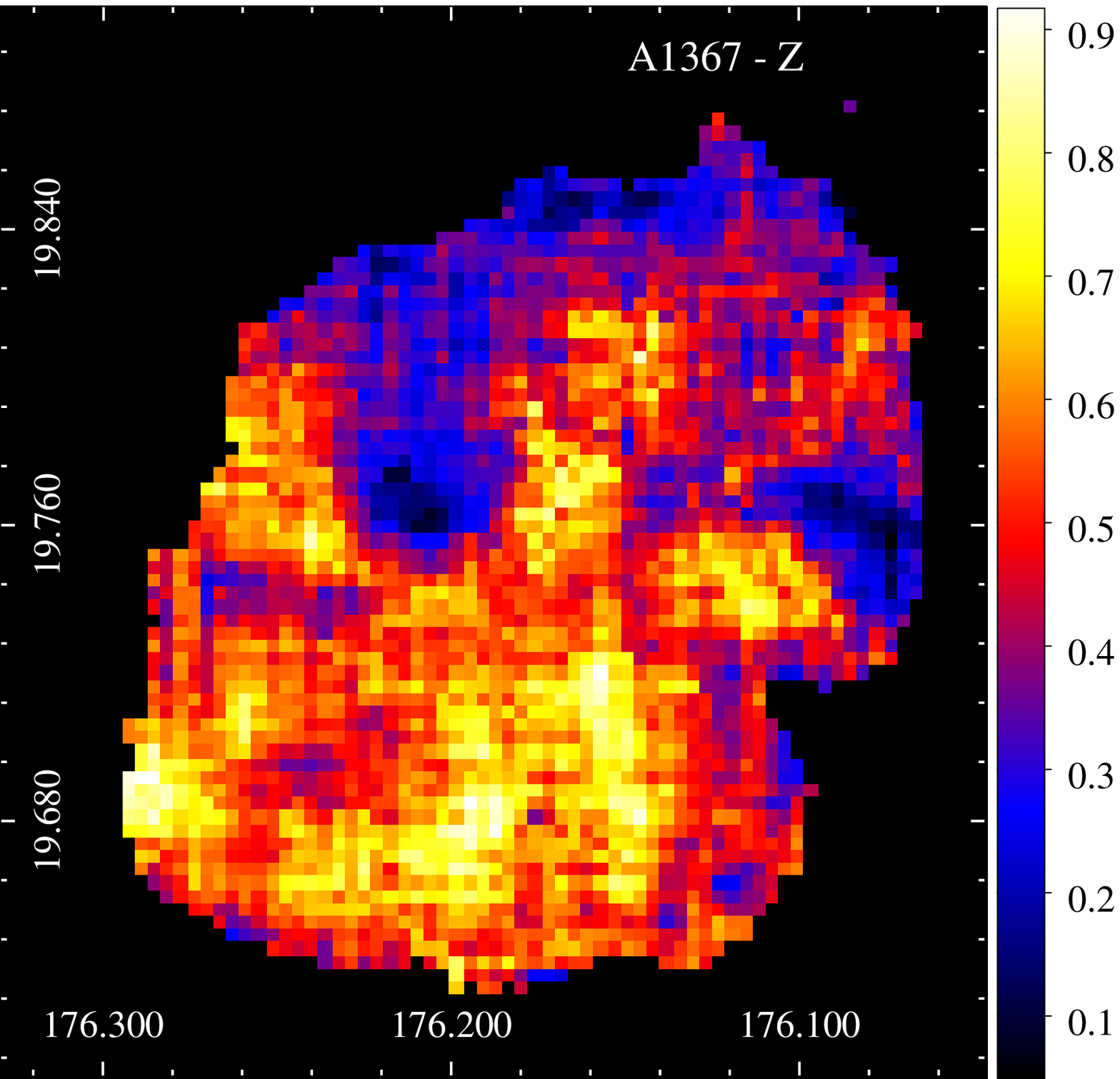}

\includegraphics[scale=0.25]{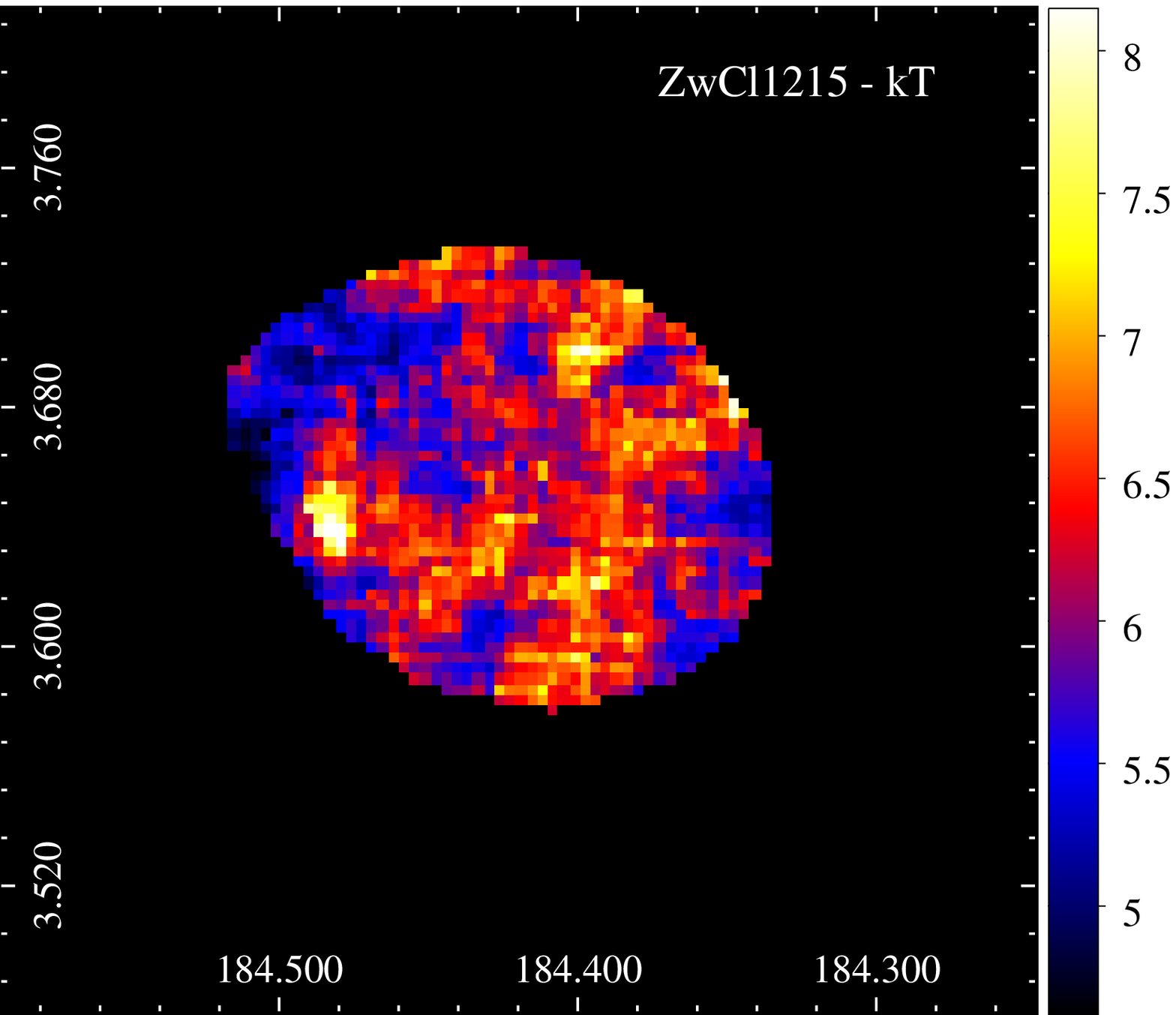}
\includegraphics[scale=0.25]{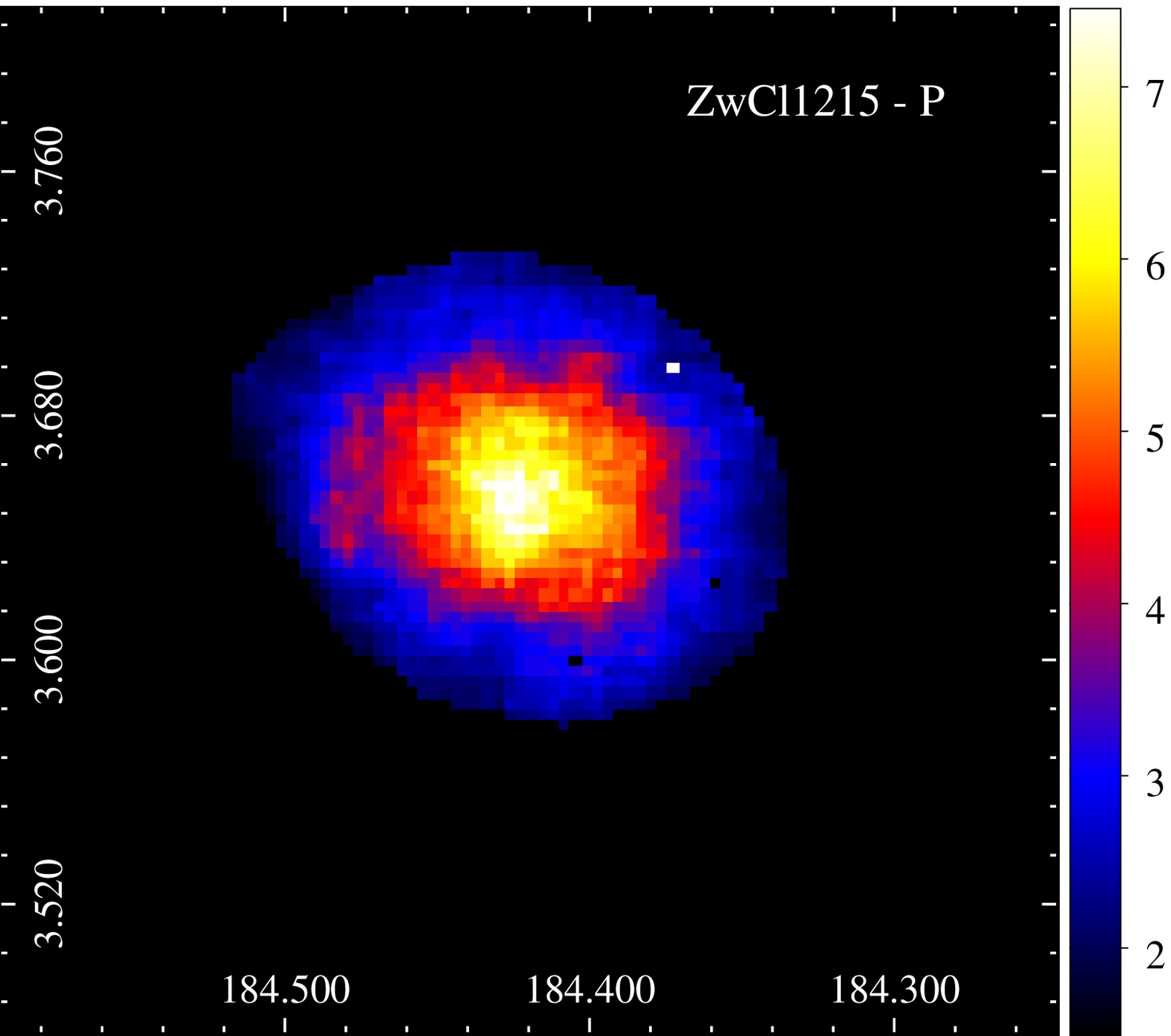}
\includegraphics[scale=0.25]{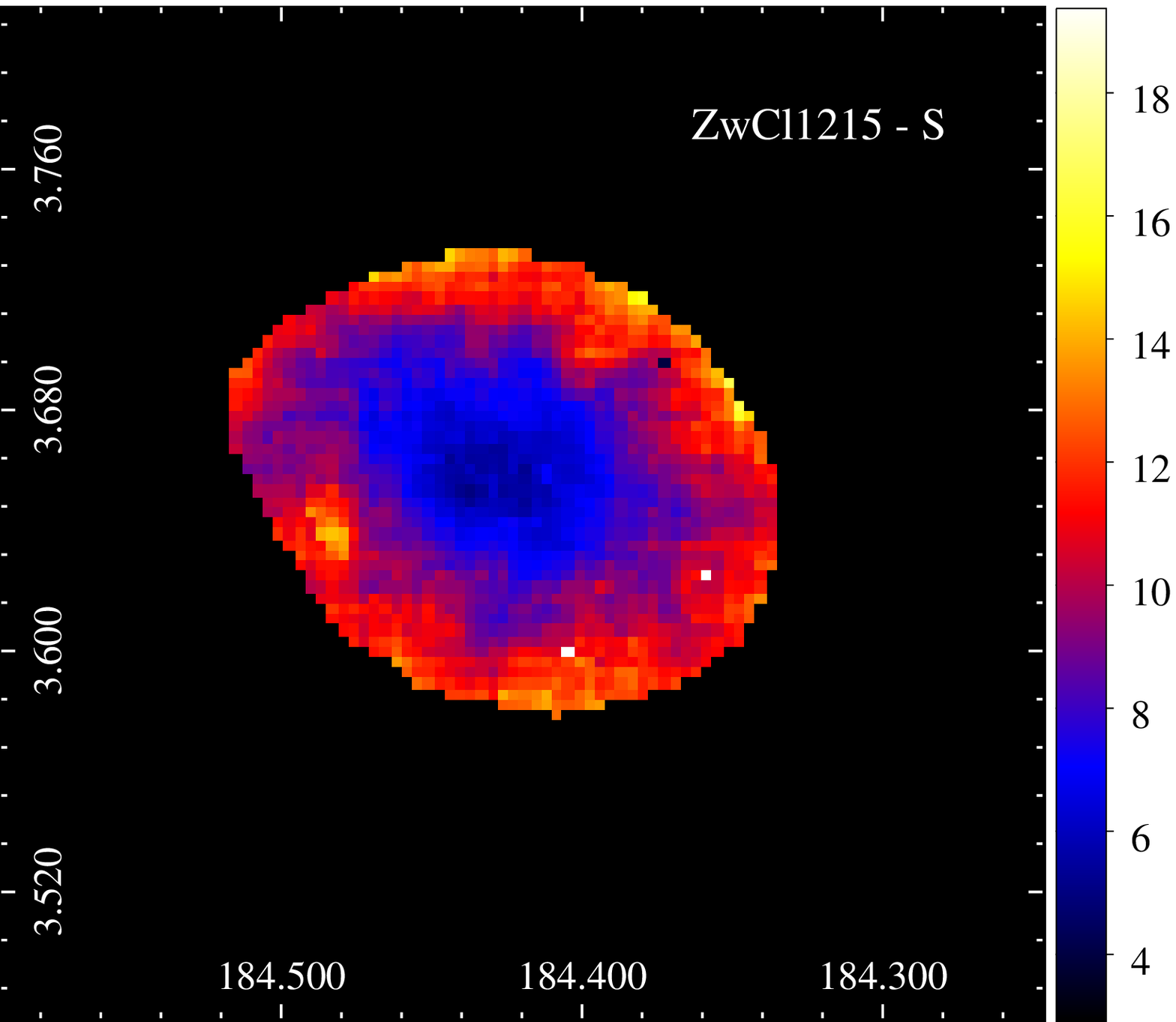}
\includegraphics[scale=0.25]{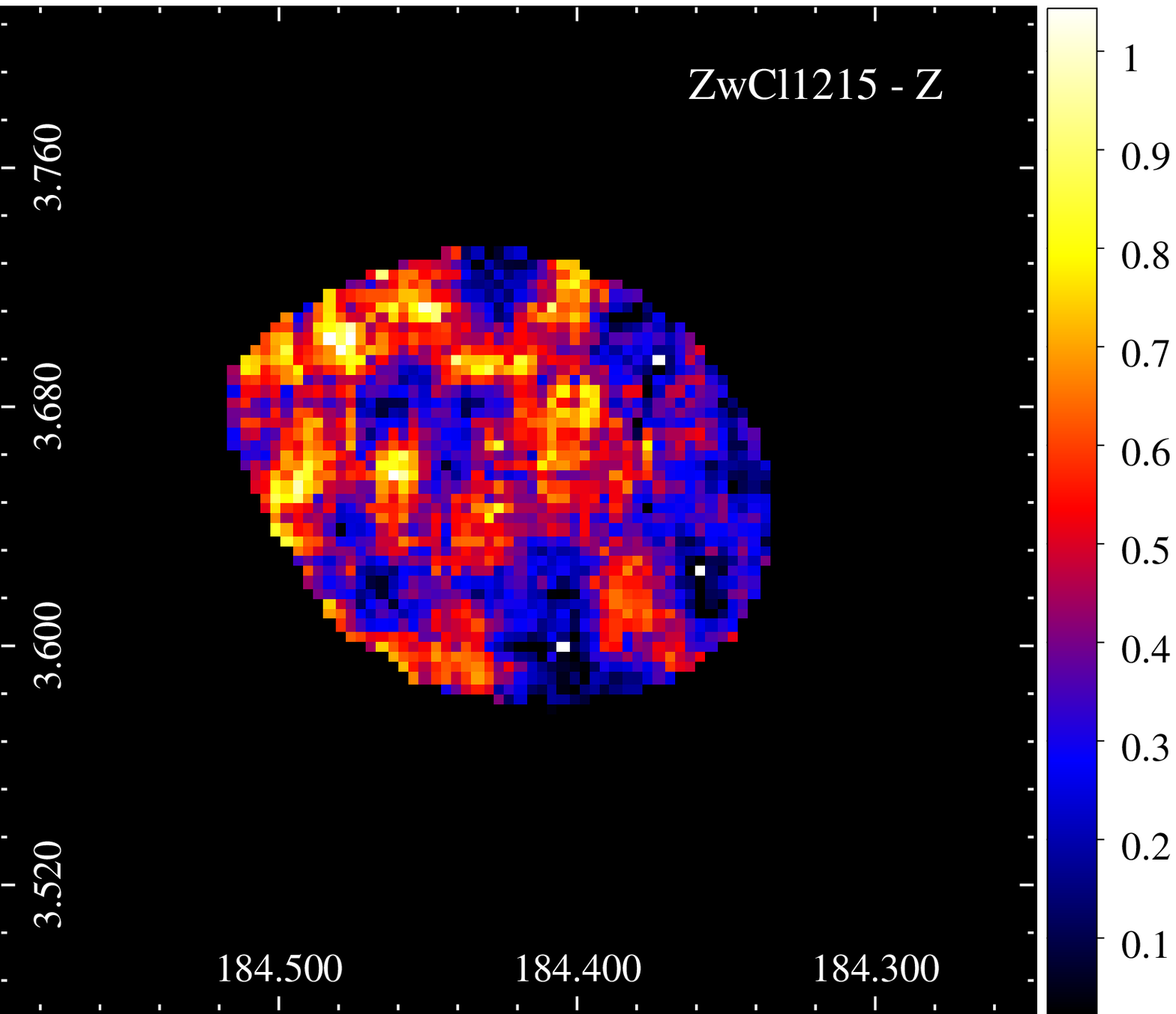}

\includegraphics[scale=0.25]{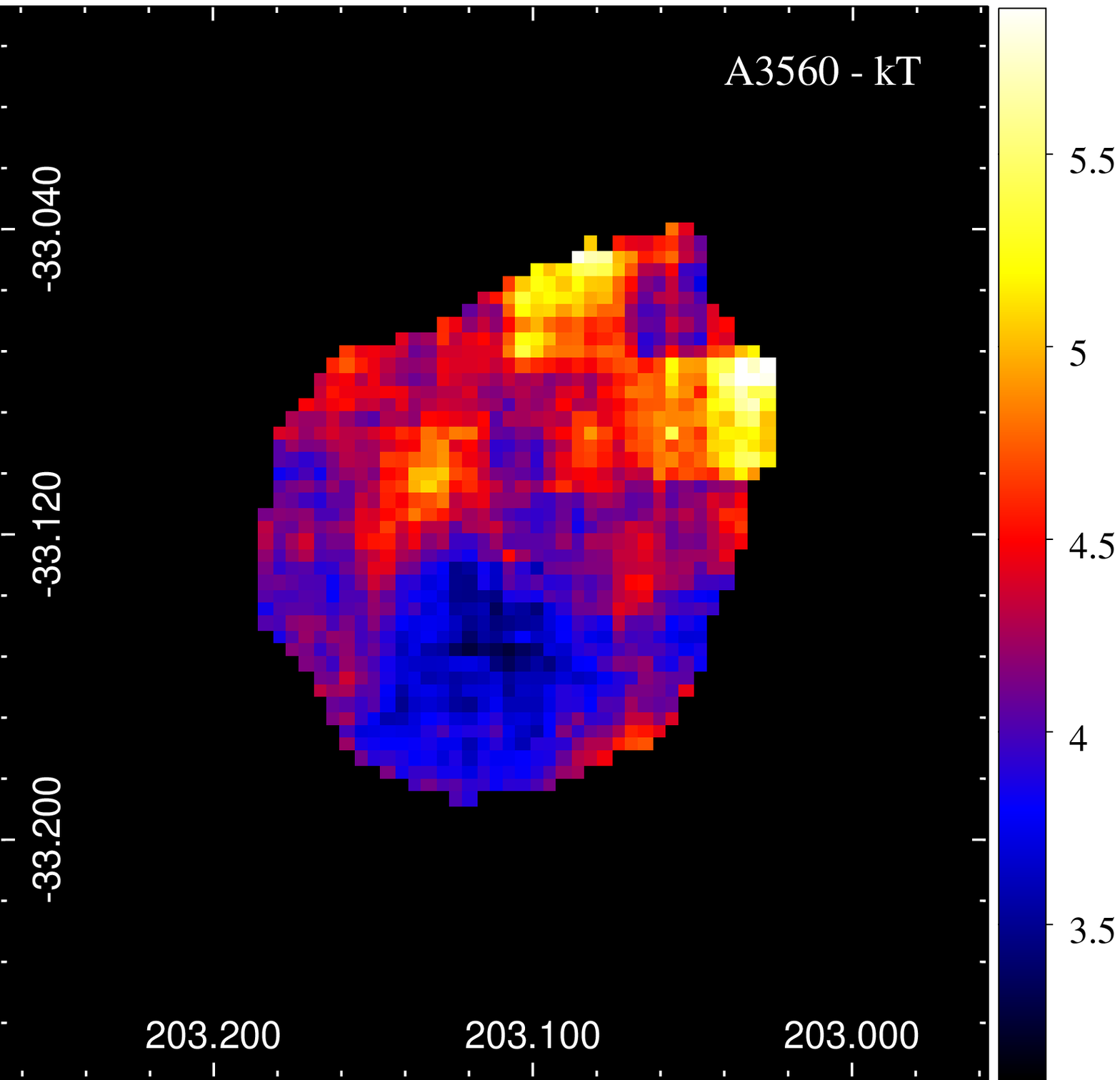}
\includegraphics[scale=0.25]{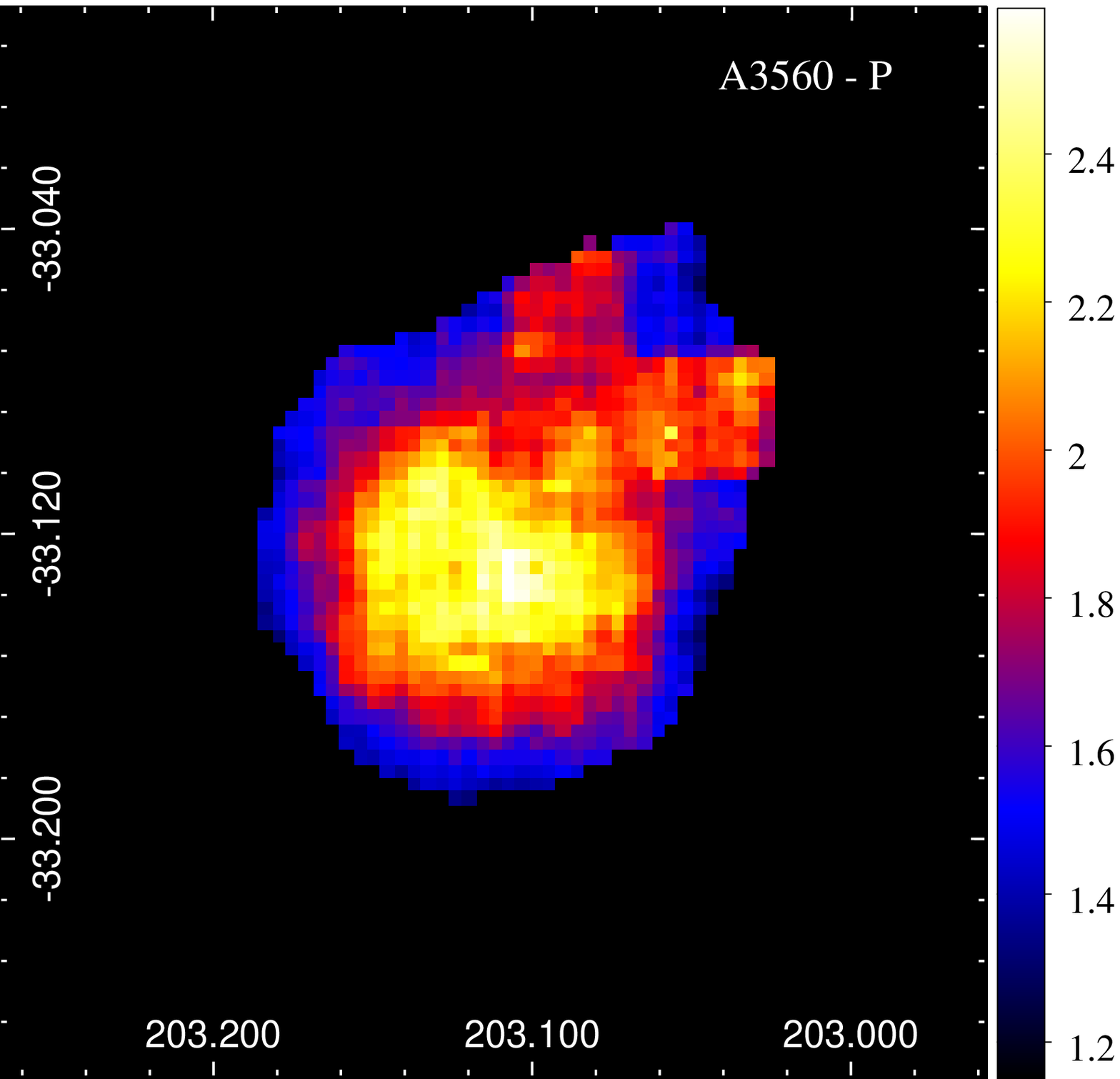}
\includegraphics[scale=0.25]{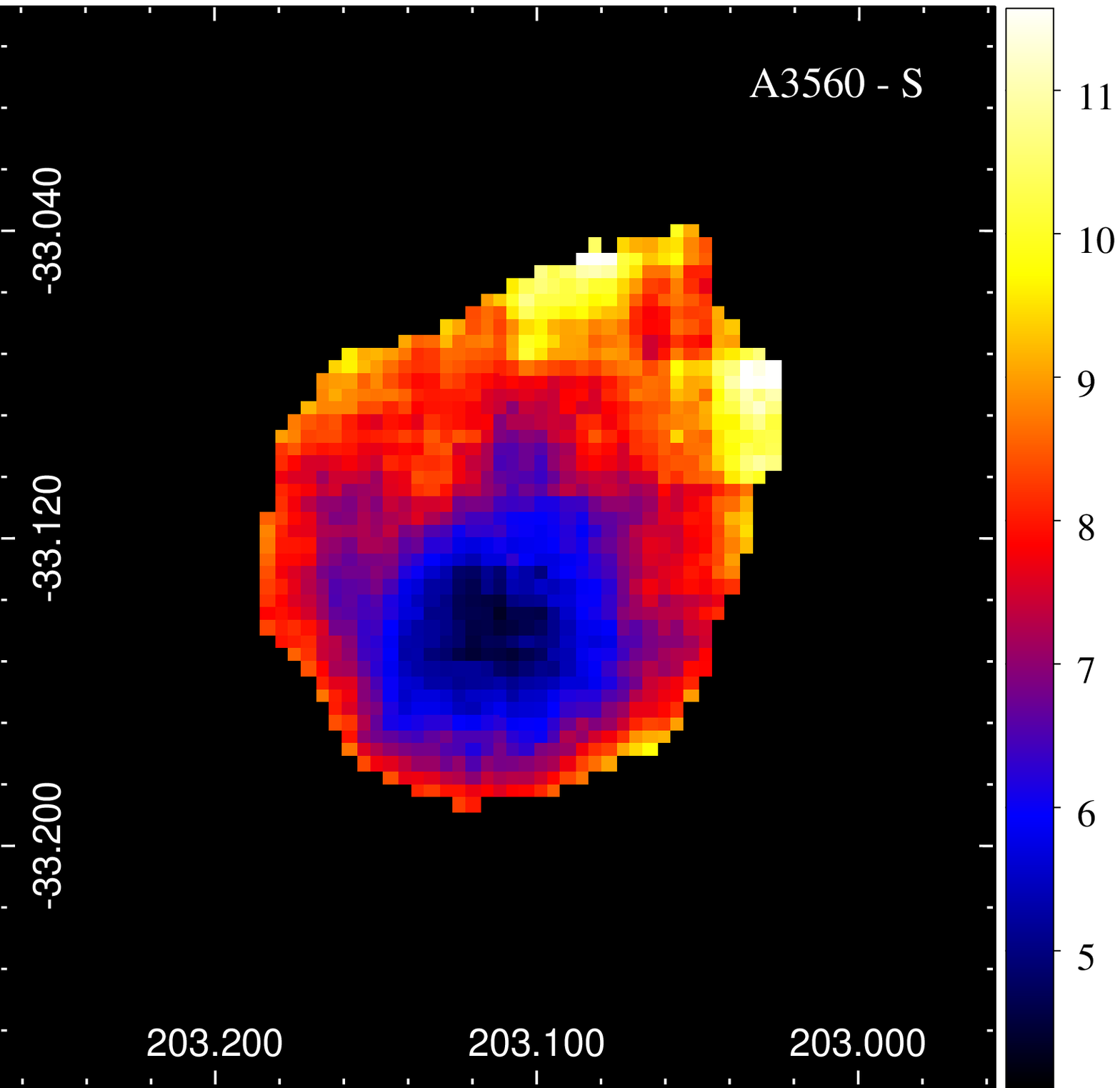}
\includegraphics[scale=0.25]{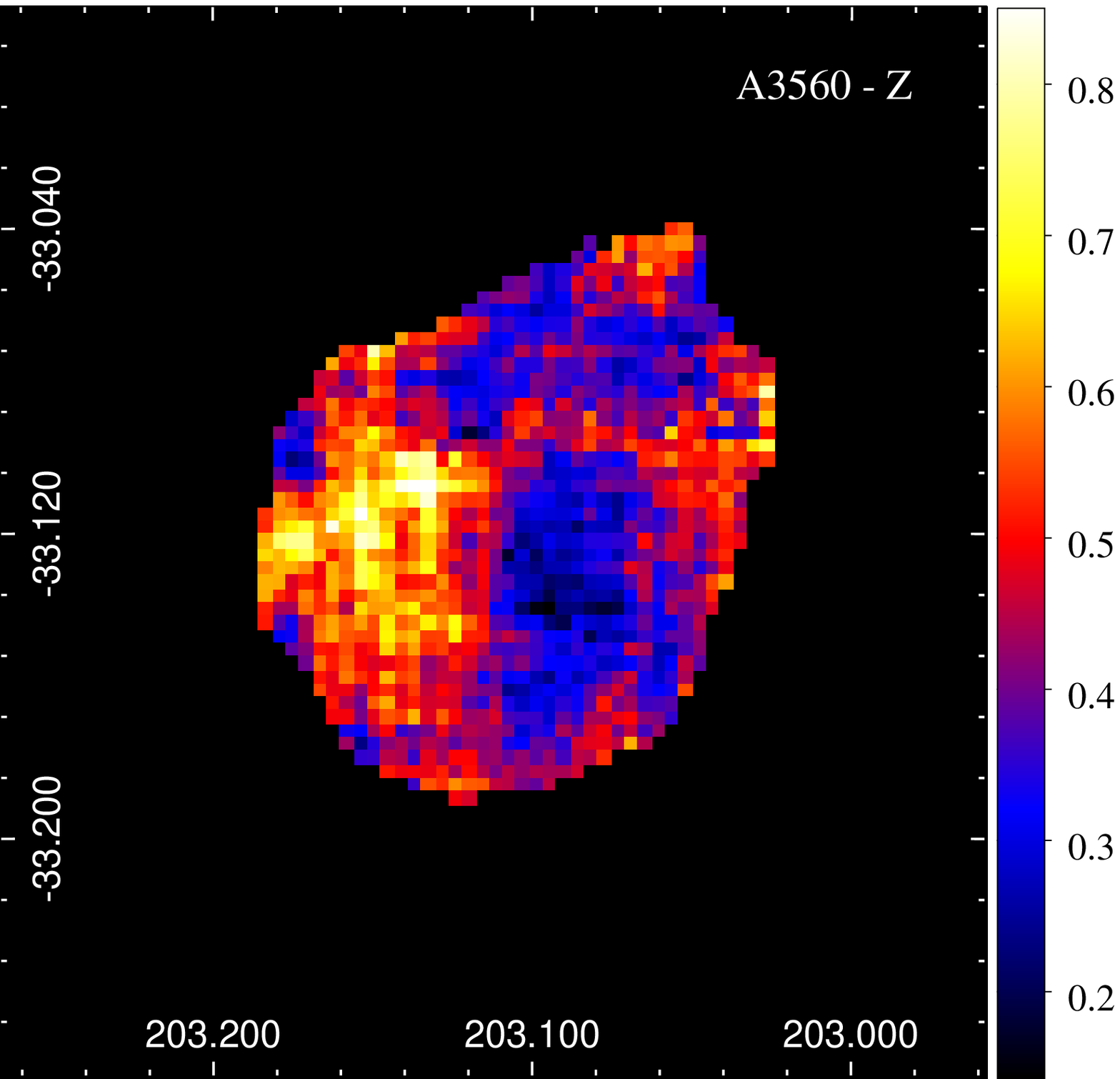}

\includegraphics[scale=0.25]{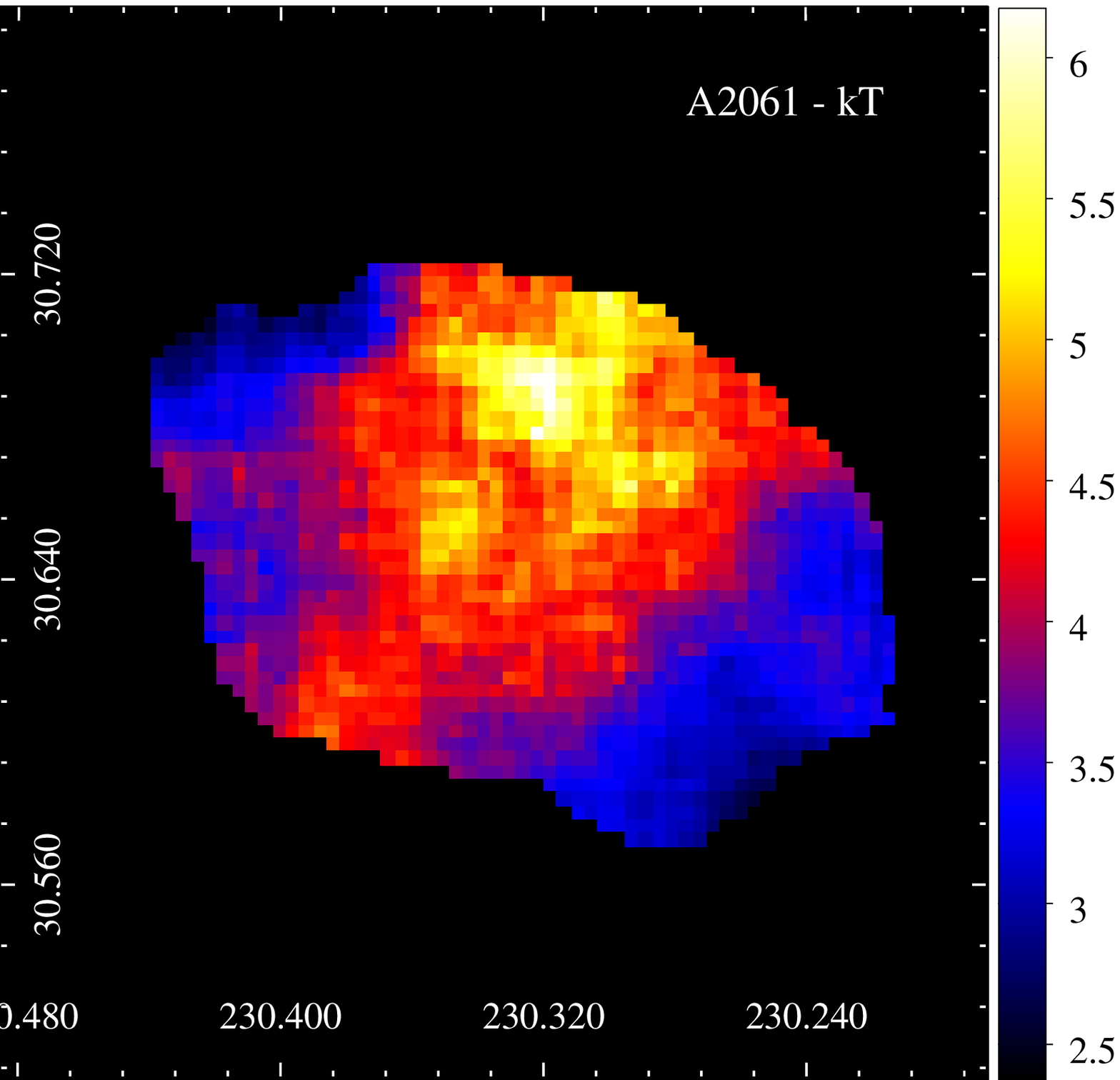}
\includegraphics[scale=0.25]{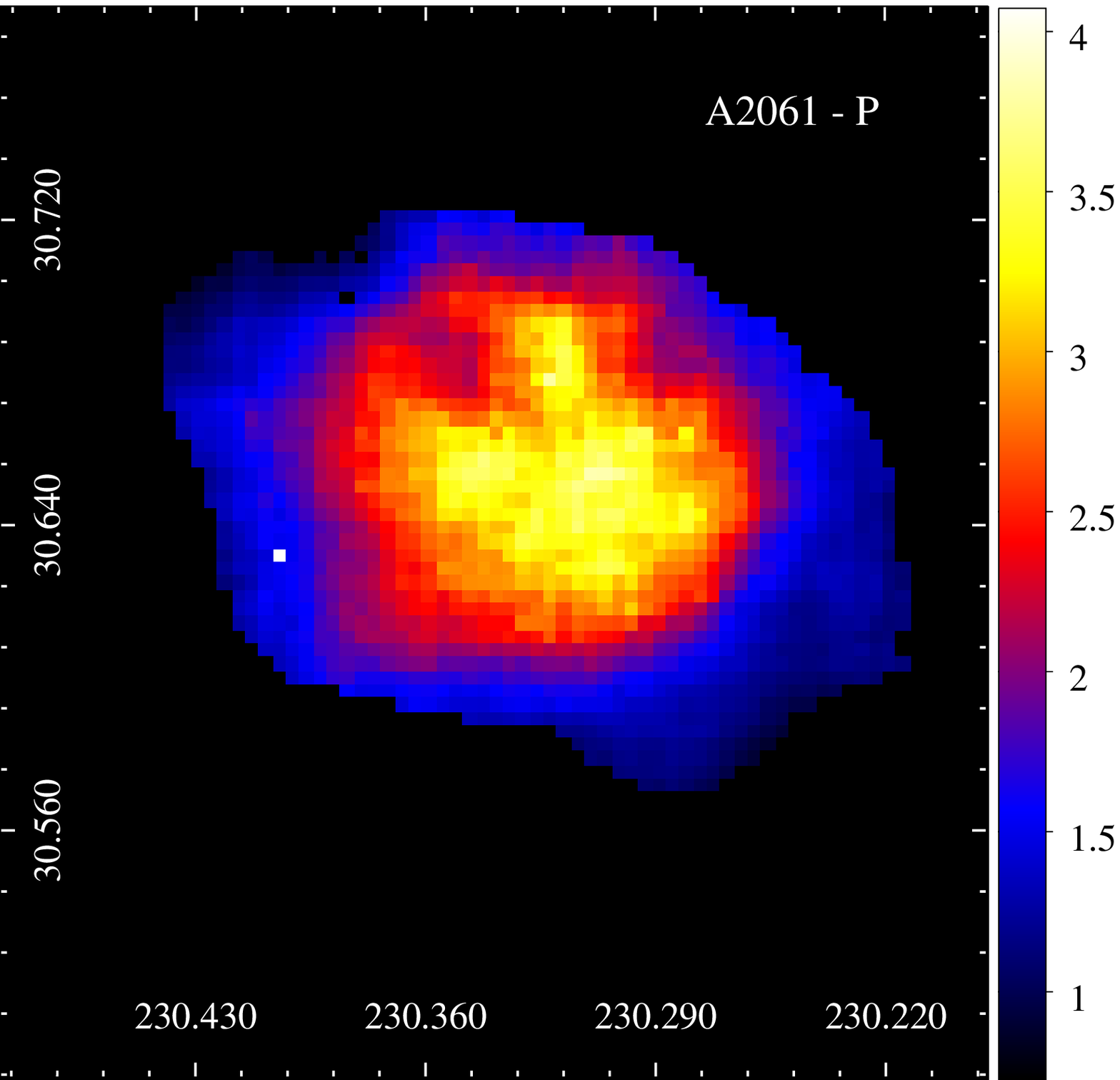}
\includegraphics[scale=0.25]{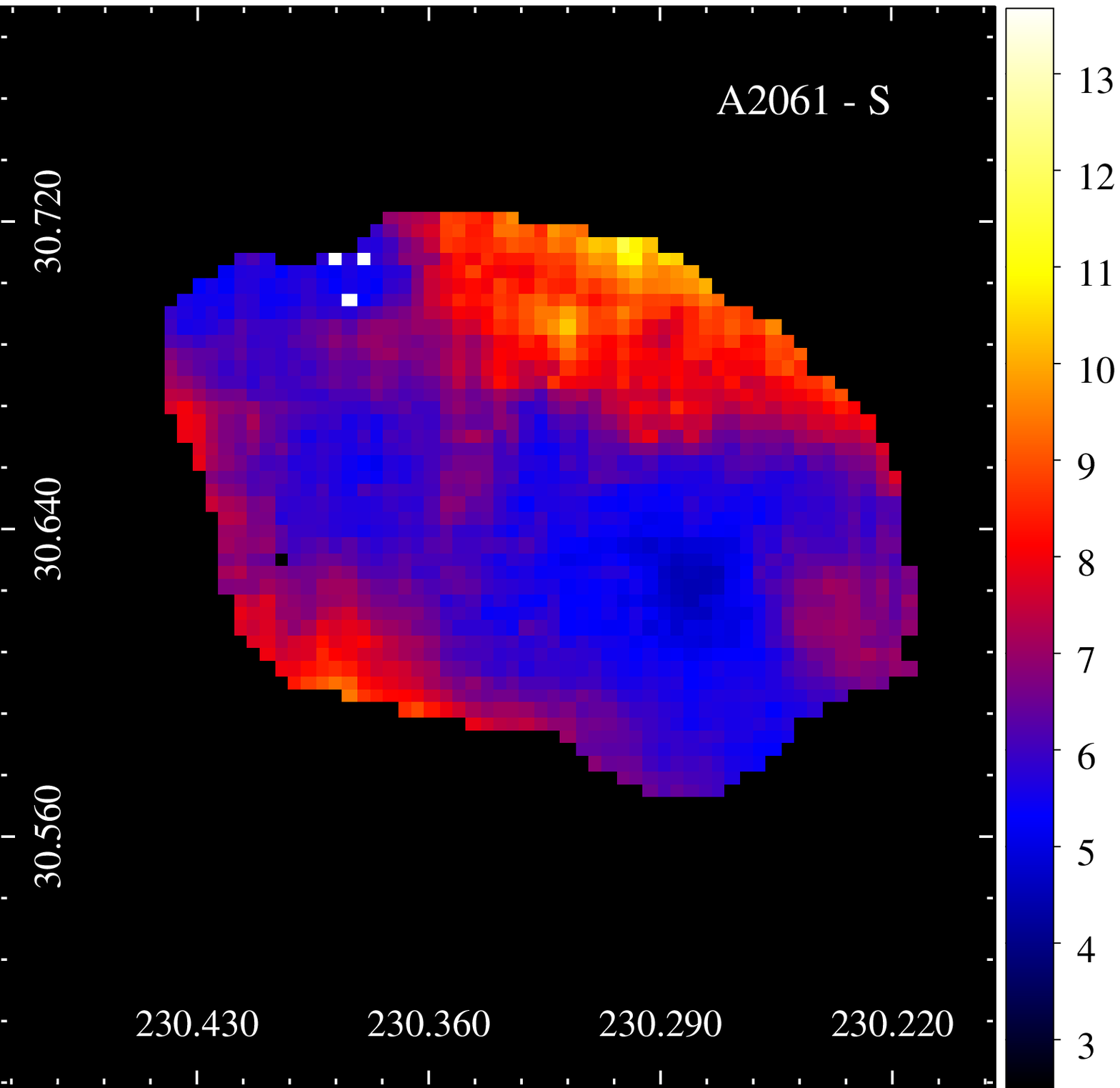}
\includegraphics[scale=0.25]{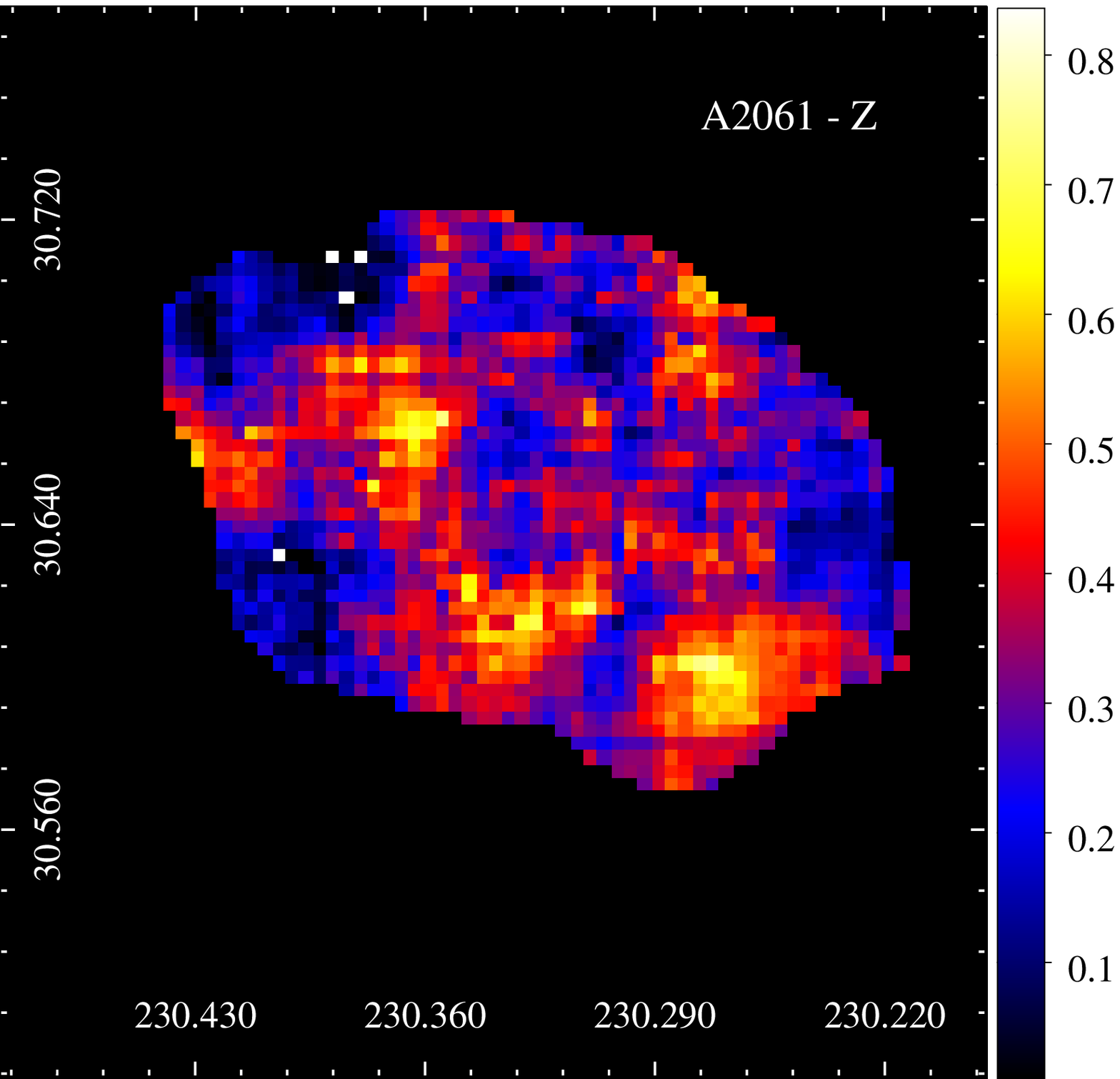}

\includegraphics[scale=0.25]{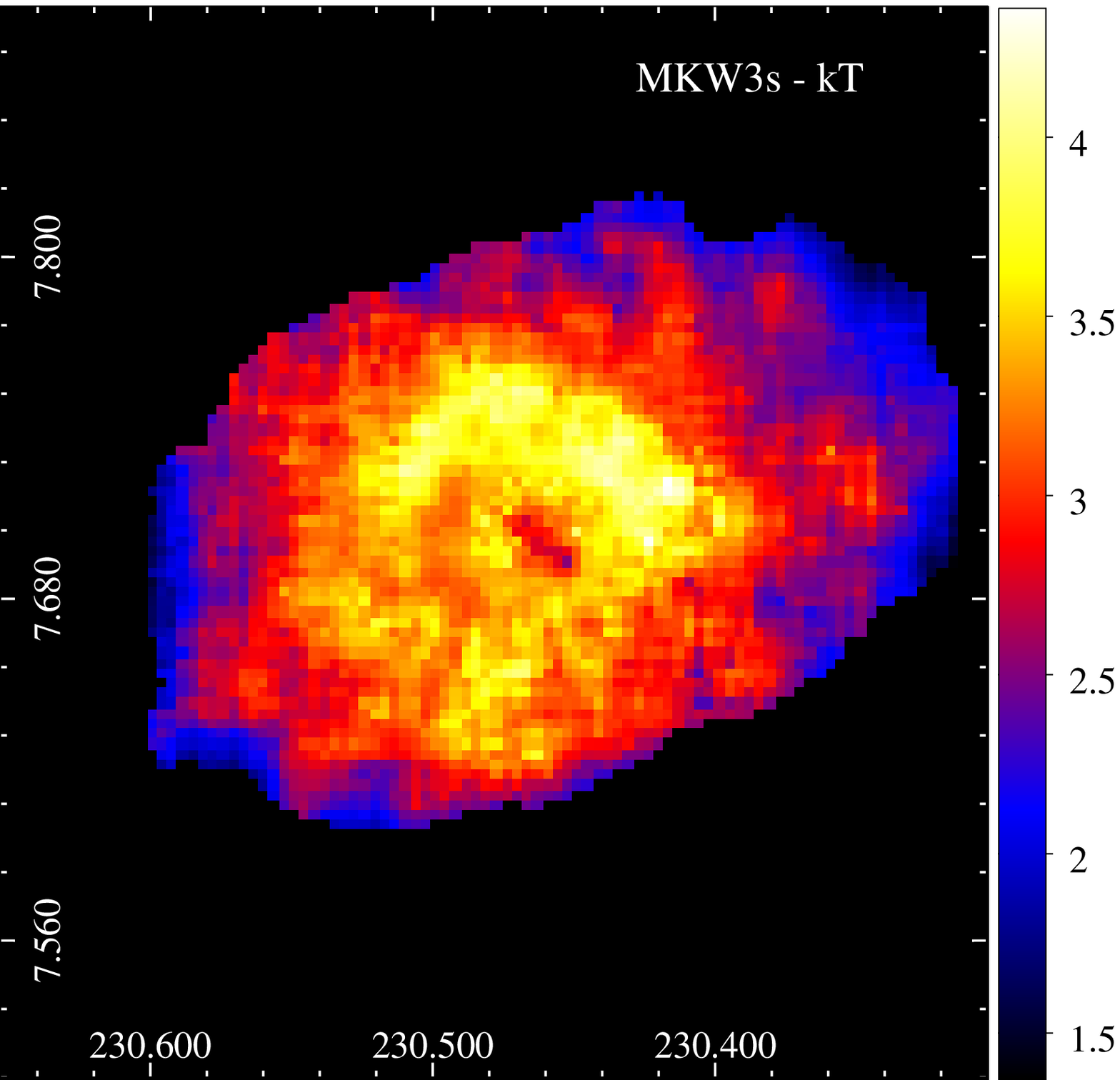}
\includegraphics[scale=0.25]{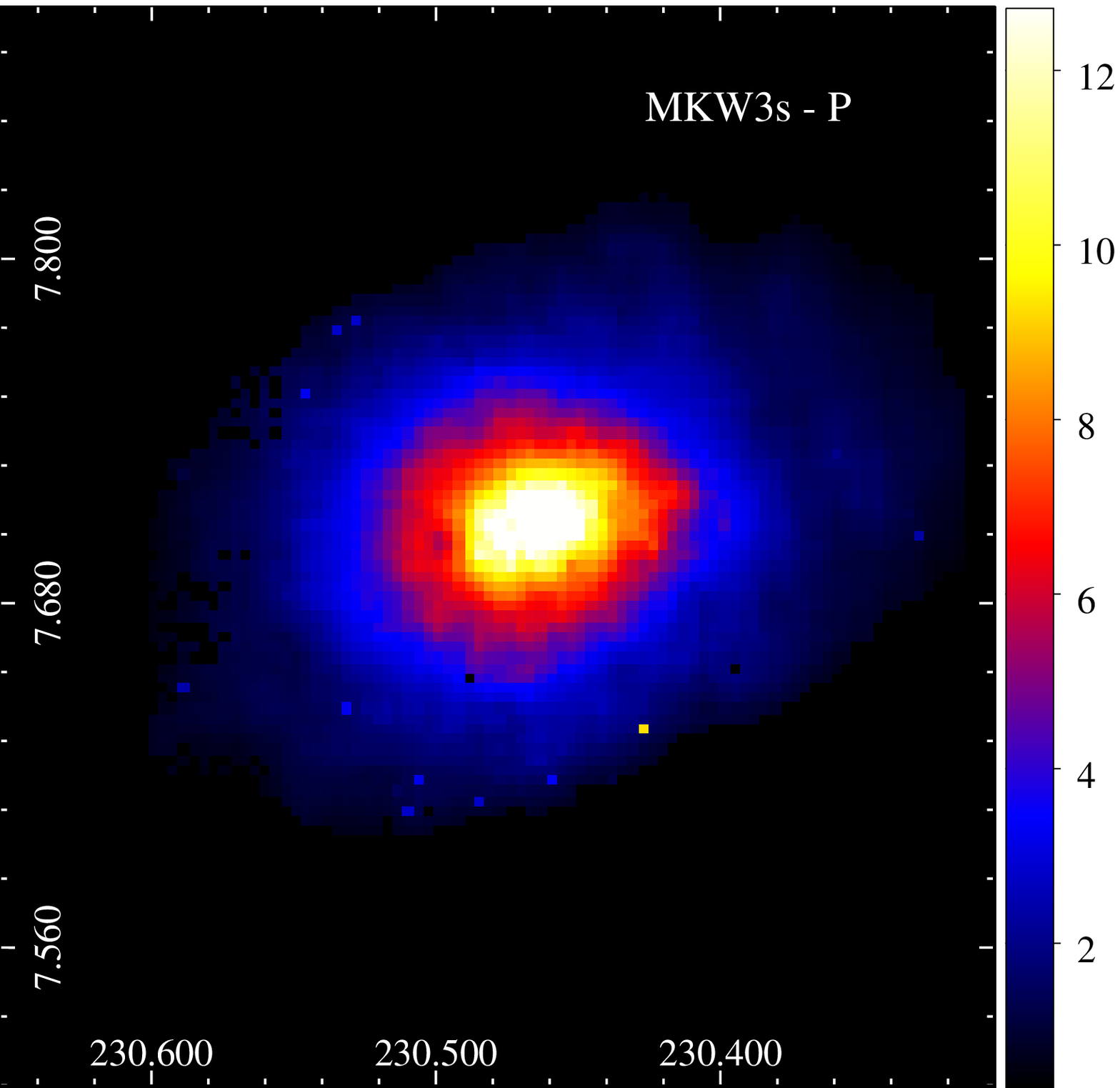}
\includegraphics[scale=0.25]{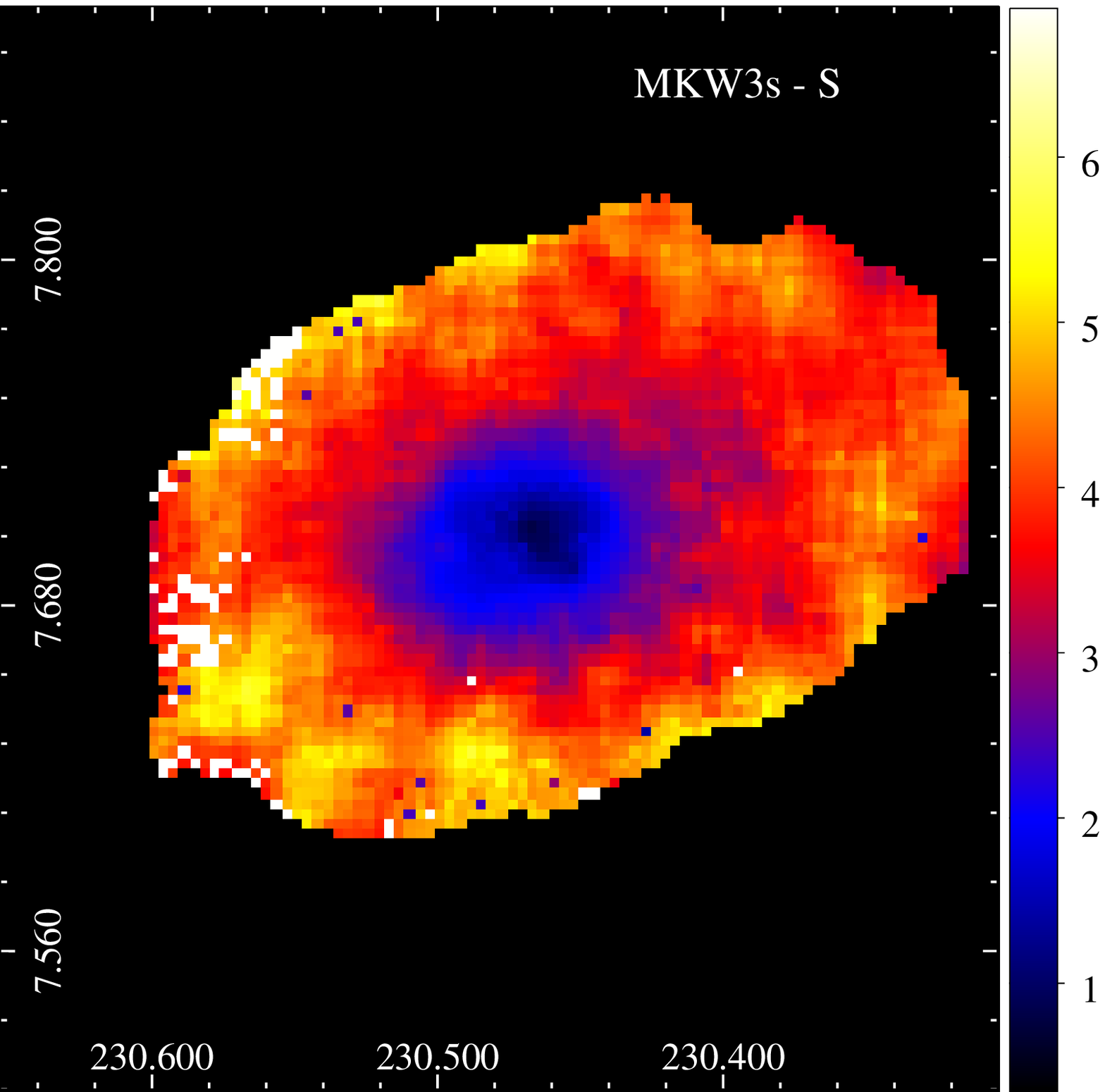}
\includegraphics[scale=0.25]{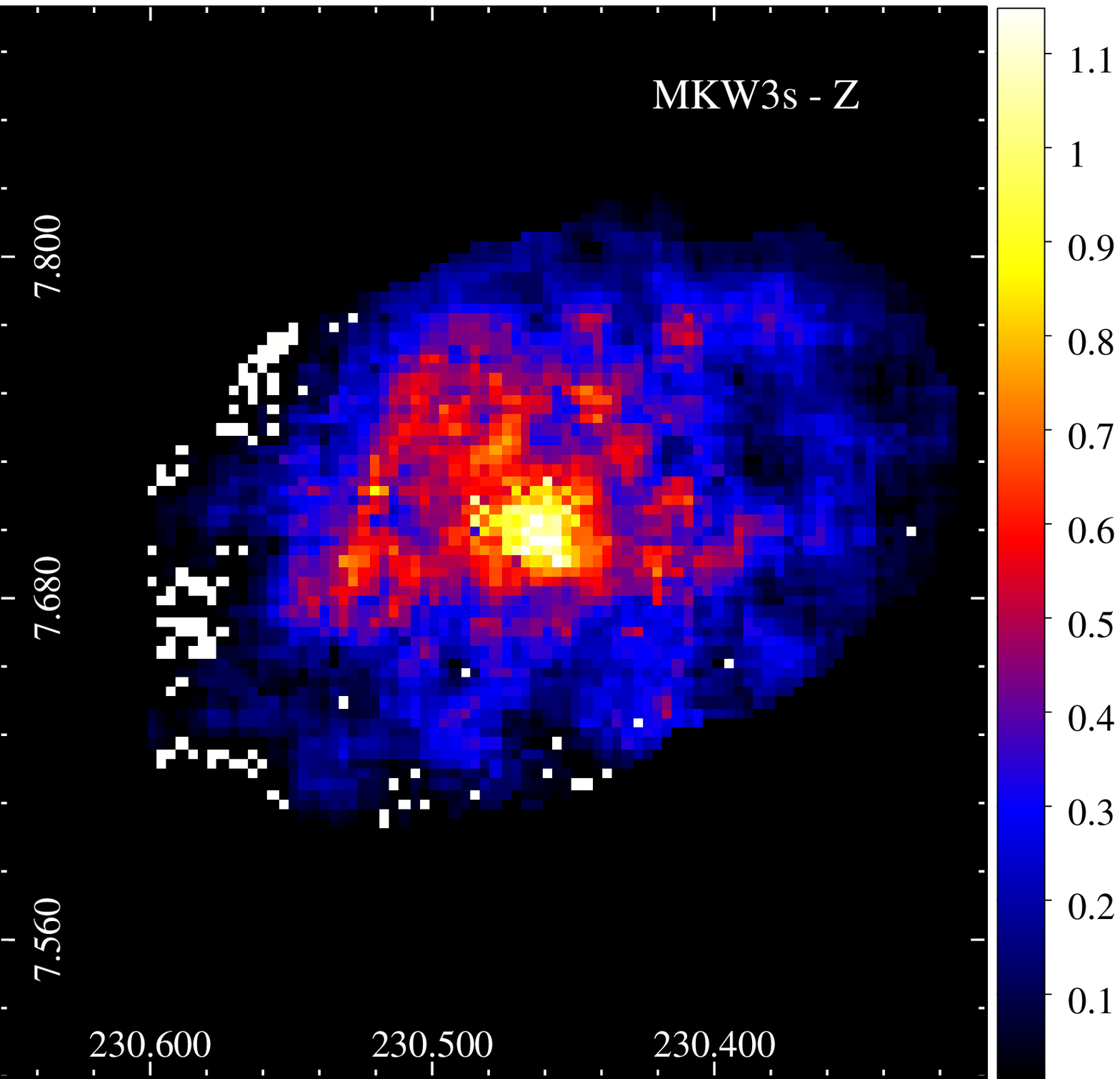}

\caption{NCC and disturbed systems. From left to right: temperature,
  pseudo-pressure, pseudo-entropy, and metallicity maps for 
 G234,  ZwCl1215, A3560, A2061, and MKW3s.}
\label{fig:NCCclusters2}
\end{figure*}

\begin{figure*}

\includegraphics[scale=0.25]{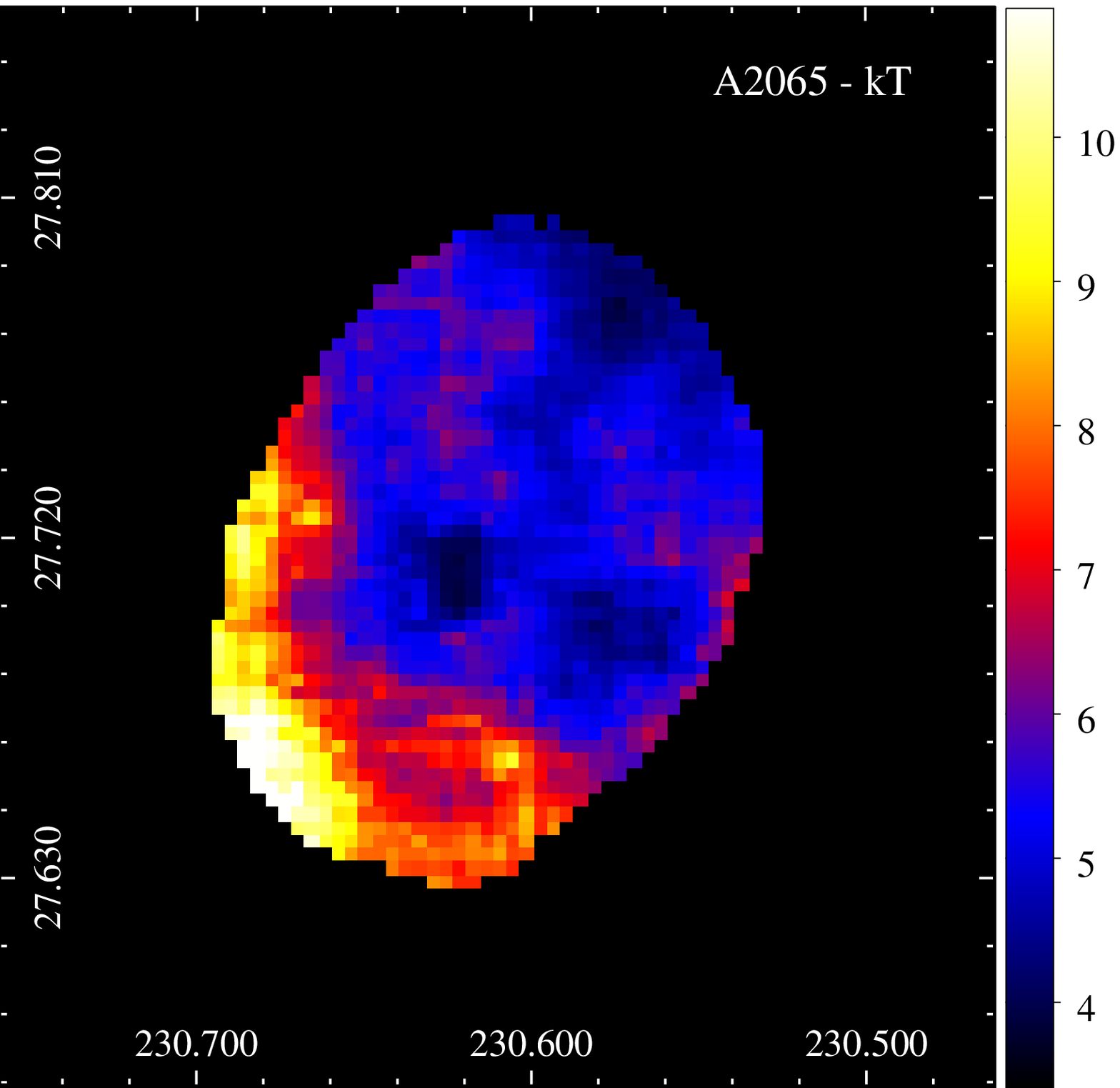}
\includegraphics[scale=0.25]{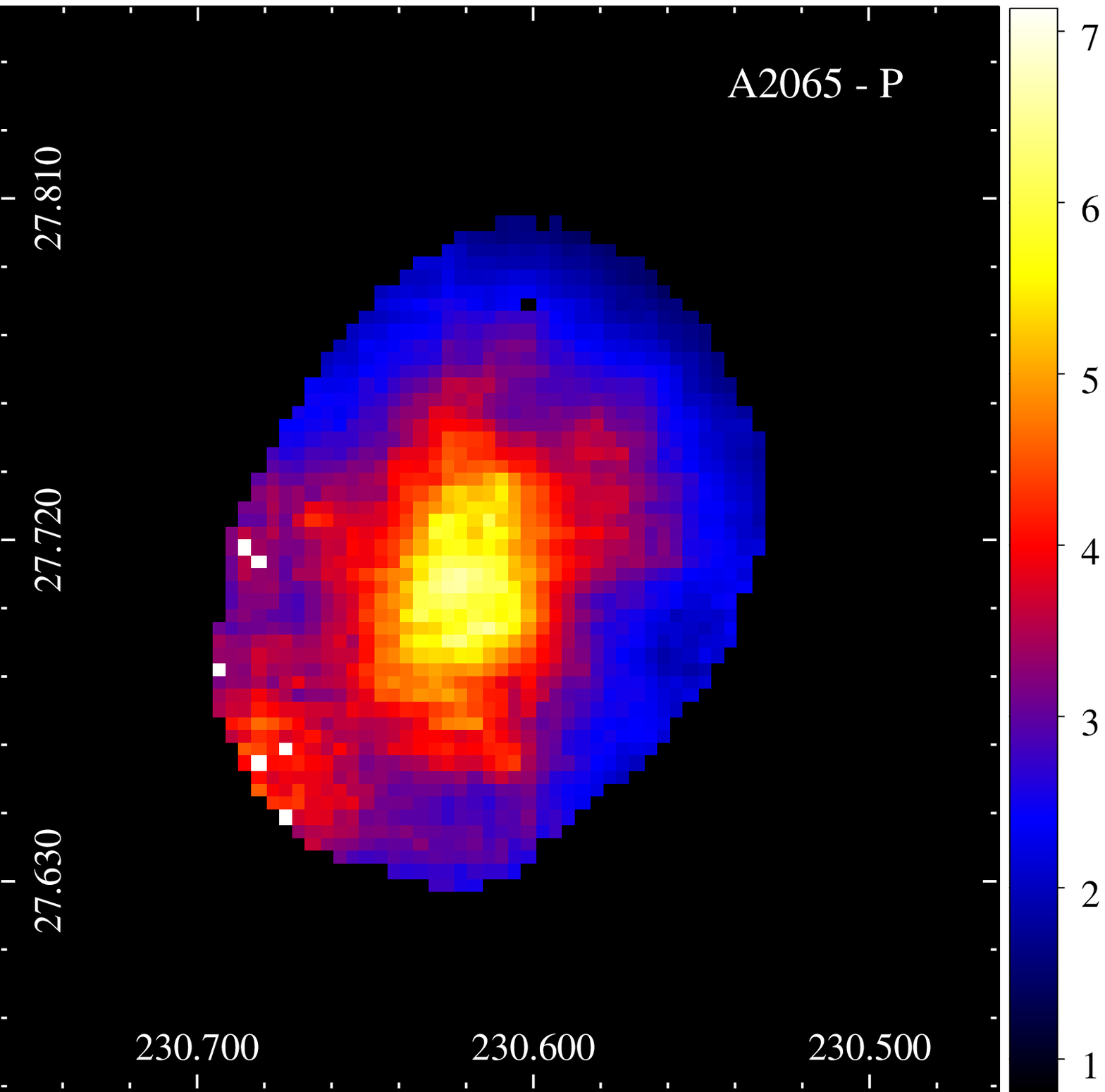}
\includegraphics[scale=0.25]{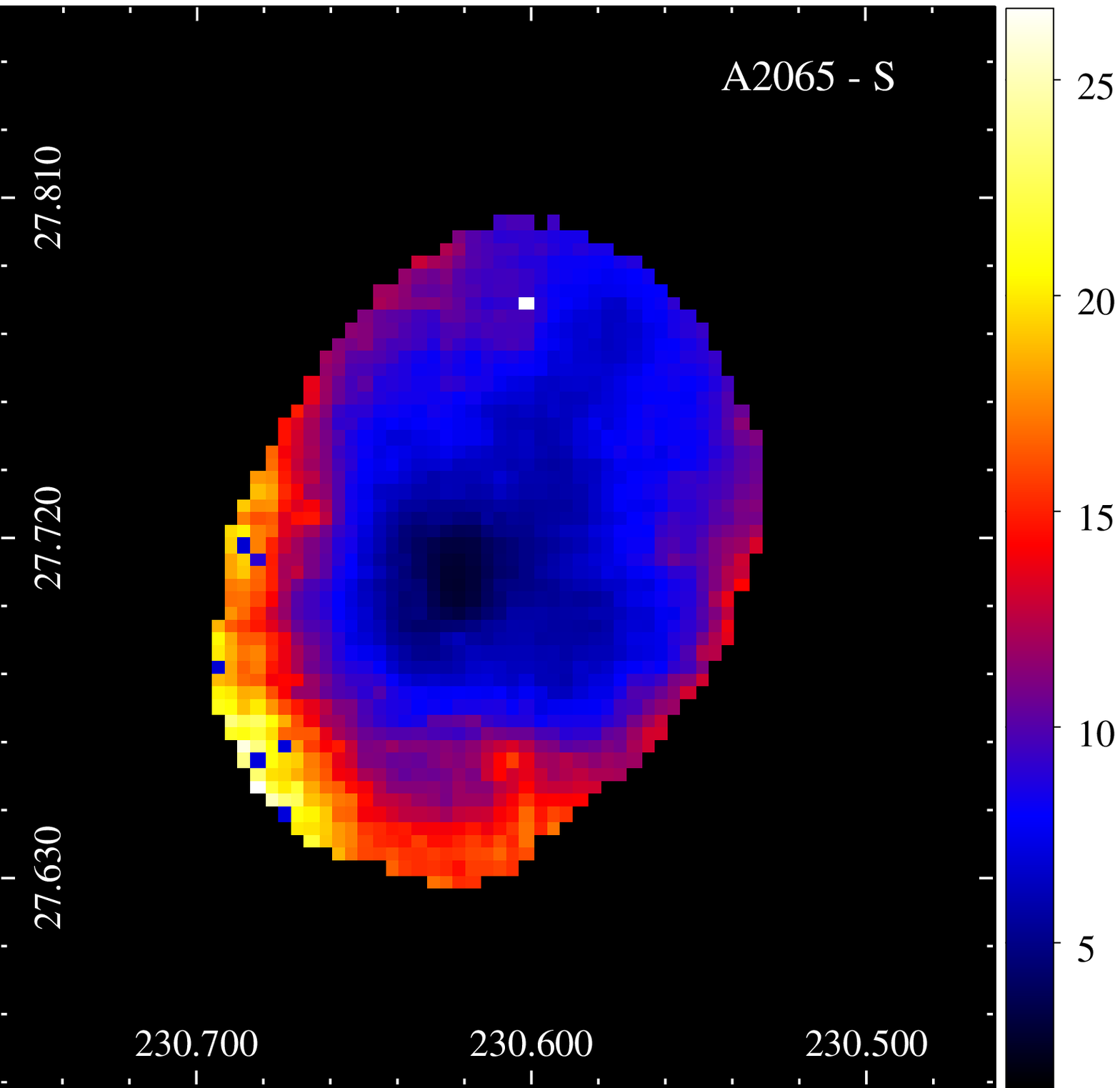}
\includegraphics[scale=0.25]{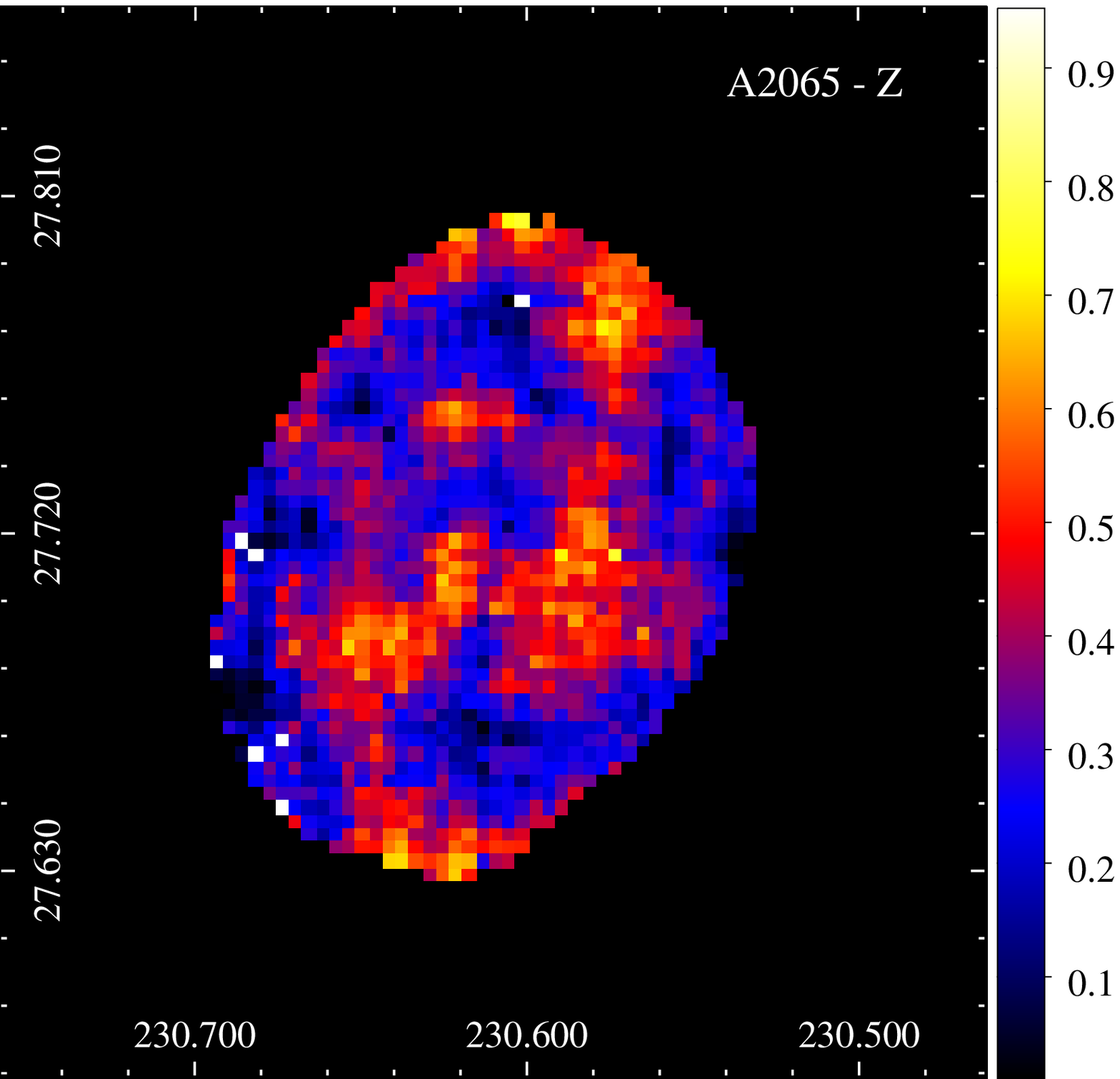}

\includegraphics[scale=0.25]{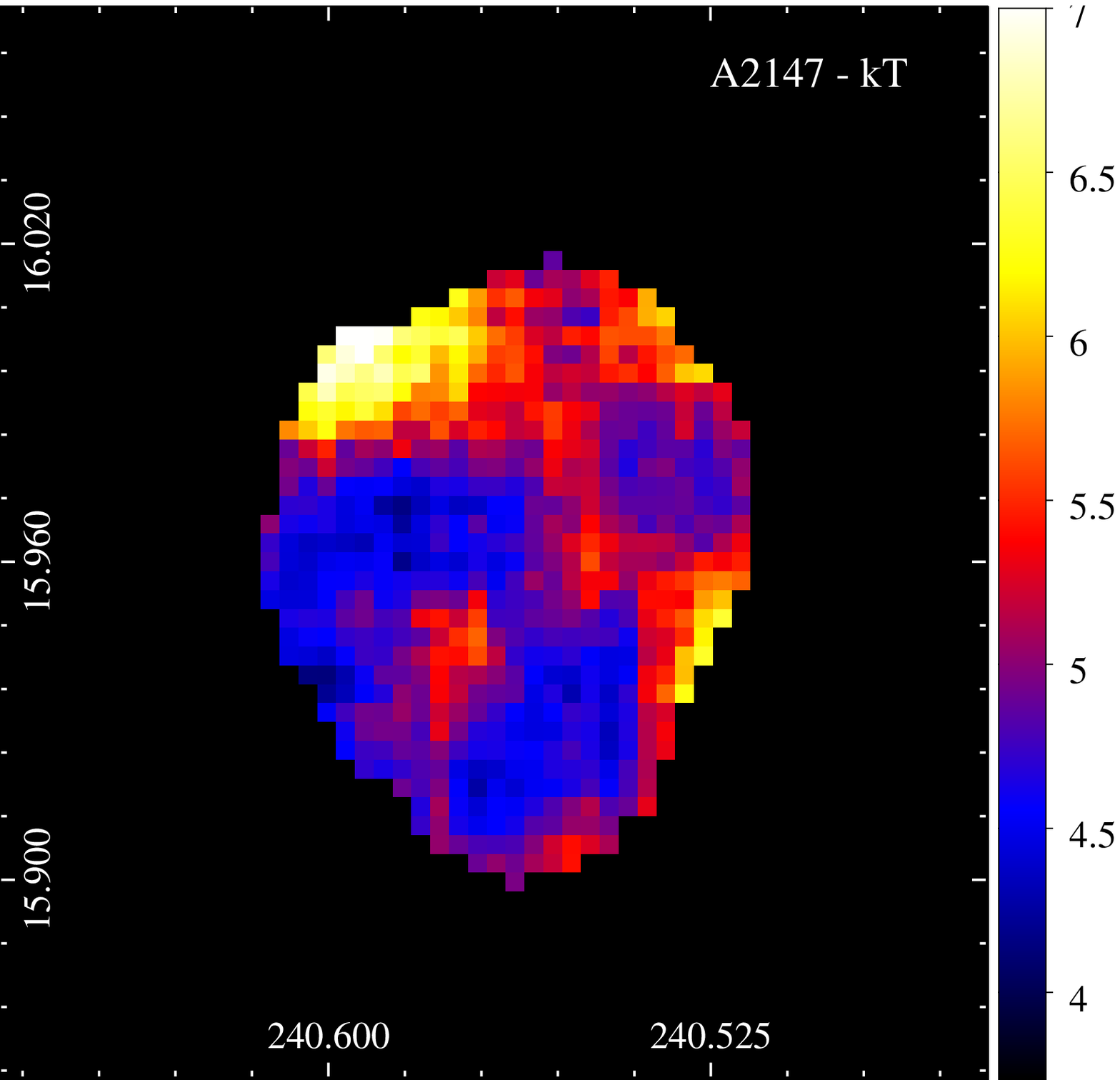}
\includegraphics[scale=0.25]{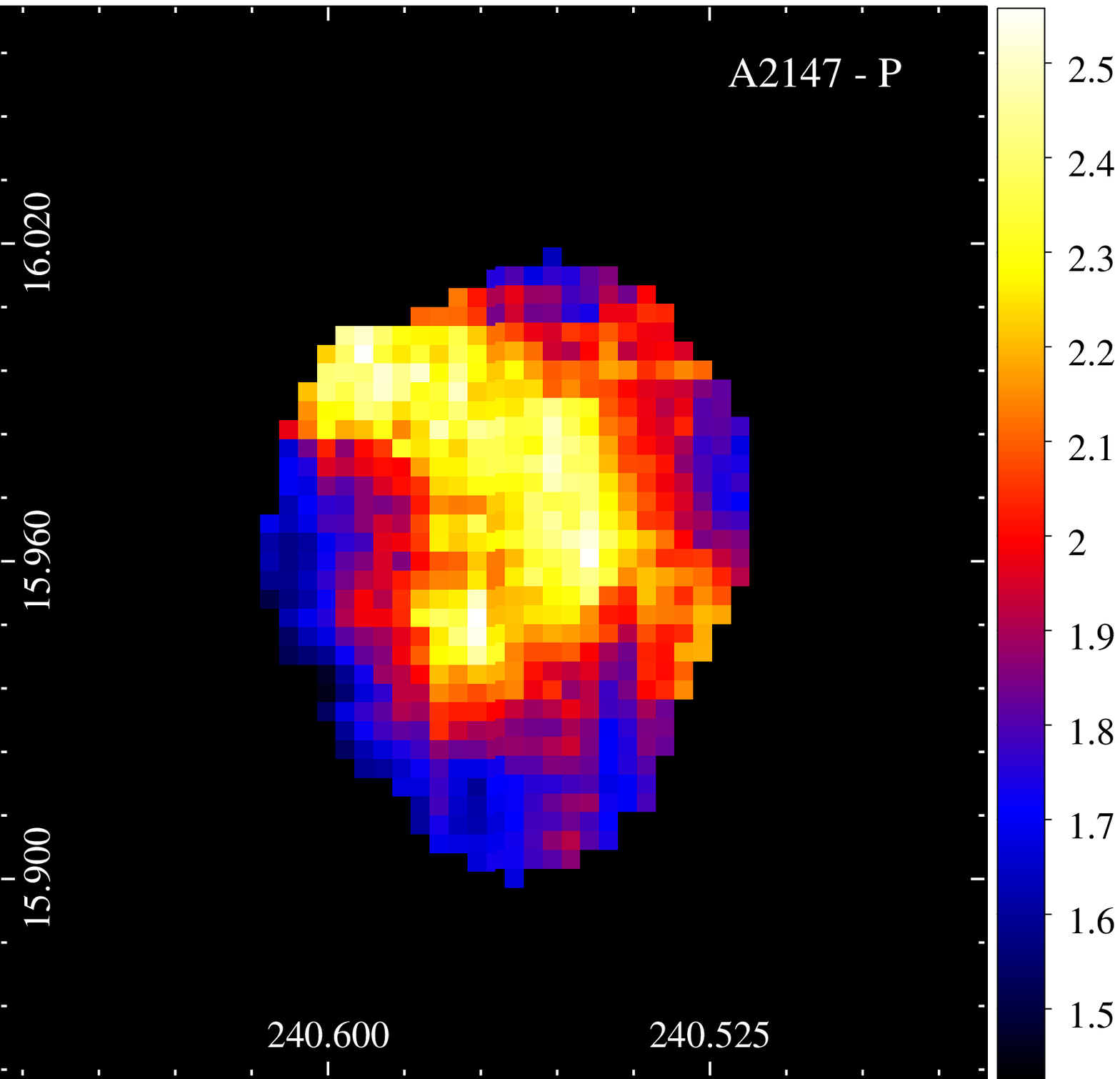}
\includegraphics[scale=0.25]{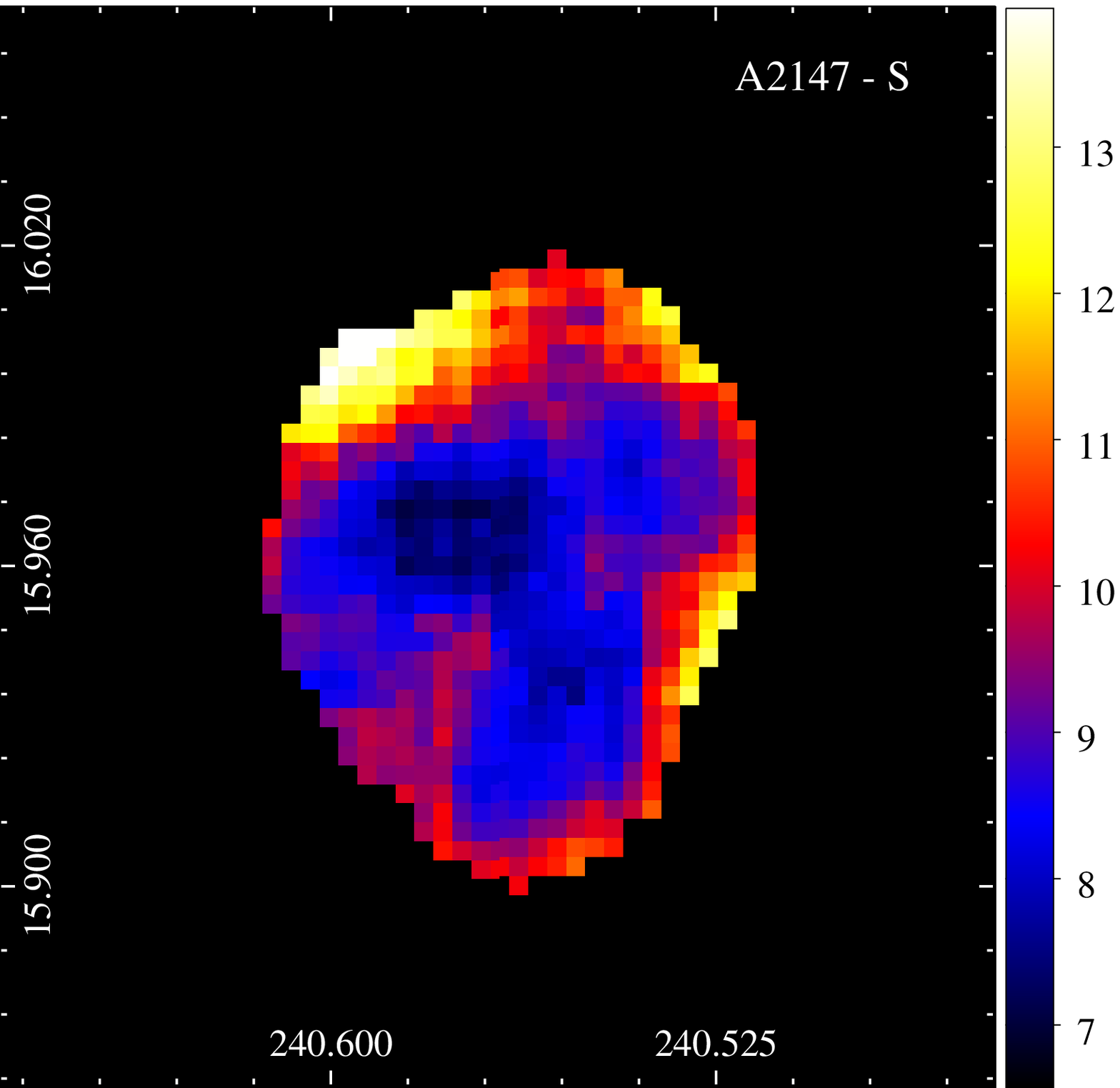}
\includegraphics[scale=0.25]{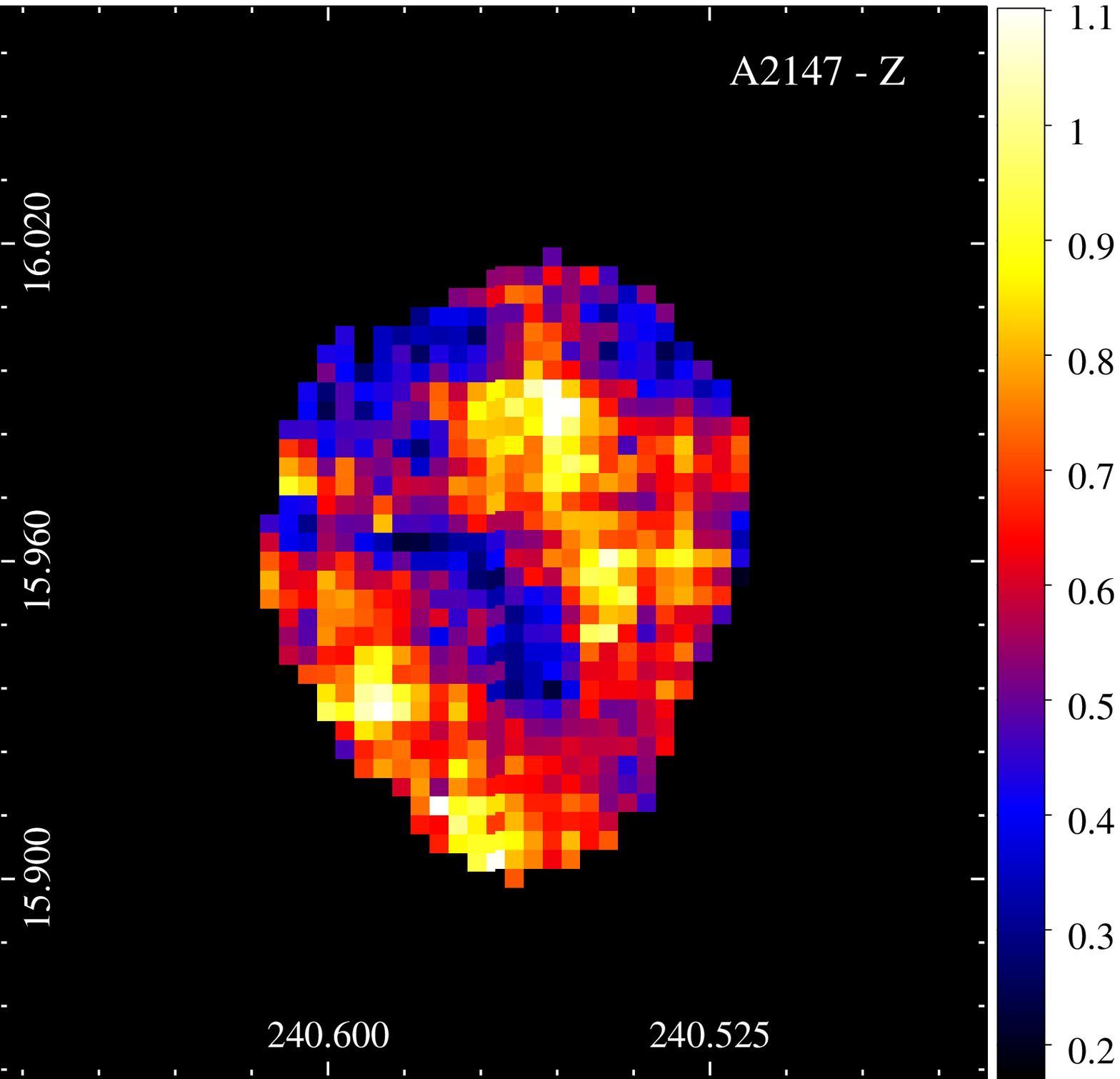}

\includegraphics[scale=0.25]{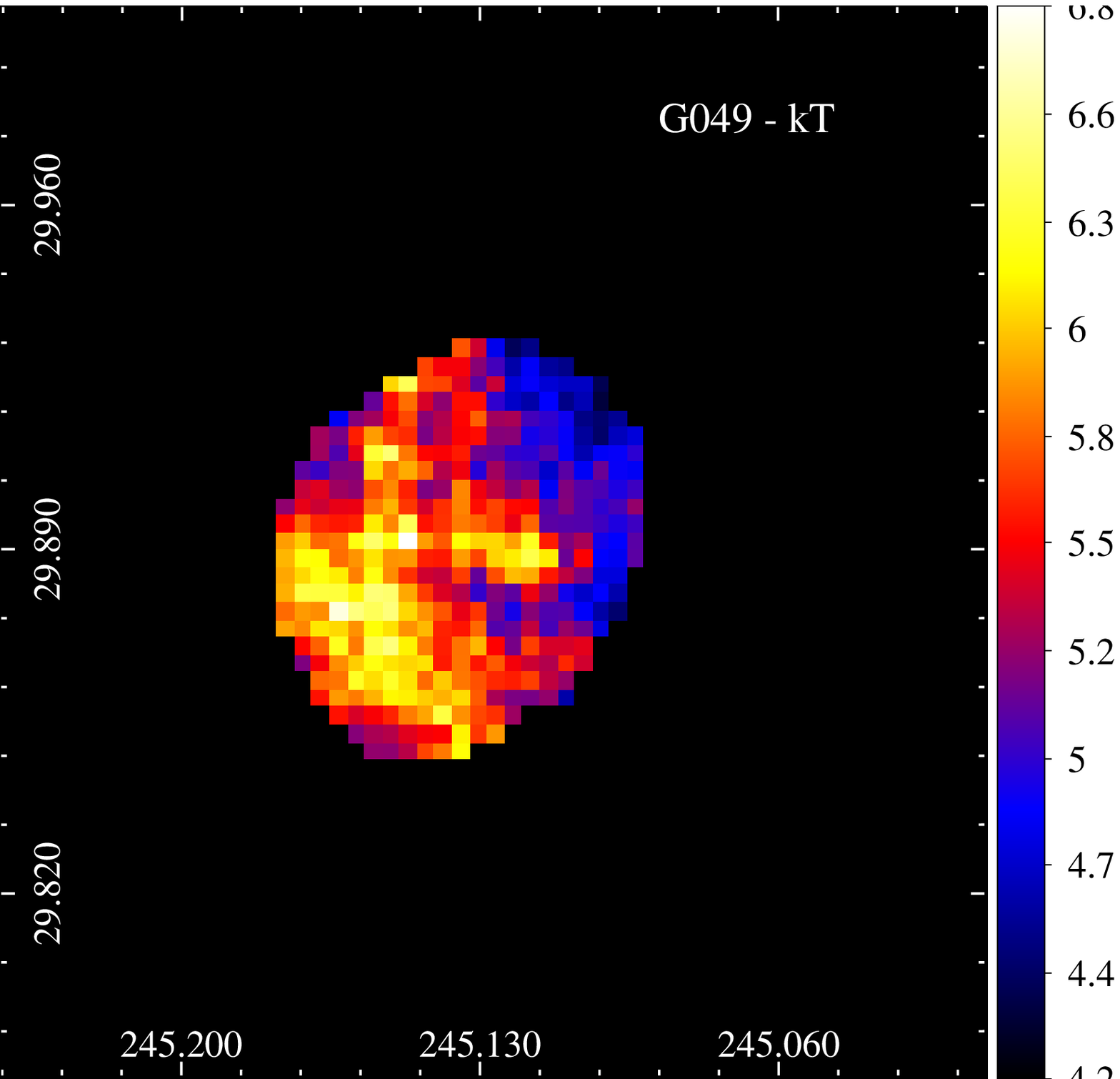}
\includegraphics[scale=0.25]{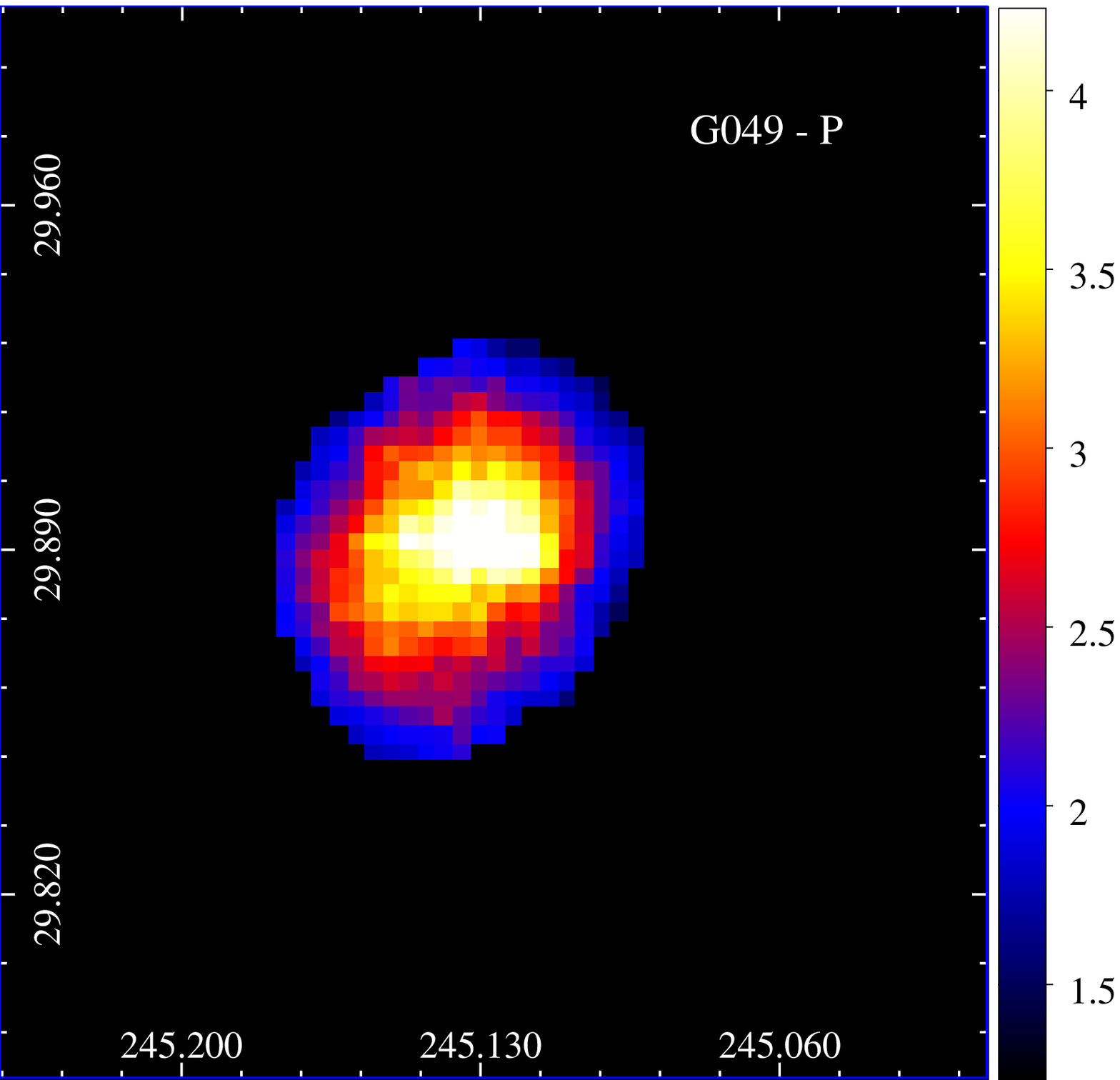}
\includegraphics[scale=0.25]{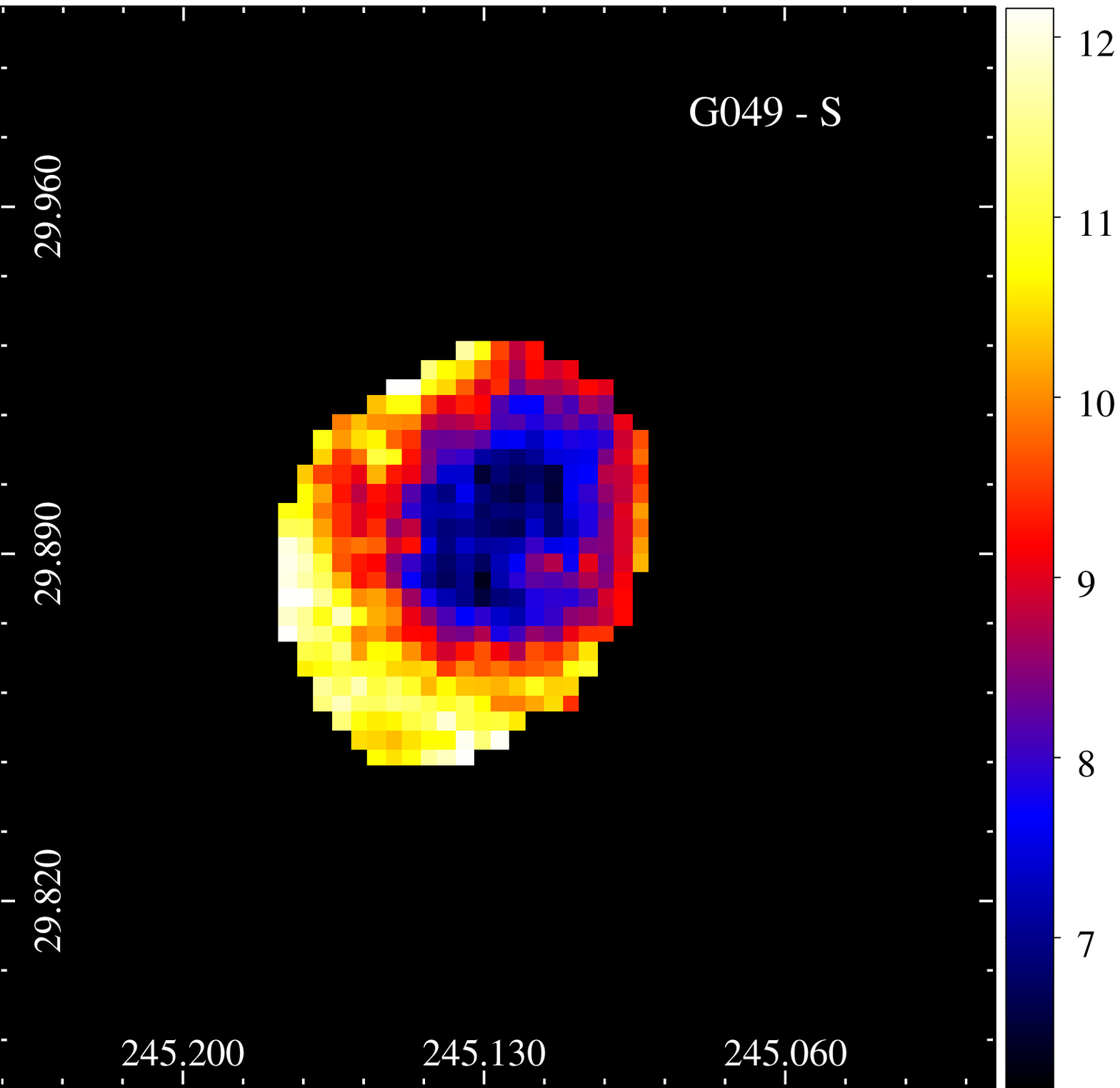}
\includegraphics[scale=0.25]{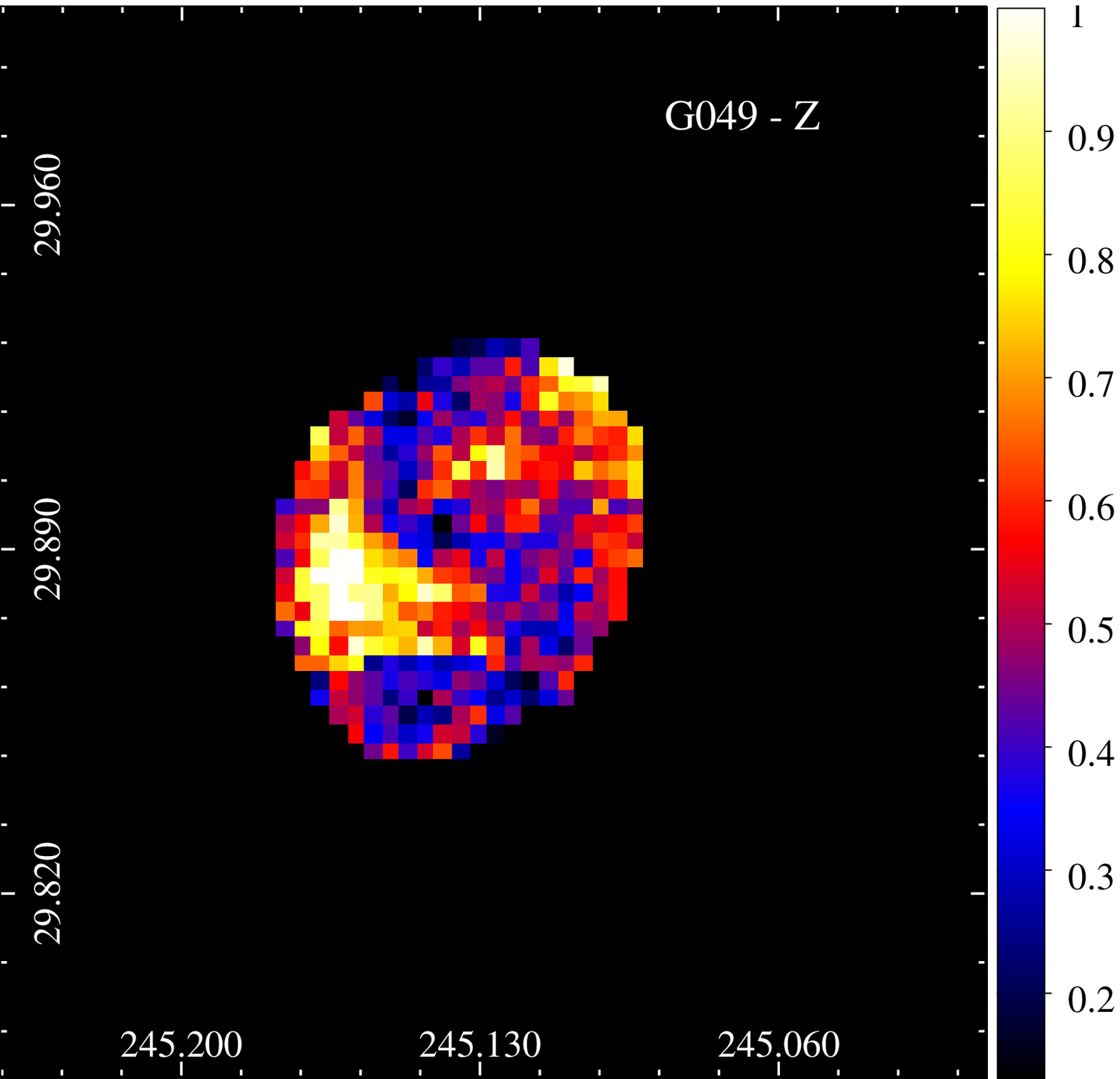}

\includegraphics[scale=0.25]{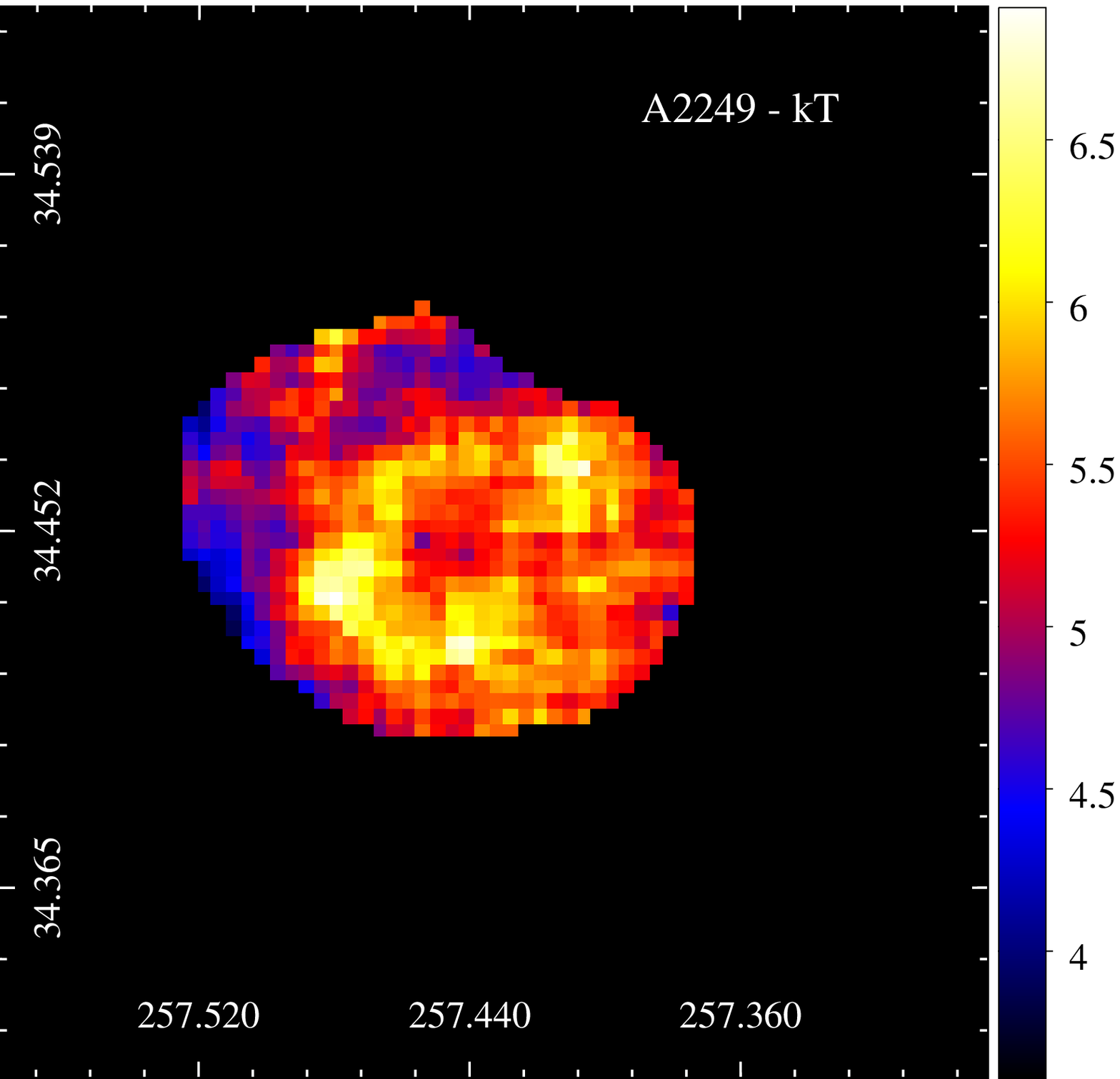}
\includegraphics[scale=0.25]{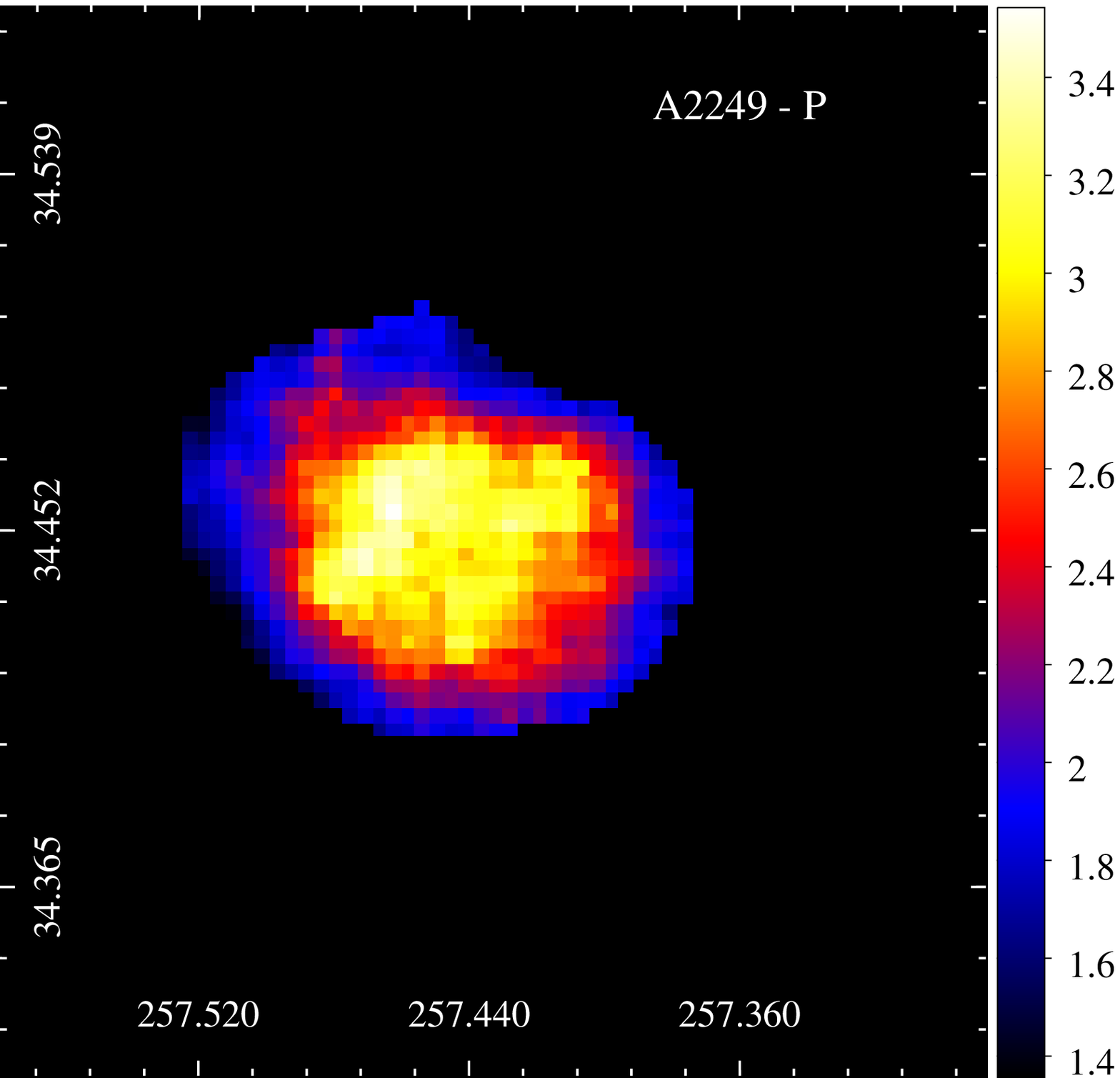}
\includegraphics[scale=0.25]{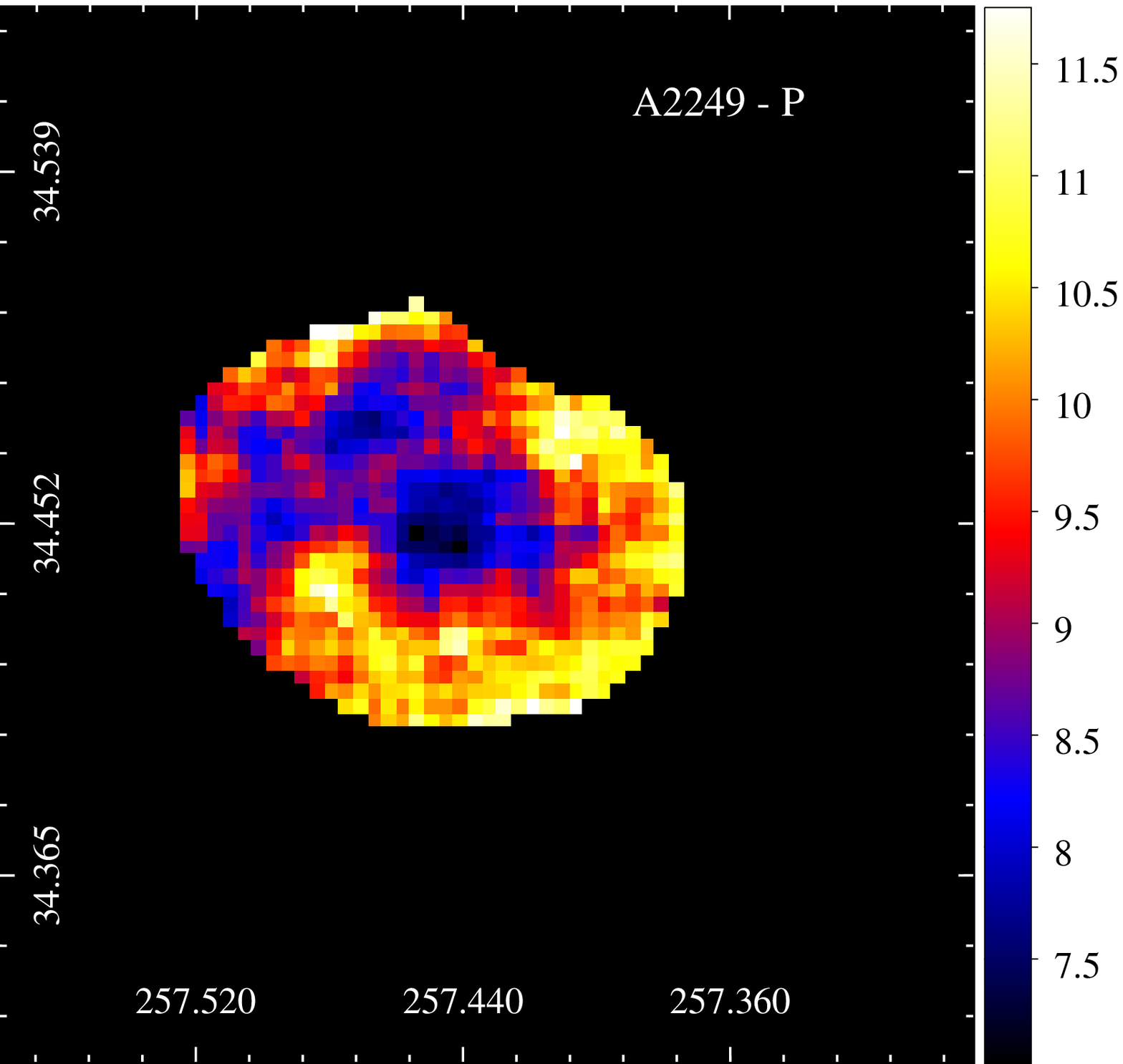}
\includegraphics[scale=0.25]{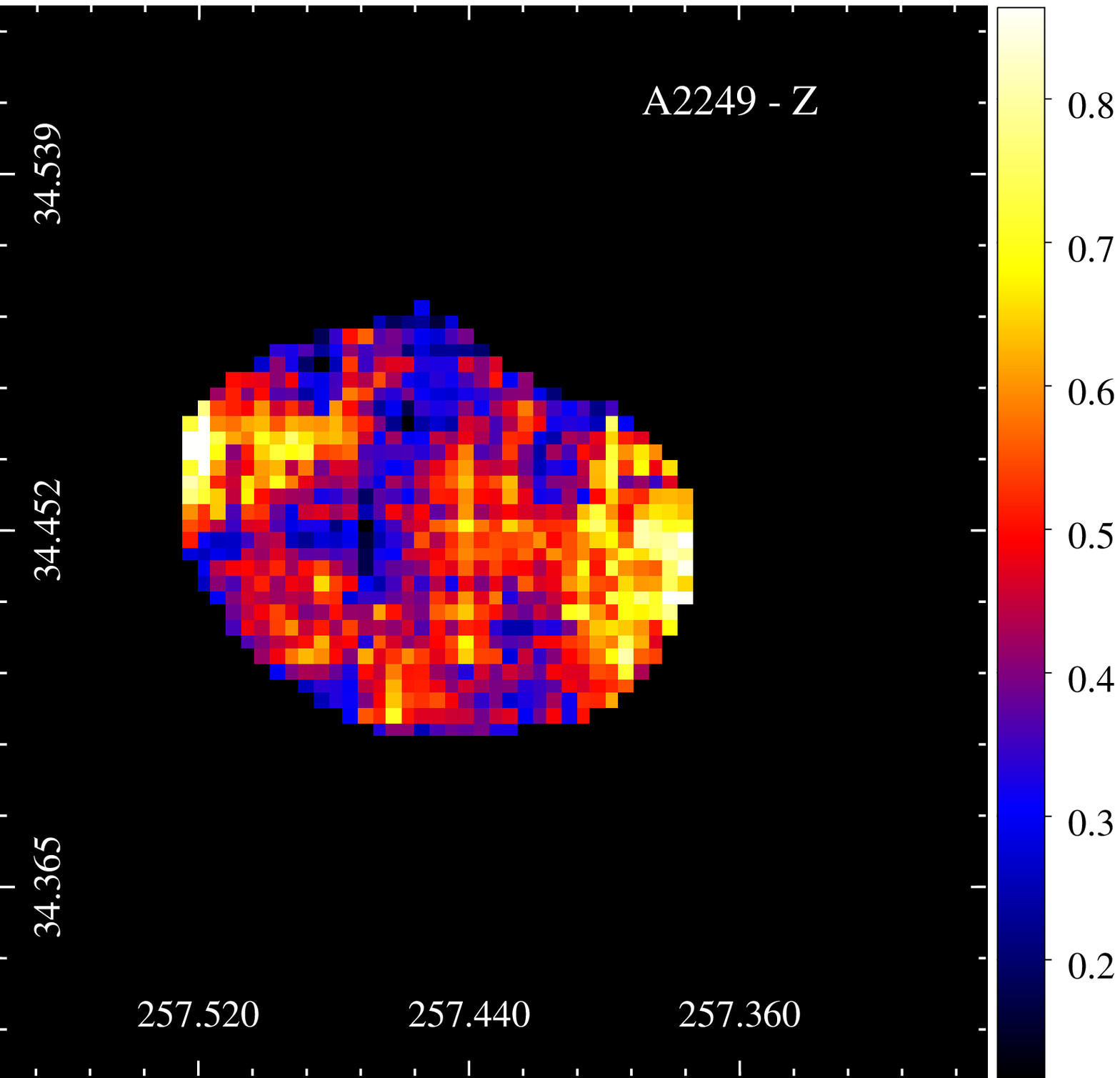}

\includegraphics[scale=0.25]{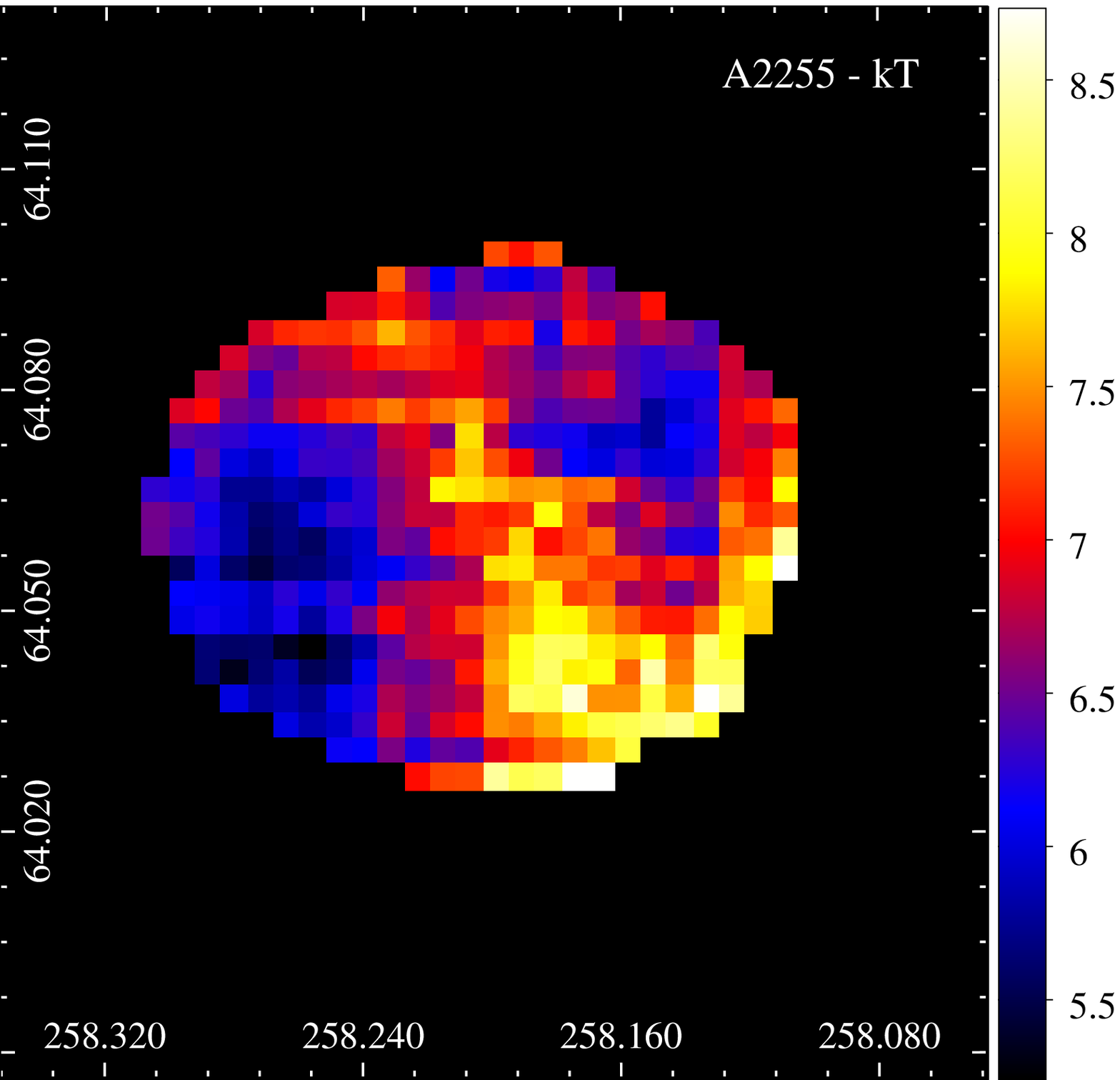}
\includegraphics[scale=0.25]{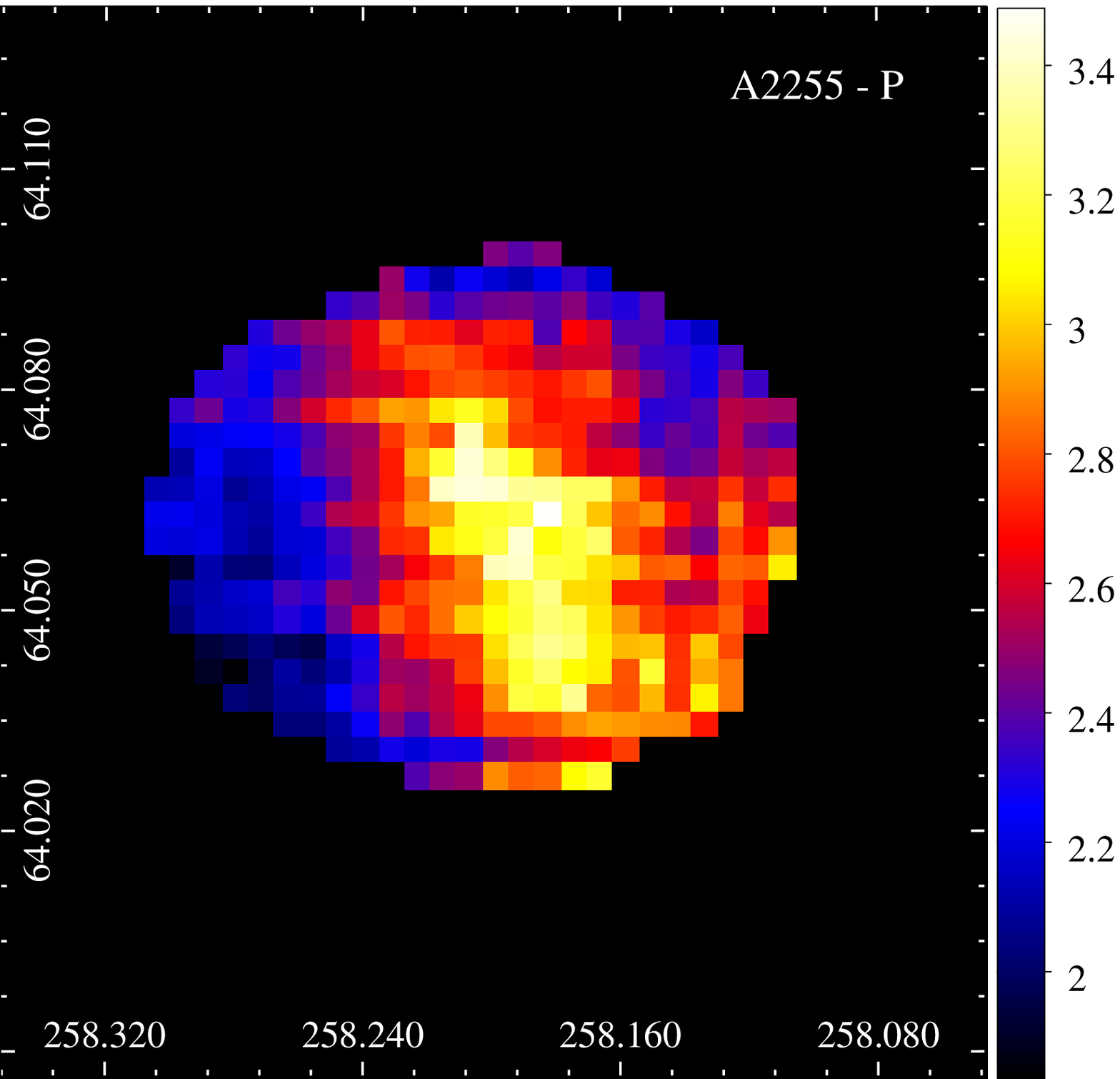}
\includegraphics[scale=0.25]{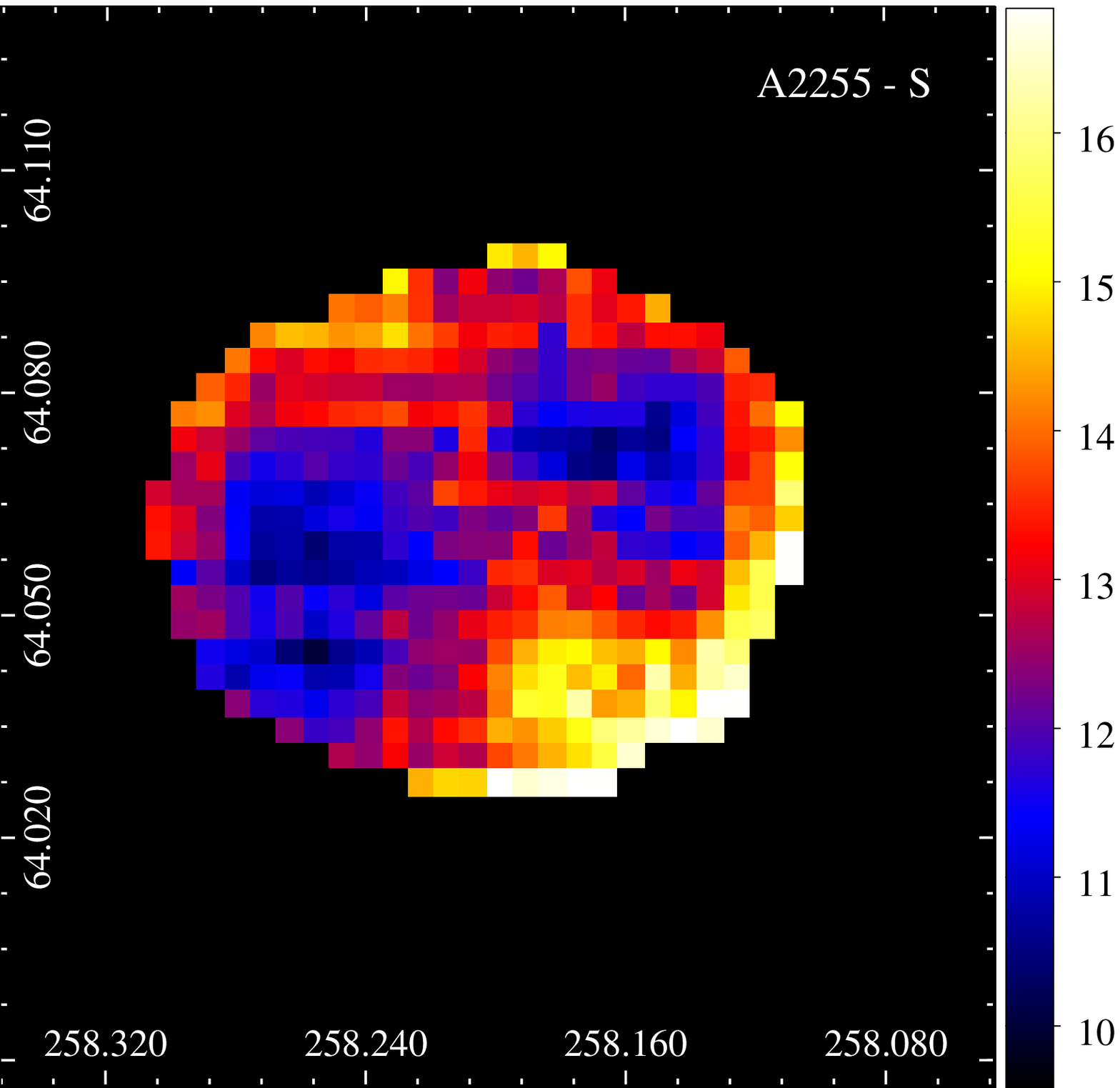}
\includegraphics[scale=0.25]{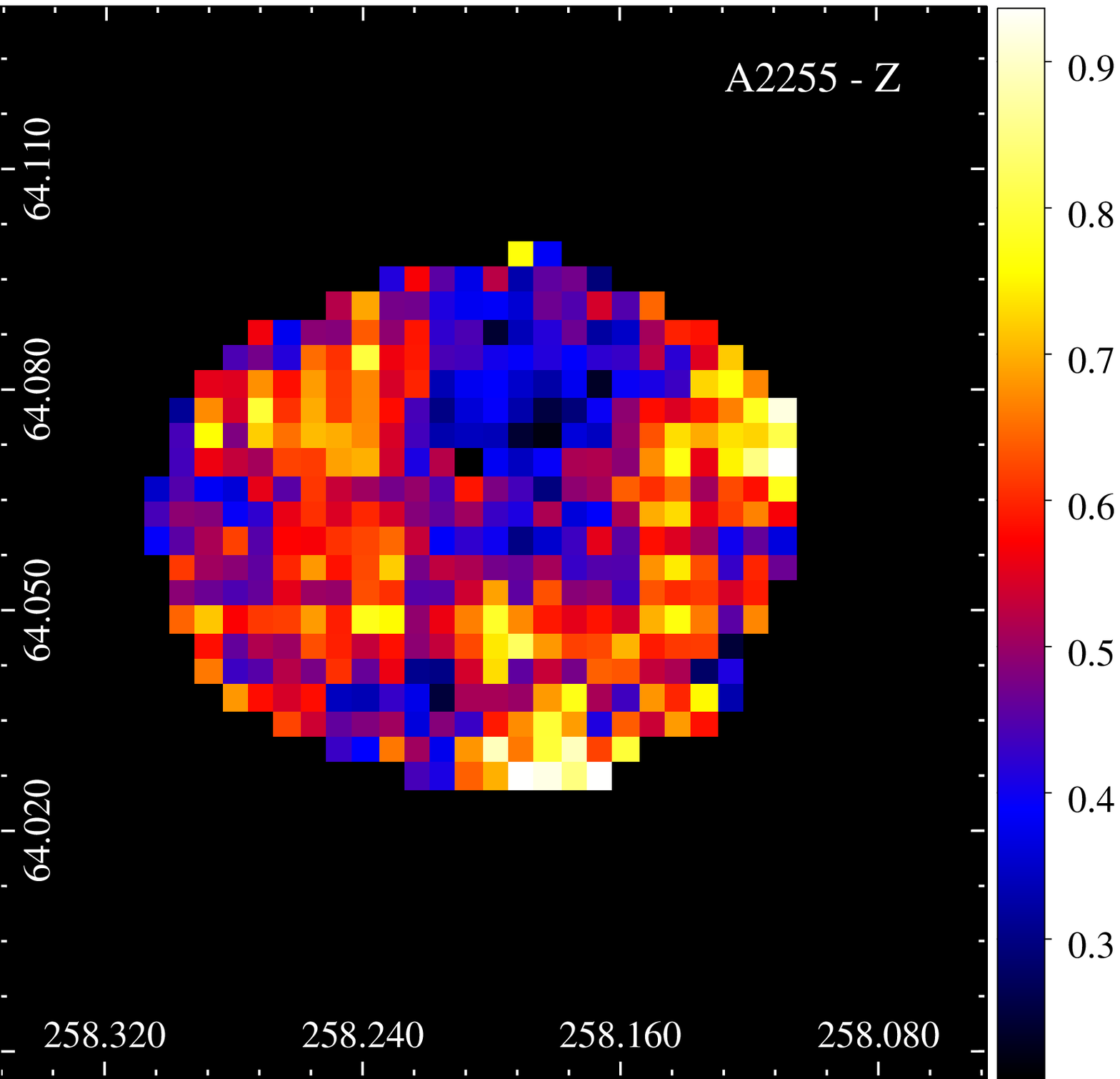}

\caption{NCC and disturbed systems. From left to right: temperature,
  pseudo-pressure, pseudo-entropy, and metallicity maps for 
  A2065, A2147, G049.33+44.38, A2249, and A2255.}
\label{fig:NCCclusters3}
\end{figure*}

\section{Discussion}
\label{sect:disc}

Based on the six criteria to
classify a system as CC, and on our 2D maps,  we discuss our results in this Section, and analyse the dynamical
states of the clusters. We give specific details for each of the 53
systems in Appendix~\ref{sect:apNotes}.

There are 17 {\scshape CC-relaxed} systems, for which the 2D maps are
shown in Figs.~\ref{fig:CCclusters}, ~\ref{fig:CCclusters2}, and
~\ref{fig:CCclusters3}.  Four of them
(G269, A3571, A2151, and A2572, ), although classified as CC, don't
show a cool centre in the kT map. For many of these clusters, the
cool centre shows a deviation from sphericity. In fact, there are only
five clusters that show almost spherical CC: A2734, A3528, A2052,
A2626, and A4059. However, looking at their P maps, A2734 and A2052
show signs of interaction. A2734 shows some elongation that is
reported in \citet{Ramella+07}. The pressure map of A2052 shows a high
pressure feature in its centre, very similar to the one presented in
\citet{Blanton11}, that may be a shock region due to the central
AGN. We highlight that in about 70\% of the local CC population the BCG is radio loud, showing non-thermal radio-jets
ejected by the central AGN, which is critical to understand the physics of the central 
region due to perturbation.  The very central region shows clear signs of perturbation, but its overall dynamics are relaxed.

There are 16 {\scshape CC-disturbed} clusters.  
The maps of these clusters are shown in
Figs.~\ref{fig:groupsA85}, ~\ref{fig:groupsA85b}, and ~\ref{fig:groupsA85c}.  
These  {\scshape CC-disturbed} clusters are very representative of
the details that 2D maps can reveal.  They represent  $\sim$30\% of
the sample, were characterised as cool-core, and indeed they have a cooler centre, 
but when analysing their global distribution, the 2D maps show that they have a very
complex structure for clusters thought to be globally relaxed, in line with recent 
numerical simulations from IllustrisTNG \citep{Barnes18}.

We classify these systems as ``disturbed'', and the perturbation is related to merger events and not to AGN feedback.
All of them except A2029 are undergoing a merger, are double systems, or have an infalling group.
For example, A85 is not spherically symmetric, and
\citet{Durret05} through temperature and metallicity maps, had already
shown results of an intense merging activity in the past, revealing
that this cluster is not relaxed, even in the central region.
A496 has a sloshing spiral arm
\citep{Ghizzardi14,Lagana10,Roediger12}, indicating a minor merger.
The temperature and entropy maps of A1775 also show a cold front as
the one reported for A496.  A1644 is a double cluster, with a main
cluster (A1644s) and a smaller and colder one (A1644n) to the
north-east \citep{Johnson+10}.  From the comparison with
hydrodynamical simulations, these authors suggest that the northern
subcluster initiated the core gas sloshing in the main cluster about
700~Myr ago.  These 16 systems in our sample show clear signs of  recent mergers,
with complex structure and geometry  but with a CC preserved. 
We give specific details for each of them, 
reporting merger events in Appendix~\ref{sect:apNotes}.

 This illustrates well the power
of 2D X-ray maps to understand in depth the dynamical state of
clusters, that can be missed in CC diagnosis tools. For merging
clusters, it is possible to go one step further and derive the
physical characteristics of the cluster merger (Mach number, age of
the merger) by comparing these maps with hydrodynamical numerical
simulations. However, to explain well the dynamical history of a
cluster, it is necessary to make specific simulations for each
object, which is very time-consuming.  This was done for
several clusters by \citet{MachadoLimaNeto13,Machado15}. In a few other
cases (see for example the discussion on A85 in the Appendix) there
happen to exist simulations that resemble our maps and allow us to
interpret the cluster merging history. However, a much larger number
of simulations than presently available is needed to account for
the variety of properties observed in this large sample. This will
be the topic of a future paper (Machado et al. in preparation). A
detailed analysis of the dynamical properties of a subsample of these
clusters with a large number of galaxy redshifts available is
also under development (Biviano et al. in preparation).

%This result goes in line with recent numerical simulations from IllustrisTNG \citep{Barnes18}.
%Examining the fraction of CCs and NCCs defined as relaxed by six common CC definitions, 
%these authors found no evidence that CC clusters are more relaxed. 
%In the merging process, the kinetic energy carried by the gas of the colliding system dissipates into thermal energy via shocks and turbulence. 
%SOMETHING SHOULD GOES HERE

There are only four {\scshape NCC-relaxed} clusters: Coma (G057.33+88.01),
A3532, bA3558 and MKW8, shown in Fig.~\ref{fig:NCCrelaxed}. They were classified as non cool-core systems
but their maps don't show any special feature indicating strong signs of perturbation.
For two of them (A3532 and bA3558) the observations were not deep enough to produce extended maps, so 
we cannot draw firm conclusions on these clusters.

There are 16 systems classified as {\scshape NCC-disturbed}, that are
presented in the last part of Tab.~\ref{tab:CC}.  We show 2D maps for
these clusters in Figs.~\ref{fig:NCCclusters1},
\ref{fig:NCCclusters2}, and \ref{fig:NCCclusters3}. 
They are irregular
in shape, elongated, and show clear signs of interaction.

For seven out of these 16 clusters, at least five criteria classify them as NCC
(A119, A3376, A3560, A2061, A2065, A2147 and A2255).  All of
these clusters are merging systems and our maps clearly confirm that
these clusters are disturbed and have undergone or are presently
undergoing one or several merging events.  However, for the other nine
clusters, at least one of the criteria give a misleading
classification, and for two of them (A754 and MKW3),  three
criteria (half of the six diagnoses considered in this work) wrongly classify them as CC.

\section{Conclusions}
\label{sect:conc}

In this section we summarize our findings.

\begin{itemize}

\item With the aim of comparing the overall dynamical and core properties of a sample of 53 galaxy clusters, the cluster dynamical 
state was investigated via ICM temperature, pressure, entropy and metallicity maps;

\item We compared six simple CC diagnoses with the results derived from our 2D maps, showing that, although very useful for large samples
and for CC characterisation, these diagnoses are somewhat too simplistic to account for the overall cluster dynamics;

\item We highlight that 2D maps reveal the detailed and complex structure of galaxy clusters that can be missed by CC diagnoses
and may affect cluster mass estimates. That is clearly seen in the 16 CC-disturbed systems we discuss in Sect.~\ref{sect:disc}.

\item An X-ray analysis provides a unique possibility to measure in detail the ICM showing complex structures;

%\item Observational  properties  arise  from  a  non-trivial  interplay  between gravitational processes, where a
% number of astrophysical processes that take place on smaller scales (AGN heating);

\item 2D maps reveal some merging galaxy clusters (A2061, A2255, NGC6338)  with shock fronts clearly visible in the sky plane 
that will be addressed in future numerical simulations with a well defined merger geometry;

\item ICM entropy  and pressure maps are of great interest because they reveal ICM global properties and record the 
thermal history of clusters. They are therefore useful quantities for studying the 
effects of feedback on the cluster environment.

\end{itemize}

\section*{Acknowledgements}
We thank the referee for his/her comments on the manuscript.
T. F. L. acknowledges financial support from FAPESP and CNPq through grants 2018/02626-8 and 303278/2015-3, respectively.
F.D. acknowledges constant support from CNES. PAAL thanks the support of CNPq, grant 308969/2014-6; and CAPES, process number 88881.120856/2016-01.

%%%%%%%%%%%%%%%%%%%%%%%%%%%%%%%%%%%%%%%%%%%%%%%%%%



%\bibliographystyle{mnras}

\bibliographystyle{mnras} 
\include{adsjournalnames} 
\bibliography{refs} 

\begin{thebibliography}{}
\makeatletter
\relax
\def\mn@urlcharsother{\let\do\@makeother \do\$\do\&\do\#\do\^\do\_\do\%\do\~}
\def\mn@doi{\begingroup\mn@urlcharsother \@ifnextchar [ {\mn@doi@}
  {\mn@doi@[]}}
\def\mn@doi@[#1]#2{\def\@tempa{#1}\ifx\@tempa\@empty \href
  {http://dx.doi.org/#2} {doi:#2}\else \href {http://dx.doi.org/#2} {#1}\fi
  \endgroup}
\def\mn@eprint#1#2{\mn@eprint@#1:#2::\@nil}
\def\mn@eprint@arXiv#1{\href {http://arxiv.org/abs/#1} {{\tt arXiv:#1}}}
\def\mn@eprint@dblp#1{\href {http://dblp.uni-trier.de/rec/bibtex/#1.xml}
  {dblp:#1}}
\def\mn@eprint@#1:#2:#3:#4\@nil{\def\@tempa {#1}\def\@tempb {#2}\def\@tempc
  {#3}\ifx \@tempc \@empty \let \@tempc \@tempb \let \@tempb \@tempa \fi \ifx
  \@tempb \@empty \def\@tempb {arXiv}\fi \@ifundefined
  {mn@eprint@\@tempb}{\@tempb:\@tempc}{\expandafter \expandafter \csname
  mn@eprint@\@tempb\endcsname \expandafter{\@tempc}}}

\bibitem[\protect\citeauthoryear{{Akamatsu}, {Hoshino}, {Ishisaki}, {Ohashi},
  {Sato}, {Takei}  \& {Ota}}{{Akamatsu} et~al.}{2011}]{Akamatsu+11}
{Akamatsu} H.,  {Hoshino} A.,  {Ishisaki} Y.,  {Ohashi} T.,  {Sato} K.,
  {Takei} Y.,   {Ota} N.,  2011, \mn@doi [\pasj] {10.1093/pasj/63.sp3.S1019},
  \href {http://adsabs.harvard.edu/abs/2011PASJ...63S1019A} {63, S1019}

\bibitem[\protect\citeauthoryear{{Akamatsu} et~al.,}{{Akamatsu}
  et~al.}{2017}]{Akamatsu17}
{Akamatsu} H.,  et~al., 2017, \mn@doi [\aap] {10.1051/0004-6361/201628400},
  \href {http://adsabs.harvard.edu/abs/2017A%26A...600A.100A} {600, A100}

\bibitem[\protect\citeauthoryear{{Alvarez}, {Randall}, {Bourdin}, {Jones}  \&
  {Holley-Bockelmann}}{{Alvarez} et~al.}{2018}]{alv18}
{Alvarez} G.~E.,  {Randall} S.~W.,  {Bourdin} H.,  {Jones} C.,
  {Holley-Bockelmann} K.,  2018, \mn@doi [\apj] {10.3847/1538-4357/aabad0},
  \href {http://adsabs.harvard.edu/abs/2018ApJ...858...44A} {858, 44}

\bibitem[\protect\citeauthoryear{{Andrade-Santos}, {Lima Neto}  \&
  {Lagan{\'a}}}{{Andrade-Santos} et~al.}{2012}]{AndradeSantos12}
{Andrade-Santos} F.,  {Lima Neto} G.~B.,   {Lagan{\'a}} T.~F.,  2012, \mn@doi
  [\apj] {10.1088/0004-637X/746/2/139}, \href
  {http://adsabs.harvard.edu/abs/2012ApJ...746..139A} {746, 139}

\bibitem[\protect\citeauthoryear{{Andrade-Santos} et~al.,}{{Andrade-Santos}
  et~al.}{2017}]{AndradeSantos17}
{Andrade-Santos} F.,  et~al., 2017, \mn@doi [\apj] {10.3847/1538-4357/aa7461},
  \href {http://adsabs.harvard.edu/abs/2017ApJ...843...76A} {843, 76}

\bibitem[\protect\citeauthoryear{{Angulo}, {Springel}, {White}, {Jenkins},
  {Baugh}  \& {Frenk}}{{Angulo} et~al.}{2012}]{Angulo12}
{Angulo} R.~E.,  {Springel} V.,  {White} S.~D.~M.,  {Jenkins} A.,  {Baugh}
  C.~M.,   {Frenk} C.~S.,  2012, \mn@doi [\mnras]
  {10.1111/j.1365-2966.2012.21830.x}, \href
  {http://adsabs.harvard.edu/abs/2012MNRAS.426.2046A} {426, 2046}

\bibitem[\protect\citeauthoryear{{Arnaud} et~al.,}{{Arnaud}
  et~al.}{2001}]{Arnaud+01}
{Arnaud} M.,  et~al., 2001, \mn@doi [\aap] {10.1051/0004-6361:20000195}, \href
  {http://adsabs.harvard.edu/abs/2001A%26A...365L..67A} {365, L67}

\bibitem[\protect\citeauthoryear{{Arnaud}, {Pratt}, {Piffaretti},
  {B{\"o}hringer}, {Croston}  \& {Pointecouteau}}{{Arnaud}
  et~al.}{2010}]{Arnaud10}
{Arnaud} M.,  {Pratt} G.~W.,  {Piffaretti} R.,  {B{\"o}hringer} H.,  {Croston}
  J.~H.,   {Pointecouteau} E.,  2010, \mn@doi [\aap]
  {10.1051/0004-6361/200913416}, \href
  {http://adsabs.harvard.edu/abs/2010A%26A...517A..92A} {517, A92}

\bibitem[\protect\citeauthoryear{{Asplund}, {Grevesse}, {Sauval}  \&
  {Scott}}{{Asplund} et~al.}{2009}]{Asplund09}
{Asplund} M.,  {Grevesse} N.,  {Sauval} A.~J.,   {Scott} P.,  2009, \mn@doi
  [\araa] {10.1146/annurev.astro.46.060407.145222}, \href
  {http://adsabs.harvard.edu/abs/2009ARA%26A..47..481A} {47, 481}

\bibitem[\protect\citeauthoryear{{Bagchi}, {Durret}, {Neto}  \&
  {Paul}}{{Bagchi} et~al.}{2006}]{Bagchi+06}
{Bagchi} J.,  {Durret} F.,  {Neto} G.~B.~L.,   {Paul} S.,  2006, \mn@doi
  [Science] {10.1126/science.1131189}, \href
  {http://adsabs.harvard.edu/abs/2006Sci...314..791B} {314, 791}

\bibitem[\protect\citeauthoryear{{Balucinska-Church} \&
  {McCammon}}{{Balucinska-Church} \& {McCammon}}{1992}]{Balucinska92}
{Balucinska-Church} M.,  {McCammon} D.,  1992, \mn@doi [\apj] {10.1086/172032},
  \href {http://adsabs.harvard.edu/abs/1992ApJ...400..699B} {400, 699}

\bibitem[\protect\citeauthoryear{{Bardelli}, {Venturi}, {Zucca}, {De Grandi},
  {Ettori}  \& {Molendi}}{{Bardelli} et~al.}{2002}]{Bardelli02}
{Bardelli} S.,  {Venturi} T.,  {Zucca} E.,  {De Grandi} S.,  {Ettori} S.,
  {Molendi} S.,  2002, \mn@doi [\aap] {10.1051/0004-6361:20021366}, \href
  {http://adsabs.harvard.edu/abs/2002A%26A...396...65B} {396, 65}

\bibitem[\protect\citeauthoryear{{Barnes} et~al.,}{{Barnes}
  et~al.}{2018}]{Barnes18}
{Barnes} D.~J.,  et~al., 2018, \mn@doi [\mnras] {10.1093/mnras/sty2078}, \href
  {http://adsabs.harvard.edu/abs/2018MNRAS.481.1809B} {481, 1809}

\bibitem[\protect\citeauthoryear{{Bauer}, {Fabian}, {Sanders}, {Allen}  \&
  {Johnstone}}{{Bauer} et~al.}{2005}]{Bauer05}
{Bauer} F.~E.,  {Fabian} A.~C.,  {Sanders} J.~S.,  {Allen} S.~W.,   {Johnstone}
  R.~M.,  2005, \mn@doi [\mnras] {10.1111/j.1365-2966.2005.08999.x}, \href
  {http://adsabs.harvard.edu/abs/2005MNRAS.359.1481B} {359, 1481}

\bibitem[\protect\citeauthoryear{{Blanton}, {Randall}, {Clarke}, {Sarazin},
  {McNamara}, {Douglass}  \& {McDonald}}{{Blanton} et~al.}{2011}]{Blanton11}
{Blanton} E.~L.,  {Randall} S.~W.,  {Clarke} T.~E.,  {Sarazin} C.~L.,
  {McNamara} B.~R.,  {Douglass} E.~M.,   {McDonald} M.,  2011, \mn@doi [\apj]
  {10.1088/0004-637X/737/2/99}, \href
  {http://adsabs.harvard.edu/abs/2011ApJ...737...99B} {737, 99}

\bibitem[\protect\citeauthoryear{{B{\"o}hringer} et~al.,}{{B{\"o}hringer}
  et~al.}{2000}]{Bohringer+00}
{B{\"o}hringer} H.,  et~al., 2000, \mn@doi [\apjs] {10.1086/313427}, \href
  {http://adsabs.harvard.edu/abs/2000ApJS..129..435B} {129, 435}

\bibitem[\protect\citeauthoryear{{Bourdin}, {Sauvageot}, {Slezak}, {Bijaoui}
  \& {Teyssier}}{{Bourdin} et~al.}{2004}]{Bourdin+04}
{Bourdin} H.,  {Sauvageot} J.-L.,  {Slezak} E.,  {Bijaoui} A.,   {Teyssier} R.,
   2004, \mn@doi [\aap] {10.1051/0004-6361:20031662}, \href
  {http://adsabs.harvard.edu/abs/2004A%26A...414..429B} {414, 429}

\bibitem[\protect\citeauthoryear{{Buote} \& {Tsai}}{{Buote} \&
  {Tsai}}{1996}]{Buote96}
{Buote} D.~A.,  {Tsai} J.~C.,  1996, \mn@doi [\apj] {10.1086/176790}, \href
  {http://adsabs.harvard.edu/abs/1996ApJ...458...27B} {458, 27}

\bibitem[\protect\citeauthoryear{{Burns}, {Roettiger}, {Ledlow}  \&
  {Klypin}}{{Burns} et~al.}{1994}]{Burns+94}
{Burns} J.~O.,  {Roettiger} K.,  {Ledlow} M.,   {Klypin} A.,  1994, \mn@doi
  [\apjl] {10.1086/187371}, \href
  {http://adsabs.harvard.edu/abs/1994ApJ...427L..87B} {427, L87}

\bibitem[\protect\citeauthoryear{{Burns}, {Hallman}, {Gantner}, {Motl}  \&
  {Norman}}{{Burns} et~al.}{2008}]{Burns08}
{Burns} J.~O.,  {Hallman} E.~J.,  {Gantner} B.,  {Motl} P.~M.,   {Norman}
  M.~L.,  2008, \mn@doi [\apj] {10.1086/526514}, \href
  {http://adsabs.harvard.edu/abs/2008ApJ...675.1125B} {675, 1125}

\bibitem[\protect\citeauthoryear{{Canizares}, {Stewart}  \&
  {Fabian}}{{Canizares} et~al.}{1983}]{Canizares83}
{Canizares} C.~R.,  {Stewart} G.~C.,   {Fabian} A.~C.,  1983, \mn@doi [\apj]
  {10.1086/161312}, \href {http://adsabs.harvard.edu/abs/1983ApJ...272..449C}
  {272, 449}

\bibitem[\protect\citeauthoryear{{Chatzikos}, {Sarazin}  \&
  {Kempner}}{{Chatzikos} et~al.}{2006}]{Chatzikos06}
{Chatzikos} M.,  {Sarazin} C.~L.,   {Kempner} J.~C.,  2006, \mn@doi [\apj]
  {10.1086/503276}, \href {http://adsabs.harvard.edu/abs/2006ApJ...643..751C}
  {643, 751}

\bibitem[\protect\citeauthoryear{{Chen}, {Reiprich}, {B{\"o}hringer}, {Ikebe}
  \& {Zhang}}{{Chen} et~al.}{2007}]{Chen07}
{Chen} Y.,  {Reiprich} T.~H.,  {B{\"o}hringer} H.,  {Ikebe} Y.,   {Zhang}
  Y.-Y.,  2007, \mn@doi [\aap] {10.1051/0004-6361:20066471}, \href
  {http://adsabs.harvard.edu/abs/2007A%26A...466..805C} {466, 805}

\bibitem[\protect\citeauthoryear{{Cortese}, {Gavazzi}, {Boselli},
  {Iglesias-Paramo}  \& {Carrasco}}{{Cortese} et~al.}{2004}]{Cortese04}
{Cortese} L.,  {Gavazzi} G.,  {Boselli} A.,  {Iglesias-Paramo} J.,   {Carrasco}
  L.,  2004, \mn@doi [\aap] {10.1051/0004-6361:20040381}, \href
  {http://adsabs.harvard.edu/abs/2004A%26A...425..429C} {425, 429}

\bibitem[\protect\citeauthoryear{{De Grandi} \& {Molendi}}{{De Grandi} \&
  {Molendi}}{2001}]{DeGrandi01}
{De Grandi} S.,  {Molendi} S.,  2001, \mn@doi [\apj] {10.1086/320098}, \href
  {http://adsabs.harvard.edu/abs/2001ApJ...551..153D} {551, 153}

\bibitem[\protect\citeauthoryear{{Donnelly}, {Forman}, {Jones}, {Quintana},
  {Ramirez}, {Churazov}  \& {Gilfanov}}{{Donnelly} et~al.}{2001}]{don01}
{Donnelly} R.~H.,  {Forman} W.,  {Jones} C.,  {Quintana} H.,  {Ramirez} A.,
  {Churazov} E.,   {Gilfanov} M.,  2001, \mn@doi [\apj] {10.1086/323521}, \href
  {http://adsabs.harvard.edu/abs/2001ApJ...562..254D} {562, 254}

\bibitem[\protect\citeauthoryear{{Durret}, {Lima Neto}, {Forman}  \&
  {Churazov}}{{Durret} et~al.}{2003}]{Durret+03}
{Durret} F.,  {Lima Neto} G.~B.,  {Forman} W.,   {Churazov} E.,  2003, \mn@doi
  [\aap] {10.1051/0004-6361:20030424}, \href
  {http://adsabs.harvard.edu/abs/2003A%26A...403L..29D} {403, L29}

\bibitem[\protect\citeauthoryear{{Durret}, {Lima Neto}  \& {Forman}}{{Durret}
  et~al.}{2005}]{Durret05}
{Durret} F.,  {Lima Neto} G.~B.,   {Forman} W.,  2005, \mn@doi [\aap]
  {10.1051/0004-6361:20041666}, \href
  {http://adsabs.harvard.edu/abs/2005A%26A...432..809D} {432, 809}

\bibitem[\protect\citeauthoryear{{Durret}, {Lagan{\'a}}, {Adami}  \&
  {Bertin}}{{Durret} et~al.}{2010}]{Durret10}
{Durret} F.,  {Lagan{\'a}} T.~F.,  {Adami} C.,   {Bertin} E.,  2010, \mn@doi
  [\aap] {10.1051/0004-6361/201014566}, \href
  {http://adsabs.harvard.edu/abs/2010A%26A...517A..94D} {517, A94}

\bibitem[\protect\citeauthoryear{{Durret}, {Lagan{\'a}}  \& {Haider}}{{Durret}
  et~al.}{2011}]{Durret11}
{Durret} F.,  {Lagan{\'a}} T.~F.,   {Haider} M.,  2011, \mn@doi [\aap]
  {10.1051/0004-6361/201015978}, \href
  {http://adsabs.harvard.edu/abs/2011A%26A...529A..38D} {529, A38}

\bibitem[\protect\citeauthoryear{{Durret}, {Perrot}, {Lima Neto}, {Adami},
  {Bertin}  \& {Bagchi}}{{Durret} et~al.}{2013}]{Durret+13}
{Durret} F.,  {Perrot} C.,  {Lima Neto} G.~B.,  {Adami} C.,  {Bertin} E.,
  {Bagchi} J.,  2013, \mn@doi [\aap] {10.1051/0004-6361/201322082}, \href
  {http://adsabs.harvard.edu/abs/2013A%26A...560A..78D} {560, A78}

\bibitem[\protect\citeauthoryear{{Ebeling}, {Mendes de Oliveira}  \&
  {White}}{{Ebeling} et~al.}{1995}]{Ebeling95}
{Ebeling} H.,  {Mendes de Oliveira} C.,   {White} D.~A.,  1995, \mn@doi
  [\mnras] {10.1093/mnras/277.3.1006}, \href
  {http://cdsads.u-strasbg.fr/abs/1995MNRAS.277.1006E} {277, 1006}

\bibitem[\protect\citeauthoryear{{Eckert}, {Molendi}  \& {Paltani}}{{Eckert}
  et~al.}{2011}]{Eckert11}
{Eckert} D.,  {Molendi} S.,   {Paltani} S.,  2011, \mn@doi [\aap]
  {10.1051/0004-6361/201015856}, \href
  {http://adsabs.harvard.edu/abs/2011A%26A...526A..79E} {526, A79}

\bibitem[\protect\citeauthoryear{{Eckert} et~al.,}{{Eckert}
  et~al.}{2017}]{Eckert+17}
{Eckert} D.,  et~al., 2017, \mn@doi [\aap] {10.1051/0004-6361/201730555}, \href
  {http://adsabs.harvard.edu/abs/2017A%26A...605A..25E} {605, A25}

\bibitem[\protect\citeauthoryear{{Ehlert}, {McDonald}, {David}, {Miller}  \&
  {Bautz}}{{Ehlert} et~al.}{2015}]{ehl15}
{Ehlert} S.,  {McDonald} M.,  {David} L.~P.,  {Miller} E.~D.,   {Bautz} M.~W.,
  2015, \mn@doi [\apj] {10.1088/0004-637X/799/2/174}, \href
  {https://ui.adsabs.harvard.edu/#abs/2015ApJ...799..174E} {799, 174}

\bibitem[\protect\citeauthoryear{{Elkholy}, {Bautz}  \& {Canizares}}{{Elkholy}
  et~al.}{2015}]{Elkholy+15}
{Elkholy} T.~Y.,  {Bautz} M.~W.,   {Canizares} C.~R.,  2015, \mn@doi [\apj]
  {10.1088/0004-637X/805/1/3}, \href
  {http://adsabs.harvard.edu/abs/2015ApJ...805....3E} {805, 3}

\bibitem[\protect\citeauthoryear{{Fabian}}{{Fabian}}{1994}]{Fabian94}
{Fabian} A.~C.,  1994, \mn@doi [\araa] {10.1146/annurev.aa.32.090194.001425},
  \href {http://adsabs.harvard.edu/abs/1994ARA%26A..32..277F} {32, 277}

\bibitem[\protect\citeauthoryear{{Fabian}, {Sanders}, {Allen}, {Crawford},
  {Iwasawa}, {Johnstone}, {Schmidt}  \& {Taylor}}{{Fabian}
  et~al.}{2003}]{Fabian03}
{Fabian} A.~C.,  {Sanders} J.~S.,  {Allen} S.~W.,  {Crawford} C.~S.,  {Iwasawa}
  K.,  {Johnstone} R.~M.,  {Schmidt} R.~W.,   {Taylor} G.~B.,  2003, \mn@doi
  [\mnras] {10.1046/j.1365-8711.2003.06902.x}, \href
  {http://adsabs.harvard.edu/abs/2003MNRAS.344L..43F} {344, L43}

\bibitem[\protect\citeauthoryear{{Finoguenov}, {Pietsch}, {Aschenbach}  \&
  {Miniati}}{{Finoguenov} et~al.}{2004}]{Finog04}
{Finoguenov} A.,  {Pietsch} W.,  {Aschenbach} B.,   {Miniati} F.,  2004,
  \mn@doi [\aap] {10.1051/0004-6361:20031672}, \href
  {http://adsabs.harvard.edu/abs/2004A%26A...415..415F} {415, 415}

\bibitem[\protect\citeauthoryear{{Frank}, {Peterson}, {Andersson}, {Fabian}  \&
  {Sanders}}{{Frank} et~al.}{2013}]{Frank+13}
{Frank} K.~A.,  {Peterson} J.~R.,  {Andersson} K.,  {Fabian} A.~C.,   {Sanders}
  J.~S.,  2013, \mn@doi [\apj] {10.1088/0004-637X/764/1/46}, \href
  {http://adsabs.harvard.edu/abs/2013ApJ...764...46F} {764, 46}

\bibitem[\protect\citeauthoryear{{Fukazawa}, {Makishima}  \&
  {Ohashi}}{{Fukazawa} et~al.}{2004}]{Fukazawa+04}
{Fukazawa} Y.,  {Makishima} K.,   {Ohashi} T.,  2004, \mn@doi [\pasj]
  {10.1093/pasj/56.6.965}, \href
  {http://adsabs.harvard.edu/abs/2004PASJ...56..965F} {56, 965}

\bibitem[\protect\citeauthoryear{{Gastaldello}, {Ettori}, {Molendi},
  {Bardelli}, {Venturi}  \& {Zucca}}{{Gastaldello}
  et~al.}{2003}]{Gastaldello+03}
{Gastaldello} F.,  {Ettori} S.,  {Molendi} S.,  {Bardelli} S.,  {Venturi} T.,
  {Zucca} E.,  2003, \mn@doi [\aap] {10.1051/0004-6361:20031348}, \href
  {http://adsabs.harvard.edu/abs/2003A%26A...411...21G} {411, 21}

\bibitem[\protect\citeauthoryear{{Ghizzardi}, {De Grandi}  \&
  {Molendi}}{{Ghizzardi} et~al.}{2014}]{Ghizzardi14}
{Ghizzardi} S.,  {De Grandi} S.,   {Molendi} S.,  2014, \mn@doi [\aap]
  {10.1051/0004-6361/201424016}, \href
  {http://adsabs.harvard.edu/abs/2014A%26A...570A.117G} {570, A117}

\bibitem[\protect\citeauthoryear{{Giovannini}, {Bonafede}, {Feretti}, {Govoni},
  {Murgia}, {Ferrari}  \& {Monti}}{{Giovannini} et~al.}{2009}]{Giovannini+09}
{Giovannini} G.,  {Bonafede} A.,  {Feretti} L.,  {Govoni} F.,  {Murgia} M.,
  {Ferrari} F.,   {Monti} G.,  2009, \mn@doi [\aap]
  {10.1051/0004-6361/200912667}, \href
  {http://adsabs.harvard.edu/abs/2009A%26A...507.1257G} {507, 1257}

\bibitem[\protect\citeauthoryear{{Gonzalez}, {Zabludoff}, {Zaritsky}  \&
  {Dalcanton}}{{Gonzalez} et~al.}{2000}]{Gonzalez00}
{Gonzalez} A.~H.,  {Zabludoff} A.~I.,  {Zaritsky} D.,   {Dalcanton} J.~J.,
  2000, \mn@doi [\apj] {10.1086/308985}, \href
  {http://adsabs.harvard.edu/abs/2000ApJ...536..561G} {536, 561}

\bibitem[\protect\citeauthoryear{{Gu} et~al.,}{{Gu} et~al.}{2009}]{Gu+09}
{Gu} L.,  et~al., 2009, \mn@doi [\apj] {10.1088/0004-637X/700/2/1161}, \href
  {http://adsabs.harvard.edu/abs/2009ApJ...700.1161G} {700, 1161}

\bibitem[\protect\citeauthoryear{{Hofmann}, {Sanders}, {Nandra}, {Clerc}  \&
  {Gaspari}}{{Hofmann} et~al.}{2016}]{Hofmann+16}
{Hofmann} F.,  {Sanders} J.~S.,  {Nandra} K.,  {Clerc} N.,   {Gaspari} M.,
  2016, \mn@doi [\aap] {10.1051/0004-6361/201526925}, \href
  {http://adsabs.harvard.edu/abs/2016A%26A...585A.130H} {585, A130}

\bibitem[\protect\citeauthoryear{{Huang} \& {Sarazin}}{{Huang} \&
  {Sarazin}}{1998}]{Huang98}
{Huang} Z.,  {Sarazin} C.~L.,  1998, \mn@doi [\apj] {10.1086/305406}, \href
  {http://adsabs.harvard.edu/abs/1998ApJ...496..728H} {496, 728}

\bibitem[\protect\citeauthoryear{{Hudson}, {Mittal}, {Reiprich}, {Nulsen},
  {Andernach}  \& {Sarazin}}{{Hudson} et~al.}{2010}]{Hudson10}
{Hudson} D.~S.,  {Mittal} R.,  {Reiprich} T.~H.,  {Nulsen} P.~E.~J.,
  {Andernach} H.,   {Sarazin} C.~L.,  2010, \mn@doi [\aap]
  {10.1051/0004-6361/200912377}, \href
  {http://adsabs.harvard.edu/abs/2010A%26A...513A..37H} {513, A37}

\bibitem[\protect\citeauthoryear{{Ignesti}, {Gitti}, {Brunetti}, {O'Sullivan},
  {Sarazin}  \& {Wong}}{{Ignesti} et~al.}{2018}]{Ignesti18}
{Ignesti} A.,  {Gitti} M.,  {Brunetti} G.,  {O'Sullivan} E.,  {Sarazin} C.,
  {Wong} K.,  2018, \mn@doi [\aap] {10.1051/0004-6361/201731380}, \href
  {http://adsabs.harvard.edu/abs/2018A%26A...610A..89I} {610, A89}

\bibitem[\protect\citeauthoryear{{Inoue}, {Hayashida}, {Ueda}, {Nagino},
  {Tsunemi}  \& {Koyama}}{{Inoue} et~al.}{2016}]{Inoue+16}
{Inoue} S.,  {Hayashida} K.,  {Ueda} S.,  {Nagino} R.,  {Tsunemi} H.,
  {Koyama} K.,  2016, \mn@doi [\pasj] {10.1093/pasj/psw027}, \href
  {http://adsabs.harvard.edu/abs/2016PASJ...68S..23I} {68, S23}

\bibitem[\protect\citeauthoryear{{Jeltema}, {Canizares}, {Bautz}  \&
  {Buote}}{{Jeltema} et~al.}{2005}]{Jeltema05}
{Jeltema} T.~E.,  {Canizares} C.~R.,  {Bautz} M.~W.,   {Buote} D.~A.,  2005,
  \mn@doi [\apj] {10.1086/428940}, \href
  {http://adsabs.harvard.edu/abs/2005ApJ...624..606J} {624, 606}

\bibitem[\protect\citeauthoryear{{Johnson}, {Markevitch}, {Wegner}, {Jones}  \&
  {Forman}}{{Johnson} et~al.}{2010}]{Johnson+10}
{Johnson} R.~E.,  {Markevitch} M.,  {Wegner} G.~A.,  {Jones} C.,   {Forman}
  W.~R.,  2010, \mn@doi [\apj] {10.1088/0004-637X/710/2/1776}, \href
  {http://adsabs.harvard.edu/abs/2010ApJ...710.1776J} {710, 1776}

\bibitem[\protect\citeauthoryear{{Jones} \& {Forman}}{{Jones} \&
  {Forman}}{1999}]{JonesForman99}
{Jones} C.,  {Forman} W.,  1999, \mn@doi [\apj] {10.1086/306646}, \href
  {http://adsabs.harvard.edu/abs/1999ApJ...511...65J} {511, 65}

\bibitem[\protect\citeauthoryear{{Kaastra} \& {Mewe}}{{Kaastra} \&
  {Mewe}}{1993}]{KM93}
{Kaastra} J.~S.,  {Mewe} R.,  1993, \aaps, \href
  {http://adsabs.harvard.edu/abs/1993A%26AS...97..443K} {97, 443}

\bibitem[\protect\citeauthoryear{{Kale} \& {Dwarakanath}}{{Kale} \&
  {Dwarakanath}}{2012}]{kal12}
{Kale} R.,  {Dwarakanath} K.~S.,  2012, \mn@doi [\apj]
  {10.1088/0004-637X/744/1/46}, \href
  {http://adsabs.harvard.edu/abs/2012ApJ...744...46K} {744, 46}

\bibitem[\protect\citeauthoryear{{Kolokotronis}, {Basilakos}, {Plionis}  \&
  {Georgantopoulos}}{{Kolokotronis} et~al.}{2001}]{Kolokotronis01}
{Kolokotronis} V.,  {Basilakos} S.,  {Plionis} M.,   {Georgantopoulos} I.,
  2001, \mn@doi [\mnras] {10.1046/j.1365-8711.2001.03924.x}, \href
  {http://adsabs.harvard.edu/abs/2001MNRAS.320...49K} {320, 49}

\bibitem[\protect\citeauthoryear{{Lagan{\'a}}, {Andrade-Santos}  \& {Lima
  Neto}}{{Lagan{\'a}} et~al.}{2010}]{Lagana10}
{Lagan{\'a}} T.~F.,  {Andrade-Santos} F.,   {Lima Neto} G.~B.,  2010, \mn@doi
  [\aap] {10.1051/0004-6361/200913180}, \href
  {http://adsabs.harvard.edu/abs/2010A%26A...511A..15L} {511, A15}

\bibitem[\protect\citeauthoryear{{Lagan{\'a}}, {Lovisari}, {Martins},
  {Lanfranchi}, {Capelato}  \& {Schellenberger}}{{Lagan{\'a}}
  et~al.}{2015}]{Lagana15}
{Lagan{\'a}} T.~F.,  {Lovisari} L.,  {Martins} L.,  {Lanfranchi} G.~A.,
  {Capelato} H.~V.,   {Schellenberger} G.,  2015, \mn@doi [\aap]
  {10.1051/0004-6361/201424821}, \href
  {http://adsabs.harvard.edu/abs/2015A%26A...573A..66L} {573, A66}

\bibitem[\protect\citeauthoryear{{Lakhchaura}, {Singh}, {Saikia}  \&
  {Hunstead}}{{Lakhchaura} et~al.}{2011}]{lak11}
{Lakhchaura} K.,  {Singh} K.~P.,  {Saikia} D.~J.,   {Hunstead} R.~W.,  2011,
  \mn@doi [\apj] {10.1088/0004-637X/743/1/78}, \href
  {http://adsabs.harvard.edu/abs/2011ApJ...743...78L} {743, 78}

\bibitem[\protect\citeauthoryear{{Lakhchaura}, {Singh}, {Saikia}  \&
  {Hunstead}}{{Lakhchaura} et~al.}{2013}]{Lakhchaura+13}
{Lakhchaura} K.,  {Singh} K.~P.,  {Saikia} D.~J.,   {Hunstead} R.~W.,  2013,
  \mn@doi [\apj] {10.1088/0004-637X/767/1/91}, \href
  {http://adsabs.harvard.edu/abs/2013ApJ...767...91L} {767, 91}

\bibitem[\protect\citeauthoryear{{Leccardi}, {Rossetti}  \&
  {Molendi}}{{Leccardi} et~al.}{2010}]{Leccardi10}
{Leccardi} A.,  {Rossetti} M.,   {Molendi} S.,  2010, \mn@doi [\aap]
  {10.1051/0004-6361/200913094}, \href
  {http://adsabs.harvard.edu/abs/2010A%26A...510A..82L} {510, A82}

\bibitem[\protect\citeauthoryear{{Lee}, {Hwang}, {Lee}, {Ko}, {Sohn}, {Shim}
  \& {Diaferio}}{{Lee} et~al.}{2015}]{Lee+15}
{Lee} G.-H.,  {Hwang} H.~S.,  {Lee} M.~G.,  {Ko} J.,  {Sohn} J.,  {Shim} H.,
  {Diaferio} A.,  2015, \mn@doi [\apj] {10.1088/0004-637X/800/2/80}, \href
  {http://adsabs.harvard.edu/abs/2015ApJ...800...80L} {800, 80}

\bibitem[\protect\citeauthoryear{{Lopes}, {Trevisan}, {Lagan{\'a}}, {Durret},
  {Ribeiro}  \& {Rembold}}{{Lopes} et~al.}{2018}]{Lopes18}
{Lopes} P.~A.~A.,  {Trevisan} M.,  {Lagan{\'a}} T.~F.,  {Durret} F.,  {Ribeiro}
  A.~L.~B.,   {Rembold} S.~B.,  2018, \mn@doi [\mnras] {10.1093/mnras/sty1374},
  \href {http://adsabs.harvard.edu/abs/2018MNRAS.478.5473L} {478, 5473}

\bibitem[\protect\citeauthoryear{{Macario}, {Markevitch}, {Giacintucci},
  {Brunetti}, {Venturi}  \& {Murray}}{{Macario} et~al.}{2011}]{Macario+11}
{Macario} G.,  {Markevitch} M.,  {Giacintucci} S.,  {Brunetti} G.,  {Venturi}
  T.,   {Murray} S.~S.,  2011, \mn@doi [\apj] {10.1088/0004-637X/728/2/82},
  \href {http://adsabs.harvard.edu/abs/2011ApJ...728...82M} {728, 82}

\bibitem[\protect\citeauthoryear{{Machado} \& {Lima Neto}}{{Machado} \& {Lima
  Neto}}{2013}]{MachadoLimaNeto13}
{Machado} R.~E.~G.,  {Lima Neto} G.~B.,  2013, \mn@doi [\mnras]
  {10.1093/mnras/stt127}, \href
  {http://adsabs.harvard.edu/abs/2013MNRAS.430.3249M} {430, 3249}

\bibitem[\protect\citeauthoryear{{Machado} \& {Lima Neto}}{{Machado} \& {Lima
  Neto}}{2015}]{Machado15}
{Machado} R.~E.~G.,  {Lima Neto} G.~B.,  2015, \mn@doi [\mnras]
  {10.1093/mnras/stu2669}, \href
  {http://adsabs.harvard.edu/abs/2015MNRAS.447.2915M} {447, 2915}

\bibitem[\protect\citeauthoryear{{Marini} et~al.,}{{Marini}
  et~al.}{2004}]{Marini+04}
{Marini} F.,  et~al., 2004, \mn@doi [\mnras]
  {10.1111/j.1365-2966.2004.08148.x}, \href
  {http://adsabs.harvard.edu/abs/2004MNRAS.353.1219M} {353, 1219}

\bibitem[\protect\citeauthoryear{{Markevitch} \& {Vikhlinin}}{{Markevitch} \&
  {Vikhlinin}}{2007}]{Mark07}
{Markevitch} M.,  {Vikhlinin} A.,  2007, \mn@doi [\physrep]
  {10.1016/j.physrep.2007.01.001}, \href
  {http://adsabs.harvard.edu/abs/2007PhR...443....1M} {443, 1}

\bibitem[\protect\citeauthoryear{{Markevitch}, {Forman}, {Sarazin}  \&
  {Vikhlinin}}{{Markevitch} et~al.}{1998}]{Mark98}
{Markevitch} M.,  {Forman} W.~R.,  {Sarazin} C.~L.,   {Vikhlinin} A.,  1998,
  \mn@doi [\apj] {10.1086/305976}, \href
  {http://adsabs.harvard.edu/abs/1998ApJ...503...77M} {503, 77}

\bibitem[\protect\citeauthoryear{{Markevitch} et~al.,}{{Markevitch}
  et~al.}{2000}]{Markevitch+00}
{Markevitch} M.,  et~al., 2000, \mn@doi [\apj] {10.1086/309470}, \href
  {http://adsabs.harvard.edu/abs/2000ApJ...541..542M} {541, 542}

\bibitem[\protect\citeauthoryear{{Markevitch}, {Gonzalez}, {David},
  {Vikhlinin}, {Murray}, {Forman}, {Jones}  \& {Tucker}}{{Markevitch}
  et~al.}{2002}]{Mark02}
{Markevitch} M.,  {Gonzalez} A.~H.,  {David} L.,  {Vikhlinin} A.,  {Murray} S.,
   {Forman} W.,  {Jones} C.,   {Tucker} W.,  2002, \mn@doi [\apjl]
  {10.1086/339619}, \href {http://adsabs.harvard.edu/abs/2002ApJ...567L..27M}
  {567, L27}

\bibitem[\protect\citeauthoryear{{Matsushita}}{{Matsushita}}{2011}]{Matsushita11}
{Matsushita} K.,  2011, \mn@doi [\aap] {10.1051/0004-6361/200913432}, \href
  {http://adsabs.harvard.edu/abs/2011A%26A...527A.134M} {527, A134}

\bibitem[\protect\citeauthoryear{{Mauduit} \& {Mamon}}{{Mauduit} \&
  {Mamon}}{2007}]{MauduitMamon07}
{Mauduit} J.-C.,  {Mamon} G.~A.,  2007, \mn@doi [\aap]
  {10.1051/0004-6361:20077721}, \href
  {http://adsabs.harvard.edu/abs/2007A%26A...475..169M} {475, 169}

\bibitem[\protect\citeauthoryear{{McNamara} et~al.,}{{McNamara}
  et~al.}{2001}]{McNamara01}
{McNamara} B.~R.,  et~al., 2001, \mn@doi [\apjl] {10.1086/338326}, \href
  {http://adsabs.harvard.edu/abs/2001ApJ...562L.149M} {562, L149}

\bibitem[\protect\citeauthoryear{{Mernier} et~al.,}{{Mernier}
  et~al.}{2017}]{Mernier17}
{Mernier} F.,  et~al., 2017, \mn@doi [\aap] {10.1051/0004-6361/201630075},
  \href {http://adsabs.harvard.edu/abs/2017A%26A...603A..80M} {603, A80}

\bibitem[\protect\citeauthoryear{{Mitsuishi} et~al.,}{{Mitsuishi}
  et~al.}{2012}]{Mitsuishi+12}
{Mitsuishi} I.,  et~al., 2012, \mn@doi [\pasj] {10.1093/pasj/64.1.18}, \href
  {http://adsabs.harvard.edu/abs/2012PASJ...64...18M} {64, 18}

\bibitem[\protect\citeauthoryear{{Molendi} \& {Pizzolato}}{{Molendi} \&
  {Pizzolato}}{2001}]{Molendi01}
{Molendi} S.,  {Pizzolato} F.,  2001, \mn@doi [\apj] {10.1086/322387}, \href
  {http://adsabs.harvard.edu/abs/2001ApJ...560..194M} {560, 194}

\bibitem[\protect\citeauthoryear{{Morris} \& {Fabian}}{{Morris} \&
  {Fabian}}{2005}]{Morris05}
{Morris} R.~G.,  {Fabian} A.~C.,  2005, \mn@doi [\mnras]
  {10.1111/j.1365-2966.2005.08822.x}, \href
  {http://adsabs.harvard.edu/abs/2005MNRAS.358..585M} {358, 585}

\bibitem[\protect\citeauthoryear{{Morrison} \& {McCammon}}{{Morrison} \&
  {McCammon}}{1983}]{Morrison83}
{Morrison} R.,  {McCammon} D.,  1983, \mn@doi [\apj] {10.1086/161102}, \href
  {http://adsabs.harvard.edu/abs/1983ApJ...270..119M} {270, 119}

\bibitem[\protect\citeauthoryear{{Neumann} et~al.,}{{Neumann}
  et~al.}{2001}]{Neumann+01}
{Neumann} D.~M.,  et~al., 2001, \mn@doi [\aap] {10.1051/0004-6361:20000182},
  \href {http://adsabs.harvard.edu/abs/2001A%26A...365L..74N} {365, L74}

\bibitem[\protect\citeauthoryear{{Nevalainen}, {Markevitch}  \&
  {Forman}}{{Nevalainen} et~al.}{2000}]{nev00}
{Nevalainen} J.,  {Markevitch} M.,   {Forman} W.,  2000, \mn@doi [\apj]
  {10.1086/308906}, \href {http://adsabs.harvard.edu/abs/2000ApJ...536...73N}
  {536, 73}

\bibitem[\protect\citeauthoryear{{Nulsen} et~al.,}{{Nulsen}
  et~al.}{2013}]{Nulsen+13}
{Nulsen} P.~E.~J.,  et~al., 2013, \mn@doi [\apj] {10.1088/0004-637X/775/2/117},
  \href {http://adsabs.harvard.edu/abs/2013ApJ...775..117N} {775, 117}

\bibitem[\protect\citeauthoryear{{O'Sullivan}, {Vrtilek}, {Harris}  \&
  {Ponman}}{{O'Sullivan} et~al.}{2007}]{OSullivan+07}
{O'Sullivan} E.,  {Vrtilek} J.~M.,  {Harris} D.~E.,   {Ponman} T.~J.,  2007,
  \mn@doi [\apj] {10.1086/511778}, \href
  {http://adsabs.harvard.edu/abs/2007ApJ...658..299O} {658, 299}

\bibitem[\protect\citeauthoryear{{O'Sullivan}, {Giacintucci}, {David},
  {Vrtilek}  \& {Raychaudhury}}{{O'Sullivan} et~al.}{2011}]{OSullivan+11}
{O'Sullivan} E.,  {Giacintucci} S.,  {David} L.~P.,  {Vrtilek} J.~M.,
  {Raychaudhury} S.,  2011, \mn@doi [\mnras]
  {10.1111/j.1365-2966.2010.17812.x}, \href
  {http://adsabs.harvard.edu/abs/2011MNRAS.411.1833O} {411, 1833}

\bibitem[\protect\citeauthoryear{{Oh} et~al.,}{{Oh} et~al.}{2016}]{Oh+16}
{Oh} S.,  et~al., 2016, \mn@doi [\apj] {10.3847/0004-637X/832/1/69}, \href
  {http://adsabs.harvard.edu/abs/2016ApJ...832...69O} {832, 69}

\bibitem[\protect\citeauthoryear{{Oh} et~al.,}{{Oh} et~al.}{2018}]{Oh18}
{Oh} S.,  et~al., 2018, \mn@doi [\apjs] {10.3847/1538-4365/aacd47}, \href
  {http://adsabs.harvard.edu/abs/2018ApJS..237...14O} {237, 14}

\bibitem[\protect\citeauthoryear{{Owers}, {Nulsen}  \& {Couch}}{{Owers}
  et~al.}{2011}]{Owers+11}
{Owers} M.~S.,  {Nulsen} P.~E.~J.,   {Couch} W.~J.,  2011, \mn@doi [\apj]
  {10.1088/0004-637X/741/2/122}, \href
  {http://adsabs.harvard.edu/abs/2011ApJ...741..122O} {741, 122}

\bibitem[\protect\citeauthoryear{{Owers} et~al.,}{{Owers}
  et~al.}{2017}]{Owers+17}
{Owers} M.~S.,  et~al., 2017, \mn@doi [\mnras] {10.1093/mnras/stx562}, \href
  {http://adsabs.harvard.edu/abs/2017MNRAS.468.1824O} {468, 1824}

\bibitem[\protect\citeauthoryear{{Pandge}, {Vagshette}, {David}  \&
  {Patil}}{{Pandge} et~al.}{2012}]{Pandge12}
{Pandge} M.~B.,  {Vagshette} N.~D.,  {David} L.~P.,   {Patil} M.~K.,  2012,
  \mn@doi [\mnras] {10.1111/j.1365-2966.2011.20358.x}, \href
  {http://adsabs.harvard.edu/abs/2012MNRAS.421..808P} {421, 808}

\bibitem[\protect\citeauthoryear{{Parekh}, {van der Heyden}, {Ferrari}, {Angus}
   \& {Holwerda}}{{Parekh} et~al.}{2015}]{Parekh+15}
{Parekh} V.,  {van der Heyden} K.,  {Ferrari} C.,  {Angus} G.,   {Holwerda} B.,
   2015, \mn@doi [\aap] {10.1051/0004-6361/201424123}, \href
  {http://adsabs.harvard.edu/abs/2015A%26A...575A.127P} {575, A127}

\bibitem[\protect\citeauthoryear{{Peres}, {Fabian}, {Edge}, {Allen},
  {Johnstone}  \& {White}}{{Peres} et~al.}{1998}]{Peres+98}
{Peres} C.~B.,  {Fabian} A.~C.,  {Edge} A.~C.,  {Allen} S.~W.,  {Johnstone}
  R.~M.,   {White} D.~A.,  1998, \mn@doi [\mnras]
  {10.1046/j.1365-8711.1998.01624.x}, \href
  {http://adsabs.harvard.edu/abs/1998MNRAS.298..416P} {298, 416}

\bibitem[\protect\citeauthoryear{{Peterson}, {Kahn}, {Paerels}, {Kaastra},
  {Tamura}, {Bleeker}, {Ferrigno}  \& {Jernigan}}{{Peterson}
  et~al.}{2003}]{Peterson03}
{Peterson} J.~R.,  {Kahn} S.~M.,  {Paerels} F.~B.~S.,  {Kaastra} J.~S.,
  {Tamura} T.,  {Bleeker} J.~A.~M.,  {Ferrigno} C.,   {Jernigan} J.~G.,  2003,
  \mn@doi [\apj] {10.1086/374830}, \href
  {http://adsabs.harvard.edu/abs/2003ApJ...590..207P} {590, 207}

\bibitem[\protect\citeauthoryear{{Piffaretti} \& {Valdarnini}}{{Piffaretti} \&
  {Valdarnini}}{2008}]{Piffaretti08}
{Piffaretti} R.,  {Valdarnini} R.,  2008, \mn@doi [\aap]
  {10.1051/0004-6361:200809739}, \href
  {http://adsabs.harvard.edu/abs/2008A%26A...491...71P} {491, 71}

\bibitem[\protect\citeauthoryear{{Planck Collaboration} et~al.,}{{Planck
  Collaboration} et~al.}{2011}]{PlanckVIII}
{Planck Collaboration} et~al., 2011, \mn@doi [\aap]
  {10.1051/0004-6361/201116459}, \href
  {http://adsabs.harvard.edu/abs/2011A%26A...536A...8P} {536, A8}

\bibitem[\protect\citeauthoryear{{Planck Collaboration} et~al.,}{{Planck
  Collaboration} et~al.}{2013}]{Planck13}
{Planck Collaboration} et~al., 2013, \mn@doi [\aap]
  {10.1051/0004-6361/201220194}, \href
  {https://ui.adsabs.harvard.edu/#abs/2013A&A...550A.134P} {550, A134}

\bibitem[\protect\citeauthoryear{{Quintana} \& {de Souza}}{{Quintana} \& {de
  Souza}}{1993}]{qui93}
{Quintana} H.,  {de Souza} R.,  1993, \aaps, \href
  {http://adsabs.harvard.edu/abs/1993A%26AS..101..475Q} {101, 475}

\bibitem[\protect\citeauthoryear{{Ramella} et~al.,}{{Ramella}
  et~al.}{2007}]{Ramella+07}
{Ramella} M.,  et~al., 2007, \mn@doi [\aap] {10.1051/0004-6361:20077245}, \href
  {http://adsabs.harvard.edu/abs/2007A%26A...470...39R} {470, 39}

\bibitem[\protect\citeauthoryear{{Read} \& {Ponman}}{{Read} \&
  {Ponman}}{2003}]{ReadPonman03}
{Read} A.~M.,  {Ponman} T.~J.,  2003, \mn@doi [\aap]
  {10.1051/0004-6361:20031099}, \href
  {http://adsabs.harvard.edu/abs/2003A%26A...409..395R} {409, 395}

\bibitem[\protect\citeauthoryear{{Roediger}, {Lovisari}, {Dupke}, {Ghizzardi},
  {Br{\"u}ggen}, {Kraft}  \& {Machacek}}{{Roediger} et~al.}{2012}]{Roediger12}
{Roediger} E.,  {Lovisari} L.,  {Dupke} R.,  {Ghizzardi} S.,  {Br{\"u}ggen} M.,
   {Kraft} R.~P.,   {Machacek} M.~E.,  2012, \mn@doi [\mnras]
  {10.1111/j.1365-2966.2011.20287.x}, \href
  {http://adsabs.harvard.edu/abs/2012MNRAS.420.3632R} {420, 3632}

\bibitem[\protect\citeauthoryear{{Roettiger}, {Stone}  \&
  {Mushotzky}}{{Roettiger} et~al.}{1998}]{Roettiger+98}
{Roettiger} K.,  {Stone} J.~M.,   {Mushotzky} R.~F.,  1998, \mn@doi [\apj]
  {10.1086/305102}, \href {http://adsabs.harvard.edu/abs/1998ApJ...493...62R}
  {493, 62}

\bibitem[\protect\citeauthoryear{{Rossetti}, {Ghizzardi}, {Molendi}  \&
  {Finoguenov}}{{Rossetti} et~al.}{2007}]{Rossetti+07}
{Rossetti} M.,  {Ghizzardi} S.,  {Molendi} S.,   {Finoguenov} A.,  2007,
  \mn@doi [\aap] {10.1051/0004-6361:20054621}, \href
  {http://adsabs.harvard.edu/abs/2007A%26A...463..839R} {463, 839}

\bibitem[\protect\citeauthoryear{{Rossetti} et~al.,}{{Rossetti}
  et~al.}{2016}]{Rossetti16}
{Rossetti} M.,  et~al., 2016, \mn@doi [\mnras] {10.1093/mnras/stw265}, \href
  {http://adsabs.harvard.edu/abs/2016MNRAS.457.4515R} {457, 4515}

\bibitem[\protect\citeauthoryear{{Sakelliou} \& {Ponman}}{{Sakelliou} \&
  {Ponman}}{2006}]{Sakelliou06}
{Sakelliou} I.,  {Ponman} T.~J.,  2006, \mn@doi [\mnras]
  {10.1111/j.1365-2966.2006.10080.x}, \href
  {http://adsabs.harvard.edu/abs/2006MNRAS.367.1409S} {367, 1409}

\bibitem[\protect\citeauthoryear{{Sanderson} \& {Ponman}}{{Sanderson} \&
  {Ponman}}{2010}]{san10}
{Sanderson} A.~J.~R.,  {Ponman} T.~J.,  2010, \mn@doi [\mnras]
  {10.1111/j.1365-2966.2009.15888.x}, \href
  {http://adsabs.harvard.edu/abs/2010MNRAS.402...65S} {402, 65}

\bibitem[\protect\citeauthoryear{{Santos}, {Tozzi}, {Rosati}  \&
  {B{\"o}hringer}}{{Santos} et~al.}{2010}]{santos10}
{Santos} J.~S.,  {Tozzi} P.,  {Rosati} P.,   {B{\"o}hringer} H.,  2010, \mn@doi
  [\aap] {10.1051/0004-6361/201015208}, \href
  {https://ui.adsabs.harvard.edu/#abs/2010A&A...521A..64S} {521, A64}

\bibitem[\protect\citeauthoryear{{Schenck}, {Datta}, {Burns}  \&
  {Skillman}}{{Schenck} et~al.}{2014}]{Schenck+14}
{Schenck} D.~E.,  {Datta} A.,  {Burns} J.~O.,   {Skillman} S.,  2014, \mn@doi
  [\aj] {10.1088/0004-6256/148/1/23}, \href
  {http://adsabs.harvard.edu/abs/2014AJ....148...23S} {148, 23}

\bibitem[\protect\citeauthoryear{{Schuecker}, {Finoguenov}, {Miniati},
  {B{\"o}hringer}  \& {Briel}}{{Schuecker} et~al.}{2004}]{Schu04}
{Schuecker} P.,  {Finoguenov} A.,  {Miniati} F.,  {B{\"o}hringer} H.,   {Briel}
  U.~G.,  2004, \mn@doi [\aap] {10.1051/0004-6361:20041039}, \href
  {http://adsabs.harvard.edu/abs/2004A%26A...426..387S} {426, 387}

\bibitem[\protect\citeauthoryear{{Simionescu}, {Roediger}, {Nulsen},
  {Br{\"u}ggen}, {Forman}, {B{\"o}hringer}, {Werner}  \&
  {Finoguenov}}{{Simionescu} et~al.}{2009}]{Simionescu09}
{Simionescu} A.,  {Roediger} E.,  {Nulsen} P.~E.~J.,  {Br{\"u}ggen} M.,
  {Forman} W.~R.,  {B{\"o}hringer} H.,  {Werner} N.,   {Finoguenov} A.,  2009,
  \mn@doi [\aap] {10.1051/0004-6361:200811071}, \href
  {http://adsabs.harvard.edu/abs/2009A%26A...495..721S} {495, 721}

\bibitem[\protect\citeauthoryear{{Slee} \& {Roy}}{{Slee} \&
  {Roy}}{1998}]{sle98}
{Slee} O.~B.,  {Roy} A.~L.,  1998, \mn@doi [\mnras]
  {10.1046/j.1365-8711.1998.01739.x}, \href
  {http://adsabs.harvard.edu/abs/1998MNRAS.297L..86S} {297, L86}

\bibitem[\protect\citeauthoryear{{Sun} \& {Murray}}{{Sun} \&
  {Murray}}{2002}]{SunMurray02}
{Sun} M.,  {Murray} S.~S.,  2002, \mn@doi [\apj] {10.1086/341756}, \href
  {http://adsabs.harvard.edu/abs/2002ApJ...576..708S} {576, 708}

\bibitem[\protect\citeauthoryear{{Sun}, {Voit}, {Donahue}, {Jones}, {Forman}
  \& {Vikhlinin}}{{Sun} et~al.}{2009}]{Sun+09}
{Sun} M.,  {Voit} G.~M.,  {Donahue} M.,  {Jones} C.,  {Forman} W.,
  {Vikhlinin} A.,  2009, \mn@doi [\apj] {10.1088/0004-637X/693/2/1142}, \href
  {http://adsabs.harvard.edu/abs/2009ApJ...693.1142S} {693, 1142}

\bibitem[\protect\citeauthoryear{{Sunyaev} \& {Zeldovich}}{{Sunyaev} \&
  {Zeldovich}}{1972}]{SZ72}
{Sunyaev} R.~A.,  {Zeldovich} Y.~B.,  1972, Comments on Astrophysics and Space
  Physics, \href {http://adsabs.harvard.edu/abs/1972CoASP...4..173S} {4, 173}

\bibitem[\protect\citeauthoryear{{Takahashi} \& {Yamashita}}{{Takahashi} \&
  {Yamashita}}{2003}]{Takahashi03}
{Takahashi} S.,  {Yamashita} K.,  2003, \mn@doi [\pasj]
  {10.1093/pasj/55.6.1105}, \href
  {http://adsabs.harvard.edu/abs/2003PASJ...55.1105T} {55, 1105}

\bibitem[\protect\citeauthoryear{{Tanaka}, {Furuzawa}, {Miyoshi}, {Tamura}  \&
  {Takata}}{{Tanaka} et~al.}{2010}]{Tanaka+10}
{Tanaka} N.,  {Furuzawa} A.,  {Miyoshi} S.~J.,  {Tamura} T.,   {Takata} T.,
  2010, \mn@doi [\pasj] {10.1093/pasj/62.3.743}, \href
  {http://adsabs.harvard.edu/abs/2010PASJ...62..743T} {62, 743}

\bibitem[\protect\citeauthoryear{{Tchernin} et~al.,}{{Tchernin}
  et~al.}{2016}]{Tchernin+16}
{Tchernin} C.,  et~al., 2016, \mn@doi [\aap] {10.1051/0004-6361/201628183},
  \href {http://adsabs.harvard.edu/abs/2016A%26A...595A..42T} {595, A42}

\bibitem[\protect\citeauthoryear{{Tittley} \& {Henriksen}}{{Tittley} \&
  {Henriksen}}{2001}]{tit01}
{Tittley} E.~R.,  {Henriksen} M.,  2001, \mn@doi [\apj] {10.1086/323955}, \href
  {http://adsabs.harvard.edu/abs/2001ApJ...563..673T} {563, 673}

\bibitem[\protect\citeauthoryear{{Tremblay} et~al.,}{{Tremblay}
  et~al.}{2012}]{Tremblay12}
{Tremblay} G.~R.,  et~al., 2012, \mn@doi [\mnras]
  {10.1111/j.1365-2966.2012.21278.x}, \href
  {http://adsabs.harvard.edu/abs/2012MNRAS.424.1042T} {424, 1042}

\bibitem[\protect\citeauthoryear{{Venturi}, {Bardelli}, {Zagaria}, {Prandoni}
  \& {Morganti}}{{Venturi} et~al.}{2002}]{ven02}
{Venturi} T.,  {Bardelli} S.,  {Zagaria} M.,  {Prandoni} I.,   {Morganti} R.,
  2002, \mn@doi [\aap] {10.1051/0004-6361:20020117}, \href
  {http://adsabs.harvard.edu/abs/2002A%26A...385...39V} {385, 39}

\bibitem[\protect\citeauthoryear{{Vikhlinin}, {Markevitch}, {Murray}, {Jones},
  {Forman}  \& {Van Speybroeck}}{{Vikhlinin} et~al.}{2005}]{Vik05}
{Vikhlinin} A.,  {Markevitch} M.,  {Murray} S.~S.,  {Jones} C.,  {Forman} W.,
  {Van Speybroeck} L.,  2005, \mn@doi [\apj] {10.1086/431142}, \href
  {http://adsabs.harvard.edu/abs/2005ApJ...628..655V} {628, 655}

\bibitem[\protect\citeauthoryear{{Wong}, {Sarazin}, {Blanton}  \&
  {Reiprich}}{{Wong} et~al.}{2008}]{Wong08}
{Wong} K.-W.,  {Sarazin} C.~L.,  {Blanton} E.~L.,   {Reiprich} T.~H.,  2008,
  \mn@doi [\apj] {10.1086/588272}, \href
  {http://adsabs.harvard.edu/abs/2008ApJ...682..155W} {682, 155}

\bibitem[\protect\citeauthoryear{{Wong}, {Irwin}, {Wik}, {Sun}, {Sarazin},
  {Fujita}  \& {Reiprich}}{{Wong} et~al.}{2016}]{Wong+16}
{Wong} K.-W.,  {Irwin} J.~A.,  {Wik} D.~R.,  {Sun} M.,  {Sarazin} C.~L.,
  {Fujita} Y.,   {Reiprich} T.~H.,  2016, \mn@doi [\apj]
  {10.3847/0004-637X/829/1/49}, \href
  {http://adsabs.harvard.edu/abs/2016ApJ...829...49W} {829, 49}

\bibitem[\protect\citeauthoryear{{Yu}, {Diaferio}, {Agulli}, {Aguerri}  \&
  {Tozzi}}{{Yu} et~al.}{2016}]{Yu+16}
{Yu} H.,  {Diaferio} A.,  {Agulli} I.,  {Aguerri} J.~A.~L.,   {Tozzi} P.,
  2016, \mn@doi [\apj] {10.3847/0004-637X/831/2/156}, \href
  {http://adsabs.harvard.edu/abs/2016ApJ...831..156Y} {831, 156}

\bibitem[\protect\citeauthoryear{{Zhang}, {Yuan}, {Yang}, {Zhang}, {Li}, {Zhou}
   \& {Jiang}}{{Zhang} et~al.}{2011}]{Zhang11}
{Zhang} L.,  {Yuan} Q.,  {Yang} Q.,  {Zhang} S.,  {Li} F.,  {Zhou} X.,
  {Jiang} Z.,  2011, \mn@doi [\pasj] {10.1093/pasj/63.3.585}, \href
  {http://adsabs.harvard.edu/abs/2011PASJ...63..585Z} {63, 585}

\bibitem[\protect\citeauthoryear{{Zhu} et~al.,}{{Zhu} et~al.}{2016}]{Zhu16}
{Zhu} Z.,  et~al., 2016, \mn@doi [\apj] {10.3847/0004-637X/816/2/54}, \href
  {http://adsabs.harvard.edu/abs/2016ApJ...816...54Z} {816, 54}

\makeatother
\end{thebibliography}
%%%%%%%%%%%%%%%%%%%%%%%%%%%%%%%%%%%%%%%%%%%%%%%%%%

%%%%%%%%%%%%%%%%% APPENDICES %%%%%%%%%%%%%%%%%%%%%

\begin{appendix}

\section{Notes on individual clusters}
\label{sect:apNotes}

We give here specific details on the clusters studied in this
work. They are presented in the same order as in Tab.~\ref{tab:CC},
to easy the comparison with the 2D maps, and for the \textit{Planck} names  the
prefix PLCKESZ is omitted.

\noindent
\begin{enumerate}
\item {\bf \scshape  CC-relaxed systems} 

\begin{itemize}

\item {\bf Abell~2734}: Based on the optical images of the WINGS
  survey, \citet{Ramella+07} give a map of isodensity contours of
  A2734 showing a strong elongation in the east-west direction
  with possibly two separate structures, the BCG being in the east
  structure. They also indicate the presence of a double substructure
  south of the main cluster.  Our maps only cover a relatively small
  part of the \citet{Ramella+07} map, but our pressure map suggests
  that this cluster is perturbed in its centre.
The maps are spherically symmetric and they don't point to an unrelaxed global structure.

\item {\bf EXO 0422-086}: This cluster was reported in the literature among
  large samples and we can highlight the abundance profile reported by
  \citet{Mernier17} that shows central values about 1$Z_{\odot}$
  decreasing to almost 0.3$Z_{\odot}$ around 0.6 $R_{500}$, 
  in agreement with our 2D metallicity map.

\item {\bf Abell S0540}: \citet {Sun+09} include this cluster (also called
  the ESO~306-G~017 group) in their sample of 43 groups observed with
  Chandra and classify it as a fossil group. They give a temperature
  of $T_{\rm 500}=2.37^{+0.12}_{-0.14}$, a mass of
  $M_{\rm 500}=10.3^{+2.1}_{-1.3}\times 10^{13}$~M$_\odot$, and an entropy
  $S_{\rm 500}=915^{+203}_{-192}$. From its mass, this object is therefore
  closer to a cluster than to a group, but according to \citet{Wong+16} it
  is a dynamically old relaxed fossil group. Our temperature
  and entropy maps  show a displacement of the cluster center
  towards the north while the metallicity map clearly exhibits
  a bimodal distribution with the western side having higher values.
  
\item {\bf G269.51+26.42}: This cluster is also known as the
  Hydra Cluster, and was previously classified as a non CC system
  \citep{san10}. The X-ray measurements listed in
  \citet{AndradeSantos17} and in \citet{Lopes18} indicate it is a CC cluster. On the
  contrary, some optical tests point to a disturbed cluster, although
  the BCG offset to the X-ray centre does not \citep{Lopes18}. Our
maps indeed suggest that this is a relaxed cluster but without a evident cool-core.

\item {\bf Abell 3528 (G303.75+33.65)}: This is a system of two interacting clusters
  Abell~3528N and Abell~3528S separated by 0.9~Mpc
  \citep{Gastaldello+03}. Based on XMM-{\it Newton}  data, these authors
  describe both subclusters as relaxed.  However, our temperature map
  shows that A3528N looks relaxed, with a cool core, but A3528S does
  not, since its cool zone is displaced relatively to the cluster
  centre and the temperature is hotter in the east and west parts of
  the cluster, indicating that at least one merger must have taken place. Our
  temperature and metallicity maps are in global agreement with the
  coarser ones of \citet{Gastaldello+03}, who suggest that the two
  subclusters are moving along the NW/SE and NE/SW directions
  respectively, the merger having had its closest encounter 1--2~Gyrs
  ago with an impact parameter of $\sim 5 r_s$.

\item {\bf Abell~1650 (G306.68+61.06)}: According to \citet{Takahashi03}, this cluster
  presents a non-uniform distribution of temperature and abundance, based on
  the analysis of XMM-{\it Newton} data, which goes in line with
  our result. However, all maps present a roundish shape.
  These authors also found cool regions with temperature
  around 4.0 keV distributed in the outer parts of the cluster, which is exactly
  what we see in the 2D temperature map.

\item {\bf Abell 3571 (G316.34+28.54)}: \citet{Lopes18}
  classify this cluster as relaxed for all except one optical
  substructure test, and it is also classified as relaxed by all the
  X-ray indicators. In the literature, this system is usually
  classified as relaxed \citep{qui93, nev00}, but an exception is
  found in the work of \citet{ven02}, suggesting (from radio data)
  that it is the final stage of a merger event. Our temperature map
  does not indicate the presence of a cool core, and our metallicity
  map is quite patchy.

\item {\bf Abell~1795 (G033.78+77.16)}: All optical substructure
classifications (except for the Lee 3D test) from \citet{Lopes18} indicate
this is a relaxed system. The same is true according to the four X-ray
CC indicators from that work and from the temperature map we show here.
This system has also been previously classified as a CC, but having a
very active central region \citep{ehl15}, with the presence of low temperature
ICM gas, sometimes coincident with H${\alpha}$ filaments.

\item { \bf Abell~2052}: \citet{Blanton11} analysed a Chandra
  observation of this cluster, revealing detailed structure in the
  inner part, including bubbles evacuated by radio lobes of the active
  galactic nucleus (AGN), compressed bubble rims, filaments, and two
  concentric shock regions.  Their 2D maps are very similar to those
  presented here.  \citet{Machado15} made simulations that could
  reproduce the sloshing feature of this cluster by two regimes: the
  first scenario has a close encounter and corresponds to a recent
  event (0.8 Gyr since pericentric passage), while the second scenario
  has a larger impact parameter and is older (almost 2.6 Gyr since
  pericentric passage).  In this second case, the simulation predicts
  that the perturbing subcluster should be located approximately 2 Mpc
  from the centre of the major cluster, where the authors were able to
  identify an optical counterpart at the same redshift. 
  Besides the inner part of the pressure map that shows instabilities probably related
  to the AGN, this cluster doesn't show strong signs of mergers.
  
\item { \bf Abell~2063}:  According to \citet{PlanckVIII}, this cluster
forms a pair with MKW3s but the SZ signal between the two clusters is
not significantly detected. Though \citet{Frank+13} classify it as
relaxed, \citet{Parekh+15} consider that this is a non-relaxed
cluster. We classify A2063 as a CC-relaxed cluster, but our maps show that it
is not fully relaxed. 
The flat temperature profile and the monotonously decreasing
metallicity profile found by \citet{Matsushita11} are not in contradiction with our
maps. 

\item {\bf Abell~2151}: This cluster is much smaller and cooler than
the other ones, and our mean temperature agrees with that of
\citet{Fukazawa+04}, who give kT=$2.14\pm 0.10$~keV, based on ASCA
data. On the other hand, these authors find a very low metallicity of
Z=$(0.14\pm 0.06)$Z$_\odot$, while we find an abundance that is in
average $\sim 0.9$ solar. All our maps suggest that this is a relaxed
cluster. There are only few papers in the literature concerning the X-ray
emission of this cluster.

\item { \bf Abell~2244 (G056.81+36.31)}: This cluster is considered by \citet{Parekh+15} as relaxed, and only
small scale temperature substructures have been detected in the
central regions of A2244 by \cite{Gu+09}, based on Chandra
data. In Table~2, we classify A2244 as a CC-relaxed cluster, 
since its overall spherical shape is preserved.

\item { \bf Abell~2572}: The core of A2572 is found to be double, consisting of two X-ray peaks of similar intensity separated by about 3 arcmim. 
The ICM between the two X-ray peaks appears undisturbed \citep{Ebeling95}. In addition to that, according to these authors, A2572 and HCG94 are falling toward each other at a velocity 
of about 1000~km~s$^{-1}$ but no clear signs of dynamical interaction are discernible in the ROSAT X-ray data of A2572.
Our 2D maps don't show strong signs of perturbation. The kT map has 
presents a small range in temperature (from 2.8 to 3.6~keV), and the Z map shows a central elongation of higher metallicity.

\item { \bf Abell~2597}: This is a CC cluster, but \citet{McNamara01} analysed the central 100~kpc 
of this cluster and found an irregular X-ray emission (their Fig. 1) very similar to the structure seen in our kT map.
\citet{Morris05} showed that EPIC fits to the central region are consistent with a cooling flow of around 100 solar masses per year.
The  Chandra observations reveal a central ($<30$~kpc) X-ray cavity that  is not detected in XMM-{\it Newton}  data due to the PSF
\citep{Tremblay12}. These authors also present evidence that in spite of the presence of a central AGN there is a residual cooling-flow.

\item {\bf Abell~2626}: This cool-core cluster has been well studied due to the
  presence of a radio source in its centre.
  \citet{Wong08} presented a detailed X-ray analysis (XMM-{\it Newton}  and
  Chandra) focused on the X-ray and radio interactions.  The cD galaxy
  IC~5338 has two nuclei and one of them has an associated hard X-ray
  point source.  \citet{Ignesti18} reported a complex system of four
  symmetric radio arcs without known correlations with the thermal X-ray
  emission.  These symmetric radio arcs together with the presence of
  two optical cores could be created by pairs of precessing radio jets
  powered by dual AGNs inside the dominant galaxy. We see a clear
  metallicity excess in the centre of this cluster in the 2D map.

\item { \bf Abell~4059}: \citet{Huang98} presented a ROSAT HRI and PSPC X-ray analysis, 
characterising this galaxy cluster as a compact cD type. The central cD galaxy hosts the strong radio source 
PKS 2354-35. In our kT, P and S maps we do not see any perturbation  produced
by this galaxy. However, in the Z map, we clearly see a central enhancement in metal
abundance that could be related to the central galaxy.

\end{itemize}

\item { \bf \scshape CC-disturbed} 

\begin{itemize}

\item {\bf Abell~85 (G115.16-72.09)}: This cluster has been known for many years to
  have a cool core, several subtructures \citep{Durret05}, and an
  extended filament detected in X-rays \citep{Durret+03}. This 
  filament was interpreted as due to groups falling on to the cluster,
  the gas in the impact region being hotter. The comparison of the
  temperature map computed from XMM-{\it Newton}  data by \citet{Durret05} (see
  their Fig.~3) with that obtained from totally independent
  hydrodynamical numerical simulations by \citet{Bourdin+04} (see
  their Fig. 9) suggests that an older merger coming from the
  northwest has also hit the cluster.  Temperature maps have also been
  obtained by \citet{Tanaka+10} and \citet{Schenck+14}. The S and SW
  subclusters described by \citet{Schenck+14} are spatially coincident
  with regions of low temperatures in our kT map.  The dynamical
  analysis by \citet{Yu+16} shows several groups passing through the
  cluster core that have disturbed the ICM, one group being associated
  with a secondary peak of X-ray emission.

\item {\bf Abell~496 (G209.56-36.49)}: This cluster is known to be an overall relaxed
  cluster \citep{Peres+98} with a cool core, though the comparison of
  its X-ray sloshing cold fronts with hydrodynamical simulations by
  \citet{Roediger12} suggest that a minor merger with an approximate
  mass of $4\times 10^{13}$~M$_\odot$ has crossed the cluster from the
  south-west to the north-east, passing the pericentre distance
  0.6--0.8~Gyr ago.  Our metallicity and entropy maps agree well with
  those of \citet{Ghizzardi14} and with the interpretation by
  \citet{Roediger12}.

\item {\bf USGC S152}: \citet{OSullivan+07} studied this cluster that
  they call the NGC 3411 group, which is also identified in NED as the
  NGC 3402 group. Based on several high resolution temperature maps
  obtained with both XMM-{\it Newton}  and Chandra data, these authors find a
  typical abundance distribution, but a clearly atypical temperature
  distribution, with a hot inner core surrounded by a cool shell of
  gas, making them conclude that the X-ray gas is not in
  hydrodynamical equilibrium. They discuss two possible scenarios to
  account for this cold halo: either the center of a cooling region
  has been reheated by an AGN outburst, or the cool shell is the
  product of a merger. Our temperature map fully agrees with theirs,
  and our metallicity map is clearly in favour of the second scenario:
  a merger in the northeast-southwest direction is the only
  explanation to the higher metallicity observed along this direction.
  
\item {\bf Abell~1644}: This is a double cluster, with a main cluster
  (A1644s) and a smaller and colder one (A1644n) to the north-east
  \citep{Johnson+10}. Surprisingly, these authors give almost similar
  masses for both clusters, though their intensity map clearly shows
  the presence of a main cluster and of a much smaller one, as also
  seen in our maps. Their contours and temperature maps agree with ours
  and show that the main cluster itself seems double, or at least
  appears strongly perturbed by a merger. From the comparison with
  hydrodynamical simulations, \citet{Johnson+10} suggest that the
  northern subcluster initiated the core gas sloshing in the main
  cluster about 700~Myr ago.
We produced the temperature map assuming a  minimum count
number of 1000 counts (see Fig.~\ref{fig:A1644})  and we can see the shock between these two systems.

\begin{figure}
\includegraphics[width=\columnwidth]{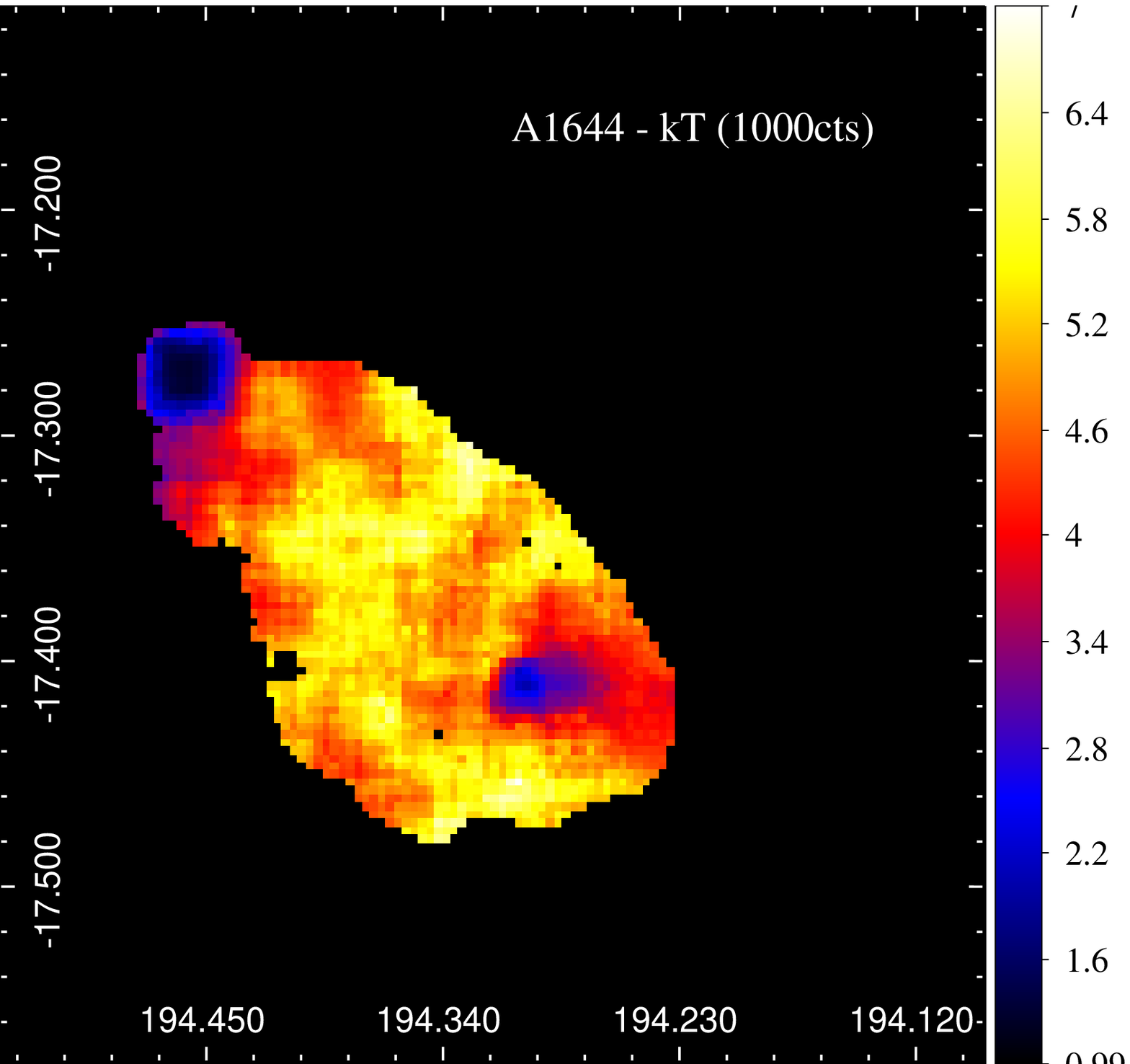}
\caption{temperature map for A1644 using a minimum count of A1644 to show the shock between the systems.}
\label{fig:A1644}
\end{figure}

\item { \bf Abell~1651 (G306.80+58.60)}: An almost isothermal profile around $\sim$ 5
  keV was found in this cluster based on ASCA data \citep{Mark98}, but our
  temperature map suggests that there is a hotter eastern region, with
  $kT \sim 8.0$ keV.  A study of the diffuse optical light around the
  BCG of Abell 1651 suggested a color and shape of the profile
  consistent with relatively little evolution in the past 5 Gyr
  \citep{Gonzalez00}. However our X-ray maps, in particular the
  temperature and metallicity maps, rather seem to suggest that at
  least one (possibly several) merger(s) has taken place.

\item {\bf Abell~3558 (G311.99+3071)}: \citet{Rossetti+07} made a detailed analysis
  of A3558 based on Chandra and XMM-{\it Newton}  data and showed that its
  cool core has survived a merger.  However, we hardly detect any
  cooler gas in the center. As stressed by \citet{Ramella+07}, this
  cluster is elongated in the southeast-northwest direction.
  According to \citet{Frank+13} and \citet{Parekh+15}, A3558 is
  unrelaxed, and linked by hot gas (2 keV) to A3556
  \citep{Mitsuishi+12}. Our maps roughly agree with the temperature
  and Fe abundance profiles calculated by \citet{Matsushita11}, and
  strongly suggest that at least one merger has taken place.

\item { \bf Abell~3562}: As A3558, this cluster is also a member of
  the Shapley supercluster.  According to \citet{Frank+13} and
  \citet{Parekh+15}, it is unrelaxed, as also suggested by its radio
  halo \citep{Giovannini+09}.  \citet{Matsushita11} classifies it as a
  non-cD cluster with a large deviation from symmetry and a rather
  large Fe abundance of $\sim 1.5$ solar, as for A3558, but we do not
  measure such a high abundance in either of the two clusters.
  \citet{Matsushita11} also finds that A3562 has a flat temperature
  profile and a uniformly decreasing metallicity profile, in
  contradiction with our maps.

\item {\bf Abell~1775}: According to \citet{Lopes18}, this cluster is
  classified as relaxed. However, our temperature map clearly shows an
  extremely sharp cold front (sloshing arm), particularly visible in
  the entropy map. At optical wavelengths, \citet{Zhang11} studied this cluster in 15 bands (SDSS + BATC),
  and the galaxy redshift histogram suggests  a bimodal structure composed by a poor subcluster at
  lower redshift, located 14 arcmin southeast of the main
  concentration. 
  This is confirmed by the luminosity function analysis of the galaxies belonging to each
  subcluster.  Probably these two clusters have merged and produced
  the spiral sloshing arm seen in the 2D maps.

\item { \bf Abell~2029 (G006.47+50.54)}: This system is classified as
disturbed by the Dressler \& Shectman (DS or $\Delta$) and Anderson-Darling
(AD) statistics in \citet{Lopes18}, but is classified as relaxed according to the
magnitude gap and to the offset between the BCG and X-ray centroid. In that paper
all four X-ray measurements indicate this is a CC cluster. This is in
agreement with previous studies, such as \citet{santos10} and \citet{Hudson10},
who classify it as a strong cool core system. On the contrary, from our
current analysis we consider it as disturbed, with a cool centre, but very
asymmetric. 

\item { \bf Abell~2142 (G044.22+48.68)}: This cluster was one of the first to show a
cold front \citep{Markevitch+00}, a phenomenon interpreted as due to a
merger by \citet{Owers+11}, based on galaxy redshifts. Though A2142 is
considered as relaxed by \citet{Frank+13}, \citet{Tchernin+16} have
shown that this cluster has rather bumpy temperature, metallicity, and
entropy profiles, while the pressure profile is quite
smooth. \citet{Akamatsu+11} obtained a temperature profile up to a
large distance of 30~arcmin from the cluster center towards the
northwest, showing a steady decrease of the temperature consistent
with our temperature map. All these results overall agree with our
maps, but illustrate again the fact that profiles do not allow to
derive the full physical properties of the X-ray gas, for which 2D
maps are necessary. The temperature and metallicity maps of A2142
clearly show that at least one merger has occurred along the southeast to northwest direction. We can note that another merger is presently
occurring, with a group falling on to A2142 \citep{Eckert+17}.

\item { \bf AWM4}: This is a poor cluster, for which
\citet{Fukazawa+04} give kT=$3.56\pm 0.07$~keV and Z=$(0.47\pm
0.09)$Z$_\odot$, based on ASCA data. \citet{OSullivan+11} computed high
spatial resolution temperature and density maps of 
AWM4, based on Chandra data, revealing a high degree of
structure. They concluded that it is unlikely that large scale sloshing
occurs in AWM4.  However, at larger scale, though our entropy and
pressure maps suggest that the cluster is relaxed, our temperature and
metallicity maps imply that at least one merging event has taken
place, probably originating from the southwest, to explain the fact
that in this zone the gas is hotter and more metal-rich. We can note
that our mean metallicity roughly agrees with that of
\citet{Fukazawa+04}, but our temperature is notably lower.

\item { \bf Abell~2199 (G062.92+43.70)}: According to \citet{Lee+15} and references therein, A2199 is at the
centre of a supercluster.  Deep Chandra observations by
\citet{Nulsen+13} have revealed evidence for a minor merger $\sim 400$
Myr ago. However, also based on Chandra data, \citet{Hofmann+16}
suggest that A2199 is an overall relaxed cluster (a result also found
by \citet{Frank+13} and \citet{Parekh+15}) with a cool core and AGN
feedback structures at its centre. These authors find a large scale
asymmetry in the temperature distribution between the north and the
south, which is also clearly visible in our temperature map. Otherwise,
\citet{Hofmann+16} find that the perturbations in entropy, pressure,
temperature and density are those expected for an overall relaxed
system.

\item{ \bf NGC6338}: This is a double cluster. 
There are X-ray cavities in the central 6 kpc detected through the analysis of Chandra data  \citep{Pandge12}.
In addition to these cavities, these authors detected a set of X-ray bright filaments which are spatially coincident
with the H$\alpha$ filaments over an extent of 15 kpc. The orientations of these filaments (NW-SE) are spatially coincident with 
the hotter regions in the kT maps and with the higher entropy values in the S map.

\item {\bf Abell 4038}: In the work of \citet{Lopes18} this cluster is
  considered as disturbed by all optical tests, except the Lee 3D and
  $\Delta m_{12}$.  Though it is considered as cool-core by the
  four X-ray substructure parameters used here, our temperature map
  does not point to a CC cluster. The unrelaxed nature of
  this cluster also agrees with the fact that the central part of the
  cluster shows the presence of two bright galaxies, and that a radio
  relic is reported by \citet{sle98} and \citet{kal12}. 

\end{itemize}

\item { \bf \scshape NCC-relaxed} 

\begin{itemize}

\item { \bf Coma Cluster  (G057.33+88.01)}: Coma is one of the most well
studied clusters, known for having two bright galaxies at its center. In
\citet{Lopes18} we derived a mass estimate of
14.6 $\times$ 10$^{14}$~M$_{\odot}$ and classifed it as non-relaxed with all
substructure tests (in the optical and X-rays), except for the n$_{\rm core}$
measurement. However, we find no indication of perturbation from the maps
derived here. Our temperature map is fully consistent (in shape and absolute values)
with the monotonously decreasing temperature profile obtained by
\citet{Matsushita11} (based on XMM-{\it Newton}  data), that show no cool
core. It also agrees with the much coarser temperature map obtained by
\citet{Arnaud+01}.  Our metallicity is also fully consistent with the
monotonously decreasing metallicity profile of \citet{Matsushita11}.
The fact that we see no trace of the influence of the infall of the
NGC 4839 group into the main cluster \citep{Neumann+01} strongly
suggests that this group is only approaching Coma for the first
time. If this group had already crossed the cluster, as it had been
suggested long ago by \citet{Burns+94}, this should had left an
imprint in the temperature and metallicity maps.

\item {\bf Abell~3532}: This cluster forms a possibly bound pair with
  Abell~3530, and both clusters belong to the Abell~3528 complex of the Shapley
  Supercluster (SSC), as presented by \citet{Lakhchaura+13} in their
  detailed introduction. Both are non cool core clusters
  \citep{Chen07}.  The radio continuum emission from galaxies in the
  SSC core seems to indicate that the Abell~3532/3530 pair of clusters
  is approaching for the first time \citep{MauduitMamon07}.  The
  Abell~3532/3530 pair of clusters was observed and analysed in detail
  in X-rays with Chandra and XMM-{\it Newton}  \citep{Lakhchaura+13}.  Their
  pressure and entropy maps agree with ours and confirm that the
  cluster is not at all relaxed.

\item {\bf bA3558}: the observation was not deep enough to produce extended maps, 
so we cannot draw firm conclusions on this cluster.
 
\item { \bf MKW8}: This cluster is classified as relaxed according
to all optical substructure tests in \citet{Lopes18}, but only C$_{\rm SB4}$
(out of the four X-ray metrics) indicates this is a CC cluster. From
our current maps we also classify it as disturbed. 
\citet{Hudson10} and \citet{Eckert11} also classify it as a NCC system, 
corroborating the discrepancy between the optical and X-ray classifications.
We have also noticed that its nearby companion (bMKW8) aligns in projection
with two background clusters, WHL J143821.9+034013 and Abell 1942, and it is not clear if the X-ray emission could be associated with those background
systems.

\end{itemize}

\item { \bf \scshape NCC-disturbed} 

\begin{itemize}

\item {\bf Abell~119 (G125.58-64.14)}: \cite{Oh+16} showed that A119 is a dynamically
  young cluster with unrelaxed groups within the virial radius
  implying a disturbed morphology.  This was confirmed by
  \citet{Owers+17}, based on a large number of galaxy redshifts from
  the SAMI cluster redshift survey. They found that A119 has a small
  substructure associated with a bright galaxy about 1~Mpc to the
  north-east of the cluster centre.  Our maps clearly confirm that
  this cluster is disturbed and has undergone or is presently
  undergoing one or several merging events. In X-rays,
  \cite{Elkholy+15} showed that the entropy profile slightly increases
  outwards, in agreement with our entropy map. On the other hand, they
  find that the metallicity profile of A119 decreases by a factor
  $2-3$ between 0.15R$_{500}$ and R$_{500}$, a result which is not
  consistent with our metallicity map.

\item {\bf Abell~3376 (G246.52-26.05)}: The bullet shape of this cluster suggests it
  has undergone a recent collision \citep{Durret+13}. The temperature
  and metallicity maps obtained by \citet{Bagchi+06} show patches of
  hotter gas and a higher metallicity in the southeast half of the
  cluster.  Hydrodynamical simulations by \citet{MachadoLimaNeto13}
  computed to match the X-ray properties suggest a collision of two
  clusters with mass ratio between 1:6 and 1:8, the subcluster
  having a gas density four times that of the main cluster. The merger
  seems to correspond to a small impact parameter ($<150$~kpc), with a
  collision axis inclined by $\sim 40^\circ$ with respect to the plane
  of the sky, and having taken place 0.5~Gyr ago, with a large Mach
  number of $\sim 4$.

\item {\bf Abell 3391}: This cluster is classified as relaxed by all
  except one optical indicator (magnitude gap) in \citet{Lopes18}, as
  there are two bright galaxies near the center of the
  system. However, all four X-ray parameters,
  point to a NCC object, and our temperature map
  indicates that this is the case. It has also been suggested that
  there is a filamentary structure linking Abell 3391 and Abell 3395,
  with the ESO-161-IG006 group between the two clusters
  \citep{tit01,alv18}.

\item {\bf Abell 3395 (G263.20-25.21)}: This cluster is an
  example of a double system, close to bA3395 (bG2263.20-25.21), 
  previously classified as a merging system \citep{lak11}. According
  to \citet{don01} the system appears to be nearly at first core
  passage.  \citet{Lopes18} found that the BCG of the second cluster
  is coincident with the X-ray centroid. As in \citet{don01} we found
  evidence that the gas between the two systems is heated.

\item {\bf Abell~754}: This cluster has been studied at various
  wavelengths \citep[][and references therein]{Inoue+16} and is
  undergoing a recent major merger that has been quantified by several
  studies. First, the comparison with hydrodynamic simulations
  \citep{Roettiger+98} has suggested a recent ($<0.3$~Gyr) off-axis
  merger between two subclusters of mass ratio 2.1:1. Then, the
  presence of a third subcluster was suggested \citep{Mark02}
  and a ``shock-SE'' region was identified by a temperature jump at
  the southeast part of the cluster coinciding with a radio relic
  \citep{Macario+11}.

%  NED: cross-id NGC 3402; NGC 3411; MCG -02-28-012; 2MASX
%  J10502610-1250422; GALEX J105026.1-125041.

\item {\bf Abell 1367 (G234.59+73.01)}: In agreement with \citet{Hudson10} and references therein, this is a
very well-studied merging cluster, observed for a long time in X-rays
\citep{SunMurray02} as well as optical wavelengths, where evidence was
found for an infalling starburst group \citep{Cortese04}.  In view of
the asymmetry of this cluster, particularly well seen in our maps,
radial profiles such as those of \citet{Matsushita11} are obviously
not well adapted, since they average values of the temperature and
metallicity in regions where these quantities are quite different.  Our
X-ray maps indeed appear rather disturbed, with mean temperature and
metallicity values agreeing with those previously derived by
\citet{Fukazawa+04} and \citet{Hudson10}, but slightly larger than
the temperature given by \citet{Frank+13}.

\item {\bf ZwCl12151$+$0400 (G282.49+65.17)}: This cluster was not studied
  individually and was only reported in the literature among large
  samples.  Our maps clearly show that it is a disturbed NCC cluster.

\item {\bf Abell~3560}: This is a rich cluster at the southern
  periphery of the A3558 complex, a chain of interacting clusters in
  the central part of the Shapley supercluster. Based on a ROSAT-PSPC
  map, \citet{Bardelli02} found that its surface brightness
  distribution is described by two components, indicating that this
  cluster is experiencing a merger event. They also mentioned that the
  main component, corresponding to the cluster, has an elongation
  towards the A3558 complex, which is located northwest of A3560. This
  is also what we found in the 2D maps presented in this work. From
  VLA radio data \citet{Bardelli02} found a peculiar bright extended
  radio source (J1332-3308).

\item { \bf Abell~2061}: This cluster forms a pair with A2067.  Based
  on BeppoSAX data, \citet{Marini+04} obtained temperature maps for
  three clusters in the Corona Borealis supercluster: A2061, A2067,
  and A2124.  Their image shows that A2061 is elongated towards A2067,
  with a ``plume'' pointing towards the second cluster, but no other
  clear sign of interaction. We also detect an elongation towards
  A2067.  \citet{Marini+04} find an overall decreasing temperature profile, with a
  hotter region (10.7 keV) in the 2-4 arcmin bin northwest of the
  cluster center. This agrees with our temperature map, where the
  hottest gas is in the northwest zone. \citet{Marini+04} suggest a
  scenario where a group is falling into A2061 and forms a shock in
  the cluster. However, the sketch of the merger that they give in
  their Fig. 17 does not seem consistent with our temperature
  map. \citet{Planck13} confirmed that no filament is detected
  between A2061 and A2067.

\item { \bf MKW3s}: According to \citet{Planck13}, this cluster
forms a pair with A2063 but the SZ signal between the two clusters is not significantly detected. 
There is a dominant central galaxy at the center of MKW3s, namely NGC~5920.
Our kT map shows that the central temperature is higher than the overall temperature, in contradiction with previous
studies that reported MKW3s as a cooling-flow system \citep{Canizares83}.

\item {\bf Abell~2065 (G042.82+56.61)}: \citet{Chatzikos06} presented the Chandra
  analysis of this merging cluster, with cool gas displaced from the
  more luminous southern cD galaxy.  They argue that A2065 is an
  unequal mass merger in which the more massive southern cluster has
  driven a shock into the ICM of the infalling northern cluster, which
  has disrupted the cool core of the latter. The spatial distribution
  of the temperature presented in Fig.~\ref{fig:NCCclusters3} is very
  similar to their Fig.~6(a), showing a cold front with temperatures
  around 5-6 keV and a shock front with temperatures reaching 10~keV.

\item {\bf Abell~2147}: This cluster forms a close pair with A2152
which is about 40~arcmin away \citep{PlanckVIII}. It is classified as
disturbed by \citet{Frank+13}, and this is obviously the case, as seen
in our maps, which are all very inhomogeneous and suggest that several
mergers must have taken place.

\item { \bf G049.33+44.38}: This cluster is part of the NORAS survey 
\citep[Northern ROSAT All-Sky Survey][]{Bohringer+00} but very little is known on it in X-rays. Our maps show that this cluster is obviously disturbed.

\item { \bf Abell~2249 (PLCKESZ G057.61+34.94)}:  The temperature profile presented in \citet{Zhu16} 
based on Chandra data suggests a 
central temperature of 8.0~keV, reaching its maximum of almost 10 keV around 1000~kpc. 
Our temperature map does not show such high values, since kT ranges from $\sim$ 4.0 keV up to 7.0 keV.
Using the KASI-Yonsei Deep Imaging Survey of Clusters (KYDISC), \citet{Oh18} found more than 400 spectroscopic 
members, and the redshift histogram is well fit by a gaussian distribution, showing no evidence of merging. 
What calls our attention is the higher temperature ``ring'' in the center, which is associated
with AGNs, although we found no indication for this in the literature.
\citet{Lopes18} also classify the velocity distribution as normal (according to the Anderson-Darling statistics), but the cluster is considered as disturbed in 3D, according to the Dressler \& Shectman (DS or $\Delta$) test.

\item { \bf Abell~2255 (G093.91+34.90)}: This is a merging cluster and our  kT and S maps show two regions of lower temperature and entropy 
while the pressure map shows a region of shock between the systems. Using 140 ks Suzaku X-ray data, \citet{Akamatsu17} 
analysed this merging system almost out to the virial radius (1.9 Mpc), confirming that the temperature drops from 6 keV around the
cluster centre to 3 keV at the outskirts. In a previous XMM-{\it Newton}  data analysis, \citet{Sakelliou06} reported an
asymmetric temperature distribution that has been assembled recently by the merging of smaller subunits that have collided some 0.1-0.2 Gyrs ago.

\end{itemize}
\end{enumerate}

\end{appendix}

%%%%%%%%%%%%%%%%%%%%%%%%%%%%%%%%%%%%%%%%%%%%%%%%%%

% Don't change these lines
\bsp	% typesetting comment
\label{lastpage}
\end{document}